\documentclass[12pt]{book}
\usepackage{amsmath,amssymb}
\usepackage{bm}
\usepackage{color}
\usepackage{cite}
\usepackage{mathtools}
\usepackage{here}

\usepackage{graphicx}
\usepackage{hyperref}

\usepackage{multirow}
\usepackage {tabularx}
\newcolumntype{C}{>{\centering\arraybackslash}X}
\newcolumntype{L}{>{\raggedright\arraybackslash}X}
\newcolumntype{R}{>{\raggedleft\arraybackslash}X}
\usepackage{longtable}

\def\slash#1{\not\!\!#1}

\begin{document}

\title{
{\Large Ph.D thesis\\*[40pt]}
{\Large \bf
The flavor structures on magnetized orbifold models and 4D modular symmetric models
\\*[20pt]}}

\author{
Shota Kikuchi
\\*[20pt]
\centerline{
\begin{minipage}{\linewidth}
\begin{center}
{\it \normalsize
Department of Physics, Hokkaido University, Sapporo 060-0810, Japan} \\*[5pt]
\end{center}
\end{minipage}}
\\*[50pt]}

\date{{\normalsize
Submitted to Department of Physics, Hokkaido University: February 2024
}}

\begin{titlepage}
\maketitle
\thispagestyle{empty}
\end{titlepage}

\newpage

\vspace{1.5 cm}
\noindent
{\large\bf Acknowledgement}\\

I would like to express my greatest appreciation to my supervisor, Tatsuo Kobayashi, for a great research environment, discussions with enthusiasm, instructive suggestions, and helpful encouragement to my study.
I am enormously grateful to many of my collaborators, Takuya H. Tatsuishi, Shintaro Takada, Hikaru Uchida, Kaito Nasu, Yuya Ogawa, Kouki Hoshiya, Shohei Takada, Hajime Otsuka, Shohei Uemura, Morimitsu Tanimoto, Kei Yamamoto, and Yusuke Yamada, for useful discussions and insightful comments.
I also thank to all members of theoretical particle and cosmological physics group at Hokkaido University and all members involved with me for supporting me continuouslly in the Ph.D. course.
The study in the Ph.D. course would not have been possible without the help of them.
I have been financially supported by Grant-in-Aid for JSPS Research Fellows No. JP22J10172 and JP22KJ0047, and JST SPRING No. JPMJSP2119.

\newpage

\vspace{1.5 cm}
\noindent
{\large\bf Abstract}\\

We study quark and lepton flavor structures on magnetized $T^2/\mathbb{Z}_2$ twisted orbifold model.
There are 6,460 number of flavor models but most of them cannot lead to realistic flavor observables because of the difficulties on realizing mass hierarchies and small (large) mixing angles of quarks (leptons).
We find that certain zero point patterns of zero-modes of fermions and Higgs fields give the flavor models being able to avoid these difficulties.
We classify such flavor models and show numerical example.
Also we study $(T^2_1\times T^2_2)/\mathbb{Z}_2^{\textrm{(per)}}$ permutation model and its $\mathbb{Z}_2$ twist orbifolding.
Additionally we study Yukawa matrices at three modular symmetric points, $\tau=i$, $\omega$ and $i\infty$, where $S$, $ST$ and $T$-symmetries remain.
We find Yukawa matrices on $T^2/\mathbb{Z}_2$ orbifold have a kind of texture structures because of the residual symmetries.

Next we study four-dimensional (4D) modular symmetric quark flavor models without fine-tuning.
Mass matrices are written in terms of the modular forms.
The modular forms become hierarchical as close to the modular symmetric points depending on its residual charges.
Actually the residual $Z_N$ symmetries with $N\geq 6$ can originate the large quark mass hierarchies.
Also the products of residual symmetries such as $Z_3\times Z_3\times Z_3$ have such possibility.
First we study the quark flavor model with $\Gamma_6$ symmetry.
We assume the vicinity of the cusp $\tau=i\infty$ where residual $Z_6$ symmetry remains.
To make our analysis simple we use only $\Gamma_6$ singlet modular forms.
Consequently we find the models realizing the order of the quark mass ratios and the absolute values of the Cabibbo-Kobayashi-Maskawa (CKM) matrix elements without fine-tuning.
In the same way, we study the quark flavor models with $A_4\times A_4\times A_4$ symmetry.
Then we discuss the CP violation caused by the value of the modulus $\tau$ and show numerical example.
Results imply the quark flavors including the CP phase require non-universal moduli.

Finally we construct the Siegel modular forms.
Zero-modes on $T^6$ at $\vec{z}=0$ are the Siegel modular forms of weight 1/2 for the subgroup of $Sp(6,\mathbb{Z})$.
They have several moduli parameters and therefore have the possibility realizing the flavor structures including the CP phases.
We study the Siegel modular forms transformed by $\widetilde{\Delta}(96)$ and show numerical example using the singlet Siegel modular forms.
We find one of moduli parameters $\omega_1$ works on the large mass hierarchies and $\omega_2$ works on the CP violation successfully in our model.

\newpage

\tableofcontents


\chapter{Introduction}
\label{Intro}

In recent years, the Standard Model (SM) of the particle physics was established by the discovery of the Higgs particle.
It gives precisely consistent predictions with almost all of the observations in experiments so far.
However, there are still unsolved mysteries in the particle physics.
The origin of the flavor structures of quarks and leptons is one of such problems.
Up-sector quarks, down-sector quarks and charged leptons have large mass hierarchies while neutrinos have extremely small masses.
Moreover quarks have small mixing angles while leptons have large one.
Also non-vanishing CP phases exist in both quark and lepton sectors.
To describe them, the SM needs to tune 22 real parameters.
Ten real parameters are for six quark masses, three mixing angles and one CP violating phase.
Remaining twelve real parameters are for six lepton masses, three mixing angles and three Dirac and Majorana CP violating phases.
Especially, it seems to be unnatural that the SM has hierarchical values of parameters to realize large mass hierarchies.
Thus explaining the flavor structures without fine-tuning is one of the challenging issues in present-day particle physics.

To understand the origin of the flavor structures, two types of approaches, bottom-up and top-down approaches have been carried out.
Non-Abelian discrete flavor models are one of the bottom-up approaches.
They have used non-Abelian discrete groups such as $S_N$, $A_N$, $\Delta(3N^2)$, $\Delta(6N^2)$ and so on as the flavor symmetries of quarks and leptons \cite{Altarelli:2010gt, Ishimori:2010au, Ishimori:2012zz, Kobayashi:2022moq, Hernandez:2012ra, King:2013eh, King:2014nza, Tanimoto:2015nfa, King:2017guk, Petcov:2017ggy, Feruglio:2019ybq}.
These symmetries are broken after the gauge singlet scalars (so-called flavons) get the vacuum expectation values (VEVs).
However the configuration of them becomes complicated in order to realize the flavor structures and it is still artificial.

Recently modular symmetric flavor models have been proposed.
In the models, the superpotentials are invariant under the modular transformation $Sp(2,\mathbb{Z})\simeq SL(2,\mathbb{Z})$.
Then three-generations of quarks and leptons are regarded as three-dimensional (reducible or irreducible) representations of the finite modular groups.
Hence, their mass matrices as well as Yukawa couplings are also the representations of the finite modular groups.
Therefore they are written in terms of the modular forms for the finite modular groups which are the holomorphic functions of the modulus $\tau$ \cite{Feruglio:2017spp} \footnote{The modular flavor symmetry was also studied from the top-down approach such as string theory \cite{Ferrara:1989bc,Ferrara:1989qb,Lerche:1989cs,Lauer:1989ax,Lauer:1990tm,Kobayashi:2018rad,Kobayashi:2018bff,Ohki:2020bpo,Kikuchi:2020frp,Kikuchi:2020nxn,
Kikuchi:2021ogn,Almumin:2021fbk,Baur:2019iai,Nilles:2020kgo,Baur:2020jwc,Nilles:2020gvu,Hoshiya:2020hki}.}.
Instead of flavons, the modular symmetry in mass matrices is violated by the VEV of the modulus $\tau$.
It is remarkable that the finite modular groups $\Gamma_N$ for $N=2,3,4$ and $5$ are isomorphic to the non-Abelian discrete groups $S_3$, $A_4$, $S_4$ and $A_5$, respectively \cite{deAdelhartToorop:2011re}.
In this sense, the modular symmetric flavor models are attractive way describing the flavor structures in the framework of the non-Abelian discrete symmetries without complicated configurations of the VEVs of the flavons.
Indeed the modular symmetric flavor models in $\Gamma_2\simeq S_3$ \cite{Kobayashi:2018vbk}, $\Gamma_3\simeq A_4$ \cite{Feruglio:2017spp}, $\Gamma_4\simeq S_4$ \cite{Penedo:2018nmg} and $\Gamma_5\simeq A_5$ \cite{Novichkov:2018nkm,Ding:2019xna} have been proposed.
Furthermore the modular symmetries including higher levels and covering groups were studied \cite{Li:2021buv,Ding:2020msi,Kobayashi:2018bff,Liu:2019khw,Novichkov:2020eep,Liu:2020akv,Liu:2020msy}.
Using them, phenomenological studies have been widely studied in many works \cite{Feruglio:2017spp,Kobayashi:2018vbk,Penedo:2018nmg,Novichkov:2018nkm,Criado:2018thu,
Kobayashi:2018scp,
Ding:2019zxk,Novichkov:2018ovf,
Kobayashi:2019mna,Wang:2019ovr,Ding:2019xna,
Liu:2019khw,Chen:2020udk,Novichkov:2020eep,Liu:2020akv,
deMedeirosVarzielas:2019cyj,
  	Asaka:2019vev,Ding:2020msi,Asaka:2020tmo,deAnda:2018ecu,Kobayashi:2019rzp,Novichkov:2018yse,Kobayashi:2018wkl,Okada:2018yrn,Okada:2019uoy,Nomura:2019jxj, Okada:2019xqk,
  	Nomura:2019yft,Nomura:2019lnr,Criado:2019tzk,
  	King:2019vhv,Gui-JunDing:2019wap,deMedeirosVarzielas:2020kji,Zhang:2019ngf,Nomura:2019xsb,Kobayashi:2019gtp,Lu:2019vgm,Wang:2019xbo,King:2020qaj,Abbas:2020qzc,Okada:2020oxh,Okada:2020dmb,Ding:2020yen,Okada:2020rjb,Okada:2020ukr,Novichkov:2021evw,Nagao:2020azf,Wang:2020lxk,
  	Okada:2020brs,Yao:2020qyy}.
(See for reviews Refs.~\cite{Kobayashi:2023zzc, Ding:2023htn}.)
Then ambiguities of coupling constants exist in the mass matrices and most of the works have used them as free parameters.
Consequently they have succeeded to realize the flavor structures by tuning these parameters.
On the other hand it is still difficult to explain the flavor structures by fewer parameters.
In addition, there remains the difficulties on realizing large hierarchies of fermion masses without fine-tuning.

The residual symmetries of the modular symmetry have the possibilities describing hierarchical fermion masses without fine-tuning.
At three modular symmetric points, $\tau=i$, $\omega\equiv e^{2\pi i\tau/3}$ and $i\infty$, $S$, $ST$ and $T$-symmetries remain \cite{Novichkov:2018ovf}.
Then the modular symmetry breaks into residual $Z_2$, $Z_3$ and $Z_N$ symmetries, respectively, where $N$ is the level of the finite modular group.
The modular forms with the residual charge $r$ take hierarchical values $\varepsilon^r$ where $\varepsilon$ stands for the deviation of the modulus $\tau$ from the modular symmetric points.
Thus, the residual symmetries make mass matrices hierarchical and can lead to large hierarchical fermion masses without hierarchical values of the coupling constants.
Along in this way, Refs.~\cite{Feruglio:2021dte,Novichkov:2021evw} have succeeded to obtain realistic lepton flavor observables without fine-tuning.
Moreover the quark flavor structures were described in $\Gamma_3\simeq A_4$ \cite{Petcov:2022fjf, Petcov:2023vws}, $\Gamma'_4\simeq S_4'$ \cite{deMedeirosVarzielas:2023crv}, $S_4'\times S_3$ \cite{Abe:2023ilq}, $\Gamma_6\simeq S_3\times A_4$ \cite{Kikuchi:2023cap} and $A_4\times A_4\times A_4$ \cite{Kikuchi:2023jap}.
Both quark and lepton flavors were studied in $S_4'$ \cite{Abe:2023qmr} and $\Gamma_6'$ \cite{Abe:2023dvr}.
On the other hand, Refs.~\cite{Petcov:2022fjf, Petcov:2023vws, Kikuchi:2023jap} have shown the difficulty on obtaining both quark hierarchical masses and the CP violating phase simultaneously.
Particularly numerical studies in Ref.~\cite{Kikuchi:2023jap} have implied that the vicinity of the modular symmetric points is favored for the large hierarchical fermion masses while it is disfavored for the CP violation.
Refs.~\cite{Petcov:2022fjf, Petcov:2023vws, Kikuchi:2023jap} have indicated that the models with multi moduli parameters can avoid such difficulty.
Multi modular symmetries were studied in Refs.~\cite{King:2021fhl,Du:2022lij,Abbas:2022slb}.
Also, the Siegel modular forms which are generalized modular forms of $Sp(2,\mathbb{Z})$, have $Sp(2g,\mathbb{Z})$ modular symmetry described by several moduli.
Therefore they are promising way explaining the flavor structures.
However, the studies of the modular symmetric flavor models using the Siegel modular forms are developing works.

In this paper, we discuss the quark flavor models with $\Gamma_6$ symmetry.
We use residual $T$ ($Z_6$) symmetry at the cusp $\tau=i\infty$ to reproduce quark mass hierarchies without fine-tuning.
As another possibility realizing such mass hierarchies, we also discuss the quark flavor models with $A_4\times A_4\times A_4$ symmetry.
We use residual $ST$ and $T$-symmetries at $\tau=\omega$ and $i\infty$, where $Z_3\times Z_3\times Z_3$ symmetry remains.
Then we will see the quark mass ratios and the absolute values of the CKM matrix elements can be realized in the vicinity of the modular symmetric points without fine-tuning.
Also we will see the inconsistency between the large quark mass hierarchies and the CP violation through numerical studies.
Furthermore, we study the construction of the Siegel modular forms for subgroups of $Sp(6,\mathbb{Z})$.
The Siegel modular forms for subgroups of $Sp(4,\mathbb{Z})$ and $Sp(6,\mathbb{Z})$ can be constructed from magnetized $T^4$ and $T^6$ models of the superstring theory, respectively \cite{Cremades:2004wa,Antoniadis:2009bg,Kikuchi:2022lfv,Kikuchi:2022psj,Kikuchi:2023awm}.
Then we will show the quark flavor observables including the CP phase can be realized without fine-tuning by deviating multi moduli parameters from the modular or CP symmetric points.

As one of the top-down approaches, higher dimensional theories are also attractive.
They can lead to non-Abelian discrete flavor symmetries as geometrical symmetries of extra-dimensions.
In particular, the supestring theory is promising candidate for the unified theory.
It predicts ten-dimensional (10D) space-time; extra six-dimensions are assumed to be compactified.
Thus six-dimensional (6D) compact space may lead to the flavor structures through its geometrical symmetry.

The torus compactification $T^2\times T^2\times T^2$ of the superstring theory is one of the simplest compactifications.
Two-dimensional (2D) torus $T^2$ has the modular symmetry $Sp(2,\mathbb{Z})\simeq SL(2,\mathbb{Z})$ as the geometrical symmetry.
Therefore it may lead to 4D modular symmetric flavor models as the low-energy effective theory.
On the other hand, we need to derive 4D chiral theory as the low-energy effective theory since the SM is a chiral theory.
It is remarkable that the torus compactification with magnetic fluxes leads to 4D chiral theories \cite{Bachas:1995ik, Blumenhagen:2000wh, Angelantonj:2000hi, Blumenhagen:2000ea, Cremades:2004wa, Abe:2008fi, Abe:2013bca, Abe:2014noa}.
The zero-mode wave functions on torus have chiral solutions depending on the signs of the fluxes.
Additionally, they have the solutions whose degeneracy number is determined by the sizes of the fluxes.
Thus there is the possibilities realizing three-generation chiral fermions on magnetized torus model.
Orbifoldings of torus have further possibilities realizing three-generation zero-modes.
Actually, several numerical studies have shown that realistic flavor observables can be realized \cite{Abe:2012fj,Abe:2014vza,Fujimoto:2016zjs,Kobayashi:2016qag,Kikuchi:2021yog,Kikuchi:2022geu,Hoshiya:2022qvr,Buchmuller:2017vho,Buchmuller:2017vut}.
In this paper, we discuss quark and lepton mass matrices derived from magnetized $T^2/\mathbb{Z}_2$ twisted orbifold model.
Zero-modes in magnetized $T^2/\mathbb{Z}_2$ orbifold model transform non-trivially under the modular symmetry 
\cite{Kobayashi:2018rad,Kobayashi:2018bff,Kariyazono:2019ehj,Ohki:2020bpo,Kikuchi:2020frp,Kikuchi:2020nxn,
Kikuchi:2021ogn,Almumin:2021fbk,Hoshiya:2020hki}.
We expect that three-generations of fermions originate from one of $T^2$s and other tori do not contribute to the generation structures.
Therefore we concentrate on zero-mode wave functions on $T^2$ and its orbifold $T^2/\mathbb{Z}_2$ in our analysis.

Yukawa couplings on magnetized torus and its orbifolding models are given by the overlap integrals of zero-mode wave functions.
As a result, they are written in terms of the modular forms and described by the modulus $\tau$.
Note that there are no ambiguities of coupling constants up to the overall factors in mass matrices contrary to ones on 4D modular symmeric flavor models.
This is because Yukawa couplings are explicitly calculated from the overlap integrals of zero-modes of fermions and Higgs fields.
Then the mass terms of Higgs fields ($\mu$ terms) are important.
In general, the superstring theory as well as magnetized orbifold models predicts multi pairs of Higgs fields.
They have the same quantum numbers under the SM gauge group $SU(3)\times SU(2)\times U(1)$ and can couple to quarks and leptons.
Hence the fermion mass matrices are given by the linear combinations of multi Higgs VEVs and Yukawa couplings.
The direction of Higgs VEVs is aligned in the lightest mass direction.
In the superstring theory, unfortunately Higgs $\mu$ terms are forbidden at the perturbative level.
Insteadly, they can be induced from nonperturbative effects such as D-brane instanton effects.
However, it is still difficult to determine the lightest mass direction of Higgs fields uniquely on the magnetized orbifold models because of the shortage of the instanton zero-mode configurations giving Higgs $\mu$ terms.
In our analysis, we regard the direction of Higgs VEVs as free parameters as in Refs.~\cite{Abe:2012fj,Abe:2014vza,Fujimoto:2016zjs,Kobayashi:2016qag}.

Similarly, D-brane instanton effects can induce Majorana masses of right-handed neutrinos \cite{Blumenhagen:2006xt, Ibanez:2006da, Ibanez:2007rs, Antusch:2007jd, Kobayashi:2015siy}.
Generating the smallness of neutrino masses is important.
Such masses are realized through the seesaw mechanism by introducing heavy Majorana mass terms of right-handed neutrino.
In Ref.~\cite{Hoshiya:2021nux}, possible Majorana masses on magnetized $T^2/\mathbb{Z}_2$ are classified.
In our analysis we assume the same D-brane instanton effects to reproduce the light neutrino masses.

As same as 4D modular symmetric flavor models, deriving hierarchical fermion masses is the key issue on the magnetized orbidold models.
On 4D modular symmetric flavor models it is solved by the deviation of the modulus from the modular symmetric points.
In place of this, we use the deviation of the Higgs VEVs here.
The mass matrices of up-sector quarks, down-sector quarks and charged leptons are approximately rank one matrices.
Hence, realistic mass matrices are realized in the vicinity of the directions of Higgs VEVs leading to rank one mass matrices.
One way to find such directions is using texture structures of Yukawa matrices at the modular symmetric points.
Suitable combinations of Yukawa textures can give rank one mass matrices.
In Ref.~\cite{Kikuchi:2021yog}, Yukawa textures at the modular symmetric points have been classified and the directions of Higgs VEVs leading to rank one mass matrices have been studied.
On the other hand, such method is available for the modulus at the modular symmetric points.
Here, we study another way using the property of wave functions on the compact space.
As we will see, such way is applied for not only quark and charged lepton mass hierarchies but also to find the directions of Higgs VEVs leading to small (large) mixing angles of quarks (leptons).
We will classify which flavor models have such directions of Higgs VEVs and show numerical example realizing quark and lepton flavor observables.

Also, we will study magnetized $(T^2_1\times T^2_2)/\mathbb{Z}_2^{\textrm{(per)}}$ and its orbifold.
Additionally we will discuss Yukawa textures on magnetized $T^2/\mathbb{Z}_2$.
Yukawa matrices have $S(\mathbb{Z}_2)$, $ST(\mathbb{Z}_3)$ and $T(\mathbb{Z}_N)$-symmetries at the modular symmetric points $\tau=i$, $\omega$ and $i\infty$, respectively.
Then Yukawa matrices are restricted to specific patterns because of the residual symmetries.
In Refs.~\cite{Novichkov:2018yse,Okada:2019uoy,Gui-JunDing:2019wap,Okada:2020rjb,Okada:2020ukr,Novichkov:2021evw}, realistic results were obtained at the vicinity of the modualr symmetric points.
(See also Ref.~\cite{Kikuchi:2022svo}.)

This paper is organized as follows.
In Chapter \ref{sec:magnetized_orbifold_models}, we discuss magnetized orbifold models.
In particular, we review torus compactification in Section \ref{subsec:Torus compactification} and study $T^2/\mathbb{Z}_2$ orbifold model in Section \ref{subsec:T2/Z2 orbifold}, $(T^2_1\times T^2_2)/\mathbb{Z}_2^{\textrm{(per)}}$ and its orbifold models in Section \ref{subsec:T2xT2} and Yukawa textures in Section \ref{subsec:Yukawa textures}.
In Chapter \ref{sec:4D modular symmetric flavor models}, we discuss 4D modular symmetric flavor models.
In particular, we study the quark flavor models with $\Gamma_6$ symmetry in Section \ref{subsec:Gamma_6} and ones with $A_4\times A_4\times A_4$ symmetry in Section \ref{subsec:A4xA4xA4}.
In Chapter \ref{sec:Constructing Siegel modular forms}, we discuss the constructing Siegel modular forms.
In Chapter \ref{sec:summary}, we conclude this paper.
In Appendix \ref{app:W-SS}, we show the equivalence of Wilson lines and SS phases.
In Appendix \ref{app:Flavor models}, we show the favorable models on $T^2/\mathbb{Z}_2$, Yukawa couplings and Majorana neutrino masses used in numerical example.
In Appendix \ref{app:D-brane instanton effects}, we review D-brane instanton effects inducing Majorana neutrino masses and Higgs $\mu$ terms.
In Appendix \ref{app:three_T2xT2dZ2xZ2}, we show possible three-generation models on $(T^2_1\times T^2_2)/(\mathbb{Z}_2^{\textrm{(t)}}\times \mathbb{Z}_2^{\textrm{(per)}})$.
In Appendix \ref{Decompositions of zero-modes with higher fluxes}, we show decompositions of zero-modes with higher fluxes.
In Appendix \ref{app:Mass matrix structures of the favorable models}, we show mass matrix structures of the favorable models on $T^2/\mathbb{Z}_2$.
In Appendix \ref{app:Gamma_6}, we review the group theory and the modular forms of $\Gamma_6$.
In Appendix \ref{app:A_4}, we review the group theory and the modular forms of $A_4$.
In Appendix \ref{app:Delta96}, we review the group theory and the Siegel modular forms of $\widetilde{\Delta}(96)$.


\chapter{Magnetized orbifold models}
\label{sec:magnetized_orbifold_models}

In this chapter, we discuss the magnetized orbifold models of the superstring theory.
We consider the orbifoldings of the magnetized torus compactification of the extra six-dimensions in the superstring theory.
Then we show that the magnetized orbifold models possess the possibilities to realize flavor structures of both quark and lepton sectors.
This chapter is along in Refs.~\cite{Kikuchi:2020frp, Kikuchi:2020nxn, Hoshiya:2020hki, Kikuchi:2021yog, Hoshiya:2022qvr}.


\section{Torus compactification}
\label{subsec:Torus compactification}

In this section, we focus on the torus compactification of the extra six-dimensions,
\begin{align}
{\cal M}^{10} = \mathbb{R}^{1,3} \times T^2_1 \times T^2_2 \times T^2_3.
\end{align}


\subsection{10D ${\cal N}=1$ $U(N)$ Super Yang-Mills theory}
\label{subsubsec:SYM}

The 10D ${\cal N}=1$ $U(N)$ Super Yang-Mills theory is a low energy effective theory of the superstring theory.
We review the theory on $\mathbb{R}^{1,3}\times T^2_1\times T^2_2\times T^2_3$ with magnetic fluxes (See Ref.~\cite{Cremades:2004wa}).
We note that magnetic fluxes can be introduced only on the compact spaces $T^2_1\times T^2_2\times T^2_3$ because of the Lorentz invariance on $\mathbb{R}^{1,3}$.
First we study Kaluza-Klein decomposition with magnetic fluxes.
Let us start from the 10D action,
\begin{align}
S = \int_{{\cal M}^4} d^4x \prod_{i=1,2,3} \int_{T^2_i} d^2z_i \left[-\frac{1}{4g} \textrm{Tr}(F^{MN}F_{MN})+\frac{i}{2g}\textrm{Tr}(\bar{\lambda}\Gamma^MD_M\lambda)\right],
\end{align}
where $F_{MN} = \partial_MA_N-\partial_NA_M-i[A_M,A_N]$ and $D_M\lambda=\partial_M\lambda-i[A_M,\lambda]$ with $M,N\in\{0,1,...,9\}$.
This action is invariant under $U(N)$ gauge transformation,
\begin{align}
\lambda \to U\lambda U^{-1}, \quad
A_M \to UA_M U^{-1} - (\partial_MU)U, \label{eq:U(N)_trans}
\end{align}
where $U$ denotes adjoint representation.
Here $\lambda$ denotes gaugino fields and $F_{MN}$ is the field strength of the 10D vector potential $A_M$.
The second term in the action contains 4D Yukawa terms,
\begin{align}
\frac{i}{2g} \textrm{Tr} (\bar{\lambda}\Gamma^MD_M\lambda)
&= \frac{i}{2g} \textrm{Tr} (-i\bar{\lambda}\Gamma^M[A_M,\lambda]) + \cdots \\
&= \frac{i}{2g} \textrm{Tr} (-i\bar{\lambda}\Gamma^\mu[A_\mu,\lambda])
+ \frac{i}{2g} \textrm{Tr} (-i\bar{\lambda}\Gamma^m[A_m,\lambda]) + \cdots , \label{eq:4DYukawa_in_Action}
\end{align}
where $\mu\in\{0,1,2,3\}$ and $m\in\{4,5,...,9\}$.
The second term on the second row in the above equation corresponds to 4D Yukawa terms.
Note that $A_m$ is regarded as 4D scalar fields.

By Kaluza-Klein decomposition, vector fields $A_M$ and Majorana-Weyl spinors $\lambda$ are decomposed into the 4D $(={\cal M}^4)$ parts and three of the 2D torus $(=T^2_i~(i=1,2,3))$ parts as
\begin{align}
A_M(x,z_1,z_2,z_3) &= \sum_{n_1,n_2,n_3} \phi_{M,n_1n_2n_3}(x) \otimes \phi_{M,n_1}(z_1) \otimes \phi_{M,n_2}(z_2) \otimes \phi_{M,n_3}(z_3), \label{eq:KK_A} \\
\lambda(x,z_1,z_2,z_3) &= \sum_{n_1,n_2,n_3} \psi_{n_1n_2n_3}(x) \otimes \psi_{n_1}(z_1) \otimes \psi_{n_2}(z_2) \otimes \psi_{n_3}(z_3). \label{eq:KK_L}
\end{align}
Here $\phi_{M,n_i}(z_i)$ is the $n_i$-th excited mode of the 2D vector fields on the $i$-th torus $T^2_i$ while $\psi_{n_i}(z_i)$ is one of the 2D Majorana-Weyl spinors.
We introduce the background magnetic fields on $T^2_1\times T^2_2\times T^2_3$,
\begin{align}
\begin{aligned}
F &= F_{z_i\bar{z}_i} dz_i\wedge d\bar{z}_i \\
&= \frac{\pi i}{\textrm{Im}\tau} 
\begin{pmatrix}
M_a^i \bm{1}_{N_a\times N_a} & & \\
& M_b^i \bm{1}_{N_b\times N_b} & \\
& & M_c^i \bm{1}_{N_c\times N_c} \\
\end{pmatrix} dz_i\wedge d\bar{z}_i, \quad i=1,2,3,
\end{aligned}
\end{align}
where $N=N_a+N_b+N_c$.
The $U(N)$ gauge symmetry breaks into $U(N_a)\times U(N_b)\times U(N_c)$ by the differences of fluxes $M^i_{a,b,c}$.
These magnetic fluxes lead to the following vector potential,
\begin{align}
A(z,\bar{z}) &=
\begin{pmatrix}
\frac{\pi M^i_a}{\textrm{Im}\tau_i} \textrm{Im} ((\bar{z}_i+\bar{\zeta}_{ia})dz_i)\bm{1}_{N_a\times N_a} & & \\
& \frac{\pi M^i_b}{\textrm{Im}\tau_i} \textrm{Im} ((\bar{z}_i+\bar{\zeta}_{ib})dz_i)\bm{1}_{N_b\times N_b} & \\
& & \frac{\pi M^i_c}{\textrm{Im}\tau_i} \textrm{Im} ((\bar{z}_i+\bar{\zeta}_{ic})dz_i)\bm{1}_{N_c\times N_c} \\
\end{pmatrix} \label{eq:vector_potential_Wilson}\\
&= -\begin{pmatrix}
\frac{\pi i M^i_a}{2\textrm{Im}\tau_i}(\bar{z}_i+\bar{\zeta}_{ia})\bm{1}_{N_a\times N_a} & & \\
& \frac{\pi i M^i_b}{2\textrm{Im}\tau_i}(\bar{z}_i+\bar{\zeta}_{ib})\bm{1}_{N_b\times N_b} & \\
& & \frac{\pi i M^i_c}{2\textrm{Im}\tau_i}(\bar{z}_i+\bar{\zeta}_{ic})\bm{1}_{N_c\times N_c}
\end{pmatrix}dz_i \notag \\
&\quad+\begin{pmatrix}
\frac{\pi iM^i_a}{2\textrm{Im}\tau_i}(z_i+\zeta_{ia})\bm{1}_{N_a\times N_a} & & \\
& \frac{\pi iM^i_b}{2\textrm{Im}\tau_i}(z_i+\zeta_{ib})\bm{1}_{N_b\times N_b} & \\
& & \frac{\pi iM^i_c}{2\textrm{Im}\tau_i}(z_i+\zeta_{ic})\bm{1}_{N_c\times N_c} \\
\end{pmatrix}d\bar{z}_i \\
&\equiv
\begin{pmatrix}
(A_{z_i})_a\bm{1}_{N_a\times N_a} & & \\
& (A_{z_i})_b\bm{1}_{N_b\times N_b} & \\
& & (A_{z_i})_c\bm{1}_{N_c\times N_c} \\
\end{pmatrix}dz_i \notag \\
&+
\begin{pmatrix}
(A_{\bar{z}_i})_a\bm{1}_{N_a\times N_a} & & \\
& (A_{\bar{z}_i})_b\bm{1}_{N_b\times N_b} & \\
& & (A_{\bar{z}_i})_c\bm{1}_{N_c\times N_c} \\
\end{pmatrix}d\bar{z}_i \\
&\equiv
\begin{pmatrix}
(A(z,\bar{z}))_a\bm{1}_{N_a\times N_a} & & \\
& (A(z,\bar{z}))_b\bm{1}_{N_b\times N_b} & \\
& & (A(z,\bar{z}))_c\bm{1}_{N_c\times N_c} \\
\end{pmatrix},
\end{align}
where $\zeta$ is so-called Wilson lines.
We can always omit Wilson lines from $A(z,\bar{z})$ by introducing Scherk-Schwarz (SS) phases instead as shown in Appendix \ref{app:W-SS}.
In the following we adopt SS phases and set Wilson lines vanishing, $\zeta=0$.

Next we study the boundary conditions on $T^2_i$.
The 2D torus can be regarded as the complex plane $\mathbb{C}$ divided by a 2D lattice $\Lambda$ spanned by two basis vectors $e_{1i}=2\pi R_i$ and $e_{2i}=2\pi R_i\tau_i$ where $R_i\equiv e_{1i}/2\pi \in\mathbb{R}$ denotes the radius of $T^2_i$ and $\tau_i\equiv e_{2i}/e_{1i} \in\mathbb{C}$ denotes the complex structure modulus of $T^2_i$.
The complex coordinates on $T^2_i$ are defined by $z_i\equiv u_i/e_{1i}$ where $u$ denotes ones on $\mathbb{C}$.
Then the 2D torus $T^2_i$ has the metric,
\begin{align}
ds^2 = 2h_{\mu\nu}^i dz_i^\mu d\bar{z}_i^\nu, \quad
h^i =|e_{1i}|^2
\begin{pmatrix}
0&\frac{1}{2}\\
\frac{1}{2}&0\\
\end{pmatrix},
\end{align}
and the gamma matrices,
\begin{align}
\Gamma^{z_i} =
\frac{1}{e_{1i}}
\begin{pmatrix}
0&2\\
0&0\\
\end{pmatrix}, \quad
\Gamma^{\bar{z}_i} =
\frac{1}{\bar{e}_{1i}}
\begin{pmatrix}
0&0\\
2&0\\
\end{pmatrix}.
\end{align}
The identifications on $T^2_i$ are written as
\begin{align}
\vec{z} \sim \vec{z} + \hat{e}_i \sim \vec{z} + \tau_i \hat{e}_i, \quad (i=1,2,3),
\end{align}
where we have used the notations,
\begin{align}
\vec{z} =
\begin{pmatrix}
z_1\\z_2\\z_3\\
\end{pmatrix}, \quad
\hat{e}_1 =
\begin{pmatrix}
1\\0\\0\\
\end{pmatrix}, \quad
\hat{e}_2 =
\begin{pmatrix}
0\\1\\0\\
\end{pmatrix}, \quad
\hat{e}_3 =
\begin{pmatrix}
0\\0\\1\\
\end{pmatrix}.
\end{align}
These identifications give the boudary conditions of the vector potential,
\begin{align}
&(A(\vec{z}+\hat{e}_i,\vec{\bar{z}}+\hat{e}_i))_{\alpha} = (A(\vec{z}_i,\vec{\bar{z}}_i))_{\alpha} + d\chi_{1i}^\alpha(\vec{z}_i,\vec{\bar{z}}_i), \\
&(A(\vec{z}+\tau_i\hat{e}_i,\vec{\bar{z}}+\tau_i\hat{e}_i))_{\alpha} = (A(\vec{z}_i,\vec{\bar{z}}_i))_{\alpha} + d\chi_{2i}^\alpha(\vec{z}_i,\vec{\bar{z}}_i), \\
&\chi_{1i}^\alpha(\vec{z}_i,\vec{\bar{z}}_i) = \left(\frac{\pi M^i_\alpha}{\textrm{Im}\tau_i}\textrm{Im}z_i\right), \quad
\chi_{2i}^\alpha(\vec{z}_i,\vec{\bar{z}}_i) = \left(\frac{\pi M^i_\alpha}{\textrm{Im}\tau_i}\textrm{Im}\bar{\tau}_iz_i\right),
\end{align}
where $\alpha\in\{a,b,c\}$.
Note that these correspond to $U(1)$ gauge transformation.
Thus the vector potential is identified up to $U(1)$ gauge transformation under the boundary conditions.
These conditions and the gauge transformation laws in Eq.~(\ref{eq:U(N)_trans}) mean that the gaugino fields,
\begin{align}
\lambda(x,\vec{z}) = \begin{pmatrix}
\lambda^{aa}(x,\vec{z}) & \lambda^{ab}(x,\vec{z}) & \lambda^{ac}(x,\vec{z}) \\
\lambda^{ba}(x,\vec{z}) & \lambda^{bb}(x,\vec{z}) & \lambda^{bc}(x,\vec{z}) \\
\lambda^{ca}(x,\vec{z}) & \lambda^{cb}(x,\vec{z}) & \lambda^{cc}(x,\vec{z}) \\
\end{pmatrix}, \label{eq:nine-bi-matter}
\end{align}
obey the boundary conditions,
\begin{align}
\begin{aligned}
&\lambda^{\alpha\alpha}(x,\vec{z}+\hat{e}_i) = \lambda^{\alpha\alpha}(x,\vec{z}+\tau_i\hat{e}_i) = \lambda^{\alpha\alpha}(x,\vec{z}), \\
&\lambda^{\alpha\beta}(x,\vec{z}+\hat{e}_i) = e^{2\pi i\alpha_{1i}^{\alpha\beta}}e^{i\chi^{\alpha\beta}_{1i}(z_i)} \lambda^{\alpha\beta}(x,\vec{z}), \quad
\lambda^{\alpha\beta}(x,\vec{z}+\tau_i\hat{e}_i) = e^{2\pi i\alpha_{2i}^{\alpha\beta}}e^{i\chi^{\alpha\beta}_{2i}(z_i)} \lambda^{\alpha\beta}(x,\vec{z}), \\
\end{aligned} \label{eq:BC_lambda}
\end{align}
where $\alpha,\beta\in\{a,b,c\}$, $\alpha\neq\beta$, $\alpha^{\alpha\beta}_{1i,2i}$ denote SS phases satisfying $\alpha^{ab}_{1i,2i}+\alpha^{bc}_{1i,2i}+\alpha^{ca}_{1i,2i}=0$ and
\begin{align}
&\chi^{\alpha\beta}_{1i}(z_i) \equiv \frac{\pi I_{\alpha\beta}^i}{\textrm{Im}\tau_i} \textrm{Im}z_i, \quad
\chi^{\alpha\beta}_{2i}(z_i) \equiv \frac{\pi I_{\alpha\beta}^i}{\textrm{Im}\tau_i} \textrm{Im}(\bar{\tau}_iz_i), \quad I_{\alpha\beta}^i \equiv M_\alpha^i-M_\beta^i. \label{eq:chi_and_M}
\end{align}
Here $\lambda^{\alpha\alpha}$ are the gaugino fields under the remaining gauge group $U(N_a)\times U(N_b)\times U(N_c)$ while $\lambda^{\alpha\beta}$ are bi-fundamental matter fields under $U(N_\alpha)\times U(N_\beta)$, $(N_\alpha,\bar{N}_\beta)$.
We note that the boundary condition after a contractible loop on $T^2_i$ must be trivial.
This requires the quantized magnetic fluxes,
\begin{align}
M_\alpha^i \in \mathbb{Z},
\end{align}
which is so-called the Dirac quantization condition.

Now we are ready to solve the Dirac equation on $T^2_i$.
We denote the 2D spinors of the gaugino fields, $\psi_{n_i}(z_i)$, as
\begin{align}
\psi_{n_i}(z_i) =
\begin{pmatrix}
\psi_{n_i+} \\ \psi_{n_i-} \\
\end{pmatrix}, \quad \psi_{n_i\pm}(z_i) =
\begin{pmatrix}
\psi^{aa}_{n_i\pm}(z_i) & \psi^{ab}_{n_i\pm}(z_i) & \psi^{ac}_{n_i\pm}(z_i) \\
\psi^{ba}_{n_i\pm}(z_i) & \psi^{bb}_{n_i\pm}(z_i) & \psi^{bc}_{n_i\pm}(z_i) \\
\psi^{ca}_{n_i\pm}(z_i) & \psi^{cb}_{n_i\pm}(z_i) & \psi^{cc}_{n_i\pm}(z_i) \\
\end{pmatrix}.
\end{align}
The boundary conditions in Eq.~(\ref{eq:BC_lambda}) are rewritten in terms of $\psi_{n_i\pm}(z_i)$ as
\begin{align}
\begin{aligned}
&\psi^{\alpha\alpha}_{n_i\pm}(z_i+1) = \psi^{\alpha\alpha}_{n_i\pm}(z_i+\tau_i) = \psi^{\alpha\alpha}_{n_i\pm}(z_i), \\
&\psi^{\alpha\beta}_{n_i\pm}(z_i+1) = e^{2\pi i\alpha_{1i}^{\alpha\beta}} e^{i\chi_{1i}^{\alpha\beta}(z_i)} \psi^{\alpha\beta}_{n_i\pm}(z_i), \quad \psi^{\alpha\beta}_{n_i\pm}(z_i+\tau_i) = e^{2\pi i\alpha_{2i}^{\alpha\beta}} e^{i\chi_{2i}^{\alpha\beta}(z_i)} \psi^{\alpha\beta}_{n_i\pm}(z_i).
\end{aligned} \label{eq:BC_psi}
\end{align}
Here we consider the Dirac equation for zero-modes, $\psi_{0_i}(z_i)=(\psi_{0_i+}, \psi_{0_i-})^T$.
The Dirac equation for zero-modes is written as
\begin{align}
i\slash{D} \psi_{0_i\pm} = (i\Gamma^{z_i}D_{z_i}+i\Gamma^{\bar{z}_i}D_{\bar{z}_i}) \psi_{0_i\pm} = i\begin{pmatrix}
0&-D^\dagger\\
D&0\\
\end{pmatrix}
\begin{pmatrix}
\psi_{0_i+}\\\psi_{0_i-}
\end{pmatrix}=0.
\end{align}
Particularly let us consider the equation for $\psi_{0_i+}$ which is given by
\begin{align}
&D\psi_{0_i+} = (\pi R_i)^{-1} (\partial_{\bar{z}_i}\psi_{0_i+} + [A_{\bar{z}_i}, \psi_{0_i+}]) \\
&= (\pi R_i)^{-1}
\begin{pmatrix}
\partial_{\bar{z}_i} \psi^{aa}_{0_i+} & (\partial_{\bar{z}_i}+\frac{\pi I_{ab}^i}{2\textrm{Im}\tau_i}z_i) \psi^{ab}_{0_i+} & (\partial_{\bar{z}_i}+\frac{\pi I_{ac}^i}{2\textrm{Im}\tau_i}z_i) \psi^{ac}_{0_i+} \\
(\partial_{\bar{z}_i}-\frac{\pi I_{ab}^i}{2\textrm{Im}\tau_i}z_i) \psi^{ba}_{0_i+} & \partial_{\bar{z}_i} \psi^{bb}_{0_i+} & (\partial_{\bar{z}_i}+\frac{\pi I_{bc}^i}{2\textrm{Im}\tau_i}z_i) \psi^{bc}_{0_i+} \\
(\partial_{\bar{z}_i}-\frac{\pi I_{ac}^i}{2\textrm{Im}\tau_i}z_i) \psi^{ca}_{0_i+} & (\partial_{\bar{z}_i}-\frac{\pi I_{bc}^i}{2\textrm{Im}\tau_i}z_i) \psi^{cb}_{0_i+} & \partial_{\bar{z}_i} \psi^{cc}_{0_i+} \\
\end{pmatrix}
= 0.
\end{align}
The boundary conditions in Eq.~(\ref{eq:BC_psi}) require the solutions of $\psi^{\alpha\alpha}_{0_i+}$ to be constants.
On the other hand, we can find that for $I_{\alpha\beta}^i>0$ only $\psi^{\alpha\beta}_{0_i+}$ have zero-mode solutions with the degeneracy number $|I_{\alpha\beta}^i|$ while for $I_{\alpha\beta}^i<0$ only $\psi^{\beta\alpha}_{0_i+}$ have such solutions.
Similarly we can find that the solutions of $\psi^{\alpha\alpha}_{0_i-}$ are constants; for $I_{\alpha\beta}^i>0$ only $\psi^{\beta\alpha}_{0_i-}$ have $|I_{\alpha\beta}^i|$ number of solutions while for $I_{\alpha\beta}^i<0$ only $\psi^{\alpha\beta}_{0_i-}$ have such solutions.
Note that these solutions have the exclusivity for the chiality.
Thus, we obtain $|I_{\alpha\beta}^i|$ generations of bi-fundamental chiral zero-modes on the magnetized $i$th torus.
The explicit formula of the $j$th zero-mode of $\psi_{0_i\pm}$ is given by
\begin{align}
&\psi^{(j+\alpha^{\alpha\beta}_{1i},\alpha^{\alpha\beta}_{2i}),|I_{\alpha\beta}^i|}_{0_i+}(z_i,\tau_i) \notag \\
&= \left(\frac{|I_{\alpha\beta}^i|}{{\cal A}^2_{i}}\right)^{1/4} 
e^{2\pi i\frac{(j+\alpha^{\alpha\beta}_{1i})\alpha^{\alpha\beta}_{2i}}{|I_{\alpha\beta}^i|}} e^{i\pi |I_{\alpha\beta}^i|z_i \frac{\textrm{Im}z_i}{\textrm{Im}\tau_i}}
\sum_{\ell\in\mathbb{Z}} e^{i\pi |I_{\alpha\beta}^i|\tau_i \left(\frac{j+\alpha^{\alpha\beta}_{1i}}{|I_{\alpha\beta}^i|}+\ell\right)^2} e^{2\pi i(|I_{\alpha\beta}^i|z_i-\alpha^{\alpha\beta}_{2i}) \left(\frac{j+\alpha^{\alpha\beta}_{1i}}{|I_{\alpha\beta}^i|}+\ell\right)} \\
&= \left(\frac{|I_{\alpha\beta}^i|}{{\cal A}^2_{i}}\right)^{1/4} 
e^{2\pi i\frac{(j+\alpha^{\alpha\beta}_{1i})\alpha^{\alpha\beta}_{2i}}{|I_{\alpha\beta}^i|}}
e^{i\pi |I_{\alpha\beta}^i|z_i \frac{\textrm{Im}z_i}{\textrm{Im}\tau_i}}
\vartheta
\begin{bmatrix}
\frac{j+\alpha^{\alpha\beta}_{1i}}{|I_{\alpha\beta}^i|} \\ -\alpha^{\alpha\beta}_{2i} \\
\end{bmatrix}
(|I_{\alpha\beta}^i|z_i, |I_{\alpha\beta}|^i\tau_i), \label{eq:zero-mode_formula} \\
&\psi^{(j+\alpha_1,\alpha_2),|I_{\alpha\beta}^i|}_{0_i-}(z_i,\tau_i) = \left(\psi^{(-j-\alpha_1,\alpha_2),|I_{\alpha\beta}^i|}_{0_i+}(z_i,\tau_i)\right)^*,
\end{align}
where $j=\{0,1,...,|I_{\alpha\beta}^i|-1\}$ and ${\cal A}_i=|e_{1i}|^2\textrm{Im}\tau_i$ stands for the area of $i$th torus.
The $\vartheta$ function is the Jacobi theta function defined by
\begin{align}
\vartheta
\begin{bmatrix}
a\\b\\
\end{bmatrix}
(\nu,\tau) =
\sum_{\ell\in\mathbb{Z}} e^{\pi i(a+\ell)^2\tau} e^{2\pi i(a+\ell)(\nu+b)}, \quad
a,b\in\mathbb{R}, \quad \nu,\tau\in\mathbb{C}.
\end{align}
They have the periodicity,
\begin{align}
&\vartheta\begin{bmatrix}a\\b\\\end{bmatrix}(\nu+n,\tau) = e^{2\pi ina}
\vartheta\begin{bmatrix}a\\b\\\end{bmatrix}(\nu,\tau), \\
&\vartheta\begin{bmatrix}a\\b\\\end{bmatrix}(\nu+n\tau,\tau) = e^{-\pi in^2\tau-2\pi in(\nu+b)}
\vartheta\begin{bmatrix}a\\b\\\end{bmatrix}(\nu,\tau),
\end{align}
where $n\in\mathbb{Z}$.
The products of the functions can be expanded as
\begin{align}
\vartheta\begin{bmatrix}\frac{r}{N_1}\\0\\\end{bmatrix}(\nu_1,N_1\tau)
\times
\vartheta\begin{bmatrix}\frac{s}{N_2}\\0\\\end{bmatrix}(\nu_2,N_2\tau)
&= \sum_{m\in\mathbb{Z}_{N_1+N_2}}
\vartheta\begin{bmatrix}\frac{r+s+N_1m}{N_1+N_2}\\0\\\end{bmatrix}(\nu_1+\nu_2,(N_1+N_2)\tau) \notag \\
\times \vartheta &\begin{bmatrix}\frac{N_2r-N_1s+N_1N_2m}{N_1N_2(N_1+N_2)}\\0\\\end{bmatrix}(\nu_1N_2-\nu_2N_1, N_1N_2(N_1+N_2)\tau).
\end{align}
Using these properties, we can show the following product expansions of zero-modes \cite{Cremades:2004wa},
\begin{align}
\psi_{0_i\pm}^{(j+\alpha_{1i}^{ab},\alpha_{2i}^{ab}),|I_{ab}^i|}(z_i,\tau_i) \cdot \psi_{0_i\pm}^{(k+\alpha_{1i}^{ca},\alpha_{2i}^{ca}),|I_{ca}^i|}(z_i,\tau_i) = \sum_{\ell\in\mathbb{Z}_{|I_{cb}^i|}} y_{T^2}^{jk\ell}(\tau_i) \cdot
\psi_{0_i\pm}^{(j+k+\ell|I_{ca}^i|+\alpha_{1i}^{cb},\alpha_{2i}^{cb}),|I_{cb}^i|}(z_i,\tau_i),
\label{eq:prod_ex}
\end{align}
where $|I_{ab}^i| + |I_{ca}^i| = |I_{cb}^i|$, $(\alpha^{ab}_{1i},\alpha^{ab}_{2i})+(\alpha^{ca}_{1i},\alpha^{ca}_{2i})=(\alpha^{cb}_{1i},\alpha^{cb}_{2i})$ and $y_{T^2}^{jk\ell}$ denotes three point couplings given by
\begin{align}
y^{jk\ell}_{T^2} &=
c_{(|I_{ab}^i|\textrm{-}|I_{ca}^i|\textrm{-}|I_{cb}^i|)} \exp \left\{
2\pi i \left( \frac{(j+\alpha^{ab}_{1i})\alpha^{ab}_{2i}}{|I_{ab}^i|}
+ \frac{(k+\alpha^{ca}_{1i})\alpha^{ca}_{2i}}{|I_{ca}^i|}
- \frac{(\ell+\alpha^{cb}_{1i})\alpha^{cb}_{2i}}{|I_{cb}^i|} \right)
\right\} \notag \\
&\times \sum_{m=0}^{|I_{ca}^i|-1} \vartheta
\begin{bmatrix}\frac{|I_{ca}^i|(j+\alpha^{ab}_{1i})-|I_{ab}^i|(k+\alpha^{ca}_{1i})+|I_{ab}^i||I_{ca}^i|m}{|I_{ab}^iI_{ca}^iI_{cb}^i|} \\ 0 \\\end{bmatrix}(|I_{ab}^i|\alpha^{ca}_{2i}-|I_{ca}^i|\alpha^{ab}_{2i}, |I_{ab}^iI_{ca}^iI_{cb}^i|\tau_i) \notag \\
&\times \delta_{(j+\alpha^{ab}_{1i})+(k+\alpha^{ca}_{1i})-(\ell+\alpha^{cb}_{1i}), |I_{cb}^i|n-|I_{ab}^i|m},
\end{align}
with
\begin{align}
n\in\mathbb{Z}, \quad
c_{(|I_{ab}^i|\textrm{-}|I_{ca}^i|\textrm{-}|I_{cb}^i|)} = (2\textrm{Im}\tau)^{-1/2} {\cal A}^{-1/2}
\left|\frac{I_{ab}^iI_{ca}^i}{I_{cb}^i}\right|^{1/4}.
\end{align}
Also zero-modes satisfy the following normalization condition,
\begin{align}
\int_{T^2_i} dz_i d\bar{z}_i \psi_{0_i\pm}^{(j+\alpha^{\alpha\beta}_{1i},\alpha^{\alpha\beta}_{2i}), |I_{\alpha\beta}^i|}(z_i,\tau_i) \left( \psi_{0_i\pm}^{(k+\alpha^{\alpha\beta}_{1i},\alpha^{\alpha\beta}_{2i}), |I_{\alpha\beta}^i|}(z_i,\tau_i) \right)^* = (2\textrm{Im}\tau_i)^{-1/2} \delta_{j,k},
\end{align}
and the periodicity,
\begin{align}
&\psi^{(j+|I_{\alpha\beta}^i|+\alpha_{1i}^{\alpha\beta},\alpha_{2i}^{\alpha\beta}), |I_{\alpha\beta}^i|}_{0_i\pm} = \psi^{(j+\alpha_{1i}^{\alpha\beta},\alpha_{2i}^{\alpha\beta}), |I_{\alpha\beta}^i|}_{0_i\pm}, \\
&\psi^{(j+\alpha_{1i}^{\alpha\beta},\alpha_{2i}^{\alpha\beta}+1), |I_{\alpha\beta}^i|}_{0_i\pm} = \psi^{(j+\alpha_{1i}^{\alpha\beta},\alpha_{2i}^{\alpha\beta}), |I_{\alpha\beta}^i|}_{0_i\pm}.
\end{align}
By use of the normalization condition, it can be found that the definition of three point couplings in Eq.~(\ref{eq:prod_ex}) is equivalent to
\begin{align}
Y^{jk\ell}_{T^2} = gy^{jk\ell}_{T^2} = g(2\textrm{Im}\tau)^{1/2}\int_{T^2} dzd\bar{z}
\psi_{0_i\pm}^{(j+\alpha_{1i}^{ab},\alpha_{2i}^{ab}),|I_{ab}^i|} \cdot
\psi_{0_i\pm}^{(k+\alpha_{1i}^{ca},\alpha_{2i}^{ca}),|I_{ca}^i|} \cdot
\left( \psi_{0_i\pm}^{(\ell+\alpha_{1i}^{cb},\alpha_{2i}^{cb}),|I_{cb}^i|} \right)^*.
\label{eq:Yukawa_overlap}
\end{align}
In Table \ref{tab:num_of_zeros_T2}, we show the number of zero-modes on magnetized $T^2_i$, which is given by a size of the flux $M_i$.
\begin{table}[H]
\centering
\begin{tabular}{c|cccccc|c} \hline
Flux & 1 & 2 & 3 & 4 & 5 & 6 & $M_i$ \\ \hline
Number of zero-modes & 1 & 2 & 3 & 4 & 5 & 6 & $M_i$ \\ \hline
\end{tabular}
\caption{The number of zero-modes on magnetized $T^2$.}
\label{tab:num_of_zeros_T2}
\end{table}

To calculate 4D Yukawa couplings, we also consider the Klein-Gordon equation for 2D parts of vector fields, $\phi_{\pm,0_i}(z_i)$, where $\pm$ is helicity on $T^2_i$.
Laplacian $\Delta$ can be constructed from Dirac operator $D$ as
\begin{align}
(i\slash{D})^2 =
\begin{pmatrix}
D^\dagger D & 0 \\
0 & D D^\dagger \\
\end{pmatrix}
= \frac{1}{2} \{D^\dagger,D\} +
\begin{pmatrix}
\frac{1}{2}[D^\dagger,D] & 0 \\
0 & \frac{1}{2}[D,D^\dagger] \\
\end{pmatrix}
= \Delta +
\begin{pmatrix}
\frac{2iF_{z_i\bar{z}_i}}{(2\pi R_i)^2} & 0 \\
0 & \frac{2iF_{\bar{z}_iz_i}}{(2\pi R_i)^2} \\
\end{pmatrix}.
\end{align}
To find functions obeying Klein-Gordon equation, first let us act $N\equiv D^\dagger D$ on the zero-modes of 2D spinors $\psi_{0_i+}$.
From $D\psi_{0_i+}=0$, we find
\begin{align}
N \psi_{0_i+}
= D^\dagger D \psi_{0_i+} =
\left(\Delta - \frac{2\pi |I_{\alpha\beta}^i|}{{\cal A}_i}\right)
\psi_{0_i+} = 0.
\end{align}
Notice that $N$ satisfies
\begin{align}
[N,D^\dagger] = \frac{4\pi |I_{\alpha\beta}^i|}{{\cal A}_i} D^\dagger.
\end{align}
Introducing
\begin{align}
a = \sqrt{\frac{{\cal A}_i}{4\pi |I_{\alpha\beta}^i|}} D, \quad
a^\dagger = \sqrt{\frac{{\cal A}_i}{4\pi |I_{\alpha\beta}^i|}} D^\dagger, \quad n = a^\dagger a,
\end{align}
we obtain the commutation relations,
\begin{align}
[a,a^\dagger] = 1, \quad [n,a^\dagger] = a^\dagger, \quad [n,a] = -a.
\end{align}
Then $(a^\dagger)^r \psi_{0_i+}$, $r\geq 0$, satisfy
\begin{align}
a \psi_{0_i+} = 0, \quad \Delta (a^\dagger)^r \psi_{0_i+} = \frac{4\pi |I_{\alpha\beta}^i|}{{\cal A}_i} \left(n+\frac{1}{2}\right) (a^\dagger)^r \psi_{0_i+} =
\frac{4\pi |I_{\alpha\beta}^i|}{{\cal A}_i} \left(r+\frac{1}{2}\right) (a^\dagger)^r \psi_{0_i+}.
\end{align}
This and the above commutation relations correspond to the harmonic oscillator quantum algebra.
Therefore the eigenstates of Laplacian $\Delta$ are given by $(a^\dagger)^r \psi_{0_i+}\propto (D^\dagger)^r \psi_{0_i+}$.
That is, the ground state of the vector fields obeying Klein-Gordon equation is written in terms of $\psi_{0_i+}$ as same as the spinor fields.
However, this ground state generally gets nonzero mass eigenvalue,
\begin{align}
m_i^2 = \frac{2\pi |I_{\alpha\beta}^i|}{{\cal A}_i} \quad (\textrm{for~scalar~fields~on~}T^2_i).
\end{align}
For vector fields on $T^2_i$, the mass squared is shifted by $\pm 4\pi |I_{\alpha\beta}^i|/{\cal A}_i$ but the ground state is given by the lowest mass state (minus sign) \cite{Cremades:2004wa}.
Note that this mass value is the contribution from one torus.
Including contributions from three tori, we find mass eigenvalue of the ground state of vector fields,
\begin{align}
m^2 = \sum_{i=1,2,3} m_i^2 = \pm \frac{2\pi |I_{\alpha\beta}^1|}{{\cal A}_1} \pm \frac{2\pi |I_{\alpha\beta}^2|}{{\cal A}_2} \pm \frac{2\pi |I_{\alpha\beta}^3|}{{\cal A}_3}, \label{eq:mass_vector}
\end{align}
where plus and minus signs correspond to scalar and vector fields on $T^2_i$, respectively.
Thus when we require supersymmetry at $\alpha\beta$ sector, mass of vector fields in Eq.~(\ref{eq:mass_vector}) must vanish.

Finally we show 4D Yukawa couplings using Kaluza-Klein decomposition.
Substituting decompositions in Eqs.~(\ref{eq:KK_A}) and (\ref{eq:KK_L}) into 4D Yukawa terms in Eq.~(\ref{eq:4DYukawa_in_Action}) and leaving only zero-modes, we can find
\begin{align}
&Y_{(1)} = \frac{1}{2g^2} \prod_{i=1,2,3} \int_{T^2_i} d^2z_i \textrm{Tr} (\psi_{0_i+} \cdot [\phi_{0_i-},\psi_{0_i+}]), \\
&Y_{(2)} = \frac{1}{2g^2} \prod_{i=1,2,3} \int_{T^2_i} d^2z_i \textrm{Tr} (\psi_{0_i-} \cdot [\phi_{0_i+},\psi_{0_i-}]),
\end{align}
where trace is regarding to gauge symbols $\alpha\beta$ and $Y_{(2)}$ is the CPT conjugates of $Y_{(1)}$.
Since the ground states of vector fields $\phi_{0_i\pm}$ are given by theta functions as $\psi_{0_i\pm}$, 4D Yukawa couplings are given by the products of three point couplings,
\begin{align}
Y_{T^2_1\times T^2_2\times T^2_3}(\tau_1,\tau_2,\tau_3) = Y_{T^2_1/\mathbb{Z}_2}(\tau_1) \times Y_{T^2_2/\mathbb{Z}_2}(\tau_2) \times Y_{T^2_3/\mathbb{Z}_2}(\tau_3). \label{eq:three-point-couplings}
\end{align}

Hereafter, we focus on the zero-modes of the bi-fundamental matter fields $\lambda^{\alpha\beta}$, $\psi^{(j+\alpha^{\alpha\beta}_{1i},\alpha^{\alpha\beta}_{2i}),|I_{\alpha\beta}^i|}_{0_i+}$, which are written by theta functions.
Then we will omit the symbols of the chirality, $\pm$, gauge representations, $\alpha\beta$, and the zero-mode, $0$, from the notation of zero-modes.
Moreover we denote $|I_{\alpha\beta}^i|$ as $M_i$.
Hence, we use the notation of zero-modes, $\psi^{(j+\alpha_{1i},\alpha_{2i}),M_i}(z_i,\tau_i)$ instead of $\psi^{(j+\alpha^{\alpha\beta}_{1i},\alpha^{\alpha\beta}_{2i}),|I_{\alpha\beta}^i|}_{0_i+}(z_i,\tau_i)$.
Also we omit $i$ meaning $i$th torus on $T^2_1\times T^2_2\times T^2_3$ if it is not necessary to distinguish three tori in the discussion.


\subsection{Modular symmetry}
\label{subsubsec:modular_symmetry}

\begin{figure}[H]
\centering
\includegraphics[width=5cm]{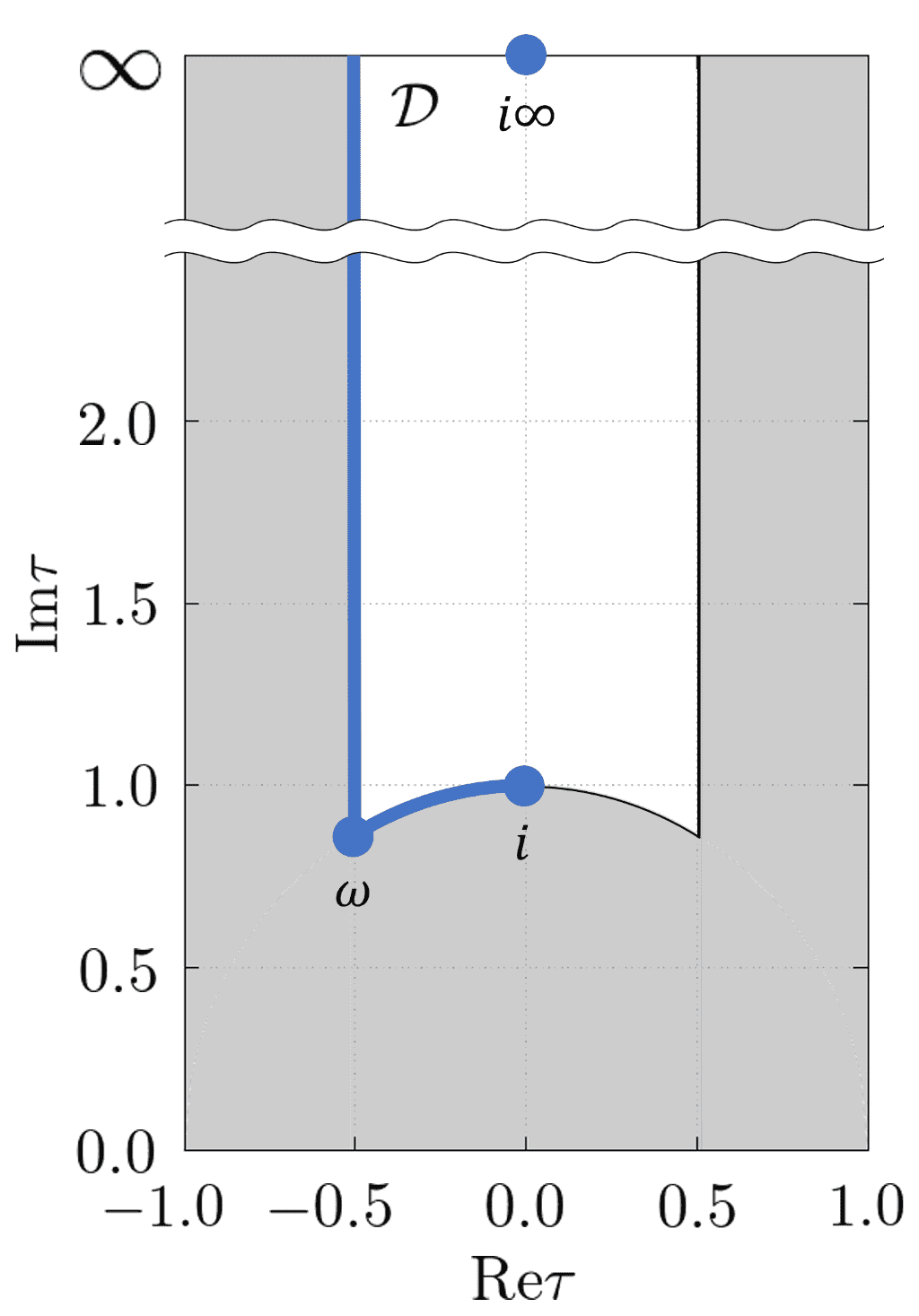}
\caption{The fundamental region ${\cal D}$ of the modulus $\tau$ under the modular group $SL(2,\mathbb{Z})\simeq \Gamma$.
White and blue corresponds to the fundamental region and gray is out of the region.
Three blue points stand for three symmetric points, $\tau=i$, $\omega$ and $i\infty$.}
\label{fig:fundamental_region}
\end{figure}

Here we give a brief review of the modular symmetry on $T^2$ and the modular forms \cite{Gunning:1962,Schoeneberg:1974,Koblitz:1984,Bruinier:2008}.
The modular transformation is defined as the basis transformation of the 2D lattice $\Lambda$ defining $T^2\simeq \mathbb{C}/\Lambda$.
It is given by
\begin{align}
\begin{pmatrix}
e_2 \\ e_1 \\
\end{pmatrix} \to
\begin{pmatrix}
a & b \\
c & d \\
\end{pmatrix}
\begin{pmatrix}
e_2 \\ e_1 \\
\end{pmatrix},
\end{align}
where
\begin{align}
\gamma =
\begin{pmatrix}
a & b \\
c & d \\
\end{pmatrix} \in Sp(2,\mathbb{Z}) \simeq SL(2,\mathbb{Z}) \equiv \Gamma,
\end{align}
satisfying
\begin{align}
\gamma J \gamma^T = J, \quad
J = \begin{pmatrix}
0 & 1 \\
-1 & 0 \\
\end{pmatrix}.
\end{align}
Under the modular transformation, the complex structure modulus and the complex coordinate on $T^2$, $\tau$ and $z$, are transformed as
\begin{align}
&\tau\equiv \frac{e_2}{e_1} \to \frac{ae_2+be_1}{ce_2+de_1} = \frac{a\tau+b}{c\tau+d}, \\
&z \equiv \frac{u}{e_1} \to \frac{u}{ce_2+de_1} = \frac{z}{c\tau+d}.
\end{align}
There are two generators, $S$ and $T$ defined by
\begin{align}
S =
\begin{pmatrix}
0 & 1 \\
-1 & 0 \\
\end{pmatrix}, \quad
T = 
\begin{pmatrix}
1 & 1 \\
0 & 1 \\
\end{pmatrix}.
\end{align}
They satisfy the algebraic relations,
\begin{align}
S^2 = -\mathbb{I}, \quad S^4 = (ST)^3 = \mathbb{I}.
\end{align}
$S$ and $T$-transformations for $\tau$ and $z$ are given by
\begin{align}
\begin{aligned}
&\tau \xrightarrow{S} -\frac{1}{\tau}, \quad z \xrightarrow{S} -\frac{z}{\tau}, \\
&\tau \xrightarrow{T} \tau+1, \quad z \xrightarrow{T} z.
\end{aligned} \label{eq:mod_t_z}
\end{align}
The modular symmetry is spontaneously broken by the VEV of $\tau$.
The modular transformations in Eq.~(\ref{eq:mod_t_z}) mean that there are no value of $\tau$ preserving full modular symmetry.
Meanwhile, there are three symmetric points where the partial modular symmetry (called residual symmetry) is preserved \cite{Novichkov:2018ovf}:
\begin{itemize}
\item $\tau=i$ is invariant under $S$-transformation;
\item $\tau=e^{2\pi i/3}\equiv \omega$ is invariant under $ST$-transformation;
\item $\tau=i\infty$ is invariant under $T$-transformation.
\end{itemize}
Figure \ref{fig:fundamental_region} shows the fundamental region ${\cal D}$ of $\tau$ under the modular group $SL(2,\mathbb{Z})\simeq\Gamma$ and the positions of three symmetric points.

Next let us study the modular forms.
We introduce the principal congruence subgroup of level $N$, $\Gamma(N)$, which is the normal subgroup of $\Gamma$,
\begin{align}
\Gamma(N) \equiv
\left\{
h=\begin{pmatrix}a&b\\c&d\\\end{pmatrix}\in\Gamma \left|
\begin{pmatrix}a&b\\c&d\\\end{pmatrix}\equiv \begin{pmatrix}1&0\\0&1\\\end{pmatrix}~(\textrm{mod}~N)
\right.\right\}.
\end{align}
In addition we introduce $\bar{\Gamma}\equiv \Gamma/\{\pm\mathbb{I}\}$ and $\bar{\Gamma}(N) \equiv \Gamma(N)/\{\pm \mathbb{I}\}$.
We note that $h=-\mathbb{I}$ is not included in $\Gamma(N)$ for $N>2$ but it is included in $\bar{\Gamma}(N)$.

The modular forms of weight $k$ are the holomorphic functions of $\tau$ transformed as
\begin{align}
f^i(\tau) \xrightarrow{\gamma} J_k(\gamma,\tau) \rho^{ij}(\gamma) f^j(\tau), \quad
\gamma \in \Gamma,
\end{align}
where $\rho$ is a unitary matrix satisfying $\rho(\gamma_1\gamma_2)=\rho(\gamma_1)\rho(\gamma_2),~\gamma_1,\gamma_2\in\Gamma$ and $J_k(\gamma,\tau)$ is the automorphy factor defined as
\begin{align}
J_k(\gamma,\tau) \equiv (c\tau+d)^k, \quad \gamma =
\begin{pmatrix}
a&b\\
c&d\\
\end{pmatrix} \in \Gamma,
\end{align}
which satisfies
\begin{align}
J_k(\gamma_1\gamma_2,\tau) = J_k(\gamma_1,\gamma_2(\tau)) J_k(\gamma_2,\tau), \quad
\gamma_1, \gamma_2 \in \Gamma.
\end{align}
Especially, the modular forms for $\Gamma(N)$ further satisfy
 \begin{align}
f^i(\tau) \xrightarrow{h} J_k(\gamma,\tau) \rho^{ij}(h)f^j(\tau) = J_k(\gamma,\tau) f^i(\tau), \quad h \in \Gamma(N).
\end{align}
Since $S^2:\tau=-\mathbb{I}:\tau=\tau$, we obtain
\begin{align}
f^i(\tau) \xrightarrow{S^2} (-1)^k \rho^{ij}(S^2)f^j(\tau) = f^i(\tau), ~\Rightarrow~ (-1)^k\rho^{ij}(-\mathbb{I}) = \mathbb{I}.
\end{align}
This implies $\rho(-\mathbb{I})=-\mathbb{I}$ for $k=2\mathbb{Z}+1$ and $\rho(-\mathbb{I})=\mathbb{I}$ for $k=2\mathbb{Z}$.
Thus, the unitary matrix $\rho(\gamma)$ is a unitary representation of
\begin{align}
\Gamma'_N\equiv\Gamma/\Gamma(N) = \langle S,T|S^4=(ST)^3=T^N=\mathbb{I},~S^2T=TS^2 \rangle,
\end{align}
for odd weights, $k\in2\mathbb{Z}+1$, while it is a unitary representation of
\begin{align}
\Gamma_N\equiv \bar{\Gamma}/\bar{\Gamma}(N) = \langle S,T|S^2=(ST)^3=T^N=\mathbb{I} \rangle,
\end{align}
for even weights, $k\in2\mathbb{Z}$.
It is remarkable that $\Gamma_2\simeq S_3$, $\Gamma_3\simeq A_4$, $\Gamma_4\simeq S_4$ and $\Gamma_5\simeq A_5$ \cite{deAdelhartToorop:2011re}.
Also $\Gamma'_N$ is the double covering group of $\Gamma_N$ \footnote{See e.g., Refs.~\cite{Liu:2019khw,Novichkov:2020eep,Liu:2020akv,Liu:2020msy,Wang:2020lxk,Yao:2020zml}.}.
The unitary representation of $\Gamma_N$, $\rho$, satisfies the algebraic relations,
\begin{align}
\rho(S)^2 = [\rho(S)\rho(T)]^3 = \rho(T)^N = \mathbb{I}.
\end{align}
Similarly one of $\Gamma_N'$ satisfies
\begin{align}
\rho(S)^4 = [\rho(S)\rho(T)]^3 = \rho(T)^N = \mathbb{I}, \quad \rho(S^2)\rho(T)=\rho(T)\rho(S^2).
\end{align}

The modular forms can be extended to the half integer weights $k/2$.
We introduce the double covering group of $\Gamma$,
\begin{align}
\widetilde{\Gamma} \equiv \{[\gamma,\epsilon] | \gamma\in\Gamma,~\epsilon\in\{\pm1\}\}.
\end{align}
There are two generators,
\begin{align}
\widetilde{S} \equiv [S,1], \quad \widetilde{T} \equiv [T,1],
\end{align}
satisfying the algebraic relations,
\begin{align}
\widetilde{S}^2 = [-\mathbb{I},1] \equiv \widetilde{Z}, ~
\widetilde{S}^4 = (\widetilde{S}\widetilde{T})^3 = [\mathbb{I},-1] = \widetilde{Z}^2, ~
\widetilde{S}^8 = (\widetilde{S}\widetilde{T})^6 = [\mathbb{I},1] \equiv \mathbb{I} = \widetilde{Z}^4,~
\widetilde{Z}\widetilde{T} = \widetilde{T}\widetilde{Z}.
\end{align}
The normal subgroup of $\widetilde{\Gamma}$, $\widetilde{\Gamma}(N)$, corresponding to $\Gamma(N)$ of $\Gamma$ is given by
\begin{align}
\widetilde{\Gamma}(N) \equiv \{ [h,\epsilon]\in\widetilde{\Gamma}|h\in\Gamma(N),~\epsilon=1 \}.
\end{align}
The automorphy factor $\widetilde{J}_{k/2}(\widetilde{\gamma},\tau)$ is given by
\begin{align}
\widetilde{J}_{k/2}(\widetilde{\gamma},\tau) \equiv \epsilon^k J_{k/2}(\gamma,\tau) = \epsilon^k (c\tau+d)^{k/2}, \quad k\in\mathbb{Z}, \quad \widetilde{\gamma} =
\begin{bmatrix}\gamma = \begin{pmatrix}a&b\\c&d\\\end{pmatrix}, \epsilon\end{bmatrix} \in \widetilde{\Gamma},
\end{align}
where we take $(-1)^{k/2}=e^{-k\pi i/2}$.
Then the modular forms of half integer weights $k/2$ are transformed as (See e.g., \cite{Koblitz:1984,shimura,Duncan:2018wbw}.)
\begin{align}
f^i(\widetilde{\gamma}(\tau)) = \widetilde{J}_{k/2}(\widetilde{\gamma},\tau) \rho^{ij}(\widetilde{\gamma}) f^j(\tau), \quad \widetilde{\gamma}\in\widetilde{\Gamma}.
\end{align}
The modular forms for $\widetilde{\Gamma}(N)$ further satisfy
\begin{align}
f^i(\widetilde{h}(\tau)) = \widetilde{J}_{k/2}(\widetilde{h},\tau) \rho^{ij}(\widetilde{h}) f^j(\tau) = \widetilde{J}_{k/2}(\widetilde{h},\tau) f^i(\tau), \quad \widetilde{h}\in\widetilde{\Gamma}(N),
\end{align}
where $\rho$ is a unitary representation of the quotient group,
\begin{align}
\widetilde{\Gamma}_N\equiv \widetilde{\Gamma}/\widetilde{\Gamma}(N)
= \langle \widetilde{S}, \widetilde{T} | \widetilde{S}^2=\widetilde{Z},~\widetilde{S}^4=(\widetilde{S}\widetilde{T})^3=\widetilde{Z}^2,~\widetilde{T}^N=\widetilde{Z}^4=\mathbb{I},~\widetilde{Z}\widetilde{T}=\widetilde{T}\widetilde{Z}\rangle.
\end{align}
Additionally we show the behaviours of the modular forms of weight 1/2 at three symmertric points $\tau=i$, $\omega$ and $i\infty$.
Since these points are invariant under $S$, $ST$ and $T$-transformations, the modular forms which are the functions of $\tau$ have the following invariances at the symmetric points,
\begin{align}
\begin{aligned}
&f^i(i) = \widetilde{J}_{1/2}(\widetilde{S},i) \rho^{ij}(\widetilde{\gamma}) f^j(i), \\
&f^i(\omega) = \widetilde{J}_{1/2}(\widetilde{ST},\omega) \rho^{ij}(\widetilde{\gamma}) f^j(\omega), \\
&f^i(i\infty) = \widetilde{J}_{1/2}(\widetilde{T},i\infty) \rho^{ij}(\widetilde{\gamma}) f^j(i\infty).
\end{aligned} \label{eq:modular_invariances_of_modular_forms}
\end{align}

Before we end this section, we show the modular transformations of the zero-modes on $T^2$.
Under $S$ and $T$-transformations, the zero-modes $\psi^{(j+\alpha_1,\alpha_2),M}(z,\tau)$ are transformed as
\begin{align}
\psi^{(j+\alpha_1,\alpha_2),M}(z,\tau) &\xrightarrow{S}
\psi^{(j+\alpha_1,\alpha_2),M}(-z/\tau,-1/\tau) \notag \\
&= (-\tau)^{1/2} \rho^{jk}_{\alpha\alpha'}(S) \psi^{(k+\alpha_1',\alpha_2'),M}(z,\tau), \\
\psi^{(j+\alpha_1,\alpha_2),M}(z,\tau) &\xrightarrow{T}
\psi^{(j+\alpha_1,\alpha_2),M}(z,\tau+1) \notag \\
&= \rho^{jk}_{\alpha\alpha'}(T)\psi^{(k+\alpha_1',\alpha_2'),M}(z,\tau),
\end{align}
where
\begin{align}
&\rho^{jk}_{\alpha\alpha'}(S) =
\frac{e^{\pi i/4}}{\sqrt{M}} e^{\frac{2\pi i}{M} (j+\alpha_1)(k+\alpha_1')} \delta_{(\alpha_2,\alpha_1),(\alpha_1',\alpha_2')}, \label{eq:rho_S} \\
&\rho^{jk}_{\alpha\alpha'}(T) =
e^{\frac{\pi i}{M}(j+\alpha_1)^2} \delta_{(\alpha_1,\alpha_2-\alpha_1+\frac{M}{2}),(\alpha_1',\alpha_2')}. \label{eq:rho_T}
\end{align}
The automorphy factor in $S$-transformation, $(-\tau)^{1/2}$, means that the zero-modes behave as the modular forms of weight 1/2.
Here we focus on the case that $(\alpha_1,\alpha_2;M)=(0,0;2\mathbb{Z})$ and $(1/2,1/2;2\mathbb{Z}+1)$.
Obviously in these cases the zero-modes are closed under $S$ and $T$-transformations.
The unitary representation $\rho$ for $(\alpha_1,\alpha_2;M)=(0,0;2\mathbb{Z})$ satisfies the algebraic relations,
\begin{align}
\begin{aligned}
&\rho(\widetilde{S})^2 = \rho(\widetilde{Z}) = e^{\pi i/2}\mathbb{I},\\
&\rho(\widetilde{S})^4 = [\rho(\widetilde{S})\rho(\widetilde{T})]^3 = \rho(\widetilde{Z})^2=e^{\pi i}\mathbb{I},\\
&\rho(\widetilde{T})^{2M} = \rho(\widetilde{Z})^4 = \mathbb{I},\\
&\rho(\widetilde{Z})\rho(\widetilde{T}) = \rho(\widetilde{T})\rho(\widetilde{Z}),
\end{aligned} \label{eq:algebra00}
\end{align}
while one for $(\alpha_1,\alpha_2;M)=(1/2,1/2;2\mathbb{Z}+1)$ satisfies,
\begin{align}
\begin{aligned}
&\rho(\widetilde{S})^2 = \rho(\widetilde{Z}) = e^{\pi i/2}\mathbb{I},\\
&\rho(\widetilde{S})^4 = [\rho(\widetilde{S})\rho(\widetilde{T})]^3 = \rho(\widetilde{Z})^2=e^{\pi i}\mathbb{I},\\
&\rho(\widetilde{T})^M = e^{\pi i/4}\mathbb{I}, \\
&\rho(\widetilde{T})^{8M} = \rho(\widetilde{Z})^4 = \mathbb{I},\\
&\rho(\widetilde{Z})\rho(\widetilde{T}) = \rho(\widetilde{T})\rho(\widetilde{Z}).
\end{aligned} \label{eq:algebra11}
\end{align}
The formers fulfill the algebraic relations of $\widetilde{\Gamma}_{2M}$.
The laters fulfill ones of $\widetilde{\Gamma}_{8M}$.
Therefore the zero-modes on $T^2$ with $(\alpha_1,\alpha_2;M)=(0,0;2\mathbb{Z})$ behave as the modular forms of weight 1/2 for $\widetilde{\Gamma}_{2M}$; ones with $(\alpha_1,\alpha_2;M)=(1/2,/1/2;2\mathbb{Z}+1)$ behave as the modular forms of weight 1/2 for $\widetilde{\Gamma}_{8M}$.


\section{$T^2/\mathbb{Z}_2$ orbifold}
\label{subsec:T2/Z2 orbifold}

In this section we study magnetized $T^2/\mathbb{Z}_2$ twisted orbifold model.
As we will see, there are further possibilities realizing three-generation zero-modes.
Furthermore we can find flavor models realizing realistic quark and lepton flavor observables.


\subsection{Zero-modes on $T^2/\mathbb{Z}_2$}
\label{subsubsec:Zero-modes_on_T2/Z2}

The $T^2/\mathbb{Z}_2$ orbifold is obtained by further identifying the $\mathbb{Z}_2$ twisted point $-z$ with $z$, i.e. $z\sim-z$ \cite{Abe:2008fi,Abe:2013bca,Abe:2014noa,Kobayashi:2017dyu}.
The zero-modes on $T^2/\mathbb{Z}_2$ orbifold are required to satisfy
\begin{align}
\psi_{T^2/\mathbb{Z}_2^m}(-z) = e^{m\pi i} \psi_{T^2/\mathbb{Z}_2^m}(z), \quad m \in \mathbb{Z}_2, \label{eq:BC_Z2}
\end{align}
in addition to the boundary conditions on $T^2$ in Eq.~(\ref{eq:BC_psi}).
Here $m\in\mathbb{Z}_2$ denotes $\mathbb{Z}_2$ parity; $m=0$ is $\mathbb{Z}_2$ even modes and $m=1$ is odd modes.
Under $\mathbb{Z}_2$ twist $z\to-z$, the zero-modes on $T^2$ are transformed as
\begin{align}
\psi^{(j+\alpha_1,\alpha_2),M}(z) \to
\psi^{(j+\alpha_1,\alpha_2),M}(-z) = e^{-2\alpha_2\pi i}\psi^{(M-(j+\alpha_1),\alpha_2),M}(z).
\end{align}
Thus, the zero-modes on $T^2/\mathbb{Z}_2$ can be defined as
\begin{align}
\psi_{T^2/\mathbb{Z}_2^m}^{(j+\alpha_1,\alpha_2),M}(z)
&= {\cal N}^{(j+\alpha_1)} \left(\psi^{(j+\alpha_1,\alpha_2),M}(z)+e^{m\pi i}\psi^{(j+\alpha_1,\alpha_2),M}(-z)\right) \\
&= {\cal N}^{(j+\alpha_1)} \left(\psi^{(j+\alpha_1,\alpha_2),M}(z)+e^{(m-2\alpha_2)\pi i}\psi^{(M-(j+\alpha_1),\alpha_2),M}(z)\right) \\
&\equiv O^{jk,\alpha_1,\alpha_2,M}_m \psi^{(k+\alpha_1,\alpha_2),M}(z), \label{eq:O}
\end{align}
where
\begin{align}
{\cal N}^{(j+\alpha_1)} = \left\{
\begin{array}{ll}
1/2 & (j+\alpha_1=0,~M/2), \\
1/\sqrt{2} & (\textrm{otherwise}), \\
\end{array}
\right.
\end{align}
and
\begin{align}
O^{jk,\alpha_1,\alpha_2,M}_m = {\cal N}^{(j+\alpha_1)}(\delta_{j,k}+e^{(m-2\alpha_2)\pi i}\delta_{M-j-2\alpha_1,k}). 
\end{align}
Note that the consistency between $\mathbb{Z}_2$ twist identification in Eq.~(\ref{eq:BC_Z2}) and the torus boundary conditions in Eq.~(\ref{eq:BC_psi}) restrict SS phases to
\begin{align}
(\alpha_1,\alpha_2) = (0,0),~ (1/2,0),~ (0,1/2),~ (1/2,1/2).
\end{align}
Table \ref{tab:num_T2/Z2} shows the number of zero-modes on $T^2/\mathbb{Z}_2$ orbifold.
\begin{table}[H]
\centering
\begin{tabular}{|c||c|c|} \hline
($\mathbb{Z}_2$ parity $m$; $\alpha_1$, $\alpha_2$) & $M=\textrm{even}$ & $M=\textrm{odd}$ \\ \hline
$(0;0,0)$ & $\frac{M}{2}+1$ & $\frac{M+1}{2}$ \\
$(1;0,0)$ & $\frac{M}{2}-1$ & $\frac{M-1}{2}$ \\
$(0;1/2,0)$ & $\frac{M}{2}$ & $\frac{M+1}{2}$ \\
$(1;1/2,0)$ & $\frac{M}{2}$ & $\frac{M-1}{2}$ \\
$(0;0,1/2)$ & $\frac{M}{2}$ & $\frac{M+1}{2}$ \\
$(1;0,1/2)$ & $\frac{M}{2}$ & $\frac{M-1}{2}$ \\
$(0;1/2,1/2)$ & $\frac{M}{2}$ & $\frac{M-1}{2}$ \\
$(1;1/2,1/2)$ & $\frac{M}{2}$ & $\frac{M+1}{2}$ \\ \hline
\end{tabular}
\caption{The number of zero-modes on magnetized $T^2/\mathbb{Z}_2$.
$\mathbb{Z}_2$ parities 0 and 1 correspond to even and odd modes, repectively.}
\label{tab:num_T2/Z2}
\end{table}
We additionally show the values of $(M;m;\alpha_1,\alpha_2)$ giving three-generation zero-modes in Table \ref{tab:three_T2/Z2}.
\begin{table}[H]
\centering
\begin{tabular}{|c|c|} \hline
$(M;m;\alpha_1,\alpha_2)$ & Independent $j$ in $\psi^{(j+\alpha_1,\alpha_2),M}_{T^2/\mathbb{Z}_2^m}$ \\ \hline
$(4;0;0,0)$ & (0,1,2) \\
$(5;0;0,0)$ & (0,1,2) \\
$(7;1;0,0)$ & (1,2,3) \\
$(8;1;0,0)$ & (1,2,3) \\
$(5;0;1/2,0)$ & (0,1,2) \\
$(6;0;1/2,0)$ & (0,1,2) \\
$(6;1;1/2,0)$ & (0,1,2) \\
$(7;1;1/2,0)$ & (0,1,2) \\
$(5;0;0,1/2)$ & (0,1,2) \\
$(6;0;0,1/2)$ & (0,1,2) \\
$(6;1;0,1/2)$ & (1,2,3) \\
$(7;1;0,1/2)$ & (1,2,3) \\
$(6;0;1/2,1/2)$ & (0,1,2) \\
$(7;0;1/2,1/2)$ & (0,1,2) \\
$(5;1;1/2,1/2)$ & (0,1,2) \\
$(6;1;1/2,1/2)$ & (0,1,2) \\ \hline
\end{tabular}
\caption{The values of $(M;m;\alpha_1,\alpha_2)$ giving three-generation zero-modes on magnetized $T^2/\mathbb{Z}_2$.}
\label{tab:three_T2/Z2}
\end{table}
Yukawa couplings on $T^2/\mathbb{Z}_2$ orbifold are given by the overlap integral of zero-modes on $T^2/\mathbb{Z}_2$:
\begin{align}
Y^{jk\ell}_{T^2/\mathbb{Z}_2} &= g(2\textrm{Im}\tau)^{1/2}
\int_{T^2/\mathbb{Z}_2} dzd\bar{z} \psi^{(j+\alpha_{1L},\alpha_{2L}),M_L}_{T^2/\mathbb{Z}_2^{m_L}}(z) \cdot 
\psi^{(k+\alpha_{1R},\alpha_{2R}),M_R}_{T^2/\mathbb{Z}_2^{m_R}}(z) \cdot 
\left( \psi^{(\ell+\alpha_{1H},\alpha_{2H}),M_H}_{T^2/\mathbb{Z}_2^{m_H}}(z) \right)^*,
\end{align}
where
\begin{align}
m_L + m_R = m_H, \label{eq:m_L+m_R=m_H}
\end{align}
is required to make Yukawa couplings $\mathbb{Z}_2$ twist invariant.
Yukawa couplings on $T^2/\mathbb{Z}_2$ can be rewritten by ones on $T^2$ as
\begin{align}
Y^{jk\ell}_{T^2/\mathbb{Z}_2} &=
O^{jj',\alpha_{1L},\alpha_{2L},M_L}_{m_L} O^{kk',\alpha_{1R},\alpha_{2R},M_R}_{m_R} \left(O^{\ell\ell',\alpha_{1H},\alpha_{2H},M_H}_{m_H}\right)^* Y^{j'k'\ell'}_{T^2},
\end{align}
where we have used Eq.~(\ref{eq:O}).
Hereafter we denote Yukawa couplings on $T^2/\mathbb{Z}_2$ as $Y^{jk\ell}$ instead of $Y^{jk\ell}_{T^2/\mathbb{Z}_2}$.

We also show the modular symmetry in the zero-modes on $T^2/\mathbb{Z}_2$.
Under the modular transformation, the zero-modes on $T^2/\mathbb{Z}_2$ are transformed as
\begin{align}
&\psi_{T^2/\mathbb{Z}_2^m}^{(j+\alpha_1,\alpha_2),M}(z,\tau) \xrightarrow{S}
\psi_{T^2/\mathbb{Z}_2^m}^{(j+\alpha_1,\alpha_2),M}(-z/\tau, -1/\tau) =
(-\tau)^{1/2} \rho^{jk}_{\alpha\alpha'}(S) \psi_{T^2/\mathbb{Z}_2^m}^{(k+\alpha_1',\alpha_2'),M}(z,\tau), \\
&\psi_{T^2/\mathbb{Z}_2^m}^{(j+\alpha_1,\alpha_2),M}(z,\tau) \xrightarrow{T}
\psi_{T^2/\mathbb{Z}_2^m}^{(j+\alpha_1,\alpha_2),M}(z, \tau+1) =
\rho^{jk}_{\alpha\alpha'}(T) \psi_{T^2/\mathbb{Z}_2^m}^{(k+\alpha_1',\alpha_2'),M}(z,\tau),
\end{align}
where
\begin{align}
&\rho^{jk}_{\alpha\alpha'}(S) = \left\{
\begin{array}{ll}
{\cal N}^{(j+\alpha_1)} {\cal N}^{(k+\alpha_1')} \frac{4e^{\pi i/4}}{\sqrt{M}} \cos \left(\frac{2\pi}{M}(j+\alpha_1)(k+\alpha_1')\right) \delta_{(\alpha_2,\alpha_1),(\alpha_1',\alpha_2')}, & (m=0), \\
{\cal N}^{(j+\alpha_1)} {\cal N}^{(k+\alpha_1')} \frac{4ie^{\pi i/4}}{\sqrt{M}} \sin \left(\frac{2\pi}{M}(j+\alpha_1)(k+\alpha_1')\right) \delta_{(\alpha_2,\alpha_1),(\alpha_1',\alpha_2')}, & (m=1), \label{eq:rho_S_T2/Z2} \\
\end{array}
\right. \\
&\rho^{jk}_{\alpha\alpha'}(T) = e^{\frac{\pi i}{M}(j+\alpha_1)^2} \delta_{(\alpha_1,\alpha_2-\alpha_1+\frac{M}{2}), (\alpha_1',\alpha_2')}. \label{eq:rho_T_T2/Z2}
\end{align}
Obviously only the zero-modes with $(\alpha_1,\alpha_2;M)=(0,0;2\mathbb{Z})$ and $(1/2,1/2;2\mathbb{Z}+1)$ are mapped into themselves, as in the case of $T^2$.
The unitary representation $\rho$ satisfies the algebraic relations in Eq.~(\ref{eq:algebra00}) for $(\alpha_1,\alpha_2;M)=(0,0;2\mathbb{Z})$ and Eq.~(\ref{eq:algebra11}) for $(1/2,1/2;2\mathbb{Z}+1)$.


\subsection{Zero points of zero-modes}
\label{subsec:Zero points of zero-modes}

Next let us see zero points of zero-modes on $T^2/\mathbb{Z}_2$.
Zero points are the coordinates such that all zero-mode wave functions vanish, $\psi^{(j+\alpha_1,\alpha_2),M}_{T^2/\mathbb{Z}_2^m}(z)=0$ for all of $j$.
Among them, we focus on zero points at the fixed points which are the invariant points under $\mathbb{Z}_2$ twist up to lattice translations of torus.
As we will see in Subsection \ref{subsubsec:favorable}, such zero points have important roles on realizing flavor structures in the magnetized $T^2/\mathbb{Z}_2$ model.
The fixed points on $T^2/\mathbb{Z}_2$ are as follows:
\begin{align}
P_F = \left\{
0,~\frac{1}{2},~\frac{\tau}{2},~\frac{1+\tau}{2}
\right\}.
\end{align}
It is easy to check that these points have $\mathbb{Z}_2$ twist invariance up to lattice translations.

In the following, we study which zero points the zero-modes $\psi^{(j+\alpha_1,\alpha_2),M}_{T^2/\mathbb{Z}_2^m}(z)$ have.
It depends on the values of SS phases $(\alpha_1,\alpha_2)$, flux $M$ and $\mathbb{Z}_2$ parity of zero-modes.
We start from the boundary conditions at the fixed points to find zero points.
It follows from zero-mode formula in Eq.~(\ref{eq:zero-mode_formula}) that
\begin{align}
\psi^{(j+\alpha_1,\alpha_2)}_{T^2}\left(z+\frac{n_1+n_2\tau}{2}\right)
= e^{\pi i(j+\alpha_1)n_1}e^{\frac{\pi iM}{4}n_1n_2} e^{\frac{\pi iM}{2}\frac{\textrm{Im}(n_1+n_2\bar{\tau})z}{\textrm{Im}\tau}}
\cdot \psi^{(j+\alpha_1+\frac{M}{2}n_2,\alpha_2+\frac{M}{2}n_1),M}_{T^2}(z),
\end{align}
where $n_1,n_2\in\mathbb{Z}$.
In addition, we can find
\begin{align}
&(-1)^{m-2\alpha_2} \psi^{(M-(j+\alpha_1),\alpha_2),M}_{T^2}\left(z+\frac{n_1+n_2\tau}{2}\right) \notag \\
&= e^{\pi i(j+\alpha_1)n_1} e^{\frac{\pi iM}{4}n_1n_2} e^{\frac{\pi iM}{2} \frac{\textrm{Im}(n_1+n_2\bar{\tau})z}{\textrm{Im}\tau}}
(-1)^{(m-Mn_1n_2-2(\alpha_1n_1+\alpha_2n_2))-2(\alpha_2+\frac{M}{2}n_1)}
\psi^{(M-(j+\alpha_1+\frac{M}{2}n_2),\alpha_2+\frac{M}{2}n_1),M}_{T^2}(z).
\end{align}
Combining these boundary conditions at the fixed points, we obtain ones of zero-modes on $T^2/\mathbb{Z}_2$:
\begin{align}
\psi^{(j+\alpha_1,\alpha_2),M}_{T^2/\mathbb{Z}_2^m}\left(z+\frac{n_1+n_2\tau}{2}\right)
= e^{\pi i(j+\alpha_1)n_1} e^{\frac{\pi iM}{4}n_1n_2} e^{\frac{\pi iM}{2} \frac{\textrm{Im}(n_1+n_2\bar{\tau})z}{\textrm{Im}\tau}}
\psi^{(j+\alpha_1+\frac{M}{2}n_2,\alpha_2+\frac{M}{2}n_1),M}_{T^2/\mathbb{Z}_2^{m'}}(z), \label{eq:BC_fixed}
\end{align}
where
\begin{align}
m' = m-Mn_1n_2-2(\alpha_1n_1+\alpha_2n_2) \quad (\textrm{mod}~2).
\end{align}
We note that again $m$ and $m'$ denote $\mathbb{Z}_2$ parity for $\mathbb{Z}_2$ twist; $m (m') = 0$ and $1$ stand for even and odd modes, respectively.
Substituting $z=0$, the relations of the zero-modes between the fixed points are obtained,
\begin{align}
\psi^{(j+\alpha_1,\alpha_2),M}_{T^2/\mathbb{Z}_2^m} \left(\frac{n_1+n_2\tau}{2}\right)
= e^{\pi i(j+\alpha_1)n_1} e^{\frac{\pi iM}{4}n_1n_2} \psi^{(j+\alpha_1+\frac{M}{2}n_2,\alpha_2+\frac{M}{2}n_1),M}_{T^2/\mathbb{Z}_2^{m'}}(0). \label{eq:fixedpoints-0}
\end{align}
This implies that the values of the zero-modes at the fixed points are given by in terms of ones at $z=0$.
Hence we should check whether the zero-modes at $z=0$ vanish or not to find which zero points zero-modes have.
From the definition of zero-modes in Eq.~(\ref{eq:O}), we find that the zero-modes at $z=0$ satisfy
\begin{align}
\psi^{(j+\alpha_1,\alpha_2),M}_{T^2/\mathbb{Z}_2^0}(0) \neq 0, \quad \psi^{(j+\alpha_1,\alpha_2),M}_{T^2/\mathbb{Z}_2^1}(0) = 0, \label{eq:zero-modes_z=0}
\end{align}
for all of $j$, $(\alpha_1,\alpha_2)$ and $M$.
Thus when the right-hand side in Eq.~(\ref{eq:fixedpoints-0}) has $\mathbb{Z}_2$ odd parity, that is,
\begin{align}
m' = m-Mn_1n_2-2(\alpha_1n_1+\alpha_2n_2) = 1~ (\textrm{mod}~2),
\end{align}
the left-hand side in Eq.~(\ref{eq:fixedpoints-0}) always vanishes.
Consequently we can find the zero points of zero-modes as shown in Table \ref{tab:zero_points}.
\begin{table}[H]
\centering
\begin{tabular}{ccc} \hline
$(m;\alpha_1,\alpha_2)$ & $M=2\mathbb{Z}$ & $M=2\mathbb{Z}+1$ \\ \hline
(0;0,0) & None & $\frac{1+\tau}{2}$ \\
(1;0,0) & $0,\frac{1}{2},\frac{\tau}{2},\frac{1+\tau}{2}$ & $0,\frac{1}{2},\frac{\tau}{2}$ \\
(0;1/2,0) & $\frac{1}{2}, \frac{1+\tau}{2}$ & $\frac{1}{2}$ \\
(1;1/2,0) & $0,\frac{\tau}{2}$ & $0,\frac{\tau}{2},\frac{1+\tau}{2}$ \\
(0;0,1/2) & $\frac{\tau}{2},\frac{1+\tau}{2}$ & $\frac{\tau}{2}$ \\
(1;0,1/2) & $0,\frac{1}{2}$ & $0,\frac{1}{2},\frac{1+\tau}{2}$ \\
(0;1/2,1/2) & $\frac{1}{2},\frac{\tau}{2}$ & $\frac{1}{2},\frac{\tau}{2},\frac{1+\tau}{2}$ \\
(1;1/2,1/2) & $0,\frac{1+\tau}{2}$ & $0$ \\ \hline
\end{tabular}
\caption{The zero points of each zero-mode at the fixed points.
$\mathbb{Z}_2$ parities 0 and 1 correspond to even and odd modes, respectively.}
\label{tab:zero_points}
\end{table}

We additionally consider the zero points of the derivative of the zero-modes on $T^2/\mathbb{Z}_2$.
Differentiating the boundary condition at the fixed points in Eq.~(\ref{eq:BC_fixed}), we obtain
\begin{align}
&\frac{\partial}{\partial z} \psi^{(j+\alpha_1,\alpha_2),M}_{T^2/\mathbb{Z}_2^m}\left(z+\frac{n_1+n_2\tau}{2}\right) \notag \\
&= e^{\pi i(j+\alpha_1)n_1} e^{\frac{\pi iM}{4}n_1n_2} e^{\frac{\pi iM}{2} \frac{\textrm{Im}(n_1+n_2\bar{\tau})z}{\textrm{Im}\tau}}
\left(\frac{\partial}{\partial z} + \frac{\pi M}{4\textrm{Im}\tau}(n_1+n_2\bar{\tau})\right) \psi^{(j+\alpha_1+\frac{M}{2}n_2,\alpha_2+\frac{M}{2}n_1),M}_{T^2/\mathbb{Z}_2^{m'}}(z).
\end{align}
Substituting $z=0$, we find
\begin{align}
\frac{\partial}{\partial z} \psi^{(j+\alpha_1,\alpha_2),M}_{T^2/\mathbb{Z}_2^m}\left(\frac{n_1+n_2\tau}{2}\right)
= e^{\pi i(j+\alpha_1)n_1} e^{\frac{\pi iM}{4}n_1n_2} 
\left(\frac{\partial}{\partial z} + \frac{\pi M}{4\textrm{Im}\tau}(n_1+n_2\bar{\tau})\right) \psi^{(j+\alpha_1+\frac{M}{2}n_2,\alpha_2+\frac{M}{2}n_1),M}_{T^2/\mathbb{Z}_2^{m'}}(0). \label{eq:derive_fixedpoints-0}
\end{align}
Notice that the derivatves of $\mathbb{Z}_2$ even and odd modes are $\mathbb{Z}_2$ odd and even modes, respectively.
Therefore the derivatves of $\mathbb{Z}_2$ even modes vanish at $z=0$ while ones of odd modes are nonzero,
\begin{align}
\frac{\partial}{\partial z} \psi^{(j+\alpha_1,\alpha_2),M}_{T^2/\mathbb{Z}_2^0}(z) = 0, \quad
\frac{\partial}{\partial z} \psi^{(j+\alpha_1,\alpha_2),M}_{T^2/\mathbb{Z}_2^1}(z) \neq 0, \label{eq:derive_zero-modes_z=0}
\end{align}
for all of $j$, $(\alpha_1,\alpha_2)$ and $M$.
Using the zero-modes at $z=0$ in Eq.~(\ref{eq:zero-modes_z=0}) and the derivatives of the zero-modes at $z=0$ in Eq.~(\ref{eq:derive_zero-modes_z=0}), it is found that either the first or second terms on the right-hand side in Eq.~(\ref{eq:derive_fixedpoints-0}) vanihies at $z=0$ but the remaining term does not vanish.
Thus only the derivatives of $\mathbb{Z}_2$ even modes vanish at $z=0$ whereas they does not vanish at $z=\frac{1}{2}, \frac{\tau}{2}$ and $\frac{1+\tau}{2}$.
The derivatives of $\mathbb{Z}_2$ odd modes does not vanish at all fixed points.
We summarize the result in Table \ref{tab:derive_zero_points}.
\begin{table}[H]
\centering
\begin{tabular}{ccc} \hline
$(m;\alpha_1,\alpha_2)$ & $M=2\mathbb{Z}$ & $M=2\mathbb{Z}+1$ \\ \hline
(0;0,0) & 0 & 0 \\
(1;0,0) & None & None \\
(0;1/2,0) & 0 & 0 \\
(1;1/2,0) & None & None \\
(0;0,1/2) & 0 & 0 \\
(1;0,1/2) & None & None \\
(0;1/2,1/2) & 0 & 0 \\
(1;1/2,1/2) & None & None \\ \hline
\end{tabular}
\caption{The zero points of the derivatives of each zero-mode at the fixed points.
$\mathbb{Z}_2$ parities 0 and 1 correspond to even and odd modes, respectively.}
\label{tab:derive_zero_points}
\end{table}
We comment on why the zero points of the derivatives of zero-modes are different from ones of the zero-modes.
This is due to the $z$ dependence of the boundary conditions in Eq.~(\ref{eq:BC_fixed}).
It leads to the second term in Eq.~(\ref{eq:derive_fixedpoints-0}) and zero points of the derivatives of zero-modes differ from ones of zero-modes.


\subsection{Three-generation models}
\label{subsubsec:three-generation-models}

Here we study three-generation models on $T^2/\mathbb{Z}_2$.
We classify all possible three-generation models with non-vanishing Yukawa couplings.
As we have seen in Subsection \ref{subsubsec:SYM}, the gauge symmetry $U(N)$ in the 10D $U(N)$ Super Yang-Mills theory is broken into $U(N_a)\times U(N_b)\times U(N_c)$ by the background magnetic fluxes.
Then nine types of the matter fields $\lambda$ appear as in Eq.~(\ref{eq:nine-bi-matter}).
We remind the following results regarding to the 2D spinors of $\lambda$.
\begin{itemize}
\item For $M_{\alpha\beta}>0$, $\alpha\beta$ sector with the positive chirality has $M_{\alpha\beta}$ degenerate solutions but $\beta\alpha$ sector with the positive chirality has no solution.
\item For $M_{\alpha\beta}>0$, $\alpha\beta$ sector with the negative chirality has no solution but $\beta\alpha$ sector with the negative chirality has $M_{\alpha\beta}$ degenerate solutions.
\item $\alpha\alpha$ sector has constant solution.
\end{itemize}
For example, when $M_{ab}\geq M_{bc}>0$, the 2D spinors $\psi_\pm$ have the solutions,
\begin{align}
&\psi_{+} = \left\{
\begin{array}{l}
\begin{pmatrix}
\psi^{aa}_{+} & \psi^{ab}_{+} & \psi^{ac}_{+} \\
\psi^{ba}_{+} & \psi^{bb}_{+} & \psi^{bc}_{+} \\
\psi^{ca}_{+} & \psi^{cb}_{+} & \psi^{cc}_{+} \\
\end{pmatrix}
=
\begin{pmatrix}
\textrm{const.} & \psi^{(j+\alpha_1^{ab},\alpha_2^{ab}),M_{ab}}_{T^2/\mathbb{Z}_2^{m_{ab}}} & \psi^{(k+\alpha_1^{ac},\alpha_2^{ac}),M_{ac}}_{T^2/\mathbb{Z}_2^{m_{ac}}} \\
\textrm{no~solution} & \textrm{const.} & \psi^{(\ell+\alpha_1^{bc},\alpha_2^{bc}),M_{bc}}_{T^2/\mathbb{Z}_2^{m_{bc}}} \\
\textrm{no~solution} & \textrm{no~solution} & \textrm{const.} \\
\end{pmatrix}
\end{array}
\right., \\
&\psi_{-} = \left\{
\begin{array}{l}
\begin{pmatrix}
\psi^{aa}_{-} & \psi^{ab}_{-} & \psi^{ac}_{-} \\
\psi^{ba}_{-} & \psi^{bb}_{-} & \psi^{bc}_{-} \\
\psi^{ca}_{-} & \psi^{cb}_{-} & \psi^{cc}_{-} \\
\end{pmatrix}
=
\begin{pmatrix}
\textrm{const.} & \textrm{no~solution} & \textrm{no~solution} \\
(\psi^{(-j-\alpha_1^{ab},\alpha_2^{ab}),M_{ab}}_{T^2/\mathbb{Z}_2^{m_{ab}}})^* & \textrm{const.} & \textrm{no~solution} \\
(\psi^{(-k-\alpha_1^{ac},\alpha_2^{ac}),M_{ac}}_{T^2/\mathbb{Z}_2^{m_{ac}}})^* & (\psi^{(-\ell-\alpha_1^{bc},\alpha_2^{bc}),M_{bc}}_{T^2/\mathbb{Z}_2^{m_{bc}}})^* & \textrm{const.} \\
\end{pmatrix}
\end{array}
\right.,
\end{align}
where $m_{ab}+m_{bc}=m_{ac}$ and $M_{ab}+M_{bc}=M_{ac}$.
In this case, $\lambda^{ab}(N_a,\bar{N}_b)$, $\lambda^{bc}(N_b,\bar{N}_c)$ and $\lambda^{ac}(\bar{N}_c,N_a)$ with positive chirality and $\lambda^{ba}(\bar{N}_a,N_b)$, $\lambda^{cb}(\bar{N}_b,N_c)$ and $\lambda^{ca}(N_c,\bar{N}_a)$ with negative chirality remain massless in the 4D effective theory after the chirality of the 10D gaugino fields is fixed.
Here we concentrate on this case and assume $\lambda$ has left (negative) chirality.

As a toy model, let us start from $U(8)$ gauge symmetry and choose $(N_a, N_b, N_c)=(4,2,2)$ to obtain the Pati-Salam gauge symmetry $SU(4)\times SU(2)_L\times SU(2)_R$ up to $U(1)$ factors, which contains the SM gauge symmetry $SU(3)_R\times SU(2)_L\times U(1)_Y$.
Note that some of the $U(1)$ part may be anomalous and massive according to the Green-Schwarz mechanism.
On this model, $\lambda^{ab}(4,\bar{2})$, $\lambda^{bc}(2,\bar{2})$ and $\lambda^{ac}(\bar{2},4)$ correspond to left-handed matter fields (quarks and leptons), (charge-conjugated) right-handed matter fields and higgsino fields.
Here we suppose supersymmetry at the $ab$, $bc$ and $ac$ sectors.
This means there are the same number of Higgs fields and higgsino fields.
Also supersymmetry ensures that tachyonic modes do not appear at tree levels.
Insteadly we can break $U(8)$ to $SU(3)_R\times SU(2)_L\times U(1)_Y$ up to $U(1)$ factors by suitable magnetic fluxes and $\mathbb{Z}_2$ twist orbifolding.
See for details of model building Refs.~ \cite{Abe:2014vza,Abe:2017gye}.

As we have seen in Subsection \ref{subsubsec:SYM}, the extra-dimensional parts of the gaugino fields are written by the products of the zero-modes on $T^2_i$, $i=1,2,3$.
When the extra-dimensions are given by $T^2_1/\mathbb{Z}_2 \times T^2_2/\mathbb{Z}_2 \times T^2_3/\mathbb{Z}_2$, they are written by ones on $T^2_i/\mathbb{Z}_2$, $i=1,2,3$.
As shown in Table \ref{tab:three_T2/Z2}, we can find various three-generation zero-modes on $T^2_i/\mathbb{Z}_2$.
When we realize three-generation zero-modes on $T^2_1/\mathbb{Z}_2$, a single zero-mode must appear on both $T^2_2$ and $T^2_3$ to obtain three-generation quarks and leptons.
Then three-generations of all fermions must be originated from $T^2_1/\mathbb{Z}_2$.
Otherwise Yukawa couplings are rank one matrix which is not favorable phenomenologically.
For example when three-generations of left- and right-handed matter fields are originated from $T^2_1/\mathbb{Z}_2$ and $T^2_2/\mathbb{Z}_2$, respectively, Yukawa couplings have the following rank one structure,
\begin{align}
Y^{jk}_{T^2_1/\mathbb{Z}_2 \times T^2_2/\mathbb{Z}_2 \times T^2_3/\mathbb{Z}_2}(\tau_1,\tau_2,\tau_3) = Y^j_{T^2_1/\mathbb{Z}_2}(\tau_1) \times Y^k_{T^2_2/\mathbb{Z}_2}(\tau_2) \times Y_{T^2_3/\mathbb{Z}_2}(\tau_3),
\end{align}
where $j$ denotes three-generation indices of left-handed matter fields, $k$ denotes ones of right-handed matter fields and we omit indices of higgsino fields.
On the other hand, when three-generations of left- and right-handed matter fields are originated from $T^2_1/\mathbb{Z}_2$, Yukawa couplings are generally rank three matrices:
\begin{align}
Y^{jk}_{T^2_1/\mathbb{Z}_2 \times T^2_2/\mathbb{Z}_2 \times T^2_3/\mathbb{Z}_2}(\tau_1,\tau_2,\tau_3) = Y^{jk}_{T^2_1/\mathbb{Z}_2}(\tau_1) \times Y_{T^2_2/\mathbb{Z}_2}(\tau_2) \times Y_{T^2_3/\mathbb{Z}_2}(\tau_3),
\end{align}
where $Y_{T^2_2/\mathbb{Z}_2}(\tau_2)$ and $Y_{T^2_3/\mathbb{Z}_2}(\tau_3)$ just contribute to the overall factor of Yukawa couplings.
Thus we should consider the case where three-generations of all fermions are originated from only $T^2_1/\mathbb{Z}_2$, and other orbifolds do not affect the generation structures.
In what follows, we concentrate on Yukawa couplings on $T^2/\mathbb{Z}_2$, $Y_{T^2_1/\mathbb{Z}_2}^{jk\ell}$,
\begin{align}
Y^{jk\ell}_{T^2_1/\mathbb{Z}_2}(\tau_1) = g(2\textrm{Im}\tau_1)^{1/2}\int_{T^2_1/\mathbb{Z}_2} dz_1d\bar{z}_1
\psi_L^j(z_1,\tau_1) \times \psi_R^k(z_1,\tau_1) \times \left(\psi_H^\ell(z_1,\tau_1)\right)^*,
\end{align}
where $\psi_L^j$, $\psi_R^k$ and $\psi_H^\ell$ denote zero-modes on $T^2_1/\mathbb{Z}_2$ of left-, right-handed matter fields and higgsino fields, respectively.

Yukawa couplings on $T^2/\mathbb{Z}_2$ are required to be $\mathbb{Z}_2$ twist even in total.
That is, Yukawa couplings must be the overlap integrals of three even modes or pairs of one even modes and two odd modes.
Otherwise they vanish.
All possible $\mathbb{Z}_2$ twist mode patterns of zero-modes are shown in Table \ref{tab:Z2twistpatterns}.
\begin{table}[H]
\centering
\begin{tabular}{c|ccc}\hline
Type & $\psi_L(\lambda^{ab})$ & $\psi_R(\lambda^{bc})$ & $\psi_H(\lambda^{ac})$ \\ \hline
I & even & even & even \\
II & odd & even & odd \\
II' & even & odd & odd \\
III & odd & odd & even \\ \hline
\end{tabular}
\caption{Possible $\mathbb{Z}_2$ twist mode patterns of zero-modes leading to non-vanishing Yukawa couplings for one torus.}
\label{tab:Z2twistpatterns}
\end{table}
Note that the type II is equivalent to type II' by flipping left- and right-handed matter fields.
These patterns are consistent with Eq.~(\ref{eq:m_L+m_R=m_H}).

Furthermore the definition of the fluxes in Eq.~(\ref{eq:chi_and_M}) requires the flux condition,
\begin{align}
|I_{ab}^i| \pm |I_{bc}^i| = |I_{ac}^i| \quad \rightarrow \quad M_{ab}^i \pm M_{bc}^i = M_{ac}^i. \label{eq:M_L+M_R=M_H}
\end{align}
Similarly, SS phases must satisfy
\begin{align}
(\alpha_1^{ab},\alpha_2^{ab}) + (\alpha_1^{bc},\alpha_2^{bc}) = (\alpha_1^{ac},\alpha_2^{ac}) \quad (\textrm{mod}~2). \label{eq:a_L+a_R=a_H}
\end{align}
Eq.~(\ref{eq:chi_and_M}) and these mean that when we choose the values of $(M^{\alpha\beta};m^{\alpha\beta},\alpha^{\alpha\beta}_1,\alpha^{\alpha\beta}_2)$ of $\psi_L$ and $\psi_R$ to make them three-generations, ones of higgsino $\psi_H$ are decided.
Hence, the generation number of higgsino as well as Higgs fields is automatically decided.
In Table \ref{tab:three-generation-models_T2/Z2}, we classify all possible three-generation models with non-vanishing Yukawa couplings satisfying $M_{ab}+M_{bc}=M_{ac}$.
\begin{table}[H]
\centering
\caption{Possible three-generation models with non-vanishing Yukawa couplings on $T^2/\mathbb{Z}_2$ twisted orbifold.
The first column shows the $\mathbb{Z}_2$ mode types shown in Table \ref{tab:Z2twistpatterns}.
The second, third and fourth columns show (flux; $\mathbb{Z}_2$ parity; SS phases) of $\psi_L$, $\psi_R$ and $\psi_H$, respectively.
The fifth column shows the number of Higgs fields.
We omit three-generation models which are equivalent to the models shown in this table by flipping left- and right-handed matter fields.
}
\label{tab:three-generation-models_T2/Z2}
\begin{tabular}{c|c|c|c|c} \hline
(type) & $(M_L;m_L;\alpha_{1L},\alpha_{2L})$ & $(M_R;m_R;\alpha_{1R},\alpha_{2R})$ & $(M_H;m_H;\alpha_{1H},\alpha_{2H})$ & $g_H$ \\ \hline
(I) & $(4;0;0,0)$ & $(4;0;0,0)$ & $(8;0;0,0)$ & 5 \\
(I) & $(4;0;0,0)$ & $(5;0;0,0)$ & $(9;0;0,0)$ & 5 \\
(II') & $(4;0;0,0)$ & $(7;1;0,0)$ & $(11;1;0,0)$ & 5 \\
(II') & $(4;0;0,0)$ & $(8;1;0,0)$ & $(12;1;0,0)$ & 5 \\
(I) & $(4;0;0,0)$ & $(5;0;1/2,0)$ & $(9;0;1/2,0)$ & 5 \\
(I) & $(4;0;0,0)$ & $(6;0;1/2,0)$ & $(10;0;1/2,0)$ & 5 \\
(II') & $(4;0;0,0)$ & $(6;1;1/2,0)$ & $(10;1;1/2,0)$ & 5 \\
(II') & $(4;0;0,0)$ & $(7;1;1/2,0)$ & $(11;1;1/2,0)$ & 5 \\
(I) & $(4;0;0,0)$ & $(5;0;0,1/2)$ & $(9;0;0,1/2)$ & 5 \\
(I) & $(4;0;0,0)$ & $(6;0;0,1/2)$ & $(10;0;0,1/2)$ & 5 \\
(II') & $(4;0;0,0)$ & $(6;1;0,1/2)$ & $(10;1;0,1/2)$ & 5 \\
(II') & $(4;0;0,0)$ & $(7;1;0,1/2)$ & $(11;1;0,1/2)$ & 5 \\
(I) & $(4;0;0,0)$ & $(6;0;1/2,1/2)$ & $(10;0;1/2,1/2)$ & 5 \\
(I) & $(4;0;0,0)$ & $(7;0;1/2,1/2)$ & $(11;0;1/2,1/2)$ & 5 \\
(II') & $(4;0;0,0)$ & $(5;1;1/2,1/2)$ & $(9;1;1/2,1/2)$ & 5 \\
(II') & $(4;0;0,0)$ & $(6;1;1/2,1/2)$ & $(10;1;1/2,1/2)$ & 5 \\
(I) & $(5;0;0,0)$ & $(5;0;0,0)$ & $(10;0;0,0)$ & 6 \\
(II') & $(5;0;0,0)$ & $(7;1;0,0)$ & $(12;1;0,0)$ & 5 \\
(II') & $(5;0;0,0)$ & $(8;1;0,0)$ & $(13;1;0,0)$ & 6 \\
(I) & $(5;0;0,0)$ & $(5;0;1/2,0)$ & $(10;0;1/2,0)$ & 5 \\
(I) & $(5;0;0,0)$ & $(6;0;1/2,0)$ & $(11;0;1/2,0)$ & 6 \\
(II') & $(5;0;0,0)$ & $(6;1;1/2,0)$ & $(11;1;1/2,0)$ & 5 \\
(II') & $(5;0;0,0)$ & $(7;1;1/2,0)$ & $(12;1;1/2,0)$ & 6 \\
(I) & $(5;0;0,0)$ & $(5;0;0,1/2)$ & $(10;0;0,1/2)$ & 5 \\
(I) & $(5;0;0,0)$ & $(6;0;0,1/2)$ & $(11;0;0,1/2)$ & 6 \\
(II') & $(5;0;0,0)$ & $(6;1;0,1/2)$ & $(11;1;0,1/2)$ & 5 \\
(II') & $(5;0;0,0)$ & $(7;1;0,1/2)$ & $(12;1;0,1/2)$ & 6 \\
(I) & $(5;0;0,0)$ & $(6;0;1/2,1/2)$ & $(11;0;1/2,1/2)$ & 5 \\
(I) & $(5;0;0,0)$ & $(7;0;1/2,1/2)$ & $(12;0;1/2,1/2)$ & 6 \\
(II') & $(5;0;0,0)$ & $(5;1;1/2,1/2)$ & $(10;1;1/2,1/2)$ & 5 \\
(II') & $(5;0;0,0)$ & $(6;1;1/2,1/2)$ & $(11;1;1/2,1/2)$ & 6 \\
(III) & $(7;1;0,0)$ & $(7;1;0,0)$ & $(14;0;0,0)$ & 8 \\
(III) & $(7;1;0,0)$ & $(8;1;0,0)$ & $(15;0;0,0)$ & 8 \\
(II) & $(7;1;0,0)$ & $(5;0;1/2,0)$ & $(12;1;1/2,0)$ & 6 \\ \hline
\end{tabular}
\end{table}
\begin{table}[H]
\centering
\begin{tabular}{c|c|c|c|c} \hline
(type) & $(M_L;m_L;\alpha_{1L},\alpha_{2L})$ & $(M_R;m_R;\alpha_{1R},\alpha_{2R})$ & $(M_H;m_H;\alpha_{1H},\alpha_{2H})$ & $g_H$ \\ \hline
(II) & $(7;1;0,0)$ & $(6;0;1/2,0)$ & $(13;1;1/2,0)$ & 6 \\
(III) & $(7;1;0,0)$ & $(6;1;1/2,0)$ & $(13;0;1/2,0)$ & 7 \\
(III) & $(7;1;0,0)$ & $(7;1;1/2,0)$ & $(14;0;1/2,0)$ & 7 \\
(II) & $(7;1;0,0)$ & $(5;0;0,1/2)$ & $(12;1;0,1/2)$ & 6 \\
(II) & $(7;1;0,0)$ & $(6;0;0,1/2)$ & $(13;1;0,1/2)$ & 6 \\
(III) & $(7;1;0,0)$ & $(6;1;0,1/2)$ & $(13;0;0,1/2)$ & 7 \\
(III) & $(7;1;0,0)$ & $(7;1;0,1/2)$ & $(14;0;0,1/2)$ & 7 \\
(II) & $(7;1;0,0)$ & $(6;0;1/2,1/2)$ & $(13;1;1/2,1/2)$ & 7 \\
(II) & $(7;1;0,0)$ & $(7;0;1/2,1/2)$ & $(14;1;1/2,1/2)$ & 7 \\
(III) & $(7;1;0,0)$ & $(5;1;1/2,1/2)$ & $(12;0;1/2,1/2)$ & 6 \\
(III) & $(7;1;0,0)$ & $(6;1;1/2,1/2)$ & $(13;0;1/2,1/2)$ & 6 \\
(III) & $(8;1;0,0)$ & $(8;1;0,0)$ & $(16;0;0,0)$ & 9 \\
(II) & $(8;1;0,0)$ & $(5;0;1/2,0)$ & $(13;1;1/2,0)$ & 6 \\
(II) & $(8;1;0,0)$ & $(6;0;1/2,0)$ & $(14;1;1/2,0)$ & 7 \\
(III) & $(8;1;0,0)$ & $(6;1;1/2,0)$ & $(14;0;1/2,0)$ & 7 \\
(III) & $(8;1;0,0)$ & $(7;1;1/2,0)$ & $(15;0;1/2,0)$ & 8 \\
(II) & $(8;1;0,0)$ & $(5;0;0,1/2)$ & $(13;1;0,1/2)$ & 6 \\
(II) & $(8;1;0,0)$ & $(6;0;0,1/2)$ & $(14;1;0,1/2)$ & 7 \\
(III) & $(8;1;0,0)$ & $(6;1;0,1/2)$ & $(14;0;0,1/2)$ & 7 \\
(III) & $(8;1;0,0)$ & $(7;1;0,1/2)$ & $(15;0;0,1/2)$ & 8 \\
(II) & $(8;1;0,0)$ & $(6;0;1/2,1/2)$ & $(14;1;1/2,1/2)$ & 7 \\
(II) & $(8;1;0,0)$ & $(7;0;1/2,1/2)$ & $(15;1;1/2,1/2)$ & 8 \\
(III) & $(8;1;0,0)$ & $(5;1;1/2,1/2)$ & $(13;0;1/2,1/2)$ & 6 \\
(III) & $(8;1;0,0)$ & $(6;1;1/2,1/2)$ & $(14;0;1/2,1/2)$ & 7 \\
(I) & $(5;0;1/2,0)$ & $(5;0;1/2,0)$ & $(10;0;0.0,0)$ & 6 \\
(I) & $(5;0;1/2,0)$ & $(6;0;1/2,0)$ & $(11;0;0.0,0)$ & 6 \\
(II') & $(5;0;1/2,0)$ & $(6;1;1/2,0)$ & $(11;1;0.0,0)$ & 5 \\
(II') & $(5;0;1/2,0)$ & $(7;1;1/2,0)$ & $(12;1;0.0,0)$ & 5 \\
(I) & $(5;0;1/2,0)$ & $(5;0;0,1/2)$ & $(10;0;1/2,1/2)$ & 5 \\
(I) & $(5;0;1/2,0)$ & $(6;0;0,1/2)$ & $(11;0;1/2,1/2)$ & 5 \\
(II') & $(5;0;1/2,0)$ & $(6;1;0,1/2)$ & $(11;1;1/2,1/2)$ & 6 \\
(II') & $(5;0;1/2,0)$ & $(7;1;0,1/2)$ & $(12;1;1/2,1/2)$ & 6 \\
(I) & $(5;0;1/2,0)$ & $(6;0;1/2,1/2)$ & $(11;0;0.0,1/2)$ & 6 \\
(I) & $(5;0;1/2,0)$ & $(7;0;1/2,1/2)$ & $(12;0;0.0,1/2)$ & 6 \\
(II') & $(5;0;1/2,0)$ & $(5;1;1/2,1/2)$ & $(10;1;0.0,1/2)$ & 5 \\
(II') & $(5;0;1/2,0)$ & $(6;1;1/2,1/2)$ & $(11;1;0.0,1/2)$ & 5 \\
(I) & $(6;0;1/2,0)$ & $(6;0;1/2,0)$ & $(12;0;0.0,0)$ & 7 \\
(II') & $(6;0;1/2,0)$ & $(6;1;1/2,0)$ & $(12;1;0.0,0)$ & 5 \\
(II') & $(6;0;1/2,0)$ & $(7;1;1/2,0)$ & $(13;1;0.0,0)$ & 6 \\ \hline
\end{tabular}
\end{table}
\begin{table}[H]
\centering
\begin{tabular}{c|c|c|c|c} \hline
(type) & $(M_L;m_L;\alpha_{1L},\alpha_{2L})$ & $(M_R;m_R;\alpha_{1R},\alpha_{2R})$ & $(M_H;m_H;\alpha_{1H},\alpha_{2H})$ & $g_H$ \\ \hline
(I) & $(6;0;1/2,0)$ & $(5;0;0,1/2)$ & $(11;0;1/2,1/2)$ & 5 \\
(II') & $(6;0;1/2,0)$ & $(7;1;0,1/2)$ & $(13;1;1/2,1/2)$ & 7 \\
(I) & $(6;0;1/2,0)$ & $(6;0;1/2,1/2)$ & $(12;0;0.0,1/2)$ & 6 \\
(I) & $(6;0;1/2,0)$ & $(7;0;1/2,1/2)$ & $(13;0;0.0,1/2)$ & 7 \\
(II') & $(6;0;1/2,0)$ & $(5;1;1/2,1/2)$ & $(11;1;0.0,1/2)$ & 5 \\
(II') & $(6;0;1/2,0)$ & $(6;1;1/2,1/2)$ & $(12;1;0.0,1/2)$ & 6 \\
(III) & $(6;1;1/2,0)$ & $(6;1;1/2,0)$ & $(12;0;0.0,0)$ & 7 \\
(III) & $(6;1;1/2,0)$ & $(7;1;1/2,0)$ & $(13;0;0.0,0)$ & 7 \\
(II) & $(6;1;1/2,0)$ & $(5;0;0,1/2)$ & $(11;1;1/2,1/2)$ & 6 \\
(II) & $(6;1;1/2,0)$ & $(6;0;0,1/2)$ & $(12;1;1/2,1/2)$ & 6 \\
(III) & $(6;1;1/2,0)$ & $(6;1;0,1/2)$ & $(12;0;1/2,1/2)$ & 6 \\
(III) & $(6;1;1/2,0)$ & $(7;1;0,1/2)$ & $(13;0;1/2,1/2)$ & 6 \\
(II) & $(6;1;1/2,0)$ & $(6;0;1/2,1/2)$ & $(12;1;0.0,1/2)$ & 6 \\
(II) & $(6;1;1/2,0)$ & $(7;0;1/2,1/2)$ & $(13;1;0.0,1/2)$ & 6 \\
(III) & $(6;1;1/2,0)$ & $(5;1;1/2,1/2)$ & $(11;0;0.0,1/2)$ & 6 \\
(III) & $(6;1;1/2,0)$ & $(6;1;1/2,1/2)$ & $(12;0;0.0,1/2)$ & 6 \\
(III) & $(7;1;1/2,0)$ & $(7;1;1/2,0)$ & $(14;0;0.0,0)$ & 8 \\
(II) & $(7;1;1/2,0)$ & $(5;0;0,1/2)$ & $(12;1;1/2,1/2)$ & 6 \\
(II) & $(7;1;1/2,0)$ & $(6;0;0,1/2)$ & $(13;1;1/2,1/2)$ & 7 \\
(III) & $(7;1;1/2,0)$ & $(6;1;0,1/2)$ & $(13;0;1/2,1/2)$ & 6 \\
(III) & $(7;1;1/2,0)$ & $(7;1;0,1/2)$ & $(14;0;1/2,1/2)$ & 7 \\
(II) & $(7;1;1/2,0)$ & $(6;0;1/2,1/2)$ & $(13;1;0.0,1/2)$ & 6 \\
(II) & $(7;1;1/2,0)$ & $(7;0;1/2,1/2)$ & $(14;1;0.0,1/2)$ & 7 \\
(III) & $(7;1;1/2,0)$ & $(5;1;1/2,1/2)$ & $(12;0;0.0,1/2)$ & 6 \\
(III) & $(7;1;1/2,0)$ & $(6;1;1/2,1/2)$ & $(13;0;0.0,1/2)$ & 7 \\
(I) & $(5;0;0,1/2)$ & $(5;0;0,1/2)$ & $(10;0;0,0.0)$ & 6 \\
(I) & $(5;0;0,1/2)$ & $(6;0;0,1/2)$ & $(11;0;0,0.0)$ & 6 \\
(II') & $(5;0;0,1/2)$ & $(6;1;0,1/2)$ & $(11;1;0,0.0)$ & 5 \\
(II') & $(5;0;0,1/2)$ & $(7;1;0,1/2)$ & $(12;1;0,0.0)$ & 5 \\
(I) & $(5;0;0,1/2)$ & $(6;0;1/2,1/2)$ & $(11;0;1/2,0.0)$ & 6 \\
(I) & $(5;0;0,1/2)$ & $(7;0;1/2,1/2)$ & $(12;0;1/2,0.0)$ & 6 \\
(II') & $(5;0;0,1/2)$ & $(5;1;1/2,1/2)$ & $(10;1;1/2,0.0)$ & 5 \\
(II') & $(5;0;0,1/2)$ & $(6;1;1/2,1/2)$ & $(11;1;1/2,0.0)$ & 5 \\
(I) & $(6;0;0,1/2)$ & $(6;0;0,1/2)$ & $(12;0;0,0.0)$ & 7 \\
(II') & $(6;0;0,1/2)$ & $(6;1;0,1/2)$ & $(12;1;0,0.0)$ & 5 \\
(II') & $(6;0;0,1/2)$ & $(7;1;0,1/2)$ & $(13;1;0,0.0)$ & 6 \\
(I) & $(6;0;0,1/2)$ & $(6;0;1/2,1/2)$ & $(12;0;1/2,0.0)$ & 6 \\
(I) & $(6;0;0,1/2)$ & $(7;0;1/2,1/2)$ & $(13;0;1/2,0.0)$ & 7 \\
(II') & $(6;0;0,1/2)$ & $(5;1;1/2,1/2)$ & $(11;1;1/2,0.0)$ & 5 \\ \hline
\end{tabular}
\end{table}
\begin{table}[H]
\centering
\begin{tabular}{c|c|c|c|c} \hline
(type) & $(M_L;m_L;\alpha_{1L},\alpha_{2L})$ & $(M_R;m_R;\alpha_{1R},\alpha_{2R})$ & $(M_H;m_H;\alpha_{1H},\alpha_{2H})$ & $g_H$ \\ \hline
(II') & $(6;0;0,1/2)$ & $(6;1;1/2,1/2)$ & $(12;1;1/2,0.0)$ & 6 \\
(III) & $(6;1;0,1/2)$ & $(6;1;0,1/2)$ & $(12;0;0,0.0)$ & 7 \\
(III) & $(6;1;0,1/2)$ & $(7;1;0,1/2)$ & $(13;0;0,0.0)$ & 7 \\
(II) & $(6;1;0,1/2)$ & $(6;0;1/2,1/2)$ & $(12;1;1/2,0.0)$ & 6 \\
(II) & $(6;1;0,1/2)$ & $(7;0;1/2,1/2)$ & $(13;1;1/2,0.0)$ & 6 \\
(III) & $(6;1;0,1/2)$ & $(5;1;1/2,1/2)$ & $(11;0;1/2,0.0)$ & 6 \\
(III) & $(6;1;0,1/2)$ & $(6;1;1/2,1/2)$ & $(12;0;1/2,0.0)$ & 6 \\
(III) & $(7;1;0,1/2)$ & $(7;1;0,1/2)$ & $(14;0;0,0.0)$ & 8 \\
(II) & $(7;1;0,1/2)$ & $(6;0;1/2,1/2)$ & $(13;1;1/2,0.0)$ & 6 \\
(II) & $(7;1;0,1/2)$ & $(7;0;1/2,1/2)$ & $(14;1;1/2,0.0)$ & 7 \\
(III) & $(7;1;0,1/2)$ & $(5;1;1/2,1/2)$ & $(12;0;1/2,0.0)$ & 6 \\
(III) & $(7;1;0,1/2)$ & $(6;1;1/2,1/2)$ & $(13;0;1/2,0.0)$ & 7 \\
(I) & $(6;0;1/2,1/2)$ & $(6;0;1/2,1/2)$ & $(12;0;0.0,0.0)$ & 7 \\
(I) & $(6;0;1/2,1/2)$ & $(7;0;1/2,1/2)$ & $(13;0;0.0,0.0)$ & 7 \\
(II') & $(6;0;1/2,1/2)$ & $(5;1;1/2,1/2)$ & $(11;1;0.0,0.0)$ & 5 \\
(II') & $(6;0;1/2,1/2)$ & $(6;1;1/2,1/2)$ & $(12;1;0.0,0.0)$ & 5 \\
(I) & $(7;0;1/2,1/2)$ & $(7;0;1/2,1/2)$ & $(14;0;0.0,0.0)$ & 8 \\
(II') & $(7;0;1/2,1/2)$ & $(5;1;1/2,1/2)$ & $(12;1;0.0,0.0)$ & 5 \\
(II') & $(7;0;1/2,1/2)$ & $(6;1;1/2,1/2)$ & $(13;1;0.0,0.0)$ & 6 \\
(III) & $(5;1;1/2,1/2)$ & $(5;1;1/2,1/2)$ & $(10;0;0.0,0.0)$ & 6 \\
(III) & $(5;1;1/2,1/2)$ & $(6;1;1/2,1/2)$ & $(11;0;0.0,0.0)$ & 6 \\
(III) & $(6;1;1/2,1/2)$ & $(6;1;1/2,1/2)$ & $(12;0;0.0,0.0)$ & 7 \\ \hline
\end{tabular}
\end{table}
In the same way, three-generation models with non-vanishing Yukawa couplings satisfying $M_{ab}-M_{bc}=M_{ac}$ can be classified.
However, such three-generation models have single Higgs field and it is difficult to realize flavor structures.
Therefore we ignore them hereafter.


\subsection{Phenomenologically favorable conditions}
\label{subsubsec:favorable}

Here we study quark and lepton flavor models on the magnetized $T^2/\mathbb{Z}_2$.
We consider all possible zero-mode assignments into left-handed quark doublets $Q=(u_L,d_L)^T$, right-handed up-sector (down-sector) quark singlets $u_R$ ($d_R$), left-handed lepton doublets $L=(\nu_L,e_L)^T$, right-handed neutrino (charged lepton) singlets $\nu_R$ ($e_R$), and up and down type Higgs fields $H_{u,d}$.
Here and hereafter we denote (flux; $\mathbb{Z}_2$ parity; SS phases) of the zero-modes assigned into each field of $f\in\{Q=(u_L,d_L)^T, u_R, d_R| L=(\nu_L,e_L)^T, \nu_R, e_R| H_u, H_d\}$ by $B_f$.
Additionally we denote the $j$th zero-mode of each field as $\psi_f^j$.
We note that the zero-modes of quarks and leptons must be three-generations; therefore (flux; $\mathbb{Z}_2$ parity; SS phases) of them are chosen from the values shown in Table \ref{tab:three_T2/Z2}.
Then mass matrices for up-sector quarks, down-sector quarks and charged leptons, $M_u$, $M_d$ and $M_e$ are given by
\begin{align}
M^{jk}_u = Y_u^{jk\ell} \langle H_u^\ell \rangle, \quad
M^{jk}_d = Y_d^{jk\ell} \langle H_d^\ell \rangle, \quad
M^{jk}_e = Y_e^{jk\ell} \langle H_d^\ell \rangle,
\end{align}
where $\langle H_{u,d}^\ell \rangle$ denote Higgs VEVs and
\begin{align}
Y^{jk\ell}_u &= g(2\textrm{Im}\tau)^{1/2} \int dzd\bar{z} \psi_{u_L}^j \cdot \psi_{u_R}^k \cdot (\psi_{H_u}^\ell)^*
= g(2\textrm{Im}\tau)^{1/2} \int dzd\bar{z} \psi_{Q}^j \cdot \psi_{u_R}^k \cdot (\psi_{H_u}^\ell)^*, \\
Y^{jk\ell}_d &= g(2\textrm{Im}\tau)^{1/2} \int dzd\bar{z} \psi_{d_L}^j \cdot \psi_{d_R}^k \cdot (\psi_{H_d}^\ell)^*
= g(2\textrm{Im}\tau)^{1/2} \int dzd\bar{z} \psi_{Q}^j \cdot \psi_{d_R}^k \cdot (\psi_{H_d}^\ell)^*, \\
Y^{jk\ell}_e &= g(2\textrm{Im}\tau)^{1/2} \int dzd\bar{z} \psi_{e_L}^j \cdot \psi_{e_R}^k \cdot (\psi_{H_d}^\ell)^*
= g(2\textrm{Im}\tau)^{1/2} \int dzd\bar{z} \psi_{L}^j \cdot \psi_{e_R}^k \cdot (\psi_{H_d}^\ell)^*.
\end{align}
On the other hand, the light neutrino mass matrix, $M_\nu$, can be induced through the seesaw mechanism:
\begin{align}
M_\nu = M_D M_{RR}^{-1} M_D^T,
\end{align}
where $M_{RR}$ is Majorana mass matrix of right-handed neutrinos and $M_D$ is Dirac mass matrix given by
\begin{align}
M_D^{jk} = Y_\nu^{jk\ell}\langle H_u^\ell \rangle,
\end{align}
with
\begin{align}
Y_\nu^{jk\ell} = g(2\textrm{Im}\tau)^{1/2} \int dzd\bar{z} \psi_{\nu_L}^j \cdot \psi_{\nu_R}^k \cdot (\psi_{H_u}^\ell)^*
= g(2\textrm{Im}\tau)^{1/2} \int dzd\bar{z} \psi_{L}^j \cdot \psi_{\nu_R}^k \cdot (\psi_{H_u}^\ell)^*.
\end{align}
In Appendix \ref{app:Majorana_D-brane}, we give a brief review of the Majorana mass terms of right-handed neutrinos induced by the D-brane instanton effects on the magnetized $T^2/\mathbb{Z}_2$ model.
Non-vanishing Yukawa couplings are obtained when (flux; $\mathbb{Z}_2$ parity; SS phases) for quarks and leptons satisfy
\begin{align}
B_Q + B_{u_R} = B_L + B_{\nu_R} = B_{H_u}, \quad
B_Q + B_{d_R} = B_L + B_{e_R} = B_{H_d}.
\end{align}
We assume the fluxes of the zero-modes of up and down type Higgs fields are larger than ones of quarks and leptons to obtain multi generation Higgs fields.
Moreover generation numbers of up and down type Higgs fields must be same to cancel the chiral anomaly in 4D supersymmetric models.
As a result, we find that 6,460 number of flavor models satisfy these conditions.
However, it is not clear which models are phenomenologically favorable among them.
Actually, there are some difficulties to realize realistic quark and lepton flavor observables and most of models are disfavored.
In what follows, we show the difficulties and find the conditions to avoid them.

\paragraph{Condition for up-sector quark masses}~\\
First we show the difficulty on realizing up-sector quark mass ratios.
Since up-sector quarks have large mass hierarchy, its mass matrix can be approximated as a rank one matrix,
\begin{align}
M_u^{jk} = Y_u^{jk\ell} \langle H_u^\ell \rangle &=
(U_L^u)^\dagger 
\begin{pmatrix}
m_u & & \\
& m_c & \\
& & m_t \\
\end{pmatrix}
U_R^u \\
&\propto (U_L^u)^\dagger
\begin{pmatrix}
{\cal O}(10^{-6}) & & \\
& {\cal O}(10^{-3}) & \\
& & 1 \\
\end{pmatrix}
U_R^u \\
&= (U_L^u)^\dagger
\begin{pmatrix}
0 & & \\
& 0 & \\
& & 1 \\
\end{pmatrix}
U_R^u + {\cal O}(10^{-3}),
\end{align}
where $U_L^u$ and $U_R^u$ are unitary matrices to diagonalize $M_u$.
This approximate rank one structure requires the direction of up type Higgs VEVs, $h_u^\ell$, such that
\begin{align}
\exists h^\ell_u~\textrm{s.t.}~Y_u^{jk\ell} \langle H_u^\ell \rangle = M_{\textrm{rank~one}} \quad (\textrm{condition~I}), \label{eq:condition_I}
\end{align}
where $M_{\textrm{rank~one}}$ denotes a rank one matrix.
If this condition is fulfilled, the mass hierarchy in the up-sector quarks can be realized by taking $\langle H_u^\ell \rangle = h_u^\ell + \varepsilon_u^\ell$ such that $\varepsilon_u/h_u \sim {\cal O}(\frac{m_c}{m_t}) \sim {\cal O}(10^{-3})$.

\paragraph{Condition for down-sector quark and charged lepton masses}~\\
Second let us see the difficulty on realizing both down-sector quark and charged lepton mass ratios.
The down-sector quark and charged lepton have large mass hierarchies and their mass matrices are approximated as rank one matrices,
\begin{align}
M_d^{jk} = Y_d^{jk\ell} \langle H_d^\ell \rangle &=
(U_L^u)^\dagger 
\begin{pmatrix}
m_d & & \\
& m_s & \\
& & m_b \\
\end{pmatrix}
U_R^d \\
&\propto (U_L^d)^\dagger
\begin{pmatrix}
{\cal O}(10^{-4}) & & \\
& {\cal O}(10^{-2}) & \\
& & 1 \\
\end{pmatrix}
U_R^d \\
&= (U_L^d)^\dagger
\begin{pmatrix}
0 & & \\
& 0 & \\
& & 1 \\
\end{pmatrix}
U_R^d + {\cal O}(10^{-2}),
\end{align}
\begin{align}
M_e^{jk} = Y_e^{jk\ell} \langle H_d^\ell \rangle &=
(U_L^e)^\dagger 
\begin{pmatrix}
m_e & & \\
& m_\mu & \\
& & m_\tau \\
\end{pmatrix}
U_R^e \\
&\propto (U_L^e)^\dagger
\begin{pmatrix}
{\cal O}(10^{-4}) & & \\
& {\cal O}(10^{-2}) & \\
& & 1 \\
\end{pmatrix}
U_R^e \\
&= (U_L^e)^\dagger
\begin{pmatrix}
0 & & \\
& 0 & \\
& & 1 \\
\end{pmatrix}
U_R^e + {\cal O}(10^{-2}),
\end{align}
where $U_L^d$ and $U_R^d$ are unitary matrices to diagonalize $M_d$; $U_L^e$ and $U_R^e$ are ones for $M_e$.
These approximate rank one structures require the direction of down type Higgs VEVs, $h_d^\ell$, such that
\begin{align}
\exists h_d^\ell~\textrm{s.t.}~&Y_d^{jk\ell} \langle H_d^\ell \rangle = M_{\textrm{Rank~one}}, \\
&Y_e^{jk\ell} \langle H_d^\ell \rangle = M_{\textrm{Rank~one}}, \quad (\textrm{condition~II}). \label{eq:condition_II}
\end{align}
If this condition is fulfilled, the mass hierarchies in the down-sector quarks and charged leptons can be realized by taking $\langle H_d^\ell \rangle = h_d^\ell + \varepsilon_d^\ell$ such that $\varepsilon_d/h_d \sim {\cal O}(\frac{m_s}{m_b}) \sim {\cal O}(10^{-2})$.

\paragraph{Condition for quark mixing}~\\
Third we study the difficulty on realizing quark mixings.
The quarks have small mixing angles.
Indeed, the absolute values of the CKM matrix elements are observed as
\begin{align}
|V_{\textrm{CKM}}| \equiv |(U_L^u)^\dagger U_L^d| =
\begin{pmatrix}
0.974 & 0.227 & 0.00361 \\
0.226 & 0.973 & 0.0405 \\
0.00854 & 0.0398 & 0.999 \\
\end{pmatrix}. \label{eq:exp_CKM}
\end{align}
This is approximately a unit matrix.
Therefore the following condition,
\begin{align}
U_L^u \sim U_L^d,
\end{align}
should be satisfied.
This can be realized when unitary matrices $u_{L,R}^{u,d}$ which diagonalize rank one matrices $Y_{u,d}^{jk\ell}\langle H_{u,d}^\ell \rangle$ as
\begin{align}
[(u_L^{u})^\dagger]^{jj'} Y^{j'k'\ell} \langle H_u^\ell \rangle [u_R^u]^{k'k} \propto
\begin{pmatrix}
0&&\\
&0&\\
&&1\\
\end{pmatrix}^{jk},
\end{align}
satisfy the following condition,
\begin{align}
u_L^u = u_L^d \quad (\textrm{condition~III}). \label{eq:condition_III}
\end{align}
Notice that when this condition is satisfied, we can take the basis where $u_L^u$, $u_L^d$, $u_R^u$ and $u_R^d$ are unit matrices.
In such basis, the quark mass matrices are given by
\begin{align}
M_u^{jk} &= Y_u^{jk\ell} \langle H_u^\ell \rangle 
= Y_u^{jk\ell}(h_u^\ell + \varepsilon_u^\ell) \propto
\begin{pmatrix}
0&&\\&0&\\&&1\\
\end{pmatrix}
+ {\cal O}\begin{pmatrix}\frac{m_c}{m_t}\end{pmatrix}, \\
M_d^{jk} &= Y_d^{jk\ell} \langle H_d^\ell \rangle 
= Y_d^{jk\ell}(h_d^\ell + \varepsilon_d^\ell) \propto
\begin{pmatrix}
0&&\\&0&\\&&1\\
\end{pmatrix}
+ {\cal O}\begin{pmatrix}\frac{m_s}{m_b}\end{pmatrix},
\end{align}
where we have assumed that the conditions I and II are satisfied.
These mass matrices imply that the unitary matrices $U_{L,R}^{u,d}$ which diagonalize $M_{u,d}$ can be estimated as
\begin{align}
&U_{L,R}^u \sim
\begin{pmatrix}
1&0&0\\
0&1&{\cal O}(\frac{m_c}{m_t})\\
0&{\cal O}(\frac{m_c}{m_t})&1\\
\end{pmatrix}
\begin{pmatrix}
1&0&{\cal O}(\frac{m_c}{m_t})\\
0&1&0\\
{\cal O}(\frac{m_c}{m_t})&0&1\\
\end{pmatrix}
\begin{pmatrix}
\ast&\ast&0\\
\ast&\ast&0\\
0&0&1\\
\end{pmatrix}
\sim
\begin{pmatrix}
\ast&\ast&{\cal O}(\frac{m_c}{m_t})\\
\ast&\ast&{\cal O}(\frac{m_c}{m_t})\\
{\cal O}(\frac{m_c}{m_t})&{\cal O}(\frac{m_c}{m_t})&1\\
\end{pmatrix}, \\
&U_{L,R}^d \sim
\begin{pmatrix}
1&0&0\\
0&1&{\cal O}(\frac{m_s}{m_b})\\
0&{\cal O}(\frac{m_s}{m_b})&1\\
\end{pmatrix}
\begin{pmatrix}
1&0&{\cal O}(\frac{m_s}{m_b})\\
0&1&0\\
{\cal O}(\frac{m_s}{m_b})&0&1\\
\end{pmatrix}
\begin{pmatrix}
\ast&\ast&0\\
\ast&\ast&0\\
0&0&1\\
\end{pmatrix}
\sim
\begin{pmatrix}
\ast&\ast&{\cal O}(\frac{m_s}{m_b})\\
\ast&\ast&{\cal O}(\frac{m_s}{m_b})\\
{\cal O}(\frac{m_s}{m_b})&{\cal O}(\frac{m_s}{m_b})&1\\
\end{pmatrix},
\end{align}
where $\ast$ stands for unestimated values.
Then the CKM matrix are estimated to be
\begin{align}
V_{\textrm{CKM}} \sim
\begin{pmatrix}
\ast&\ast&{\cal O}(\frac{m_s}{m_b})\\
\ast&\ast&{\cal O}(\frac{m_s}{m_b})\\
{\cal O}(\frac{m_s}{m_b})&{\cal O}(\frac{m_s}{m_b})&1\\
\end{pmatrix}
\sim
\begin{pmatrix}
\ast&\ast&{\cal O}(10^{-2})\\
\ast&\ast&{\cal O}(10^{-2})\\
{\cal O}(10^{-2})&{\cal O}(10^{-2})&1\\
\end{pmatrix}.
\end{align}
This estimation is consistent with the experimental results in Eq.~(\ref{eq:exp_CKM}).
Thus the condition III in Eq.~(\ref{eq:condition_III}) can partially realize small mixing angles in the CKM matrix.

\paragraph{Condition for lepton mixing}~\\
Fourth we study the difficulty on realizing lepton mixings.
The leptons have large mixing angles.
Indeed, the absolute values of the PMNS matrix elements are observed as
\begin{align}
|V_{\textrm{PMNS}}| =
\begin{pmatrix}
0.801\to0.845 & 0.513\to0.579 & 0.143\to0.156\\
0.232\to0.507 & 0.459\to0.694 & 0.629\to0.779\\
0.260\to0.526 & 0.470\to0.702 & 0.609\to0.763\\
\end{pmatrix}.
\end{align}
To find the condition to realize these values, we consider the light neutrino mass matrix under the condition I in Eq.~(\ref{eq:condition_I}),
\begin{align}
M_\nu^{jk} &= Y_\nu^{jj'm} \langle H_u^{m} \rangle [M_{RR}^{-1}]^{j'k'} (Y_\nu^{k'kn})^T \langle H_u^n\rangle \\
&= Y_\nu^{jj'm} h_u^m [M_{RR}^{-1}]^{j'k'} (Y_\nu^{k'kn})^T h_u^n
+ {\cal O}(\varepsilon_u) + {\cal O}(\varepsilon_u^2). \label{eq:mass_nu_ex}
\end{align}
Since we assume $\varepsilon_u/h_u\sim{\cal O}(\frac{m_c}{m_t})\sim{\cal O}(10^{-3})$, the first term is dominant in above.
The direction of $h_u$ is determined to fulfill the condition I and there are no parameters to be used for realizing lepton mixings.
Thus it is difficult to realize realistic lepton mixings by the first term in Eq.~(\ref{eq:mass_nu_ex}) unless $M_{RR}$ and $Y_\nu^{jk\ell}h_u^\ell$ have ideal structures.
This difficulty can be avoided when the following condition is satisfied,
\begin{align}
Y_u^{jk\ell} h_u^\ell = M_{\textrm{Rank~one}} \quad \Rightarrow \quad Y_\nu^{jk\ell} h_u^\ell = 0 \quad (\textrm{condition~IV}). \label{eq:condition_IV}
\end{align}
When this condition is fulfilled, the light neutrino mass matrix is given by
\begin{align}
M_\nu^{jk} = Y_\nu^{jj'm} \varepsilon_u^m [M_{RR}^{-1}]^{j'k'} (Y_\nu^{j'jn})^T \varepsilon_u^n.
\end{align}
The structure of this mass matrix can be controled by the directions of $\varepsilon_u$.
Therefore there are the possibilities to realize realistic lepton mixings by taking suitable directions of $\varepsilon_u$.

In Table \ref{tab:favorable_conditions}, we summarize all conditions we have proposed to realize realistic flavor observables.
\begin{table}[H]
\centering
\begin{tabular}{c|c} \hline
& Conditions \\ \hline
I & $\exists h_u^\ell~\textrm{s.t.}~Y_u^{jk\ell}h_u^\ell = \textrm{rank~one~matrix}$ \\
II & $\exists h_d^\ell~\textrm{s.t.}~Y_d^{jk\ell}h_d^\ell = \textrm{rank~one~matrix}$, and $Y_e^{jk\ell}h_d^\ell = \textrm{rank~one~matrix}$ \\
III & $u_L^u=u_L^d$ \\
IV & $Y_u^{jk\ell} h_u^\ell = \textrm{rank~one~matrix}~\Rightarrow~Y_\nu^{jk\ell}h_u^\ell=0$ \\ \hline
\end{tabular}
\caption{The conditions I, II, III and IV.}
\label{tab:favorable_conditions}
\end{table}
We regard these conditions as phenomenologically favorable conditions.
In the following, we study which models can fulfill these conditions.


\subsection{Zero point analysis}
\label{subsubsec:Zero_point_analysis}

To find which flavor models are phenomenologically favorable, we study the model constraints to satisfy the conditions I, II, III and IV which require the directions of Higgs VEVs, $h_{u,d}$, leading to rank one (for quarks and charged leptons) and vanishing mass matrices (for neutrinos).
Here we show such directions can be obtained in several cases by checking the zero points of zero-modes of each field.
In more detail, patterns of zero points which zero-modes of each field have determine whether the directions of Higgs VEVs $h_{u,d}$ satisfying the conditions I, II, III and IV exist or not.
The procedure is as follows.
First we start from Yukawa couplings between left-handed fermion zero-modes $\psi_L^j$, right-handed fermion zero-modes $\psi_R^k$, and Higgs field zero-modes $\psi_H^\ell$,
\begin{align}
Y^{jk\ell} = gy^{jk\ell} = g(2\textrm{Im}\tau)^{1/2} \int dzd\bar{z} \psi_L^j(z) \cdot \psi_R^k(z) \cdot (\psi_H^\ell(z))^*.
\end{align}
Then we consider the structures of Yukawa couplings for each zero point pattern of zero-modes.
As we will see, the zero point patterns have the information which linear combinations of Yukawa matrices lead to rank one or vanishing mass matrices.
Second we will construct unitary transformation matrices for Higgs field zero-modes which correspond to this linear combination.
Finally we classify the structures of mass matrices in each zero point pattern.
Here we denote sets of the zero points at the fixed points of $\psi_L^j$, $\psi_R^k$ and $\psi_H^\ell$ as $P_{\psi_L}$, $P_{\psi_R}$ and $P_{\psi_H}$; ones of the derivatives of $\psi_L^j$, $\psi_R^k$ and $\psi_H^\ell$ as $P'_{\psi_L}$, $P'_{\psi_R}$ and $P'_{\psi_H}$.

Yukawa couplings between $\psi_L^j$, $\psi_R^k$ and $\psi_H^\ell$ lead to the product expansion,
\begin{align}
\psi_L^j(z) \cdot \psi_R^k(z) = y^{jk\ell} \psi_H^\ell(z).
\end{align}
Let us focus on one point $z=p$ on $T^2/\mathbb{Z}_2$ (not necessary to be the fixed points),
\begin{align}
\psi_L^j(p) \cdot \psi_R^k(p) = y^{jk\ell} \psi_H^\ell(p).
\end{align}
We consider a unitary transformation for $\psi_L^j$ such as
\begin{align}
&\psi_L^j \to \hat{\psi}_L^j = U^{jj'}_{\psi_L}(p) \psi_L^{j'}, \\
&U_{\psi_L}^{jj'}(p) =
\begin{pmatrix}
\cos\theta_2 & 0 & -\sin\theta_2 \\
0 & 1 & 0 \\
\sin\theta_2 & 0 & \cos\theta_2 \\
\end{pmatrix}
\begin{pmatrix}
1 & 0 & 0 \\
0 & \cos\theta_1 & -\sin\theta_1 \\
0 & \sin\theta_1 & \cos\theta_1 \\
\end{pmatrix}
\begin{pmatrix}
e^{-i\alpha_0} & 0 & 0 \\
0 & e^{-i\alpha_1} & 0 \\
0 & 0 & e^{-i\alpha_2} & 0 \\
\end{pmatrix}, \label{eq:unitary_p}
\end{align}
where
\begin{align}
&\alpha_i = \left\{
\begin{array}{ll}
\textrm{arg} (\psi_L^j(p)) & \textrm{for}~p \notin P_{\psi_L}, \\
\textrm{arg} \left(\frac{\partial}{\partial z}\psi_L^j(p)\right) & \textrm{for}~p \in P_{\psi_L}, \\
\end{array}
\right. \\
&\theta_1 = \left\{
\begin{array}{ll}
\tan^{-1} \left|\frac{\psi^1_L(p)}{\psi^2_L(p)}\right| & \textrm{for}~p \notin P_{\psi_L}, \\
\tan^{-1} \left|\frac{\frac{\partial}{\partial z}\psi_L^1(p)}{\frac{\partial}{\partial z}\psi_L^2(p)}\right| & \textrm{for}~p \in P_{\psi_L}, \\
\end{array}
\right. \\
&\theta_2 = \left\{
\begin{array}{ll}
\tan^{-1} \frac{\left|\psi_L^0(p)\right|}{\left|\psi_L^1(p)\right|\sin\theta_1+\left|\psi_L^2(p)\right|\cos\theta_1} & \textrm{for}~p \notin P_{\psi_L}, \\
\tan^{-1} \frac{\left|\frac{\partial}{\partial z}\psi_L^0(p)\right|}{\left|\frac{\partial}{\partial z}\psi_L^1(p)\right|\sin\theta_1+\left|\frac{\partial}{\partial z}\psi_L^2(p)\right|\cos\theta_1} & \textrm{for}~p \in P_{\psi_L}. \\
\end{array}
\right.
\end{align}
After the unitary transformation for $p\notin P_{\psi_L}$, redefined zero-modes $\hat{\psi}_L^j(z)$ satisfy
\begin{align}
\hat{\psi}_L^0(p) = \hat{\psi}_L^1(p) = 0, \quad \hat{\psi}_L^2(p) \neq 0.
\end{align}
For $p\in P_{\psi_L}$, the derivatives of redefined zero-modes $\hat{\psi}_L^j(z)$ satisfy
\begin{align}
\frac{\partial}{\partial z}\hat{\psi}_L^0(p) = \frac{\partial}{\partial z}\hat{\psi}_L^1(p) = 0, \quad \frac{\partial}{\partial z}\hat{\psi}_L^2(p) \neq 0.
\end{align}
Note that when $p\notin P_{\psi_L}$, it is unknown whether the derivatives of redefined zero-modes $\frac{\partial}{\partial z}\hat{\psi}_L^j(z)$ vanish or not.
Similarly when $p\in P_{\psi_L}$, it is unknown whether redefined zero-modes $\hat{\psi}_L^j(z)$ vanish or not.
In the same way, we can obtain redefined zero-modes, $\hat{\psi}_R$ and $\hat{\psi}_H$, for $\psi_R$ and $\psi_H$ by unitary transformations $U_{\psi_R}(p)$ and $U_{\psi_H}(p)$.
They satisfy
\begin{align}
&\left\{
\begin{array}{ll}
\hat{\psi}_R^0(p) = \hat{\psi}_R^1(p) = 0, \quad \hat{\psi}_R^2(p) \neq 0 & \textrm{for}~p \notin P_{\psi_R}, \\
\frac{\partial}{\partial z}\hat{\psi}_R^0(p) = \frac{\partial}{\partial z}\hat{\psi}_R^1(p) = 0, \quad \frac{\partial}{\partial z}\hat{\psi}_R^2(p) \neq 0 & \textrm{for}~p \in P_{\psi_R}, \\
\end{array}
\right. \\
&\left\{
\begin{array}{ll}
\hat{\psi}_H^0(p) = \hat{\psi}_H^1(p) = \cdots = \hat{\psi}_H^{g_H-2}(p) = 0, \quad \hat{\psi}_H^{g_H-1}(p) \neq 0 & \textrm{for}~p \notin P_{\psi_H}, \\
\frac{\partial}{\partial z}\hat{\psi}_H^0(p) = \frac{\partial}{\partial z}\hat{\psi}_H^1(p) = \cdots = \frac{\partial}{\partial z}\hat{\psi}_H^{g_H-2}(p) = 0, \quad \frac{\partial}{\partial z}\hat{\psi}_H^{g_H-1}(p) \neq 0 & \textrm{for}~p \in P_{\psi_H}, \\
\end{array}
\right. \label{eq:redefined_Higgs}
\end{align}
where $g_H$ denotes generation number of Higgs fields.

Now we are ready to discuss the structures of Yukawa couplings for each zero point patterns.
We consider the product expansions of redefined zero-modes,
\begin{align}
\hat{\psi}_L^j(z) \cdot \hat{\psi}_R^k(z) = \hat{y}^{jk\ell} \hat{\psi}_H^\ell(z), \label{eq:PERZ}
\end{align}
where
\begin{align}
\hat{Y}^{jk\ell} = g\hat{y}^{jk\ell} = U_L^{jj'}(p) U_R^{kk'}(p) (U_H^{\ell\ell'}(p))^* Y^{j'k'\ell'}
= g(2\textrm{Im}\tau)^{1/2} \int dzd\bar{z}
\hat{\psi}_L^j(z) \cdot \hat{\psi}_R^k(z) \cdot (\hat{\psi}_H^\ell(z))^*.
\end{align}
Let us consider all possible zero point patterns for $\hat{\psi}_L^j$, $\hat{\psi}_R^k$ and $\hat{\psi}_H^\ell$.
Non-vanishing Yukawa coupling conditions in Eqs.~(\ref{eq:m_L+m_R=m_H}), (\ref{eq:M_L+M_R=M_H}) and (\ref{eq:a_L+a_R=a_H}) mean that when we choose the values of (flux; $\mathbb{Z}_2$ parity; SS phases) of $\hat{\psi}_L^j$ and $\hat{\psi}_R^k$ to make them three-generations, ones of $\hat{\psi}_H^\ell$ are determined.
Since zero points of zero-modes at the fixed points depend on the values of (flux; $\mathbb{Z}_2$ parity; SS phases) as classified in Table \ref{tab:zero_points}, we can find all possible zero point patterns from non-vanishing Yukawa coupling conditions.
Additionally, Table \ref{tab:zero_points} and \ref{tab:derive_zero_points} imply that when $p$ is in $P_{\psi_f}$, it is not in $P'_{\psi_f}$ for $f\in\{L,R,H\}$.
Consequently, we find the following zero point patterns are possible:
\begin{enumerate}
\item[(1)] $p\notin P_{\psi_L}$, $p\notin P_{\psi_R}$, $p\notin P_{\psi_H}$;
\item[(2)] $p\in P_{\psi_L}$, $p\in P_{\psi_R}$, $p\notin P_{\psi_H}$;
\item[(3)] $p\in P_{\psi_L}$, $p\notin P_{\psi_R}$, $p\in P_{\psi_H}$ ($p\notin P'_{\psi_L}$, $p\notin P'_{\psi_H}$);
\item[(4)] $p\notin P_{\psi_L}$, $p\in P_{\psi_R}$, $p\in P_{\psi_H}$ ($p\notin P'_{\psi_R}$, $p\notin P'_{\psi_H}$).
\end{enumerate}
In each pattern, we consider the structures of $\hat{Y}^{jk(g_H-1)}=U_{\psi_L}^{jj'}(p)U_{\psi_R}^{kk'}(p)(U_{\psi_H}^{(g_H-1)\ell}(p))^* Y^{j'k'\ell'}$ because we take the Higgs mode basis such that $(g_H-1)$th zero-mode of Higgs fields is non-vanishing at $p$ and others vanish.

\paragraph{Pattern (1) $\bm{p\notin P_{\psi_L}}$, $\bm{p\notin P_{\psi_R}}$, $\bm{p\notin P_{\psi_H}}$}~\\
Table \ref{tab:zero_points_pattern_(1)} shows the zero points of redefined zero-modes in this pattern.
\begin{table}[H]
\centering
\renewcommand{\arraystretch}{1.2}
\begin{tabularx}{150mm}{c|CCCCCCC} \hline
& $j=0$ & $1$ & $2$ & $3$ & $\cdots$ & $g_H-2$ & $g_H-1$ \\ \hline
$\hat{\psi}_L^j=U_{\psi_L}^{jk}\psi_L^k$ & $P_{\psi_L}$, $p$ & $P_{\psi_L}$, $p$ & $P_{\psi_L}$ & - & $\cdots$ & - & - \\
$\hat{\psi}_R^j=U_{\psi_R}^{jk}\psi_R^k$ & $P_{\psi_R}$, $p$ & $P_{\psi_R}$, $p$ & $P_{\psi_R}$ & - & $\cdots$ & - & - \\
$\hat{\psi}_H^j=U_{\psi_H}^{jk}\psi_H^k$ & $P_{\psi_H}$, $p$ & $P_{\psi_H}$, $p$ & $P_{\psi_H}$, $p$ & $P_{\psi_H}$, $p$ & $\cdots$ & $P_{\psi_H}$, $p$ & $P_{\psi_H}$ \\ \hline
\end{tabularx}
\caption{Zero points of redefined zero modes in pattern (1).}
\label{tab:zero_points_pattern_(1)}
\end{table}
Then the product expansion in Eq.~(\ref{eq:PERZ}) at $z=p$ leads to
\begin{align}
\underbrace{\hat{\psi}_L^j(p)}_{\propto \delta^{j,2}} \cdot \underbrace{\hat{\psi}_R^k(p)}_{\propto \delta^{k,2}} = \hat{y}^{jk\ell} \underbrace{\hat{\psi}_H^\ell(p)}_{\propto \delta^{\ell,(g_H-1)}} \quad &\Leftrightarrow \quad
\hat{Y}^{jk(g_H-1)} \propto \delta^{j,2}\delta^{k,2} \quad (\textrm{rank=one~matrix}) \\
&\Leftrightarrow \quad Y^{jk\ell} (U_{\psi_H}^{(g_H-1)}(p))^* = M_{\textrm{Rank~one}}. \label{eq:pattern_(1)}
\end{align}

\paragraph{Pattern (2) $\bm{p\in P_{\psi_L}}$, $\bm{p\in P_{\psi_R}}$, $\bm{p\notin P_{\psi_H}}$}~\\
Table \ref{tab:zero_points_pattern_(2)} shows the zero points of redefined zero-modes in this pattern.
\begin{table}[H]
\centering
\renewcommand{\arraystretch}{1.2}
\begin{tabularx}{150mm}{c|CCCCCCC} \hline
& $j=0$ & $1$ & $2$ & $3$ & $\cdots$ & $g_H-2$ & $g_H-1$ \\ \hline
$\hat{\psi}_L^j=U_{\psi_L}^{jk}\psi_L^k$ & $P_{\psi_L}$ & $P_{\psi_L}$ & $P_{\psi_L}$ & - & $\cdots$ & - & - \\
$\hat{\psi}_R^j=U_{\psi_R}^{jk}\psi_R^k$ & $P_{\psi_R}$ & $P_{\psi_R}$ & $P_{\psi_R}$ & - & $\cdots$ & - & - \\
$\hat{\psi}_H^j=U_{\psi_H}^{jk}\psi_H^k$ & $P_{\psi_H}$, $p$ & $P_{\psi_H}$, $p$ & $P_{\psi_H}$, $p$ & $P_{\psi_H}$, $p$ & $\cdots$ & $P_{\psi_H}$, $p$ & $P_{\psi_H}$ \\ \hline
\end{tabularx}
\caption{Zero points of redefined zero modes in pattern (2).}
\label{tab:zero_points_pattern_(2)}
\end{table}
Then the product expansion in Eq.~(\ref{eq:PERZ}) at $z=p$ leads to
\begin{align}
\underbrace{\hat{\psi}_L^j(p)}_{=0} \cdot \underbrace{\hat{\psi}_R^k(p)}_{=0} = \hat{y}^{jk\ell} \underbrace{\hat{\psi}_H^\ell(p)}_{\propto \delta^{\ell,(g_H-1)}} \quad &\Leftrightarrow \quad
\hat{Y}^{jk(g_H-1)} =0 \\
&\Leftrightarrow \quad Y^{jk\ell} (U_{\psi_H}^{(g_H-1)}(p))^* = 0.
\end{align}

\paragraph{Pattern (3) $\bm{p\in P_{\psi_L}}$, $\bm{p\notin P_{\psi_R}}$, $\bm{p\in P_{\psi_H}}$ ($\bm{p\notin P'_{\psi_L}}$, $\bm{p\notin P'_{\psi_H}}$)}~\\
Table \ref{tab:zero_points_pattern_(3)} shows the zero points of redefined zero-modes and their derivatives in this pattern.
\begin{table}[H]
\centering
\renewcommand{\arraystretch}{1.2}
\begin{tabularx}{162mm}{c|CCCCCCC} \hline
& $j=0$ & $1$ & $2$ & $3$ & $\cdots$ & $g_H-2$ & $g_H-1$ \\ \hline
$\hat{\psi}_L^j=U_{\psi_L}^{jk}\psi_L^k$ & $P_{\psi_L}$ & $P_{\psi_L}$ & $P_{\psi_L}$ & - & $\cdots$ & - & - \\
$\hat{\psi}_R^j=U_{\psi_R}^{jk}\psi_R^k$ & $P_{\psi_R}$, $p$ & $P_{\psi_R}$, $p$ & $P_{\psi_R}$ & - & $\cdots$ & - & - \\
$\hat{\psi}_H^j=U_{\psi_H}^{jk}\psi_H^k$ & $P_{\psi_H}$ & $P_{\psi_H}$ & $P_{\psi_H}$ & $P_{\psi_H}$ & $\cdots$ & $P_{\psi_H}$ & $P_{\psi_H}$ \\
$\frac{\partial}{\partial z}\hat{\psi}_L^j=U_{\psi_L}^{jk}\frac{\partial}{\partial z}\psi_L^k$ & $P'_{\psi_L}$, $p$ & $P'_{\psi_L}$, $p$ & $P'_{\psi_L}$ & - & $\cdots$ & - & - \\
$\frac{\partial}{\partial z}\hat{\psi}_H^j=U_{\psi_H}^{jk}\frac{\partial}{\partial z}\psi_H^k$ & $P'_{\psi_H}$, $p$ & $P'_{\psi_H}$, $p$ & $P'_{\psi_H}$, $p$ & $P'_{\psi_H}$, $p$ & $\cdots$ & $P'_{\psi_H}$, $p$ & $P'_{\psi_H}$ \\ \hline
\end{tabularx}
\caption{Zero points of redefined zero modes in pattern (3).}
\label{tab:zero_points_pattern_(3)}
\end{table}
Then the product expansion in Eq.~(\ref{eq:PERZ}) at $z=p$ give no information for $\hat{Y}^{jk(g_H-1)}$ since both left and righ-hand sides trivially vanish,
\begin{align}
\underbrace{\hat{\psi}_L^j(p)}_{=0} \cdot \underbrace{\hat{\psi}_R^k(p)}_{\propto \delta^{k,2}} = \hat{y}^{jk\ell} \underbrace{\hat{\psi}_H^\ell(p)}_{=0}.
\end{align}
Instead we consider the derivative of the product expansion in Eq.~(\ref{eq:PERZ}).
At $z=0$, it leads to
\begin{align}
&\underbrace{\frac{\partial}{\partial z} \hat{\psi}_L^j(p)}_{\propto \delta^{j,2}} \cdot \underbrace{\hat{\psi}_R^k(p)}_{\propto \delta^{k,2}}
+ \underbrace{\hat{\psi}_L^j(p)}_{=0} \cdot \frac{\partial}{\partial z} \hat{\psi}_R^k(p)
= \hat{y}^{jk\ell} \underbrace{\frac{\partial}{\partial z} \hat{\psi}_H^\ell(p)}_{\propto \delta^{\ell,(g_H-1)}} \notag \\
&\Leftrightarrow \quad \hat{Y}^{jk(g_H-1)} \propto \delta^{j,2} \delta^{k,2} \quad (\textrm{rank~one~matrix}) \\
&\Leftrightarrow \quad Y^{jk\ell} (U_{\psi_H}^{(g_H-1)\ell}(p))^* = M_{\textrm{Rank~one}}. \label{eq:pattern_(3)}
\end{align}

\paragraph{Pattern (4) $\bm{p\notin P_{\psi_L}}$, $\bm{p\in P_{\psi_R}}$, $\bm{p\in P_{\psi_H}}$ ($\bm{p\notin P'_{\psi_R}}$, $\bm{p\notin P'_{\psi_H}}$)}~\\
This pattern is flipping between $\psi_L$ and $\psi_R$ in the pattern (3).
Thus we obtain the same result as the pattern (3),
\begin{align}
Y^{jk\ell} (U_{\psi_H}^{(g_H-1)\ell}(p))^* = M_{\textrm{Rank~one}}.
\end{align}

In Eqs.~(\ref{eq:pattern_(1)}) and (\ref{eq:pattern_(3)}), we note that unitary transformations $U_{\psi_L}$ and $U_{\psi_R}$ do not affect the rank of the matrix.
Consequently we find the direction of Higgs VEVs, $h_{u,d}^\ell=v_{u,d}(U_{\psi_{H_{u,d}}}^{(g_H-1)\ell})^*$, leading to rank one fermion mass matrices in three patterns (1), (3) and (4).
Additionaly, it leads to vanishing fermion mass matrices in the pattern (2).
Reminds that the favorable condition I and II require the rank one fermion mass matrices while the condition IV requires the vanishing fermion mass matrices.
Therefore the rank one and vanishing mass matrices in four patterns can be used for realizing these conditions.


\subsection{Phenomenologically favorable models}

Here we classify all of quark and lepton flavor models satisfying the favorable conditions I, II, III and IV.
In what follows, we denote sets of the zero points at the fixed points of each field $f$ as $P_f$ for
\begin{align}
f \in \{Q=(u_L,d_L)^T,u_R,d_R|L=(\nu_L,e_L)^T,\nu_R,e_R|H_u,H_d\}.
\end{align}
As we will see soon, the sets of zero points of quarks, leptons and Higgs fields are restricted to realize the favorable conditions.

\paragraph{Condition I}~\\
The condition I requires the directions of up type Higgs VEVs leading to the rank one mass matrix for the up-sector quark,
\begin{align}
M_u = Y^{jk\ell}_u h_u^\ell = M_{\textrm{Rank~one}}.
\end{align}
According to Subsection \ref{subsubsec:Zero_point_analysis}, this is realized when $p_u$ in the following patterns exists,
\begin{align}
\exists p_u~\textrm{s.t.}~\left\{
\begin{array}{l}
\textrm{(1)}~p_u \notin P_{Q},~p_u \notin P_{u_R},~p_u\notin P_{H_u}, \\
\textrm{(3)}~p_u \in P_{Q},~p_u \notin P_{u_R},~p_u\in P_{H_u},~(p_u\notin P'_{Q},~p_u\notin P'_{H_u}), \\
\textrm{(4)}~p_u \notin P_{Q},~p_u \in P_{u_R},~p_u\in P_{H_u},~(p_u\notin P'_{u_R},~p_u\notin P'_{H_u}), \\
\end{array}
\right. \quad (\textrm{constraint~I}).
\end{align}

\paragraph{Condition II}~\\
The condition II requires the directions of down type Higgs VEVs leading to the rank one mass matrices for both down-sector quark and charged lepton,
\begin{align}
M_d = Y^{jk\ell}_d h_d^\ell = M_{\textrm{Rank~one}}, \quad
M_e = Y^{jk\ell}_e h_d^\ell = M_{\textrm{Rank~one}}.
\end{align}
According to Subsection \ref{subsubsec:Zero_point_analysis}, this is realized when $p_d$ in the following patterns exists,
\begin{align}
\exists p_d~\textrm{s.t.}~&\left\{
\begin{array}{l}
\textrm{(1)}~p_d \notin P_{Q},~p_d \notin P_{d_R},~p_d\notin P_{H_d}, \\
\textrm{(3)}~p_d \in P_{Q},~p_d \notin P_{d_R},~p_d\in P_{H_d},~(p_d\notin P'_{Q},~p_d\notin P'_{H_d}), \\
\textrm{(4)}~p_d \notin P_{Q},~p_d \in P_{d_R},~p_d\in P_{H_d},~(p_d\notin P'_{d_R},~p_d\notin P'_{H_d}), \\
\end{array}
\right. \quad (\textrm{constraint~II}_1), \\
\textrm{and}~&\left\{
\begin{array}{l}
\textrm{(1)}~p_d \notin P_{L},~p_d \notin P_{e_R},~p_d\notin P_{H_d}, \\
\textrm{(3)}~p_d \in P_{L},~p_d \notin P_{e_R},~p_d\in P_{H_d},~(p_d\notin P'_{L},~p_d\notin P'_{H_d}), \\
\textrm{(4)}~p_d \notin P_{L},~p_d \in P_{e_R},~p_d\in P_{H_d},~(p_d\notin P'_{e_R},~p_d\notin P'_{H_d}), \\
\end{array}
\right. \quad (\textrm{constraint~II}_2).
\end{align}
Note that it is not necessary to fulfill identical patterns in constraint II$_1$ and II$_2$.

\paragraph{Condition III}~\\
The condition III requires that unitary matrices $u_L^{u,d}$ which diagonalize rank one mass matrices $Y^{jk\ell}_{u,d}h_{u,d}^\ell$ are equal, $u_L^u=u_L^d$.
Assuming the condition I (constraint I) is satisfied, we can find rank one up-sector quark mass matrix,
\begin{align}
Y_u^{jk\ell} h_u^\ell = v_uY_u^{jk\ell} (U_{\psi_H}^{(g_H-1)\ell}(p_u))^*.
\end{align}
This can be diagonalized by
\begin{align}
U_{\psi_L}^{jj'}(p_u) [v_uY_u^{j'k'\ell}(U_{\psi_H}^{(g_H-1)\ell}(p_u))^*] U_{\psi_R}^{kk'}(p_u) \propto \delta^{j,2}\delta^{k,2} =
\begin{pmatrix}
0&&\\
&0&\\
&&1\\
\end{pmatrix},
\end{align}
for $(\psi_L,\psi_R,\psi_H)=(Q,u_L,H_u)$.
Here $U_{\psi_L}$, $U_{\psi_R}$ and $U_{\psi_H}$ are defined in Eq.~(\ref{eq:unitary_p}) and the sentence below.
Obviously $U_{\psi_L}(p_u)=U_{Q}(p_u)$ is equivalent to $u_L^u$ since it diagonalizes rank one mass matrix $Y_u^{jk\ell} h_u^\ell$.
In the same way, assuming the condition II (constraint II$_1$ and II$_2$) is satisfied, we can show $U_{\psi_L}(p_d)=U_Q(p_d)$ is equivalent to $u_L^d$.
Thus, the condition III is satisfied when $U_Q(p_u)=U_Q(p_d)$, that is,
\begin{align}
p_u = p_d \quad (\textrm{constraint~III}).
\end{align}

\paragraph{Condition IV}~\\
The condition IV requires that when the up-sector quark mass matrix is rank one matrix, the neutrino Dirac mass matrix vanishes.
That is, when the condition I is satisfied, the neutrino Dirac mass matrix must vanish,
\begin{align}
Y^{jk\ell}_u h_u^\ell = M_{\textrm{Rank~one}} \quad \Rightarrow \quad Y_\nu^{jk\ell} h_u^\ell = 0.
\end{align}
According to Subsection \ref{subsubsec:Zero_point_analysis}, this is realized when $p_d$ in the pattern (2) exists,
\begin{align}
\exists p_u~\textrm{s.t.}~\textrm{(2)}~p_u \in P_L,~p \in P_{\nu_R},~p \notin P_{H_u} \quad (\textrm{constraint~IV}).
\end{align}

In Table \ref{tab:Constraints}, we summarize all constraints.
\begin{table}[h]
  \begin{center}
  \renewcommand{\arraystretch}{1.2}
    \begin{tabular}{|c|c|c|c|c|c|c|c|c|} \hline
      & $P_Q$ & $P_{u_R}$ & $P_{d_R}$ & $P_{L}$ & $P_{\nu_R}$ & $P_{e_R}$ & $P_{H_u}$ & $P_{H_d}$ \\ \hline
      \multirow{3}{*}{I: $p_u$ is $\left\{ \begin{matrix} (1)\\(3)\\(4)\\\end{matrix}\right.$} & {\bf not in} & {\bf not in} & - & - & - & - & {\bf not in} & - \\ 
      & in & not in & - & - & - & - & in & - \\
      & not in & in & - & - & - & - & in & - \\ \hline
      \multirow{3}{*}{II$_1$: $p_d$ is $\left\{ \begin{matrix} (1)\\(3)\\(4)\\ \end{matrix} \right.$} & not in & - & not in & - & - & - & - & not in \\
      & in & - & not in & - & - & - & - & in \\
      & {\bf not in} & - & {\bf in} & - & - & - & - & {\bf in} \\ \hline
      \multirow{3}{*}{II$_2$: $p_d$ is $\left\{ \begin{matrix} (1)\\(3)\\(4)\\ \end{matrix} \right.$} & - & - & - & not in & - & not in & - & not in \\
      & - & - & - & {\bf in} & - & {\bf not in} & - & {\bf in} \\
      & - & - & - & not in & - & in & - & in \\ \hline
      III: $p_u=p_d$ & - & - & - & - & - & - & - & - \\ \hline
      IV: $p_u$ is & - & - & - & {\bf in} & {\bf in} & - & {\bf not in} & - \\ \hline
    \end{tabular}
  \end{center}
  \caption{The constraints I, II$_1$, II$_2$, III and IV.
For example, if $p_u$ corresponds to $(1)$, it is not included in $P_Q$.
If $p_u$ corresponds to $(3)$, it is included in $P_Q$.
  The bold texts denote the choices in Eq.~(\ref{eq:consistent_p}) which are consistent with all constraints.}
  \label{tab:Constraints}
\end{table}

Now we are ready to classify all of quark and lepton flavor models satisfying the constraints.
See Table \ref{tab:Constraints}.
Firstly the constraint III imposes $p_u=p_d \equiv p$.
Second the constraint IV requires
\begin{align}
p \in P_L, \quad p\in P_{\nu_R}, \quad p \notin P_{H_u}, \label{eq:constraint_IV_for_p}
\end{align}
and $p$ must be in $P_F$ since $P_L$, $P_{\nu_R}\subset P_F$.
Third, the constraint I which is consistent with the constraint IV requires that $p$ is in the pattern (1),
\begin{align}
p \notin P_Q, \quad p \notin P_{u_R}, \quad p \notin P_{H_u}, \label{eq:constraint_I_for_p}
\end{align}
because $p\notin P_{H_u}$ from Eq.~(\ref{eq:constraint_IV_for_p}).
Fourth, the constraint II$_2$ which is consistent with the constraint IV requires that $p$ is in the pattern (3),
\begin{align}
p \in P_L, \quad p \notin P_{e_R}, \quad p \in P_{H_d}, \label{eq:constraint_II2_for_p}
\end{align}
because $p\in P_L$ from Eq.~(\ref{eq:constraint_IV_for_p}).
Finally the constraint II$_1$ which is consistent with other constraints requires that $p$ is in the pattern (4),
\begin{align}
p \notin P_Q, \quad p \in P_{d_R}, \quad p \in P_{H_d}, 
\end{align}
because $p\notin P_Q$ from Eq.~(\ref{eq:constraint_I_for_p}) and $p\in P_{H_d}$ from Eq.~(\ref{eq:constraint_II2_for_p}).
As a result, the point $p$ which is consistent with all constraints must satisfy
\begin{align}
p \notin P_L \cup P_{d_R} \cup P_{\nu_R} \cup P_{H_d} \subset P_F, \quad
p \notin P_Q \cup P_{u_R} \cup P_{e_R} \cup P_{H_u} \subset P_F. \label{eq:consistent_p}
\end{align}
Thus, we can find quark and lepton flavor models satisfying the favorable conditions I, II, III and IV by checking whether zero points of the zero-modes of each field fulfill this condition, Eq.~(\ref{eq:consistent_p}), or not.
In Appendix \ref{app:Favorable_models}, we classify all quark and lepton flavor models satisfying Eq.~(\ref{eq:consistent_p}).
There are 408 flavor models in total.
Note that these models also satisfy non-vanishing Yukawa coupling conditions, Eqs.~(\ref{eq:m_L+m_R=m_H}), (\ref{eq:M_L+M_R=M_H}) and (\ref{eq:a_L+a_R=a_H}), and the anomaly cancellation condition which makes the generation number of up and down type Higgs fields the same.


\subsection{Modular symmetric models}

Here we study the flavor models which have a specific property under the modular transformation.
To calculate theoritical values of the flavor observables, we must identify two types of VEVs; one is the VEV of the modulus and another one is the VEVs of Higgs fields.
In the former, we assume the vacuum where the modulus lies on either of three modular symmetric points, $\tau=i$, $\omega$ and $i\infty$; as shown in Subsection \ref{subsubsec:modular_symmetry}, these points are invariant under $S$, $ST$ and $T$-transformations, respectively.
In the latter, we consider the direction of Higgs VEVs aligned in eigenbasis of $S$, $ST$ or $T$-transformations corresponding to each symmetric point.
In the following, we will show that some flavor models have the possibility to reproduce realistic flavor observables in a vicinity of $S$-symmetric vacuum but there are no consistent models for $ST$ and $T$-symmetric vacua.

First we consider the direction of Higgs VEVs under an assumption that the value of the modulus lies on either of $\tau=i$, $\omega$ and $i\infty$.
The direction of Higgs VEVs is aligned in the lightest mass direction.
It is known that supersymmetric mass term ($\mu$ term) of Higgs fields can be generated by D-brane instanton effects \cite{Blumenhagen:2006xt,Ibanez:2006da,Ibanez:2007rs,Antusch:2007jd,Kobayashi:2015siy}.
As we show in Appendix \ref{app:Higgs_mu_terms}, actually D-brane instanton effects yield the following Higgs $\mu$ term in a leading order:
\begin{align}
\mu^{jk} \varepsilon_{nm} H_{um}^j H_{dn}^k = \Lambda e^{-S_{\textrm{inst}}} (2\textrm{Im}\tau)^{-1} (Y_u^jY_d^k) \varepsilon_{nm} H_{um}^j H_{dn}^k \equiv d(\tau) (Y_u^jY_d^k) \varepsilon_{nm} H_{um}^j H_{dn}^k, \label{eq:mu_sec}
\end{align}
where $\Lambda$ stands for a typical scale such as the compactification scale and $S_{\textrm{inst}}$ is the instanton action.
Here $Y_u^j$ ($Y_d^k$) are three point couplings between instanton zero-modes $\alpha$, $\beta$ ($\gamma$) and Higgs fields $H_u^j$ ($H_d^k$),
\begin{align}
&Y_u^j = g(\textrm{Im}\tau)^{1/2} \int dzd\bar{z} \psi_{\alpha}(z) \cdot \psi_{\beta}(z) \cdot (\psi_{H_u}^j(z))^*, \\
&Y_d^k = g(\textrm{Im}\tau)^{1/2} \int dzd\bar{z} \psi_{\alpha}(z) \cdot \psi_{\gamma}(z) \cdot (\psi_{H_d}^k(z))^*,
\end{align}
where $\psi_\alpha$, $\psi_\beta$ and $\psi_\gamma$ are the zero-mode wave functions of instanton zero-modes $\alpha$, $\beta$ and $\gamma$; $\psi_{H_u}$ and $\psi_{H_d}$ are ones of up and down type Higgs fields $H_u$ and $H_d$.

As we have seen in Subsection \ref{subsubsec:modular_symmetry}, the zero-modes on $T^2/\mathbb{Z}_2$ behave as the modular forms of weight 1/2.
Hence, under the modular transformation, the zero-modes $\psi_{\alpha,\beta,\gamma}$ and $\psi_{H_{u,d}}$ are transformed as
\begin{align}
&\psi_{\alpha,\beta,\gamma} \to \widetilde{J}_{1/2}(\widetilde{\gamma},\tau) \rho_{\alpha,\beta,\gamma}(\widetilde{\gamma}) \psi_{\alpha,\beta,\gamma}, \\
&\psi_{H_{u,d}}^j \to \widetilde{J}_{1/2}(\widetilde{\gamma},\tau) \rho^{jk}_{H_{u,d}}(\widetilde{\gamma}) \psi^{k}_{H_{u,d}},
\end{align}
where $\rho_{\alpha,\beta,\gamma}$ and $\rho_{H_{u,d}}$ denote $1\times 1$ and $g_H\times g_H$ unitary representation matrices for $\alpha$, $\beta$, $\gamma$ and $H_{u,d}$.
Using these transformations, we obtain the modular transformation of three point couplings $Y_{u,d}$,
\begin{align}
&Y_u^j(\tau) \to \widetilde{J}_{1/2}(\widetilde{\gamma},\tau) \rho_\alpha(\widetilde{\gamma}) \cdot \rho_\beta(\widetilde{\gamma}) \cdot (\rho_{H_u}^{jk}(\widetilde{\gamma}))^* Y_u^k(\tau), \\
&Y_d^j(\tau) \to \widetilde{J}_{1/2}(\widetilde{\gamma},\tau) \rho_\alpha(\widetilde{\gamma}) \cdot \rho_\gamma(\widetilde{\gamma}) \cdot (\rho_{H_d}^{jk}(\widetilde{\gamma}))^* Y_d^k(\tau).
\end{align}
From the modular invariances shown in Eq.~(\ref{eq:modular_invariances_of_modular_forms}), at three symmetric points $\tau=i$, $\omega$ and $i\infty$, we find
\begin{align}
&Y_{u,d}^j(i) = \widetilde{J}_{1/2}(\widetilde{S},i) \rho_\alpha(\widetilde{S}) \cdot \rho_{\beta,\gamma}(\widetilde{S}) \cdot (\rho_{H_{u,d}}^{jk}(\widetilde{S}))^* Y_{u,d}^k(i), \\
&Y_{u,d}^j(\omega) = \widetilde{J}_{1/2}(\widetilde{ST},\omega) \rho_\alpha(\widetilde{T}) \cdot \rho_{\beta,\gamma}(\widetilde{T}) \cdot (\rho_{H_{u,d}}^{jk}(\widetilde{T}))^* Y_{u,d}^k(\omega), \\
&Y_{u,d}^j(i\infty) = \widetilde{J}_{1/2}(\widetilde{T},i\infty) \rho_\alpha(\widetilde{T}) \cdot \rho_{\beta,\gamma}(\widetilde{T}) \cdot (\rho_{H_{u,d}}^{jk}(\widetilde{T}))^* Y_{u,d}^k(i\infty).
\end{align}
Taking three point couplings at $\tau=i$, $\omega$ and $i\infty$ into eigenstates of $S$, $ST$ and $T$-transformations, respectively, the above relations are rewitten as
\begin{align}
&\hat{Y}_{u,d}^j(i) = \textrm{diag}\left(\widetilde{J}_{1/2}(\widetilde{S},i) \rho_\alpha(\widetilde{S}) \cdot \rho_{\beta,\gamma}(\widetilde{S}) \cdot (\rho_{H_{u,d}}^{jk}(\widetilde{S}))^*\right) \hat{Y}_{u,d}^k(i), \\
&\hat{Y}_{u,d}^j(\omega) = \textrm{diag}\left(\widetilde{J}_{1/2}(\widetilde{ST},\omega) \rho_\alpha(\widetilde{T}) \cdot \rho_{\beta,\gamma}(\widetilde{T}) \cdot (\rho_{H_{u,d}}^{jk}(\widetilde{T}))^*\right) \hat{Y}_{u,d}^k(\omega), \\
&\hat{Y}_{u,d}^j(i\infty) = \textrm{diag}\left(\widetilde{J}_{1/2}(\widetilde{T},i\infty) \rho_\alpha(\widetilde{T}) \cdot \rho_{\beta,\gamma}(\widetilde{T}) \cdot (\rho_{H_{u,d}}^{jk}(\widetilde{T}))^*\right) \hat{Y}_{u,d}^k(i\infty),
\end{align}
where $\hat{Y}_{u,d}(i)$, $\hat{Y}_{u,d}(\omega)$ and $\hat{Y}_{u,d}(i\infty)$ stand for three-point couplings in eigenstates of $S$, $ST$ and $T$-transformations, respectively.
These relations mean that only $j$th three point coupling satisfying
\begin{align}
\left\{
\begin{array}{ll}
\textrm{diag} \left(\widetilde{J}_{1/2}(\widetilde{S},i) \rho_\alpha(\widetilde{S}) \cdot \rho_{\beta,\gamma}(\widetilde{S}) \cdot (\rho_{H_{u,d}}^{jk}(\widetilde{S}))^*\right)^{jj} = 1 & \textrm{for}~\tau=i, \\
\textrm{diag} \left(\widetilde{J}_{1/2}(\widetilde{ST},\omega) \rho_\alpha(\widetilde{ST}) \cdot \rho_{\beta,\gamma}(\widetilde{ST}) \cdot (\rho_{H_{u,d}}^{jk}(\widetilde{ST}))^*\right)^{jj} = 1 & \textrm{for}~\tau=\omega, \\
\textrm{diag} \left(\widetilde{J}_{1/2}(\widetilde{T},i\infty) \rho_\alpha(\widetilde{T}) \cdot \rho_{\beta,\gamma}(\widetilde{T}) \cdot (\rho_{H_{u,d}}^{jk}(\widetilde{T}))^*\right)^{jj} = 1 & \textrm{for}~\tau=i\infty, \\
\end{array}
\right.
\end{align}
does not vanish at each symmetric point.
On the other hand, one satisfying
\begin{align}
\left\{
\begin{array}{ll}
\textrm{diag} \left(\widetilde{J}_{1/2}(\widetilde{S},i) \rho_\alpha(\widetilde{S}) \cdot \rho_{\beta,\gamma}(\widetilde{S}) \cdot (\rho_{H_{u,d}}^{jk}(\widetilde{S}))^*\right)^{jj} \neq 1 & \textrm{for}~\tau=i, \\
\textrm{diag} \left(\widetilde{J}_{1/2}(\widetilde{ST},\omega) \rho_\alpha(\widetilde{ST}) \cdot \rho_{\beta,\gamma}(\widetilde{ST}) \cdot (\rho_{H_{u,d}}^{jk}(\widetilde{ST}))^*\right)^{jj} \neq 1 & \textrm{for}~\tau=\omega, \\
\textrm{diag} \left(\widetilde{J}_{1/2}(\widetilde{T},i\infty) \rho_\alpha(\widetilde{T}) \cdot \rho_{\beta,\gamma}(\widetilde{T}) \cdot (\rho_{H_{u,d}}^{jk}(\widetilde{T}))^*\right)^{jj} \neq 1 & \textrm{for}~\tau=i\infty, \\
\end{array}
\right.
\end{align}
vanishes because it leads to
\begin{align}
\hat{Y}_{u,d} \neq \hat{Y}_{u,d}.
\end{align}
We call the formers as the invariant Yukawa couplings and the latters as the variant Yukawa couplings.
Notice that only the directions of Higgs fields corresponding to the invariant Yukawa couplings can have nonzero masses because $\mu$ term elements corresponding to the variant Yukawa couplings vanish.
In other words, nonzero masses are generated only for the Higgs fields which are invariant under $S$, $ST$ or $T$-transformations.
Thus the mass eigenstates of the leading order mass term at three symmetric points are given by $S$, $ST$ and $T$-transformations eigenstates, respectively.

In general, there would exist some configurations yielding a single instanton zero-mode; therefore the leading order mass term is generalized as
\begin{align}
\mu^{jk}(\tau) = \sum_a d_a Y_u^{ja}Y_d^{ka} \equiv \sum_a c_a \mu^{jk}_a(\tau),
\end{align}
where $a$ runs all possible configurations of instanton zero-mode.
Note that the mass eigenstates at three symmetric points are given by the eigenstates of the modular transformations as one configuration case.
Thus in the leading order, the Higgs VEVs which are aligned in the lightest mass direction lie on the eigenstates of $S$, $ST$ and $T$-transformations at three symmetric points $\tau=i$, $\omega$ and $i\infty$, respectively.

However, on the magnetized $T^2/\mathbb{Z}_2$ model, unfortunately we cannot find the leading order Higgs $\mu$ term being able to fix the lightest mass direction uniquely.
This is due to the shortage of number of instanton zero-mode configurations which couple to Higgs fields.
In what follows, we assume the directions of Higgs VEVs aligned along eigenstates of residual modular symmetry as the leading order although it is unknown the full order $\mu$ term structure.

Next, let us study the conditions to realize the Higgs VEVs which correspond to the eigenstates of the residual modular transformations at the symmetric points.
Then we ignore $T$-symmetric vacuum, $\tau=i\infty$, since the values of elements of Yukawa couplings at $\tau=i\infty$ are strictly restricted by $T$-symmetry and it is difficult to realize realistic flavor observables.
Furthermore, the symmetric point $\tau=i\infty$ corresponds to the decompactification limit, and it is not valid from the veiwpoint of 4D effective theory.

As shown in Subsection \ref{subsubsec:Zero-modes_on_T2/Z2}, the zero-modes on $T^2/\mathbb{Z}_2$ are mapped into themselves under the modular transformation when (flux; SS phases) of the zero-modes are ($2\mathbb{Z}$; 0,0) and ($2\mathbb{Z}+1$; 1/2,1/2).
Here we consider what zero-modes are closed under $S$ or $ST$-transformations to find the conditions realizing $S$ or $ST$-invariant Higgs VEVs.
Note that the modular transformation for Higgs VEVs are given by
\begin{align}
\langle H_{u,d}^\ell \rangle \to \widetilde{J}_{1/2}^*(\widetilde{\gamma},\tau) \rho^{\ell\ell'}(\widetilde{\gamma}) \langle H_{u,d}^{\ell'} \rangle,
\end{align}
since $(\psi_{H_{u,d}}^\ell)^* \langle H_{u,d}^\ell\rangle$ is modular invariant.
From the unitary transformation matrices in Eqs.~(\ref{eq:rho_S}) and (\ref{eq:rho_T}), $S$ and $ST$-transformations for zero-modes are closed when
\begin{align}
\left\{
\begin{array}{ll}
(\alpha_1, \alpha_2) = (0,0) \textrm{~or~} (1/2,1/2) & \textrm{for~}S\textrm{-transformation}, \\
(\alpha_1, \alpha_2) = (M/2,M/2)~(\textrm{mod}~1) & \textrm{for~}ST\textrm{-transformation}. \\
\end{array}
\right.
\end{align}
Thus, when SS phases of the zero-modes of Higgs fields satisfy the above conditions, Higgs VEVs can lie on the eigenstates of the residual modular transformation at the symmetric points.
However, we should consider the directions of Higgs VEVs, $h_{u,d}^\ell = v_{u,d}(U_{H_{u,d}}^{(g_H-1)\ell}(p))^*$, to realize realistic flavor observables.
That is, we should consider whether $h_{u,d}^\ell$ can be eigenstates of the residual modular transformations at the symmetric points or not.
The conditions for the modular eigenstates $h_{u,d}^\ell$ are given by
\begin{align}
\left\{
\begin{array}{ll}
p = 0 \textrm{~or~} \frac{1+i}{2} & \textrm{for~}S\textrm{-symmetric vaccum}, \\
p = 0 & \textrm{for~}ST\textrm{-symmetric vaccum}. \\
\end{array}
\right.
\end{align}
Let us prove this condition for $S$-symmetric vacuum.
When the non-vanishing redefined zero-modes of Higgs fields satisfying Eq.~(\ref{eq:redefined_Higgs}), $\hat{\psi}_{H_{u,d}}^{(g_H-1)}(z,\tau)=U_{H_{u,d}}^{(g_H-1)\ell}(p)\psi_{H_{u,d}}^\ell(z,\tau)$, are eigenstates of $S$-transformation, the direction $h_{u,d}^\ell = v_{u,d}(U_{H_{u,d}}^{(g_H-1)\ell}(p))^*$ becomes $S$-eigenstate at $\tau=i$.
This can be checked by calculating $S$-transformation of $\hat{\psi}_{H_{u,d}}^{(g_H-1)}(p,i)$ for $p\notin P_{H_{u,d}}$ and $\frac{\partial}{\partial z}\hat{\psi}_{H_{u,d}}^{(g_H-1)}(p,i)$ for $p\in P_{H_{u,d}}$.
Remind Eq.~(\ref{eq:redefined_Higgs}).
$S$-transformation for $\hat{\psi}_{H_{u,d}}^{(g_H-1)}(p,i)$ is given by
\begin{align}
\hat{\psi}_{H_{u,d}}^{(g_H-1)}(p,i) \xrightarrow{S} \hat{\psi}_{H_{u,d}}^{(g_H-1)}(S:(p,i)) &= \left\{
\begin{array}{ll}
\hat{\psi}_{H_{u,d}}^{(g_H-1)}(0,i) & \textrm{for}~p=0, \\
\hat{\psi}_{H_{u,d}}^{(g_H-1)}(\frac{-1+i}{2},i) & \textrm{for}~p=\frac{1+i}{2}, \\
\end{array}\right. \\
&= \left\{
\begin{array}{ll}
\hat{\psi}_{H_{u,d}}^{(g_H-1)}(0,i) & \textrm{for}~p=0, \\
e^{-2\pi i\alpha_{1H_{u,d}}}e^{-\pi i\frac{M_{H_{u,d}}}{2}}\hat{\psi}_{H_{u,d}}^{(g_H-1)}(\frac{1+i}{2},i) & \textrm{for}~p=\frac{1+i}{2}, \\
\end{array}\right.
\end{align}
where we have used the boundary condition in Eq.~(\ref{eq:BC_psi}) in the second row.
On the other hand, one for $\frac{\partial}{\partial z}\hat{\psi}_{H_{u,d}}^{(g_H-1)}(p,i)$ is given by
\begin{align}
  \textstyle\frac{\partial}{\partial z} \hat{\psi}^{(g_H-1)}_{H_{u,d}}(p,i) &\xrightarrow{S} (-i)\textstyle\frac{\partial}{\partial z} \hat{\psi}^{(g_H-1)}_{H_{u,d}}(S:(p,i)) \notag \\
  &=
  \left\{
  \begin{array}{l}
    (-i)\frac{\partial}{\partial z}\hat{\psi}^{(g_H-1)}_{H_{u,d}}(0,i) \quad {\rm for}~p=0, \\
    (-i)\frac{\partial}{\partial z}\hat{\psi}^{(g_H-1)}_{H_{u,d}}(\frac{-1+i}{2},i) \quad {\rm for}~p=\frac{1+i}{2}, \\
  \end{array}
  \right. \\
  &=
  \left\{
  \begin{array}{l}
    (-i)\frac{\partial}{\partial z}\hat{\psi}^{(g_H-1)}_{H_{u,d}}(0,i) \quad {\rm for}~p=0, \\
    (-i)e^{-2\pi i\alpha_{1H_{u,d}}}e^{-\pi i\frac{M_{H_{u,d}}}{2}} \frac{\partial}{\partial z}\hat{\psi}^{(g_H-1)}_{H_{u,d}}(\frac{1+i}{2},i) \quad {\rm for}~p=\frac{1+i}{2}, \\
  \end{array}
  \right.
\end{align}
where we have used the boundary condition in Eq.~(\ref{eq:BC_fixed}) in the third row.
Since the transformation law does not depend on $z$, these also consist for $z\neq p$,
\begin{align}
  &\hat{\psi}^{(g_H-1)}_{H_{u,d}}(z,i) \xrightarrow{S} 
  \left\{
  \begin{array}{l}
    \hat{\psi}^{(g_H-1)}_{H_{u,d}}(z,i) \quad {\rm for}~p=0\notin P_{H_{u,d}}, \\
    e^{-2\pi i\alpha_{1H_{u,d}}}e^{-\pi i\frac{M_{H_{u,d}}}{2}}\hat{\psi}^{(g_H-1)}_{H_{u,d}}(z,i) \quad {\rm for}~p=\frac{1+i}{2} \notin P_{H_{u,d}}, \\
  \end{array}
  \right.
  \\
  &\textstyle\frac{\partial}{\partial z} \hat{\psi}^{(g_H-1)}_{H_{u,d}}(z,i) \xrightarrow{S} 
  \left\{
  \begin{array}{l}
    (-i)\frac{\partial}{\partial z}\hat{\psi}^{(g_H-1)}_{H_{u,d}}(z,i) \quad {\rm for}~p=0 \in P_{H_{u,d}}, \\
    (-i)e^{-2\pi i\alpha_{1H_{u,d}}}e^{-\pi i\frac{M_{H_{u,d}}}{2}} \frac{\partial}{\partial z}\hat{\psi}^{(g_H-1)}_{H_{u,d}}(z,i) \quad {\rm for}~p=\frac{1+i}{2} \in P_{H_{u,d}}. \\
  \end{array} \label{eq:derivative_psi_S}
  \right.
\end{align}
Using $\frac{\partial}{\partial z}\xrightarrow{S}(-i)\frac{\partial}{\partial z}$, we can obtain $S$-transformation for $\hat{\psi}^{(g_H-1)}_{H_{u,d}}(z,i)$ instead of $\frac{\partial}{\partial z}\hat{\psi}^{(g_H-1)}_{H_{u,d}}(z,i)$ in Eq.~(\ref{eq:derivative_psi_S}).
Consequently we obtain
\begin{align}
  \hat{\psi}^{(g_H-1)}_{H_{u,d}}(z,i) \xrightarrow{S} 
  \left\{
  \begin{array}{l}
    \hat{\psi}^{(g_H-1)}_{H_{u,d}}(z,i) \quad {\rm for}~p=0\notin P_{H_{u,d}}, \\
    e^{-2\pi i\alpha_{1H_{u,d}}}e^{-\pi i\frac{M_{H_{u,d}}}{2}}\hat{\psi}^{(g_H-1)}_{H_{u,d}}(z,i) \quad {\rm for}~p=\frac{1+i}{2} \notin P_{H_{u,d}}, \\
    \hat{\psi}^{(g_H-1)}_{H_{u,d}}(z,i) \quad {\rm for}~p=0\in P_{H_{u,d}}, \\
    e^{-2\pi i\alpha_{1H_{u,d}}}e^{-\pi i\frac{M_{H_{u,d}}}{2}}\hat{\psi}^{(g_H-1)}_{H_{u,d}}(z,i) \quad {\rm for}~p=\frac{1+i}{2} \in P_{H_{u,d}}. \\
  \end{array}
  \right.
\end{align}
Thus $\hat{\psi}^{(g_H-1)}_{H_{u,d}}(z,i)=U_{H_{u,d}}^{(g_H-1)\ell}(p)\psi_{H_{u,d}}^\ell(z,i)$ with $p=0$ and $\frac{1+i}{2}$ are eigenstates of $S$-transformation.
That is, the direction $h^\ell_{u,d}=v_{u,d}(U_{H_{u,d}}^{(g_H-1)\ell}(p))^*$ with $p=0$ and $\frac{1+i}{2}$ is $S$-symmetric vacuum.
This is because that $z=0$ and $\frac{1+i}{2}$ are invariant under $S$-transformation up to the lattice transformation of torus.
On the other hand, it is difficult in the same way to show whether $\hat{\psi}^{(g_H-1)}_{H_{u,d}}(z,i)$ for $p=\frac{1}{2}$ and $\frac{i}{2}$ are eigenstates of $S$-transformation or not.
This is because that $z=\frac{1}{2}$ and $\frac{i}{2}$ are not invariant under $S$-transformation.
Instead, we adopt the direct calculation using $S$-transformation in Eq.~(\ref{eq:rho_S}).
For flavor models shown in Table \ref{tab:classification}, we check whether $\hat{\psi}^{(g_H-1)}_{H_{u,d}}(z,i)=U_{H_{u,d}}^{(g_H-1)\ell}(p)\psi_{H_{u,d}}^\ell(z,i)$ is an eigenstate of $S$-transformation or not.
As a result, there are no models where $\hat{\psi}^{(g_H-1)}_{H_{u,d}}(z,i)$ with $p=\frac{1}{2}$ and $\frac{i}{2}$ is eigenstates of $S$-tranformation.

So far we have discussed Higgs VEVs in $S$-symmetric vacuum.
Ones in $ST$-symmetric vaccum can be studied in the same way.
The direction $h^\ell_{u,d}$ at $ST$-invariant point $p=0$ is eigenstates of $ST$-transformation but one at $ST$-variant points $p\in \{\frac{1}{2},~\frac{\omega}{2},~\frac{1+\omega}{2}\}$ is not eigenstates of $ST$-transformation.
Thus the direction $h^\ell_{u,d}=v_{u,d}(U_{H_{u,d}}^{(g_H-1)\ell}(p))^*$ with $p=0$ is $ST$-symmetric vacuum.

Now we have two conditions for $S$ and $ST$-symmetric vacua.
The conditions for $S$-symmetric vacuum are
\begin{align}
(\alpha_1,\alpha_2) = (0,0) \textrm{~or~} (1/2,1/2), \quad p = 0 \textrm{~or~}\frac{1+i}{2}. \label{eq:conditions_S}
\end{align}
Ones for $ST$-symmetric vacuum are
\begin{align}
(\alpha_1,\alpha_2) = (M/2,M/2) \textrm{~(mod~1)}, \quad p = 0. \label{eq:conditions_ST}
\end{align}
We classify all flavor models satisfying these conditions in addition to the favorable condition in Eq.~(\ref{eq:consistent_p}).
Consequently we cannot find the flavor models satisfying the conditions for $ST$-symmetric vacuum in Eq.~(\ref{eq:conditions_ST}) while we can find the models satisfying the conditions for $S$-symmetric vacuum in Eq.~(\ref{eq:conditions_S}).
We show the results for $S$-symmetric vacuum in Table \ref{tab:classification_S_eig}.
There are 24 flavor models in total.
\begin{table}[H]
\begin{align}
  \begin{array}{c|c|c|c|c|c|c|c|c|c} \hline
    B_Q & B_{u_R} & B_{d_R} & B_L & B_{\nu_R} & B_{e_R} & B_{H_u} & B_{H_d} & g_H & p \\ 
\hline \hline
5,0,0,0 & 7,0,\frac{1}{2},\frac{1}{2} & 6,1,\frac{1}{2},\frac{1}{2} & 6,1,\frac{1}{2},0 & 6,1,0,\frac{1}{2} & 5,0,0,\frac{1}{2} & 12,0,\frac{1}{2},\frac{1}{2} & 11,1,\frac{1}{2},\frac{1}{2} & 6 & 0 \\
5,0,0,0 & 7,0,\frac{1}{2},\frac{1}{2} & 6,1,\frac{1}{2},\frac{1}{2} & 6,1,0,\frac{1}{2} & 6,1,\frac{1}{2},0 & 5,0,\frac{1}{2},0 & 12,0,\frac{1}{2},\frac{1}{2} & 11,1,\frac{1}{2},\frac{1}{2} & 6 & 0 \\
5,0,\frac{1}{2},0 & 6,0,\frac{1}{2},0 & 6,1,0,\frac{1}{2} & 6,1,\frac{1}{2},\frac{1}{2} & 5,1,\frac{1}{2},\frac{1}{2} & 5,0,0,0 & 11,0,0,0 & 11,1,\frac{1}{2},\frac{1}{2} & 6 & 0 \\
5,0,\frac{1}{2},0 & 6,1,0,\frac{1}{2} & 6,0,\frac{1}{2},0 & 6,1,\frac{1}{2},\frac{1}{2} & 5,0,0,0 & 5,1,\frac{1}{2},\frac{1}{2} & 11,1,\frac{1}{2},\frac{1}{2} & 11,0,0,0 & 6 & \frac{1+i}{2} \\
5,0,0,\frac{1}{2} & 6,0,0,\frac{1}{2} & 6,1,\frac{1}{2},0 & 6,1,\frac{1}{2},\frac{1}{2} & 5,1,\frac{1}{2},\frac{1}{2} & 5,0,0,0 & 11,0,0,0 & 11,1,\frac{1}{2},\frac{1}{2} & 6 & 0 \\
5,0,0,\frac{1}{2} & 6,1,\frac{1}{2},0 & 6,0,0,\frac{1}{2} & 6,1,\frac{1}{2},\frac{1}{2} & 5,0,0,0 & 5,1,\frac{1}{2},\frac{1}{2} & 11,1,\frac{1}{2},\frac{1}{2} & 11,0,0,0 & 6 & \frac{1+i}{2} \\
5,1,\frac{1}{2},\frac{1}{2} & 7,1,0,0 & 6,1,\frac{1}{2},\frac{1}{2} & 6,0,\frac{1}{2},0 & 6,0,0,\frac{1}{2} & 5,0,\frac{1}{2},0 & 12,0,\frac{1}{2},\frac{1}{2} & 11,0,0,0 & 6 & \frac{1+i}{2} \\
5,1,\frac{1}{2},\frac{1}{2} & 7,1,0,0 & 6,1,\frac{1}{2},\frac{1}{2} & 6,0,0,\frac{1}{2} & 6,0,\frac{1}{2},0 & 5,0,0,\frac{1}{2} & 12,0,\frac{1}{2},\frac{1}{2} & 11,0,0,0 & 6 & \frac{1+i}{2} \\
6,0,\frac{1}{2},0 & 5,0,\frac{1}{2},0 & 7,1,\frac{1}{2},0 & 6,1,\frac{1}{2},\frac{1}{2} & 5,1,\frac{1}{2},\frac{1}{2} & 7,0,\frac{1}{2},\frac{1}{2} & 11,0,0,0 & 13,1,0,0 & 6 & 0 \\
6,0,\frac{1}{2},0 & 6,0,0,\frac{1}{2} & 6,1,0,\frac{1}{2} & 6,1,\frac{1}{2},0 & 6,1,0,\frac{1}{2} & 6,0,0,\frac{1}{2} & 12,0,\frac{1}{2},\frac{1}{2} & 12,1,\frac{1}{2},\frac{1}{2} & 6 & 0 \\
6,0,\frac{1}{2},0 & 6,0,0,\frac{1}{2} & 6,1,0,\frac{1}{2} & 6,1,0,\frac{1}{2} & 6,1,\frac{1}{2},0 & 6,0,\frac{1}{2},0 & 12,0,\frac{1}{2},\frac{1}{2} & 12,1,\frac{1}{2},\frac{1}{2} & 6 & 0 \\
6,0,0,\frac{1}{2} & 5,0,0,\frac{1}{2} & 7,1,0,\frac{1}{2} & 6,1,\frac{1}{2},\frac{1}{2} & 5,1,\frac{1}{2},\frac{1}{2} & 7,0,\frac{1}{2},\frac{1}{2} & 11,0,0,0 & 13,1,0,0 & 6 & 0 \\
6,0,0,\frac{1}{2} & 6,0,\frac{1}{2},0 & 6,1,\frac{1}{2},0 & 6,1,\frac{1}{2},0 & 6,1,0,\frac{1}{2} & 6,0,0,\frac{1}{2} & 12,0,\frac{1}{2},\frac{1}{2} & 12,1,\frac{1}{2},\frac{1}{2} & 6 & 0 \\
6,0,0,\frac{1}{2} & 6,0,\frac{1}{2},0 & 6,1,\frac{1}{2},0 & 6,1,0,\frac{1}{2} & 6,1,\frac{1}{2},0 & 6,0,\frac{1}{2},0 & 12,0,\frac{1}{2},\frac{1}{2} & 12,1,\frac{1}{2},\frac{1}{2} & 6 & 0 \\
6,0,\frac{1}{2},\frac{1}{2} & 7,0,\frac{1}{2},\frac{1}{2} & 7,1,0,0 & 7,1,\frac{1}{2},0 & 6,1,\frac{1}{2},0 & 6,0,0,\frac{1}{2} & 13,0,0,0 & 13,1,\frac{1}{2},\frac{1}{2} & 7 & 0 \\
6,0,\frac{1}{2},\frac{1}{2} & 7,0,\frac{1}{2},\frac{1}{2} & 7,1,0,0 & 7,1,0,\frac{1}{2} & 6,1,0,\frac{1}{2} & 6,0,\frac{1}{2},0 & 13,0,0,0 & 13,1,\frac{1}{2},\frac{1}{2} & 7 & 0 \\
6,0,\frac{1}{2},\frac{1}{2} & 7,1,0,0 & 7,0,\frac{1}{2},\frac{1}{2} & 7,1,\frac{1}{2},0 & 6,0,0,\frac{1}{2} & 6,1,\frac{1}{2},0 & 13,1,\frac{1}{2},\frac{1}{2} & 13,0,0,0 & 7 & \frac{1+i}{2} \\
6,0,\frac{1}{2},\frac{1}{2} & 7,1,0,0 & 7,0,\frac{1}{2},\frac{1}{2} & 7,1,0,\frac{1}{2} & 6,0,\frac{1}{2},0 & 6,1,0,\frac{1}{2} & 13,1,\frac{1}{2},\frac{1}{2} & 13,0,0,0 & 7 & \frac{1+i}{2} \\
6,1,\frac{1}{2},0 & 5,0,0,\frac{1}{2} & 7,1,0,\frac{1}{2} & 6,1,\frac{1}{2},\frac{1}{2} & 5,0,0,0 & 7,1,0,0 & 11,1,\frac{1}{2},\frac{1}{2} & 13,0,\frac{1}{2},\frac{1}{2} & 6 & \frac{1+i}{2} \\
6,1,\frac{1}{2},0 & 6,1,0,\frac{1}{2} & 6,0,0,\frac{1}{2} & 6,0,\frac{1}{2},0 & 6,0,0,\frac{1}{2} & 6,1,0,\frac{1}{2} & 12,0,\frac{1}{2},\frac{1}{2} & 12,1,\frac{1}{2},\frac{1}{2} & 6 & \frac{1+i}{2} \\
6,1,\frac{1}{2},0 & 6,1,0,\frac{1}{2} & 6,0,0,\frac{1}{2} & 6,0,0,\frac{1}{2} & 6,0,\frac{1}{2},0 & 6,1,\frac{1}{2},0 & 12,0,\frac{1}{2},\frac{1}{2} & 12,1,\frac{1}{2},\frac{1}{2} & 6 & \frac{1+i}{2} \\
6,1,0,\frac{1}{2} & 5,0,\frac{1}{2},0 & 7,1,\frac{1}{2},0 & 6,1,\frac{1}{2},\frac{1}{2} & 5,0,0,0 & 7,1,0,0 & 11,1,\frac{1}{2},\frac{1}{2} & 13,0,\frac{1}{2},\frac{1}{2} & 6 & \frac{1+i}{2} \\
6,1,0,\frac{1}{2} & 6,1,\frac{1}{2},0 & 6,0,\frac{1}{2},0 & 6,0,\frac{1}{2},0 & 6,0,0,\frac{1}{2} & 6,1,0,\frac{1}{2} & 12,0,\frac{1}{2},\frac{1}{2} & 12,1,\frac{1}{2},\frac{1}{2} & 6 & \frac{1+i}{2} \\
6,1,0,\frac{1}{2} & 6,1,\frac{1}{2},0 & 6,0,\frac{1}{2},0 & 6,0,0,\frac{1}{2} & 6,0,\frac{1}{2},0 & 6,1,\frac{1}{2},0 & 12,0,\frac{1}{2},\frac{1}{2} & 12,1,\frac{1}{2},\frac{1}{2} & 6 & \frac{1+i}{2} \\ \hline
  \end{array} \notag
\end{align}
\caption{All quark and lepton flavor models satisfying $S$-symmetric vacuum conditions in Eq.~(\ref{eq:conditions_S}).
The first to eighth columns show the flux $M$ , $\mathbb{Z}_2$ parity $m$ (even, odd = 0, 1) and SS phases $(\alpha_1,\alpha_2)$ of the zero-modes of the fields.
$g_H$ denotes the number of Higgs fields.}
  \label{tab:classification_S_eig}
\end{table}

Finally we comment on the realization of the flavor models in $S$-symmetric vacuum.
As we have discussed in Subsection \ref{subsubsec:favorable}, we need to slightly deviate the direction of Higgs VEVs from $h_{u,d}^\ell$ to realize small but nonzero masses of the first and second generation quarks and charged leptons.
Then the deviations may not be $S$-symmetric direction.
Therefore it may be difficult to realize flavor observables in exact $S$-symmetric vacuum.
Actually in the following, we will show in an numerical example the deviation from $S$-symmetric directions is required to obtain realistic flavor observables.


\subsection{Numerical example}
\label{subsubsec:numerical_example_T2dZ2}

Here, we show a numerical example deriving a realistic quark and lepton flavor structures in the vicinity of $S$-eigenstates.
We use the flavor model shown in Table \ref{tab:ex-model}.
\begin{table}[H]
\begin{align}
  \begin{array}{c|c|c|c|c|c|c|c|c} \hline
    B_Q & B_{u_R} & B_{d_R} & B_L & B_{\nu_R} & B_{e_R} & B_{H_u} & B_{H_d} & g_H \\ \hline\hline
    6,0,\frac{1}{2},0 & 6,0,0,\frac{1}{2} & 6,1,0,\frac{1}{2} & 6,1,0,\frac{1}{2} & 6,1,\frac{1}{2},0 & 6,0,\frac{1}{2},0 & 12,0,\frac{1}{2},\frac{1}{2} & 12,1,\frac{1}{2},\frac{1}{2} & 6 \\ \hline
  \end{array} \notag
\end{align}
\caption{Flux, $\mathbb{Z}_2$ parity (even, ood = 0, 1), SS phases $(\alpha_1,\alpha_2)$ of quarks, leptons and Higgs fields in the model.
$g_H$ denotes the number of Higgs fields.}
  \label{tab:ex-model}
\end{table}
In this model, quark doublets $Q$ have (flux, $\mathbb{Z}_2$ parity, SS phases $\alpha_1,\alpha_2$) = ($6,0,0,\frac{1}{2}$); right-handed up-sector quarks $u_R$ have $(5,0,0,\frac{1}{2})$; right-handed down-sector quarks $d_R$ have $(6,0,\frac{1}{2},0)$; lepton doublets $L$ have $(6,0,\frac{1}{2},0)$; right-handed neutrinos $\nu_R$ have $(5,0,\frac{1}{2},0)$; right-handed charged leptons $e_R$ have $(6,0,0,\frac{1}{2})$; up type Higgs fields $H_u$ have $(11,0,0,0)$; down type Higgs fields $H_d$ have $(12,0,\frac{1}{2},\frac{1}{2})$.
The number of both up and down types Higgs fields are six.
Yukawa couplings $Y_u^{ijk}$, $Y_{d}^{ijk}$, $Y_\nu^{ijk}$ and $Y_e^{ijk}$ are shown in Appendix \ref{App:Yukawa-couplings}; the Majorana mass matrix of the right-handed neutrinos induced by D-brane instanton effect is shown in Appendix \ref{App:Majorana}.

First of all, we fix the value of modulus by $\tau=i$ and use the slight deviations of Higgs VEVs from $h_{u,d}$ as parameters.
Higgs VEV directions satisfying the favorable conditions I-IV, $h_{u,d}^\ell$, in this model are given by
\begin{align}
  &h_u^\ell = v_u(0.8464, 0.5014, 0.1759, 0.03657, 0.004504, 0.0003144), \\
  &h_d^\ell = v_d(0.4330, 0.7696, 0.4501, 0.1310, 0.02074, 0.001945),
\end{align}
where $h_u^\ell$ and $h_d^\ell$ are $S$-eigenstates with eigenvalues $+1$ and $+i$, respectively.
Thus the modulus $\tau$ is $S$-symmetric vacuum, while these directions of Higgs VEVs 
are $S$-eigenstates.
Before we consider the deviation from $S$-eigenstate directions, we try to realize flavor observables in exact $S$-eigenstate directions of the Higgs VEVs.
Six pairs of up (down) type Higgs fields contain three $S$-eigenstates with eigenvalue $+1$ ($+i$) in total.
Therefore there are three of degree of freedom in $S$-eigenstate directions of up and down type Higgs VEVs respectively.
We use them as free parameters.
To obtain realistic flavors, let us choose the following directions of Higgs VEVs,
\begin{align}
  &\langle H_u^\ell \rangle = v_u(0.8466, 0.5009, 0.1762, 0.03715, 0.004794, 0.0003797), \label{eq:VEVHu_S} \\
  &\langle H_d^\ell \rangle = v_d(0.5006, 0.7890, 0.3521, 0.05382, -0.003787, -0.003709). \label{eq:VEVHd_S}
\end{align}
Note that again these directions are $S$-eigenstates with eigenvalues $+1$ and $+i$, respectively.
They lead to the following up-sector quark, down-sector quark and charged lepton mass ratios,
\begin{align}
  &(m_u,m_c,m_t)/m_t = (2.96\times 10^{-5},5.35\times 10^{-4},1), \\
  &(m_d,m_s,m_b)/m_b = (4.36\times 10^{-4},1.17\times 10^{-2},1), \\
  &(m_e,m_\mu,m_\tau)/m_\tau = (4.36\times 10^{-4},1.17\times 10^{-2},1),
\end{align}
and a ratio of the differences of the squares of the neutrino masses,
\begin{align}
  \sqrt{\frac{\Delta m_{\nu 12}^2}{\Delta m_{\nu 13}^2}}
  = \sqrt{\frac{|m_{\nu_1}^2-m_{\nu_2}^2|}{|m_{\nu_1}^2-m_{\nu_3}^2|}} = 0.179,
\end{align}
for normal ordering (NO), $m_{\nu_1}<m_{\nu_2}<m_{\nu_3}$.
The absolute values of the CKM matrix, $|V_{\rm CKM}|$, and the Pontecorvo-Maki-Nakagawa-Sakata (PMNS) matrix, $|V_{\rm PMNS}|$, are obtained as follows,
\begin{align}
  &|V_{\rm CKM}| = 
  \begin{pmatrix}
    0.972 & 0.235 & 0.00134 \\
    0.233 & 0.964 & 0.126 \\
    0.0309 & 0.122 & 0.992 \\
      \end{pmatrix}, \quad
  |V_{\rm PMNS}| = 
  \begin{pmatrix}
    0.990 & 0.137 & 0.0134 \\
    0.129 & 0.957 & 0.261 \\
    0.0487 & 0.257 & 0.965 \\
      \end{pmatrix}.
\end{align}
The large hierarchies in the mass ratios of quarks and charged lepton are realized.
Also the ratio of the differences of the squares of the neutrino masses is realistic.
Moreover the absolute values of the CKM matrix can be realized up to ${\cal O}(1)$.
On the other hand, the absolute values of the PMNS matrix are nearly an unit matrix and they cannot reproduce large mixings in lepton flavors.
Consequently, in this model it is difficult to realize both quark and lepton flavor observables in the exact $S$-eigenstate directions of the Higgs VEVs.

Next, we try the realization in the vicinity of above $S$-eigenstate directions of Higgs VEVs.
We use all six pairs of Higgs VEVs as free parameters for both up and down types but fixing the modulus at $\tau=i$ is continued to simplify the analysis.
To realize flavor structures in the vicinity of $h_{u,d}^\ell$, we have chosen the following directions of Higgs VEVs,
\begin{align}
  &\langle H_u^\ell \rangle = v_u(0.8509, 0.4970, 0.1679, 0.02805, -0.006762, -0.003731), \label{eq:VEVHu} \\
  &\langle H_d^\ell \rangle = v_d(0.4340, 0.7688, 0.4499, 0.1283, 0.02538, 0.03302). \label{eq:VEVHd}
\end{align}
The norm of $h_u^\ell$ in $\langle H_{u}^k \rangle$ is 0.9998 and one of $h_d^\ell$ in $\langle H_{d}^k \rangle$ is 0.9995.
Thus these directions lie on the vicinity of $h_{u,d}^\ell$.
Then the mass matrices for up-sector quark $M_u$, down-sector quark $M_d$, neutrino $M_\nu$ and charged lepton $M_e$ are given by
\begin{align}
&M_u/m_t =
\begin{pmatrix}
0.7202 & 0.5992 & 0.1214 \\
0.2492 & 0.2063 & 0.03922 \\
0.03057 & 0.02249 & -0.002550 \\
\end{pmatrix}, \quad
M_{d}/m_b =
\begin{pmatrix}
0.8675 & 0.3620 & 0.05514 \\
0.3053 & 0.1303 & 0.02287 \\
0.03861 & 0.03580 & 0.03967 \\
\end{pmatrix}, \\
&M_\nu/m_{\nu_3} =
\begin{pmatrix}
-0.3614 & -0.09456 & -0.3323 \\
-0.09456 & -0.1345 & -0.4077 \\
-0.3323 & -0.4077 & -0.5819 \\
\end{pmatrix}, \quad
M_e/m_\tau =
\begin{pmatrix}
0.8675 & 0.3053 & 0.03861 \\
0.3620 & 0.1303 & 0.03580 \\
0.05514 & 0.02287 & 0.03967 \\
\end{pmatrix}.
\end{align}
They lead to the following up-sector quark, down-sector quark and charged lepton mass ratios,
\begin{align}
  &(m_u,m_c,m_t)/m_t = (3.13\times 10^{-5},8.14\times 10^{-3},1), \\
  &(m_d,m_s,m_b)/m_b = (8.46\times 10^{-4},4.10\times 10^{-2},1), \\
  &(m_e,m_\mu,m_\tau)/m_\tau = (8.46\times 10^{-4},4.10\times 10^{-2},1),
\end{align}
and a ratio of the differences of the squares of the neutrino masses,
\begin{align}
  \sqrt{\frac{\Delta m_{\nu 12}^2}{\Delta m_{\nu 13}^2}}
  = \sqrt{\frac{|m_{\nu_1}^2-m_{\nu_2}^2|}{|m_{\nu_1}^2-m_{\nu_3}^2|}} = 0.162,
\end{align}
for NO.
For inverted ordering (IO), $m_{\nu_3}<m_{\nu_1}<m_{\nu_2}$, we cannot find the directions of Higgs VEVs leading to realistic results.
The absolute values of the CKM matrix, $|V_{\rm CKM}|$, and the PMNS matrix, $|V_{\rm PMNS}|$, are obtained as follows,
\begin{align}
  &|V_{\rm CKM}| = 
  \begin{pmatrix}
    0.973 & 0.232 & 0.00234 \\
    0.232 & 0.973 & 0.0162 \\
    0.00603 & 0.0152 & 1.00 \\
      \end{pmatrix}, \quad
  |V_{\rm PMNS}| = 
  \begin{pmatrix}
    0.841 & 0.522 & 0.147 \\
    0.246 & 0.608 & 0.755 \\
    0.483 & 0.598 & 0.639 \\
      \end{pmatrix}.
\end{align}
The results are summarized in Table \ref{MassandCKMandPMNS}.
As a result, in this model we could realize both quark and lepton flavor structures in the vicinity of $S$-eigenstate direction of Higgs VEVs.
\begin{table}[H]
  \begin{center}
    \renewcommand{\arraystretch}{1.3}
    $\begin{array}{c|c|c} \hline
      & {\rm Obtained\ values} & {\rm Reference\ values} \\ \hline
      (m_u,m_c,m_t)/m_t & (3.13\times 10^{-5},8.14\times 10^{-3},1) & (5.58\times 10^{-6},2.69\times 10^{-3},1) \\ \hline
      (m_d,m_s,m_b)/m_b & (8.46\times 10^{-4},4.10\times 10^{-2},1) & (6.86\times 10^{-4},1.37\times 10^{-2},1) \\ \hline
      |V_{\rm CKM}| 
      &
      \begin{pmatrix}
        0.973 & 0.232 & 0.00234 \\
    0.232 & 0.973 & 0.0162 \\
    0.00603 & 0.0152 & 1.00 \\
      \end{pmatrix}
      & 
      \begin{pmatrix}
        0.974 & 0.227 & 0.00361 \\
        0.226 & 0.973 & 0.0405 \\
        0.00854 & 0.0398 & 0.999 
      \end{pmatrix}\\ \hline
      \sqrt{\Delta m_{\nu 12}^2/\Delta m_{\nu 13}^2} & 0.162~{\rm (NO)} & 0.173 \\ \hline
      (m_e,m_\mu,m_\tau)/m_\tau & (8.46\times 10^{-4},4.10\times 10^{-2},1) & (2.78\times 10^{-4},5.88\times 10^{-2},1) \\ \hline
      |V_{\rm PMNS}| 
      &
      \begin{pmatrix}
        0.841 & 0.522 & 0.147 \\
    0.246 & 0.608 & 0.755 \\
    0.483 & 0.598 & 0.639 \\
      \end{pmatrix}
      & 
      \begin{pmatrix}
        0.801{\text -}0.845 & 0.513{\text -}0.579 & 0.143{\text -}0.156 \\
        0.232{\text -}0.507 & 0.459{\text -}0.694 & 0.629{\text -}0.779 \\
        0.260{\text -}0.526 & 0.470{\text -}0.702 & 0.609{\text -}0.763 \\
      \end{pmatrix}\\ \hline
    \end{array}$
    \caption{The mass ratios of the quarks and leptons, and the absolute values of the CKM matrix and the PMNS matrix elements at $\tau=i$ under the vacuum alignments of Higgs fields in Eqs.~(\ref{eq:VEVHu}) and (\ref{eq:VEVHd}).
    Reference values of mass ratios are shown in Refs.~\cite{Antusch:2013jca,Bjorkeroth:2015ora}.
    Those of the CKM matrix and PMNS matrix elements are shown in Refs.~\cite{Zyla:2020zbs} and \cite{NuFIT:2021}.}
    \label{MassandCKMandPMNS}
  \end{center}
\end{table}


\section{$(T^2_1\times T^2_2)/\mathbb{Z}_2^{\textrm{(per)}}$ orbifolds}
\label{subsec:T2xT2}

In the previous section, we have found the possibility to realize both quark and lepton flavor strcutures on the magnetized $T^2/\mathbb{Z}_2^{\textrm{(per)}}$ model by checking the zero points of the zero-modes of each field.
In this section, we study $(T^2_1\times T^2_2)/\mathbb{Z}_2^{\textrm{(per)}}$ permutation orbifold model as the further possibility realizing flavor structures.
Also we study the orbifolding by both $\mathbb{Z}_2$ twist ($\mathbb{Z}_2^{\textrm{(t)}}$) and $\mathbb{Z}_2$ permutation ($\mathbb{Z}_2^{\textrm{(per)}}$), $(T^2_1\times T^2_2)/(\mathbb{Z}_2^{\textrm{(t)}}\times\mathbb{Z}_2^{\textrm{(per)}})$.
$\mathbb{Z}_2$ permutation identifies two tori, $T^2_1$ and $T^2_2$.
On $(T^2_1\times T^2_2)/\mathbb{Z}_2^{\textrm{(per)}}$ orbifold, the complex structure modulus of $T_1^2$, $\tau_1$, and one of $T_2^2$, $\tau_2$, are identified as
\begin{align}
\tau_1 = \tau_2 \equiv \tau,
\end{align}
because $T^2_1$ and $T^2_2$ are identified.


\subsection{Zero-modes on $(T^2_1\times T^2_2)/\mathbb{Z}_2^{\textrm{(per)}}$}

The $(T^2_1\times T^2_2)/\mathbb{Z}_2^{\textrm{(per)}}$ permutation orbifold is obtained by further identifying the $\mathbb{Z}_2$ permutation point $z_1$ with $z_2$, i.e. $z_1\sim z_2$, under the moduli stabilization $\tau_1 = \tau_2 \equiv \tau$.
$\mathbb{Z}_2$ permutation is given by $(z_1,z_2;\tau_1,\tau_2) \rightarrow (z_2,z_1;\tau_2,\tau_1)$ for $\tau_1=\tau_2\equiv \tau$; therefore it corresponds to the permutation of $T^2_1$ and $T^2_2$, $T^2_1 \leftrightarrow T^2_2$.
The zero-modes on $(T^2_1\times T^2_2)/\mathbb{Z}_2^{\textrm{(per)}}$ orbifold are required to satisfy
\begin{align}
\psi_{(T^2_1\times T^2_2)/\mathbb{Z}_2^{\textrm{(per)}n}}(z_1,z_2) 
\xrightarrow{\mathbb{Z}_2^{\textrm{(per)}}} \psi_{(T^2_1\times T^2_2)/\mathbb{Z}_2^{\textrm{(per)}n}}(z_2,z_1) 
= e^{n\pi i} \psi_{(T^2_1\times T^2_2)/\mathbb{Z}_2^{\textrm{(per)}n}}(z_1,z_2) , \quad n \in \mathbb{Z}_2, \label{eq:BC_Z2per}
\end{align}
in addition to the boundary conditions on $T^2_1$ and $T_2^2$ in Eq.~(\ref{eq:BC_psi}).
Here $n\in\mathbb{Z}_2$ denotes $\mathbb{Z}_2^{\textrm{(per)}}$ parity; $n=0$ is $\mathbb{Z}_2^{\textrm{(per)}}$ even modes and $n=1$ is odd modes.
Under $\mathbb{Z}_2$ permutation $(z_1,z_2)\to(z_2,z_1)$, the products of the zero-modes on $T_1^2$ and $T_2^2$ are tranformed as
\begin{align}
\psi^{(j_1+\alpha_{11},\alpha_{21}),M_1}_{T^2_1}(z_1) \cdot \psi^{(j_2+\alpha_{12},\alpha_{22}),M_2}_{T^2_2}(z_2) \to \psi^{(j_1+\alpha_{11},\alpha_{21}),M_1}_{T^2_1}(z_2) \cdot \psi^{(j_2+\alpha_{12},\alpha_{22}),M_2}_{T^2_2}(z_1).
\end{align}
Thus, the zero-modes on $(T^2_1\times T^2_2)/\mathbb{Z}_2^{\textrm{(per)}}$ satisfying Eq.~(\ref{eq:BC_Z2per}) can be defined as \cite{Kikuchi:2020nxn}
\begin{align}
&\psi^{(\vec{j}+\vec{\alpha}_1,\vec{\alpha}_2),M}_{(T^2_1\times T^2_2)/\mathbb{Z}_2^{\textrm{(per)}n}}(z_1,z_2) \notag \\
&= {\cal N}^{\vec{j}}_{\textrm{(per)}} \left(
\psi^{(j_1+\alpha_{1},\alpha_{2}),M}_{T^2_1}(z_1) \cdot \psi^{(j_2+\alpha_{1},\alpha_{2}),M}_{T^2_2}(z_2)
+ e^{n\pi i} \psi^{(j_1+\alpha_{1},\alpha_{2}),M}_{T^2_2}(z_2) \cdot \psi^{(j_2+\alpha_{1},\alpha_{2}),M}_{T^2_1}(z_1)
\right),
\end{align}
where $n\in\mathbb{Z}_2$ and
\begin{align}
&\vec{j} = (j_1, j_2), \quad j_1,j_2 \in \mathbb{Z}_M, \quad j_1 \geq j_2, \\
&\vec{\alpha} = (\vec{\alpha}_1, \vec{\alpha}_2) = ((\alpha_{11},\alpha_{12}), (\alpha_{21},\alpha_{22})) = ((\alpha_1,\alpha_1),(\alpha_2,\alpha_2)), \\
&M_1 = M_2 \equiv M, \\
&{\cal N}^{\vec{j}}_{\textrm{(per)}} = \left\{
\begin{array}{ll}
1/2 & (j_1=j_2), \\
1/\sqrt{2} & (j_1 \neq j_2). \\
\end{array}
\right.
\end{align}
Note that $M_1=M_2\equiv M$, $\alpha_{11}=\alpha_{12}$ and $\alpha_{21}=\alpha_{22}$ are required because $T^2_1$ and $T^2_2$ are identified.
In Table \ref{tab:num_of_zeros_T2xT2/Z2}, we show the number of zero-modes on magnetized $(T^2_1\times T^2_2)/\mathbb{Z}_2^{\textrm{(per)}}$ up to $M\leq 6$.
\begin{table}[H]
\centering
\begin{tabular}{c|cccccc|c} \hline
Flux & 1 & 2 & 3 & 4 & 5 & 6 & $M$ \\ \hline
Even & 1 & 3 & 6 & 10 & 15 & 21 & $\frac{M(M+1)}{2}$ \\
Odd & 0 & 1 & 3 & 6 & 10 & 15 & $\frac{M(M-1)}{2}$ \\ \hline
\end{tabular}
\caption{The number of zero-modes on magnetized $(T^2_1\times T^2_2)/\mathbb{Z}_2^{\textrm{(per)}}$.
The second and third rows show the number of even- and odd-modes under $\mathbb{Z}_2$ permutation, respectively.}
\label{tab:num_of_zeros_T2xT2/Z2}
\end{table}
We can find three-generations for even-modes at $M=2$ and for odd-modes at $M=3$.

Yukawa couplings on $(T^2_1\times T^2_2)/\mathbb{Z}_2^{\textrm{(per)}}$ are given by the overlap integral of the zero-modes on $(T^2_1\times T^2_2)/\mathbb{Z}_2^{\textrm{(per)}}$:
\begin{align}
&Y^{\vec{j}\vec{k}\vec{\ell}}_{(T^2_1\times T^2_2)/\mathbb{Z}_2^{\textrm{(per)}}} \notag \\
&= g^2(2\textrm{Im}\tau) \int_{(T^2_1\times T^2_2)/\mathbb{Z}_2^{\textrm{(per)}}} d^2z_1d^2z_2
\psi^{(\vec{j}+\vec{\alpha}_{1L},\vec{\alpha}_{2L}),M_L}_{(T^2_1\times T^2_2)/\mathbb{Z}_2^{\textrm{(per)}n_L}}
\cdot \psi^{(\vec{k}+\vec{\alpha}_{1R},\vec{\alpha}_{2R}),M_R}_{(T^2_1\times T^2_2)/\mathbb{Z}_2^{\textrm{(per)}n_R}}
\cdot \left(\psi^{(\vec{\ell}+\vec{\alpha}_{1H},\vec{\alpha}_{2H}),M_H}_{(T^2_1\times T^2_2)/\mathbb{Z}_2^{\textrm{(per)}n_H}}\right)^*,
\end{align}
where
\begin{align}
\begin{aligned}
&M_L + M_R = M_H, \\
&\vec{\alpha}_{1L} + \vec{\alpha}_{1R} = \vec{\alpha}_{1H}, \\
&\vec{\alpha}_{2L} + \vec{\alpha}_{2R} = \vec{\alpha}_{2H}, \\
&n_L + n_R = n_H~ (\textrm{mod}~2), \\
\end{aligned} \label{eq:non-vanishing_perm}
\end{align}
are required to obtain non-vanishing Yukawa couplings.

We additionally show the modular transformation for zero-modes on $(T^2_1\times T^2_2)/\mathbb{Z}_2^{\textrm{(per)}}$.
Under $S$ and $T$-transformations, the zero-modes are transformed as
\begin{align}
\psi^{(\vec{j}+\vec{\alpha}_1,\vec{\alpha}_2),M}_{(T^2_1\times T^2_2)/\mathbb{Z}_2^{\textrm{(per)}n}}(z_1,z_2,\tau) \xrightarrow{S}
J_1(\gamma,\tau) \rho_{(T^2_1\times T^2_2)/\mathbb{Z}_2^{\textrm{(per)}n}}^{\vec{j}\vec{j}'\vec{\alpha}_1\vec{\alpha}_1'\vec{\alpha}_2\vec{\alpha}_2'}(\gamma) \psi^{(\vec{j}'+\vec{\alpha}'_1,\vec{\alpha}'_2),M}_{(T^2_1\times T^2_2)/\mathbb{Z}_2^{\textrm{(per)}n}}(z_1,z_2,\tau),
\end{align}
where
\begin{align}
&\rho_{(T^2_1\times T^2_2)/\mathbb{Z}_2^{\textrm{(per)}n}}^{\vec{j}\vec{j}'\vec{\alpha}_1\vec{\alpha}_1'\vec{\alpha}_2\vec{\alpha}_2'}(\gamma)
= 2{\cal N}_{\textrm{(per)}}^{\vec{j}} {\cal N}_{\textrm{(per)}}^{\vec{j}'} 
\left(
\rho_{T^2_1\times T^2_2}^{(j_1j_2)(j_1'j_2')\vec{\alpha}_1\vec{\alpha}_1'\vec{\alpha}_2\vec{\alpha}_2'}(\gamma)
+ e^{n\pi i} \rho_{T^2_1\times T^2_2}^{(j_1j_2)(j_2'j_1')\vec{\alpha}_1\vec{\alpha}_1'\vec{\alpha}_2\vec{\alpha}_2'}(\gamma)
\right), \\
&\rho_{T^2_1\times T^2_2}^{(j_1j_2)(j_1'j_2')\vec{\alpha}_1\vec{\alpha}_1'\vec{\alpha}_2\vec{\alpha}_2'}(\gamma)
= \rho^{j_1j_1'\alpha_1\alpha_1'\alpha_2\alpha_2'}_{T^2_1}(\widetilde{\gamma}) \cdot \rho^{j_2j_2'\alpha_1\alpha_1'\alpha_2\alpha_2'}_{T^2_2}(\widetilde{\gamma}),
\end{align}
and $\rho^{j_ij_i'\alpha_1\alpha_1'\alpha_2\alpha_2'}_{T^2_i}$, $i=1,2$, are given by Eqs.~(\ref{eq:rho_S}) and (\ref{eq:rho_T}).
Unitary transformation matrices $\rho_{T^2_1\times T^2_2/\mathbb{Z}_2^{\textrm{(per)}n}}$ as well as $\rho_{T^2_1\times T^2_2}$ satisfy the algebraic relations of $\Gamma'_{2M}$:
\begin{align}
\rho(S)^4 = [\rho(S)\rho(T)]^3 = \rho(T)^{2M} = \mathbb{I}, \quad \rho(S^2)\rho(T) = \rho(T)\rho(S^2).
\end{align}
Thus, the zero-modes on $(T^2_1\times T^2_2)/\mathbb{Z}_2^{\textrm{(per)}}$ behave as the modular forms of weight 1 for $\Gamma(2M)$.


\subsection{Zero-modes on $(T^2_1\times T^2_2)/(\mathbb{Z}_2^{\textrm{(t)}}\times\mathbb{Z}_2^{\textrm{(per)}})$}

Next, we discuss the zero-modes on $(T^2_1\times T^2_2)/(\mathbb{Z}_2^{\textrm{(t)}}\times\mathbb{Z}_2^{\textrm{(per)}})$.
The zero-modes on $(T^2_1\times T^2_2)/(\mathbb{Z}_2^{\textrm{(t)}}\times\mathbb{Z}_2^{\textrm{(per)}})$ orbifold are required to satisfy Eq.~(\ref{eq:BC_Z2per}) in addition to the boundary conditions on $T^2_1/\mathbb{Z}_2$ and $T^2_2/\mathbb{Z}_2$.
They are given by
\begin{align}
&\psi^{(\vec{j}+\vec{\alpha}_1,\vec{\alpha}_2),M}_{m,n}(z_1,z_2) \notag \\
&= {\cal N}^{\vec{j}}_{\textrm{(per)}} \left(
\psi^{(j_1+\alpha_{1},\alpha_{2}),M}_{T^2_1/\mathbb{Z}_2^m}(z_1) \cdot \psi^{(j_2+\alpha_{1},\alpha_{2}),M}_{T^2_2/\mathbb{Z}_2^m}(z_2)
+ e^{n\pi i} \psi^{(j_1+\alpha_{1},\alpha_{2}),M}_{T^2_2/\mathbb{Z}_2^m}(z_2) \cdot \psi^{(j_2+\alpha_{1},\alpha_{2}),M}_{T^2_1/\mathbb{Z}_2^m}(z_1)
\right),
\end{align}
where $m,n\in\mathbb{Z}_2$.
Note that $m=0$ and $1$ denote $\mathbb{Z}_2$ twist even and odd modes, respectively; $n=0$ and $n=1$ denote $\mathbb{Z}_2$ permutation even and odd modes, respectively.
In Table \ref{tab:num_T2xT2/Z2xZ2}, we show the number of zero-modes on magnetized $(T^2_1\times T^2_2)/(\mathbb{Z}_2^{\textrm{(t)}}\times\mathbb{Z}_2^{\textrm{(per)}})$.
\begin{table}[H]
\centering
\begin{tabular}{|c||c|c|} \hline
($\mathbb{Z}_2^{\textrm{(t)}}$ parity $m$; $\mathbb{Z}_2^{\textrm{(per)}}$ parity $n$; $\alpha_1$, $\alpha_2$) & $M=\textrm{even}$ & $M=\textrm{odd}$ \\ \hline
$(0;0;0,0)$ & $\frac{1}{2}(\frac{M}{2}+1)(\frac{M}{2}+2)$ & $\frac{1}{8}(M+1)(M+3)$ \\
$(1;0;0,0)$ & $\frac{M}{4}(\frac{M}{2}-1)$ & $\frac{1}{8}(M-1)(M+1)$ \\
$(0;0;1/2,0)$ & $\frac{M}{4}(\frac{M}{2}+1)$ & $\frac{1}{8}(M+1)(M+3)$ \\
$(1;0;1/2,0)$ & $\frac{M}{4}(\frac{M}{2}+1)$ & $\frac{1}{8}(M-1)(M+1)$ \\
$(0;0;0,1/2)$ & $\frac{M}{4}(\frac{M}{2}+1)$ & $\frac{1}{8}(M+1)(M+3)$ \\
$(1;0;0,1/2)$ & $\frac{M}{4}(\frac{M}{2}+1)$ & $\frac{1}{8}(M-1)(M+1)$ \\
$(0;0;1/2,1/2)$ & $\frac{M}{4}(\frac{M}{2}+1)$ & $\frac{1}{8}(M-1)(M+1)$ \\
$(1;0;1/2,1/2)$ & $\frac{M}{4}(\frac{M}{2}+1)$ & $\frac{1}{8}(M+1)(M+3)$ \\ 
$(0;1;0,0)$ & $\frac{M}{4}(\frac{M}{2}+1)$ & $\frac{1}{8}(M-1)(M+1)$ \\
$(1;1;0,0)$ & $\frac{1}{2}(\frac{M}{2}-1)(\frac{M}{2}-2)$ & $\frac{1}{8}(M-1)(M-3)$ \\
$(0;1;1/2,0)$ & $\frac{M}{4}(\frac{M}{2}-1)$ & $\frac{1}{8}(M-1)(M+1)$ \\
$(1;1;1/2,0)$ & $\frac{M}{4}(\frac{M}{2}-1)$ & $\frac{1}{8}(M-1)(M-3)$ \\
$(0;1;0,1/2)$ & $\frac{M}{4}(\frac{M}{2}-1)$ & $\frac{1}{8}(M-1)(M+1)$ \\
$(1;1;0,1/2)$ & $\frac{M}{4}(\frac{M}{2}-1)$ & $\frac{1}{8}(M-1)(M-3)$ \\
$(0;1;1/2,1/2)$ & $\frac{M}{4}(\frac{M}{2}-1)$ & $\frac{1}{8}(M-1)(M-3)$ \\
$(1;1;1/2,1/2)$ & $\frac{M}{4}(\frac{M}{2}-1)$ & $\frac{1}{8}(M-1)(M+1)$ \\\hline
\end{tabular}
\caption{The number of zero-modes on magnetized $(T^2_1\times T^2_2)/(\mathbb{Z}_2^{\textrm{(t)}}\times\mathbb{Z}_2^{\textrm{(per)}})$.
$\mathbb{Z}_2^{\textrm{(t)}}$ parities 0 and 1 correspond to $\mathbb{Z}_2$ twist even and odd modes, respectively; $\mathbb{Z}_2^{\textrm{(per)}}$ parities 0 and 1 correspond to $\mathbb{Z}_2$ permutation even and odd modes, respectively.}
\label{tab:num_T2xT2/Z2xZ2}
\end{table}
We additionally show the values of $(M;m;n;\alpha_1,\alpha_2)$ giving three-generation zero-modes in Table \ref{tab:three_T2xT2/Z2xZ2}.
We can find 32 three-generation zero-modes in total.
\begin{table}[H]
\centering
\begin{tabular}{|c|c|} \hline
($M$; $m$; $n$; $\alpha_1$, $\alpha_2$) & ($M$; $m$; $n$; $\alpha_1$, $\alpha_2$) \\
$M=\textrm{even}$ & $M=\textrm{odd}$ \\ \hline
$(2;0;0;0,0)$ & $(3;0;0;0,0)$ \\
$(6;1;0;0,0)$ & $(5;1;0;0,0)$ \\
$(4;0;0;1/2,0)$ & $(3;0;0;1/2,0)$ \\
$(4;1;0;1/2,0)$ & $(5;1;0;1/2,0)$ \\
$(4;0;0;0,1/2)$ & $(3;0;0;0,1/2)$ \\
$(4;1;0;0,1/2)$ & $(5;1;0;0,1/2)$ \\
$(4;0;0;1/2,1/2)$ & $(5;0;0;1/2,1/2)$ \\
$(4;1;0;1/2,1/2)$ & $(3;1;0;1/2,1/2)$ \\ 
$(4;0;1;0,0)$ & $(5;0;1;0,0)$ \\
$(8;1;1;0,0)$ & $(7;1;1;0,0)$ \\
$(6;0;1;1/2,0)$ & $(5;0;1;1/2,0)$ \\
$(6;1;1;1/2,0)$ & $(7;1;1;1/2,0)$ \\
$(6;0;1;0,1/2)$ & $(5;0;1;0,1/2)$ \\
$(6;1;1;0,1/2)$ & $(7;1;1;0,1/2)$ \\
$(6;0;1;1/2,1/2)$ & $(7;0;1;1/2,1/2)$ \\
$(6;1;1;1/2,1/2)$ & $(5;1;1;1/2,1/2)$ \\\hline
\end{tabular}
\caption{The values of $(M;m;n;\alpha_1,\alpha_2)$ giving three-generation zero-modes on magnetized $(T^2_1\times T^2_2)/(\mathbb{Z}_2^{\textrm{(t)}}\times\mathbb{Z}_2^{\textrm{(per)}})$.}
\label{tab:three_T2xT2/Z2xZ2}
\end{table}
Yukawa couplings on $(T^2_1\times T^2_2)/(\mathbb{Z}_2^{\textrm{(t)}}\times\mathbb{Z}_2^{\textrm{(per)}})$ are given by the overlap integral of the zero-modes on $(T^2_1\times T^2_2)/(\mathbb{Z}_2^{\textrm{(t)}}\times\mathbb{Z}_2^{\textrm{(per)}})$:
\begin{align}
&Y^{\vec{j}\vec{k}\vec{\ell}}_{(T^2_1\times T^2_2)/(\mathbb{Z}_2^{\textrm{(t)}}\times\mathbb{Z}_2^{\textrm{(per)}})} \notag \\
&= g^2(2\textrm{Im}\tau) \int_{(T^2_1\times T^2_2)/(\mathbb{Z}_2^{\textrm{(t)}}\times\mathbb{Z}_2^{\textrm{(per)}})} d^2z_1d^2z_2
\psi^{(\vec{j}+\vec{\alpha}_{1L},\vec{\alpha}_{2L}),M_L}_{m_L,n_L}
\cdot \psi^{(\vec{k}+\vec{\alpha}_{1R},\vec{\alpha}_{2R}),M_R}_{m_R,n_R}
\cdot \left(\psi^{(\vec{\ell}+\vec{\alpha}_{1H},\vec{\alpha}_{2H}),M_H}_{m_H,n_H}\right)^*,
\end{align}
where
\begin{align}
&m_L + m_R = m_H~ (\textrm{mod}~2), \label{eq:non-vanishing_twist}
\end{align}
is required in addition to Eq.~(\ref{eq:non-vanishing_perm}).

Also we show the modular transformation for zero-modes on $(T^2_1\times T^2_2)/(\mathbb{Z}_2^{\textrm{(t)}}\times\mathbb{Z}_2^{\textrm{(per)}})$.
Under $S$ and $T$-transformations, the zero-modes are transformed as
\begin{align}
\psi^{(\vec{j}+\vec{\alpha}_1,\vec{\alpha}_2),M}_{m,n}(z_1,z_2,\tau) \xrightarrow{S}
J_1(\gamma,\tau) \rho_{m,n}^{\vec{j}\vec{j}'\vec{\alpha}_1\vec{\alpha}_1'\vec{\alpha}_2\vec{\alpha}_2'}(\gamma) \psi^{(\vec{j}'+\vec{\alpha}'_1,\vec{\alpha}'_2),M}_{m,n}(z_1,z_2,\tau),
\end{align}
where
\begin{align}
&\rho_{m,n}^{\vec{j}\vec{j}'\vec{\alpha}_1\vec{\alpha}_1'\vec{\alpha}_2\vec{\alpha}_2'}(\gamma)
= 2{\cal N}_{\textrm{(per)}}^{\vec{j}} {\cal N}_{\textrm{(per)}}^{\vec{j}'} 
\left(
\rho_{(T^2_1\times T^2_2)/\mathbb{Z}_2^{\textrm{(t)}}}^{(j_1j_2)(j_1'j_2')\vec{\alpha}_1\vec{\alpha}_1'\vec{\alpha}_2\vec{\alpha}_2'}(\gamma)
+ e^{n\pi i} \rho_{(T^2_1\times T^2_2)/\mathbb{Z}_2^{\textrm{(t)}}}^{(j_1j_2)(j_2'j_1')\vec{\alpha}_1\vec{\alpha}_1'\vec{\alpha}_2\vec{\alpha}_2'}(\gamma)
\right), \\
&\rho_{(T^2_1\times T^2_2)/\mathbb{Z}_2^{\textrm{(t)}}}^{(j_1j_2)(j_1'j_2')\vec{\alpha}_1\vec{\alpha}_1'\vec{\alpha}_2\vec{\alpha}_2'}(\gamma)
= \rho^{j_1j_1'\alpha_1\alpha_1'\alpha_2\alpha_2'}_{T^2_1/\mathbb{Z}_2^{\textrm{(t)}}}(\widetilde{\gamma}) \cdot \rho^{j_2j_2'\alpha_1\alpha_1'\alpha_2\alpha_2'}_{T^2_2/\mathbb{Z}_2^{\textrm{(t)}}}(\widetilde{\gamma}),
\end{align}
and $\rho^{j_ij_i'\alpha_1\alpha_1'\alpha_2\alpha_2'}_{T^2_i/\mathbb{Z}_2^{\textrm{(t)}}}$, $i=1,2$, are given by Eqs.~(\ref{eq:rho_S_T2/Z2}) and (\ref{eq:rho_T_T2/Z2}).
$\rho_{m,n}$ satisfies same algebraic relations as $\rho_{T^2_1\times T^2_2/\mathbb{Z}_2^{\textrm{(per)}n}}$.
Thus, the zero-modes on $(T^2_1\times T^2_2)/(\mathbb{Z}_2^{\textrm{(t)}}\times\mathbb{Z}_2^{\textrm{(per)}})$ behave as the modular forms of weight 1 for $\Gamma(2M)$.


\subsection{Three-generation models}

Here let us study three-generation models on magnetized $(T^2_1\times T^2_2)/\mathbb{Z}_2^{\textrm{(per)}}$ and $(T^2_1\times T^2_2)/(\mathbb{Z}_2^{\textrm{(t)}}\times\mathbb{Z}_2^{\textrm{(per)}})$.
In the same way as $T_2/\mathbb{Z}_2^{\textrm{(t)}}$ studied in Subsection \ref{subsubsec:three-generation-models}, we can find three-generation models on $(T^2_1\times T^2_2)/\mathbb{Z}_2^{\textrm{(per)}}$ and $(T^2_1\times T^2_2)/(\mathbb{Z}_2^{\textrm{(t)}}\times\mathbb{Z}_2^{\textrm{(per)}})$.
Then we assume the following structures of 4D Yukawa couplings,
\begin{align}
&Y^{jk}_{(T^2_1\times T^2_2)/\mathbb{Z}_2^{\textrm{(per)}}\times T^2_3}(\tau,\tau_3)
= Y^{jk}_{(T^2_1\times T^2_2)/\mathbb{Z}_2^{\textrm{(per)}}}(\tau) \times
Y_{T^2_3}(\tau_3), \\
&Y^{jk}_{(T^2_1\times T^2_2)/(\mathbb{Z}_2^{\textrm{(t)}}\times\mathbb{Z}_2^{\textrm{(per)}})\times T^2_3/\mathbb{Z}_2}(\tau,\tau_3)
= Y^{jk}_{(T^2_1\times T^2_2)/(\mathbb{Z}_2^{\textrm{(t)}}\times\mathbb{Z}_2^{\textrm{(per)}})}(\tau) \times
Y_{T^2_3/\mathbb{Z}_2}(\tau_3).
\end{align}
That is, three-generations are originated from the zero-modes on $(T^2_1\times T^2_2)/\mathbb{Z}_2^{\textrm{(per)}}$ and $(T^2_1\times T^2_2)/(\mathbb{Z}_2^{\textrm{(t)}}\times\mathbb{Z}_2^{\textrm{(per)}})$.
Note that to obtain non-vanishing Yukawa couplings, Eqs.~(\ref{eq:non-vanishing_perm}) and (\ref{eq:non-vanishing_twist}) are required.
In Table \ref{tab:three-generation-models_T2xT2/Z2}, we classify all possible three-generation models on $(T^2_1\times T^2_2)/\mathbb{Z}_2^{\textrm{(per)}}$ with non-vanishing Yukawa couplings.
We show ones on $(T^2_1\times T^2_2)/(\mathbb{Z}_2^{\textrm{(t)}}\times\mathbb{Z}_2^{\textrm{(per)}})$ in Appendix \ref{app:three_T2xT2dZ2xZ2}.
\begin{table}[H]
\centering
\begin{tabular}{c|c|c|c} \hline
$M_L;n_L$ & $M_R;n_R$ & $M_H;n_H$ & $g_H$ \\ \hline\hline
$2;0$ & $2;0$ & $4;0$ & 10 \\
$3;1$ & $2;0$ & $5;1$ & 10 \\
$3;1$ & $3;1$ & $6;0$ & 21 \\ \hline
\end{tabular}
\caption{Possible three-generation models with non-vanishing Yukawa couplings on $(T^2_1\times T^2_2)/\mathbb{Z}_2^{\textrm{(per)}}$.
The first, second and third columns show (flux; $\mathbb{Z}_2^{\textrm{(per)}}$ parity) of $\psi_L$, $\psi_R$ and $\psi_H$, respectively.
The fourth column shows the number of Higgs fields.
We omit three-generation models which is equivalent to the model shown in this table by flipping left- and right-handed matter fields.
}
\label{tab:three-generation-models_T2xT2/Z2}
\end{table}


\subsection{Numerical example on $(T^2_1\times T^2_2)/(\mathbb{Z}_2^{\textrm{(t)}}\times\mathbb{Z}_2^{\textrm{(per)}})$}

Here we show a numerical example on $(T^2_1\times T^2_2)/(\mathbb{Z}_2^{\textrm{(t)}}\times\mathbb{Z}_2^{\textrm{(per)}})$.
We focus on the quark flavors to make our analysis simple.
We use the following model on $(T^2_1\times T^2_2)/(\mathbb{Z}_2^{\textrm{(t)}}\times\mathbb{Z}_2^{\textrm{(per)}})$,
\begin{align}
\begin{aligned}
&(M_Q;m_Q;n_Q;\alpha_{1Q},\alpha_{2Q}) = (2;0;0;0,0), \\
&(M_{u_R};m_{u_R};n_{u_R};\alpha_{1u_R},\alpha_{2u_R}) = (2;0;0;0,0), \\
&(M_{d_R};m_{d_R};n_{d_R};\alpha_{1d_R},\alpha_{2d_R}) = (2;0;0;0,0), \\
&(M_{H_u};m_{H_u};n_{H_u};\alpha_{1H_u},\alpha_{2H_u}) = (4;0;0;0,0), \\
&(M_{H_d};m_{H_d};n_{H_d};\alpha_{1H_d},\alpha_{2H_d}) = (4;0;0;0,0), \\
\end{aligned}
\end{align}
which has six zero-modes of Higgs fields.
The zero-modes of left-handed quark doublets $Q=(u_L,d_L)^T$, right-handed up-sector (down-sector) quarks $u_R$ ($d_R$) and up (down) type Higgs fields are shown in Table \ref{tab:zero-modes_numerical_example_on_T2xT2dZ2xZ2}.
\begin{table}[H]
\centering
\renewcommand{\arraystretch}{1.2}
\begin{tabular}{c|c|c} \hline
$\vec{j}$ & $Q$, $u_R$, $d_R$ & $H_u$, $H_d$ \\ \hline
(0,0) & $\psi_{0,0}^{((0,0),(0,0),2)}(z_1,z_2,\tau)$ & $\psi_{0,0}^{((0,0),(0,0),2)}(z_1,z_2,\tau)$ \\
(1,0) & $\psi_{0,0}^{((1,0),(0,0),2)}(z_1,z_2,\tau)$ & $\psi_{0,0}^{((1,0),(0,0),2)}(z_1,z_2,\tau)$ \\
(1,1) & $\psi_{0,0}^{((1,1),(0,0),2)}(z_1,z_2,\tau)$ & $\psi_{0,0}^{((1,1),(0,0),2)}(z_1,z_2,\tau)$ \\
(2,0) & - & $\psi_{0,0}^{((2,0),(0,0),2)}(z_1,z_2,\tau)$ \\
(2,1) & - & $\psi_{0,0}^{((2,1),(0,0),2)}(z_1,z_2,\tau)$ \\
(2,2) & - & $\psi_{0,0}^{((2,2),(0,0),2)}(z_1,z_2,\tau)$ \\ \hline
\end{tabular}
\caption{Zero-modes of left-handed quark doublets $Q=(u_L,d_L)^T$, right-handed up-sector (down-sector) quarks $u_R$ ($d_R$) and up (down) type Higgs fields.}
\label{tab:zero-modes_numerical_example_on_T2xT2dZ2xZ2}
\end{table}

Yukawa couplings for up and down-sectors on this model, $Y^{\vec{j}\vec{k}\vec{\ell}}_u$ and $Y^{\vec{j}\vec{k}\vec{\ell}}_d$, are given by
\begin{align}
  \begin{array}{ll}
    Y^{\vec{j}\vec{k}(0,0)}_{u,d}= c_{(2\textrm{-}2\textrm{-}4)}^2
    \begin{pmatrix}
      y_{a}&0&0\\
      0&y_{b}&0\\
      0&0&y_{c}\\
    \end{pmatrix},
    &
    Y^{\vec{j}\vec{k}(1,0)}_{u,d}= c_{(2\textrm{-}2\textrm{-}4)}^2
    \begin{pmatrix}
      0&y_{d}&0\\
      y_{d}&0&y_{e}\\
      0&y_{e}&0\\
    \end{pmatrix},\\
    Y^{\vec{j}\vec{k}(1,1)}_{u,d}= c_{(2\textrm{-}2\textrm{-}4)}^2
    \begin{pmatrix}
      0&0&y_{f}\\
      0&y_{f}&0\\
      y_{f}&0&0\\
    \end{pmatrix},
    &
    Y^{\vec{j}\vec{k}(2,0)}_{u,d}= c_{(2\textrm{-}2\textrm{-}4)}^2
    \begin{pmatrix}
      \sqrt{2}y_{b}&0&0\\
      0&\frac{1}{\sqrt{2}}(y_{a}+y_{c})&0\\
      0&0&\sqrt{2}y_{b}\\
    \end{pmatrix},\\
    Y^{\vec{j}\vec{k}(2,1)}_{u,d}= c_{(2\textrm{-}2\textrm{-}4)}^2
    \begin{pmatrix}
      0&y_{e}&0\\
      y_{e}&0&y_{d}\\
      0&y_{d}&0\\
    \end{pmatrix},
    &
    Y^{\vec{j}\vec{k}(2,2)}_{u,d}= c_{(2\textrm{-}2\textrm{-}4)}^2
    \begin{pmatrix}
      y_{c}&0&0\\
      0&y_{b}&0\\
      0&0&y_{a}\\
    \end{pmatrix},
  \end{array}
\end{align}
where
\begin{align}
  \begin{array}{ll}
    y_{a}=(\eta_0^{(16)}+\eta_8^{(16)})^2, & y_{b}=2(\eta_0^{(16)}+\eta_8^{(16)})\eta_4^{(16)},\\
    y_{c}=4(\eta_4^{(16)})^2, & y_{d}=\sqrt{2}(\eta_0^{(16)}+\eta_8^{(16)})(\eta_2^{(16)}+\eta_6^{(16)}),\\
    y_{e}=2\sqrt{2}\eta_4^{(16)}(\eta_2^{(16)}+\eta_6^{(16)}), & y_{f}=2(\eta_2^{(16)}+\eta_6^{(16)})^2.
  \end{array} \notag
\end{align}
Here we have used the notation,
\begin{align}
\eta_N^{(M)} = \vartheta
\begin{bmatrix}\frac{N}{M}\\0\\\end{bmatrix}(0,M\tau).
\end{align}

As same as the way in Subsection \ref{subsubsec:numerical_example_T2dZ2}, we regard the directions of Higgs VEVs as free parameters.
First let us choose the following direction of Higgs VEVs,
\begin{align}
  \begin{array}{c}
    \langle H_u^{(2,2)} \rangle \neq 0, \ ({\rm other\ VEVs})=0, \\
    \langle H_d^{(2,1)} \rangle = 0.4 \langle H_d^{(2,2)} \rangle \neq 0, \ ({\rm other\ VEVs})=0.
  \end{array}
  \label{VEVs1}
\end{align}
Then mass matrices for up-sector and down-sector quarks are given by
\begin{align}
  M_u = Y_u^{\vec{j}\vec{k}\vec{\ell}}\langle H_u^{\vec{\ell}} \rangle = c_{(2\textrm{-}2\textrm{-}4)}^2
  \begin{pmatrix}
    y_c & & \\
    & y_b & \\
    & & y_a 
  \end{pmatrix}
  \langle H_u^{(2,2)} \rangle,
\end{align}
\begin{align}
  M_d = Y_d^{\vec{j}\vec{k}\vec{\ell}}\langle H_d^{\vec{\ell}} \rangle = c_{(2\textrm{-}2\textrm{-}4)}^2
  \begin{pmatrix}
    y_c & 0.4y_e & 0 \\
    0.4 y_e & y_b & 0.4y_d \\
    0 & 0.4y_d & y_a 
  \end{pmatrix}
  \langle H_d^{(2,2)} \rangle.
\end{align}
At $\tau = 1.5i$, we can obtain the mass ratios of the quarks and the absolute values of the CKM matrix elements as shown in Table \ref{MassandCKM1}.

\begin{table}[H]
  \begin{center}
    \renewcommand{\arraystretch}{1.2}
    $\begin{array}{c|c|c} \hline
      & {\rm Obtained\ values} & {\rm Observed\ values} \\ \hline
      (m_u,m_c,m_t)/m_t & (3.22\times 10^{-4},1.80\times 10^{-2},1) & (1.26\times 10^{-5},7.38\times 10^{-3},1) \\ \hline
      (m_d,m_s,m_b)/m_b & (1.02\times 10^{-3},1.24\times 10^{-2},1) & (1.12\times 10^{-3},2.22\times 10^{-2},1) \\ \hline
      |V_{\rm CKM}| \equiv |{(U_L^u)}^{\dagger}U_L^d|
      &
      \begin{pmatrix}
        0.973 & 0.230 & 0.000515 \\
        0.226 & 0.959 & 0.170 \\
        0.0395 & 0.165 & 0.986 
      \end{pmatrix}
      & 
      \begin{pmatrix}
        0.974 & 0.227 & 0.00361 \\
        0.226 & 0.973 & 0.0405 \\
        0.00854 & 0.0398 & 0.999 
      \end{pmatrix}\\ \hline
    \end{array}$
    \caption{The mass ratios of the quarks and the absolute values of the CKM matrix elements at $\tau=1.5i$ for the direction of Higgs VEVs in Eq.~(\ref{VEVs1}).
    Observed values are shown in Ref \cite{Zyla:2020zbs}.}
    \label{MassandCKM1}
  \end{center}
\end{table}

Assuming further non-vanishing VEVs of Higgs fields, more realistic results can be obtained.
Let us choose the following direction of Higgs VEVs,
\begin{align}
  \begin{array}{c}
    \langle H_u^{(2,1)} \rangle = 0.22 \langle H_u^{(2,2)} \rangle \neq 0, \ ({\rm other\ VEVs})=0, \\
    \langle H_d^{(2,0)} \rangle = -0.10 \langle H_d^{(2,2)} \rangle \neq 0,\ \langle H_d^{(2,1)} \rangle = 0.34 \langle H_d^{(2,2)} \rangle \neq 0, \ ({\rm other\ VEVs})=0.
  \end{array}
  \label{VEVs2}
\end{align}
Then mass matrices for up-sector and down-sector quarks are given by
\begin{align}
  M_u = Y_u^{\vec{j}\vec{k}\vec{\ell}}\langle H_u^{\vec{\ell}} \rangle = c_{(2\textrm{-}2\textrm{-}4)}^2
  \begin{pmatrix}
    y_c & 0.22y_e & 0 \\
    0.22y_e & y_b & 0.22y_d \\
    0 & 0.22y_d & y_a 
  \end{pmatrix}
  H_u^{(2,2)},
\end{align}
\begin{align}
  M_d = Y_d^{\vec{j}\vec{k}\vec{\ell}}\langle H_d^{\vec{\ell}} \rangle = c_{(2\textrm{-}2\textrm{-}4)}^2
  \begin{pmatrix}
    y_c-0.10\times \sqrt{2}y_b & 0.34y_e & 0 \\
    0.34 y_e & y_b-\frac{0.10}{\sqrt{2}}(y_a+y_c) & 0.34y_d \\
    0 & 0.34y_d & y_a-0.10 \times \sqrt{2}y_b 
  \end{pmatrix}
  H_d^{(2,2)}.
\end{align}
At $\tau = 1.5i$, we can obtain the mass ratios of the quarks and the absolute values of the CKM matrix elements as shown in Table \ref{MassandCKM2}.

\begin{table}[H]
  \begin{center}
  \renewcommand{\arraystretch}{1.2}
    $\begin{array}{c|c|c} \hline
      & {\rm Obtained\ values} & {\rm Observed\ values} \\ \hline
      (m_u,m_c,m_t)/m_t & (1.35\times 10^{-5},8.96\times 10^{-3},1) & (1.26\times 10^{-5},7.38\times 10^{-3},1) \\ \hline
      (m_d,m_s,m_b)/m_b & (2.08\times 10^{-3},7.20\times 10^{-2},1) & (1.12\times 10^{-3},2.22\times 10^{-2},1) \\ \hline
      |V_{\rm CKM}| \equiv |{(U_L^u)}^{\dagger}U_L^d|
      &
      \begin{pmatrix}
        0.973 & 0.229 & 0.00772 \\
        0.229 & 0.973 & 0.0403 \\
        0.00171 & 0.0410 & 0.999 
      \end{pmatrix}
      & 
      \begin{pmatrix}
        0.974 & 0.227 & 0.00361 \\
        0.226 & 0.973 & 0.0405 \\
        0.00854 & 0.0398 & 0.999 
      \end{pmatrix}\\ \hline
    \end{array}$
    \caption{The mass ratios of the quarks and the absolute values of the CKM matrix elements at $\tau=1.5i$ for the direction of Higgs VEVs in Eq.~(\ref{VEVs2}).
Observed values are shown in Ref \cite{Zyla:2020zbs}.}
    \label{MassandCKM2}
  \end{center}
\end{table}
Thus, we can also realize realistic quark flavor structures on magnetized $(T^2_1\times T^2_2)/(\mathbb{Z}_2^{\textrm{(t)}}\times\mathbb{Z}_2^{\textrm{(per)}})$.
When we consider the realization of both quark and lepton flavor structures, we should use the zero point analysis studied in Subsection \ref{subsubsec:Zero_point_analysis}.
Then we can classify the phenomenologically favorable models on $(T^2_1\times T^2_2)/(\mathbb{Z}_2^{\textrm{(t)}}\times\mathbb{Z}_2^{\textrm{(per)}})$ but we do not touch to it in this paper.


\section{Yukawa textures}
\label{subsec:Yukawa textures}

In the two previous sections, we have seen the possibilities realizing flavor structures on magnetized orbifold models.
We have obtained various three-generation models where multi Higgs fields appear as shown in Tables \ref{tab:three-generation-models_T2/Z2}, \ref{tab:three-generation-models_T2xT2/Z2} and \ref{tab:three-generation-models_T2xT2/Z2xZ2}.
To realize realistic flavor observables we have used the directions of multi Higgs VEVs as free parameters.
Then it is possible to find the phenomenologically favorable flavor models and direction of Higgs VEVs through the zero point analysis studied in Subsection \ref{subsubsec:Zero_point_analysis}.
Thus the zero point analysis is one way to realize flavor structures on magnetized orbifold models.

As an another way to approach to the flavor structures, Yukawa textures which restrict some of entries in the fermion mass matrices to zero have been studied on the bottom-up approach.
In this section, we will show that Yukawa textures also can be obtained at the modular symmetric points on magnetized orbifold models.
Furthermore, we will illustrate numerical studies for quark flavor structures using Yukawa textures.


\subsection{Yukawa textures on $T^2/\mathbb{Z}_2$}

Here, we study Yukawa textures on magnetized $T^2/\mathbb{Z}_2$.
As we will see, Yukawa textures can be realized at three modular symmetric points $\tau=i$, $\omega$ and $i\infty$ because of the residual $S$, $ST$ and $T$-symmetris.

\paragraph{$S$-symmetry}~\\
First we study Yukawa textures at $S$-symmetric point $\tau=i$.
Only if $\tau=i$, the wave functions on $T^2/\mathbb{Z}_2$ can be expanded by eigenstates of the $\mathbb{Z}_4$ twist (for eigenstates of the $\mathbb{Z}_4$ twist, see Refs.~\cite{Abe:2013bca,Abe:2014noa,Kikuchi:2020frp,Kobayashi:2017dyu}).
The $\mathbb{Z}_4$ twist is the transformation of the complex coordinate on $T^2$,
\begin{align}
z \to iz.
\end{align}
At $\tau=i$, $S$-transformation is given by
\begin{align}
S:(z,\tau) \to \left(-\frac{z}{\tau}, -\frac{1}{\tau}\right) \quad \Rightarrow \quad
S:(z,i) \to \left(-\frac{z}{i}, -\frac{1}{i}\right) = (iz, i).
\end{align}
Note that $\tau=i$ is invariant under $S$-transformation; therefore, the $\mathbb{Z}_4$ twist at $\tau=i$ corresponds to $S$-transformation.
This means that eigenstates and eigenvalues of $S$-transformation are equivalent to ones of the $\mathbb{Z}_4$ twist at $\tau=i$.
In Table \ref{tab:numZ4}, we show the number of each $\mathbb{Z}_4$ ($S$) eigenstate in the zero-modes on $T^2/\mathbb{Z}_2$ at $\tau=i$.
\begin{table}[H]
\centering
\begin{tabular}{c|cccc} \hline
\multirow{2}{*}{$\mathbb{Z}_2$ parity; SS phases; \# of generations} & \multicolumn{4}{c}{number of each $\mathbb{Z}_4$ $(S)$ eigenstate} \\
 & $\eta=1$ & $\eta=-1$ & $\eta=i$ & $\eta=-i$ \\ \hline
0; 0,0; $2n$ & $n$ & $n$ & 0 & 0 \\
0; 0,0; $2n+1$ & $n+1$ & $n$ & 0 & 0 \\
0; 1/2,1/2; $2n$ & $n$ & $n$ & 0 & 0 \\
0; 1/2,1/2; $2n+1$ & $n+1$ & $n$ & 0 & 0 \\
1; 0,0; $2n$ & 0 & 0 & $n$ & $n$ \\
1; 0,0; $2n+1$ & 0 & 0 & $n+1$ & $n$ \\
1; 1/2,1/2; $2n$ & 0 & 0 & $n$ & $n$ \\
1; 1/2,1/2; $2n+1$ & 0 & 0 & $n+1$ & $n$ \\ \hline
\end{tabular}
\caption{The number of each $\mathbb{Z}_4$ eigenstate in zero-modes on $T^2/\mathbb{Z}_2$ at $\tau=i$.
$\mathbb{Z}_2$ parity 0 and 1 denote even and odd modes, respectively.
$\eta$ denotes the eigenvalues of the $\mathbb{Z}_4$ twist.}
\label{tab:numZ4}
\end{table}
Note that only the zero-modes with SS phases (0,0) and (1/2,1/2) are mapped to themselves under the $\mathbb{Z}_4$ twist ($S$-transformation); therefore we ignore the zero-modes with $(0,1/2)$ and $(1/2,0)$.
Table \ref{tab:numZ4} means that any three-generation zero-modes with even parity consist of two $\eta=1$ and one $\eta=-1$ eigenstates.
Similarly, any three-generation zero-modes with odd parity consist of two $\eta=i$ and one $\eta=-i$ eigenstates.
Notice that $\mathbb{Z}_2$ even modes do not contain $\eta=i$ and $-i$ eigenstates since $\eta^2$ is equivalent to $\mathbb{Z}_2$ eigenvalue.
Also $\mathbb{Z}_2$ odd modes do not contain $\eta=1$ and $-1$ eigenstates.

According to Eq.~(\ref{eq:modular_invariances_of_modular_forms}), Yukawa couplings which are the modular forms satisfy
\begin{align}
Y^{jk\ell} = \widetilde{J}_{1/2}(\widetilde{S},i) \rho_L^{jj'}(\widetilde{S})
\widetilde{J}_{1/2}(\widetilde{S},i) \rho_R^{kk'}(\widetilde{S})
(\widetilde{J}_{1/2}(\widetilde{S},i) \rho_H^{\ell\ell'}(\widetilde{S}))^* Y^{j'k'\ell'}, \label{eq:ResidualS_Y}
\end{align}
at $\tau=i$ because of the residual $S$-symmetry.
Here $\rho_L$, $\rho_R$ and $\rho_H$ are unitary matrices defined by Eq.~(\ref{eq:rho_S_T2/Z2}) and $\widetilde{J}_{1/2}(\widetilde{S},i)=(-i)^{1/2}$.
On the $\mathbb{Z}_4$ eigenstates, $\widetilde{J}_{1/2}(\widetilde{S},i)\rho_{L,R,H}(\widetilde{S})$ is given by a diagonalized matix composed of $\mathbb{Z}_4$ eigenvalues.
The number of $\mathbb{Z}_4$ eigenvalues in the diagonalized matrix can be obtained from Table \ref{tab:numZ4}.
Then we can classify the structures of Yukawa matrices on $S$-eigenbasis at $\tau=i$ by use of the residual $S$-symmetry in Eq.~(\ref{eq:ResidualS_Y}).
The results are shown in Table \ref{tab:Yukawa_textures_S}.
\begin{table}[H]
\centering
\begin{tabular}{c|cccc} \hline
$\mathbb{Z}_2$ parities of & \multicolumn{4}{c}{Structures of Yukawa matrices for each $S$-eigenstate Higgs mode} \\
$(L,R,H)$ & 1 & -1 & $i$ & $-i$ \\ \hline
(0,0,0) & $\quad\begin{pmatrix}\ast&\ast&0\\\ast&\ast&0\\0&0&\ast\\\end{pmatrix}\quad$ & $\begin{pmatrix}0&0&\ast\\0&0&\ast\\\ast&\ast&0\\\end{pmatrix}$ & None & None \\
(0,1,1) & None & None & $\begin{pmatrix}\ast&\ast&0\\\ast&\ast&0\\0&0&\ast\\\end{pmatrix}$ & $\begin{pmatrix}0&0&\ast\\0&0&\ast\\\ast&\ast&0\\\end{pmatrix}$ \\
(1,0,1) & None & None & $\begin{pmatrix}\ast&\ast&0\\\ast&\ast&0\\0&0&\ast\\\end{pmatrix}$ & $\begin{pmatrix}0&0&\ast\\0&0&\ast\\\ast&\ast&0\\\end{pmatrix}$ \\
(1,1,0) & $\quad\begin{pmatrix}0&0&\ast\\0&0&\ast\\\ast&\ast&0\\\end{pmatrix}\quad$ & $\begin{pmatrix}\ast&\ast&0\\\ast&\ast&0\\0&0&\ast\\\end{pmatrix}$ & None & None \\ \hline
\end{tabular}
\caption{The structures of Yukawa matrices for each $S$-eigenstate Higgs mode.
Yukawa matrices are $S$-eigenstates.
The symbol “$\ast$” stands for nonzero matrix elements.}
\label{tab:Yukawa_textures_S}
\end{table}
As shown in Table \ref{tab:Yukawa_textures_S}, we can find two types of Yukawa textures,
\begin{align}
\begin{pmatrix}\ast&\ast&0\\\ast&\ast&0\\0&0&\ast\\\end{pmatrix}, \quad
\begin{pmatrix}0&0&\ast\\0&0&\ast\\\ast&\ast&0\\\end{pmatrix},
\end{align}
at $\tau=i$.

As a simple example of Yukawa textures, we show Yukawa matrices on the following three-generation model,
\begin{align}
\begin{aligned}
&(M_L;m_L;\alpha_{1L},\alpha_{2L}) = (4;0;0,0), \\
&(M_R;m_R;\alpha_{1R},\alpha_{2R}) = (4;0;0,0), \\
&(M_H;m_H;\alpha_{1H},\alpha_{2H}) = (8;0;0,0), \\
\end{aligned} \label{eq:448}
\end{align}
which has five zero-modes of Higgs fields.
Three-generation zero-modes of left-handed fermions are same as ones of right-handed fermions.
On $S$-eigenstates, they are transformed by
\begin{align}
\textrm{diag} \left(\widetilde{J}_{1/2}(\widetilde{S},i) \rho_{L}(\widetilde{S})\right)
= \textrm{diag} \left(\widetilde{J}_{1/2}(\widetilde{S},i) \rho_{R}(\widetilde{S})\right)
= \begin{pmatrix}
1&&\\&1&\\&&-1\\
\end{pmatrix},
\end{align}
under $S$-transformation.
On the other hand, five-generation zero-modes of Higgs fields are transformed by
\begin{align}
\textrm{diag} \left(\widetilde{J}_{1/2}(\widetilde{S},i) \rho_{H}(\widetilde{S})\right)
= \begin{pmatrix}
1&&&&\\&1&&&\\&&1&&\\&&&-1&\\&&&&-1\\
\end{pmatrix}.
\end{align}
Then the residual $S$-symmetry in Eq.~(\ref{eq:ResidualS_Y}) is given by
\begin{align}
Y^{jk\ell} = 
\begin{pmatrix}
1&&\\&1&\\&&-1\\
\end{pmatrix}^{jj'}
\begin{pmatrix}
1&&\\&1&\\&&-1\\
\end{pmatrix}^{kk'}
\begin{pmatrix}
1&&&&\\&1&&&\\&&1&&\\&&&-1&\\&&&&-1\\
\end{pmatrix}^{\ell\ell'} Y^{j'k'\ell'}.
\end{align}
This leads to the following Yukawa textures,
\begin{align}
&Y^{jk0},~Y^{jk1},~Y^{jk2} = \begin{pmatrix}\ast&\ast&0\\\ast&\ast&0\\0&0&\ast\\\end{pmatrix}, \quad
Y^{jk3},~Y^{jk4} = 
\begin{pmatrix}0&0&\ast\\0&0&\ast\\\ast&\ast&0\\\end{pmatrix}.
\end{align}

\paragraph{$ST$-symmetry}~\\
Second we study Yukawa textures at $ST$-symmetric point $\tau=\omega$.
Only if $\tau=\omega$, the wave functions on $T^2/\mathbb{Z}_2$ can be expanded by eigenstates of the $\mathbb{Z}_6$ twist (for eigenstates of the $\mathbb{Z}_6$ twist, see Refs.~\cite{Abe:2013bca,Abe:2014noa,Kobayashi:2017dyu,Kikuchi:2020frp}).
The $\mathbb{Z}_6$ twist is the transformation of the complex coordinate on $T^2$,
\begin{align}
z \to e^{\pi i/3}z = \omega^{1/2}z.
\end{align}
At $\tau=\omega$, $ST$-transformation is given by
\begin{align}
ST:(z,\tau) \to \left(-\frac{z}{\tau+1}, -\frac{1}{\tau+1}\right) \quad \Rightarrow \quad
ST:(z,\omega) \to \left(-\frac{z}{\omega+1}, -\frac{1}{\omega+1}\right) = (\omega z, \omega).
\end{align}
Note that $\tau=\omega$ is invariant under $ST$-transformation; therefore, the $\mathbb{Z}_6^2=\mathbb{Z}_3$ twist at $\tau=\omega$ corresponds to $ST$-transformation.
This means that eigenstates of $ST$-transformation are equivalent to ones of the $\mathbb{Z}_6$ twist at $\tau=\omega$.
Similarly, eigenvalues of $ST$-transformation are equivalent to ones of the $\mathbb{Z}_6^2=\mathbb{Z}_3$ twist at $\tau=\omega$.
In Table \ref{tab:numZ6}, we show the number of each $\mathbb{Z}_6$ eigenstate in the zero-modes on $T^2/\mathbb{Z}_2$ at $\tau=\omega$.
\begin{table}[H]
\centering
\begin{tabular}{c|cccccc} \hline
\multirow{2}{*}{$\mathbb{Z}_2$ parity; SS phases; \# of generations} & \multicolumn{6}{c}{number of each $\mathbb{Z}_6$ eigenstate} \\
 & $\eta=1$ & $\eta=\omega^{1/2}$ & $\eta=\omega$ & $\eta=\omega^{3/2}$ & $\eta=\omega^2$ & $\eta=\omega^{5/2}$ \\ \hline
0; 0,0; $3n$ & $n$ & 0 & $n$ & 0 & $n$ & 0 \\
0; 0,0; $3n+1$ & $n+1$ & 0 & $n$ & 0 & $n$ & 0 \\
0; 0,0; $3n+2$ & $n+1$ & 0 & $n+1$ & 0 & $n$ & 0 \\
0; 1/2,1/2; $3n$ & $n$ & 0 & $n$ & 0 & $n$ & 0 \\
0; 1/2,1/2; $3n+1$ & $n+1$ & 0 & $n$ & 0 & $n$ & 0 \\
0; 1/2,1/2; $3n+2$ & $n+1$ & 0 & $n+1$ & 0 & $n$ & 0 \\
1; 0,0; $3n$ & 0 & $n$ & 0 & $n$ & 0 & $n$ \\
1; 0,0; $3n+1$ & 0 & $n+1$ & 0 & $n$ & 0 & $n$ \\
1; 0,0; $3n+2$ & 0 & $n+1$ & 0 & $n+1$ & 0 & $n$ \\
1; 1/2,1/2; $3n$ & 0 & $n$ & 0 & $n$ & 0 & $n$ \\
1; 1/2,1/2; $3n+1$ & 0 & $n+1$ & 0 & $n$ & 0 & $n$ \\
1; 1/2,1/2; $3n+2$ & 0 & $n+1$ & 0 & $n+1$ & 0 & $n$ \\ \hline
\end{tabular}
\caption{The number of each $\mathbb{Z}_6$ eigenstate in zero-modes on $T^2/\mathbb{Z}_2$ at $\tau=\omega$.
$\mathbb{Z}_2$ parity 0 and 1 denote even and odd modes, respectively.
$\eta$ denotes the eigenvalues of the $\mathbb{Z}_6$ twist.}
\label{tab:numZ6}
\end{table}
Note that only the zero-modes with SS phases ($M/2$, $M/2$) (mod 1) are mapped to themselves under the $\mathbb{Z}_6$ twist (and $ST$-transformation); therefore we ignore the zero-modes with other SS phases.
Table \ref{tab:numZ6} means that any three-generation zero-modes with even parity consist of one $\eta=1$, one $\eta=\omega$ and one $\eta=\omega^2$ eigenstates.
Similarly, any three-generation zero-modes with odd parity consist of one $\eta=\omega^{1/2}$, one $\eta=\omega^{3/2}$ and one $\eta=\omega^{5/2}$ eigenstates.
Notice that $\mathbb{Z}_2$ even modes do not contain $\eta=\omega^{1/2}$, $\omega^{3/2}$ and $\omega^{5/2}$ eigenstates since $\eta^3$ is equivalent to $\mathbb{Z}_2$ eigenvalue.
Also $\mathbb{Z}_2$ odd modes do not contain $\eta=1$ and $\omega$ and $\omega^2$ eigenstates.

According to Eq.~(\ref{eq:modular_invariances_of_modular_forms}), Yukawa couplings which are the modular forms satisfy
\begin{align}
Y^{jk\ell} = \widetilde{J}_{1/2}(\widetilde{ST},\omega) \rho_L^{jj'}(\widetilde{ST})
\widetilde{J}_{1/2}(\widetilde{ST},\omega) \rho_R^{kk'}(\widetilde{ST})
(\widetilde{J}_{1/2}(\widetilde{ST},\omega) \rho_H^{\ell\ell'}(\widetilde{ST}))^* Y^{j'k'\ell'}, \label{eq:ResidualST_Y}
\end{align}
at $\tau=\omega$ because of the residual $ST$-symmetry.
Here $\rho_L$, $\rho_R$ and $\rho_H$ are unitary matrices defined by Eqs.~(\ref{eq:rho_S_T2/Z2}) and (\ref{eq:rho_T_T2/Z2}), and $\widetilde{J}_{1/2}(\widetilde{ST},\omega)=(-(\omega+1))^{1/2}$.
On the $\mathbb{Z}_6$ eigenstates, $\widetilde{J}_{1/2}(\widetilde{ST},\omega)\rho_{L,R,H}(\widetilde{ST})$ is given by a diagonalized matix composed of $\mathbb{Z}_6^2=\mathbb{Z}_3$ eigenvalues.
The number of $\mathbb{Z}_6^2=\mathbb{Z}_3$ eigenvalues in the diagonalized matrix can be obtained from Table \ref{tab:numZ6}.
Then we can classify the structures of Yukawa matrices on $ST$-eigenbasis at $\tau=\omega$ by use of the residual $ST$-symmetry in Eq.~(\ref{eq:ResidualST_Y}).
The results are shown in Table \ref{tab:Yukawa_textures_ST}.
\begin{table}[H]
\centering
\begin{tabular}{c|ccc} \hline
$\mathbb{Z}_2$ parities of & \multicolumn{3}{c}{Structures of Yukawa matrices for each $ST$-eigenstate Higgs mode} \\
$(L,R,H)$ & 1 & $\omega^2$ & $\omega$ \\ \hline
All patterns & $\qquad\begin{pmatrix}\ast&0&0\\0&0&\ast\\0&\ast&0\\\end{pmatrix}\qquad$ & $\qquad\begin{pmatrix}0&\ast&0\\\ast&0&0\\0&0&\ast\\\end{pmatrix}\qquad$ & $\qquad\begin{pmatrix}0&0&\ast\\0&\ast&0\\\ast&0&0\\\end{pmatrix}\qquad$ \\ \hline
\end{tabular}
\caption{The structures of Yukawa matrices for each $ST$-eigenstate Higgs mode.
Yukawa matrices are $ST$-eigenstates.
The symbol “$\ast$” stands for nonzero matrix elements.}
\label{tab:Yukawa_textures_ST}
\end{table}
As shown in Table \ref{tab:Yukawa_textures_ST}, we can find three types of Yukawa textures,
\begin{align}
\begin{pmatrix}\ast&0&0\\0&0&\ast\\0&\ast&0\\\end{pmatrix}, \quad
\begin{pmatrix}0&\ast&0\\\ast&0&0\\0&0&\ast\\\end{pmatrix}, \quad
\begin{pmatrix}0&0&\ast\\0&\ast&0\\\ast&0&0\\\end{pmatrix}, \quad
\end{align}
at $\tau=\omega$.

As a simple example of Yukawa textures, we show Yukawa matrices on the three-generation model in Eq.~(\ref{eq:448}).
On $ST$-eigenstates, three-generation zero-modes of left- and right-handed fermions are transformed by
\begin{align}
\textrm{diag} \left(\widetilde{J}_{1/2}(\widetilde{ST},\omega) \rho_L(\widetilde{ST})\right)
= \textrm{diag} \left(\widetilde{J}_{1/2}(\widetilde{ST},\omega) \rho_R(\widetilde{ST})\right)
= \begin{pmatrix}1&&\\&\omega^2&\\&&\omega\\\end{pmatrix},
\end{align}
under $ST$-transformation.
On the other hand, five-generation zero-modes of Higgs fields are transformed by
\begin{align}
\textrm{diag} \left(\widetilde{J}_{1/2}(\widetilde{ST},\omega) \rho_H(\widetilde{ST})\right)
= \begin{pmatrix}1&&&&\\&1&&&\\&&\omega^2&&\\&&&\omega^2&\\&&&&\omega\\\end{pmatrix}.
\end{align}
Then the residual $ST$-symmetry in Eq.~(\ref{eq:ResidualST_Y}) is given by
\begin{align}
Y^{jk\ell} =
\begin{pmatrix}1&&\\&\omega^2&\\&&\omega\\\end{pmatrix}^{jj'}
\begin{pmatrix}1&&\\&\omega^2&\\&&\omega\\\end{pmatrix}^{kk'}
\begin{pmatrix}1&&&&\\&1&&&\\&&\omega^2&&\\&&&\omega^2&\\&&&&\omega\\\end{pmatrix}^{\ell\ell'} Y^{j'k'\ell'}.
\end{align}
This leads to the following Yukawa textures,
\begin{align}
Y^{jk0},~Y^{jk1} = \begin{pmatrix}\ast&0&0\\0&0&\ast\\0&\ast&0\\\end{pmatrix}, \quad
Y^{jk2},~Y^{jk3} = \begin{pmatrix}0&\ast&0\\\ast&0&0\\0&0&\ast\\\end{pmatrix}, \quad
Y^{jk4} = \begin{pmatrix}0&0&\ast\\0&\ast&0\\\ast&0&0\\\end{pmatrix}.
\end{align}

\paragraph{$T$-symmetry}~\\
Third we study Yukawa textures at $T$-symmetric point $\tau=i\infty$.
Only the zero-modes with SS phases $(M/2,M/2)$ (mod 1) are mapped to themselves under $T$-transformation; therefore we ignore the zero-modes with other SS phases.
Then the zero-modes can be expanded by $T$-transformation eigenstates.

According to Eq.~(\ref{eq:modular_invariances_of_modular_forms}), Yukawa couplings which are the modular forms satisfy
\begin{align}
Y^{jk\ell} = \widetilde{J}_{1/2}(\widetilde{T},i\infty) \rho_L^{jj'}(\widetilde{T})
\widetilde{J}_{1/2}(\widetilde{T},i\infty) \rho_R^{kk'}(\widetilde{T})
(\widetilde{J}_{1/2}(\widetilde{T},i\infty) \rho_H^{\ell\ell'}(\widetilde{T}))^* Y^{j'k'\ell'}, \label{eq:ResidualT_Y}
\end{align}
at $\tau=i\infty$ because of the residual $T$-symmetry.
Here $\rho_L$, $\rho_R$ and $\rho_H$ are unitary matrices defined by Eq.~(\ref{eq:rho_T_T2/Z2}) and $\widetilde{J}_{1/2}(\widetilde{T},i\infty)=1$.
This leads to
\begin{align}
Y^{jk\ell} = Y^{jk\ell}
\exp\left[\pi i\left(\frac{(j+\alpha_{1L})^2}{M_L} + \frac{(k+\alpha_{1R})^2}{M_R} - \frac{(\ell+\alpha_{1H})^2}{M_H}\right) \right],
\end{align}
and the non-vanishing condition of the elements of Yukawa matrices,
\begin{align}
\left(\frac{(j+\alpha_{1L})^2}{M_L} + \frac{(k+\alpha_{1R})^2}{M_R} - \frac{(\ell+\alpha_{1H})^2}{M_H}\right) ~\textrm{mod}~2 = 0, \quad \textrm{otherwise}~Y^{jk\ell} = 0.
\end{align}
This condition makes almost all of elements of Yukawa matrices vanish.
As an example let us see Yukawa matrices on the three-generation model in Eq.~(\ref{eq:448}).
On $T$-eigenstates, three-generation zero-modes of left- and right-handed fermions are transformed by
\begin{align}
\textrm{diag} \left(\widetilde{J}_{1/2}(\widetilde{T},i\infty) \rho_L(\widetilde{T})\right)
= \textrm{diag} \left(\widetilde{J}_{1/2}(\widetilde{T},i\infty) \rho_R(\widetilde{T})\right)
= \begin{pmatrix}1&&\\&e^{\pi i/4}&\\&&-1\\\end{pmatrix},
\end{align}
under $T$-transformation.
On the other hand, five-generation zero-modes of Higgs fields are transformed by
\begin{align}
\textrm{diag} \left(\widetilde{J}_{1/2}(\widetilde{T},i\infty) \rho_H(\widetilde{T})\right)
= \begin{pmatrix}1&&&&\\&e^{\pi i/8}&&&\\&&e^{\pi i/2}&&\\&&&e^{9\pi i/8}&\\&&&&1\\\end{pmatrix}.
\end{align}
Then the residual $T$-symmetry in Eq.~(\ref{eq:ResidualT_Y}) is given by
\begin{align}
Y^{jk\ell} =
\begin{pmatrix}1&&\\&e^{\pi i/4}&\\&&-1\\\end{pmatrix}^{jj'}
\begin{pmatrix}1&&\\&e^{\pi i/4}&\\&&-1\\\end{pmatrix}^{kk'}
\begin{pmatrix}1&&&&\\&e^{\pi i/8}&&&\\&&e^{\pi i/2}&&\\&&&e^{9\pi i/8}&\\&&&&1\\\end{pmatrix}^{\ell\ell'} Y^{j'k'\ell'}.
\end{align}
This leads to the following Yukawa textures,
\begin{align}
Y^{jk0},~Y^{jk4} = \begin{pmatrix}\ast&0&0\\0&0&0\\0&0&\ast\\\end{pmatrix}, \quad
Y^{jk1},~Y^{jk3} = \begin{pmatrix}0&0&0\\0&0&0\\0&0&0\\\end{pmatrix}, \quad
Y^{jk2} = \begin{pmatrix}0&0&0\\0&\ast&0\\0&0&0\\\end{pmatrix}.
\end{align}
Obviously we cannot realize flavor mixings from these Yukawa textures since they are diagonalized.
Similarly, it is difficult to obtain realistic flavor observables from Yukawa textures on other three-generation models.
Therefore in what follows we avoid the discussion of Yukawa textures at $T$-symmetric point $\tau=i\infty$. 


\subsection{The Fritzch and Fritzch-Xing mass matrices}

Here we give a brief review of the Fritzch and Fritzch-Xing mass matrices.
In Ref.~\cite{Fritzsch:1979zq}, Fritzch proposed the quark mass marices with texture zeros.
It was extended in Ref.~\cite{Fritzsch:1995dj}.
(See for a review Ref.~\cite{Xing:2020ijf}.)
Additionally, various structures of texture zeros were studied \cite{Ramond:1993kv}.
Actually, phenomenologically viable four zero textures of Hermitian quark mass matrices have been investigated and it has been found that there are several possibilities.
(See e.g. Ref.~ \cite{Bagai:2021nsl} and references therein.)

The Fritzch mass matrices for up and down-sector quarks, $M_{u,d}^{(F)}$, are defined as the following Hermitian matrices,
\begin{align}
M_u^{(F)} =
\begin{pmatrix}
0 & C_u & 0 \\
C_u^* & 0 & B_u \\
0 & B_u^* & A_u \\
\end{pmatrix}, \quad
M_d^{(F)} =
\begin{pmatrix}
0 & C_d & 0 \\
C_d^* & 0 & B_d \\
0 & B_d^* & A_d \\
\end{pmatrix},
\end{align}
where $A_{u,d}$ are taken to be real and positive parameters; $B_{u,d}$ and $C_{u,d}$ are complex parameters.
In addition, $A_{u,d}\gg|B_{u,d}|\gg|C_{u,d}|$ is assumed.
These matrices are extended to the Fritzch-Xing mass matrices, $M_{u,d}^{(F\textrm{-}X)}$, by introducing one more parameter in (1,1) entry in mass matrices as
\begin{align}
M_u^{(F\textrm{-}X)} =
\begin{pmatrix}
0 & C_u & 0 \\
C_u^* & B'_u & B_u \\
0 & B_u^* & A_u \\
\end{pmatrix}, \quad
M_d^{(F\textrm{-}X)} =
\begin{pmatrix}
0 & C_d & 0 \\
C_d^* & B'_d & B_d \\
0 & B_d^* & A_d \\
\end{pmatrix},
\end{align}
where $A_{u,d}$ are taken to be real and positive parameters; $B'_{u,d}$ are real parameters; $B_{u,d}$ and $C_{u,d}$ are complex parameters.
In addition, $A_{u,d}\gg|B_{u,d}|\gg|B'_{u,d}|\gg|C_{u,d}|$ is assumed.

In what follows, we focus on three-generation model in Eq.~(\ref{eq:448}) and illustrate numerical studies for quark flavor structures.
Then we will show that the Fritzch and Fritzch-Xing mass matrices can be obtained from Yukawa textures at $ST$ and $S$-symmetric points, respectively.
Furthermore, we will obtain realistic quark flavor observables by taking appropriate directions of Higgs VEVs.
In this sense, Yukawa textures at $S$ and $ST$-symmetric points have the further possibility to realize flavor structures.
The approach using Yukawa textures can be applied to lepton flavor structures as well as quarks.
However in this paper we do not touch to it.


\subsection{Numerical example at $\tau=i$}

Here we study three-generation model in Eq.~(\ref{eq:448}) at $\tau=i$.
We will show that the Fritzch-Xing mass matrices can be realized from Yukawa textures at $S$-symmetric point $\tau=i$ on this model.

Yukawa couplings $Y^{jk\ell}$ on this model are given by
\begin{align}
  \begin{array}{ll}
    Y^{jk0} = c_{(4\textrm{-}4\textrm{-}8)} \begin{pmatrix}
    X_0 &  &  \\
    & X_1 &  \\
    &  & X_2 \\
  \end{pmatrix}, &
  Y^{jk1} = c_{(4\textrm{-}4\textrm{-}8)} \begin{pmatrix}
  & X_3 &  \\
  X_3 &  & X_4 \\
  & X_4 &  \\
  \end{pmatrix}, \\
  Y^{jk2} = c_{(4\textrm{-}4\textrm{-}8)} \begin{pmatrix}
 &  & \sqrt{2}X_1 \\
 & \frac{1}{\sqrt{2}}(X_0 + X_2) &  \\
\sqrt{2}X_1 &  &  \\
\end{pmatrix}, &
  Y^{jk3} = c_{(4\textrm{-}4\textrm{-}8)} \begin{pmatrix}
 & X_4 &  \\
X_4 &  & X_3 \\
 & X_3 &  \\
\end{pmatrix}, \\
  Y^{jk4} = c_{(4\textrm{-}4\textrm{-}8)} \begin{pmatrix}
X_2 &  &  \\
 & X_1 &  \\
 &  & X_0 \\
\end{pmatrix},
& 
\end{array} \label{YukawaMatrix448}
\end{align}
with
\begin{align}
\begin{aligned}
  &X_0 = \eta_{0}^{(128)} + 2\eta_{32}^{(128)} + \eta_{64}^{(128)}, \\
  &X_1 = \eta_{8}^{(128)} + \eta_{24}^{(128)} + \eta_{40}^{(128)} + \eta_{56}^{(128)}, \\
  &X_2 = 2(\eta_{16}^{(128)} + \eta_{48}^{(128)}), \notag \\
  &X_3 = \eta_{4}^{(128)} + \eta_{28}^{(128)} + \eta_{36}^{(128)} + \eta_{60}^{(128)}, \\
  &X_4 = \eta_{12}^{(128)} + \eta_{20}^{(128)} + \eta_{44}^{(128)} + \eta_{52}^{(128)}. \\
\end{aligned}
\end{align}
Here, we have used the notation,
\begin{align}
  \eta_N^{(M)} = \vartheta
  \begin{bmatrix}
    \frac{N}{M} \\
    0 \\
  \end{bmatrix}
  (0,M\tau).
\end{align}
On $S$-eigenstates, Yukawa matrices at $\tau=i$ are calculated as
\begin{align}
  \begin{array}{ll}
  Y^{jk0} =c_{(4\textrm{-}4\textrm{-}8)} 
  \begin{pmatrix}
    1.00 & -0.0839 & 0 \\
    -0.0839 & 0.00704 & 0 \\
    0 & 0 & 0 \\
  \end{pmatrix}, &
  Y^{jk1} =c_{(4\textrm{-}4\textrm{-}8)} 
  \begin{pmatrix}
    -0.0572 & -0.248 & 0 \\
    -0.248 & -0.943 & 0 \\
    0 & 0 & 0 \\
  \end{pmatrix}, \\
  Y^{jk2} =c_{(4\textrm{-}4\textrm{-}8)} 
  \begin{pmatrix}
    0.0683 & -0.301 & 0 \\
    -0.301 & 0.281 & 0 \\
    0 & 0 & 0.844 \\
  \end{pmatrix}, &
  Y^{jk3} =c_{(4\textrm{-}4\textrm{-}8)} 
  \begin{pmatrix}
    0 & 0 & 0 \\
    0 & 0 & -0.636 \\
    0 & -0.636 & 0 \\
  \end{pmatrix}, \\
  Y^{jk4} =c_{(4\textrm{-}4\textrm{-}8)} 
  \begin{pmatrix}
    0 & 0 & 0.602 \\
    0 & 0 & -0.158 \\
    0.602 & -0.158 & 0 \\
  \end{pmatrix}. & \\
  \end{array} \label{eq:SYukawa5}
\end{align}
Notice that the linear combination of $Y^{jk0}$, $Y^{jk1}$, $Y^{jk2}$ and $Y^{jk3}$ can form the structure of the Fritzch-Xing mass matrices.
For example, let us choose the following directions of Higgs VEVs,
\begin{align}
\begin{aligned}
&\langle H^\ell_u\rangle = v_u(1.00, 0.00838, 0, -0.000211, 0), \\
&\langle H^\ell_d\rangle = v_d(0.999, -0.0402, 0, -0.0145, 0). \\
\end{aligned} \label{eq:VEVs_S}
\end{align}
They leads to
\begin{align}
  M_u^{jk} &=
  Y^{jk\ell} \langle H^\ell_u\rangle 
  =c_{(4\textrm{-}4\textrm{-}8)} 
  \begin{pmatrix}
    1.00 & -8.60\times 10^{-2} & 0 \\
    -8.60\times 10^{-2}  & -8.53\times 10^{-4} & 1.34\times 10^{-4} \\
    0 & 1.34\times 10^{-4} & 0 \\
  \end{pmatrix},
\end{align}
\begin{align}
  M_d^{jk} &= 
  Y^{jk\ell} \langle H^\ell_d\rangle 
  =c_{(4\textrm{-}4\textrm{-}8)} 
  \begin{pmatrix}
    1.00 & -7.39\times 10^{-2} & 0 \\
    -7.39\times 10^{-2} & 4.49\times 10^{-2} & 9.20 \times 10^{-3} \\
    0 & 9.20 \times 10^{-3} & 0 \\
  \end{pmatrix}.
\end{align}
These mass matrices become the structures of the Fritzch-Xing mass matrices by the basis transformations,
\begin{align}
M_{u,d} \to 
\begin{pmatrix}0&0&1\\0&1&0\\1&0&0\\\end{pmatrix} M_{u,d} \begin{pmatrix}0&0&1\\0&1&0\\1&0&0\\\end{pmatrix}.
\end{align}
Thus the Fritzch-Xing mass matrices can be realized on three-generation model in Eq.~(\ref{eq:448}) at $\tau=i$ \footnote{The Fritzch-Xing mass matrix can be obtained by another type of string compactifictaion \cite{Kobayashi:1995ft,Kobayashi:1996ib,Kobayashi:1997kk}.}.
These mass matrices give the mass ratios of the quarks and the absolute values of the CKM matrix elements shown in Table \ref{tab:MassandCKM_S}.
\begin{table}[H]
  \begin{center}
    \renewcommand{\arraystretch}{1.2}
    $\begin{array}{c|c|c} \hline
      & {\rm Obtained\ values} & {\rm Comparison\ values} \\ \hline
      (m_u,m_c,m_t)/m_t & (2.16 \times 10^{-6},8.13\times 10^{-3},1) & (5.58\times 10^{-6},2.69\times 10^{-3},1) \\ \hline
      (m_d,m_s,m_b)/m_b & (2.02 \times 10^{-3},4.10\times 10^{-2},1) & (6.86\times 10^{-4},1.37\times 10^{-2},1) \\ \hline
      |V_{\rm CKM}| \equiv |{(U_L^u)}^{\dagger}U_L^d|
      &
      \begin{pmatrix}
        0.973 & 0.233 & 0.000550 \\
        0.233 & 0.973 & 0.00848 \\
        0.00251 & 0.00812 & 1.00
      \end{pmatrix}
      & 
      \begin{pmatrix}
        0.974 & 0.227 & 0.00361 \\
        0.226 & 0.973 & 0.0405 \\
        0.00854 & 0.0398 & 0.999 
      \end{pmatrix}\\ \hline
    \end{array}$
    \caption{The mass ratios of the quarks and the absolute values of the CKM matrix elements at $\tau=i$ under the directions of Higgs VEVs in Eq.~(\ref{eq:VEVs_S}).
    Comparison values of mass ratios are shown in Ref \cite{Bjorkeroth:2015ora}.
    Ones of the CKM matrix elements are shown in Ref \cite{Zyla:2020zbs}.}
    \label{tab:MassandCKM_S}
  \end{center}
\end{table}


\subsection{Numerical example at $\tau=\omega$}

Here we study three-generation model in Eq.~(\ref{eq:448}) at $\tau=\omega$.
We will show that the Fritzch mass matrices can be realized from Yukawa textures at $ST$-symmetric point $\tau=\omega$ on this model.

Yukawa couplings $Y^{jk\ell}$ on this model are given by Eq.~(\ref{YukawaMatrix448}).
On $ST$-eigenstates, Yukawa matrices at $\tau=\omega$ are calculated as
\begin{align}
  \begin{array}{l}
  Y^{jk0} =c_{(4\textrm{-}4\textrm{-}8)} 
  \begin{pmatrix}
    0.9535+0.04357i & 0 & 0 \\
    0 & 0 & 0 \\
    0 & 0 & 0 \\
  \end{pmatrix}, \\
  Y^{jk1} =c_{(4\textrm{-}4\textrm{-}8)} 
  \begin{pmatrix}
    0.2852-0.1027i & 0 & 0 \\
    0 & 0 & 0.8093-0.0005968i \\
    0 & 0.8093-0.0005968i & 0 \\
  \end{pmatrix}, \\
  Y^{jk2} =c_{(4\textrm{-}4\textrm{-}8)} 
  \begin{pmatrix}
    0 & -0.6454-0.06436i & 0 \\
    -0.6454-0.06436i & 0 & 0 \\
    0 & 0 & 0 \\
  \end{pmatrix}, \\
  Y^{jk3} =c_{(4\textrm{-}4\textrm{-}8)} 
  \begin{pmatrix}
    0 & 0.1615+0.1576i & 0 \\
    0.1615+0.1576i & 0 & 0 \\
    0 & 0 & -0.6802-0.5248i \\
  \end{pmatrix}, \\
  Y^{jk4} =c_{(4\textrm{-}4\textrm{-}8)} 
  \begin{pmatrix}
    0 & 0 & 0.4039+0.08034i \\
    0 & 0.1607-0.8077i & 0 \\
    0.4039+0.08034i & 0 & 0 \\
  \end{pmatrix}. \\
  \end{array} \label{eq:STYukawa5}
\end{align}
Notice that the linear combination of $Y^{jk0}$, $Y^{jk1}$ and $Y^{jk2}$ can form the structure of the Fritzch mass matrices.
For example, let us choose the following directions of Higgs VEVs,
\begin{align}
\begin{aligned}
&\langle H^\ell_u\rangle = v_u(0.9969, -0.0002465, 0.07846, 0, 0), \\
&\langle H^\ell_d\rangle = v_d(0.9900, 0.001771, 0.1409, 0, 0). \\
\end{aligned} \label{eq:VEVs_ST}
\end{align}
They leads to
\begin{align}
  &M_u^{jk} = Y^{jk\ell}\langle H^\ell_u \rangle \notag \\
  &=c_{(4\textrm{-}4\textrm{-}8)} 
  \begin{pmatrix}
    0.9505+0.04346i & -0.05064-0.005050i & 0 \\
    -0.05064-0.005050i & 0 & -(1.995-0.001471i)\times 10^{-4} \\
    0 & -(1.995-0.001471i)\times 10^{-4} & 0 \\
  \end{pmatrix}, \\
  &M_d^{jk} = Y^{jk\ell}\langle H^\ell_d \rangle \notag \\
  &=c_{(4\textrm{-}4\textrm{-}8)} 
  \begin{pmatrix}
    0.9445+0.04296i & -0.09093-0.009068i & 0 \\
    -0.09093-0.009068i  & 0 & (1.433-0.001057)\times 10^{-3} \\
    0 & (1.433-0.001057)\times 10^{-3} & 0 \\
  \end{pmatrix}.
\end{align}
These mass matrices become the structures of the Fritzch mass matrices by the basis transformations,
\begin{align}
M_{u,d} \to 
\begin{pmatrix}0&0&1\\0&1&0\\1&0&0\\\end{pmatrix} M_{u,d} \begin{pmatrix}0&0&1\\0&1&0\\1&0&0\\\end{pmatrix}.
\end{align}
Thus the Fritzch mass matrices can be realized on three-generation model in Eq.~(\ref{eq:448}) at $\tau=\omega$.
These mass matrices give the mass ratios of the quarks and the absolute values of the CKM matrix elements shown in Table \ref{tab:MassandCKM_ST}.
\begin{table}[H]
  \begin{center}
    \renewcommand{\arraystretch}{1.2}
    $\begin{array}{c|c|c} \hline
      & {\rm Obtained\ values} & {\rm Comparison\ values} \\ \hline
      (m_u,m_c,m_t)/m_t & (1.52 \times 10^{-5}, 2.86\times 10^{-3},1) & (5.58\times 10^{-6},2.69\times 10^{-3},1) \\ \hline
      (m_d,m_s,m_b)/m_b & (2.37 \times 10^{-4}, 9.41\times 10^{-3},1) & (6.86\times 10^{-4},1.37\times 10^{-2},1) \\ \hline
      |V_{\rm CKM}| \equiv |{(U_L^u)}^{\dagger}U_L^d|
      &
      \begin{pmatrix}
        0.974 & 0.228 & 0.00292 \\
        0.228 & 0.973 & 0.0421 \\
        0.00677 & 0.0416 & 0.999
      \end{pmatrix}
      & 
      \begin{pmatrix}
        0.974 & 0.227 & 0.00361 \\
        0.226 & 0.973 & 0.0405 \\
        0.00854 & 0.0398 & 0.999 
      \end{pmatrix}\\ \hline
    \end{array}$
    \caption{The mass ratios of the quarks and the absolute values of the CKM matrix elements at $\tau=\omega$ under the directions of Higgs VEVs in Eq.~(\ref{eq:VEVs_ST}).
    Comparison values of mass ratios are shown in Ref \cite{Bjorkeroth:2015ora}.
    Ones of the CKM matrix elements are shown in Ref \cite{Zyla:2020zbs}.}
    \label{tab:MassandCKM_ST}
  \end{center}
\end{table}


\chapter{4D modular symmetric flavor models}
\label{sec:4D modular symmetric flavor models}

In this chapter, we consider 4D modular symmetric flavor models.
Especially, we study the quark flavor structures on the models.
This chapter is along in Refs.~\cite{Kikuchi:2023cap,Kikuchi:2023jap}.


\section{Framework}
\label{sec:framework}

Here we consider the 4D modular invariant supersymmetric theory.
In the theory, the superpotential $W(\tau)$ is modular invariant.
The superpotential relevant to quark mass terms are written as
\begin{align}
W(\tau) = \sum_{j,k=1,2,3}\left[\alpha^{jk}Y_u^{jk}(\tau) Q^j u_R^k H_u + \beta^{jk}Y_d^{jk}(\tau)Q^jd_R^k H_d\right], \label{eq:superpotential}
\end{align}
where $Q^j=(u_L^j,d_L^j)^T$, $u_R^k$ and $d_R^j$ are superfields corresponding to three-generations of left-handed quark doublets, right-handed up-sector quarks and right-handed down-sector quarks, respectively; $H_u^\ell$ and $H_d^\ell$ are up type and down type Higgs superfields, respectively; $\alpha^{jk}$ and $\beta^{jk}$ are coupling constants.
The 4D modular invariant supersymmetric theory may be derived from the higher-dimensional theories such as the superstring theory as the low-energy effective theory.
Indeed, the magnetized orbifold models we have studied in Chapter \ref{sec:magnetized_orbifold_models} have the modular symmetry as the geometrical symmetry of the compact space.
We note that Higgs fields $H_u$ and $H_d$ in this theory are assumed to be single generations although multi generation Higgs fields appear in the magnetized orbifold models.
This is because only the lightest mass eigenstate of multi Higgs fields is expected to be not decoupled at the low-enegy scale.
Thus, the 4D modular invariant supersymmetric theory we will study in this chapter can be the candidates of the low-energy effective theory of the magnetized orbifold models.

Under the modular transformation, the 4D superfields $\Phi \in \{Q,u_R,d_R,H_u,H_d\}$ are transformed as
\begin{align}
\Phi^j \to (c\tau+d)^{-k_\Phi^j} \rho^{jk}_\Phi(\gamma)^* \Phi^j,
\end{align}
where $-k_j\leq 0$ stands for the modular weights of 4D superfields and $\rho(\gamma)$ is a unitary representation matrix of the finite modular group.
Hence, quarks are assigned into three-dimensional (redusible or irreducible) representation of a finite modular group with weight $-k_\Phi^j$ while Higgs fields are assigned into one-dimensional irreducible representation.
Then Yukawa couplings for up and down-sector quarks, $Y_u$ and $Y_d$, are transformed as
\begin{align}
&Y_u^{jk} \to (c\tau+d)^{k_{Y_u}^{jk}} \rho_Q^{jj'}(\gamma) \rho_{u_R}^{kk'}(\gamma) \rho_{H_u}(\gamma) Y_u^{j'k'}, \\
&Y_d^{jk} \to (c\tau+d)^{k_{Y_d}^{jk}} \rho_Q^{jj'}(\gamma) \rho_{d_R}^{kk'}(\gamma) \rho_{H_d}(\gamma) Y_d^{j'k'},
\end{align}
where $k_{Y_u}^{jk} = k_Q^j+k_{u_R}^k+k_{H_u}$ and $k_{Y_d}^{jk} = k_Q^j+k_{d_R}^k+k_{H_d}$.
This means Yukawa couplings are given by the modular forms of weight $k_{Y_{u,d}}^{jk}$ for the finite modular group, which are holomorphic functions of the modulus $\tau$.
The fundamental region ${\cal D}$ of the modulus $\tau$ is shown in Table \ref{fig:fundamental_region}.

Since the superpotential is modular invariant, it must be singlet under the modular transformation.
That is, the most general form of the superpotential is given by
\begin{align}
W(\tau) = &\sum_{\bm{r}_i} \left[Y_{\bm{r}_i}(\tau)
\begin{pmatrix}Q^1 & Q^2 & Q^3 \\\end{pmatrix}
\begin{pmatrix}
\alpha_{\bm{r}_i}^{11} & \alpha_{\bm{r}_i}^{12} & \alpha_{\bm{r}_i}^{13} \\
\alpha_{\bm{r}_i}^{21} & \alpha_{\bm{r}_i}^{22} & \alpha_{\bm{r}_i}^{23} \\
\alpha_{\bm{r}_i}^{31} & \alpha_{\bm{r}_i}^{32} & \alpha_{\bm{r}_i}^{33} \\
\end{pmatrix}
\begin{pmatrix}
u_R^1 \\ u_R^2 \\ u_R^3 \\
\end{pmatrix} H_u
\right]_{\bm{1}} \notag \\
&+\sum_{\bm{r}_i} \left[Y_{\bm{r}_i}(\tau)
\begin{pmatrix}Q^1 & Q^2 & Q^3 \\\end{pmatrix}
\begin{pmatrix}
\beta_{\bm{r}_i}^{11} & \beta_{\bm{r}_i}^{12} & \beta_{\bm{r}_i}^{13} \\
\beta_{\bm{r}_i}^{21} & \beta_{\bm{r}_i}^{22} & \beta_{\bm{r}_i}^{23} \\
\beta_{\bm{r}_i}^{31} & \beta_{\bm{r}_i}^{32} & \beta_{\bm{r}_i}^{33} \\
\end{pmatrix}
\begin{pmatrix}
d_R^1 \\ d_R^2 \\ d_R^3 \\
\end{pmatrix} H_d
\right]_{\bm{1}},
\end{align}
where $Y_{\bm{r}_i}(\tau)$ are modular forms belonging to all possible irreducible representations $\bm{r}_i$ of the finite modular group and choosen to make the superpotential $W(\tau)$ singlet.
Coupling constants $\alpha^{jk}_{\bm{r}_i}$ ($\beta^{jk}_{\bm{r}_i}$) may be related each other when quarks belong to multiplets.
They lead to the up and down-sector quark mass matrices, $M_u$ and $M_d$,
\begin{align}
&M_u = \sum_{\bm{r}_i} \left[Y_{\bm{r}_i}(\tau)
\begin{pmatrix}Q^1 & Q^2 & Q^3 \\\end{pmatrix}
\begin{pmatrix}
\alpha_{\bm{r}_i}^{11} & \alpha_{\bm{r}_i}^{12} & \alpha_{\bm{r}_i}^{13} \\
\alpha_{\bm{r}_i}^{21} & \alpha_{\bm{r}_i}^{22} & \alpha_{\bm{r}_i}^{23} \\
\alpha_{\bm{r}_i}^{31} & \alpha_{\bm{r}_i}^{32} & \alpha_{\bm{r}_i}^{33} \\
\end{pmatrix}
\begin{pmatrix}
u_R^1 \\ u_R^2 \\ u_R^3 \\
\end{pmatrix} \langle H_u\rangle
\right]_{\bm{1}}, \\
&M_d =\sum_{\bm{r}_i} \left[Y_{\bm{r}_i}(\tau)
\begin{pmatrix}Q^1 & Q^2 & Q^3 \\\end{pmatrix}
\begin{pmatrix}
\beta_{\bm{r}_i}^{11} & \beta_{\bm{r}_i}^{12} & \beta_{\bm{r}_i}^{13} \\
\beta_{\bm{r}_i}^{21} & \beta_{\bm{r}_i}^{22} & \beta_{\bm{r}_i}^{23} \\
\beta_{\bm{r}_i}^{31} & \beta_{\bm{r}_i}^{32} & \beta_{\bm{r}_i}^{33} \\
\end{pmatrix}
\begin{pmatrix}
d_R^1 \\ d_R^2 \\ d_R^3 \\
\end{pmatrix} \langle H_d\rangle
\right]_{\bm{1}}.
\end{align}

We comment on the normalization of modular forms.
There is an ambiguity on the normalization of modular forms.
However we expect naturally that such normalization would give ${\cal O}(1)$ contributions and not be the origin of hierarchical structures.
As mentioned above, our models may originate from the compactification of higher-dimensional theory such as the magnetized orbifold models of superstring theory.
Then obtained values in our models appear in high energy scale such as the GUT scale, $2\times 10^{16}$ GeV, and they are affected by the renormalization group effects through some factors, although those effects depend on the scenario.
For example, renormalization group effects in the minimal supersymmetric scenario were studied in Refs.~\cite{Antusch:2013jca,Bjorkeroth:2015ora}.
Therefore, we should realize the values of the quark flavor structures at high energy scale up to ${\cal O}(1)$.
In the following analysis, we use the values at the GUT scale, $2\times 10^{16}$ GeV, for $\tan\beta=5$ in Refs.~\cite{Antusch:2013jca,Bjorkeroth:2015ora} as reference values. 


\section{Quark flavor models without fine-tuning}

Next let us review the modular symmetric quark flavor models without fine-tuning.
We assume ${\cal O}(1)$ sizes of coupling constants $\alpha^{jk}_{\bm{r}_i}$ and $\beta^{jk}_{\bm{r}_i}$ on the quark mass matrices because we do not explain large mass hierarchies of quarks by use of them.
We expect that large mass hierarchies of quarks originate from only the value of the modulus $\tau$.

In the vicinity of the modular symmetric points, the modular forms take hierarchical values depending on their residual charges.
This can be understood as follows.
At the modular symmetric points, $\tau=i$, $\omega$ and $i\infty$, residual $Z_2$, $Z_3$ and $Z_N$ symmetries exist respectively, where $N$ denotes a level of the finite modular group.
For example, $A_4$ modular symmetry is broken to $Z_3$ residual symmetry at $\tau=i\infty$ since $T^3=\mathbb{I}$.
To study the hierarchical structures of the modular forms in the vicinity of the symmetric points, let us introduce following three quantities,
\begin{align}
\varepsilon_S \equiv \frac{\tau-i}{\tau+i}, \quad
\varepsilon_{ST} \equiv \frac{\tau-\omega}{\tau-\omega^2}, \quad
\varepsilon_T \equiv e^{2\pi i\tau/N}.
\end{align}
Notice that they represent the deviations from the symmetric points.
They are transformed as
\begin{align}
\varepsilon_S \xrightarrow{S} -\varepsilon_S, \quad
\varepsilon_{ST} \xrightarrow{ST} \omega^2 \varepsilon_{ST}, \quad
\varepsilon_T \xrightarrow{T} e^{2\pi i/N} \varepsilon_T,
\end{align}
under $S$, $ST$ and $T$-transformations, respectively.
We define residual $S$, $ST$ and $T$-charges as $r$ in eigenvalues $(-1)^r$, $\omega^{2r}$ and $e^{2\pi ir/N}$ for $S$, $ST$ and $T$-transformations.
Hence, $\varepsilon_S$, $\varepsilon_{ST}$ and $\varepsilon_T$ have residual $S$-charge 1, $ST$-charge 1 and $T$-charge 1, respectively.
Therefore, the modular forms $f^j(\tau)$ with residual $S$-charge $r_S$ in the vicinity of the symmetric point $\tau=i$ can be expanded by $\varepsilon_S$ as \cite{Novichkov:2021evw}
\begin{align}
&f^j(\tau) \simeq \sum_{n=0}^\infty c_n\varepsilon_S^{r_S+2n} \simeq c_0 \varepsilon_S^{r_S}, \quad \tau\sim i,
\end{align}
where $c_n$ denote constant coefficients and $\varepsilon_S \ll 1$ at $\tau\simeq i$.
Similarly, the modular forms $f^j(\tau)$ with residual $ST$-charge $r_{ST}$ in the vicinity of the symmetric point $\tau=\omega$ can be expanded by $\varepsilon_{ST}$ as
\begin{align}
&f^j(\tau) \simeq \sum_{n=0}^\infty c_n\varepsilon_{ST}^{r_{ST}+3n} \simeq c_0 \varepsilon_{ST}^{r_{ST}}, \quad \tau \sim \omega, \label{eq:ex_by_q_ST}
\end{align}
where $\varepsilon_{ST} \ll 1$ at $\tau\simeq \omega$.
The modular forms $f^j(\tau)$ with residual $T$-charge $r_{T}$ in the vicinity of the symmetric point $\tau=i\infty$ can be expanded by $\varepsilon_{T}$ as
\begin{align}
&f^j(\tau) \simeq \sum_{n=0}^\infty c_n\varepsilon_{T}^{r_T+Nn} \simeq c_0 \varepsilon_{T}^{r_{T}}, \quad \tau \sim i\infty, \label{eq:f expanded by eT}
\end{align}
where $\varepsilon_T \ll 1$ at $\tau\simeq i\infty$.
Thus the modular forms in the vicinity of the symmetric points can generate large hierarchies and become the origin of the quark mass hierarchies.

As an illustlating example, we suppose that quark doublets $Q$, up-sector quark singlets $u_R$ and up type Higgs field $H_u$ with the residual $ST$-charges,
\begin{align}
Q:(1,1,0), \quad u_R:(0,1,0), \quad H_u:0.
\end{align}
Then the elements of the mass matrix of up-sector quarks $M_u$ have the following $ST$-charges,
\begin{align}
M_u:
\begin{pmatrix}
2 & 1 & 2 \\
2 & 1 & 2 \\
0 & 2 & 0 \\
\end{pmatrix}.
\end{align}
According to Eq.~(\ref{eq:ex_by_q_ST}), the sizes of elements of $M_u$ are approximately given by
\begin{align}
|M_u| \simeq
\begin{pmatrix}
|\varepsilon_{ST}|^2 & |\varepsilon_{ST}| & |\varepsilon_{ST}|^2 \\
|\varepsilon_{ST}|^2 & |\varepsilon_{ST}| & |\varepsilon_{ST}|^2 \\
1 & |\varepsilon_{ST}|^2 & 1 \\
\end{pmatrix},
\end{align}
at $\tau\simeq\omega$.

In this way, hierarchical structures of mass matrices can be obtained in the vicinity of the symmetric points.
Nevertheless the realization of the quark flavor structure is not straightforward.
We need to obtain the quark mass ratios, $m_u/m_t\sim 10^{-5}$, $m_c/m_t\sim 10^{-2}$, $m_d/m_b\sim 10^{-3}$ and $m_s/m_b\sim 10^{-2}$.
When we take $\varepsilon = {\cal O}(0.1)$, $m_u/m_t\sim 10^{-5}$ can be realized by the residual charge 5 as $\varepsilon^5\sim{\cal O}(10^{-5})$.
This means the residual $Z_N$ symmetries with $N\geq 6$ are required to realize quark mass hierarchies.
Such residual symmetries are yielded only at the cusp, $\tau=i\infty$, for the finite modular groups satisfying $T^N=\mathbb{I}$ with $N\geq 6$.
When we consider the products of the finite modular groups, we can relax this requirement.
For example, let us consider the products of the finite modular group $A_4\times A_4\times A_4$ controled by three moduli $\tau_1$, $\tau_2$ and $\tau_3$.
In this case, each $A_4$ generates residual $Z_3$ symmetry at both symmetric points, $\omega$ and $i\infty$.
Thus the residual $Z_3\times Z_3 \times Z_3$ symmetry and the hierarchical value $\varepsilon^6$ are realized in the vicinity of $\omega$ and $i\infty$.


\section{The models with $\Gamma_6$ modular symmetry}
\label{subsec:Gamma_6}

As a finite modular group breaking into residual $Z_N$ symmetry with $N\geq 6$ at $\tau=i\infty$, first we consider the finite modular symmetry at level 6, $\Gamma_6\simeq S_3\times A_4$.
$\Gamma_6$ modular group has two generators, $S$ and $T$-transformations, and they satisfy the algebraic relations,
\begin{align}
S^2 = (ST)^3 = T^6 = ST^2ST^3ST^4ST^3 = \mathbb{I}.
\end{align}
Thus it can be expected that the hierarchical value $\varepsilon^5$ are realized in the vicinity of $\tau=i\infty$.

To make our analysis simple, we use only $\Gamma_6$ singlets and assign Higgs fields $H_u$ and $H_d$ into $\Gamma_6$ trivial singlet $\bm{1}_0^0$ with weight 0.
In addition, we restrict all coupling constants $\alpha_{\bm{r}_i}^{jk}$ and $\beta_{\bm{r}_i}^{jk}$ to $\pm 1$ in quark mass matrices to avoid the fine-tuning by them.
Since we require the hierarchical value $\varepsilon^5$, we focus on the vicinity of the cusp, $\tau=i\infty$.
We use only the deviation of the modulus from the cusp (and the choices of $+1$ or $-1$ in $\alpha_{\bm{r}_i}^{jk}$ and $\beta_{\bm{r}_i}^{jk}$) as a free parameter.

As reviewed in Appendix \ref{app:gt_Gamma_6}, in $\Gamma_6$ modular group, six singlets, $\bm{1}_0^0$, $\bm{1}_1^0$, $\bm{1}_2^0$, $\bm{1}_0^1$, $\bm{1}_1^1$ and $\bm{1}_2^1$ exist.
They have six different $T$-charges up to 5.
Therefore $\varepsilon$ to the powers up to 5 can appear in mass matrices.
Table \ref{tab:T-charges_of_six_singlets} shows $T$-charges and the orders of the modular forms of each $\Gamma_6$ singlet.
\begin{table}[H]
\centering
\begin{tabular}{c|cccccc} \hline
Singlets & $\bm{1}_0^0$ & $\bm{1}_2^1$ & $\bm{1}_1^0$ & $\bm{1}_0^1$ & $\bm{1}_2^0$ & $\bm{1}_1^1$ \\ \hline
$T$-charges & 0 & 1 & 2 & 3 & 4 & 5 \\
Orders & $1$ & $\varepsilon$ & $\varepsilon^2$ & $\varepsilon^3$ & $\varepsilon^4$ & $\varepsilon^5$ \\ \hline
\end{tabular}
\caption{$T$-charges of six $\Gamma_6$ singlets and their orders in the vicinity of $\tau=i\infty$.}
\label{tab:T-charges_of_six_singlets}
\end{table}
To realize large mass hierarchies of quarks, we concentrate on the following four types of mass matrices,
\begin{align}
  &\textrm{Type~I:} \quad M_u \propto
  \begin{pmatrix}
    \varepsilon^5 & \varepsilon^{3-a+b} & \varepsilon^b \\
    \pm\varepsilon^{5+a-b} & \pm\varepsilon^3 & \pm\varepsilon^a \\
    \pm\varepsilon^{5-b} & \pm\varepsilon^{3-a} & \pm1 \\
  \end{pmatrix}, \quad
  M_d \propto
  \begin{pmatrix}
    \varepsilon^3 & \varepsilon^{2-a+b} & \varepsilon^b \\
    \pm\varepsilon^{3+a-b} & \pm\varepsilon^2 & \pm\varepsilon^a \\
    \pm\varepsilon^{3-b} & \pm\varepsilon^{2-a} & \pm1 \\
  \end{pmatrix}, \label{eq:firstmassmtx} \\
  &\textrm{Type~II:} \quad M_u \propto
  \begin{pmatrix}
    \varepsilon^5 & \varepsilon^{3-a+b} & \varepsilon^b \\
    \pm\varepsilon^{5+a-b} & \pm\varepsilon^3 & \pm\varepsilon^a \\
    \pm\varepsilon^{5-b} & \pm\varepsilon^{3-a} & \pm1 \\
  \end{pmatrix}, \quad
  M_d \propto
  \begin{pmatrix}
    \varepsilon^4 & \varepsilon^{2-a+b} & \varepsilon^b \\
    \pm\varepsilon^{4+a-b} & \pm\varepsilon^2 & \pm\varepsilon^a \\
    \pm\varepsilon^{4-b} & \pm\varepsilon^{2-a} & \pm1 \\
  \end{pmatrix}, \label{eq:secondmassmtx} \\
  &\textrm{Type~III:} \quad M_u \propto
  \begin{pmatrix}
    \varepsilon^5 & \varepsilon^{2-a+b} & \varepsilon^b \\
    \pm\varepsilon^{5+a-b} & \pm\varepsilon^2 & \pm\varepsilon^a \\
    \pm\varepsilon^{5-b} & \pm\varepsilon^{2-a} & \pm1 \\
  \end{pmatrix}, \quad
  M_d \propto
  \begin{pmatrix}
    \varepsilon^3 & \varepsilon^{2-a+b} & \varepsilon^b \\
    \pm\varepsilon^{3+a-b} & \pm\varepsilon^2 & \pm\varepsilon^a \\
    \pm\varepsilon^{3-b} & \pm\varepsilon^{2-a} & \pm1 \\
  \end{pmatrix}, \label{eq:thirdmassmtx} \\
  &\textrm{Type~IV:} \quad M_u \propto
  \begin{pmatrix}
    \varepsilon^5 & \varepsilon^{2-a+b} & \varepsilon^b \\
    \pm\varepsilon^{5+a-b} & \pm\varepsilon^2 & \pm\varepsilon^a \\
    \pm\varepsilon^{5-b} & \pm\varepsilon^{2-a} & \pm1 \\
  \end{pmatrix}, \quad
  M_d \propto
  \begin{pmatrix}
    \varepsilon^4 & \varepsilon^{2-a+b} & \varepsilon^b \\
    \pm\varepsilon^{4+a-b} & \pm\varepsilon^2 & \pm\varepsilon^a \\
    \pm\varepsilon^{4-b} & \pm\varepsilon^{2-a} & \pm1 \\
  \end{pmatrix}, \label{eq:fourthmassmtx}
\end{align}
where $\pm$ corresponds any possible combinations of signs and $a,b \in \{0,1,2,3,4,5\}$.
Without loss of generality it is possible to fix the signs of (1,1), (1,2) and (1,3) elements of $M_u$ and $M_d$ to $+1$ by the basis transformation for right-handed quarks $u_R$ and $d_R$.
The powers of $\varepsilon$ on diagonal elements in $M_u$ and $M_d$ are (5,3,0) and (3,2,0) for type I, (5,3,0) and (4,2,0) for type II, (5,2,0) and (3,2,0) for type III and (5,2,0) and (4,2,0) for type IV.
These powers seem to be suitable in order to realize hierarchical quark mass ratios.
Four types of mass matrices can be led by the following singlet assignments of quarks,
\begin{align}
  &\textrm{Type~I}: \notag \\
  &Q = (\bm{1}^{b~\textrm{mod~2}}_{b~\textrm{mod~3}},\bm{1}^{a~\textrm{mod~2}}_{a~\textrm{mod~3}},\bm{1^0_0}), ~ u_R = (\bm{1}^{5-b~\textrm{mod~2}}_{5-b~\textrm{mod~3}},\bm{1}^{3-a~\textrm{mod~2}}_{3-a~\textrm{mod~3}},\bm{1^0_0}), ~ d_R = (\bm{1}^{3-b~\textrm{mod~2}}_{3-b~\textrm{mod~3}},\bm{1}^{2-a~\textrm{mod~2}}_{2-a~\textrm{mod~3}},\bm{1^0_0}), \\
  &\textrm{Type~II}: \notag \\
  &Q = (\bm{1}^{b~\textrm{mod~2}}_{b~\textrm{mod~3}},\bm{1}^{a~\textrm{mod~2}}_{a~\textrm{mod~3}},\bm{1^0_0}), ~ u_R = (\bm{1}^{5-b~\textrm{mod~2}}_{5-b~\textrm{mod~3}},\bm{1}^{3-a~\textrm{mod~2}}_{3-a~\textrm{mod~3}},\bm{1^0_0}), ~ d_R = (\bm{1}^{4-b~\textrm{mod~2}}_{4-b~\textrm{mod~3}},\bm{1}^{2-a~\textrm{mod~2}}_{2-a~\textrm{mod~3}},\bm{1^0_0}), \\
  &\textrm{Type~III}: \notag \\
  &Q = (\bm{1}^{b~\textrm{mod~2}}_{b~\textrm{mod~3}},\bm{1}^{a~\textrm{mod~2}}_{a~\textrm{mod~3}},\bm{1^0_0}), ~ u_R = (\bm{1}^{5-b~\textrm{mod~2}}_{5-b~\textrm{mod~3}},\bm{1}^{2-a~\textrm{mod~2}}_{2-a~\textrm{mod~3}},\bm{1^0_0}), ~ d_R = (\bm{1}^{3-b~\textrm{mod~2}}_{3-b~\textrm{mod~3}},\bm{1}^{2-a~\textrm{mod~2}}_{2-a~\textrm{mod~3}},\bm{1^0_0}), \\
  &\textrm{Type~IV}: \notag \\
  &Q = (\bm{1}^{b~\textrm{mod~2}}_{b~\textrm{mod~3}},\bm{1}^{a~\textrm{mod~2}}_{a~\textrm{mod~3}},\bm{1^0_0}), ~ u_R = (\bm{1}^{5-b~\textrm{mod~2}}_{5-b~\textrm{mod~3}},\bm{1}^{2-a~\textrm{mod~2}}_{2-a~\textrm{mod~3}},\bm{1^0_0}), ~ d_R = (\bm{1}^{4-b~\textrm{mod~2}}_{4-b~\textrm{mod~3}},\bm{1}^{2-a~\textrm{mod~2}}_{2-a~\textrm{mod~3}},\bm{1^0_0}).
\end{align}
However, there are the cases that these assignments cannot lead to the mass matrices in four types.
This is because all of the singlet modular forms of $\Gamma_6$ with certain $T$-charges do not exist for weights less than 14.
For example, the modular forms of weight 12 belong to $\bm{1}^1_1$ do not exist as shown in Appendix \ref{app:modular_forms_6}.
Thus, some of mass matrix elements may vanish due to the shortage of the singlet modular forms for weights less than 14.
We study the case of Yukawa couplings of the weight 14 in Subsection \ref{subsubsec:weight14} and one of weights less than 14 in Subsection \ref{subsubsec:weight<14}.


\subsection{Weight 14}
\label{subsubsec:weight14}

Here we study the models with Yukawa couplings of weight 14 where all of mass matrix elements do not vanish.
As a benchmark point of the modulus, let us choose $\tau=3.2i$ around the cusp.
As shown in Appendix \ref{app:modular_forms_6}, seven singlet modular forms of weight 14, $Y_{\bm{1}^0_0}^{(14)}$, $Y_{\bm{1}^1_2i}^{(14)}$, $Y_{\bm{1}^0_1}^{(14)}$, $Y_{\bm{1}^1_0}^{(14)}$, $Y_{\bm{1}^0_2}^{(14)}$, $Y_{\bm{1}^1_1}^{(14)}$ and $Y_{\bm{1}^1_2ii}^{(14)}$, exist. At the benchmark point $\tau=3.2i$, they are evaluated as
\begin{align}
\begin{aligned}
  &Y_{\bm{1}^0_0}^{(14)}/Y_{\bm{1}^0_0}^{(14)} = 1 \rightarrow 1, \quad
  Y_{\bm{1}^1_2i}^{(14)}/Y_{\bm{1}^0_0}^{(14)} = 0.172 \rightarrow \varepsilon, \\
  &Y_{\bm{1}^0_1}^{(14)}/Y_{\bm{1}^0_0}^{(14)} = 2.08\times 10^{-2} \rightarrow \varepsilon^2, \quad
  Y_{\bm{1}^1_0}^{(14)}/Y_{\bm{1}^0_0}^{(14)} = 3.58\times 10^{-3} \rightarrow \varepsilon^3, \\
  &Y_{\bm{1}^0_2}^{(14)}/Y_{\bm{1}^0_0}^{(14)} = 4.35\times 10^{-4} \rightarrow \varepsilon^4, \quad
  Y_{\bm{1}^1_1}^{(14)}/Y_{\bm{1}^0_0}^{(14)} = 7.46\times 10^{-5} \rightarrow \varepsilon^5, \\
  &Y_{\bm{1}^1_2ii}^{(14)}/Y_{\bm{1}^0_0}^{(14)} = 1.56\times 10^{-6} \rightarrow \varepsilon^7.
\end{aligned} \label{eq:Y_weight14_3.2i}
\end{align}
Here $Y_{\bm{1}^1_2ii}^{(14)} \sim \varepsilon^7$ originates from $Y_{\bm{1}^1_0}^{(6)} \cdot Y_{\bm{1}^0_2}^{(8)} \sim \varepsilon^3\cdot \varepsilon^4$ while $Y_{\bm{1}^1_2i}^{(14)} \sim \varepsilon$ originates from $Y_{\bm{1}^1_2}^{(6)} \cdot Y_{\bm{1}^0_0}^{(8)} \sim \varepsilon \cdot 1$.
The modular forms of order $\varepsilon^{n+6}$ for $0\leq n<6$ can appear when the modular forms of order $\varepsilon^n$ belonging to same irreducible representation exist.
Actually, at weight 14 we find the modular forms $Y_{\bm{1}^1_2i}^{(14)}\sim \varepsilon$ and $Y_{\bm{1}^1_2ii}^{(14)}\sim \varepsilon^7$.
In what follows, we ignore $Y_{\bm{1}^1_2ii}^{(14)}$ because it is ignorable for $Y_{\bm{1}^1_2i}^{(14)}$ as shown in Eq.~(\ref{eq:Y_weight14_3.2i}).
Also we assume left-handed quark doublets $Q$, right-handed up-sector quark singlets $u_R$ and right-handed down-sector quark singlets $d_R$ have weight -7 in order to make Yukawa couplings have weight 14.

\paragraph{Type I: (5,3,0) and (3,2,0)}~\\
First we study the quark flavor model in type I.
The quark mass matrices of type I are given by
\begin{align}
  &M_u =
  \begin{pmatrix}
    \alpha^{11} Y_{\bm{1}^1_1}^{(14)} & \alpha^{12} Y_{\bm{1}^{3+a-b~\textrm{mod~2}}_{3+a-b~\textrm{mod~3}}}^{(14)} & \alpha^{13} Y_{\bm{1}^{6-b~\textrm{mod~2}}_{6-b~\textrm{mod~3}}}^{(14)} \\
    \alpha^{21} Y_{\bm{1}^{1-a+b~\textrm{mod~2}}_{1-a+b~\textrm{mod~3}}}^{(14)} & \alpha^{22} Y_{\bm{1}^1_0}^{(14)} & \alpha^{23} Y_{\bm{1}^{6-a~\textrm{mod~2}}_{6-a~\textrm{mod~3}}}^{(14)} \\
    \alpha^{31} Y_{\bm{1}^{1+b~\textrm{mod~2}}_{1+b~\textrm{mod~3}}}^{(14)} & \alpha^{32} Y_{\bm{1}^{3+a~\textrm{mod~2}}_{3+a~\textrm{mod~3}}}^{(14)} & \alpha^{33} Y_{\bm{1}^0_0}^{(14)} \\
  \end{pmatrix}, \label{eq:M_u530} \\
  &M_d =
  \begin{pmatrix}
    \beta^{11} Y_{\bm{1}^1_0}^{(14)} & \beta^{12} Y_{\bm{1}^{4+a-b~\textrm{mod~2}}_{4+a-b~\textrm{mod~3}}}^{(14)} & \beta^{13} Y_{\bm{1}^{6-b~\textrm{mod~2}}_{6-b~\textrm{mod~3}}}^{(14)} \\
    \beta^{21} Y_{\bm{1}^{3-a+b~\textrm{mod~2}}_{3-a+b~\textrm{mod~3}}}^{(14)} & \beta^{22} Y_{\bm{1}^0_1}^{(14)} & \beta^{23} Y_{\bm{1}^{6-a~\textrm{mod~2}}_{6-a~\textrm{mod~3}}}^{(14)} \\
    \beta^{31} Y_{\bm{1}^{3+b~\textrm{mod~2}}_{3+b~\textrm{mod~3}}}^{(14)} & \beta^{32} Y_{\bm{1}^{4+a~\textrm{mod~2}}_{4+a~\textrm{mod~3}}}^{(14)} & \beta^{33} Y_{\bm{1}^0_0}^{(14)} \\
  \end{pmatrix}, \label{eq:M_d320}
\end{align}
where $\alpha^{jk}$ and $\beta^{jk}$ are coupling constants which we have restricted to $\pm1$.
When we choose unsuitable signs $\pm 1$, the mass matrices are approximately rank-deficient and they would lead to extremely small mass eigenvalues.
To avoid rank-deficient, we must choose appropriate signs $\pm 1$ in coupling constants.
Moreover, we need to choose the values of $a$ and $b$ which determine $T$-charges of Yukawa couplings.
Consequently, we find best-fit choices at $\tau=3.2i$,
\begin{align}
a = 2, \quad b = 3,
\end{align}
and
\begin{align}
\begin{pmatrix}
\alpha^{11} & \alpha^{12} & \alpha^{13} \\
\alpha^{21} & \alpha^{22} & \alpha^{23} \\
\alpha^{31} & \alpha^{32} & \alpha^{33} \\
\end{pmatrix} =
\begin{pmatrix}
1 & 1 & 1 \\
1 & 1 & 1 \\
-1 & -1 & 1 \\
\end{pmatrix}, \quad
\begin{pmatrix}
\beta^{11} & \beta^{12} & \beta^{13} \\
\beta^{21} & \beta^{22} & \beta^{23} \\
\beta^{31} & \beta^{32} & \beta^{33} \\
\end{pmatrix} =
\begin{pmatrix}
1 & 1 & 1 \\
1 & -1 & -1 \\
-1 & 1 & -1 \\
\end{pmatrix}. \label{eq:Choise_typeI}
\end{align}
Then $T$-charges of $Q$, $u_R$ and $d_R$ are
\begin{align}
Q = (\bm{1}^1_0,\bm{1}^0_2,\bm{1}^0_0), \quad u_R = (\bm{1}^0_2,\bm{1}^1_1,\bm{1}^0_0), \quad d_R = (\bm{1}^0_0,\bm{1}^0_0,\bm{1}^0_0). \label{eq:Assign_typeI}
\end{align}
They lead to the hierarchical quark mass matrices,
\begin{align}
M_u/Y_{\bm{1}^0_0}^{(14)} =
\begin{pmatrix}
Y_{\bm{1}^1_1}^{(14)} & Y_{\bm{1}^0_2}^{(14)} & Y_{\bm{1}^1_0}^{(14)} \\
Y_{\bm{1}^0_2}^{(14)} & Y_{\bm{1}^1_0}^{(14)} & Y_{\bm{1}^0_1}^{(14)} \\
-Y_{\bm{1}^0_1}^{(14)} & -Y_{\bm{1}^1_2i}^{(14)} & Y_{\bm{1}^0_0}^{(14)} \\
\end{pmatrix}/Y_{\bm{1}^0_0}^{(14)}
&=
\begin{pmatrix}
0.0000746 & 0.000435 & 0.00358 \\
0.000435 & 0.00358 & 0.0208 \\
-0.0208 & -0.172 & 1 \\
\end{pmatrix} \notag \\
&\sim
\begin{pmatrix}
\varepsilon^5 & \varepsilon^4 & \varepsilon^3 \\
\varepsilon^4 & \varepsilon^3 & \varepsilon^2 \\
-\varepsilon^2 & -\varepsilon & 1 \\
\end{pmatrix}, \\
M_d/Y_{\bm{1^0_0}}^{(14)} =
\begin{pmatrix}
Y_{\bm{1}^1_0}^{(14)} & Y_{\bm{1}^1_0}^{(14)} & Y_{\bm{1}^1_0}^{(14)} \\
Y_{\bm{1}^0_1}^{(14)} & -Y_{\bm{1}^0_1}^{(14)} & -Y_{\bm{1}^0_1}^{(14)} \\
-Y_{\bm{1}^0_0}^{(14)} & Y_{\bm{1}^0_0}^{(14)} & -Y_{\bm{1}^0_0}^{(14)} \\
\end{pmatrix}/Y_{\bm{1}^0_0}^{(14)}
&=
\begin{pmatrix}
0.00358 & 0.00358 & 0.00358 \\
0.0208 & -0.0208 & -0.0208 \\
-1 & 1 & -1 \\
\end{pmatrix} \notag \\
&\sim
\begin{pmatrix}
\varepsilon^3 & \varepsilon^3 & \varepsilon^3 \\
\varepsilon^2 & -\varepsilon^2 & -\varepsilon^2 \\
-1 & 1 & -1 \\
\end{pmatrix}.
\end{align}
These mass matrices give the following quark mass ratios,
\begin{align}
&(m_u,m_c,m_t)/m_t = (2.11\times 10^{-5}, 7.07\times 10^{-3},1), \\
&(m_d,m_s,m_b)/m_b = (2.91\times 10^{-3}, 1.97\times 10^{-2},1),
\end{align}
and the absolute values of the CKM matrix elements,
\begin{align}
|V_{\textrm{CKM}}| =
\begin{pmatrix}
0.973 & 0.231 & 0.000681 \\
0.231 & 0.973 & 0.0270 \\
0.00690 & 0.0261 & 1.00 \\
\end{pmatrix}.
\end{align}
Results are summarized in Table \ref{tab:type-I}.
\begin{table}[H]
  \begin{center}
    \renewcommand{\arraystretch}{1.3}
    \begin{tabular}{c|ccccccc} \hline
      & $\frac{m_u}{m_t}\times10^{6}$ & $\frac{m_c}{m_t}\times10^3$ & $\frac{m_d}{m_b}\times10^4$ & $\frac{m_s}{m_b}\times10^2$ & $|V_{\textrm{CKM}}^{us}|$ & $|V_{\textrm{CKM}}^{cb}|$ & $|V_{\textrm{CKM}}^{ub}|$ \\ \hline
      obtained values & 21.1 & 7.07 & 29.1 & 1.97 & 0.231 & 0.0270 & 0.000681 \\
      observed values & 12.6 & 7.38 & 11.2 & 2.22 & 0.227 & 0.0405 & 0.00361 \\
      GUT scale values & 5.39 & 2.80 & 9.21 & 1.82 & 0.225 & 0.0400 & 0.00353 \\ \hline
    \end{tabular}
  \end{center}
  \caption{The mass ratios of the quarks and the absolute values of the CKM matrix elements at the benchmark point $\tau=3.2i$ in the best-fit model by Eqs.~(\ref{eq:Assign_typeI}) and (\ref{eq:Choise_typeI}) of type I with Yukawa couplings of weight 14.
  Observed values  Ref.~\cite{Zyla:2020zbs} and GUT scale values with $\tan \beta=5$ \cite{Antusch:2013jca,Bjorkeroth:2015ora} are shown.}
\label{tab:type-I}
\end{table}
Again we note that our purpose is to realize the order of quark mass ratios and mixing angles without fine-tuning by coupling constants.
For this purpose, we have restricted the coupling constants $\alpha^{jk}, \beta^{jk}$ to $\pm 1$ to make our point clear.
Varying $\alpha^{jk}, \beta^{jk}={\cal O}(1)$, more realistic results can be obtained.
For example, we set
\begin{align}
\begin{aligned}
&\begin{pmatrix}
\alpha^{11} & \alpha^{12} & \alpha^{13} \\
\alpha^{21} & \alpha^{22} & \alpha^{23} \\
\alpha^{31} & \alpha^{32} & \alpha^{33} \\
\end{pmatrix} =
\begin{pmatrix}
2.547 & 1.987 & 1.052 \\
1.124 & 1.000 & 2.998 \\
-2.511 & -1.001 & 2.754 \\
\end{pmatrix}, \\
&\begin{pmatrix}
\beta^{11} & \beta^{12} & \beta^{13} \\
\beta^{21} & \beta^{22} & \beta^{23} \\
\beta^{31} & \beta^{32} & \beta^{33} \\
\end{pmatrix} =
\begin{pmatrix}
1.149 & 1.000 & 1.405 \\
2.997 & -2.999 & -1.504 \\
-2.961 & 1.664 & -1.494 \\
\end{pmatrix}.
\end{aligned} \label{eq:Choise_typeI_order_one}
\end{align}
Then, we obtain the following quark mass ratios,
\begin{align}
&(m_u,m_c,m_t)/m_t = (5.39\times 10^{-5}, 2.80\times 10^{-3},1), \\
&(m_d,m_s,m_b)/m_b = (9.21\times 10^{-3}, 1.82\times 10^{-2},1),
\end{align}
and the absolute values of the CKM matrix elements,
\begin{align}
|V_{\textrm{CKM}}| =
\begin{pmatrix}
0.974 & 0.225 & 0.00353 \\
0.225 & 0.974 & 0.0400 \\
0.00556 & 0.0398 & 0.999 \\
\end{pmatrix}.
\end{align}
Results are summarized in Table \ref{tab:type-I order one}.
Similarly, other models in this type could be realistic when we vary $\alpha^{jk}, \beta^{jk}={\cal O}(1)$.
\begin{table}[H]
  \begin{center}
    \renewcommand{\arraystretch}{1.3}
    \begin{tabular}{c|ccccccc} \hline
      & $\frac{m_u}{m_t}\times10^{6}$ & $\frac{m_c}{m_t}\times10^3$ & $\frac{m_d}{m_b}\times10^4$ & $\frac{m_s}{m_b}\times10^2$ & $|V_{\textrm{CKM}}^{us}|$ & $|V_{\textrm{CKM}}^{cb}|$ & $|V_{\textrm{CKM}}^{ub}|$ \\ \hline
      obtained values & 5.39 & 2.80 & 9.21 & 1.82 & 0.225 & 0.0400 & 0.00353 \\
      observed values & 12.6 & 7.38 & 11.2 & 2.22 & 0.227 & 0.0405 & 0.00361 \\
      GUT scale values & 5.39 & 2.80 & 9.21 & 1.82 & 0.225 & 0.0400 & 0.00353 \\ \hline
    \end{tabular}
  \end{center}
  \caption{The mass ratios of the quarks and the absolute values of the CKM matrix elements at the benchmark point $\tau=3.2i$ in the best-fit model by Eqs.~(\ref{eq:Assign_typeI}) and (\ref{eq:Choise_typeI_order_one}) of type I with Yukawa couplings of weight 14.
  Observed values  Ref.~\cite{Zyla:2020zbs} and GUT scale values with $\tan \beta=5$ \cite{Antusch:2013jca,Bjorkeroth:2015ora} are shown.}
\label{tab:type-I order one}
\end{table}

\paragraph{Type II: (5,3,0) and (4,2,0)}~\\
Second we study the quark flavor model in type II.
The quark mass matrices of type II are given by Eq.~(\ref{eq:M_u530}) and
\begin{align}
  M_d =
  \begin{pmatrix}
    \beta^{11} Y_{\bm{1}^0_2}^{(14)} & \beta^{12} Y_{\bm{1}^{4+a-b~\textrm{mod~2}}_{4+a-b~\textrm{mod~3}}}^{(14)} & \beta^{13} Y_{\bm{1}^{6-b~\textrm{mod~2}}_{6-b~\textrm{mod~3}}}^{(14)} \\
    \beta^{21} Y_{\bm{1}^{2-a+b~\textrm{mod~2}}_{2-a+b~\textrm{mod~3}}}^{(14)} & \beta^{22} Y_{\bm{1}^0_1}^{(14)} & \beta^{23} Y_{\bm{1}^{6-a~\textrm{mod~2}}_{6-a~\textrm{mod~3}}}^{(14)} \\
    \beta^{31} Y_{\bm{1}^{2+b~\textrm{mod~2}}_{2+b~\textrm{mod~3}}}^{(14)} & \beta^{32} Y_{\bm{1}^{4+a~\textrm{mod~2}}_{4+a~\textrm{mod~3}}}^{(14)} & \beta^{33} Y_{\bm{1}^0_0}^{(14)} \\
  \end{pmatrix}. \label{eq:M_d420}
\end{align}
To calculate the quark flavor observables, we need to choose the signs $\pm 1$ in coupling constants and the values of $a$ and $b$.
Consequently, we find best-fit choices at $\tau=3.2i$,
\begin{align}
a = 2, \quad b = 3,
\end{align}
and
\begin{align}
\begin{pmatrix}
\alpha^{11} & \alpha^{12} & \alpha^{13} \\
\alpha^{21} & \alpha^{22} & \alpha^{23} \\
\alpha^{31} & \alpha^{32} & \alpha^{33} \\
\end{pmatrix} =
\begin{pmatrix}
1 & 1 & 1 \\
1 & 1 & -1 \\
-1 & -1 & -1 \\
\end{pmatrix}, \quad
\begin{pmatrix}
\beta^{11} & \beta^{12} & \beta^{13} \\
\beta^{21} & \beta^{22} & \beta^{23} \\
\beta^{31} & \beta^{32} & \beta^{33} \\
\end{pmatrix} =
\begin{pmatrix}
1 & 1 & 1 \\
-1 & 1 & 1 \\
1 & 1 & -1 \\
\end{pmatrix}. \label{eq:Choise_typeII}
\end{align}
Then $T$-charges of $Q$, $u_R$ and $d_R$ are
\begin{align}
Q = (\bm{1}^1_0,\bm{1}^0_2,\bm{1}^0_0), \quad u_R = (\bm{1}^0_2,\bm{1}^1_1,\bm{1}^0_0), \quad d_R = (\bm{1}^1_1,\bm{1}^0_0,\bm{1}^0_0). \label{eq:Assign_typeII}
\end{align}
They lead to the hierarchical quark mass matrices,
\begin{align}
M_u/Y_{\bm{1^0_0}}^{(14)} =
\begin{pmatrix}
Y_{\bm{1}^1_1}^{(14)} & Y_{\bm{1}^0_2}^{(14)} & Y_{\bm{1}^1_0}^{(14)} \\
Y_{\bm{1}^0_2}^{(14)} & Y_{\bm{1}^1_0}^{(14)} & -Y_{\bm{1}^0_1}^{(14)} \\
-Y_{\bm{1}^0_1}^{(14)} & -Y_{\bm{1}^1_2i}^{(14)} & -Y_{\bm{1}^0_0}^{(14)} \\
\end{pmatrix}/Y_{\bm{1}^0_0}^{(14)}
&=
\begin{pmatrix}
0.0000746 & 0.000435 & 0.00358 \\
0.000435 & 0.00358 & -0.0208 \\
-0.0208 & -0.172 & -1 \\
\end{pmatrix} \notag \\
&\sim
\begin{pmatrix}
\varepsilon^5 & \varepsilon^4 & \varepsilon^3 \\
\varepsilon^4 & \varepsilon^3 & -\varepsilon^2 \\
-\varepsilon^2 & -\varepsilon & -1 \\
\end{pmatrix}, \\
M_d/Y_{\bm{1^0_0}}^{(14)} =
\begin{pmatrix}
Y_{\bm{1}^0_2}^{(14)} & Y_{\bm{1}^1_0}^{(14)} & Y_{\bm{1}^1_0}^{(14)} \\
-Y_{\bm{1}^1_0}^{(14)} & Y_{\bm{1}^0_1}^{(14)} & Y_{\bm{1}^0_1}^{(14)} \\
Y_{\bm{1}^1_2i}^{(14)} & Y_{\bm{1}^0_0}^{(14)} & -Y_{\bm{1}^0_0}^{(14)} \\
\end{pmatrix}/Y_{\bm{1}^0_0}^{(14)}
&=
\begin{pmatrix}
0.000435 & 0.00358 & 0.00358 \\
-0.00358 & 0.0208 & 0.0208 \\
0.172 & 1 & -1 \\
\end{pmatrix} \notag \\
&\sim
\begin{pmatrix}
\varepsilon^4 & \varepsilon^3 & \varepsilon^3 \\
-\varepsilon^3 & \varepsilon^2 & \varepsilon^2 \\
\varepsilon & 1 & -1 \\
\end{pmatrix}.
\end{align}
These mass matrices give the following quark mass ratios,
\begin{align}
&(m_u,m_c,m_t)/m_t = (2.14\times 10^{-5}, 7.00\times 10^{-3},1), \\
&(m_d,m_s,m_b)/m_b = (7.16\times 10^{-4}, 2.11\times 10^{-2},1),
\end{align}
and the absolute values of the CKM matrix elements,
\begin{align}
|V_{\textrm{CKM}}| =
\begin{pmatrix}
0.982 & 0.190 & 0.00309 \\
0.190 & 0.982 & 0.0200 \\
0.00683 & 0.0191 & 1.00 \\
\end{pmatrix}.
\end{align}
Results are summarized in Table \ref{tab:type-II}.
\begin{table}[H]
  \begin{center}
    \renewcommand{\arraystretch}{1.3}
    \begin{tabular}{c|ccccccc} \hline
      & $\frac{m_u}{m_t}\times10^{6}$ & $\frac{m_c}{m_t}\times10^3$ & $\frac{m_d}{m_b}\times10^4$ & $\frac{m_s}{m_b}\times10^2$ & $|V_{\textrm{CKM}}^{us}|$ & $|V_{\textrm{CKM}}^{cb}|$ & $|V_{\textrm{CKM}}^{ub}|$ \\ \hline
      obtained values & 21.4 & 7.00 & 7.16 & 2.11 & 0.190 & 0.0200 & 0.00309 \\
      observed values & 12.6 & 7.38 & 11.2 & 2.22 & 0.227 & 0.0405 & 0.00361 \\
      GUT scale values & 5.39 & 2.80 & 9.21 & 1.82 & 0.225 & 0.0400 & 0.00353 \\ \hline
    \end{tabular}
  \end{center}
  \caption{The mass ratios of the quarks and the absolute values of the CKM matrix elements at the benchmark point $\tau=3.2i$ in the best-fit model by Eqs.~(\ref{eq:Assign_typeII}) and (\ref{eq:Choise_typeII}) of type II with Yukawa couplings of weight 14.
  Observed values  Ref.~\cite{Zyla:2020zbs} and GUT scale values with $\tan \beta=5$ \cite{Antusch:2013jca,Bjorkeroth:2015ora} are shown.}
\label{tab:type-II}
\end{table}

\paragraph{Type III: (5,2,0) and (3,2,0)}~\\
Third we study the quark flavor model in type III.
The mass matrices of type III are given by Eq.~(\ref{eq:M_d320}) and
\begin{align}
  &M_u =
  \begin{pmatrix}
    \alpha^{11} Y_{\bm{1}^1_1}^{(14)} & \alpha^{12} Y_{\bm{1}^{4+a-b~\textrm{mod~2}}_{4+a-b~\textrm{mod~3}}}^{(14)} & \alpha^{13} Y_{\bm{1}^{6-b~\textrm{mod~2}}_{6-b~\textrm{mod~3}}}^{(14)} \\
    \alpha^{21} Y_{\bm{1}^{1-a+b~\textrm{mod~2}}_{1-a+b~\textrm{mod~3}}}^{(14)} & \alpha^{22} Y_{\bm{1}^0_1}^{(14)} & \alpha^{23} Y_{\bm{1}^{6-a~\textrm{mod~2}}_{6-a~\textrm{mod~3}}}^{(14)} \\
    \alpha^{31} Y_{\bm{1}^{1+b~\textrm{mod~2}}_{1+b~\textrm{mod~3}}}^{(14)} & \alpha^{32} Y_{\bm{1}^{4+a~\textrm{mod~2}}_{4+a~\textrm{mod~3}}}^{(14)} & \alpha^{33} Y_{\bm{1}^0_0}^{(14)} \\
  \end{pmatrix}. \label{eq:M_u520}
\end{align}
To calculate the quark flavor observables, we need to choose the signs $\pm 1$ in coupling constants and the values of $a$ and $b$.
Consequently, we find best-fit choices at $\tau=3.2i$,
\begin{align}
a = 2, \quad b = 3,
\end{align}
and
\begin{align}
\begin{pmatrix}
\alpha^{11} & \alpha^{12} & \alpha^{13} \\
\alpha^{21} & \alpha^{22} & \alpha^{23} \\
\alpha^{31} & \alpha^{32} & \alpha^{33} \\
\end{pmatrix} =
\begin{pmatrix}
1 & 1 & 1 \\
1 & -1 & -1 \\
1 & 1 & -1 \\
\end{pmatrix}, \quad
\begin{pmatrix}
\beta^{11} & \beta^{12} & \beta^{13} \\
\beta^{21} & \beta^{22} & \beta^{23} \\
\beta^{31} & \beta^{32} & \beta^{33} \\
\end{pmatrix} =
\begin{pmatrix}
1 & 1 & 1 \\
1 & 1 & -1 \\
1 & -1 & 1 \\
\end{pmatrix}. \label{eq:Choise_typeIII}
\end{align}
Then $T$-charges of $Q$, $u_R$ and $d_R$ are
\begin{align}
Q = (\bm{1}^1_0,\bm{1}^0_2,\bm{1}^0_0), \quad u_R = (\bm{1}^0_2,\bm{1}^0_0,\bm{1}^0_0), \quad d_R = (\bm{1}^0_0,\bm{1}^0_0,\bm{1}^0_0). \label{eq:Assign_typeIII}
\end{align}
They lead to the hierarchical quark mass matrices,
\begin{align}
M_u/Y_{\bm{1}^0_0}^{(14)} =
\begin{pmatrix}
Y_{\bm{1}^1_1}^{(14)} & Y_{\bm{1}^1_0}^{(14)} & Y_{\bm{1}^1_0}^{(14)} \\
Y_{\bm{1}^0_2}^{(14)} & -Y_{\bm{1}^0_1}^{(14)} & -Y_{\bm{1}^0_1}^{(14)} \\
Y_{\bm{1}^0_1}^{(14)} & Y_{\bm{1}^0_0}^{(14)} & -Y_{\bm{1}^0_0}^{(14)} \\
\end{pmatrix}/Y_{\bm{1}^0_0}^{(14)}
&=
\begin{pmatrix}
0.0000746 & 0.00358 & 0.00358 \\
0.000435 & -0.0208 & -0.0208 \\
0.0208 & 1 & -1 \\
\end{pmatrix} \notag \\
&\sim
\begin{pmatrix}
\varepsilon^5 & \varepsilon^3 & \varepsilon^3 \\
\varepsilon^4 & -\varepsilon^2 & -\varepsilon^2 \\
\varepsilon^2 & 1 & -1 \\
\end{pmatrix}, \\
M_d/Y_{\bm{1}^0_0}^{(14)} =
\begin{pmatrix}
Y_{\bm{1}^1_0}^{(14)} & Y_{\bm{1}^1_0}^{(14)} & Y_{\bm{1}^1_0}^{(14)} \\
Y_{\bm{1}^0_1}^{(14)} & Y_{\bm{1}^0_1}^{(14)} & -Y_{\bm{1}^0_1}^{(14)} \\
Y_{\bm{1}^0_0}^{(14)} & -Y_{\bm{1}^0_0}^{(14)} & Y_{\bm{1}^0_0}^{(14)} \\
\end{pmatrix}/Y_{\bm{1}^0_0}^{(14)}
&=
\begin{pmatrix}
0.00358 & 0.00358 & 0.00358 \\
0.0208 & 0.0208 & -0.0208 \\
1 & -1 & 1 \\
\end{pmatrix} \notag \\
&\sim
\begin{pmatrix}
\varepsilon^3 & \varepsilon^3 & \varepsilon^3 \\
\varepsilon^2 & \varepsilon^2 & -\varepsilon^2 \\
1 & -1 & 1 \\
\end{pmatrix}.
\end{align}
These mass matrices give the following quark mass ratios,
\begin{align}
&(m_u,m_c,m_t)/m_t = (1.04\times 10^{-4}, 2.12\times 10^{-2},1), \\
&(m_d,m_s,m_b)/m_b = (2.91\times 10^{-3}, 1.97\times 10^{-2},1),
\end{align}
and the absolute values of the CKM matrix elements,
\begin{align}
|V_{\textrm{CKM}}| =
\begin{pmatrix}
0.967 & 0.255 & 0.00000171 \\
0.255 & 0.967 & 0.00706 \\
0.00180 & 0.00682 & 1.00 \\
\end{pmatrix}.
\end{align}
Results are shown in Table \ref{tab:type-III}.
\begin{table}[H]
  \begin{center}
    \renewcommand{\arraystretch}{1.3}
    \begin{tabular}{c|ccccccc} \hline
      & $\frac{m_u}{m_t}\times10^{6}$ & $\frac{m_c}{m_t}\times10^3$ & $\frac{m_d}{m_b}\times10^4$ & $\frac{m_s}{m_b}\times10^2$ & $|V_{\textrm{CKM}}^{us}|$ & $|V_{\textrm{CKM}}^{cb}|$ & $|V_{\textrm{CKM}}^{ub}|$ \\ \hline
      obtained values & 104 & 21.2 & 29.1 & 1.97 & 0.255 & 0.00706 & 0.00000171 \\
      observed values & 12.6 & 7.38 & 11.2 & 2.22 & 0.227 & 0.0405 & 0.00361 \\
      GUT scale values & 5.39 & 2.80 & 9.21 & 1.82 & 0.225 & 0.0400 & 0.00353 \\ \hline
    \end{tabular}
  \end{center}
  \caption{The mass ratios of the quarks and the absolute values of the CKM matrix elements at the benchmark point $\tau=3.2i$ in the best-fit model by Eqs.~(\ref{eq:Assign_typeIII}) and (\ref{eq:Choise_typeIII}) of type III with Yukawa couplings of weight 14.
  Observed values  Ref.~\cite{Zyla:2020zbs} and GUT scale values with $\tan \beta=5$ \cite{Antusch:2013jca,Bjorkeroth:2015ora} are shown.}
\label{tab:type-III}
\end{table}

\paragraph{Type IV: (5,2,0) and (4,2,0)}~\\
Finally we study the quark flavor model in type IV.
The mass matrices of type IV are given by Eqs.~(\ref{eq:M_u520}) and (\ref{eq:M_d420}).
To calculate the quark flavor observables, we need to choose the signs $\pm 1$ in coupling constants and the values of $a$ and $b$.
Consequently, we find best-fit choices at $\tau=3.2i$,
\begin{align}
a = 5, \quad b = 0,
\end{align}
and
\begin{align}
\begin{pmatrix}
\alpha^{11} & \alpha^{12} & \alpha^{13} \\
\alpha^{21} & \alpha^{22} & \alpha^{23} \\
\alpha^{31} & \alpha^{32} & \alpha^{33} \\
\end{pmatrix} =
\begin{pmatrix}
1 & 1 & 1 \\
1 & 1 & 1 \\
1 & -1 & -1 \\
\end{pmatrix}, \quad
\begin{pmatrix}
\beta^{11} & \beta^{12} & \beta^{13} \\
\beta^{21} & \beta^{22} & \beta^{23} \\
\beta^{31} & \beta^{32} & \beta^{33} \\
\end{pmatrix} =
\begin{pmatrix}
1 & 1 & 1 \\
1 & -1 & -1 \\
1 & 1 & -1 \\
\end{pmatrix}. \label{eq:Choise_typeIV}
\end{align}
Then $T$-charges of $Q$, $u_R$ and $d_R$ are
\begin{align}
Q = (\bm{1}^0_0,\bm{1}^1_2,\bm{1}^0_0), \quad u_R = (\bm{1}^1_2,\bm{1}^1_0,\bm{1}^0_0), \quad d_R = (\bm{1}^0_1,\bm{1}^1_0,\bm{1}^0_0). \label{eq:Assign_typeIV}
\end{align}
They lead to the hierarchical quark mass matrices,
\begin{align}
M_u/Y_{\bm{1}^0_0}^{(14)} =
\begin{pmatrix}
Y_{\bm{1}^1_1}^{(14)} & Y_{\bm{1}^1_0}^{(14)} & Y_{\bm{1}^0_0}^{(14)} \\
Y_{\bm{1}^0_2}^{(14)} & Y_{\bm{1}^0_1}^{(14)} & Y_{\bm{1}^1_1}^{(14)} \\
Y_{\bm{1}^1_1}^{(14)} & -Y_{\bm{1}^1_0}^{(14)} & -Y_{\bm{1}^0_0}^{(14)} \\
\end{pmatrix}/Y_{\bm{1}^0_0}^{(14)}
&=
\begin{pmatrix}
0.0000746 & 0.00358 & 1 \\
0.000435 & 0.0208 & 0.0000746 \\
0.0000746 & -0.00358 & -1 \\
\end{pmatrix} \notag \\
&\sim
\begin{pmatrix}
\varepsilon^5 & \varepsilon^3 & 1 \\
\varepsilon^4 & \varepsilon^2 & \varepsilon^5 \\
\varepsilon^5 & -\varepsilon^3 & -1 \\
\end{pmatrix}, \\
M_d/Y_{\bm{1}^0_0}^{(14)} =
\begin{pmatrix}
Y_{\bm{1}^0_2}^{(14)} & Y_{\bm{1}^1_0}^{(14)} & Y_{\bm{1}^0_0}^{(14)} \\
Y_{\bm{1}^1_0}^{(14)} & -Y_{\bm{1}^0_1}^{(14)} & -Y_{\bm{1}^1_1}^{(14)} \\
Y_{\bm{1}^0_2}^{(14)} & Y_{\bm{1}^1_0}^{(14)} & -Y_{\bm{1}^0_0}^{(14)} \\
\end{pmatrix}/Y_{\bm{1}^0_0}^{(14)}
&=
\begin{pmatrix}
0.000435 & 0.00358 & 1 \\
0.00358 & -0.0208 & -0.0000746 \\
0.000435 & 0.00358 & -1 \\
\end{pmatrix} \notag \\
&\sim
\begin{pmatrix}
\varepsilon^4 & \varepsilon^3 & 1 \\
\varepsilon^3 & -\varepsilon^2 & -\varepsilon^5 \\
\varepsilon^4 & \varepsilon^3 & -1 \\
\end{pmatrix}.
\end{align}
These mass matrices give the following quark mass ratios,
\begin{align}
&(m_u,m_c,m_t)/m_t = (7.46\times 10^{-5}, 1.47\times 10^{-2},1), \\
&(m_d,m_s,m_b)/m_b = (1.01\times 10^{-3}, 1.54\times 10^{-2},1),
\end{align}
and the absolute values of the CKM matrix elements,
\begin{align}
|V_{\textrm{CKM}}| =
\begin{pmatrix}
0.974 & 0.226 & 0.0000000194 \\
0.226 & 0.974 & 0.000158 \\
0.0000358 & 0.000154 & 1.00 \\
\end{pmatrix}.
\end{align}
Results are shown in Table \ref{tab:type-IV}.
\begin{table}[H]
  \begin{center}
    \renewcommand{\arraystretch}{1.3}
    \begin{tabular}{c|ccccccc} \hline
      & $\frac{m_u}{m_t}\times10^{6}$ & $\frac{m_c}{m_t}\times10^3$ & $\frac{m_d}{m_b}\times10^4$ & $\frac{m_s}{m_b}\times10^2$ & $|V_{\textrm{CKM}}^{us}|$ & $|V_{\textrm{CKM}}^{cb}|$ & $|V_{\textrm{CKM}}^{ub}|$ \\ \hline
      obtained values & 74.6 & 14.7 & 10.1 & 1.54 & 0.226 & 0.000158 & $1.94\times 10^{-8}$ \\
      observed values & 12.6 & 7.38 & 11.2 & 2.22 & 0.227 & 0.0405 & 0.00361 \\
      GUT scale values & 5.39 & 2.80 & 9.21 & 1.82 & 0.225 & 0.0400 & 0.00353 \\ \hline
    \end{tabular}
  \end{center}
  \caption{The mass ratios of the quarks and the absolute values of the CKM matrix elements at the benchmark point $\tau=3.2i$ in the best-fit model by Eqs.~(\ref{eq:Assign_typeIV}) and (\ref{eq:Choise_typeIV}) of type IV with Yukawa couplings of weight 14.
  Observed values  Ref.~\cite{Zyla:2020zbs} and GUT scale values with $\tan \beta=5$ \cite{Antusch:2013jca,Bjorkeroth:2015ora} are shown.}
\label{tab:type-IV}
\end{table}


\subsection{Weights less than 14}
\label{subsubsec:weight<14}

Next we study the models with Yukawa couplings of weights less than 14 where some of mass matrix elements vanish because there do not exist modular forms of proper weights and representations.

\paragraph{Type III: (5,2,0) and (3,2,0) with weights 8 and 10}~\\
First we study the model where Yukawa couplings for up-sector have weight 8 and ones for down-sector have weight 10.
In addition, we consider the model in type III.
As a benchmark point of the modulus, we choose $\tau=3.7i$ around the cusp.
As shown in Appendix \ref{app:modular_forms_6}, four singlet modular forms of weight 8, $Y^{(8)}_{{\bm{1}}^0_0}$, $Y^{(8)}_{{\bm{1}}^1_2}$, $Y^{(8)}_{{\bm{1}}^0_1}$ and $Y^{(8)}_{{\bm{1}}^0_2}$ exist.
At the benchmark point $\tau=3.7i$, they are evaluated as
\begin{align}
\begin{aligned}
  &Y_{{\bm{1}^0_0}}^{(8)}/Y_{{\bm{1}^0_0}}^{(8)} = 1 \rightarrow 1, \quad
  Y_{{\bm{1}^1_2}}^{(8)}/Y_{{\bm{1}^0_0}}^{(8)} = -7.19\times 10^{-2} \rightarrow \varepsilon, \\
  &Y_{{\bm{1}^0_1}}^{(8)}/Y_{{\bm{1}^0_0}}^{(8)} = 7.32\times 10^{-3} \rightarrow \varepsilon^2, \quad
  Y_{{\bm{1}^0_2}}^{(8)}/Y_{{\bm{1}^0_0}}^{(8)} = 5.35\times 10^{-5} \rightarrow \varepsilon^4.
\end{aligned} \label{eq:modular form weight 8}
\end{align}
At weight 10, five singlet modular forms, $Y^{(10)}_{{\bm{1}}^0_0}$, $Y^{(10)}_{{\bm{1}}^1_2}$, $Y^{(10)}_{{\bm{1}}^0_1}$, $Y^{(10)}_{{\bm{1}}^1_0}$ and $Y^{(10)}_{{\bm{1}}^1_2}$ exist.
At $\tau=3.7i$, they are evaluated as
\begin{align}
\begin{aligned}
  &Y_{{\bm{1}^0_0}}^{(10)}/Y_{{\bm{1}^0_0}}^{(10)} = 1 \rightarrow 1, \quad
  Y_{{\bm{1}^1_2}}^{(10)}/Y_{{\bm{1}^0_0}}^{(10)} = 0.102 \rightarrow \varepsilon, \\
  &Y_{{\bm{1}^0_1}}^{(10)}/Y_{{\bm{1}^0_0}}^{(10)} = 7.32\times 10^{-3} \rightarrow \varepsilon^2, \quad
  Y_{{\bm{1}^1_0}}^{(10)}/Y_{{\bm{1}^0_0}}^{(10)} = 7.44\times 10^{-4} \rightarrow \varepsilon^3, \\
  &Y_{{\bm{1}^1_1}}^{(10)}/Y_{{\bm{1}^0_0}}^{(10)} = 5.44\times 10^{-6} \rightarrow \varepsilon^5.
\end{aligned} \label{eq:modular form weight 10}
\end{align}
Here we assume left-handed quark doublets $Q$, right-handed up-sector quark singlets $u_R$ and right-handed down-sector quark singlets $d_R$ have weights -4, -4 and -6, respectively.
In this case Yukawa couplings for up-sector have weight 8 and ones for down-sector have weight 10.

The mass matrices in type III are given by
\begin{align}
  &M_u =
  \begin{pmatrix}
    0 & \alpha^{12} Y_{{\bm{1}}^{4+a-b~\textrm{mod~2}}_{4+a-b~\textrm{mod~3}}}^{(8)} & \alpha^{13} Y_{{\bm{1}}^{6-b~\textrm{mod~2}}_{6-b~\textrm{mod~3}}}^{(8)} \\
    \alpha^{21} Y_{{\bm{1}}^{1-a+b~\textrm{mod~2}}_{1-a+b~\textrm{mod~3}}}^{(8)} & \alpha^{22} Y_{{\bm{1}}^0_1}^{(8)} & \alpha^{23} Y_{{\bm{1}}^{6-a~\textrm{mod~2}}_{6-a~\textrm{mod~3}}}^{(8)} \\
    \alpha^{31} Y_{{\bm{1}}^{1+b~\textrm{mod~2}}_{1+b~\textrm{mod~3}}}^{(8)} & \alpha^{32} Y_{{\bm{1}}^{4+a~\textrm{mod~2}}_{4+a~\textrm{mod~3}}}^{(8)} & \alpha^{33} Y_{{\bm{1}}^0_0}^{(8} \\
  \end{pmatrix}, \label{eq:mass_mtx_w8_III} \\
  &M_d =
  \begin{pmatrix}
    \beta^{11} Y_{{\bm{1}}^1_0}^{(10)} & \beta^{12} Y_{{\bm{1}}^{4+a-b~\textrm{mod~2}}_{4+a-b~\textrm{mod~3}}}^{(10)} & \beta^{13} Y_{{\bm{1}}^{6-b~\textrm{mod~2}}_{6-b~\textrm{mod~3}}}^{(10)} \\
    \beta^{21} Y_{{\bm{1}}^{3-a+b~\textrm{mod~2}}_{3-a+b~\textrm{mod~3}}}^{(10)} & \beta^{22} Y_{{\bm{1}}^0_1}^{(10)} & \beta^{23} Y_{{\bm{1}}^{6-a~\textrm{mod~2}}_{6-a~\textrm{mod~3}}}^{(10)} \\
    \beta^{31} Y_{{\bm{1}}^{3+b~\textrm{mod~2}}_{3+b~\textrm{mod~3}}}^{(10)} & \beta^{32} Y_{{\bm{1}}^{4+a~\textrm{mod~2}}_{4+a~\textrm{mod~3}}}^{(10)} & \beta^{33} Y_{{\bm{1}}^0_0}^{(10)} \\
  \end{pmatrix}.
\end{align}
Here some of mass matrix elements can vanish depending on their $T$-charges, that is, values of $a$ and $b$.
To calculate the quark flavor observables, we need to choose the signs $\pm 1$ in coupling constants and the values of $a$ and $b$.
Consequently, we find best-fit choices at $\tau=3.7i$,
\begin{align}
a = 1, \quad b = 2,
\end{align}
and
\begin{align}
\begin{pmatrix}
\textrm{-} & \textrm{-} & \alpha^{13} \\
\alpha^{21} & \alpha^{22} & \alpha^{23} \\
\textrm{-} & \alpha^{32} & \alpha^{33} \\
\end{pmatrix} =
\begin{pmatrix}
\textrm{-} & \textrm{-} & 1 \\
1 & 1 & 1 \\
\textrm{-} & -1 & -1 \\
\end{pmatrix}, \quad
\begin{pmatrix}
\beta^{11} & \beta^{12} & \beta^{13} \\
\beta^{21} & \beta^{22} & \beta^{23} \\
\beta^{31} & \beta^{32} & \beta^{33} \\
\end{pmatrix} =
\begin{pmatrix}
1 & 1 & 1 \\
-1 & -1 & 1 \\
1 & -1 & 1 \\
\end{pmatrix}. \label{eq:Choise_typeIII108}
\end{align}
Then $T$-charges of $Q$, $u_R$ and $d_R$ are
\begin{align}
Q = ({\bm{1}^0_2},{\bm{1}^1_1},{\bm{1}^0_0}), \quad u_R = ({\bm{1}^1_0},{\bm{1}^1_1},{\bm{1}^0_0}), \quad d_R = ({\bm{1}^1_1},{\bm{1}^1_1},{\bm{1}^0_0}). \label{eq:Assign_typeIII108}
\end{align}
They lead to the hierarchical quark mass matrices,
\begin{align}
M_u/Y_{{\bm{1}^0_0}}^{(8)} =
\begin{pmatrix}
0 & 0 & Y_{{\bm{1}^0_1}}^{(8)} \\
Y_{{\bm{1}^0_2}}^{(8)} & Y_{{\bm{1}^0_1}}^{(8)} & Y_{{\bm{1}^1_2}}^{(8)} \\
0 & -Y_{{\bm{1}^1_2}}^{(8)} & -Y_{{\bm{1}^0_0}}^{(8)} \\
\end{pmatrix}/Y_{{\bm{1}^0_0}}^{(8)}
&=
\begin{pmatrix}
0 & 0 & 0.00732 \\
0.0000535 & 0.00732 & -0.0719 \\
0 & 0.0719 & -1 \\
\end{pmatrix} \notag \\
&\sim
\begin{pmatrix}
0 & 0 & \varepsilon^2 \\
\varepsilon^4 & \varepsilon^2 & -\varepsilon \\
0 & \varepsilon & -1 \\
\end{pmatrix}, \label{eq:Upweight8} \\
M_d/Y_{{\bm{1}^0_0}}^{(10)} =
\begin{pmatrix}
Y_{{\bm{1}^1_0}}^{(10)} & Y_{{\bm{1}^1_0}}^{(10)} & Y_{{\bm{1}^0_1}}^{(10)} \\
-Y_{{\bm{1}^0_1}}^{(10)} & -Y_{{\bm{1}^0_1}}^{(10)} & Y_{{\bm{1}^1_2}}^{(10)} \\
Y_{{\bm{1}^1_2}}^{(10)} & -Y_{{\bm{1}^1_2}}^{(10)} & Y_{{\bm{1}^0_0}}^{(10)} \\
\end{pmatrix}/Y_{{\bm{1}^0_0}}^{(10)}
&=
\begin{pmatrix}
0.000744 & 0.000744 & 0.00732 \\
-0.00732 & -0.00732 & 0.102 \\
0.102 & -0.102 & 1 \\
\end{pmatrix} \notag \\
&\sim
\begin{pmatrix}
\varepsilon^3 & \varepsilon^3 & \varepsilon^2 \\
-\varepsilon^2 & -\varepsilon^2 & \varepsilon \\
\varepsilon & -\varepsilon & 1 \\
\end{pmatrix}.
\end{align}
These mass matrices give the following quark mass ratios,
\begin{align}
&(m_u,m_c,m_t)/m_t = (1.27\times 10^{-5}, 2.18\times 10^{-3},1), \\
&(m_d,m_s,m_b)/m_b = (1.44\times 10^{-3}, 1.74\times 10^{-2},1),
\end{align}
and the absolute values of the CKM matrix elements,
\begin{align}
|V_{\textrm{CKM}}| =
\begin{pmatrix}
0.974 & 0.227 & 0.00741 \\
0.227 & 0.973 & 0.0300 \\
0.0140 & 0.0276 & 1.00 \\
\end{pmatrix}.
\end{align}
Results are shown in Table \ref{tab:small-weight}.
\begin{table}[H]
  \begin{center}
    \renewcommand{\arraystretch}{1.3}
    \begin{tabular}{c|ccccccc} \hline
      & $\frac{m_u}{m_t}\times10^{6}$ & $\frac{m_c}{m_t}\times10^3$ & $\frac{m_d}{m_b}\times10^4$ & $\frac{m_s}{m_b}\times10^2$ & $|V_{\textrm{CKM}}^{us}|$ & $|V_{\textrm{CKM}}^{cb}|$ & $|V_{\textrm{CKM}}^{ub}|$ \\ \hline
      obtained values & 12.7 & 2.18 & 14.4 & 1.74 & 0.227 & 0.0300 & 0.00741 \\
      observed values & 12.6 & 7.38 & 11.2 & 2.22 & 0.227 & 0.0405 & 0.00361 \\
      GUT scale values & 5.39 & 2.80 & 9.21 & 1.82 & 0.225 & 0.0400 & 0.00353 \\ \hline
    \end{tabular}
  \end{center}
  \caption{The mass ratios of the quarks and the absolute values of the CKM matrix elements at the benchmark point $\tau=3.7i$ in the best-fit model by Eqs.~(\ref{eq:Assign_typeIII108}) and (\ref{eq:Choise_typeIII108}) of type III with up-sector Yukawa couplings of weight 8 and down-sector Yukawa couplings of weight 10.
  Observed values  Ref.~\cite{Zyla:2020zbs} and GUT scale values with $\tan \beta=5$ \cite{Antusch:2013jca,Bjorkeroth:2015ora} are shown.}
\label{tab:small-weight}
\end{table}

\paragraph{Type III: (5,2,0) and (3,2,0) with weights 8 and 12}~\\
Next we study the model where Yukawa couplings for up-sector have weight 8 and ones for down-sector have weight 12.
In addition, we consider the model in type III.
As a benchmark point of the modulus, we choose $\tau=3.7i$ around the cusp.
As shown in Appendix \ref{app:modular_forms_6}, four singlet modular forms of weight 8, $Y^{(8)}_{{\bm{1}}^0_0}$, $Y^{(8)}_{{\bm{1}}^1_2}$, $Y^{(8)}_{{\bm{1}}^0_1}$ and $Y^{(8)}_{{\bm{1}}^0_2}$ exist.
At the benchmark point $\tau=3.7i$, they are evaluated as in Eq.~(\ref{eq:modular form weight 8}).
At weight 12, six singlet modular forms, $Y^{(12)}_{{\bm{1}}^0_0i}$, $Y^{(12)}_{{\bm{1}}^1_2}$, $Y^{(12)}_{{\bm{1}}^0_1}$, $Y^{(12)}_{{\bm{1}}^1_0}$, $Y^{(12)}_{{\bm{1}}^0_2}$ and $Y^{(12)}_{{\bm{1}}^0_0ii}$ exist.
At $\tau=3.7i$, they are evaluated as
\begin{align}
\begin{aligned}
  &Y_{{\bm{1}^0_0i}}^{(12)}/Y_{{\bm{1}^0_0i}}^{(12)} = 1 \rightarrow 1, \quad
  Y_{{\bm{1}^1_2}}^{(12)}/Y_{{\bm{1}^0_0i}}^{(12)} = 0.102 \rightarrow \varepsilon, \\
  &Y_{{\bm{1}^0_1}}^{(12)}/Y_{{\bm{1}^0_0i}}^{(12)} = 1.03\times 10^{-2} \rightarrow \varepsilon^2, \quad
  Y_{{\bm{1}^1_0}}^{(12)}/Y_{{\bm{1}^0_0i}}^{(12)} = 7.44\times 10^{-4} \rightarrow \varepsilon^3, \\
  &Y_{{\bm{1}^0_2}}^{(12)}/Y_{{\bm{1}^0_0i}}^{(12)} = 7.57\times 10^{-5} \rightarrow \varepsilon^4, \quad
  Y_{{\bm{1}^0_0ii}}^{(12)}/Y_{{\bm{1}^0_0i}}^{(12)} = 5.54\times 10^{-7} \rightarrow \varepsilon^6. 
\end{aligned}
\end{align}
Here $Y_{{\bm{1}^0_0ii}}^{(12)} \sim \varepsilon^6$ originates from $Y_{{\bm{1}^1_0}}^{(6)} Y_{{\bm{1}^1_0}}^{(6)} \sim \varepsilon^3\cdot \varepsilon^3$ while $Y_{{\bm{1}^0_0i}}^{(12)} \sim 1$ originates from $Y_{{\bm{1}^0_0}}^{(6)} Y_{{\bm{1}^0_0}}^{(6)} \sim 1\cdot 1$.
In what follows, we ignore $Y_{{\bm{1}^0_0ii}}^{(12)}$ because it belongs to the same representation as $Y_{{\bm{1}^0_0i}}^{(12)}$ and $Y_{{\bm{1}^0_0i}}^{(12)}>>Y_{{\bm{1}^0_0ii}}^{(12)}$.
Also we assume left-handed quark doublets $Q$, right-handed up-sector quark singlets $u_R$ and right-handed down-sector quark singlets $d_R$ have weights -4, -4 and -8, respectively.
In this case Yukawa couplings for up-sector have weight 8 and ones for down-sector have weight 12.

The mass matrices in type III are given by Eq.~(\ref{eq:mass_mtx_w8_III}) and
\begin{align}
  &M_d =
  \begin{pmatrix}
    \beta^{11} Y_{{{\bm{1}}}^1_0}^{(12)} & \beta^{12} Y_{{{\bm{1}}}^{4+a-b~\textrm{mod~2}}_{4+a-b~\textrm{mod~3}}}^{(12)} & \beta^{13} Y_{{{\bm{1}}}^{6-b~\textrm{mod~2}}_{6-b~\textrm{mod~3}}}^{(12)} \\
    \beta^{21} Y_{{{\bm{1}}}^{3-a+b~\textrm{mod~2}}_{3-a+b~\textrm{mod~3}}}^{(12)} & \beta^{22} Y_{{{\bm{1}}}^0_1}^{(12)} & \beta^{23} Y_{{{\bm{1}}}^{6-a~\textrm{mod~2}}_{6-a~\textrm{mod~3}}}^{(12)} \\
    \beta^{31} Y_{{{\bm{1}}}^{3+b~\textrm{mod~2}}_{3+b~\textrm{mod~3}}}^{(12)} & \beta^{32} Y_{{{\bm{1}}}^{4+a~\textrm{mod~2}}_{4+a~\textrm{mod~3}}}^{(12)} & \beta^{33} Y_{{{\bm{1}}}^0_0}^{(12)} \\
  \end{pmatrix}.
\end{align}
Here some of mass matrix elements can vanish depending on their $T$-charges, that is, values of $a$ and $b$.
To calculate the quark flavor observables, we need to choose the signs $\pm 1$ in coupling constants and the values of $a$ and $b$.
Consequently, we find best-fit choices at $\tau=3.7i$,
\begin{align}
a = 1, \quad b = 2,
\end{align}
and
\begin{align}
\begin{pmatrix}
\alpha^{11} & \alpha^{12} & \alpha^{13} \\
\alpha^{21} & \alpha^{22} & \alpha^{23} \\
\alpha^{31} & \alpha^{32} & \alpha^{33} \\
\end{pmatrix} =
\begin{pmatrix}
1 & 1 & 1 \\
1 & 1 & 1 \\
1 & -1 & -1 \\
\end{pmatrix}, \quad
\begin{pmatrix}
\beta^{11} & \beta^{12} & \beta^{13} \\
\beta^{21} & \beta^{22} & \beta^{23} \\
\beta^{31} & \beta^{32} & \beta^{33} \\
\end{pmatrix} =
\begin{pmatrix}
1 & 1 & 1 \\
-1 & -1 & 1 \\
1 & -1 & 1 \\
\end{pmatrix}. \label{eq:Choise_typeIII128}
\end{align}
Then $T$-charges of $Q$, $u_R$ and $d_R$ are
\begin{align}
Q = ({\bm{1}^0_2},{\bm{1}^1_1},{\bm{1}^0_0}), \quad u_R = ({\bm{1}^1_0},{\bm{1}^1_1},{\bm{1}^0_0}), \quad d_R = ({\bm{1}^1_1},{\bm{1}^1_1},{\bm{1}^0_0}). \label{eq:Assign_typeIII128}
\end{align}
They lead to the hierarchical quark mass matrices,
\begin{align}
M_u/Y_{{\bm{1}^0_0}}^{(8)} =
\begin{pmatrix}
0 & 0 & Y_{{\bm{1}^0_1}}^{(8)} \\
Y_{{\bm{1}^0_2}}^{(8)} & Y_{{\bm{1}^0_1}}^{(8)} & Y_{{\bm{1}^1_2}}^{(8)} \\
0 & -Y_{{\bm{1}^1_2}}^{(8)} & -Y_{{\bm{1}^0_0}}^{(8)} \\
\end{pmatrix}/Y_{{\bm{1}^0_0}}^{(8)}
&=
\begin{pmatrix}
0 & 0 & 0.00732 \\
0.0000535 & 0.00732 & -0.0719 \\
0 & 0.0719 & -1 \\
\end{pmatrix} \notag \\
&\sim
\begin{pmatrix}
0 & 0 & \varepsilon^2 \\
\varepsilon^4 & \varepsilon^2 & -\varepsilon \\
0 & \varepsilon & -1 \\
\end{pmatrix}, \\
M_d/Y_{{\bm{1}^0_0i}}^{(12)} =
\begin{pmatrix}
Y_{{\bm{1}^1_0}}^{(12)} & Y_{{\bm{1}^1_0}}^{(12)} & Y_{{\bm{1}^0_1}}^{(12)} \\
-Y_{{\bm{1}^0_1}}^{(12)} & -Y_{{\bm{1}^0_1}}^{(12)} & Y_{{\bm{1}^1_2}}^{(12)} \\
Y_{{\bm{1}^1_2}}^{(12)} & -Y_{{\bm{1}^1_2}}^{(12)} & Y_{{\bm{1}^0_0i}}^{(12)} \\
\end{pmatrix}/Y_{{\bm{1}^0_0i}}^{(12)}
&=
\begin{pmatrix}
0.000744 & 0.000744 & 0.0103 \\
-0.0103 & -0.0103 & 0.102 \\
0.102 & -0.102 & 1 \\
\end{pmatrix} \notag \\
&\sim
\begin{pmatrix}
\varepsilon^3 & \varepsilon^3 & \varepsilon^2 \\
-\varepsilon^2 & -\varepsilon^2 & \varepsilon \\
\varepsilon & -\varepsilon & 1 \\
\end{pmatrix}.
\end{align}
These mass matrices give the following quark mass ratios,
\begin{align}
&(m_u,m_c,m_t)/m_t = (1.27\times 10^{-5}, 2.18\times 10^{-3},1), \\
&(m_d,m_s,m_b)/m_b = (1.76\times 10^{-3}, 2.02\times 10^{-2}, 1),
\end{align}
and the absolute values of the CKM matrix elements,
\begin{align}
|V_{\textrm{CKM}}| =
\begin{pmatrix}
0.974 & 0.227 & 0.0103 \\
0.226 & 0.974 & 0.0308 \\
0.0170 & 0.0276 & 0.999 \\
\end{pmatrix}.
\end{align}
Results are shown in Table \ref{tab:small-weight812}.
\begin{table}[H]
  \begin{center}
    \renewcommand{\arraystretch}{1.3}
    \begin{tabular}{c|ccccccc} \hline
      & $\frac{m_u}{m_t}\times10^{6}$ & $\frac{m_c}{m_t}\times10^3$ & $\frac{m_d}{m_b}\times10^4$ & $\frac{m_s}{m_b}\times10^2$ & $|V_{\textrm{CKM}}^{us}|$ & $|V_{\textrm{CKM}}^{cb}|$ & $|V_{\textrm{CKM}}^{ub}|$ \\ \hline
      obtained values & 12.7 & 2.18 & 17.6 & 2.02 & 0.227 & 0.0308 & 0.0103 \\
      observed values & 12.6 & 7.38 & 11.2 & 2.22 & 0.227 & 0.0405 & 0.00361 \\
      GUT scale values & 5.39 & 2.80 & 9.21 & 1.82 & 0.225 & 0.0400 & 0.00353 \\ \hline
    \end{tabular}
  \end{center}
  \caption{The mass ratios of the quarks and the absolute values of the CKM matrix elements at the benchmark point $\tau=3.7i$ in the best-fit model by Eqs.~(\ref{eq:Assign_typeIII128}) and (\ref{eq:Choise_typeIII128}) of type III with up-sector Yukawa couplings of weight 8 and down-sector Yukawa couplings of weight 12.
  Observed values  Ref.~\cite{Zyla:2020zbs} and GUT scale values with $\tan \beta=5$ \cite{Antusch:2013jca,Bjorkeroth:2015ora} are shown.}
\label{tab:small-weight812}
\end{table}


\subsection{The origin of $\Gamma_6$ modular symmetry}

As we have seen throughout this section, the values of the quark flavor structures can be realized up to ${\cal O}(1)$ without fine-tuning in $\Gamma_6$ modular symmetry.
Then residual $T$-symmetry ($Z_6$-symmetry) of $\Gamma_6$ on the cusp $\tau=i\infty$ can originate hierarchical quark mass ratios.
Before we end this section, we also comment on a plausible origin of $\Gamma_6$ modular symmetry of the theories.
As we have seen in Chapter \ref{sec:magnetized_orbifold_models}, some modular forms are derived from the torus compactification $T^2_1\times T^2_2\times T^2_3$ of the low-energy effective theory of the superstring theory with magnetic flux background.
For example the modular forms of weight 1/2 transformed by $\widetilde{\Gamma}_{N}$ are obtained on magnetized $T^2$; the modular forms of weight 1 transformed by $\Gamma'_N$ are obtained on magnetized $T^2_1\times T^2_2$ with the moduli stabilization $\tau_1=\tau_2$.
In addition, non-factorizable toroidal compactifications $T^4$ and $T^6$ can lead to more various modular forms.
Therefore it may be expected that $\Gamma_6$ modular symmetry originates from magnetized troidal compactification models.
As a simple case, $\Gamma_6\simeq S_3\times A_4\simeq \Gamma_2\times \Gamma_3$ may originate from $\Gamma_2$ modular symmetry on $T^2_1$, $\Gamma_3$ modular symmetry on $T^2_2$ and trivial modular symmetry on $T^2_3$ under the moduli stabilization $\tau_1=\tau_2$.


\section{The models with $A_4\times A_4 \times A_4$ modular symmetry}
\label{subsec:A4xA4xA4}

As another promissing example for the quark flavor models without fine-tuning, we study the models with $A_4\times A_4 \times A_4$ modular symmetry, which is controled by three moduli $\tau_1$, $\tau_2$ and $\tau_3$.
$A_4$ modular group has two generators, $S$ and $T$-transformations, and they satisfy the algebraic relations,
\begin{align}
S^2 = (ST)^3 = T^3 = \mathbb{I}.
\end{align}
Since $A_4$ modular symmetry breaks to residual $Z_3$ symmetry at both $\tau=\omega$ and $i\infty$, $A_4\times A_4 \times A_4$ modular symmetry breaks to residual $Z_3\times Z_3 \times Z_3$ symmetry which is able to generate the hierarchical value $\varepsilon^6$ in the vicinity of $\tau=\omega$ and $i\infty$.
Hereafter, we assume the moduli stabilization $\tau_1=\tau_2=\tau_3\equiv \tau$ and the vicinity of the symmetric points $\omega$ or $i\infty$ in order to generate the hierarchical values of the modular forms.
Note that the non-universal moduli such as $\tau_1\sim \omega$, $\tau_2\sim \omega$ and $\tau_3\sim i\infty$ are possible to generate the hierarchical value $\varepsilon^6$ as well as the universal moduli $\tau_1=\tau_2=\tau_3\equiv \tau\sim \omega$ or $i\infty$.
Also we will study the models with non-universal moduli $\tau_1=\tau_2\neq\tau_3$ in Subsection \ref{subsubsec:non-universal}.

The superpotentials for up and down-sectors, $W_u$ and $W_d$, with $A_4\times A_4 \times A_4$ modular symmetry are given by
\begin{equation}
\label{eq: up_superpotential_(A4)^3}
    W_u = \sum_{{\bm{r}}_1, {\bm{r}}_2, {\bm{r}}_3}
    \left[ Y_{{\bm{r}}_1}(\tau)
    Y_{{\bm{r}}_2}(\tau)
    Y_{{\bm{r}}_3}(\tau)
    (Q^1\ Q^2\ Q^3) 
    \begin{pmatrix}
    \alpha_{{\bm{r}}_1{\bm{r}}_2{\bm{r}}_3}^{11} &  \alpha_{{\bm{r}}_1{\bm{r}}_2{\bm{r}}_3}^{12} &  \alpha_{{\bm{r}}_1{\bm{r}}_2{\bm{r}}_3}^{13} \\
    \alpha_{{\bm{r}}_1{\bm{r}}_2{\bm{r}}_3}^{21} &  \alpha_{{\bm{r}}_1{\bm{r}}_2{\bm{r}}_3}^{22} &  \alpha_{{\bm{r}}_1{\bm{r}}_2{\bm{r}}_3}^{23} \\
     \alpha_{{\bm{r}}_1{\bm{r}}_2{\bm{r}}_3}^{31} &  \alpha_{{\bm{r}}_1{\bm{r}}_2{\bm{r}}_3}^{32} &  \alpha_{{\bm{r}}_1{\bm{r}}_2{\bm{r}}_3}^{33}
    \end{pmatrix}
    \begin{pmatrix}
    u_R^1 \\ u_R^2 \\ u_R^3
    \end{pmatrix}
    H_u
    \right]_{\bf{1}},
\end{equation}
\begin{equation}
\label{eq: down_superpotential}
     W_d = \sum_{{\bm{r}}_1, {\bm{r}}_2, {\bm{r}}_3}
    \left[ Y_{{\bm{r}}_1}(\tau)
    Y_{{\bm{r}}_2}(\tau)
    Y_{{\bm{r}}_3}(\tau)
    (Q^1\ Q^2\ Q^3) 
    \begin{pmatrix}
    \beta_{{\bm{r}}_1{\bm{r}}_2{\bm{r}}_3}^{11} &  \beta_{{\bm{r}}_1{\bm{r}}_2{\bm{r}}_3}^{12} &  \beta_{{\bm{r}}_1{\bm{r}}_2{\bm{r}}_3}^{13} \\
    \beta_{{\bm{r}}_1{\bm{r}}_2{\bm{r}}_3}^{21} &  \beta_{{\bm{r}}_1{\bm{r}}_2{\bm{r}}_3}^{22} &  \beta_{{\bm{r}}_1{\bm{r}}_2{\bm{r}}_3}^{23} \\
    \beta_{{\bm{r}}_1{\bm{r}}_2{\bm{r}}_3}^{31} &  \beta_{{\bm{r}}_1{\bm{r}}_2{\bm{r}}_3}^{32} &  \beta_{{\bm{r}}_1{\bm{r}}_2{\bm{r}}_3}^{33}
    \end{pmatrix}
    \begin{pmatrix}
    d_R^1 \\ d_R^2 \\ d_R^3
    \end{pmatrix}
    H_d
    \right]_{\bm{1}},
\end{equation}
where $Y_{{\bm{r}}_n}(\tau), (n=1,2,3)$ are the modular forms belonging to all possible irreducible representations ${\bm{r}}_n$ with respect to $n$-th $A_4$.
Coupling constants $\alpha_{{\bm{r}}_1{\bm{r}}_2{\bm{r}}_3}^{jk}$ ($\beta_{{\bm{r}}_1{\bm{r}}_2{\bm{r}}_3}^{jk}$) may be related each other when quarks belong to multiplets.
Yukawa couplings are given by the products of three modular forms $Y_{{\bm{r}}_1}(\tau) Y_{{\bm{r}}_2}(\tau) Y_{{\bm{r}}_3}(\tau)$.
Notice that this corresponds to Yukawa couplings on the magnetized orbifold models in Eq.~(\ref{eq:three-point-couplings}) under the moduli stabilization $\tau_1=\tau_2=\tau_3$.

After the Higgs fields get the VEVs, these superpotentials lead to the mass terms,
\begin{align}
\begin{aligned}
\label{eq: up_mass_(A4)^3}
    (Q^1\ Q^2\ & Q^3) M_u
    \begin{pmatrix}
      u_R^1 \\ u_R^2 \\ u_R^3
    \end{pmatrix}  \\
    &= \sum_{{\bm{r}}_1, {\bm{r}}_2, {\bm{r}}_3}
    \left[ \prod_{n=1}^3 Y_{{\bm{r}}_n}
    (Q^1\ Q^2\ Q^3) 
    \begin{pmatrix}
    \alpha_{{\bm{r}}_1{\bm{r}}_2{\bm{r}}_3}^{11} &  \alpha_{{\bm{r}}_1{\bm{r}}_2{\bm{r}}_3}^{12} &  \alpha_{{\bm{r}}_1{\bm{r}}_2{\bm{r}}_3}^{13} \\
    \alpha_{{\bm{r}}_1{\bm{r}}_2{\bm{r}}_3}^{21} &  \alpha_{{\bm{r}}_1{\bm{r}}_2{\bm{r}}_3}^{22} &  \alpha_{{\bm{r}}_1{\bm{r}}_2{\bm{r}}_3}^{23} \\
     \alpha_{{\bm{r}}_1{\bm{r}}_2{\bm{r}}_3}^{31} &  \alpha_{{\bm{r}}_1{\bm{r}}_2{\bm{r}}_3}^{32} &  \alpha_{{\bm{r}}_1{\bm{r}}_2{\bm{r}}_3}^{33}
    \end{pmatrix}
    \begin{pmatrix}
    u_R^1 \\ u_R^2 \\ u_R^3
    \end{pmatrix}
   \langle H_u \rangle
    \right]_{\bm{1}},
    \end{aligned}
\end{align}
\begin{align}
\begin{aligned}
\label{eq: down_mass_(A4)^3}
    (Q^1\ Q^2\ & Q^3) M_d
    \begin{pmatrix}
      d_R^1 \\ d_R^2 \\ d_R^3
    \end{pmatrix} \\
    &= \sum_{{\bm{r}}_1, {\bm{r}}_2, {\bm{r}}_3}
    \left[ \prod_{n=1}^3 Y_{{\bm{r}}_n}
    (Q^1\ Q^2\ Q^3) 
 \begin{pmatrix}
    \beta_{{\bm{r}}_1{\bm{r}}_2{\bm{r}}_3}^{11} &  \beta_{{\bm{r}}_1{\bm{r}}_2{\bm{r}}_3}^{12} &  \beta_{{\bm{r}}_1{\bm{r}}_2{\bm{r}}_3}^{13} \\
    \beta_{{\bm{r}}_1{\bm{r}}_2{\bm{r}}_3}^{21} &  \beta_{{\bm{r}}_1{\bm{r}}_2{\bm{r}}_3}^{22} &  \beta_{{\bm{r}}_1{\bm{r}}_2{\bm{r}}_3}^{23} \\
    \beta_{{\bm{r}}_1{\bm{r}}_2{\bm{r}}_3}^{31} &  \beta_{{\bm{r}}_1{\bm{r}}_2{\bm{r}}_3}^{32} &  \beta_{{\bm{r}}_1{\bm{r}}_2{\bm{r}}_3}^{33}
    \end{pmatrix}
    \begin{pmatrix}
    d_R^1 \\ d_R^2 \\ d_R^3
    \end{pmatrix}
   \langle H_d \rangle
    \right]_{\bm{1}}.
\end{aligned}
\end{align}
To make our analysis simple, we use only $A_4$ singlets and assign Higgs fields $H_u$ and $H_d$ into $A_4\times A_4\times A_4$ trivial singlets $\bm{1}\times \bm{1}\times \bm{1}$ with weight 0.
In addition, we restrict the coupling constants $\alpha_{{\bm{r}}_1{\bm{r}}_2{\bm{r}}_3}^{jk}$ and $\beta_{{\bm{r}}_1{\bm{r}}_2{\bm{r}}_3}^{jk}$ to $\pm 1$ in quark mass matrices as same as the analysis on $\Gamma_6$ modular symmetry in the previous section.
To avoid fine-tuning by coupling constants, we use only the deviation of the modulus from the symmetric points $\omega$ or $i\infty$ (and the choices of $+1$ or $-1$ in $\alpha$ and $\beta$) as a free parameter.

As reviewed in Appendix \ref{app:gt_A_4}, in $A_4$ modular group, three singlet irreducible representations $\bm{1}$, $\bm{1}'$ and $\bm{1}''$ exist.
Under $S$, $ST$ and $T$-transformations, the modular forms of weight $k$ belonging to $\bm{1}$, $\bm{1}'$ and $\bm{1}''$, $Y_{\bm{1}}(\tau)$, $Y_{\bm{1}'}(\tau)$ and $Y_{\bm{1}''}(\tau)$ are transformed as
\begin{align}
\begin{aligned}
&S:Y_{\bm{1}}(\tau) = (-\tau)^kY_{\bm{1}}(\tau), \quad ST:Y_{\bm{1}}(\tau) = (-1-\tau)^kY_{\bm{1}}(\tau), \quad T:Y_{\bm{1}}(\tau) = Y_{\bm{1}}(\tau), \\
&S:Y_{\bm{1}'}(\tau) = (-\tau)^kY_{\bm{1}'}(\tau), \quad ST:Y_{\bm{1}'}(\tau) = (-1-\tau)^k\omega Y_{\bm{1}'}(\tau), \quad T:Y_{\bm{1}'}(\tau) = \omega Y_{\bm{1}'}(\tau), \\
&S:Y_{\bm{1}''}(\tau) = (-\tau)^k Y_{\bm{1}''}(\tau), \quad ST: Y_{\bm{1}''}(\tau) = (-1-\tau)^k\omega^2 Y_{\bm{1}''}(\tau), \quad T: Y_{\bm{1}''}(\tau) = \omega^2 Y_{\bm{1}''}(\tau).
\end{aligned} \label{eq:S ST T-trans for Y8}
\end{align}
As shown in Appendix \ref{app:modular_forms_A_4}, at weight $k=8$, we can find the modular forms belonging to $\bm{1}$, $\bm{1}'$ and $\bm{1}''$, $Y_{\bm{1}}^{(8)}(\tau)$, $Y_{\bm{1}'}^{(8)}(\tau)$ and $Y_{\bm{1}''}^{(8)}(\tau)$, while at weights less than 8 some singlet modular forms do not exist.
For example, at weight 6 the modular forms belonging to $\bm{1}'$ and $\bm{1}''$ do not exist.
Therefore we consider the assignments of weights,
\begin{align}
k_Q = k_{u_R} = k_{d_R} = 4,
\end{align}
to make Yukawa couplings have weights 8.

First let us consider the modular forms in the vicinity of the cusp $\tau\sim i\infty$.
Using Eqs.~(\ref{eq:f expanded by eT}) and (\ref{eq:S ST T-trans for Y8}), we obtain the hierarchical structures of the modular forms,
\begin{align}
&Y_{\bm{1}}^{(8)}(\tau) \simeq \varepsilon_T^0 = 1, \\
&Y_{\bm{1}'}^{(8)}(\tau) \simeq \varepsilon_T = e^{2\pi i\tau/3}, \\
&Y_{\bm{1}''}^{(8)}(\tau) \simeq \varepsilon_T^2 = e^{4\pi i\tau/3},
\end{align}
at $\tau\sim i\infty$.

Next let us consider the modular forms in the vicinity of left cusp $\tau\sim \omega$.
Using Eqs.~(\ref{eq:ex_by_q_ST}) and (\ref{eq:S ST T-trans for Y8}), we obtain the hierarchical structures of the modular forms,
\begin{align}
&Y_{\bm{1}}^{(8)}(\tau) \simeq \varepsilon_{ST}^2 = \left(\frac{\tau-\omega}{\tau-\omega^2}\right)^2, \\
&Y_{\bm{1}'}^{(8)}(\tau) \simeq \varepsilon_{ST} = \frac{\tau-\omega}{\tau-\omega^2}, \\
&Y_{\bm{1}''}^{(8)}(\tau) \simeq \varepsilon_{ST}^0 = 1,
\end{align}
at $\tau\sim \omega$.
Here we have used $(-1-\omega)=\omega^2$.


\subsection{Favorable models}
\label{subsubsec:Favorable models}

Here we study favorable models realizing quark flavor observables at $\tau\sim \omega$ and $i\infty$.
To realize hierarchical quark mass ratios, we consider the mass matrices of the forms,
\begin{align}
\label{eq: hierarchy_mass}
    M_u &\propto 
    \begin{pmatrix}
    \mathcal{O}(\varepsilon^6) & \ast & \ast \\
    \ast & \mathcal{O}(\varepsilon^3) & \ast \\
    \ast & \ast & \mathcal{O}(1)
    \end{pmatrix},\quad
     M_d \propto 
    \begin{pmatrix}
    \mathcal{O}(\varepsilon^4) & \ast & \ast \\
    \ast & \mathcal{O}(\varepsilon^2) & \ast \\
    \ast & \ast &  \mathcal{O}(1)
    \end{pmatrix},
\end{align}
where we assume $\varepsilon \sim 0.15$ and $*$ denotes unfixed values at this stage.
Each $A_4$ can yield the modular forms of the orders 1, $\varepsilon$ and $\varepsilon^2$ at $\tau\sim \omega$ and $i\infty$.
We denote the orders of the modular forms of the $n$-th $A_4$ as 1, $\varepsilon_n$ and $\varepsilon^2_n$.
Then up-sector mass matrix $M_u$ of the above forms is realized in the following two types:
\begin{align}
   \text{Type $123$}: M_u &\propto
    \begin{pmatrix}
    \mathcal{O}({\varepsilon_1}^2{\varepsilon_2}^2{\varepsilon_3}^2) & & \\
    & \mathcal{O}({\varepsilon_1}{\varepsilon_2}{\varepsilon_3}) & \\
    & & \mathcal{O}(1)
    \end{pmatrix},\\
    \label{eq: 1^22}
    \text{Type $1^22$}: M_u &\propto
    \begin{pmatrix}
    \mathcal{O}({\varepsilon_1}^2{\varepsilon_2}^2{\varepsilon_3}^2) & & \\
    & \mathcal{O}({\varepsilon_1}^2{\varepsilon_2}) & \\
    & & \mathcal{O}(1)
    \end{pmatrix},
\end{align}
where the notation of the type, $\ell^im^jn^k$, corresponds to up-sector mass matrix with $M_u^{22}\propto {\cal O}(\varepsilon_\ell^i \varepsilon_m^j \varepsilon_n^k)$.
Notice that type $123$ has a permutation symmetry of three $A_4$'s while type $1^22$ does not has such symmetry.
On the other hand, other possible types, $1^23$, $12^2$, $2^23$, $13^2$ and $23^2$ are equivalent to type $1^22$ since each $A_4$ cannot be distinguished under the moduli stabilization $\tau_1=\tau_2=\tau_3$.

Next we classify the possible types of down-sector mass matrix $M_d$.
When $M_u$ is in type $123$, the possible types of down-sector mass matrix $M_d$ are given by
\begin{align}
\begin{aligned}
    \text{$1^22^2$-$12$}:M_d &\propto
    \begin{pmatrix}
    \mathcal{O}({\varepsilon_1}^2{\varepsilon_2}^2) & & \\
    & \mathcal{O}(\varepsilon_1\varepsilon_2) & \\
    & & \mathcal{O}(1)
    \end{pmatrix}, \quad
\text{$1^22^2$-$23$}:M_d \propto
    \begin{pmatrix}
    \mathcal{O}({\varepsilon_1}^2{\varepsilon_2}^2) & & \\
    & \mathcal{O}(\varepsilon_2\varepsilon_3) & \\
    & & \mathcal{O}(1)
    \end{pmatrix}, \\
\text{$1^22^2$-$1^2$}:M_d &\propto
    \begin{pmatrix}
    \mathcal{O}({\varepsilon_1}^2{\varepsilon_2}^2) & & \\
    & \mathcal{O}(\varepsilon_1^2) & \\
    & & \mathcal{O}(1)
    \end{pmatrix}, \quad
\text{$1^22^2$-$3^2$}:M_d \propto
    \begin{pmatrix}
    \mathcal{O}({\varepsilon_1}^2{\varepsilon_2}^2) & & \\
    & \mathcal{O}(\varepsilon_3^2) & \\
    & & \mathcal{O}(1)
    \end{pmatrix}, \\
    \text{$123^2$-$12$}:M_d &\propto
    \begin{pmatrix}
    \mathcal{O}(\varepsilon_1\varepsilon_2{\varepsilon_3}^2) & & \\
    & \mathcal{O}(\varepsilon_1\varepsilon_2) & \\
    & & \mathcal{O}(1)
    \end{pmatrix}, \quad
\text{$123^2$-$23$}:M_d \propto
    \begin{pmatrix}
    \mathcal{O}(\varepsilon_1\varepsilon_2{\varepsilon_3}^2) & & \\
    & \mathcal{O}(\varepsilon_2\varepsilon_3) & \\
    & & \mathcal{O}(1)
    \end{pmatrix}, \\
\text{$123^2$-$1^2$}:M_d &\propto
    \begin{pmatrix}
    \mathcal{O}(\varepsilon_1\varepsilon_2{\varepsilon_3}^2) & & \\
    & \mathcal{O}(\varepsilon_1^2) & \\
    & & \mathcal{O}(1)
    \end{pmatrix}, \quad
\text{$123^2$-$3^2$}:M_d \propto
    \begin{pmatrix}
    \mathcal{O}(\varepsilon_1\varepsilon_2{\varepsilon_3}^2) & & \\
    & \mathcal{O}(\varepsilon_3^2) & \\
    & & \mathcal{O}(1)
    \end{pmatrix},
\end{aligned}
\end{align}
where the notation of the type, $\ell^im^jn^k$-$r^ps^q$, corresponds to down-sector mass matrix with $M_u^{11}\propto {\cal O}(\varepsilon_\ell^i \varepsilon_m^j \varepsilon_n^k)$ and $M_u^{22}\propto {\cal O}(\varepsilon_r^p \varepsilon_s^q)$.
Other types such as $1^22^2$-$13$ are equivalent to any of the above types.

In the same way, we can find the possible types of down-sector mass matrix when $M_u$ is in type $1^22$.
The possible types of $M_d^{11}$ are
\begin{align}
    \begin{aligned}
\text{$1^22^2$}:\ &  \mathcal{O}({\varepsilon_1}^2 {\varepsilon_2}^2),\quad 
\text{$2^23^2$}:\ 
\mathcal{O}({\varepsilon_2}^2{\varepsilon_3}^2),\quad 
\text{$1^23^2$}:\  \mathcal{O}({\varepsilon_1}^2{\varepsilon_3}^2), \\
\text{$1^223$}:\ &
\mathcal{O}({\varepsilon_1}^2{\varepsilon_2}{\varepsilon_3}),\quad
\text{$12^23$}:\  \mathcal{O}({\varepsilon_1}{\varepsilon_2}^2{\varepsilon_3}),\quad 
\text{$123^2$}:\  \mathcal{O}({\varepsilon_1} {\varepsilon_2}{\varepsilon_3}^2),
    \end{aligned}
\end{align}
and ones of $M_d^{22}$ are
\begin{align}
\begin{aligned}
\text{$1^2$}:\ &  \mathcal{O}({\varepsilon_1}^2),
\quad 
\text{$2^2$}:\ 
\mathcal{O}({\varepsilon_2}^2),
\quad 
\text{$3^2$}:\  \mathcal{O}({\varepsilon_3}^2),\\
\text{$12$}:\ &  \mathcal{O}({\varepsilon_1} {\varepsilon_2}),\quad \text{$23$}:\ 
\mathcal{O}({\varepsilon_2}{\varepsilon_3}),\quad 
\text{$13$}:\  \mathcal{O}({\varepsilon_1}{\varepsilon_3}).
\end{aligned}
\end{align}
In total, 44 types realizing hierarchical structures in Eq.~(\ref{eq: hierarchy_mass}) exist.

In these types we investigate phenomenologically favorable models.
As benchmark points of modulus $\tau$, we choose $\tau=2.1i$ and $\omega+0.051i$ in the vicinity of $i\infty$ and $\omega$.
The values of the modular forms of weight 8 at the benchmark points are evaluated as
\begin{align}
&\tau=2.1i:~
Y^{(8)}_{{\bf{1}}}/ Y^{(8)}_{{\bf{1}}} = 1 \rightarrow 1,\quad 
Y^{(8)}_{{\bf{1}}'}/ Y^{(8)}_{{\bf{1}}} = -0.148 \rightarrow \varepsilon,\quad
Y^{(8)}_{{\bf{1}}''}/ Y^{(8)}_{{\bf{1}}} = 0.0218 \rightarrow \varepsilon^2, \\
&\tau=\omega+0.051i:~
| Y^{(8)}_{{\bf{1}}''}/ Y^{(8)}_{{\bf{1}}''}| = 1 \rightarrow 1,\quad 
|Y^{(8)}_{{\bf{1}}'}/ Y^{(8)}_{{\bf{1}}''} | = 0.148 \rightarrow \varepsilon,\quad
|Y^{(8)}_{{\bf{1}}}/ Y^{(8)}_{{\bf{1}}''}| = 0.0218  \rightarrow \varepsilon^2.
\end{align}
We enumerate the models for each choice of the signs $\pm 1$ in coupling constants $\alpha$ and $\beta$ for 44 type.
Then we pick up the models realizing the orders of the quark mass ratios and mixing angles.
Therefore we require the following hierarchical results as phenomenologically favorable conditions,
\begin{align}
\begin{aligned}
\label{eq: mass_ratio_order}
    &1/3< \frac{(m_u/m_t)_{\textrm{model}}}{(m_u/m_t)_{\textrm{GUT}}} < 3,\quad  
    1/3 < \frac{(m_c/m_t)_{\textrm{model}}}{(m_c/m_t)_{\textrm{GUT}}} < 3, \\
    &1/3< \frac{(m_d/m_b)_{\textrm{model}}}{(m_d/m_b)_{\textrm{GUT}}} < 3,\quad  
    1/3 < \frac{(m_s/m_b)_{\textrm{model}}}{(m_s/m_b)_{\textrm{GUT}}} < 3, \\
    &2/3 < \frac{|V_{\textrm{CKM}}^x|_{\textrm{model}}}{|V_{\textrm{CKM}}^x|_{\textrm{GUT}}} < 3/2, \quad (x\in\{us,cb,ub\}).
\end{aligned}
\end{align}
As a result, we find 1,584 number of models satisfying these conditions at both benchmark points $\tau = 2.1i$ and $\omega+0.051i$.
Table \ref{tab:chi<0.01atinfinite} shows the results at $\tau=2.1i$ and Table \ref{tab:chi<0.01atomega} shows ones at $\tau=\omega+0.051i$.
\begin{table}[H]
  \centering
  \begin{tabular}{cc|cc}
    \hline
Type & Number of models  & Type & Number of models  \\
    \hline \hline
$123\textrm{-}1^22^2\textrm{-}12$ & 64 & $1^22\textrm{-}1^23^2\textrm{-}3^2$ & 64 \\
$123\textrm{-}1^22^2\textrm{-}23$ & 64 & $1^22\textrm{-}1^23^2\textrm{-}12$ & 32 \\
$123\textrm{-}1^22^2\textrm{-}1^2$ & 96 & $1^22\textrm{-}1^23^2\textrm{-}23$ & 32 \\
$123\textrm{-}1^22^2\textrm{-}3^2$ & 96 & $1^22\textrm{-}1^23^2\textrm{-}13$ & 64 \\
$123\textrm{-}123^2\textrm{-}12$ & 32 & $1^22\textrm{-}1^223\textrm{-}1^2$ & 16 \\
$123\textrm{-}123^2\textrm{-}23$ & 16 & $1^22\textrm{-}1^223\textrm{-}2^2$ & 48 \\
$123\textrm{-}123^2\textrm{-}1^2$ & 32 & $1^22\textrm{-}1^223\textrm{-}3^2$ & 48 \\
$123\textrm{-}123^2\textrm{-}3^2$ & 32 & $1^22\textrm{-}1^223\textrm{-}12$ & 32 \\
$1^22\textrm{-}1^22^2\textrm{-}1^2$ & 32 & $1^22\textrm{-}1^223\textrm{-}23$ & 16 \\
$1^22\textrm{-}1^22^2\textrm{-}2^2$ & 64 & $1^22\textrm{-}1^223\textrm{-}13$ & 48 \\
$1^22\textrm{-}1^22^2\textrm{-}3^2$ & 64 & $1^22\textrm{-}12^23\textrm{-}1^2$ & 16 \\
$1^22\textrm{-}1^22^2\textrm{-}12$ & 32 & $1^22\textrm{-}12^23\textrm{-}2^2$ & 32 \\
$1^22\textrm{-}1^22^2\textrm{-}23$ & 32 & $1^22\textrm{-}12^23\textrm{-}3^2$ & 32 \\
$1^22\textrm{-}1^22^2\textrm{-}13$ & 64 & $1^22\textrm{-}12^23\textrm{-}12$ & 16 \\
$1^22\textrm{-}2^23^2\textrm{-}1^2$ & 32 & $1^22\textrm{-}12^23\textrm{-}23$ & 16 \\
$1^22\textrm{-}2^23^2\textrm{-}2^2$ & 32 & $1^22\textrm{-}12^23\textrm{-}13$ & 32 \\
$1^22\textrm{-}2^23^2\textrm{-}3^2$ & 32 & $1^22\textrm{-}123^2\textrm{-}1^2$ & 0 \\
$1^22\textrm{-}2^23^2\textrm{-}12$ & 0 & $1^22\textrm{-}123^2\textrm{-}2^2$ & 16 \\
$1^22\textrm{-}2^23^2\textrm{-}23$ & 32 & $1^22\textrm{-}123^2\textrm{-}3^2$ & 16 \\
$1^22\textrm{-}2^23^2\textrm{-}13$ & 32 & $1^22\textrm{-}123^2\textrm{-}12$ & 16 \\
$1^22\textrm{-}1^23^2\textrm{-}1^2$ & 32 & $1^22\textrm{-}123^2\textrm{-}23$ & 0 \\
$1^22\textrm{-}1^23^2\textrm{-}2^2$ & 64 & $1^22\textrm{-}123^2\textrm{-}13$ & 16 \\
    \hline
  \end{tabular}
    \caption{Number of models satisfying hierarchy conditions in Eq. (\ref{eq: mass_ratio_order}) at the benchmark point $\tau = 2.1i$.
The first and third columns denote the type of $M_u^{11}$-$M_d^{22}$-$M_d^{11}$.}
    \label{tab:chi<0.01atinfinite}
\end{table}
\begin{table}[H]
  \centering
  \begin{tabular}{cc|cc}
    \hline
Type & Number of models  & Type & Number of models  \\
    \hline \hline
$123\textrm{-}1^22^2\textrm{-}12$ & 64 & $1^22\textrm{-}1^23^2\textrm{-}3^2$ & 64 \\
$123\textrm{-}1^22^2\textrm{-}23$ & 64 & $1^22\textrm{-}1^23^2\textrm{-}12$ & 32 \\
$123\textrm{-}1^22^2\textrm{-}1^2$ & 96 & $1^22\textrm{-}1^23^2\textrm{-}23$ & 32 \\
$123\textrm{-}1^22^2\textrm{-}3^2$ & 96 & $1^22\textrm{-}1^23^2\textrm{-}13$ & 64 \\
$123\textrm{-}123^2\textrm{-}12$ & 32 & $1^22\textrm{-}1^223\textrm{-}1^2$ & 16 \\
$123\textrm{-}123^2\textrm{-}23$ & 16 & $1^22\textrm{-}1^223\textrm{-}2^2$ & 48 \\
$123\textrm{-}123^2\textrm{-}1^2$ & 32 & $1^22\textrm{-}1^223\textrm{-}3^2$ & 48 \\
$123\textrm{-}123^2\textrm{-}3^2$ & 32 & $1^22\textrm{-}1^223\textrm{-}12$ & 32 \\
$1^22\textrm{-}1^22^2\textrm{-}1^2$ & 32 & $1^22\textrm{-}1^223\textrm{-}23$ & 16 \\
$1^22\textrm{-}1^22^2\textrm{-}2^2$ & 64 & $1^22\textrm{-}1^223\textrm{-}13$ & 48 \\
$1^22\textrm{-}1^22^2\textrm{-}3^2$ & 64 & $1^22\textrm{-}12^23\textrm{-}1^2$ & 16 \\
$1^22\textrm{-}1^22^2\textrm{-}12$ & 32 & $1^22\textrm{-}12^23\textrm{-}2^2$ & 32 \\
$1^22\textrm{-}1^22^2\textrm{-}23$ & 32 & $1^22\textrm{-}12^23\textrm{-}3^2$ & 32 \\
$1^22\textrm{-}1^22^2\textrm{-}13$ & 64 & $1^22\textrm{-}12^23\textrm{-}12$ & 16 \\
$1^22\textrm{-}2^23^2\textrm{-}1^2$ & 32 & $1^22\textrm{-}12^23\textrm{-}23$ & 16 \\
$1^22\textrm{-}2^23^2\textrm{-}2^2$ & 32 & $1^22\textrm{-}12^23\textrm{-}13$ & 32 \\
$1^22\textrm{-}2^23^2\textrm{-}3^2$ & 32 & $1^22\textrm{-}123^2\textrm{-}1^2$ & 0 \\
$1^22\textrm{-}2^23^2\textrm{-}12$ & 0 & $1^22\textrm{-}123^2\textrm{-}2^2$ & 16 \\
$1^22\textrm{-}2^23^2\textrm{-}23$ & 32 & $1^22\textrm{-}123^2\textrm{-}3^2$ & 16 \\
$1^22\textrm{-}2^23^2\textrm{-}13$ & 32 & $1^22\textrm{-}123^2\textrm{-}12$ & 16 \\
$1^22\textrm{-}1^23^2\textrm{-}1^2$ & 32 & $1^22\textrm{-}123^2\textrm{-}23$ & 0 \\
$1^22\textrm{-}1^23^2\textrm{-}2^2$ & 64 & $1^22\textrm{-}123^2\textrm{-}13$ & 16 \\
\hline
  \end{tabular}
  \caption{Number of models satisfying hierarchy conditions in Eq. (\ref{eq: mass_ratio_order}) at the benchmark point $\tau = \omega+0.051i$.
The first and third columns denote the type of $M_u^{11}$-$M_d^{22}$-$M_d^{11}$.}
  \label{tab:chi<0.01atomega}
\end{table}


\subsection{Numerical example}
\label{subsection:Numerical example}

To illustrate the numerical calculations of quark flavors without fine-tuning, here we show some examples using the favorable models classified in Tables \ref{tab:chi<0.01atinfinite} and \ref{tab:chi<0.01atomega}.


\paragraph{Type $1^22$-$1^22^2$-$1^2$ at $\tau=2.1i \sim i\infty$}~\\
In type $1^22$-$1^22^2$-$1^2$, the $T$-charge assignments to quark fields are given by
\begin{align}
&\{Q^1,Q^2,Q^3\}:~\{(a_1,a_2,a_3),(b_1,b_2,b_3),(0,0,0)\}, \\
&\{u_R^1,u_R^2,u_R^3\}:~\{(1-a_1,1-a_2,1-a_3)_{\textrm{mod~3}},(1-b_1,2-b_2,-b_3)_{\textrm{mod~3}},(0,0,0)\}, \\
&\{d_R^1,d_R^2,d_R^3\}:~\{(1-a_1,1-a_2,-a_3)_{\textrm{mod~3}},(1-b_1,-b_2,-b_3)_{\textrm{mod~3}},(0,0,0)\},
\end{align}
where $a_i \in \{0,1,2 \}$ and $b_i \in \{0,1,2 \}$ are $T$-charges of the $i$-th $A_4$ for $Q^1$ and $Q^2$, respectively.
We find the best-fit choices of $a_i$ and $b_i$,
\begin{align}
(a_1,a_2,a_3) = (1,1,1), \quad (b_1,b_2,b_3) = (2,2,0),
\end{align}
and signs $\pm 1$ in the coupling constants $\alpha$ and $\beta$,
\begin{equation}
\begin{pmatrix}
\alpha^{11} & \alpha^{12} & \alpha^{13} \\
\alpha^{21} & \alpha^{22} & \alpha^{23} \\
\alpha^{31} & \alpha^{32} & \alpha^{33} \\
\end{pmatrix}
=
\begin{pmatrix}
1 & 1 & 1 \\
1 & 1 & -1 \\
1 & -1 & -1 \\
\end{pmatrix}, \quad\begin{pmatrix}
\beta^{11} & \beta^{12} & \beta^{13} \\
\beta^{21} & \beta^{22} & \beta^{23} \\
\beta^{31} & \beta^{32} & \beta^{33} \\
\end{pmatrix}
=
\begin{pmatrix}
1 & 1 & 1 \\
1 & -1 & -1 \\
1 & 1 & 1 \\
\end{pmatrix}.
\end{equation}
In these choices, we obtain the quark mass matrices,
\begin{align}
&M_u = \langle H_u \rangle\begin{pmatrix}
Y^{(8)}_{\bm{1''}}Y^{(8)}_{\bm{1''}}Y^{(8)}_{\bm{1''}} & Y^{(8)}_{\bm{1}}Y^{(8)}_{\bm{1''}}Y^{(8)}_{\bm{1''}} & Y^{(8)}_{\bm{1''}}Y^{(8)}_{\bm{1''}}Y^{(8)}_{\bm{1''}} \\
Y^{(8)}_{\bm{1'}}Y^{(8)}_{\bm{1'}}Y^{(8)}_{\bm{1}} & Y^{(8)}_{\bm{1''}}Y^{(8)}_{\bm{1'}}Y^{(8)}_{\bm{1}} & -Y^{(8)}_{\bm{1'}}Y^{(8)}_{\bm{1'}}Y^{(8)}_{\bm{1}} \\
Y^{(8)}_{\bm{1}}Y^{(8)}_{\bm{1}}Y^{(8)}_{\bm{1}} & -Y^{(8)}_{\bm{1'}}Y^{(8)}_{\bm{1}}Y^{(8)}_{\bm{1}} & -Y^{(8)}_{\bm{1}}Y^{(8)}_{\bm{1}}Y^{(8)}_{\bm{1}} \\
\end{pmatrix}, \\
&M_d = \langle H_d \rangle\begin{pmatrix}
Y^{(8)}_{\bm{1''}}Y^{(8)}_{\bm{1''}}Y^{(8)}_{\bm{1}} & Y^{(8)}_{\bm{1}}Y^{(8)}_{\bm{1'}}Y^{(8)}_{\bm{1''}} & Y^{(8)}_{\bm{1''}}Y^{(8)}_{\bm{1''}}Y^{(8)}_{\bm{1''}} \\
Y^{(8)}_{\bm{1'}}Y^{(8)}_{\bm{1'}}Y^{(8)}_{\bm{1'}} & -Y^{(8)}_{\bm{1''}}Y^{(8)}_{\bm{1}}Y^{(8)}_{\bm{1}} & -Y^{(8)}_{\bm{1'}}Y^{(8)}_{\bm{1'}}Y^{(8)}_{\bm{1}} \\
Y^{(8)}_{\bm{1}}Y^{(8)}_{\bm{1}}Y^{(8)}_{\bm{1'}} & Y^{(8)}_{\bm{1'}}Y^{(8)}_{\bm{1''}}Y^{(8)}_{\bm{1}} & Y^{(8)}_{\bm{1}}Y^{(8)}_{\bm{1}}Y^{(8)}_{\bm{1}} \\
\end{pmatrix}.
\end{align}
At $\tau=2.1i$, they take the hierarchical structures,
\begin{align}
|M_u/M_u^{33}| &=
\begin{pmatrix}
1.03\times 10^{-5} & 4.74\times 10^{-4} & 1.03\times 10^{-5} \\
2.18\times 10^{-2} & 3.21\times 10^{-3} & 2.18\times 10^{-2} \\
1.00 & 1.48\times 10^{-1} & 1.00 \\
\end{pmatrix} \\
&\sim
\begin{pmatrix}
{\cal O}(\varepsilon^6) & {\cal O}(\varepsilon^4) & {\cal O}(\varepsilon^6) \\
{\cal O}(\varepsilon^2) & {\cal O}(\varepsilon^3) & {\cal O}(\varepsilon^2) \\
{\cal O}(1) & {\cal O}(\varepsilon) & {\cal O}(1) \\
\end{pmatrix}, \\
|M_d/M_d^{33}| &=
\begin{pmatrix}
4.74\times 10^{-4} & 3.21\times 10^{-3} & 1.03\times 10^{-5} \\
3.21\times 10^{-3} & 2.18\times 10^{-2} & 2.18\times 10^{-2} \\
1.48\times 10^{-1} & 3.21\times 10^{-3} & 1.00 \\
\end{pmatrix} \\
&\sim
\begin{pmatrix}
{\cal O}(\varepsilon^4) & {\cal O}(\varepsilon^3) & {\cal O}(\varepsilon^6) \\
{\cal O}(\varepsilon^3) & {\cal O}(\varepsilon^2) & {\cal O}(\varepsilon^2) \\
{\cal O}(\varepsilon) & {\cal O}(\varepsilon^3) & {\cal O}(1) \\
\end{pmatrix}.
\end{align}
From these mass matrices, we can calculate the quark mass ratios and the absolute values of the CKM matrix elements.
The results are shown in Table \ref{tab: 1^22-123^2-13}.
\begin{table}[H]
  \begin{center}
    \renewcommand{\arraystretch}{1.3}
    \begin{tabular}{c|ccccccc} \hline
      & $\frac{m_u}{m_t}\times10^{6}$ & $\frac{m_c}{m_t}\times10^3$ & $\frac{m_d}{m_b}\times10^4$ & $\frac{m_s}{m_b}\times10^2$ & $|V_{\textrm{CKM}}^{us}|$ & $|V_{\textrm{CKM}}^{cb}|$ & $|V_{\textrm{CKM}}^{ub}|$ \\ \hline
      obtained values & 10.22 & 4.50 & 13.22 & 2.27 & 0.202 & 0.0419 & 0.00318 \\ \hline
      GUT scale values & 5.39 & 2.80 & 9.21 & 1.82 & 0.225 & 0.0400 & 0.00353 \\
      $1\sigma$ errors & $\pm 1.68$ & $\pm 0.12$ & $\pm 1.02$ & $\pm 0.10$ & $\pm 0.0007$ & $\pm 0.0008$ & $\pm 0.00013$ \\ \hline
    \end{tabular}
  \end{center}
  \caption{The mass ratios of the quarks and the absolute values of the CKM matrix elements at the benchmark point $\tau=2.1i$.
GUT scale values at $2\times 10^{16}$ GeV with $\tan \beta=5$ \cite{Antusch:2013jca,Bjorkeroth:2015ora} and $1\sigma$ errors are shown.}
\label{tab: 1^22-123^2-13}
\end{table}

As same as the analysis of quark flavor models with $\Gamma_6$ modular symmetry studied in Section \ref{subsec:Gamma_6}, we have fixed the coupling constants $\alpha^{jk}$ and $\beta^{jk}$ to $\pm 1$.
Then more realistic results can be obtained by varying $\alpha^{jk}$, $\beta^{jk}={\cal O}(1)$.
In the above model, let us choose the following ${\cal O}(1)$ size of coupling constants,
\begin{equation}
\begin{pmatrix}
\alpha^{11} & \alpha^{12} & \alpha^{13} \\
\alpha^{21} & \alpha^{22} & \alpha^{23} \\
\alpha^{31} & \alpha^{32} & \alpha^{33} \\
\end{pmatrix}
=
\begin{pmatrix}
2.71 & 1.94 & 2.67 \\
2.53 & 1.99 & -2.23 \\
2.82 & -1.39 & -2.44 \\
\end{pmatrix}, \quad\begin{pmatrix}
\beta^{11} & \beta^{12} & \beta^{13} \\
\beta^{21} & \beta^{22} & \beta^{23} \\
\beta^{31} & \beta^{32} & \beta^{33} \\
\end{pmatrix}
=
\begin{pmatrix}
1.24 & 1.96 & 3.00 \\
2.45 & -1.88 & -2.26 \\
1.00 &  1.20 & 2.35 \\
\end{pmatrix}.
\end{equation}
In these choices, we obtain the quark mass ratios,
\begin{align}
&(m_u,m_c,m_t)/m_t = (5.39\times 10^{-6}, 2.80\times 10^{-3}, 1), \\
&(m_d,m_s,m_b)/m_b = (9.21\times 10^{-4}, 1.82\times 10^{-2}, 1),
\end{align}
and the absolute values of the CKM matrix elements,
\begin{align}
|V_{\textrm{CKM}}| =
\begin{pmatrix}
0.974 & 0.225 & 0.00353 \\
0.225 & 0.974 & 0.0400 \\
0.00556 & 0.0397 & 0.999 \\
\end{pmatrix}.
\end{align}
The results are summarized in Table \ref{tab: orderone}.
\begin{table}[H]
  \begin{center}
    \renewcommand{\arraystretch}{1.3}
    \begin{tabular}{c|ccccccc} \hline
      & $\frac{m_u}{m_t}\times10^{6}$ & $\frac{m_c}{m_t}\times10^3$ & $\frac{m_d}{m_b}\times10^4$ & $\frac{m_s}{m_b}\times10^2$ & $|V_{\textrm{CKM}}^{us}|$ & $|V_{\textrm{CKM}}^{cb}|$ & $|V_{\textrm{CKM}}^{ub}|$ \\ \hline
      obtained values & 5.39 & 2.80 & 9.21 & 1.82 & 0.225 & 0.0400 & 0.00353 \\ \hline
      GUT scale values & 5.39 & 2.80 & 9.21 & 1.82 & 0.225 & 0.0400 & 0.00353 \\
      $1\sigma$ errors & $\pm 1.68$ & $\pm 0.12$ & $\pm 1.02$ & $\pm 0.10$ & $\pm 0.0007$ & $\pm 0.0008$ & $\pm 0.00013$ \\ \hline
    \end{tabular}
  \end{center}
  \caption{The mass ratios of the quarks and the absolute values of the CKM matrix elements at the benchmark point $\tau=2.1i$.
GUT scale values at $2\times 10^{16}$ GeV with $\tan \beta=5$ \cite{Antusch:2013jca,Bjorkeroth:2015ora} and $1\sigma$ errors are shown.}
\label{tab: orderone}
\end{table}


\paragraph{Type $123$-$1^22^2$-$12$ at $\tau=\omega+0.051i \sim \omega$}~\\
In type $123$-$1^22^2$-$12$, the $ST$-charge assignments to quark fields are given by
\begin{align}
&\{Q^1,Q^2,Q^3\}:~\{(a_1,a_2,a_3),(b_1,b_2,b_3),(0,0,0)\}, \\
&\{u_R^1,u_R^2,u_R^3\}:~\{(1-a_1,1-a_2,1-a_3)_{\textrm{mod~3}},(2-b_1,2-b_2,2-b_3)_{\textrm{mod~3}},(0,0,0)\}, \\
&\{d_R^1,d_R^2,d_R^3\}:~\{(1-a_1,1-a_2,-a_3)_{\textrm{mod~3}},(2-b_1,2-b_2,-b_3)_{\textrm{mod~3}},(0,0,0)\},
\end{align}
where $a_i\in\{0,1,2\}$ and $b_i\in\{0,1,2\}$ are $ST$-charges of the $i$-th $A_4$ for $Q^1$ and $Q^2$, respectively.
We find the best-fit choices of $a_i$ and $b_i$,
\begin{align}
(a_1,a_2,a_3) = (1,1,1), \quad (b_1,b_2,b_3) = (2,0,2),
\end{align}
and signs $\pm 1$ in the coupling constants $\alpha$ and $\beta$,
\begin{equation}
\begin{pmatrix}
\alpha^{11} & \alpha^{12} & \alpha^{13} \\
\alpha^{21} & \alpha^{22} & \alpha^{23} \\
\alpha^{31} & \alpha^{32} & \alpha^{33} \\
\end{pmatrix}
=
\begin{pmatrix}
1 & 1 & 1 \\
1 & 1 & -1 \\
1 & -1 & -1 \\
\end{pmatrix}, \quad\begin{pmatrix}
\beta^{11} & \beta^{12} & \beta^{13} \\
\beta^{21} & \beta^{22} & \beta^{23} \\
\beta^{31} & \beta^{32} & \beta^{33} \\
\end{pmatrix}
=
\begin{pmatrix}
1 & 1 & 1 \\
1 & -1 & 1 \\
1 & -1 & -1 \\
\end{pmatrix}.
\end{equation}
In these choices, we obtain the quark mass matrices,
\begin{align}
&M_u = \langle H_u \rangle\begin{pmatrix}
Y^{(8)}_{\bm{1}}Y^{(8)}_{\bm{1}}Y^{(8)}_{\bm{1}} & Y^{(8)}_{\bm{1}}Y^{(8)}_{\bm{1''}}Y^{(8)}_{\bm{1}} & Y^{(8)}_{\bm{1}}Y^{(8)}_{\bm{1}}Y^{(8)}_{\bm{1}} \\
Y^{(8)}_{\bm{1'}}Y^{(8)}_{\bm{1''}}Y^{(8)}_{\bm{1'}} & Y^{(8)}_{\bm{1'}}Y^{(8)}_{\bm{1'}}Y^{(8)}_{\bm{1'}} & -Y^{(8)}_{\bm{1'}}Y^{(8)}_{\bm{1''}}Y^{(8)}_{\bm{1'}} \\
Y^{(8)}_{\bm{1''}}Y^{(8)}_{\bm{1''}}Y^{(8)}_{\bm{1''}} & -Y^{(8)}_{\bm{1''}}Y^{(8)}_{\bm{1'}}Y^{(8)}_{\bm{1''}} & -Y^{(8)}_{\bm{1''}}Y^{(8)}_{\bm{1''}}Y^{(8)}_{\bm{1''}} \\
\end{pmatrix}, \\
&M_d = \langle H_d \rangle\begin{pmatrix}
Y^{(8)}_{\bm{1}}Y^{(8)}_{\bm{1}}Y^{(8)}_{\bm{1''}} & Y^{(8)}_{\bm{1}}Y^{(8)}_{\bm{1''}}Y^{(8)}_{\bm{1'}} & Y^{(8)}_{\bm{1}}Y^{(8)}_{\bm{1}}Y^{(8)}_{\bm{1}} \\
Y^{(8)}_{\bm{1'}}Y^{(8)}_{\bm{1''}}Y^{(8)}_{\bm{1}} & -Y^{(8)}_{\bm{1'}}Y^{(8)}_{\bm{1'}}Y^{(8)}_{\bm{1''}} & Y^{(8)}_{\bm{1'}}Y^{(8)}_{\bm{1''}}Y^{(8)}_{\bm{1'}} \\
Y^{(8)}_{\bm{1''}}Y^{(8)}_{\bm{1''}}Y^{(8)}_{\bm{1'}} & -Y^{(8)}_{\bm{1''}}Y^{(8)}_{\bm{1'}}Y^{(8)}_{\bm{1}} & -Y^{(8)}_{\bm{1''}}Y^{(8)}_{\bm{1''}}Y^{(8)}_{\bm{1''}} \\
\end{pmatrix}.
\end{align}
At $\tau=\omega+0.051i$, they take the hierarchical structures,
\begin{align}
|M_u/M_u^{33}| &=
\begin{pmatrix}
1.04\times 10^{-5} & 4.76\times 10^{-4} & 1.04\times 10^{-5} \\
2.18\times 10^{-2} & 3.22\times 10^{-3} & 2.18\times 10^{-2} \\
1.00 & 1.48\times 10^{-1} & 1.00 \\
\end{pmatrix} \\
&\sim
\begin{pmatrix}
{\cal O}(\varepsilon^6) & {\cal O}(\varepsilon^4) & {\cal O}(\varepsilon^6) \\
{\cal O}(\varepsilon^2) & {\cal O}(\varepsilon^3) & {\cal O}(\varepsilon^2) \\
{\cal O}(1) & {\cal O}(\varepsilon) & {\cal O}(1) \\
\end{pmatrix}, \\
|M_d/M_d^{33}| &=
\begin{pmatrix}
4.76\times 10^{-4} & 3.22\times 10^{-3} & 1.04\times 10^{-5} \\
3.22\times 10^{-3} & 2.18\times 10^{-2} & 2.18\times 10^{-2} \\
1.48\times 10^{-1} & 3.22\times 10^{-3} & 1.00 \\
\end{pmatrix} \\
&\sim
\begin{pmatrix}
{\cal O}(\varepsilon^4) & {\cal O}(\varepsilon^3) & {\cal O}(\varepsilon^6) \\
{\cal O}(\varepsilon^3) & {\cal O}(\varepsilon^2) & {\cal O}(\varepsilon^2) \\
{\cal O}(\varepsilon) & {\cal O}(\varepsilon^3) & {\cal O}(1) \\
\end{pmatrix}.
\end{align}
From these mass matrices, we can calculate the quark mass ratios and the absolute values of the CKM matrix elements.
The results are shown in Table \ref{tab: 123-123^2-12_omega}.
\begin{table}[H]
  \begin{center}
    \renewcommand{\arraystretch}{1.3}
    \begin{tabular}{c|ccccccc} \hline
      & $\frac{m_u}{m_t}\times10^{6}$ & $\frac{m_c}{m_t}\times10^3$ & $\frac{m_d}{m_b}\times10^4$ & $\frac{m_s}{m_b}\times10^2$ & $|V_{\textrm{CKM}}^{us}|$ & $|V_{\textrm{CKM}}^{cb}|$ & $|V_{\textrm{CKM}}^{ub}|$ \\ \hline
      obtained values & 10.3 & 4.52 & 13.29 & 2.27 & 0.202 & 0.0420 & 0.00319 \\ \hline
      GUT scale values & 5.39 & 2.80 & 9.21 & 1.82 & 0.225 & 0.0400 & 0.00353 \\ 
      $1\sigma$ errors & $\pm 1.68$ & $\pm 0.12$ & $\pm 1.02$ & $\pm 0.10$ & $\pm 0.0007$ & $\pm 0.0008$ & $\pm 0.00013$ \\ \hline
    \end{tabular}
  \end{center}
  \caption{The mass ratios of the quarks and the absolute values of the CKM matrix elements at the benchmark point $\tau=\omega + 0.051i$.
GUT scale values at $2\times 10^{16}$ GeV with $\tan \beta=5$ \cite{Antusch:2013jca,Bjorkeroth:2015ora} and $1\sigma$ errors are shown.}
\label{tab: 123-123^2-12_omega}
\end{table}


\subsection{CP violation}
\label{subsubsec:CP violation}

So far, we have studied the quark mass ratios and the absolute values of the CKM matrix elements on the $A_4\times A_4\times A_4$ modular symmetric flavor models without fine-tuning.
Here we additionally study the possibility of the realization of the CP violation.
We assume that the CP violation is only induced by the VEV of the modulus $\tau$.
The independent values of the modulus $\tau$ under the modular transformation (so-called the fundamental region) are shown in Figure \ref{fig:fundamental_region}.

Under the CP transformation, the modulus $\tau$ is transformed as \cite{Baur:2019kwi,Novichkov:2019sqv,Baur:2019iai}
\begin{align}
\tau \to -\tau^*,
\end{align}
i.e.,
\begin{align}
\textrm{Re}\tau \to -\textrm{Re}\tau, \quad \textrm{Im}\tau \to \textrm{Im}\tau.
\end{align}
The CP invariant points of the modulus are obviously given by following three,
\begin{itemize}
\item the imaginary axis, Re$\tau=0$;
\item the boundary, Re$\tau=\pm 1/2$;
\item the unit arc, $|\tau|=1$.
\end{itemize}
Note that the imaginary axis Re$\tau=0$ is CP invariant due to $T$-symmetry,
\begin{align}
\tau = \frac{1}{2}+i\textrm{Im}\tau \xrightarrow{CP} -\frac{1}{2}+i\textrm{Im}\tau
\xrightarrow{T} \frac{1}{2}+i\textrm{Im}\tau = \tau,
\end{align}
and the unit arc $|\tau|=1$ is CP invariant due to $S$-symmetry,
\begin{align}
\tau = e^{i\textrm{Arg}\tau} \xrightarrow{CP} -e^{-i\textrm{Arg}\tau}
\xrightarrow{S} e^{i\textrm{Arg}\tau} = \tau.
\end{align}
Therefore, we cannot obtain non-vanishing CP phase in our numerical examples studied in Subsection \ref{subsection:Numerical example} since we have fixed the modulus at $\tau=2.1i$ (the imaginary axis) or $\omega+0.051i$ (the boundary).
In order to find the models leading to non-vanishing CP phase, here we study the necessary conditions for the CP violation.

In the vicinity of the symmetric points, the mass matrices can be classified to two types.
In one type the CP violation is not induced while in another type it can be occured.
To understand this, first we study the mass matrices in the vicinity of $\tau=i\infty$ and $A_4$ modular symmetry instead of $A_4\times A_4\times A_4$.
As shown in Eq.~(\ref{eq:f expanded by eT}), the mass matrix elements as well as the modular forms with $T$-charge $r_T$ can be expanded by $\varepsilon_T^{r_T}=e^{2\pi i\tau r_T/N}=e^{2\pi i\tau r_T/3}$ in the first order approximation.
Thus, the up-sector mass matrix $M_u$ with the following $T$-charges,
\begin{align}
M_u:
\begin{pmatrix}
2 & 2 & 1 \\
1 & 1 & 0 \\
1 & 1 & 0 \\
\end{pmatrix},
\end{align}
can be estimated as
\begin{align}
M_u \sim
\begin{pmatrix}
\varepsilon_T^2 & \varepsilon_T^2 & \varepsilon_T \\
\varepsilon_T & \varepsilon_T & 1 \\
\varepsilon_T & \varepsilon_T & 1 \\
\end{pmatrix}.
\end{align}
This charge pattern can be realized by $T$-charge assignments of the fields,
\begin{align}
Q:(-1,0,0), \quad u_R:(-1,-1,0), \quad H_u:0.
\end{align}
Note that the second order contribution to the first order is ${\cal O}(\varepsilon_T^3)$ and it is ignorable in the vicinity of $\tau=i\infty$ where $\varepsilon_T\ll 1$.
Actually we have assumed $\varepsilon_T\sim 0.15$ in Subsection \ref{subsubsec:Favorable models} to generate large quark mass hierarchies.
This mass matrix has the phase factors,
\begin{align}
  M_u &\sim 
  \begin{pmatrix}
    |\varepsilon_T|^2e^{4\pi i\textrm{Re}\tau/3} & |\varepsilon_T|^2e^{4\pi i\textrm{Re}\tau/3} & |\varepsilon_T|e^{2\pi i\textrm{Re}\tau/3} \\
    |\varepsilon_T|e^{2\pi i\textrm{Re}\tau/3} & |\varepsilon_T|e^{2\pi i\textrm{Re}\tau/3}  & 1 \\
    |\varepsilon_T|e^{2\pi i\textrm{Re}\tau/3} & |\varepsilon_T|e^{2\pi i\textrm{Re}\tau/3} & 1 \\
  \end{pmatrix} \\
  &=
  \begin{pmatrix}
    |\varepsilon_T|^2e^{2i\alpha} & |\varepsilon_T|^2e^{2i\alpha} & |\varepsilon_T|e^{i\alpha} \\
    |\varepsilon_T|e^{i\alpha} & |\varepsilon_T|e^{i\alpha}  & 1 \\
    |\varepsilon_T|e^{i\alpha} & |\varepsilon_T|e^{i\alpha} & 1 \\
  \end{pmatrix},
\end{align}
where $\alpha = 2\pi\textrm{Re}\tau/3$.
For the phase transformations $u_L$ and $u_R$,
\begin{align}
u_L =
\begin{pmatrix}
e^{i\alpha} & & \\
& 1 & \\
& & 1 \\
\end{pmatrix}, \quad
u_R =
\begin{pmatrix}
e^{-i\alpha} & & \\
& e^{-i\alpha} & \\
& & 1 \\
\end{pmatrix},
\end{align}
the phase factors in the mass matrix are completely vanished as
\begin{align}
M_u &\to u_L^\dagger M_u u_R \notag \\
&\sim u_L^\dagger
\begin{pmatrix}
    |\varepsilon_T|^2e^{2i\alpha} & |\varepsilon_T|^2e^{2i\alpha} & |\varepsilon_T|e^{i\alpha} \\
    |\varepsilon_T|e^{i\alpha} & |\varepsilon_T|e^{i\alpha}  & 1 \\
    |\varepsilon_T|e^{i\alpha} & |\varepsilon_T|e^{i\alpha} & 1 \\
  \end{pmatrix}
u_R \\
&=
\begin{pmatrix}
e^{-i\alpha} & & \\
& 1 & \\
& & 1 \\
\end{pmatrix}
\begin{pmatrix}
    |\varepsilon_T|^2e^{2i\alpha} & |\varepsilon_T|^2e^{2i\alpha} & |\varepsilon_T|e^{i\alpha} \\
    |\varepsilon_T|e^{i\alpha} & |\varepsilon_T|e^{i\alpha}  & 1 \\
    |\varepsilon_T|e^{i\alpha} & |\varepsilon_T|e^{i\alpha} & 1 \\
  \end{pmatrix}
\begin{pmatrix}
e^{-i\alpha} & & \\
& e^{-i\alpha} & \\
& & 1 \\
\end{pmatrix} \\
&=
\begin{pmatrix}
    |\varepsilon_T|^2 & |\varepsilon_T|^2 & |\varepsilon_T| \\
    |\varepsilon_T| & |\varepsilon_T| & 1 \\
    |\varepsilon_T| & |\varepsilon_T| & 1 \\
  \end{pmatrix}.
\end{align}
This is because the powers of the phase factors in the mass matrix are determined by their $T$-charges and the phase transformations $u_L$ and $u_R$ correspond to them as
\begin{align}
  &u_L^\dagger =
  \begin{pmatrix}
    e^{-i\phi^1\alpha} & & \\
    & e^{-i\phi^2\alpha} & \\
    & & e^{-i\phi^3\alpha} \\
  \end{pmatrix}, \quad \phi^i = [-(T\textrm{-charge~of~}Q^i)]_{\textrm{mod~3}}, \label{eq:u_L} \\
  &u_R =
  \begin{pmatrix}
    e^{-i\psi^1\alpha} & & \\
    & e^{-i\psi^2\alpha} & \\
    & & e^{-i\psi^3\alpha} \\
  \end{pmatrix}, \quad \psi^i = [-(T\textrm{-charge~of~}u_R^i)]_{\textrm{mod~3}}. \label{eq:u_R}
\end{align}
Here we have used the notation $[q]_{\textrm{mod~3}}=r$ when $q=3n+r$ with the maximum integer $n$ such that $r=0,1,2$.
This is one example of the type where the CP violation is not induced \footnote{Similar behaviors at the modular symmetric points were studied in Refs.~\cite{Kobayashi:2019uyt,Kikuchi:2022geu}.}.

On the other hand, the mass matrix with the following $T$-charges,
\begin{align}
  M_u:~
  \begin{pmatrix}
    0 & 2 & 1 \\
    2 & 1 & 0 \\
    2 & 1 & 0 \\
  \end{pmatrix},
\end{align}
has non-vanishing phase factors.
This charge pattern can be realized by $T$-charge assignments of the fields,
\begin{align}
Q:(-1,0,0), \quad u_R:(-2,-1,0), \quad H_u:0.
\end{align}
This mass matrix can be estimated as
\begin{align}
  M_u &\sim
  \begin{pmatrix}
    1 & \varepsilon_T^2 & \varepsilon_T \\
    \varepsilon_T^2 & \varepsilon_T & 1 \\
    \varepsilon_T^2 & \varepsilon_T & 1 \\
  \end{pmatrix} \\
  &=
  \begin{pmatrix}
    1 & |\varepsilon_T|^2e^{4\pi i\textrm{Re}\tau/3} & |\varepsilon_T|e^{2\pi i\textrm{Re}\tau/3} \\
    |\varepsilon_T|^2e^{4\pi i\textrm{Re}\tau/3} & |\varepsilon_T|e^{2\pi i\textrm{Re}\tau/3} & 1 \\
    |\varepsilon_T|^2e^{4\pi i\textrm{Re}\tau/3} & |\varepsilon_T|e^{2\pi i\textrm{Re}\tau/3} & 1 \\
  \end{pmatrix} \\
  &=
  \begin{pmatrix}
    1 & |\varepsilon_T|^2e^{2i\alpha} & |\varepsilon_T|e^{i\alpha} \\
    |\varepsilon_T|^2e^{2i\alpha} & |\varepsilon_T|e^{i\alpha} & 1 \\
    |\varepsilon_T|^2e^{2i\alpha} & |\varepsilon_T|e^{i\alpha} & 1 \\
  \end{pmatrix},
\end{align}
where $\alpha = 2\pi \textrm{Re}\tau/3$.
Under the basis transformations in Eqs.~(\ref{eq:u_L}) and (\ref{eq:u_R}), the phase factors in this matrix are partially vanished,
\begin{align}
  M_u &\rightarrow u_L^\dagger M_u u_R \notag \\
  &\sim
  u_L^\dagger
  \begin{pmatrix}
    1 & |\varepsilon_T|^2e^{2i\alpha} & |\varepsilon_T|e^{i\alpha} \\
    |\varepsilon_T|^2e^{2i\alpha} & |\varepsilon_T|e^{i\alpha} & 1 \\
    |\varepsilon_T|^2e^{2i\alpha} & |\varepsilon_T|e^{i\alpha} & 1 \\
  \end{pmatrix}
  u_R \\
  &=
  \begin{pmatrix}
    e^{-i\alpha} & & \\
    & 1 & \\
    & & 1 \\
  \end{pmatrix}
  \begin{pmatrix}
    1 & |\varepsilon_T|^2e^{2i\alpha} & |\varepsilon_T|e^{i\alpha} \\
    |\varepsilon_T|^2e^{2i\alpha} & |\varepsilon_T|e^{i\alpha} & 1 \\
    |\varepsilon_T|^2e^{2i\alpha} & |\varepsilon_T|e^{i\alpha} & 1 \\
  \end{pmatrix}
  \begin{pmatrix}
    e^{-2i\alpha} & & \\
    & e^{-i\alpha} & \\
    & & 1 \\
  \end{pmatrix} \\
  &=
  \begin{pmatrix}
    e^{-3i\alpha} & |\varepsilon_T|^2 & |\varepsilon_T| \\
    |\varepsilon_T|^2 & |\varepsilon_T| & 1 \\
    |\varepsilon_T|^2 & |\varepsilon_T| & 1 \\
  \end{pmatrix}.
\end{align}
In this case, the phase factor $e^{-3i\alpha}$ in the (1,1) matrix element survives.
This is because $T$-charge is identified up to $N=3$ in mass matrix elements.
Notice that if the (1,1) matrix element in $M_u$ has $T$-charge 3, it has the phase factor $e^{3i\alpha}$ and the mass matrix in the first approximation becomes completely real after the phase transformation.
Thus the identification of $T$-charge in the mass matrix leads to non-vanishing phase factors.
This is one example of the type where the CP violation can be induced.

This mechanism is generallized as follows.
The $T$-charge of the $M_u^{jk}$ element is given by one of $Q^j$ and $u_R^k$,
\begin{align}
  T\textrm{-charge~of~}M_u^{jk} = [-(T\textrm{-charge~of~}Q^j) - (T\textrm{-charge~of~}u_R^k)]_{\textrm{mod~3}}.
\end{align}
In the first order approximation, $M^{jk}_u$ element has the phase factor,
\begin{align}
  \textrm{exp}\left[i\alpha(T\textrm{-charge~of~}M_u^{jk})\right] =
  \textrm{exp}\left[i\alpha[-(T\textrm{-charge~of~}Q^j) - (T\textrm{-charge~of~}u_R^k)]_{\textrm{mod~3}}\right].
\end{align}
On the other hand, the mass matrix element after the phase transformation, $[u_L^\dagger M_u u_R]^{jk}$, has the phase factor,
\begin{align}
&\textrm{exp}\left[i\alpha[-(T\textrm{-charge~of~}Q^j) - (T\textrm{-charge~of~}u_R^k)]_{\textrm{mod~3}}\right] \notag \\
&\times
  \textrm{exp}\left[-i\alpha([-(T\textrm{-charge~of~}Q^j)]_{\textrm{mod~3}} + [-(T\textrm{-charge~of~}u_R^k)]_{\textrm{mod~3}})\right].
\end{align}
This is cancelled only if
\begin{align}
&[-(T\textrm{-charge~of~}Q^j) - (T\textrm{-charge~of~}u_R^k)]_{\textrm{mod~3}} \notag \\
&=
[-(T\textrm{-charge~of~}Q^j)]_{\textrm{mod~3}} + [-(T\textrm{-charge~of~}u_R^k)]_{\textrm{mod~3}}, \label{eq:[Q+u]mod3=[Q]mod3+[u]mod3}
\end{align}
is satisfied.
Notice that the left-hand side in Eq.~(\ref{eq:[Q+u]mod3=[Q]mod3+[u]mod3}) is less than 3.
Thus when
\begin{align}
  [-(T\textrm{-charge~of~}Q^j)]_{\textrm{mod~3}} +[- ( T\textrm{-charge~of~}u_R^k)]_{\textrm{mod~3}} \geq 3, \label{eq:CP_condition_up}
\end{align}
is satisfied, $M_u^{jk}$ gets non-vanishing phase factor $e^{-3i\alpha}$ after the basis transformations in Eqs.~(\ref{eq:u_L}) and (\ref{eq:u_R}).
Similarly, we find the condition for down-sector quarks,
\begin{align}
  [-(T\textrm{-charge~of~}Q^j)]_{\textrm{mod~3}} +[- ( T\textrm{-charge~of~}d_R^k)]_{\textrm{mod~3}} \geq 3. \label{eq:CP_condition_down}
\end{align}
Remind that the residual charges determine the hierarchical structures of the mass matrices.
These conditions imply that residual charges further determine the phase factors in mass matrices.
Therefore, the CP violation is strongly related to the hierarchical quark mass ratios through the residual charges.
This is also true in other modular symmetric flavor models since above analysis only depends on the residual charges of the modular group.

When either of the conditions in Eqs.~(\ref{eq:CP_condition_up}) and (\ref{eq:CP_condition_down}) are satisfied, the quark mass matrices can be complex and the CP violation can be induced depending on the value of the modulus $\tau$.
Otherwise, the quark mass matrices are real in the first order approximation and the sufficient CP violation never occur.
Note that again the second order contribution to the first order is estimated as ${\cal O}(10^{-3})$ and it is ignorable.
Thus we can regard Eqs.~(\ref{eq:CP_condition_up}) and (\ref{eq:CP_condition_down}) as the necessary conditions for the CP violation.

Next we consider the necessary conditions in the vicinity of $\tau=\omega$.
Since $A_4$ modular group satisfies $(ST)^3=\mathbb{I}$, we can regard the same conditions in Eqs.~(\ref{eq:CP_condition_up}) and (\ref{eq:CP_condition_down}),
\begin{align}
  &[-(ST\textrm{-charge~of~}Q^j)]_{\textrm{mod~3}} +[- ( ST\textrm{-charge~of~}u_R^k)]_{\textrm{mod~3}} \geq 3, \\
  &[-(ST\textrm{-charge~of~}Q^j)]_{\textrm{mod~3}} +[- ( ST\textrm{-charge~of~}d_R^k)]_{\textrm{mod~3}} \geq 3,
\end{align}
as the necessary conditions for the CP violation.

Now we extend the necessary conditions to $A_4\times A_4\times A_4$ modular symmetry.
We have three necessary conditions for three $A_4$ symmetry.
When at least one necessary condition for $A_4$ symmetry is satisfied, the CP violation can be induced.
Consequently we find that all of favorable models shown in Tables \ref{tab:chi<0.01atinfinite} and \ref{tab:chi<0.01atomega} satisfy the necessary conditions for the CP violation.
In Appendix \ref{app:Mass matrix structures of the favorable models}, we classify the phase factors after the basis transformations in Eqs.~(\ref{eq:u_L}) and (\ref{eq:u_R}), and the hierarchical structures of the mass matrices of the favorable models.
Indeed these mass matrix structures contain non-vanishing phase factors.
Nevertheless the numerical analysis shows that all of the favorable models cannot obtain sufficient CP phase in the vicinity of the symmetric points.
To understand this, let us consider the CKM matrix at the first order approximation of $\varepsilon$.
It is calculated from the massmatrix structures in Table \ref{tab:mass_matrix_structures}.
As a result, we find all of the favorable models have either of the following two structures of the CKM matrix at the first order approximation,
\begin{align}
  V_{\textrm{CKM}} =
&\begin{pmatrix}
1 & 1.5 |\varepsilon|p^* & - |\varepsilon|^{3}p^* \\
- 1.5 |\varepsilon| p & 1 & - 2 |\varepsilon|^{2} \\
- 2 |\varepsilon|^{3} p & 2 |\varepsilon|^{2} & 1 \\
\end{pmatrix} , \quad
\begin{pmatrix}
1 & - 1.5 |\varepsilon| p^* & |\varepsilon|^{3}p^* \\
1.5 |\varepsilon| p & 1 & - 2 |\varepsilon|^{2} \\
2 |\varepsilon|^{3} p & 2 |\varepsilon|^{2} & 1 \\
\end{pmatrix}, \label{eq:CKM_at_first_order}
\end{align}
where $p$ is given by $\varepsilon/|\varepsilon|$.
It follows from this that Jarlskog invariant $J_\textrm{CP}$ vanishes at the first order approximation,
\begin{align}
  J_{\textrm{CP}} = |\textrm{Im}(V_{\textrm{CKM}}^{us}V_{\textrm{CKM}}^{cb}(V_{\textrm{CKM}}^{ub}V_{\textrm{CKM}}^{cs})^*)| = \textrm{Im}(3p^*p|\varepsilon|^{6}) = 0. \label{eq:first_p_cancellation}
\end{align}
When we consider the second order contribution to the CKM matrix, we obtain ${\cal O}(\varepsilon^2)$ deviation compared to the first order.
Hence, we expect
\begin{align}
J_\textrm{CP} \lesssim 3 \times |\varepsilon|^8, \label{eq:J_CP_second}
\end{align}
at the second order approximation.
Since we have assumed $|\varepsilon|\sim0.15$ to realize quark mass hierarchies, we obtain $J_\textrm{CP} \lesssim 7.7\times 10^{-7}$.
On the other hand, this is extremely small compared with the observed value $J_\textrm{CP} = 2.8\times 10^{-5}$.
Thus, the favorable models with $A_4\times A_4\times A_4$ modular symmetry summarized in Tables \ref{tab:chi<0.01atinfinite} and \ref{tab:chi<0.01atomega} cannot realize the sufficient CP phase although realistic values of the quark mass ratios and the mixing angles are obtained in the vicinity of the symmetric points.
In the following, we will confirm this through the numerical examples using the some of the favorable models.


\subsection{Numerical example of the CP violation}

\begin{figure}[H]
  \centering
  \includegraphics[width=8cm]{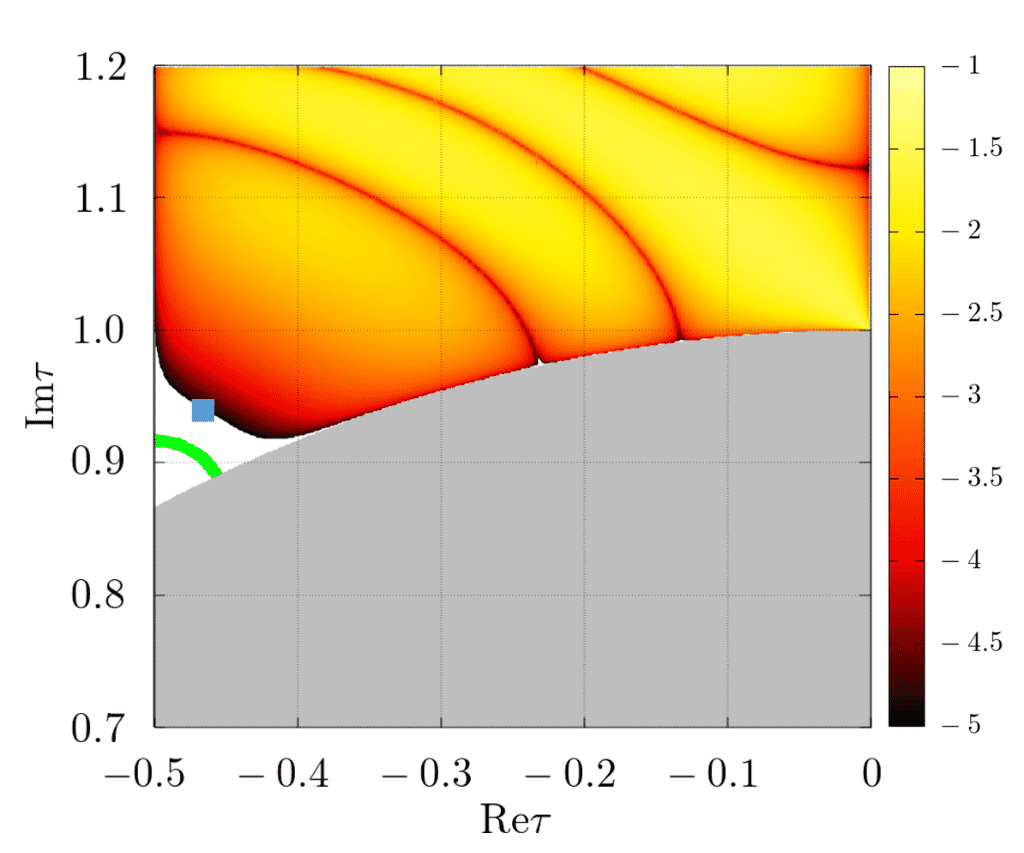}
  \caption{Allowed regions and Jarlskog invariant in the $\tau$ plane around $\tau=\omega$ for the model in type $1^22\textrm{-}1^23^2\textrm{-}3^2$.
  Green shows the region satisfying hierarchy conditions in Eq.~(\ref{eq: mass_ratio_order}), and black, red and yellow colors correspond to $\textrm{log}_{10}J_{\textrm{CP}}$.
  White shows the region with $\textrm{log}_{10}J_{\textrm{CP}}<-5$.
  Blue square denotes the point $\tau=\omega+(0.0326+0.0753i)$ on numerical example in Table \ref{tab:J>000001_omega}.
  Gray is out of fundamental region.}
  \label{fig:chi_J_omega}
\end{figure}

Here we will try to realize both the CP phase and the quark hierarchical structures in the vicinity of the symmetric points $\tau=i\infty$ or $\omega$ using the favorable models.
To illustrate the inconsistency between the CP violation and the quark hierarchical structures, we study the model in type $1^22\textrm{-}1^23^2\textrm{-}3^2$ at $\tau\sim\omega$.
In type $1^22\textrm{-}1^23^2\textrm{-}3^2$, the $ST$-charge assignments to quark fields are given by
\begin{align}
  &\{Q^1,Q^2,Q^3\}:~\{(a_1,a_2,a_3),(b_1,b_2,b_3),(0,0,0)\}, \\
  &\{u_R^1,u_R^2,u_R^3\}:~\{(2-a_1,2-a_2,2-a_3)_{\textrm{mod~3}},(2-b_1,1-b_2,-b_3)_{\textrm{mod~3}},(0,0,0)\}, \\
  &\{d_R^1,d_R^2,d_R^3\}:~\{(2-a_1,-a_2,2-a_3)_{\textrm{mod~3}},(-b_1,-b_2,2-b_3)_{\textrm{mod~3}},(0,0,0)\},
\end{align}
where $a_i\in\{0,1,2\}$ and $b_i\in\{0,1,2\}$ are $ST$-charges of the $i$-th $A_4$ for $Q^1$ and $Q^2$, respectively.
We find the best-fit choices of $a_i$ and $b_i$,
\begin{align}
(a_1,a_2,a_3) = (1,1,1), \quad (b_1,b_2,b_3) = (1,0,0), \label{eq:J_model_charge}
\end{align}
and signs $\pm 1$ in the coupling constants $\alpha$ and $\beta$,
\begin{align}
\begin{pmatrix}
    \alpha^{11} & \alpha^{12} & \alpha^{13} \\
    \alpha^{21} & \alpha^{22} & \alpha^{23} \\
    \alpha^{31} & \alpha^{32} & \alpha^{33} \\
  \end{pmatrix}
  =
  \begin{pmatrix}
    1 & 1 & 1 \\
    1 & -1 & -1 \\
    1 & 1 & -1 \\
  \end{pmatrix}, \quad
  \begin{pmatrix}
    \beta^{11} & \beta^{12} & \beta^{13} \\
    \beta^{21} & \beta^{22} & \beta^{23} \\
    \beta^{31} & \beta^{32} & \beta^{33} \\
  \end{pmatrix}
  =
  \begin{pmatrix}
    1 & 1 & 1 \\
   - 1 & 1 & -1 \\
    -1 & -1 & 1 \\
  \end{pmatrix}. \label{eq:J_model_sign}
\end{align}
In these choices, we obtain the quark mass matrices,
\begin{align}
  &M_u = \langle H_u \rangle
  \begin{pmatrix}
    Y^{(8)}_{\bm{1}}Y^{(8)}_{\bm{1}}Y^{(8)}_{\bm{1}} & Y^{(8)}_{\bm{1}}Y^{(8)}_{\bm{1''}}Y^{(8)}_{\bm{1}} & Y^{(8)}_{\bm{1}}Y^{(8)}_{\bm{1}}Y^{(8)}_{\bm{1}} \\
    Y^{(8)}_{\bm{1}}Y^{(8)}_{\bm{1''}}Y^{(8)}_{\bm{1''}} & -Y^{(8)}_{\bm{1}}Y^{(8)}_{\bm{1'}}Y^{(8)}_{\bm{1''}} & -Y^{(8)}_{\bm{1}}Y^{(8)}_{\bm{1''}}Y^{(8)}_{\bm{1''}} \\
    Y^{(8)}_{\bm{1''}}Y^{(8)}_{\bm{1''}}Y^{(8)}_{\bm{1''}} & Y^{(8)}_{\bm{1''}}Y^{(8)}_{\bm{1'}}Y^{(8)}_{\bm{1''}} & -Y^{(8)}_{\bm{1''}}Y^{(8)}_{\bm{1''}}Y^{(8)}_{\bm{1''}} \\
  \end{pmatrix}, \\
  &M_d = \langle H_d \rangle
  \begin{pmatrix}
    Y^{(8)}_{\bm{1}}Y^{(8)}_{\bm{1''}}Y^{(8)}_{\bm{1}} & Y^{(8)}_{\bm{1''}}Y^{(8)}_{\bm{1}}Y^{(8)}_{\bm{1'}} & Y^{(8)}_{\bm{1}}Y^{(8)}_{\bm{1}}Y^{(8)}_{\bm{1}} \\
    -Y^{(8)}_{\bm{1}}Y^{(8)}_{\bm{1'}}Y^{(8)}_{\bm{1''}} & Y^{(8)}_{\bm{1''}}Y^{(8)}_{\bm{1''}}Y^{(8)}_{\bm{1}} & -Y^{(8)}_{\bm{1}}Y^{(8)}_{\bm{1''}}Y^{(8)}_{\bm{1''}} \\
    -Y^{(8)}_{\bm{1''}}Y^{(8)}_{\bm{1'}}Y^{(8)}_{\bm{1''}} & -Y^{(8)}_{\bm{1'}}Y^{(8)}_{\bm{1''}}Y^{(8)}_{\bm{1}} & Y^{(8)}_{\bm{1''}}Y^{(8)}_{\bm{1''}}Y^{(8)}_{\bm{1''}} \\
  \end{pmatrix}.
\end{align}
This is one of the models in Table \ref{tab:chi<0.01atomega} and can satisfy hierarchy conditions in Eq.~(\ref{eq: mass_ratio_order}) at the benchmark point $\tau=\omega+0.051i$.
To check whether the CP violation can occur or not in the vicinity of $\tau=\omega$, we calculate Jarlskog invariant $J_{\textrm{CP}}=\textrm{Im}(V_{\textrm{CKM}}^{us}V_{\textrm{CKM}}^{cb}(V_{\textrm{CKM}}^{ub}V_{\textrm{CKM}}^{cs})^*)$ in the $\tau$ plane around $\tau=\omega$.
The results are shown in Figure \ref{fig:chi_J_omega}.
As a result, we cannot find the region of the modulus where both the sufficient Jarlskog invariant and realistic quark hierarchical structures can be realized.
Suitable value of Jarlskog invariant $J_{\textrm{CP}}>10^{-5}$ can be realized at $|\tau-\omega|> 0.080$ ($|\varepsilon|> 0.23$).
This is consistent with the estimation by Eq.~(\ref{eq:J_CP_second}).
At $|\varepsilon|\sim0.23$, it gives $J_{\textrm{CP}}\lesssim 2.3\times 10^{-5}\sim 10^{-5}$ but it is too large compared with $|\varepsilon|\sim 0.15$ which is required to realize hierarchical quark mass ratios.
Thus, it is difficult to realize both the CP phase and quark mass ratios simultaneously.
As a numerical example, in Table \ref{tab:J>000001_omega}, we show the results at $\tau=\omega+(0.0326+0.0753i)$ where $J_{\textrm{CP}}>10^{-5}$ is realized.
\begin{table}[H]
\small
  \begin{center}
    \renewcommand{\arraystretch}{1.3}
    \begin{tabular}{c|cccccccc} \hline
      & $\frac{m_u}{m_t}{\times10^{6}}$ & $\frac{m_c}{m_t}{\times10^3}$ & $\frac{m_d}{m_b}{\times10^4}$ & $\frac{m_s}{m_b}{\times10^2}$ & $|V_{\textrm{CKM}}^{us}|$ & $|V_{\textrm{CKM}}^{cb}|$ & $|V_{\textrm{CKM}}^{ub}|$ & $J_{\textrm{CP}}{\times 10^5}$ \\ \hline
      obtained values & 162 & 17.8 & 76.9 & 5.93 & 0.287 & 0.100 & 0.0128 & 1.01 \\ \hline
      GUT scale values & 5.39 & 2.80 & 9.21 & 1.82 & 0.225 & 0.0400 & 0.00353 & 2.80 \\
      $1\sigma$ errors & $\pm 1.68$ & $\pm 0.12$ & $\pm 1.02$ & $\pm 0.10$ & $\pm 0.0007$ & $\pm 0.0008$ & $\pm 0.00013$ & $^{+0.14}_{-0.12}$ \\ \hline
    \end{tabular}
  \end{center}
  \caption{The values of the quark mass ratios and the absolute values of the CKM matrix elements at the benchmark point $\tau=\omega+(0.0326+0.0753i)$.
GUT scale values at $2\times 10^{16}$ GeV with $\tan \beta=5$ \cite{Antusch:2013jca,Bjorkeroth:2015ora} and $1\sigma$ errors are shown.}
\label{tab:J>000001_omega}
\normalsize
\end{table}
The results show that the value of the up quark mass ratio is too large compared to observed value.
Hence, large quark mass hierarchies are not realized in this example although $J_{\textrm{CP}}>10^{-5}$ is realized.
Similar results can be obtained for the favorable models in the vicinity of $\tau=i\infty$.


\subsection{Non-universal moduli}
\label{subsubsec:non-universal}

As we have studied in the two previous subsections, the sufficient amount of the CP violation does not occur in the favorable models with $A_4\times A_4\times A_4$ modular symmetry.
This is because the structures of the CKM matrix in the favorable models are commonly given by Eq.~(\ref{eq:CKM_at_first_order}) and they lead to vanishing Jarlskog invariant at the first order approximation.
Indeed, the numerical example shows that the CP violation and the quark mass hierarchies are not realized simultaneously.
In this sense, the relation of the trade-off between the CP violation and the quark mass hierarchies exist.

To improve the relation of the trade-off, let us study the models with non-universal moduli.
So far, we have assumed the moduli stabilization $\tau_1=\tau_2=\tau_3\equiv \tau$.
Then the value of the phase factors in the mass matrices can be controlled by the value of the modulus $\tau$.
However, such phase factors are cancelled in Jarlskog invariant of the favorable models.
We expect that this cancellation is due to the strong restriction of the modulus by the moduli stabilization $\tau_1=\tau_2=\tau_3\equiv \tau$.
When we have the phase factors controlled by multi moduli parameters, $\tau_1$, $\tau_2$ and $\tau_3$, in mass matrices, they would give non-vanishing contributions to Jarlskog invariant.
Here we concentrate on non-universal moduli in the vicinity of $\omega$,
\begin{align}
\tau_1 = \tau_2 \equiv \tau \neq \tau_3, \quad
|\tau_1-\omega| = |\tau_2-\omega| = |\tau_3-\omega| \ll 1,
\end{align}
and study the CP violation and the quark mass hierarchies by a concrete model.

As an example, we consider the model in type $123$-$1^22^2$-$1^2$.
In type $123$-$1^22^2$-$1^2$, the $ST$-charge assignments to quark fields are given by
\begin{align}
&\{Q^1,Q^2,Q^3\}:~\{(a_1,a_2,a_3),(b_1,b_2,b_3),(0,0,0)\}, \\
&\{u_R^1,u_R^2,u_R^3\}:~\{(1-a_1,1-a_2,1-a_3)_{\textrm{mod~3}},(2-b_1,2-b_2,2-b_3)_{\textrm{mod~3}},(0,0,0)\}, \\
&\{d_R^1,d_R^2,d_R^3\}:~\{(1-a_1,1-a_2,-a_3)_{\textrm{mod~3}},(1-b_1,-b_2,-b_3)_{\textrm{mod~3}},(0,0,0)\},
\end{align}
where $a_i\in\{0,1,2\}$ and $b_i\in\{0,1,2\}$ are $Z_3$-charges of the $i$-th $A_4$ for $Q^1$ and $Q^2$, respectively.
We find the best-fit choices of $a_i$ and $b_i$,
\begin{align}
(a_1,a_2,a_3) = (1,1,1), \quad (b_1,b_2,b_3) = (2,2,0),
\end{align}
and signs $\pm 1$ in the coupling constants $\alpha$ and $\beta$,
\begin{align}
&\begin{pmatrix}
\alpha^{11} & \alpha^{12} & \alpha^{13} \\
\alpha^{21} & \alpha^{22} & \alpha^{23} \\
\alpha^{31} & \alpha^{32} & \alpha^{33} \\
\end{pmatrix}
=
\begin{pmatrix}
1 & 1 & 1 \\
1 & 1 & -1 \\
1 & -1 & -1 \\
\end{pmatrix}, \quad\begin{pmatrix}
\beta^{11} & \beta^{12} & \beta^{13} \\
\beta^{21} & \beta^{22} & \beta^{23} \\
\beta^{31} & \beta^{32} & \beta^{33} \\
\end{pmatrix}
=
\begin{pmatrix}
1 & 1 & 1 \\
-1 & -1 & 1 \\
1 & 1 & -1 \\
\end{pmatrix}.
\end{align}
In these choices, we obtain the quark mass matrices,
\begin{align}
&\frac{M_u}{ \langle H_u \rangle} =  \begin{pmatrix}
Y^{(8)}_{\bm{1''}}(\tau)Y^{(8)}_{\bm{1''}}(\tau)Y^{(8)}_{\bm{1''}}(\tau_3) & Y^{(8)}_{\bm{1''}}(\tau)Y^{(8)}_{\bm{1''}}(\tau)Y^{(8)}_{\bm{1}}(\tau_3) & Y^{(8)}_{\bm{1''}}(\tau)Y^{(8)}_{\bm{1''}}(\tau)Y^{(8)}_{\bm{1''}}(\tau_3) \\
Y^{(8)}_{\bm{1'}}(\tau)Y^{(8)}_{\bm{1'}}(\tau)Y^{(8)}_{\bm{1}}(\tau_3) & Y^{(8)}_{\bm{1'}}(\tau)Y^{(8)}_{\bm{1'}}(\tau)Y^{(8)}_{\bm{1'}}(\tau_3) & -Y^{(8)}_{\bm{1'}}(\tau)Y^{(8)}_{\bm{1'}}(\tau)Y^{(8)}_{\bm{1}}(\tau_3) \\
Y^{(8)}_{\bm{1}}(\tau)Y^{(8)}_{\bm{1}}(\tau)Y^{(8)}_{\bm{1}}(\tau_3) & -Y^{(8)}_{\bm{1}}(\tau)Y^{(8)}_{\bm{1}}(\tau)Y^{(8)}_{\bm{1'}}(\tau_3) & -Y^{(8)}_{\bm{1}}(\tau)Y^{(8)}_{\bm{1}}(\tau)Y^{(8)}_{\bm{1}}(\tau_3) \\
\end{pmatrix}, \\
&\frac{M_d }{ \langle H_d \rangle}= \begin{pmatrix}
Y^{(8)}_{\bm{1''}}(\tau)Y^{(8)}_{\bm{1''}}(\tau)Y^{(8)}_{\bm{1}}(\tau_3) & Y^{(8)}_{\bm{1}}(\tau)Y^{(8)}_{\bm{1'}}(\tau)Y^{(8)}_{\bm{1''}}(\tau_3) & Y^{(8)}_{\bm{1''}}(\tau)Y^{(8)}_{\bm{1''}}(\tau)Y^{(8)}_{\bm{1''}}(\tau_3) \\
-Y^{(8)}_{\bm{1'}}(\tau)Y^{(8)}_{\bm{1'}}(\tau)Y^{(8)}_{\bm{1'}}(\tau_3) & -Y^{(8)}_{\bm{1''}}(\tau)Y^{(8)}_{\bm{1}}(\tau)Y^{(8)}_{\bm{1}}(\tau_3) & Y^{(8)}_{\bm{1'}}(\tau)Y^{(8)}_{\bm{1'}}(\tau)Y^{(8)}_{\bm{1}}(\tau_3) \\
Y^{(8)}_{\bm{1}}(\tau)Y^{(8)}_{\bm{1}}(\tau)Y^{(8)}_{\bm{1'}}(\tau_3) & Y^{(8)}_{\bm{1'}}(\tau)Y^{(8)}_{\bm{1''}}(\tau)Y^{(8)}_{\bm{1}}(\tau_3) & -Y^{(8)}_{\bm{1}}(\tau)Y^{(8)}_{\bm{1}}(\tau)Y^{(8)}_{\bm{1}}(\tau_3) \\
\end{pmatrix}.
\end{align}
This is one of the models in Table \ref{tab:chi<0.01atomega} and can satisfy hierarchy conditions in Eq.~(\ref{eq: mass_ratio_order}) at the benchmark point $\tau=\tau_3=\omega+0.051i$ although the CP violation does not occur there.
The phase factors after the basis transformation in Eqs.~(\ref{eq:u_L}) and (\ref{eq:u_R}), and the hierarchical structures of the mass matrices are given by
\begin{align}
  M_u \sim
  \begin{pmatrix}
    |\varepsilon|^6 & |\varepsilon|^4p_3^{-1} & |\varepsilon|^6 \\
    |\varepsilon|^2 & |\varepsilon|^3 & -|\varepsilon|^2 \\
    1 & -|\varepsilon| & -1 \\
  \end{pmatrix}, \quad
  M_d \sim
  \begin{pmatrix}
    |\varepsilon|^4p_3^{-1} & |\varepsilon|^3p^{-2} & |\varepsilon|^6 \\
    -|\varepsilon|^3 & -|\varepsilon|^2p^{-1} & |\varepsilon|^2 \\
    |\varepsilon| & |\varepsilon|^3 & -1 \\
  \end{pmatrix},
\end{align}
at the first order approximation.
Here we have used $p \equiv \varepsilon/|\varepsilon|$ for $\varepsilon=(\tau-\omega)/(\tau-\omega^2)$ and $p_3 \equiv \varepsilon_3/|\varepsilon_3|$ for $\varepsilon_3=(\tau_3-\omega)/(\tau_3-\omega^2)$.
These mass matrices lead to the CKM matrix,
\begin{align}
  V_{\textrm{CKM}} =
  \begin{pmatrix}
1 & -|\varepsilon|\left(p^* + 0.5p_3^*\right) & |\varepsilon|^{3}p_3^* \\
|\varepsilon| \left(p + 0.5 p_3\right) & 1 & - 2 |\varepsilon|^{2} \\
2 |\varepsilon|^{3} p & 2 |\varepsilon|^{2} & 1 \\
  \end{pmatrix},
\end{align}
at the first order approximation.
Note that when $\tau_3=\tau$, that is, $p_3=p$, this CKM matrix is equivalent to one shown in Eq.~(\ref{eq:CKM_at_first_order}).
This CKM matrix gives non-vanishing Jarlskog invariant,
\begin{align}
  J_{\textrm{CP}} &= |\textrm{Im}(- 2|\varepsilon|^{6} \left(- p^* - 0.5p_3^*\right) p_3)| = 2|\varepsilon|^6\cdot |\textrm{Im}(p^*p_3)|.
\end{align}
To illustrate the realization of both Jarlskog invariant and quark mass hierarchies, let us choose
\begin{align}
  \tau = \omega+0.055i, \quad \tau_3 = \omega+0.055e^{2\pi i/5},
\end{align}
and show the results in Table \ref{tab:non-universal_result}.
\begin{table}[H]
\small
  \begin{center}
    \renewcommand{\arraystretch}{1.3}
    \begin{tabular}{c|cccccccc} \hline
      & $\frac{m_u}{m_t}{\times10^{6}}$ & $\frac{m_c}{m_t}{\times10^3}$ & $\frac{m_d}{m_b}{\times10^4}$ & $\frac{m_s}{m_b}{\times10^2}$ & $|V_{\textrm{CKM}}^{us}|$ & $|V_{\textrm{CKM}}^{cb}|$ & $|V_{\textrm{CKM}}^{ub}|$ & $J_{\textrm{CP}}{\times 10^5}$ \\ \hline
      obtained values & 16.0 & 5.63 & 6.16 & 2.52 & 0.214 & 0.0498 & 0.00411 & 2.53 \\ \hline
      GUT scale values & 5.39 & 2.80 & 9.21 & 1.82 & 0.225 & 0.0400 & 0.00353 & 2.80 \\ 
      $1\sigma$ errors & $\pm 1.68$ & $\pm 0.12$ & $\pm 1.02$ & $\pm 0.10$ & $\pm 0.0007$ & $\pm 0.0008$ & $\pm 0.00013$ & $^{+0.14}_{-0.12}$ \\ \hline
    \end{tabular}
  \end{center}
  \caption{The mass ratios of the quarks and the absolute values of the CKM matrix elements at $\tau=\omega+0.055i$ and $\tau_3=\omega+0.055e^{2\pi i/5}$.
GUT scale values at $2\times 10^{16}$ GeV with $\tan \beta=5$ \cite{Antusch:2013jca,Bjorkeroth:2015ora} and $1\sigma$ errors are shown.}
\label{tab:non-universal_result}
\normalsize
\end{table}
This results satisfy the hierarchy conditions in Eq.~(\ref{eq: mass_ratio_order}).
Thus, realistic quark mass ratios, absolute values of CKM matrix elements and Jarlskog invariant are simultaneously realized in the $A_4\times A_4\times A_4$ modular symmetric flavor models with the non-universal moduli.


\chapter{Constructing Siegel modular forms}
\label{sec:Constructing Siegel modular forms}

In Chapter \ref{sec:magnetized_orbifold_models}, we have studied the magnetized $T^2$ and orbifold models as one of the top-down approaches to the flavor structures.
Then the zero-mode wave functions on magnetized $T^2$ and its orbifold behave as the modular forms of $Sp(2,\mathbb{Z})\simeq SL(2,\mathbb{Z})$.
Also Yukawa couplings are written by the modular forms.
In Chapter \ref{sec:4D modular symmetric flavor models}, we have studied 4D modular symmetric flavor models as one of the bottom-up approaches to the flavor structures.
Then we have used some modular forms of finite modular groups to calculate Yukawa couplings.
Thus, in these two approaches, the modular forms are important.

Magnetized $T^6$ model has several moduli and the modular symmetry $Sp(6,\mathbb{Z})$ as the geometrical symmetry.
Then Siegel modular forms for the subgroup of $Sp(6,\mathbb{Z})$ can be derived from them.
Siegel modular forms have several moduli.
Therefore the magnetized $T^6$ models and the modular symmetric flavor models based on Siegel modular forms have rich possibilities realizing flavor structures.
Indeed, we have seen that the relation of the trade-off between the CP violation and the quark mass hierarchies in $A_4\times A_4\times A_4$ modular symmetric flavor models can be improved by non-universal moduli.
Hence, Siegel modular forms described by multi moduli parameters are attractive way to explaining the flavor structures.
In this chapter, we will construct some examples of Siegel modular forms from the zero-mode wave functions on magnetized $T^6$.

This chapter is along in Ref.~\cite{Kikuchi:2023dow}.


\section{Zero-modes on $T^6$}

First we review zero-modes on magnetized $T^6$ \cite{Cremades:2004wa,Antoniadis:2009bg,Kikuchi:2023awm}.
6D torus $T^6$ is defined by $T^6\simeq \mathbb{C}/\Lambda$ where $\Lambda$ is a 6D lattice spanned by following six basis vectors,
\begin{align}
\vec{e}_1 = \begin{pmatrix}1\\0\\0\\\end{pmatrix}, \quad
\vec{e}_4 = \Omega\vec{e}_1 = \begin{pmatrix}\omega_1\\\omega_4\\\omega_6\\\end{pmatrix}, \\
\vec{e}_2 = \begin{pmatrix}0\\1\\0\\\end{pmatrix}, \quad
\vec{e}_5 = \Omega\vec{e}_2 = \begin{pmatrix}\omega_4\\\omega_2\\\omega_5\\\end{pmatrix}, \\
\vec{e}_3 = \begin{pmatrix}0\\0\\1\\\end{pmatrix}, \quad
\vec{e}_6 = \Omega\vec{e}_3 = \begin{pmatrix}\omega_6\\\omega_5\\\omega_3\\\end{pmatrix}.
\end{align}
Here $\Omega$ is the complex structure moduli defined by
\begin{align}
\Omega =
\begin{pmatrix}
\omega_1 & \omega_4 & \omega_6 \\
\omega_4 & \omega_2 & \omega_5 \\
\omega_6 & \omega_5 & \omega_3 \\
\end{pmatrix}.
\end{align}
The complex coordinate on $T^6$, $\vec{z}$, satisfies the identifications on $\mathbb{C}^3$,
\begin{align}
\vec{z} \sim \vec{z} + \vec{e}_i, \quad i=1,2,...,6.
\end{align}
The background magnetic flux on $T^6$ is written by \cite{Antoniadis:2009bg}
\begin{align}
F = \frac{1}{2} p_{x^ix^j}dx^i \wedge dx^j + \frac{1}{2}p_{y^iy^j}dy^i\wedge dy^j + p_{x^iy^j}dx^i\wedge dy^j,
\end{align}
where $x$ and $y$ are coordinate vectors on $T^6$ indirectly given by
\begin{align}
z^i = x^i + \Omega^i_j y^j , \quad \bar{z}^i = x^i + \bar{\Omega}^i_j y^j.
\end{align}
In terms of $z$ and $\bar{z}$, the flux $F$ is rewritten as
\begin{align}
F = \frac{1}{2} p_{z^iz^j} dz^i \wedge dz^j + \frac{1}{2} p_{\bar{z}^i\bar{z}^j} d\bar{z}^i \wedge d\bar{z}^j + p_{z^i\bar{z}^j} (idz^i \wedge d\bar{z}^j),
\end{align}
where
\begin{align}
&p_{z^iz^j} = (\bar{\Omega}-\Omega)^{-1} (\bar{\Omega}p_{xx}\bar{\Omega}+p_{yy}+p_{xy}^T\bar{\Omega}-\bar{\Omega}p_{xy}) (\bar{\Omega}-\Omega)^{-1}, \\
&p_{\bar{z}^i\bar{z}^j} = (\bar{\Omega}-\Omega)^{-1} (\Omega p_{xx}\Omega+p_{yy}+p_{xy}^T\Omega-\Omega p_{xy}) (\bar{\Omega}-\Omega)^{-1}, \\
&p_{z^i\bar{z}^j} = i(\bar{\Omega}-\Omega)^{-1} (\bar{\Omega}p_{xx}\Omega+p_{yy}+p_{xy}^T\Omega-\bar{\Omega}p_{xy}) (\bar{\Omega}-\Omega)^{-1}.
\end{align}
When we require supersymmetry, the flux $F$ must be a (1,1)-form.
This is achieved by $F_{zz}=F_{\bar{z}\bar{z}}=0$, which is so-called the $F$-term condition.
To make our analysis simple, we assume $p_{xx}=p_{yy}=0$.
Then the $F$-term condition $F_{zz}=F_{\bar{z}\bar{z}}=0$ is rewritten as
\begin{align}
p_{xy}^T\Omega = \Omega p_{xy}.
\end{align}
In addition, the Dirac quantization condition make $p_{xy}/2\pi$ integral,
\begin{align}
p_{xy} = 2\pi N^T,
\end{align}
where $N$ must be a $3\times 3$ integer matrix.
In this chapter we concentrate on symmetric $N$ matrix.
We note that symmetric $N$ matrix is required by the consistency of $S$-symmetry as we will see in the following.
Eventually, we obtain the $F$-term condition,
\begin{align}
N\Omega = \Omega N,
\end{align}
and the flux $F$,
\begin{align}
F = \pi (N(\textrm{Im}\Omega)^{-1})_{ij} (idz^i \wedge d\bar{z}^j).
\end{align}
The vector potential $A$ is led from $F=dA$ as
\begin{align}
A =& \pi \textrm{Im} \left(N(\vec{\bar{z}}+\vec{\bar{\zeta}})(\textrm{Im}\Omega)^{-1} d\vec{z}\right) \notag \\
=& -\frac{i\pi}{2}\left(N(\vec{\bar{z}}+\vec{\bar{\zeta}})(\textrm{Im}\Omega)^{-1}\right)dz^i
+ \frac{i\pi}{2}\left(N(\vec{z}+\vec{\zeta})(\textrm{Im}\Omega)^{-1}\right)d\bar{z}^i \\
\equiv& A_{z^i}dz^i + A_{\bar{z}^i}d\bar{z}^i, \notag
\end{align}
where $\vec{\zeta}$ is the Wilson line.
In what follows we assume the vanishing Wilson line, $\vec{\zeta} = 0$, for simplicity.

Now we are ready to derive the zero-modes on magnetized $T^6$.
We consider fermion massless modes (zero-modes) of the spinor,
\begin{align}
\Psi(\vec{z},\vec{\bar{z}}) = (
\psi_{(+,+,+)},
\psi_{(+,+,-)},
\psi_{(+,-,+)},
\psi_{(+,-,-)},
\psi_{(-,+,+)},
\psi_{(-,+,-)},
\psi_{(-,-,+)},
\psi_{(-,-,-)}
)^T,
\end{align}
where $(\pm,\pm,\pm)$ denotes the chiralities for each complex plane on $T^6$.
They obey the massless Dirac equation,
\begin{align}
i \slash{D} \Psi(\vec{z}, \vec{\bar{z}}) &= \frac{i}{\pi R}
\begin{pmatrix}
0 & D_{z^3} & D_{z^2} & 0 & D_{z^1} & 0 & 0 & 0 \\
\bar{D}_{\bar{z}^3} & 0 & 0 & -D_{z^2} & 0 & -D_{z^1} & 0 & 0 \\
\bar{D}_{\bar{z}^2} & 0 & 0 & D_{z^3} & 0 & 0 & -D_{z^1} & 0 \\
0 & -\bar{D}_{\bar{z}^2} & \bar{D}_{\bar{z}^3} & 0 & 0 & 0 & 0 & D_{z^1} \\
\bar{D}_{\bar{z}^1} & 0 & 0 & 0 & 0 & D_{z^3} & D_{z^2} & 0 \\
0 & -\bar{D}_{\bar{z}^1} & 0 & 0 & \bar{D}_{\bar{z}^3} & 0 & 0 & -D_{z^2} \\
0 & 0 & -\bar{D}_{\bar{z}^1} & 0 & \bar{D}_{\bar{z}^2} & 0 & 0 & D_{z^3} \\
0 & 0 & 0 & \bar{D}_{\bar{z}^1} & 0 & -\bar{D}_{\bar{z}^2} & \bar{D}_{\bar{z}^3} & 0 \\
\end{pmatrix}
\begin{pmatrix}
\psi_{(+,+,+)} \\
\psi_{(+,+,-)} \\
\psi_{(+,-,+)} \\
\psi_{(+,-,-)} \\
\psi_{(-,+,+)} \\
\psi_{(-,+,-)} \\
\psi_{(-,-,+)} \\
\psi_{(-,-,-)} \\
\end{pmatrix} = 0,
\end{align}
where
\begin{align}
D_{z^i} = \partial_{z^i} -iA_{z^i}, \quad \bar{D}_{\bar{z}^i} = \bar{\partial}_{\bar{z}^i} -iA_{\bar{z}^i}.
\end{align}
The boundary conditions of $\Psi(\vec{z},\vec{\bar{z}})$ are given by
\begin{align}
\Psi(\vec{z}+\vec{e}_k) = e^{i\xi_{\vec{e}_k}(\vec{z})} \Psi(\vec{z}), \quad
\Psi(\vec{z}+\Omega\vec{e}_k) = e^{i\xi_{\Omega\vec{e}_k}(\vec{z})} \Psi(\vec{z}),
\end{align}
where $k=1,2,3$ and
\begin{align}
\xi_{\vec{e}_k}(\vec{z}) = \pi(N(\textrm{Im}\Omega)^{-1}\textrm{Im}\vec{z})_k, \quad
\xi_{\Omega\vec{e}_k}(\vec{z}) = \pi\textrm{Im}(N\bar{\Omega}(\textrm{Im}\Omega)^{-1}\vec{z})_k.
\end{align}
We focus on the case where only the spinor component with chiralities $(+,+,+)$, $\psi_{(+,+,+)}$, do not vanish.
Then $\psi_{(+,+,+)}$ obeys the masslesss Dirac equation,
\begin{align}
\bar{D}_{\bar{z}^i} \psi_{(+,+,+)} = 0.
\end{align}
The solution to this equation is given by \cite{Cremades:2004wa}
\begin{align}
\psi_N^{\vec{j}}(\vec{z},\Omega) = {\cal N} e^{i\pi (N\vec{z})^T(\textrm{Im}\Omega)^{-1}\textrm{Im}\vec{z}} \cdot \vartheta
\begin{bmatrix}
\vec{j}N^{-1} \\ 0 \\
\end{bmatrix}
(N\vec{z},N\Omega), \label{eq;zero-modes_on_T6}
\end{align}
where $\vartheta\begin{bmatrix}
\vec{j}N^{-1} \\ 0 \\
\end{bmatrix}
(N\vec{z},N\Omega)$ is the Riemann-theta function with characteristics defined by
\begin{align}
\vartheta
\begin{bmatrix}
\vec{a} \\ \vec{b}
\end{bmatrix}(\vec{z},\Omega') =
\sum_{\vec{m}\in\mathbb{Z}^3}
e^{\pi i(\vec{m}+\vec{a})^T\Omega'(\vec{m}+\vec{a})} e^{2\pi i(\vec{m}+\vec{a})^T(\vec{z}+\vec{b})}, \quad \Omega'\in {\cal H}_3, \quad \vec{a},\vec{b}\in\mathbb{R}^3.
\end{align}
The independent indices of $\vec{J}\in\mathbb{Z}^3$ label the degeneracy of zero-modes.
Since zero-modes have the periodicity,
\begin{align}
\psi_N^{\vec{j}+N\vec{e}_k} = \psi_N^{\vec{j}}, \quad k=1,2,3,
\end{align}
the independent $\vec{J}$ are given by $|\det N|$ number of lattice points in $\Lambda_N$ which is spanned by basis vectors $N\vec{e}_k$, $k=1,2,3$.
Note that zero-modes in Eq.~(\ref{eq;zero-modes_on_T6}) are well-defined only if $\Omega'=N\Omega$ is an element of the Siegel upper-half plane ${\cal H}_3$ defined as \cite{Siegel:1943}
\begin{align}
{\cal H}_3 = \{\Omega'\in GL(3,\mathbb{C})|{\Omega'}^T = \Omega',~\textrm{Im}\Omega'>0 \}.
\end{align}
Thus three eigenvalues of $\textrm{Im}N\Omega$ must be positive and this requirement restrict the consistent region of the moduli $\Omega$.


\section{$Sp(6,\mathbb{Z})$ modular symmetry}

Next we study $Sp(6,\mathbb{Z})$ modular symmetry on zero-modes on magnetized $T^6$.
We will show zero-modes on magnetized $T^6$ behave as the Siegel modular forms for subgroups of $Sp(6,\mathbb{Z})$.


\subsection{Modular transformation and Siegel modular forms}

We start from the review of the modular symmetry on $T^6$ and the Siegel modular forms \cite{Ding:2020zxw,Siegel:1943,Igusa:1972,Freitag:1983,Freitag:1991,Klingen:1990,Geer:2008,Fay:1973,Mumford:1984,Kikuchi:2023awe}.
The modular transformation is the basis transformation of the lattice $\Lambda$ defining $T^6\simeq \mathbb{C}^3/\Lambda$.
It is given by
\begin{align}
&\begin{pmatrix}
\vec{e}_4 \\
\vec{e}_1 \\
\end{pmatrix}
\to
\begin{pmatrix}
A & B \\
C & D \\
\end{pmatrix}
\begin{pmatrix}
\vec{e}_4 \\
\vec{e}_1 \\
\end{pmatrix}, \\
&\begin{pmatrix}
\vec{e}_5 \\
\vec{e}_2 \\
\end{pmatrix}
\to
\begin{pmatrix}
A & B \\
C & D \\
\end{pmatrix}
\begin{pmatrix}
\vec{e}_5 \\
\vec{e}_2 \\
\end{pmatrix},  \\
&\begin{pmatrix}
\vec{e}_6 \\
\vec{e}_3 \\
\end{pmatrix}
\to
\begin{pmatrix}
A & B \\
C & D \\
\end{pmatrix}
\begin{pmatrix}
\vec{e}_6 \\
\vec{e}_3 \\
\end{pmatrix},
\end{align}
where
\begin{align}
\gamma =
\begin{pmatrix}
A & B \\
C & D \\
\end{pmatrix} \in Sp(6,\mathbb{Z}),
\end{align}
satisfying
\begin{align}
\gamma J \gamma^T = J, \quad J = \begin{pmatrix}\bm{0}_3 & \bm{1}_3 \\ -\bm{1}_3 & \bm{0}_3 \\\end{pmatrix}.
\end{align}
Under the modular transformation, the complex structure moduli of $T^6$, $\Omega$, are transformed as
\begin{align}
\Omega \equiv \begin{pmatrix}\vec{e}_4&\vec{e}_5&\vec{e}_6\end{pmatrix}\begin{pmatrix}\vec{e}_1&\vec{e}_2&\vec{e}_3\end{pmatrix}^{-1}
&\to
(A\Omega+B)\Omega^{-1}
\begin{pmatrix}\vec{e}_4&\vec{e}_5&\vec{e}_6\end{pmatrix}\begin{pmatrix}\vec{e}_1&\vec{e}_2&\vec{e}_3\end{pmatrix}^{-1}(C\Omega+D)^{-1} \notag \\
&= (A\Omega+B)(C\Omega+D)^{-1}.
\end{align}
The modular transformation is generated by seven generators, $S$ and $T_i$ $(i=1,2,...,6)$, defined by
\begin{align}
S = 
\begin{pmatrix}
\bm{0}_3 & \bm{1}_3 \\
-\bm{1}_3 & \bm{0}_3 \\
\end{pmatrix}, \quad
T_i =
\begin{pmatrix}
\bm{1}_3 & B_i \\
\bm{0}_3 & \bm{1}_3 \\
\end{pmatrix},
\quad i = 1,2,...,6,
\end{align}
where
\begin{align}
&B_1 =
\begin{pmatrix}
1 & 0 & 0 \\
0 & 0 & 0 \\
0 & 0 & 0 \\
\end{pmatrix}, \quad
B_2 =
\begin{pmatrix}
0 & 0 & 0 \\
0 & 1 & 0 \\
0 & 0 & 0 \\
\end{pmatrix}, \quad
B_3 =
\begin{pmatrix}
0 & 0 & 0 \\
0 & 0 & 0 \\
0 & 0 & 1 \\
\end{pmatrix}, \\
&B_4 =
\begin{pmatrix}
0 & 1 & 0 \\
1 & 0 & 0 \\
0 & 0 & 0 \\
\end{pmatrix}, \quad
B_5 =
\begin{pmatrix}
0 & 0 & 0 \\
0 & 0 & 1 \\
0 & 1 & 0 \\
\end{pmatrix}, \quad
B_6 =
\begin{pmatrix}
0 & 0 & 1 \\
0 & 0 & 0 \\
1 & 0 & 0 \\
\end{pmatrix}.
\end{align}

The Siegel modular forms $f^j(\Omega)$ are holomorphic functions of $\Omega$.
Under the modular transformation, they are transformed as
\begin{align}
f^j(\Omega) \rightarrow f^j(\gamma:\Omega) = [\det(C\Omega+D)]^{k} \rho(\gamma)^{jk} f^k(\Omega),
\end{align}
where $k$ is the so-called modular weight and $\rho$ is a unitary representation of subgroup of $Sp(6,\mathbb{Z})$.
In particular, the Siegel modular forms obey
\begin{align}
&f^j(\Omega) \xrightarrow{S} f^j(S:\Omega) = [\det (-\Omega)]^k \rho(S)^{jk} f^k(\Omega), \\
&f^j(\Omega) \xrightarrow{T} f^j(T_i:\Omega) = \rho(T_i)^{jk} f^k(\Omega), \quad i=1,2,...,6.
\end{align}
Generally $\rho$ fulfills $\rho(h)=\mathbb{I}$ for elements $h\in G \subset Sp(6,\mathbb{Z})$ and forms a finite group.


\subsection{Modular symmetry in zero-modes}

Next we discuss the modular transformation for zero-modes on $T^6$ \footnote{Modular symmetry in magnetized $T^{2g}$ was classified in Ref.~\cite{Kikuchi:2023awe}.}.
Since zero-modes on $T^6$ in Eq.~(\ref{eq;zero-modes_on_T6}) are the functions of $\Omega$ and $\vec{z}$, we need to find the modular transformation for $\vec{z}$.
Under the modular transformation, the complex coordinate on $T^6$, $\vec{z}$, is transformed as
\begin{align}
\vec{z} \to {(C\Omega + D)^{-1}}^T \vec{z}.
\end{align}
Then we can find the modular transformation for zero-modes and the consistency conditions:
\begin{itemize}
\item $S$-transformation\\
When the consisteny conditions,
\begin{align}
(N\Omega)^T=N\Omega, \quad N=N^T, \quad \Omega=\Omega^T,
\end{align}
are satisfied, $S$-transformation for zero-modes are given by
\begin{align}
S: \psi^{\vec{j}}_N(\vec{z},\Omega) &= \psi^{\vec{j}}_N(-\Omega^{-1}\vec{z}, -\Omega^{-1}) 
= \sqrt{\det (-\Omega)} \sum_{\vec{k}\in\Lambda_N}\rho(S)^{\vec{j}\vec{k}} \psi^{\vec{k}}_N(\vec{z},\Omega), 
\end{align}
where
\begin{align}
\rho(S)^{\vec{j}\vec{k}} = \frac{e^{3\pi i/4}}{\sqrt{\det N}} e^{2\pi i \vec{j}^T N^{-1} \vec{k}},
\quad \vec{j}, \vec{k}\in \Lambda_N. \label{eq:rho_S}
\end{align}
Note that the consistency condition for $S$-transformation is equivalent to the $F$-term condition for symmetric $N$ and $\Omega$.
Then $\Omega$ must commute to $N$ matrix.
\item $T$-transformations\\
When the consistency conditions,
\begin{align}
(N\Omega)^T = N\Omega, \quad (NB)^T = NB, \quad (NB)^{kk}\in2\mathbb{Z}~(k=1,2,3), \label{eq:consistency_T}
\end{align}
are satisfied, $T$-transformations for zero-modes are given by
\begin{align}
T_i: \psi^{\vec{j}}_N(\vec{z},\Omega) &= \psi^{\vec{j}}_N(\vec{z}, \Omega+B_i)
= \sum_{\vec{k}\in\Lambda_N}\rho(T_i)^{\vec{j}\vec{k}} \psi^{\vec{k}}_N(\vec{z},\Omega),
\end{align}
where
\begin{align}
\rho(T_i)^{\vec{j}\vec{k}} = e^{\pi i \vec{j}^T N^{-1}B_i \vec{j}} \delta_{\vec{j},\vec{k}},
\quad \vec{j}, \vec{k}\in \Lambda_N. \label{eq:rho_T}
\end{align}
Note that what $T$-transformation can be defined for zero-modes depend on the existence of $B$ matrix commuting to $N$ matrix.
In other words, some of $T$-symmetries of $Sp(6,\mathbb{Z})$ can be violated in zero-modes because of the structure of N matrix.
\end{itemize}
Notice that these transformations correspond to ones for the Siegel modular forms of weight 1/2.
Thus, zero-modes on $T^6$ behave as the Siegel modular forms of weight 1/2 for subgroup of $Sp(6,\mathbb{Z})$.
Actually, taking $\vec{z}=0$, zero-modes on $T^6$ become exactly the Siegel modular forms.

In this chapter, we focus on zero-modes with $N$ matrix possessing three different eigenvalues.
When $N$ matrix has three different eigenvalues, it cannot be expanded by $\bm{1}_3$ and $N^{-1}$.
This can be understood as follows.
If we can find $\alpha,\beta\in\mathbb{R}$ such that $N=\alpha\bm{1}_3+\beta N^{-1}$, we obtain
\begin{align}
N^2 - \alpha N - \beta \bm{1}_3 = 0 ~\to~
\begin{pmatrix}\lambda_1^2\\\lambda_2^2\\\lambda_3^2\\\end{pmatrix}
-\alpha \begin{pmatrix}\lambda_1\\\lambda_2\\\lambda_3\\\end{pmatrix}
-\beta \begin{pmatrix}1\\1\\1\\\end{pmatrix} = 0,
\end{align}
where $\lambda_1$, $\lambda_2$ and $\lambda_3$ are eigenvalues of $N$ matrix.
These quadratic equations for $\lambda_1$, $\lambda_2$ and $\lambda_3$ give two solutions.
Therefore at least two of $\lambda_1$, $\lambda_2$ and $\lambda_3$ have same solution and $N$ matrix is degenerate if $N=\alpha\bm{1}_3+\beta N^{-1}$ consists.

Assuming $N$ matrix with three different eigenvalues, we can find the structures of $\Omega$ and $B$ matrix commuting to $N$ matrix.
When we write
\begin{align}
\begin{pmatrix}\lambda_1&&\\&\lambda_2&\\&&\lambda_3\\\end{pmatrix}=O^TNO, \quad
O^T
\begin{pmatrix}
\omega_1 & \omega_4 & \omega_6 \\
\omega_4 & \omega_2 & \omega_5 \\
\omega_6 & \omega_5 & \omega_3 \\
\end{pmatrix}
O
=
\begin{pmatrix}
\omega'_1 & \omega'_4 & \omega'_6 \\
\omega'_4 & \omega'_2 & \omega'_5 \\
\omega'_6 & \omega'_5 & \omega'_3 \\
\end{pmatrix},
\end{align}
where $O$ is a orthogonal matrix, $N\Omega=\Omega N$ leads to
\begin{align}
&\begin{pmatrix}
\lambda_1&&\\
&\lambda_2&\\
&&\lambda_3\\
\end{pmatrix}
\begin{pmatrix}
\omega'_1 & \omega'_4 & \omega'_6 \\
\omega'_4 & \omega'_2 & \omega'_5 \\
\omega'_6 & \omega'_5 & \omega'_3 \\
\end{pmatrix}
=
\begin{pmatrix}
\omega'_1 & \omega'_4 & \omega'_6 \\
\omega'_4 & \omega'_2 & \omega'_5 \\
\omega'_6 & \omega'_5 & \omega'_3 \\
\end{pmatrix}
\begin{pmatrix}
\lambda_1&&\\
&\lambda_2&\\
&&\lambda_3\\
\end{pmatrix}
\to \omega_4' = \omega_5' = \omega_6' = 0,
\end{align}
for three different eigenvalues $(\lambda_1,\lambda_2,\lambda_3)$.
Thus there are three independent symmetric matrices commuting to $N$ matrix.
Since $N\neq \alpha\bm{1}_3+\beta N^{-1}$, we can regard $\bm{1}_3$, $N$ and $N^{-1}$ as three independent symmetric matrices commuting to $N$ matrix.
This means that $\Omega$ satisfying the $F$ term comdition $N\Omega=\Omega N$ can be expanded as
\begin{align}
\Omega = k_1\bm{1}_3 + k_2N + k_3N^{-1},
\end{align}
where $k_i$ ($i=1,2,3$) are any complex values.
In the same way, $B$ matrix satisfying the consistency condition $NB=BN$ can be written as
\begin{align}
B = n_1\bm{1}_3 + n_2 \frac{1}{\textrm{lcm}(N)} N + n_3 \frac{\det N}{\textrm{lcm}(N)^2}N^{-1},
\end{align}
where $\textrm{lcm}(N)$ are the least common multiple of $N$ matrix elements, and $n_i$ ($i=1,2,3$) are any real values but $B$ must be an integral matrix.
Obviously we can find three independent $B$ matrices, that is, three independent $T$-transformations which are well-defined in zero-modes.
Therefore $S$-symmetry and three $T$-symmetries remain in zero-modes with $N$ matrix possessing three different eigenvalues.


\section{Examples of the Siegel modular forms}
\label{subsec:Examples of the Siegel modular forms}

In this section we show some examples of the Siegel modular forms constructed from zero-modes on $T^6$.
As mentioned in the previous section, we focus on zero-modes with $N$ matrix possessing three different eigenvalues.
Then $S$-symmetry and three $T$-symmetries remain in zero-modes.
Thus taking $\vec{z}=0$ we can obtain the Siegel modular forms transformed by subgroup of $Sp(6,\mathbb{Z})$ which are generated by $S$-transformation and three $T$-transformations.
We will show five Siegel modular forms at $\det N = 2,3,4,5$ and 6.


\begin{itemize}
\item $\det N=2$

We consider the $N$ matrix,
\begin{align}
N=
\begin{pmatrix}
-2 & 1 & -1 \\
1 & -2 & 1 \\
-1 & 1 & 0 \\
\end{pmatrix},
\end{align}
whose determinant is 2.
Hence there are two independent degenerate zero-modes,
\begin{align}
\psi^{\vec{j}}_N(\vec{z},\Omega) =
\begin{pmatrix}
\psi_N^{\begin{psmallmatrix}0\\0\\0\\\end{psmallmatrix}},
\psi_N^{\begin{psmallmatrix}-1\\0\\0\\\end{psmallmatrix}}
\end{pmatrix}.
\end{align}
These zero-modes are even modes under $\vec{z}\to -\vec{z}$ and do not vanish at $\vec{z}=0$.
Therefore even modes at $\vec{z}=0$,
\begin{align}
\label{eq:N=2}
\hat{\psi}^j_{\bm{2}}(\Omega) =
\begin{pmatrix}
\hat{\psi}^0_{\bm{2}}(\Omega) \\
\hat{\psi}^1_{\bm{2}}(\Omega) \\
\end{pmatrix}
\equiv
\begin{pmatrix}
\psi_N^{\begin{psmallmatrix}0\\0\\0\\\end{psmallmatrix}}(0,\Omega) \\
\psi_N^{\begin{psmallmatrix}-1\\0\\0\\\end{psmallmatrix}}(0,\Omega) \\
\end{pmatrix},
\end{align}
are the Siegel modular forms of weight 1/2.
They are mapped into themselves under $S$-transformation and following three $T$-transformations,
\begin{align}
T_I =
\begin{pmatrix}
\bm{1}_3 & B_I \\
\bm{0}_3 & \bm{1}_3 \\
\end{pmatrix}, \quad
T_{II} =
\begin{pmatrix}
\bm{1}_3 & B_{II} \\
\bm{0}_3 & \bm{1}_3 \\
\end{pmatrix}, \quad
T_{III} =
\begin{pmatrix}
\bm{1}_3 & B_{III} \\
\bm{0}_3 & \bm{1}_3 \\
\end{pmatrix},
\end{align}
where
\begin{align}
&B_I = \bm{1}_3 =
\begin{pmatrix}
1 & 0 & 0 \\
0 & 1 & 0 \\
0 & 0 & 1 \\
\end{pmatrix}, \\
&B_{II} = 2\bm{1}_3 + N =
\begin{pmatrix}
0 & 1 & -1 \\
1 & 0 & 1 \\
-1 & 1 & 2 \\
\end{pmatrix}, \\
&B_{III} = -\bm{1}_3 -2N^{-1} =
\begin{pmatrix}
0 & 1 & 1 \\
1 & 0 & -1 \\
1 & -1 & -4 \\
\end{pmatrix}.
\end{align}
It can be checked that these $B$ matrices satisfy the consistency conditions for $T$-transformations in Eq.~(\ref{eq:consistency_T}).
The unitary representation matrices of $S$, $T_I$, $T_{II}$ and $T_{III}$-transformations, $\rho(S)$, $\rho(T_I)$, $\rho(T_{II})$ and $\rho(T_{III})$, form a group, $\widetilde{S}_4 \simeq T'\rtimes Z_4$, whose order is 96 \footnote{The group isomorphism $\widetilde{S}_4 \simeq T'\rtimes Z_4$ is pointed out in Refs.~\cite{Liu:2020msy,Uchida:2023yaj}}.
Then, $\hat{\psi}^j_{\bm{2}}(\Omega) $ in Eq.~(\ref{eq:N=2}) are a doublet under $\widetilde{S}_4$.


\item $\det N=3$

We consider the following $N$ matrix,
\begin{align}
N = 
\begin{pmatrix}
0 & 3 & 4 \\
3 & 0 & -4 \\
4 & -4 & -11 \\
\end{pmatrix},
\end{align}
whose determinant is 3.
Hence there are three independent degenerate zero-modes,
\begin{align}
\psi^{\vec{j}}_N(\vec{z},\Omega) =
\begin{pmatrix}
\psi_N^{\begin{psmallmatrix}2\\2\\0\\\end{psmallmatrix}},
\psi_N^{\begin{psmallmatrix}1\\1\\0\\\end{psmallmatrix}},
\psi_N^{\begin{psmallmatrix}0\\0\\0\\\end{psmallmatrix}}
\end{pmatrix}.
\end{align}
These zero-modes are decomposed into even and odd modes under $\vec{z}\to-\vec{z}$.
At $\vec{z}=0$, only even modes do not vanish.
Therefore, even modes at $\vec{z}=0$,
\begin{align}
\label{eq:N=3}
\hat{\psi}^{\vec{j}}_{\bm{2}}(\Omega) =
\begin{pmatrix}
\hat{\psi}^0_{\bm{2}}(\Omega) \\
\hat{\psi}^1_{\bm{2}}(\Omega) \\
\end{pmatrix}
\equiv
\begin{pmatrix}
\frac{1}{\sqrt{2}}\begin{pmatrix}\psi_N^{\begin{psmallmatrix}2\\2\\0\\\end{psmallmatrix}}(0,\Omega) + \psi_N^{\begin{psmallmatrix}1\\1\\0\\\end{psmallmatrix}}(0,\Omega)\end{pmatrix} \\
\psi_N^{\begin{psmallmatrix}0\\0\\0\\\end{psmallmatrix}}(0,\Omega) \\
\end{pmatrix},
\end{align}
are the Siegel modular forms of weight 1/2.
They are mapped into themselves under $S$-transformation and following three $T$-transformations,
\begin{align}
T_I =
\begin{pmatrix}
\bm{1}_3 & B_I \\
\bm{0}_3 & \bm{1}_3 \\
\end{pmatrix}, \quad
T_{II} =
\begin{pmatrix}
\bm{1}_3 & B_{II} \\
\bm{0}_3 & \bm{1}_3 \\
\end{pmatrix}, \quad
T_{III} =
\begin{pmatrix}
\bm{1}_3 & B_{III} \\
\bm{0}_3 & \bm{1}_3 \\
\end{pmatrix},
\end{align}
where
\begin{align}
&B_I = 2\bm{1}_3 =
\begin{pmatrix}
2 & 0 & 0 \\
0 & 2 & 0 \\
0 & 0 & 2 \\
\end{pmatrix}, \\
&B_{II} = \frac{16}{13}\bm{1}_3 + \frac{3}{13}N + \frac{3}{13}N^{-1} =
\begin{pmatrix}
0 & 2 & 0 \\
2 & 0 & 0 \\
0 & 0 & -2 \\
\end{pmatrix}, \\
&B_{III} = \frac{6}{13}\bm{1}_3 - \frac{17}{104}N + \frac{9}{104}N^{-1} =
\begin{pmatrix}
0 & 0 & -1 \\
0 & 0 & 1 \\
-1 & 1 & 2 \\
\end{pmatrix}.
\end{align}
It can be checked that these $B$ matrices satisfy the consistency conditions for $T$-transformations in Eq.~(\ref{eq:consistency_T}).
The unitary representation matrices of $S$, $T_I$, $T_{II}$ and $T_{III}$-transformations, $\rho(S)$, $\rho(T_I)$, $\rho(T_{II})$ and $\rho(T_{III})$, form a group, $(Z_8 \times Z_2) \rtimes Z_2 \rtimes Z_3$, whose order is 96.
Then, $\hat{\psi}^{\vec{j}}_{\bm{2}}(\Omega) $ in Eq.~(\ref{eq:N=3}) are a doublet of this group.


\item $\det N=4$

We consider the following $N$ matrix,
\begin{align}
N = 
\begin{pmatrix}
-1&1&-1\\
1&-1&-1\\
-1&-1&2\\
\end{pmatrix}, \label{eq:N_matrix}
\end{align}
whose determinant is 4.
Hence there are four independent degenerate zero-modes,
\begin{align}
\psi^{\vec{j}}_N(\vec{z},\Omega) =
\begin{pmatrix}
\psi_N^{\begin{psmallmatrix}0\\-1\\0\\\end{psmallmatrix}},
\psi_N^{\begin{psmallmatrix}0\\0\\0\\\end{psmallmatrix}},
\psi_N^{\begin{psmallmatrix}0\\0\\-1\\\end{psmallmatrix}},
\psi_N^{\begin{psmallmatrix}-1\\0\\0\\\end{psmallmatrix}}
\end{pmatrix}.
\end{align}
These zero-modes are decomposed into even and odd modes under $\vec{z}\to-\vec{z}$.
At $\vec{z}=0$, only even modes do not vanish.
Therefore, even modes at $\vec{z}=0$,
\begin{align}
\hat{\psi}^{\vec{j}}_{\bm{3}}(\Omega) =
\begin{pmatrix}
\hat{\psi}^0_{\bm{3}}(\Omega) \\
\hat{\psi}^1_{\bm{3}}(\Omega) \\
\hat{\psi}^2_{\bm{3}}(\Omega) \\
\end{pmatrix}
\equiv
\begin{pmatrix}
\frac{1}{\sqrt{2}}\psi_N^{\begin{psmallmatrix}0\\-1\\0\\\end{psmallmatrix}}(0,\Omega)
+ \frac{1}{\sqrt{2}} \psi_N^{\begin{psmallmatrix}-1\\0\\0\\\end{psmallmatrix}}(0,\Omega) \\
\psi_N^{\begin{psmallmatrix}0\\0\\0\\\end{psmallmatrix}}(0,\Omega) \\
-\psi_N^{\begin{psmallmatrix}0\\0\\-1\\\end{psmallmatrix}}(0,\Omega) \\
\end{pmatrix}, \label{eq:modular_form_1/2_Delta96}
\end{align}
are the Siegel modular forms of weight 1/2.
They are mapped into themselves under $S$-transformation and following three $T$-transformations,
\begin{align}
T_I =
\begin{pmatrix}
\bm{1}_3 & B_I \\
\bm{0}_3 & \bm{1}_3 \\
\end{pmatrix}, \quad
T_{II} =
\begin{pmatrix}
\bm{1}_3 & B_{II} \\
\bm{0}_3 & \bm{1}_3 \\
\end{pmatrix}, \quad
T_{III} =
\begin{pmatrix}
\bm{1}_3 & B_{III} \\
\bm{0}_3 & \bm{1}_3 \\
\end{pmatrix},
\end{align}
where
\begin{align}
&B_I = -\frac{4}{3}\bm{1}_3 + \frac{2}{3}N -\frac{4}{3}N^{-1} =
\begin{pmatrix}
-1 & 1 & 0 \\
1 & -1 & 0 \\
0 & 0 & 0 \\
\end{pmatrix}, \label{eq:BI} \\
&B_{II} = -\bm{1}_3 + N =
\begin{pmatrix}
-2 & 1 & -1 \\
1 & -2 & -1 \\
-1 & -1 & 1 \\
\end{pmatrix}, \label{eq:BII} \\
&B_{III} = -\frac{2}{3}\bm{1}_3-\frac{2}{3}N+\frac{4}{3}N^{-1} =
\begin{pmatrix}
-1 & -1 & 0 \\
-1 & -1 & 0 \\
0 & 0 & -2 \\
\end{pmatrix}. \label{eq:BIII}
\end{align}
It can be checked that these $B$ matrices satisfy the consistency conditions for $T$-transformations in Eq.~(\ref{eq:consistency_T}).
The unitary representation matrices of $S$, $T_I$, $T_{II}$ and $T_{III}$-transformations, $\rho(S)$, $\rho(T_I)$, $\rho(T_{II})$ and $\rho(T_{III})$, form a group, $\widetilde{\Delta}(96)\simeq \Delta(48)\rtimes Z_8$, whose order is 384.
Then, $\hat{\psi}^{\vec{j}}_{\bm{3}}(\Omega) $ in Eq.~(\ref{eq:modular_form_1/2_Delta96}) are a triplet of this group.


\item $\det N=5$

We consider the following $N$ matrix,
\begin{align}
N = 
\begin{pmatrix}
-3 & -2 &  2 \\
-2 & -3 & -2 \\
 2 & -2 & -7 \\
\end{pmatrix},
\end{align}
whose determinant is 5.
Hence there are five independent degenerate zero-modes,
\begin{align}
\psi^{\vec{j}}_N(\vec{z},\Omega) =
\begin{pmatrix}
\psi_N^{\begin{psmallmatrix}0\\0\\0\\\end{psmallmatrix}},
\psi_N^{\begin{psmallmatrix}-1\\-1\\0\\\end{psmallmatrix}},
\psi_N^{\begin{psmallmatrix}-2\\-2\\0\\\end{psmallmatrix}},
\psi_N^{\begin{psmallmatrix}-3\\-3\\0\\\end{psmallmatrix}},
\psi_N^{\begin{psmallmatrix}-4\\-4\\0\\\end{psmallmatrix}}
\end{pmatrix}.
\end{align}
These zero-modes are decomposed into even and odd modes under $\vec{z}\to-\vec{z}$.
At $\vec{z}=0$, only even modes do not vanish.
Therefore, even modes at $\vec{z}=0$,
\begin{align}
\label{eq:N=5}
\hat{\psi}^{\vec{j}}_{\bm{3}}(\Omega) =
\begin{pmatrix}
\hat{\psi}^0_{\bm{3}}(\Omega) \\
\hat{\psi}^1_{\bm{3}}(\Omega) \\
\hat{\psi}^2_{\bm{3}}(\Omega) \\
\end{pmatrix}
\equiv
\begin{pmatrix}
\frac{1}{\sqrt{2}}\begin{pmatrix}\psi_N^{\begin{psmallmatrix}-2\\-2\\0\\\end{psmallmatrix}}(0,\Omega) + \psi_N^{\begin{psmallmatrix}-3\\-3\\0\\\end{psmallmatrix}}(0,\Omega)\end{pmatrix} \\
\frac{1}{\sqrt{2}}\begin{pmatrix}\psi_N^{\begin{psmallmatrix}-1\\-1\\0\\\end{psmallmatrix}}(0,\Omega) + \psi_N^{\begin{psmallmatrix}-4\\-4\\0\\\end{psmallmatrix}}(0,\Omega)\end{pmatrix} \\
\psi_N^{\begin{psmallmatrix}0\\0\\0\\\end{psmallmatrix}}(0,\Omega)
\end{pmatrix},
\end{align}
are the Siegel modular forms of weight 1/2.
They are mapped into themselves under $S$-transformation and following three $T$-transformations,
\begin{align}
T_I =
\begin{pmatrix}
\bm{1}_3 & B_I \\
\bm{0}_3 & \bm{1}_3 \\
\end{pmatrix}, \quad
T_{II} =
\begin{pmatrix}
\bm{1}_3 & B_{II} \\
\bm{0}_3 & \bm{1}_3 \\
\end{pmatrix}, \quad
T_{III} =
\begin{pmatrix}
\bm{1}_3 & B_{III} \\
\bm{0}_3 & \bm{1}_3 \\
\end{pmatrix},
\end{align}
where
\begin{align}
&B_I = 2\bm{1}_3 =
\begin{pmatrix}
2 & 0 & 0 \\
0 & 2 & 0 \\
0 & 0 & 2 \\
\end{pmatrix}, \\
&B_{II} = -\frac{3}{2}\bm{1}_3 - \frac{1}{2}N =
\begin{pmatrix}
0 & 1 & -1 \\
1 & 0 & 1 \\
-1 & 1 & 2 \\
\end{pmatrix}, \\
&B_{III} = \frac{19}{2}\bm{1}_3 + \frac{7}{4}N - \frac{5}{4}N^{-1} =
\begin{pmatrix}
0 & 1 & 1 \\
1 & 0 & -1 \\
1 & -1 & -4 \\
\end{pmatrix}.
\end{align}
It can be checked that these $B$ matrices satisfy the consistency conditions for $T$-transformations in Eq.~(\ref{eq:consistency_T}).
The unitary representation matrices of $S$, $T_I$, $T_{II}$ and $T_{III}$-transformations, $\rho(S)$, $\rho(T_I)$, $\rho(T_{II})$ and $\rho(T_{III})$, form a group, $A_5\times Z_8$, whose order is 480.
Then, $\hat{\psi}^{\vec{j}}_{\bm{3}}(\Omega)$ in Eq.~(\ref{eq:N=5}) are a triplet of this group.


\item $\det N=6$

We consider the following $N$ matrix,
\begin{align}
N = 
\begin{pmatrix}
-2 & -1 & -1 \\
-1 & -2 &  1 \\
-1 &  1 &  0 \\
\end{pmatrix},
\end{align}
whose determinant is 6.
Hence there are six independent degenerate zero-modes,
\begin{align}
\psi^{\vec{j}}_N(\vec{z},\Omega) =
\begin{pmatrix}
\psi_N^{\begin{psmallmatrix}0\\0\\0\\\end{psmallmatrix}},
\psi_N^{\begin{psmallmatrix}-1\\0\\0\\\end{psmallmatrix}},
\psi_N^{\begin{psmallmatrix}-1\\-1\\0\\\end{psmallmatrix}},
\psi_N^{\begin{psmallmatrix}-2\\-1\\0\\\end{psmallmatrix}},
\psi_N^{\begin{psmallmatrix}-2\\-2\\0\\\end{psmallmatrix}},
\psi_N^{\begin{psmallmatrix}-3\\-2\\0\\\end{psmallmatrix}}
\end{pmatrix}.
\end{align}
These zero-modes are decomposed into even and odd modes under $\vec{z}\to-\vec{z}$.
At $\vec{z}=0$, only even modes do not vanish.
Therefore, even modes at $\vec{z}=0$,
\begin{align}
\label{eq:N=6}
\hat{\psi}^{\vec{j}}_{\bm{4}}(\Omega) =
\begin{pmatrix}
\hat{\psi}^0_{\bm{4}}(\Omega) \\
\hat{\psi}^1_{\bm{4}}(\Omega) \\
\hat{\psi}^2_{\bm{4}}(\Omega) \\
\hat{\psi}^3_{\bm{4}}(\Omega) \\
\end{pmatrix}
\equiv
\begin{pmatrix}
\frac{1}{\sqrt{2}}\begin{pmatrix}\psi_N^{\begin{psmallmatrix}-1\\-1\\0\\\end{psmallmatrix}}(0,\Omega) + \psi_N^{\begin{psmallmatrix}-2\\-2\\0\\\end{psmallmatrix}}(0,\Omega)\end{pmatrix} \\
\psi_N^{\begin{psmallmatrix}-2\\-1\\0\\\end{psmallmatrix}}(0,\Omega) \\
\frac{1}{\sqrt{2}}\begin{pmatrix}\psi_N^{\begin{psmallmatrix}-1\\0\\0\\\end{psmallmatrix}}(0,\Omega) + \psi_N^{\begin{psmallmatrix}-3\\-2\\0\\\end{psmallmatrix}}(0,\Omega)\end{pmatrix} \\
\psi_N^{\begin{psmallmatrix}0\\0\\0\\\end{psmallmatrix}}(0,\Omega)
\end{pmatrix},
\end{align}
are the Siegel modular forms of weight 1/2.
They are mapped into themselves under $S$-transformation and following three $T$-transformations,
\begin{align}
T_I =
\begin{pmatrix}
\bm{1}_3 & B_I \\
\bm{0}_3 & \bm{1}_3 \\
\end{pmatrix}, \quad
T_{II} =
\begin{pmatrix}
\bm{1}_3 & B_{II} \\
\bm{0}_3 & \bm{1}_3 \\
\end{pmatrix}, \quad
T_{III} =
\begin{pmatrix}
\bm{1}_3 & B_{III} \\
\bm{0}_3 & \bm{1}_3 \\
\end{pmatrix},
\end{align}
where
\begin{align}
&B_I = \bm{1}_3 =
\begin{pmatrix}
1 & 0 & 0 \\
0 & 1 & 0 \\
0 & 0 & 1 \\
\end{pmatrix}, \\
&B_{II} = -3\bm{1}_3 -2N+6N^{-1} =
\begin{pmatrix}
0 & 1 & -1 \\
1 & 0 & 1 \\
-1 & 1 & 0 \\
\end{pmatrix}, \\
&B_{III} = -2\bm{1}_3 - N =
\begin{pmatrix}
0 & 1 & 1 \\
1 & 0 & -1 \\
1 & -1 & -2 \\
\end{pmatrix}.
\end{align}
It can be checked that these $B$ matrices satisfy the consistency conditions for $T$-transformations in Eq.~(\ref{eq:consistency_T}).
The unitary representation matrices of $S$, $T_I$, $T_{II}$ and $T_{III}$-transformations, $\rho(S)$, $\rho(T_I)$, $\rho(T_{II})$ and $\rho(T_{III})$, form a group, $\widetilde{S}_4\rtimes A_4$, whose order is 1152.
Then, $\hat{\psi}^{\vec{j}}_{\bm{4}}(\Omega) $ in Eq.~(\ref{eq:N=6}) are a quartet of this group.

\end{itemize}

Table \ref{tab:examples_SiegelModularForms}, summarizes the results.
\begin{table}[H]
\centering
\renewcommand{\arraystretch}{1.5}
\begin{tabular}{c|c|c|c} \hline
$N$ & $\det N$ & $B_I,B_{II},B_{III}$ & Groups \\ \hline
$\begin{psmallmatrix}-2&1&-1\\1&-2&1\\-1&1&0\\\end{psmallmatrix}$
& 2 & $
\begin{psmallmatrix}
1 & 0 & 0 \\
0 & 1 & 0 \\
0 & 0 & 1 \\
\end{psmallmatrix},
\begin{psmallmatrix}
0 & 1 & -1 \\
1 & 0 & 1 \\
-1 & 1 & 2 \\
\end{psmallmatrix},
\begin{psmallmatrix}
0 & 1 & 1 \\
1 & 0 & -1 \\
1 & -1 & -4 \\
\end{psmallmatrix}
$ & $\widetilde{S}_4$ \\
$\begin{psmallmatrix}0&3&4\\3&0&-4\\4&-4&-11\\\end{psmallmatrix}$
& 3 & $
\begin{psmallmatrix}
2 & 0 & 0 \\
0 & 2 & 0 \\
0 & 0 & 2 \\
\end{psmallmatrix},
\begin{psmallmatrix}
0 & 2 & 0 \\
2 & 0 & 0 \\
0 & 0 & -2 \\
\end{psmallmatrix},
\begin{psmallmatrix}
0 & 0 & -1 \\
0 & 0 & 1 \\
-1 & 1 & 2 \\
\end{psmallmatrix}
$ & $(Z_8 \times Z_2) \rtimes Z_2 \rtimes Z_3$ \\
$\begin{psmallmatrix}-1&1&-1\\1&-1&-1\\-1&-1&2\\\end{psmallmatrix}$
& 4 & $
\begin{psmallmatrix}
-1 & 1 & 0 \\
1 & -1 & 0 \\
0 & 0 & 0 \\
\end{psmallmatrix},
\begin{psmallmatrix}
-2 & 1 & -1 \\
1 & -2 & -1 \\
-1 & -1 & 1 \\
\end{psmallmatrix},
\begin{psmallmatrix}
-1 & -1 & 0 \\
-1 & -1 & 0 \\
0 & 0 & -2 \\
\end{psmallmatrix}
$ & $\widetilde{\Delta}(96)$ \\
$\begin{psmallmatrix}
-3 & -2 &  2 \\
-2 & -3 & -2 \\
 2 & -2 & -7 \\
\end{psmallmatrix}$ & 5 & 
$\begin{psmallmatrix}
2 & 0 & 0 \\
0 & 2 & 0 \\
0 & 0 & 2 \\
\end{psmallmatrix},
\begin{psmallmatrix}
0 & 1 & -1 \\
1 & 0 & 1 \\
-1 & 1 & 2 \\
\end{psmallmatrix},
\begin{psmallmatrix}
0 & 1 & 1 \\
1 & 0 & -1 \\
1 & -1 & -4 \\
\end{psmallmatrix}$
& $A_5\times Z_8$ \\
$\begin{psmallmatrix}-2 & -1 & -1 \\-1 & -2 &  1 \\-1 &  1 &  0 \\\end{psmallmatrix}$ & 6 &
$\begin{psmallmatrix}
1 & 0 & 0 \\
0 & 1 & 0 \\
0 & 0 & 1 \\
\end{psmallmatrix},
\begin{psmallmatrix}
0 & 1 & -1 \\
1 & 0 & 1 \\
-1 & 1 & 0 \\
\end{psmallmatrix},
\begin{psmallmatrix}
0 & 1 & 1 \\
1 & 0 & -1 \\
1 & -1 & -2 \\
\end{psmallmatrix}$
 & $\widetilde{S}_4\rtimes A_4$ \\ \hline
\end{tabular}
\caption{Examples of the Siegel modular forms of weight 1/2 obtained from the zero-modes on magnetized $T^6$.}
\label{tab:examples_SiegelModularForms}
\end{table}


\section{The Siegel modular forms for $\widetilde{\Delta}(96)$}

In the previous section, we have constructed some of Siegel modular forms of weight 1/2 from zero-modes on $T^6$.
In this section, we give a further study of the Siegel modular forms of weight 1/2 for $\widetilde{\Delta}(96)$ in Eq.~(\ref{eq:modular_form_1/2_Delta96}).
Also we construct the Siegel modular forms of higher weights.


\subsection{Weight 1/2}

The Siegel modular forms of weight 1/2 for $\widetilde{\Delta}(96)$ are given by Eq.~(\ref{eq:modular_form_1/2_Delta96}).
For convenience, we redefine them as
\begin{align}
\begin{pmatrix}
\zeta(\Omega) \\
\eta(\Omega) \\
\theta(\Omega) \\
\end{pmatrix}
\equiv
\begin{pmatrix}
\frac{1}{\sqrt{2}}\psi_N^{\begin{psmallmatrix}0\\-1\\0\\\end{psmallmatrix}}(0,\Omega)
+ \frac{1}{\sqrt{2}} \psi_N^{\begin{psmallmatrix}-1\\0\\0\\\end{psmallmatrix}}(0,\Omega) \\
\psi_N^{\begin{psmallmatrix}0\\0\\0\\\end{psmallmatrix}}(0,\Omega) \\
-\psi_N^{\begin{psmallmatrix}0\\0\\-1\\\end{psmallmatrix}}(0,\Omega) \\
\end{pmatrix}
=
\begin{pmatrix}
\sqrt{2}\psi_N^{\begin{psmallmatrix}0\\-1\\0\\\end{psmallmatrix}}(0,\Omega) \\
\psi_N^{\begin{psmallmatrix}0\\0\\0\\\end{psmallmatrix}}(0,\Omega) \\
-\psi_N^{\begin{psmallmatrix}0\\0\\-1\\\end{psmallmatrix}}(0,\Omega) \\
\end{pmatrix}, \label{eq:zeetth}
\end{align}
where the minus sign in the third row is merely the convension.
The independent $B$ matrices satisfying the consistency conditions for $T$-transformation in Eq.~(\ref{eq:consistency_T}) are given by Eqs.~(\ref{eq:BI}), (\ref{eq:BII}) and (\ref{eq:BIII}).
Then the moduli $\Omega$ commuting to $N$ matrix is generally written as
\begin{align}
\Omega = \omega_1B_I + \omega_2B_{II} + \omega_3B_{III}, \label{eq:omega_para}
\end{align}
where $(\omega_1,\omega_2,\omega_3)$ are the moduli parameters.
Under $T$-transformations, $(\omega_1,\omega_2,\omega_3)$ are transformed as
\begin{align}
\begin{aligned}
&(\omega_1, \omega_2, \omega_3) \xrightarrow{T_I} (\omega_1+1, \omega_2, \omega_3), \\
&(\omega_1, \omega_2, \omega_3) \xrightarrow{T_{II}} (\omega_1, \omega_2+1, \omega_3), \\
&(\omega_1, \omega_2, \omega_3) \xrightarrow{T_{III}} (\omega_1, \omega_2,\omega_3+1).
\end{aligned} \label{eq:omega123toT123}
\end{align}

Under $S$, $T_I$, $T_{II}$ and $T_{III}$-transformations, the triplet Siegel modular forms $(\zeta,\eta,\theta)$ are transformed as
\begin{align}
\begin{pmatrix}
\zeta \\ \eta \\ \theta \\
\end{pmatrix}
\xrightarrow{S}
\sqrt{\det (-\Omega)} \rho(S)
\begin{pmatrix}
\zeta \\ \eta \\ \theta \\
\end{pmatrix}, \quad
\begin{pmatrix}
\zeta \\ \eta \\ \theta \\
\end{pmatrix}
\xrightarrow{T_{I,II,III}}
\rho(T_{I,II,III})
\begin{pmatrix}
\zeta \\ \eta \\ \theta \\
\end{pmatrix},
\end{align}
where
\begin{align}
\begin{aligned}
&\rho(S) = e^{-7\pi i/4}
\begin{pmatrix}
0 & \frac{1}{\sqrt{2}}i & \frac{1}{\sqrt{2}}i \\
\frac{1}{\sqrt{2}}i & \frac{1}{2}i & -\frac{1}{2}i \\
\frac{1}{\sqrt{2}}i & -\frac{1}{2}i & \frac{1}{2}i \\
\end{pmatrix}, \quad
\rho(T_{I}) =
\begin{pmatrix}
i & 0 & 0 \\
0 & 1 & 0 \\
0 & 0 & 1 \\
\end{pmatrix}, \\
&\rho(T_{II}) =
\begin{pmatrix}
e^{7\pi i/4} & 0 & 0 \\
0 & 1 & 0 \\
0 & 0 & -1 \\
\end{pmatrix}, \quad
\rho(T_{III}) =
\begin{pmatrix}
-1 & 0 & 0 \\
0 & 1 & 0 \\
0 & 0 & 1 \\
\end{pmatrix}.
\end{aligned} \label{eq:rhos}
\end{align}
One can check $\rho(T_I)=\rho(T_{II}^{6})$ and $\rho(T_{III})=\rho(T_{II}^{4})$.
Hence we can regard $\rho(S)$ and $\rho(T_{II})$ as generators of unitary representations although $T_I$ and $T_{III}$ cannot be generated by $S$ and $T_{II}$.
$\rho(S)$ and $\rho(T_{II})$ form a group $\widetilde{\Delta}(96)\simeq \Delta(48)\rtimes Z_8$ and satisfy the algebraic relations,
\begin{align}
\begin{aligned}
&\rho(S)^2=-i\bm{1}_3, ~(\rho(S)\rho(T_{II}))^3 = \rho(T_{II})^{8} = (\rho(S)^{-1}\rho(T_{II})^{-1}\rho(S)\rho(T_{II}))^3 = \bm{1}_3, \\
&\rho(S)^2\rho(T_{II}) = \rho(T_{II})\rho(S)^2.
\end{aligned}
\end{align}
In Appendix \ref{app:gt_Delta96}, we summarize the group theory of $\widetilde{\Delta}(96)$.
Thus, $(\zeta,\eta,\theta)$ are the Siegel modular forms of weight 1/2 for $\widetilde{\Delta}(96)$.

Introducing $q_i\equiv e^{\pi i\omega_i/4}$ $(i=1,2,3)$, $(\zeta,\eta,\theta)$ can be expanded by the powers of $q_i$.
Notice that
\begin{align}
\begin{aligned}
(q_1,q_2,q_3) \xrightarrow{T_I} (e^{\frac{\pi i}{4}}q_1,q_2,q_3), \\
(q_1,q_2,q_3) \xrightarrow{T_{II}} (q_1,e^{\frac{\pi i}{4}}q_2,q_3), \\
(q_1,q_2,q_3) \xrightarrow{T_{III}} (q_1,q_2,e^{\frac{\pi i}{4}}q_3), \\
\end{aligned}
\end{align}
and
\begin{align}
\begin{aligned}
&(\zeta, \eta, \theta) \xrightarrow{T_I} (e^{\frac{\pi i}{2}}\zeta, \eta, \theta), \\
&(\zeta, \eta, \theta) \xrightarrow{T_{II}} (e^{\frac{7\pi i}{4}}\zeta, \eta, e^{\pi i}\theta), \\
&(\zeta, \eta, \theta) \xrightarrow{T_{III}} (e^{\pi i}\zeta, \eta, \theta).
\end{aligned} \label{eq:T-charges}
\end{align}
$T$-charges of $(\zeta, \eta, \theta)$ mean that they are expanded by $q_i$ as
\begin{align}
&\zeta = q_1^2q_2^7q_3^4\sum_{\vec{m}\in\mathbb{Z}^3} c^1_{m_1m_2m_3} q_1^{8m_1} q_2^{8m_2} q_3^{8m_3}, \\
&\eta = \sum_{\vec{m}\in\mathbb{Z}^3} c^2_{m_1m_2m_3} q_1^{8m_1} q_2^{8m_2} q_3^{8m_3}, \\
&\theta = q_2^4\sum_{\vec{m}\in\mathbb{Z}^3} c^3_{m_1m_2m_3} q_1^{8m_3} q_2^{8m_1} q_3^{8m_2},
\end{align}
where $c^1$, $c^2$ and $c^3$ denote constant order 3 tensors.
Substituting $\Omega$ in Eq.~(\ref{eq:omega_para}) to the definitions of $(\zeta, \eta, \theta)$ in Eq.~(\ref{eq:zeetth}), we actually obtain following $q$-expansions,
\begin{align}
\begin{pmatrix}
\zeta \\ \eta \\ \theta \\
\end{pmatrix}
=
\begin{pmatrix}
\sqrt{2}(2q_1^{2}q_2^{7}q_3^{4}
+2q_1^{2}q_2^{7}q_3^{-4}
+2q_1^{2}q_2^{15}q_3^{-12}
+2q_1^{2}q_2^{15}q_3^{12}
+2q_1^{18}q_2^{31}q_3^{-4}
+2q_1^{18}q_2^{31}q_3^{4} + \cdots) \\
1
+2q_2^{16}
+4q_1^{8}q_2^{16}
+2q_2^{16}q_3^{16}
+2q_2^{16}q_3^{-16}
+4q_1^{8}q_2^{24} + \cdots \\
-(2q_2^{4}
+2q_2^{12}
+2q_1^{8}q_2^{12}
+4q_1^{8}q_2^{28}
+4q_1^{8}q_2^{28}q_3^{16}
+4q_1^{8}q_2^{28}q_3^{-16} + \cdots) \\
\end{pmatrix}. \label{eq:q-ex_zethet}
\end{align}
It is remarkable that only non-negative powers of $q_1$ and $q_2$ appear in $(\zeta,\eta,\theta)$ while it is not true for $q_3$.
This is because diag$(NB_I)\geq 0$ and diag$(NB_{II}) \geq 0$ but diag$(NB_{III})$ includes a negative eigenvalue.
It follows from these $q$-expansions that at two cusps $\omega_1= i\infty$ and $\omega_2= i\infty$ where $q_1$ and $q_2$ vanish respectively, $(\zeta,\eta,\theta)$ have the following behaviours,
\begin{align}
\begin{array}{llll}
\omega_1=i\infty: &
\zeta = 0, & \eta = 1+2q_2^{16}+\cdots, & \theta = -2q_2^4-2q_2^{12}+\cdots, \\
\omega_2=i\infty: &
\zeta = 0, & \eta = 1, & \theta = 0.
\end{array}
\label{eq:1/2 at cusps}
\end{align}

Finally we study the Siegel modular forms of higher weights.
Taking tensor products of $(\zeta,\eta,\theta)$, the Siegel modular forms of higher weights are constructed.
In Appendix \ref{app:mf_Delta96}, we show the Siegel modular forms up to weight 5.
The dimension of the Siegel modular forms of weight $k$ is $_{2k+2}C_2=\frac{1}{2}(2k+2)(2k+1)$.
This can be shown from the behaviours of $(\zeta,\eta,\theta)$ at the cusps in Eq.~(\ref{eq:1/2 at cusps}).
We use a mathematical induction.
Here we introduce a vector $Y_n$ containing the products of $n$ number of $(\zeta,\eta,\theta)$:
\begin{align}
Y_n \equiv
\begin{pmatrix}
\zeta^n \\
\zeta^{n-1}\eta \\
\vdots \\
\zeta\eta\theta^{n-2} \\
\zeta\theta^{n-1} \\
\eta^n \\
\eta^{n-1}\theta \\
\vdots \\
\eta \theta^{n-1} \\
\theta^n \\
\end{pmatrix}
\begin{array}{l}
\left.
\begin{matrix}
\\ \\ \\ \\ \\
\end{matrix}
\right\} \textrm{Products with one or more $\zeta$} \\
\left.
\begin{matrix}
\\ \\ \\ \\ \\
\end{matrix}
\right\} \textrm{Products without $\zeta$} \\
\end{array}.
\end{align}
Let us assume that $Y_{n-1}$ spans $_{n+1}C_2=\frac{1}{2}(n+1)n$ dimensional spaces.
Note that $Y_{n-1}$ contains $\frac{1}{2}(n+1)n$ number of products.
Hence this assumption means all products of $n-1$ number of $(\zeta,\eta,\theta)$ are linearly independent each other.
This is true for $Y_1=(\zeta,\eta,\theta)$.
Thus we should show that $Y_n$ spans $_{n+2}C_2=\frac{1}{2}(n+2)(n+1)$ dimensional spaces under this assumption.

First let us consider the products with one or more $\zeta$ in $Y_n$.
They are given by $\zeta Y_{n-1}$ and span $\frac{1}{2}(n+1)n$ dimensional spaces.
This means
\begin{align}
\sum_{j=0,1,...,n-1} \sum_{k=0,1,...,j} c_{jk} \zeta^{n-j} \eta^{j-k} \theta^{k} = 0 \to c_{jk}=0. \label{eq:c_jk=0}
\end{align}

Second we consider the products without $\zeta$.
As shown in Eq.~(\ref{eq:q-ex_zethet}), the lowest orders of $q_2$ in $\eta$ and $\theta$ are $q_2^0$ and $q_2^4$, respectively.
This means
\begin{align}
\eta^n \neq \sum_{j=1,2,...,n} c_j\eta^{n-j} \theta^{j}, \quad \forall c_j \in \mathbb{C},
\end{align}
because the lowest order in the left-hand side is $q_2^0$ while one in the right-hand side is $q_2^{4}$.
Similarly, we find
\begin{align}
\eta^{n-1}\theta \neq \sum_{j=2,3,...,n} c_j\eta^{n-j} \theta^{j}, \quad \forall c_j \in \mathbb{C},
\end{align}
because the lowest order in the left-hand side is $q_2^{4}$ while one in the right-hand side is $q_2^{8}$.
Repeating this procedure, we obtain
\begin{align}
\sum_{j=0,1,...,n} c_j\eta^{n-j} \theta^{j} = 0 \to c_j = 0. \label{eq:c_j=0}
\end{align}

Third, we consider the equation,
\begin{align}
\sum_{j=0,1,...,n-1} \sum_{k=0,1,...,j} c_{jk} \zeta^{n-j} \eta^{j-k} \theta^{k}
+
\sum_{j=0,1,...,n} c_j\eta^{n-j} \theta^{j} = 0.
\end{align}
As shown in Eq.~(\ref{eq:1/2 at cusps}), the first term vanishes at the cusp $\omega_1=i\infty$.
This means that this equation is devided to two equations,
\begin{align}
\sum_{j=0,1,...,n-1} \sum_{k=0,1,...,j} c_{jk} \zeta^{n-j} \eta^{j-k} \theta^{k} = 0, \quad
\sum_{j=0,1,...,n} c_j\eta^{n-j} \theta^{j} = 0.
\end{align}
These two equations consist only if $c_{jk}=0$ and $c_j=0$ as shown in Eqs.~(\ref{eq:c_jk=0}) and (\ref{eq:c_j=0}).
Thus we obtain
\begin{align}
\sum_{j=0,1,...,n-1} \sum_{k=0,1,...,j} c_{jk} \zeta^{n-j} \eta^{j-k} \theta^{k}
+
\sum_{j=0,1,...,n} c_j\eta^{n-j} \theta^{j} = 0 \to c_{jk}=0,~c_j=0.
\end{align}
This means that all products in $Y_n$ are linearly independent each other.
Therefore we have proven that $Y_n$ spans $_{n+2}C_2=\frac{1}{2}(n+2)(n+1)$ dimensional spaces when $Y_{n-1}$ spans $_{n+1}C_2=\frac{1}{2}(n+1)n$ dimensional spaces.


\subsection{Residual symmetries}

Next we study residual symmetries at the modular symmetric points.
$S$, $T_I$, $T_{II}$ and $T_{III}$-transformations satisfy the following algebraic relations,
\begin{align}
S^4 =
(ST_I^{-1}T_{II})^{12} =
(ST_{I}^{-2}T_{II})^{12} =
\mathbb{I}. \label{eq:algebra_N}
\end{align}
We can find three invariant moduli (modular symmetric points) corresponding to these algebraic relations.
The generators of algebraic relations, $S$, $ST^{-1}_IT_{II}$ and $ST_I^{-2}T_{II}$, act on the moduli $\Omega$ as
\begin{align}
&S:\Omega = -\Omega^{-1}, \\
&(ST_I^{-1}T_{II}):\Omega = -(\Omega-B_I+B_{II})^{-1}, \\
&(ST_I^{-2}T_{II}):\Omega = -(\Omega-2B_I+B_{II})^{-1}.
\end{align}
Solving the equations,
\begin{align}
&-\Omega_S = \Omega_S, \\
&-(\Omega_{ST_I^{-1}T_{II}}-B_I+B_{II})^{-1} = \Omega_{ST_I^{-1}T_{II}}, \\
&-(\Omega_{ST_I^{-2}T_{II}}-2B_I+B_{II})^{-1} = \Omega_{ST_I^{-2}T_{II}},
\end{align}
we obtain the invariant moduli $\Omega_S$, $\Omega_{ST_I^{-1}T_{II}}$ and $\Omega_{ST_I^{-2}T_{II}}$ under $S$, $ST_I^{-1}T_{II}$ and $ST_I^{-2}T_{II}$-transformations, respectively.
In addition there are three invariant moduli $\Omega_{T_I}$, $\Omega_{T_{II}}$ and $\Omega_{T_{III}}$ under $T_I$, $T_{II}$ and $T_{III}$-transformations, respectively.
In Table \ref{tab:residuals}, we show these invariant moduli.
Note that the structures of the moduli are restricted by $N\Omega=\Omega N$.
Hence, it is diagonalized by the orthogonal matrix $O$ which diagonalize $N$ matrix in Eq.~(\ref{eq:N_matrix}) as
\begin{align}
O^T N O =
\begin{pmatrix}
\sqrt{3}+1 & 0 & 0 \\
0 & -2 & 0 \\
0 & 0 & -\sqrt{3}+1 \\
\end{pmatrix},
\end{align}
where
\begin{align}
O =
\begin{pmatrix}
-\frac{1}{\sqrt{6+2\sqrt{3}}} & \frac{1}{\sqrt{2}} & \frac{1}{\sqrt{6-2\sqrt{3}}} \\
-\frac{1}{\sqrt{6+2\sqrt{3}}} & -\frac{1}{\sqrt{2}} & \frac{1}{\sqrt{6-2\sqrt{3}}} \\
\frac{1+\sqrt{3}}{\sqrt{6+2\sqrt{3}}} & 0 & -\frac{1-\sqrt{3}}{\sqrt{6-2\sqrt{3}}} \\
\end{pmatrix}. \label{eq:orthogonal}
\end{align}
\begin{table}[H]
\centering
\begin{tabular}{ccc} \hline
$\gamma$ & $\gamma:\Omega$ & Invariant moduli \\ \hline
$S$ & $-\Omega^{-1}$ & 
$\Omega_{S}\equiv O\begin{pmatrix}e^{\pm\frac{\pi i}{2}}&0&0\\0&e^{\pm\frac{\pi i}{2}}&0\\0&0&e^{\pm\frac{\pi i}{2}}\\\end{pmatrix}O^T$ \\
$ST_I^{-1}T_{II}$ & $-(\Omega-B_I+B_{II})^{-1}$ & 
$\Omega_{ST_I^{-1}T_{II}}\equiv O\begin{pmatrix}e^{\pm\frac{\pi i}{6}}&0&0\\0&e^{\pm\frac{2\pi i}{3}}&0\\0&0&e^{\pm\frac{5\pi i}{6}}\\\end{pmatrix}O^T$ \\
$ST_{I}^{-2}T_{II}$ & $-(\Omega-2B_I+B_{II})^{-1}$ & 
$\Omega_{ST_I^{-2}T_{II}}\equiv O\begin{pmatrix}e^{\pm\frac{\pi i}{6}}&0&0\\0&e^{\pm\frac{\pi i}{3}}&0\\0&0&e^{\pm\frac{5\pi i}{6}}\\\end{pmatrix}O^T$ \\
$T_I$ & $\Omega+B_I$ & 
$\Omega_{T_I}\equiv i\infty B_I$ ($\omega_1=i\infty$) \\
$T_{II}$ & $\Omega+B_{II}$ & 
$\Omega_{T_{II}}\equiv i\infty B_{II}$ ($\omega_2=i\infty$) \\
$T_{III}$ & $\Omega+B_{III}$ & 
$\Omega_{T_{III}}\equiv i\infty B_{III}$ ($\omega_3=i\infty$) \\
\hline
\end{tabular}
\caption{Invariant moduli corresponding to the algebraic relations in Eq.~(\ref{eq:algebra_N}) and $T$-transformations.
$\pm$ in the third column means any double sign.}
\label{tab:residuals}
\end{table}


\section{Numerical example}
\label{subsec:Num}

In this section, we study quark flavor models using the Siegel modular forms for $\widetilde{\Delta}(96)$ which we have studied in the previous section.
To make our analysis simple, we use the Siegel modular forms belonging to $\widetilde{\Delta}(96)$ singlets.
There are eight singlets $\bm{1}_q$, $q=0,1,...,7$, in $\widetilde{\Delta}(96)$.
When left-handed matter fields $L=(L_1,L_2,L_3)$ belong to singlets $(\bm{1}_{q_{L1}}, \bm{1}_{q_{L2}}, \bm{1}_{q_{L3}})$ with weights $(-k_{L1},-k_{L2},-k_{L3})$ and right-handed matter fields $R=(R_1,R_2,R_3)$ belong to $(\bm{1}_{q_{R1}}, \bm{1}_{q_{R2}}, \bm{1}_{q_{R3}})$ with $(-k_{L1},-k_{L2},-k_{L3})$, fermion mass matrices are given by
\begin{align}
M &\propto \alpha_{\bm{1}_{q_{11}}}^{(k_{11})}\begin{pmatrix}
C^{\frac{k_{11}}{2}} Y_{\bm{1}_{q_{11}}}^{(k_{11})} & 0 & 0 \\
0 & 0 & 0 \\
0 & 0 & 0 \\
\end{pmatrix}
+ \alpha_{\bm{1}_{q_{12}}}^{(k_{12})}\begin{pmatrix}
0 & C^{\frac{k_{12}}{2}} Y_{\bm{1}_{q_{12}}}^{(k_{12})} & 0 \\
0 & 0 & 0 \\
0 & 0 & 0 \\
\end{pmatrix}
+ \alpha_{\bm{1}_{q_{13}}}^{(k_{13})}\begin{pmatrix}
0 & 0 & C^{\frac{k_{13}}{2}} Y_{\bm{1}_{q_{13}}}^{(k_{13})} \\
0 & 0 & 0 \\
0 & 0 & 0 \\
\end{pmatrix} \notag \\
&+ \alpha_{\bm{1}_{q_{21}}}^{(k_{21})}\begin{pmatrix}
0 & 0 & 0 \\
C^{\frac{k_{21}}{2}}Y_{\bm{1}_{q_{21}}}^{(k_{21})} & 0 & 0 \\
0 & 0 & 0 \\
\end{pmatrix}
+ \alpha_{\bm{1}_{q_{22}}}^{(k_{22})}\begin{pmatrix}
0 & 0 & 0 \\
0 & C^{\frac{k_{22}}{2}}Y_{\bm{1}_{q_{22}}}^{(k_{22})} & 0 \\
0 & 0 & 0 \\
\end{pmatrix}
+ \alpha_{\bm{1}_{q_{23}}}^{(k_{23})}\begin{pmatrix}
0 & 0 & 0 \\
0 & 0 & C^{\frac{k_{23}}{2}}Y_{\bm{1}_{q_{23}}}^{(k_{23})} \\
0 & 0 & 0 \\
\end{pmatrix} \notag \\
&+ \alpha_{\bm{1}_{q_{31}}}^{(k_{31})}\begin{pmatrix}
0 & 0 & 0 \\
0 & 0 & 0 \\
C^{\frac{k_{31}}{2}} Y_{\bm{1}_{q_{31}}}^{(k_{31})} & 0 & 0 \\
\end{pmatrix}
+ \alpha_{\bm{1}_{q_{32}}}^{(k_{32})}\begin{pmatrix}
0 & 0 & 0 \\
0 & 0 & 0 \\
0 & C^{\frac{k_{32}}{2}} Y_{\bm{1}_{q_{32}}}^{(k_{32})} & 0 \\
\end{pmatrix}
+ \alpha_{\bm{1}_{q_{33}}}^{(k_{33})}\begin{pmatrix}
0 & 0 & 0 \\
0 & 0 & 0 \\
0 & 0 & C^{\frac{k_{33}}{2}}Y_{\bm{1}_{q_{33}}}^{(k_{33})} \\
\end{pmatrix},
\end{align}
where $\alpha$ denotes coupling constants and
\begin{align}
q_{jk} = (-q_{Lj}-q_{Ek})~\textrm{mod}~8, \quad k_{jk} = k_{Lj} + k_{Rk}.
\end{align}
Here we denote $C=(2^3\det \textrm{Im}\Omega)$ and its powers are originated from the ratio of Kahler metric.
Note that when all of mass matrix elements have same weights, the powers of $C$ contribute to the overall factor.

As we have seen throughout Chapter \ref{sec:4D modular symmetric flavor models}, hierarchical values of the modular forms are required to reproduce large quark mass hierarchies.
For this purpose we concentrate on the vicinity of the cusp,
\begin{align}
\Omega \sim \Omega_{T_I} = i\infty B_I,
\end{align}
where $T_I$ symmetry remains.
When $\Omega$ lies on the vicinity of $\Omega_{T_I}$, the Siegel modular forms $f(\Omega)$ with the residual charge $r$ are expanded by powers of $q_1=e^{\pi i\omega_1/4}$ as
\begin{align}
f(\Omega) \sim q_1^r \ll 1, \quad \omega_1 \sim i\infty. \label{eq:fsimq1}
\end{align}
Actually $q$-expansions in Eq.~(\ref{eq:q-ex_zethet}) are written as
\begin{align}
\begin{pmatrix}
\zeta \\ \eta \\ \theta \\
\end{pmatrix}
\sim
\begin{pmatrix}
\sqrt{2} q_1^2 (2q_2^7q_3^4+2q_2^7q_3^{-4}+2q_2^{15}q_3^{12}+2q_2^{15}q_3^{-12}+\cdots) \\
1+2q_2^{16} + 2q_2^{16}q_3^{16}+\cdots \\
-2q_2^4-2q_2^{12}+\cdots \\
\end{pmatrix},
\end{align}
in the first order approximation of $q_1$.
$(\zeta,\eta,\theta)$ have $T_I$-charge $(2,0,0)$ and this result is consistent.
Thus the modular forms become hierarchical as close to $\Omega_{T_I}$ depending on their $T_I$-charges.
To obtain more realistic mass hierarchies of quarks, we further assume $\omega_2\sim i\infty$ and $|\omega_2| < |\omega_1|$, where $|q_1|<|q_2|<1$.
Then $(\zeta, \eta, \theta)$ is evaluated as
\begin{align}
\begin{pmatrix}
\zeta \\ \eta \\ \theta \\
\end{pmatrix}
\sim
\begin{pmatrix}
4\sqrt{2} q_1^2 q_2^7 \\
1 \\
-2q_2^4 \\
\end{pmatrix}, \label{eq:zetaetatheta_q1q2}
\end{align}
in the first order approximation of $q_1$ and $q_2$.
We have the hierarchy $\zeta,\theta \ll \eta\sim 1$ while the hierarchy between $\zeta$ and $\theta$ is controlled by the values of $\omega_1$ and $\omega_2$.

Next, let us study the behaviors of singlet Siegel modular forms in the assumption $\omega_1,\omega_2\sim i\infty$, $|\omega_2|<|\omega_1|$.
That is, we use the approximation in Eq.~(\ref{eq:zetaetatheta_q1q2}).
In Table \ref{tab:T-charges_of_singlets}, we show $T_I$, $T_{II}$ and $T_{III}$-charges of eight $\widetilde{\Delta}(96)$ singlets.
\begin{table}[H]
\centering
\begin{tabular}{c|cccccccc} \hline
& $\bm{1}_0$ & $\bm{1}_1$ & $\bm{1}_2$ & $\bm{1}_3$ & $\bm{1}_4$ & $\bm{1}_5$ & $\bm{1}_6$ & $\bm{1}_7$ \\ \hline
$T_I$-charges & 0 & 6 & 4 & 2 & 0 & 6 & 4 & 2 \\
$T_{II}$-charges & 0 & 1 & 2 & 3 & 4 & 5 & 6 & 7 \\
$T_{III}$-charges & 0 & 4 & 0 & 4 & 0 & 4 & 0 & 4 \\ \hline
\end{tabular}
\caption{$T_I$, $T_{II}$ and $T_{III}$-charges of $\widetilde{\Delta}(96)$ singlets.}
\label{tab:T-charges_of_singlets}
\end{table}
Up to weight 5, we can find eight modular forms belonging to singlets,
\begin{align}
Y^{(3/2)}_{\bm{1}_7}, ~Y^{(2)}_{\bm{1}_4}, ~Y^{(3)}_{\bm{1}_6}, ~Y^{(7/2)}_{\bm{1}_3}, ~Y^{(4)}_{\bm{1}_{0a}}, ~Y^{(4)}_{\bm{1}_{0b}}, ~Y^{(9/2)}_{\bm{1}_5}, ~Y^{(5)}_{\bm{1}_2}.
\end{align}
Note that the Siegel modular forms belonging to the singlet $\bm{1}_1$ do not appear up to weight 5.
When $\zeta,\theta \ll \eta\sim 1$, they are evaluated as
\begin{align}
\begin{aligned}
&Y^{(3/2)}_{\bm{1}_7} \simeq
\frac{\sqrt{6}}{2}\zeta\simeq 4\sqrt{3}q_1^2q_2^7, \quad
Y^{(2)}_{\bm{1}_4} \simeq
\frac{2\sqrt{3}}{3}\theta\simeq -\frac{4\sqrt{3}}{3}q_2^4, \quad
Y^{(3)}_{\bm{1}_6} \simeq
\frac{3}{2}\zeta^{2} \simeq 48q_1^4q_2^{14}, \\
&Y^{(7/2)}_{\bm{1}_3} \simeq
\sqrt{2}\zeta\theta \simeq -16q_1^2q_2^{11}, \quad
Y^{(4)}_{\bm{1}_{0a}} \simeq
\frac{1}{4\sqrt{6}}, \quad
Y^{(4)}_{\bm{1}_{0b}} \simeq
\frac{4}{3}\theta^{2} \simeq \frac{16}{3}q_2^8, \\
&Y^{(9/2)}_{\bm{1}_5} \simeq
-\frac{3}{2\sqrt{6}}\zeta^{3} \simeq -64\sqrt{3}q_1^6q_2^{21}, \quad
Y^{(5)}_{\bm{1}_2} \simeq
\frac{1}{\sqrt{3}}\zeta^{2}\theta \simeq -\frac{64}{\sqrt{3}}q_1^4q_2^{18},
\end{aligned}
\end{align}
in the first order approximation of $\zeta$ and $\theta$.
In what follows we ignore $Y^{(4)}_{\bm{1}_{0b}}$ because it is negligible comparing with $Y^{(4)}_{\bm{1}_{0a}}$ since $\theta\ll 1$.

Using these singlet Siegel modular forms, we build the quark flavor model in the vicinity of $\Omega_{T_I}$.
Let us study the model with the assignments in Table \ref{eq:assignments}.
\begin{table}[H]
\centering
\begin{tabular}{c|ccccc} \hline
& $Q$ & $u_R$ & $d_R$ & $H_u$ & $H_d$ \\ \hline
Weights & $(-1/2, -5/2, -3/2)$ & $(0, -1/2, -5/2)$ & $(-5/2, -1, -5/2)$ & 0 & 0 \\
Irr. reps. & $(\bm{1}_2, \bm{1}_6, \bm{1}_0)$ & $(\bm{1}_1, \bm{1}_4, \bm{1}_0)$ & $(\bm{1}_0, \bm{1}_7, \bm{1}_0)$ & $\bm{1}_0$ & $\bm{1}_0$ \\ \hline
\end{tabular}
\caption{Assignments in our model.}
\label{eq:assignments}
\end{table}
In this model, up and down-sector quark mass matrices $M_u$ and $M_d$ are given by
\begin{align}
M_u &\propto
\begin{pmatrix}
0 & 0 & C^{-1/2}\alpha^{13}Y_{\bm{1}_6}^{(3)} \\
0 &C^{-1/2}\alpha^{22}Y_{\bm{1}_6}^{(3)} & C^{1/2}\alpha^{23}Y_{\bm{1}_2}^{(5)} \\
C^{-5/4}\alpha^{31}Y_{\bm{1}_7}^{(3/2)} & C^{-1}\alpha^{32}Y_{\bm{1}_4}^{(2)} & \alpha^{33}Y_{\bm{1}_{0a}}^{(4)} \\
\end{pmatrix} \\
&\simeq 
\begin{pmatrix}
0 & 0 & C^{-1/2}\alpha^{13}48q_1^4q_2^{14} \\
0 & C^{-1/2}\alpha^{22}48q_1^4q_2^{14} & -C^{1/2}\alpha^{23}\frac{64}{\sqrt{3}}q_1^4q_2^{18} \\
C^{-5/4}\alpha^{31}4\sqrt{3}q_1^2q_2^7 & -C^{-1}\alpha^{32}\frac{4\sqrt{3}}{3}q_2^4 & \alpha^{33}\frac{1}{4\sqrt{6}} \\
\end{pmatrix}, \label{eq:up_mass_mtx} \\
 M_d &\propto
\begin{pmatrix}
C^{-1/2}\beta^{11}Y_{\bm{1}_6}^{(3)} &C^{-5/4}\beta^{12}Y_{\bm{1}_7}^{(3/2)} & C^{-1/2}\beta^{13}Y_{\bm{1}_6}^{(3)} \\
C^{1/2}\beta^{21}Y_{\bm{1}_2}^{(5)} &C^{-1/4}\beta^{22}Y_{\bm{1}_3}^{(7/2)} & C^{1/2}\beta^{23}Y_{\bm{1}_2}^{(5)} \\
\beta^{31}Y_{\bm{1}_{0a}}^{(4)} & 0 &\beta^{33}Y_{\bm{1}_{0a}}^{(4)} \\
\end{pmatrix} \\
&\simeq 
\begin{pmatrix}
C^{-1/2}\beta^{11}48q_1^4q_2^{14} & C^{-5/4}\beta^{12}4\sqrt{3}q_1^2q_2^7 & C^{-1/2}\beta^{13}48q_1^4q_2^{14} \\
-C^{1/2}\beta^{21}\frac{64}{\sqrt{3}}q_1^4q_2^{18} & -C^{-1/4}\beta^{22}16q_1^2q_2^{11} & -C^{1/2}\beta^{23}\frac{64}{\sqrt{3}}q_1^4q_2^{18} \\
\beta^{31}\frac{1}{4\sqrt{6}} & 0 &\beta^{33}\frac{1}{4\sqrt{6}} \\
\end{pmatrix}, \label{eq:down_mass_mtx}
\end{align}
where $\alpha^{ij}$ and $\beta^{ij}$ are the coupling constants.
Zero textures in mass matrices are due to the shortage of the singlet modular forms.
We expect that large quark mass hierarchies do not originate from the values of the coupling constants.
Therefore we assume ${\cal O}(1)$ sizes of $\alpha^{ij}$ and $\beta^{ij}$.
Also we regard values of moduli $(\omega_1, \omega_2,\omega_3)$ as free parameters.

To generate large mass hierarchies, in the vicinity of $\omega_1\sim i\infty$, we choose the moduli,
\begin{align}
(\omega_1, \omega_2, \omega_3) = (1.3i, 0.6i+0.86, 0),
\end{align}
where
\begin{align}
(q_1,q_2,q_3) = (0.360, 0.624e^{0.674i}, 1).
\end{align}
For CP violation, we have deviated $\omega_2$ from the imaginary axis\footnote{CP violation does not occur without $\textrm{Re}\Omega$ since the CP transformation is given by $\Omega\to -\Omega^*$ \cite{Baur:2019iai,Baur:2019kwi,Novichkov:2019sqv}.}.
Additonally we assume the following ${\cal O}(1)$ sizes of $\alpha^{ij}$ and $\beta^{ij}$,
\begin{align}
&\begin{pmatrix}
\text{-} & \text{-} & \alpha^{13} \\
\text{-} & \alpha^{22} & \alpha^{23} \\
\alpha^{31} & \alpha^{32} & \alpha^{33} \\
\end{pmatrix}
=
\begin{pmatrix}
\text{-} & \text{-} & 1.16 \\
\text{-} & 1.72 & -3.03 \\
1.00 & -1.36 & 1.00 \\
\end{pmatrix}, \quad
\begin{pmatrix}
\beta^{11} & \beta^{12} & \beta^{13} \\
\beta^{21} & \beta^{22} & \beta^{23} \\
\beta^{31} & \text{-} & \beta^{33} \\
\end{pmatrix}
=
\begin{pmatrix}
1.46 & 3.39 & 1.66 \\
3.44 & -1.00 & -2.89 \\
2.08 & \text{-} & -1.27 \\
\end{pmatrix}.
\end{align}
They lead to the quark mass ratios,
\begin{align}
&(m_u, m_c, m_t)/m_t = (8.25\times 10^{-6}, 2.70\times 10^{-3}, 1), \\
&(m_d, m_s, m_b)/m_b = (1.01\times 10^{-3}, 2.02\times 10^{-2}, 1),
\end{align}
absolute values of the CKM matrix elements,
\begin{align}
|V_{\textrm{CKM}}| =
\begin{pmatrix}
0.975 & 0.224 & 0.00340 \\
0.224 & 0.974 & 0.0392 \\
0.00809 & 0.0385 & 0.999 \\
\end{pmatrix},
\end{align}
and the value of Jarlskog invariant,
\begin{align}
J_{\textrm{CP}} = |\textrm{Im}(V_{\textrm{CKM}}^{us}V_{\textrm{CKM}}^{cb}(V_{\textrm{CKM}}^{ub}V_{\textrm{CKM}}^{cs})^*)| = 2.68\times 10^{-5}.
\end{align}
Results are summarized in Table \ref{tab:fitting}.
\begin{table}[H]
\small
  \begin{center}
    \renewcommand{\arraystretch}{1.3}
    \begin{tabular}{c|cccccccc} \hline
      & $\frac{m_u}{m_t}{\times10^{6}}$ & $\frac{m_c}{m_t}{\times10^3}$ & $\frac{m_d}{m_b}{\times10^4}$ & $\frac{m_s}{m_b}{\times10^2}$ & $|V_{\textrm{CKM}}^{us}|$ & $|V_{\textrm{CKM}}^{cb}|$ & $|V_{\textrm{CKM}}^{ub}|$ & $J_{\textrm{CP}}{\times 10^5}$ \\ \hline
      obtained values & 8.25 & 2.70 & 10.1 & 2.02 & 0.224 & 0.0392 & 0.00340 & 2.68 \\ \hline
      GUT scale values & 5.39 & 2.80 & 9.21 & 1.82 & 0.225 & 0.0400 & 0.00353 & 2.80 \\
      $1\sigma$ errors & $\pm 1.68$ & $\pm 0.12$ & $\pm 1.02$ & $\pm 0.10$ & $\pm 0.0007$ & $\pm 0.0008$ & $\pm 0.00013$ & $^{+0.14}_{-0.12}$ \\ \hline
    \end{tabular}
  \end{center}
  \caption{The mass ratios of the quarks and the absolute values of the CKM matrix elements at $(\omega_1, \omega_2, \omega_3) = (1.3i, 0.6i+0.86, 0)$.
GUT scale values at $2\times 10^{16}$ GeV with $\tan \beta=5$ \cite{Antusch:2013jca,Bjorkeroth:2015ora} and $1\sigma$ errors are shown.}
\label{tab:fitting}
\normalsize
\end{table}
Thus the quark flavor observables can be realized without fine-tuning of the coupling constants in the vicinity of invariant moduli $\Omega_{T_I}$.

Finally, we comment on the CP violation in our model.
At $\omega_1\sim i\infty$, $\omega_2\sim i\infty$ and $\omega_3=0$, up and down-sector quark mass matrices $M_u$ and $M_d$ are evaluated by Eqs.~(\ref{eq:up_mass_mtx}) and (\ref{eq:down_mass_mtx}), respectively.
Then phase factors of $q_1$ in mass matrices are comletely vanished under the following basis transformations,
\begin{align}
M_u \to U_Q^\dagger M_u U_{u_R}, \quad M_d \to U_Q^\dagger M_d U_{d_R},
\end{align}
where
\begin{align}
&U_Q =
\begin{pmatrix}
 e^{4i\textrm{Arg}(q_1)} &&\\
&  e^{4i\textrm{Arg}(q_1)} &\\
&& 1 \\
\end{pmatrix}, \\
&U_{u_R} =
\begin{pmatrix}
e^{-2i\textrm{Arg}(q_1)} &&\\
& 1 &\\
&& 1 \\
\end{pmatrix}, \\
&U_{d_R} =
\begin{pmatrix}
1 &&\\
& e^{2i\textrm{Arg}(q_1)} &\\
&& 1 \\
\end{pmatrix}.
\end{align}
After the basis transformation, the mass matrices are written as
\begin{align}
&U_Q^\dagger M_u U_{u_R} \notag \\
&\simeq 
\begin{pmatrix}
0 & 0 & C^{-1/2}\alpha^{13}48|q_1|^4q_2^{14} \\
0 & C^{-1/2}\alpha^{22}48|q_1|^4q_2^{14} & -C^{1/2}\alpha^{23}\frac{64}{\sqrt{3}}|q_1|^4q_2^{18} \\
C^{-5/4}\alpha^{31}4\sqrt{3}|q_1|^2q_2^7 & -C^{-1}\alpha^{32}\frac{4\sqrt{3}}{3}q_2^4 & \alpha^{33}\frac{1}{4\sqrt{6}} \\
\end{pmatrix}, \\
&U_Q^\dagger M_d U_{d_R} \notag \\
&\simeq 
\begin{pmatrix}
C^{-1/2}\beta^{11}48|q_1|^4q_2^{14} & C^{-5/4}\beta^{12}4\sqrt{3}|q_1|^2q_2^7 & C^{-1/2}\beta^{13}48|q_1|^4q_2^{14} \\
-C^{1/2}\beta^{21}\frac{64}{\sqrt{3}}|q_1|^4q_2^{18} & -C^{-1/4}\beta^{22}16|q_1|^2q_2^{11} & -C^{1/2}\beta^{23}\frac{64}{\sqrt{3}}|q_1|^4q_2^{18} \\
\beta^{31}\frac{1}{4\sqrt{6}} & 0 &\beta^{33}\frac{1}{4\sqrt{6}} \\
\end{pmatrix},
\end{align}
up to the overall factors.
The phase factors of $q_1$ in mass matrices are completely canceled \footnote{Similar behaviors at the modular symmetric points were studied in Refs.~\cite{Kobayashi:2019uyt,Kikuchi:2022geu}.}.
Therefore $\textrm{Re}\omega_1$ cannot contribute to the CP violation in our model.
In the same way, we check the effectiveness of the phase factor of $q_2$ in mass matrices.
We consider the further basis transformations,
\begin{align}
U_Q^\dagger M_u U_{u_R} \to \hat{U}_Q^\dagger U_Q^\dagger M_u U_{u_R} \hat{U}_{u_R}, \quad U_Q^\dagger M_d U_{d_R} \to \hat{U}_Q^\dagger U_Q^\dagger M_d U_{d_R} \hat{U}_{d_R},
\end{align}
where
\begin{align}
&\hat{U}_Q =
\begin{pmatrix}
e^{14i\textrm{Arg}(q_2)} &&\\
& e^{18i\textrm{Arg}(q_2)} &\\
&& 1 \\
\end{pmatrix}, \\
&\hat{U}_{u_R} =
\begin{pmatrix}
e^{-7i\textrm{Arg}(q_2)} &&\\
& e^{4i\textrm{Arg}(q_2)} &\\
&& 1 \\
\end{pmatrix}, \\
&\hat{U}_{d_R} =
\begin{pmatrix}
1 &&\\
& e^{7i\textrm{Arg}(q_2)} &\\
&& 1 \\
\end{pmatrix}.
\end{align}
Then mass matrices are written as
\begin{align}
&\hat{U}_Q^\dagger U_Q^\dagger M_u U_{u_R} \hat{U}_{u_R} \notag \\
&\simeq
\begin{pmatrix}
0 & 0 & C^{-1/2}\alpha^{13}48|q_1|^4|q_2|^{14} \\
0 & C^{-1/2}\alpha^{22}48|q_1|^4|q_2|^{14} & -C^{1/2}\alpha^{23}\frac{64}{\sqrt{3}}|q_1|^4|q_2|^{18} \\
C^{-5/4}\alpha^{31}4\sqrt{3}|q_1|^2|q_2|^7 & -C^{-1}\alpha^{32}\frac{4\sqrt{3}}{3}|q_2|^4 e^{8i\textrm{Ang}(q_2)} &\alpha^{33}\frac{1}{4\sqrt{6}} \\
\end{pmatrix}, \\
&\hat{U}_Q^\dagger U_Q^\dagger M_d U_{d_R} \hat{U}_{d_R} \notag \\
&\simeq
\begin{pmatrix}
C^{-1/2}\beta^{11}48|q_1|^4|q_2|^{14} & C^{-5/4}\beta^{12}4\sqrt{3}|q_1|^2|q_2|^7 & C^{-1/2}\beta^{13}48|q_1|^4|q_2|^{14} \\
-C^{1/2}\beta^{21}\frac{64}{\sqrt{3}}|q_1|^4|q_2|^{18} & -C^{-1/4}\beta^{22}16|q_1|^2|q_2|^{11} & -C^{1/2}\beta^{23}\frac{64}{\sqrt{3}}|q_1|^4|q_2|^{18} \\
\beta^{31}\frac{1}{4\sqrt{6}} & 0 & \beta^{33}\frac{1}{4\sqrt{6}} \\
\end{pmatrix},
\end{align}
up to the overall factors.
After the basis transformations, the phase factor $e^{8i\textrm{Arg}(q_2)}=e^{2\pi i\textrm{Re}\omega_2}$ survives in (3,2) element of $M_u$.
Therefore $\textrm{Re}\omega_2$ in the region $1/2>|\textrm{Re}\omega_2|>0$ can contribute to the CP violation.
Actually, sufficient CP violation can be obtained by the deviation of $\omega_2$ from the imaginary axis (CP symmetric points) as we have seen in the above numerical example.
In this sense $\omega_1$ works on mass hierarchies while $\omega_2$ works on the CP violation in our model.
Thus, the Siegel modular forms which are described by the multi moduli parameters are the promising way realizing both mass hierarchies and the CP violation without fine-tuning of the coupling constants.


\chapter{Summary}
\label{sec:summary}

In this paper, we have seen two approaches to the flavor structures.
One is the magnetized orbifold models of the superstring theory; another one is 4D modular symmetric flavor models.
Furthermore, we have constructed some examples of Siegel modular forms from magnetized $T^6$ model.


\section{Summary of magnetized orbifold models}

In Chapter \ref{sec:magnetized_orbifold_models}, we have discussed the magnetized orbifold models along in Refs.~\cite{Kikuchi:2020frp, Kikuchi:2020nxn, Hoshiya:2020hki, Kikuchi:2021yog, Hoshiya:2022qvr}.
The 10D ${\cal N}=1$ $U(N)$ Super Yang-Mills theory is a low-energy effective theory of the superstring theory.
In the theory, the torus compactification of the extra six-dimensions with magnetic flux leads to the chiral zero-mode solutions of the matter fields.
In addition, zero-modes are degenerated depending on the size of flux.
Hence, magnetized torus compactification model naturally leads to the chiral matter fields with the generation structure.
In this sense, magnetized torus compactification model is an attractive way realizing flavor structures of fermions.
Then the modular symmetry of the torus may describe the origin of the flavor structures.
Also the orbifoldings of torus have further possibilities.

First we have studied zero-modes on magnetized $T^2/\mathbb{Z}_2$ twisted orbifold.
Zero-modes with flux $M$ behave as the modular forms of weight 1/2 for $\widetilde{\Gamma}_{2M}$ and number of degeneracy (generation) is determined by their flux, $\mathbb{Z}_2$ twist parity and SS phases.
Then we have found various three-generation models as shown in Table \ref{tab:three-generation-models_T2/Z2}.
Using them, we have found 6,460 of quark and lepton flavor models.
On the other hand, all of these flavor models cannot give realistic flavor observables because of the difficulties on realizing hierarchical structures of quarks and leptons.
To avoid the difficulties, we have found four conditions I, II, III and IV.
To realize large up-sector quark mass hierarchies, the condition I requires the directions of up type Higgs VEVs $h^k_u$ leading to rank one mass matrix of up-sector quarks.
To realize large down-sector quark mass and charged lepton mass hierarchies, the condition II requires the directions of down type Higgs VEVs $h^k_d$ leading to rank one mass matrices of down-sector quarks and charged leptons.
To realize small quark mixings, the condition III requires the equality of $u_L^u$ and $u_L^d$ which are unitary matrices diagonalizing rank one mass matrices realized in the conditions I and II.
To realize large lepton mixings, the condition IV requires $h_u^k$ realized in the condition I is also the direction leading to vanishing neutrino Dirac mass matrix.
The flavor models satisfying these conditions have the possibility realizing large mass hierarchies of up-sector quarks, down-sector quarks and charged leptons, and small and large mixing angles of quarks and leptons in the vicinity of $h_{u,d}^k$.
Therefore we have regarded the flavor models satisfying these conditions as the phenomenologically favorable models.
To find such models, we have conducted the zero point analysis in Subsection \ref{subsubsec:Zero_point_analysis}.
Then zero point analysis provides the phenomenologically favorable condition in Eq.~(\ref{eq:consistent_p}) which corresponds to the conditions I, II, III and IV.
That is, whether the conditions I, II, III and IV to obtain phenomenologically favarable models are satisfied or not depends on only zero points of the zero-modes appearing the flavor models.
Consequently 408 of flavor models satisfying such condition have been found.

Additionally, we have classified the flavor models where $h_{u,d}^k$ are along in the modular symmetric directions when the modulus lies on the modular symmetric points.
Actually leading order Higgs $\mu$ term induced by the D-brane instanton effects at $\tau=i,\omega$ and $i\infty$ gives the Higgs VEVs along in the modular symmetric directions.
We have found that 24 of flavor models satisfying phenomenologically favorable condition can have $h_{u,d}^k$ along in $S$-eigenstates at $\tau=i$.
Thus these flavor models have the possibility realizing flavor structures in the vicinity of $S$-eigenstate direction of Higgs VEVs, $h_{u,d}^k$.
Indeed the numerical example in Subsection \ref{subsubsec:numerical_example_T2dZ2} have shown that one of 24 flavor model leads to the realistic flavor observables in the vicinity of $h_{u,d}^k$, that is, the vicinity of $S$-eigenstate directions.

We note that zero point analysis we have conducted can be applied for other orbifold models such as $T^2/\mathbb{Z}_3$, $T^2/\mathbb{Z}_4$ and $T^2/\mathbb{Z}_6$.
This is because zero point analysis depends on only the zero points of the zero-modes.
Thus it would be intersting to investigate the possibilities realizing flavor structures through zero point analysis in other orbifold models.
Also studying Higgs $\mu$ term is an important issue to identify the direction of Higgs VEVs although we have assumed the vicinity of $h_{u,d}^k$.

In addition, we have studied $(T^2_1\times T^2_2)/\mathbb{Z}_2^{\textrm{(per)}}$ and $(T^2_1\times T^2_2)/(\mathbb{Z}_2^{\textrm{(t)}} \times \mathbb{Z}_2^{\textrm{(per)}})$ orbifold models.
Zero-modes behave as the modular forms of weight 1 for $\Gamma'_{2M}$ and number of degeneracy (generation) is determined by their flux, $\mathbb{Z}_2$ twist (permutation) parity and SS phases.
We have classified three-generation models and shown numerical example realizing quark flavors.
Thus these orbifolds also have the possibility realizing flavor structures.

As another way approaching to the flavor structures on the magnetized orbifold models, we have studied Yukawa textures on $T^2/\mathbb{Z}_2$.
At the modular symmetric points $\tau=i$ and $\omega$, the structures of Yukawa matrices are restricted by the residual symmetries.
We have found two structures of Yukawa matrices for $\tau=i$ and three structures of Yukawa matrices for $\tau=\omega$ as shown in Tables \ref{tab:Yukawa_textures_S} and \ref{tab:Yukawa_textures_ST}, respectively.
Using these structures, we have shown Fritzch-Xing and Fritzch mass matrices are realized for $\tau=i$ and $\omega$, respectively.
In numerical studies, we have obtained realistic results of quark flavors for both cases.
Therefore Yukawa textures at the modular symmetric points are an attractive way realizing the flavor structures.


\section{Summary of 4D modular symmetric flavor models}

In Chapter \ref{sec:4D modular symmetric flavor models}, we have discussed 4D modular symmetric flavor models along in Refs.~\cite{Kikuchi:2023cap, Kikuchi:2023jap}.
In the theory, the superpotential is modular invariant.
Then superfields are transformed by the finite modular group and have negative weights.
Therefore fermion mass matrices as well as Yukawa couplings are written in terms of the modular forms of the finite modular group.
Note that we have ambiguilties of coupling constants in mass matrices.

In this paper, we have studied the possibilities realizing quark flavor structures without fine-tuning by coupling constants.
In the vicinity of the modular symmetric points, the modular forms with residual charge $r$ are approximately given by $\varepsilon^r$, which is the deviation of the modulus from the modular symmetric points.
Since the mass matrices are written by the modular forms, they are also written by the powers of $\varepsilon$.
To obtain large mass hierarchies of up-sector quarks, we need the residual $Z_N$ symmetries with $N\geq 6$ which can yield hierarchical value $\varepsilon^{N-1}$.
Instead, the products of the residual symmetries such as $Z_3\times Z_3\times Z_3$ can reproduce such hierarchical values.

Firstly we have studied the quark flavor models with $\Gamma_6$ modular symmetry.
$\Gamma_6$ satisfies $T^6=\mathbb{I}$; therefore residual $Z_6$ symmetry remains at $T$-symmetric point, $\tau=i\infty$.
To reproduce hierarchical values of the modular forms, we have assumed the vicinity of $\tau=i\infty$ where hierarchical values up to $\varepsilon^5$ appear.
To find phenomenologically favorable models, we have considered four types of quark mass matrices.
In four types, diagonal components in mass matrices have suitable powers of $\varepsilon$ to realize quark mass ratios.
We have treated the degree of freedom to choose the powers of $\varepsilon$ in nondiagonal components as model depending values.
For simplicity we have used only $\Gamma_6$ singlets, and restricted the coupling constants to $\pm 1$ to avoid fine-tuning by them.
When both up and down-sector Yukawa couplings have weight 14 where the modular forms of six $\Gamma_6$ singlets exist, we have obtained favorable models at $\tau=3.2i\sim i\infty$.
When Yukwa couplings have weight less than 14, some of components of mass matrices vanish because of the shortage of the modular forms belonging to $\Gamma_6$ singlets.
We have found the favorable models in the case that up-sector Yukawa couplings have weight 8 and down-sector Yukawa couplings have weight 10 as well as weight 12 at $\tau=3.7i\sim i\infty$.
These models can realize the order of the quark mass ratios and the absolute values of the CKM matrix elements without fine-tuning.
Thus, the modular symmetirc flavor models with $\Gamma_6$ symmetry have the possibility realizing quark flavor structures without fine-tuning.

Second we have studied the quark flavor models with $A_4\times A_4\times A_4$ modular symmetry which are described by three moduli $\tau_1$, $\tau_2$ and $\tau_3$.
To make our analysis simple, we have assumed the moduli stabilization $\tau_1=\tau_2=\tau_3\equiv\tau$.
$A_4$ satisfies $(ST)^3=T^3=\mathbb{I}$; therefore residual $Z_3$ symmetries remain at $ST$ and $T$-symmetric points, $\tau=\omega$ and $i\infty$.
To reproduce hierarchical values of the modular forms, we have assumed the vicinity of $\tau=\omega$ or $i\infty$ where hierarchical values up to $\varepsilon^2\times \varepsilon^2\times \varepsilon^2=\varepsilon^6$ appear.
As same as the analysis of the models with $\Gamma_6$ symmetry, we have considered the mass matrices whose diagonal components have suitable powers of $\varepsilon$.
Also we have used only $A_4$ singlets, and restricted the coupling constants to $\pm 1$ to avoid fine-tuning by them.
When both up and down-sector Yukawa couplings have weight 8 where the modular forms of three $A_4$ singlets exist, we have obtained favorable models at $\tau=2.1i\sim i\infty$ and $\omega+0.051i\sim\omega$ as shown in Tables \ref{tab:chi<0.01atinfinite} and \ref{tab:chi<0.01atomega}.
These models can realize the order of the quark mass ratios and the absolute values of the CKM matrix elements without fine-tuning.
However, the first order approximation tells that the structures of the CKM matrix in these favorable models lead to vanishing CP phase.
Numerical example in the vicinity of $\tau\sim i\infty$ actually have shown that the sufficient CP violation needs $|\varepsilon|\sim 0.23$ while hierarchical quark mass ratios need $|\varepsilon|\sim0.15$.
Thus the relation of the trade-off between the CP violation and the quark mass hierarchies exist.

The relation of the trade-off can be improved by introducing non-universal moduli.
So far we have considered the moduli stabilization $\tau_1=\tau_2=\tau_3\equiv\tau$.
Instead we have assumed the non-universal moduli $\tau_1=\tau_2\equiv\tau\neq\tau_3$, $|\tau-\omega|=|\tau_3-\omega|\ll 1$.
Then numerical example have shown that quark mass ratios, the absolute values of the CKM matrix elements and the CP phase can be realized simultaneously at non universal moduli $\tau=\omega+0.055i$, $\tau_3=\omega+0.055e^{2\pi i/5}$.
This result implies that the models with multi moduli parameters are an attractive way explaining the flavor structures including the CP phase.


\section{Summary of constructing Siegel modular forms}

In Chapter \ref{sec:Constructing Siegel modular forms}, we have studied the construction of Siegel modular forms along in Ref.~\cite{Kikuchi:2023dow}.
As mentioned in the end of Chapter \ref{sec:4D modular symmetric flavor models}, multi moduli parameters have the possibility realizing the CP phase and large quark mass hierarchies simultaneously.
Siegel modular forms have several moduli and therefore they are promising approach to the quark flavor structures.

6D torus $T^6$ has $Sp(6,\mathbb{Z})$ modular symmetry as the geometrical symmetry and zero-modes on magnetized $T^6$ behave as the Siegel modular forms of weight 1/2 for the subgroup of $Sp(6,\mathbb{Z})$.
Taking $\vec{z}=0$, they become exactly the Siegel modular forms.
On the other hand, $T$-symmetries on zero-modes are partially violated depending on the structure of $N$ matrix.
Moreover the consistency condition for $S$-transformation requires the moduli commuting to $N$ matrix, $N\Omega=\Omega N$, which is equivalent to the $F$-term condition for symmetric $\Omega$ and $N$.
In this paper we have focused on $N$ matrix possessing three different eigenvalues, where three $T$-symmetries denoted as $T_I$, $T_{II}$ and $T_{III}$ remain.
Then we have shown five examples of the Siegel modular forms of weight 1/2 in Section \ref{subsec:Examples of the Siegel modular forms}.

As a constructing example, we have studied the Siegel modular forms transformed by $\widetilde{\Delta}(96)\simeq \Delta(48)\rtimes Z_8$.
From zero-modes on $T^6$ with $N$ matrix in Eq.~(\ref{eq:N_matrix}), we have obtained the Siegel modular forms of weight 1/2 transformed by unitary matrices in Eq.~(\ref{eq:rhos}).
These unitary matrices form a group $\widetilde{\Delta}(96)\simeq \Delta(48)\rtimes Z_8$.
Thus $\hat{\psi}^{\vec{j}}_{\bm{3}}(\Omega) $ in Eq.~(\ref{eq:modular_form_1/2_Delta96}) are a triplet of this group.
Higher weight Siegel modular forms have been constructed by tensor products of them.
The dimension of the Siegel modular forms of weight $k$ is given by $_{2k+2}C_2$.

Using the Siegel modular forms of $\widetilde{\Delta}(96)$, we have given the numerical studies for quark flavor structures.
To make our analysis simple, we have used singlet Siegel modular forms.
To reproduce quark mass hierarchies, we have assumed the vicinity of the cusp $\Omega_{T_I}$.
When the moduli lie on the vicinity of $\Omega_{T_I}$, the mass matrix elements become hierarchical depending on its $T_I$ charges.
In the numerical studies, we have chosen $(\omega_1,\omega_2,\omega_3)=(1.3i,0.6i+0.86,0)$.
To obtain non-vanishing CP phase, we have deviated $\omega_2$ from the imaginary axis.
In addition we have deviated the coupling constants by ${\cal O}(1)$.
Then we have obtained realistic quark flavor observables including sufficient amount of CP phase as shown in Table \ref{tab:fitting}.

In the numerical example, one of the moduli parameter, $\omega_1$, works on generating large quark mass hierarchies but cannot contribute to CP violation.
This is because the phase factors of $q_1=e^{\pi i\omega_1/4}$ in mass matrices are completely canceled by the basis transformation.
Meanwhile the phase factors of $q_2=e^{\pi i\omega_2/4}$ in mass matrices survive after any basis transformation.
That is, $\omega_2$ can contribute to the CP violation.
In our numerical example, the deviation of $\omega_1$ from the cusp ($T_I$-symmetric point) contribute to the large mass hierarchies while the deviation of $\omega_2$ from the imaginary axis (CP symmetric points) contribute to the CP violation.
Thus the Siegel modular forms with multi moduli parameters have the possibility realizing flavor structures including CP phases without fine-tuning.

We note that the deviation of the moduli from the modular or CP symmetric points are important in our model.
The moduli stabilization leading to such deviations is a key issue\footnote{See for moduli stabilization in moduli flavor models Refs.~\cite{Kobayashi:2019xvz,Ishiguro:2020tmo,Abe:2020vmv,Novichkov:2022wvg,Ishiguro:2022pde, Knapp-Perez:2023nty, King:2023snq, Kobayashi:2023spx}. See also Ref.~\cite{Kikuchi:2023uqo}}.


\section{Discussion}

Here we discuss the further possibilities realizing the flavor structures.
Let us start from the comparison of magnetized orbifold models and 4D modular symmetric flavor models.
As mentioned in Section \ref{sec:framework}, 4D modular symmetric flavor models may be derived from the higher dimensional theories such as the magnetized orbifold models as the low-energy effective theory.
Actually, the modular forms play important and similar roles for both models; Yukawa couplings are written in terms of the modular forms.
Hence, they would be related strongly through the modular forms.
In the following, we show some facts on both models regarding to the modulus, Higgs fields and the origins of large mass hierarchies and CP violation in numerical examples.
\begin{itemize}
\item Magnetized orbifold models (quark and lepton flavor model on $T^2/\mathbb{Z}_2$ twisted orbifold in Subsection \ref{subsubsec:numerical_example_T2dZ2})
\begin{itemize}
\item The modulus $\tau$ is assumed to lie on $S$-symmetric point.
\item There are six pairs of Higgs fields.
Higgs VEVs are aligned in the lightest mass direction but we have used its direction as free parameters.
\item Yukawa couplings have no ambiguilties except for the overall factors since they are calculated by overlap integral of zero-mode wave functions.
\item Large mass hierarchies originate from the deviation of Higgs VEVs from $S$-eigenstate directions.
\item CP violation is expected to be induced from the complex direction of Higgs VEVs or the deviation of the modulus from CP symmetric points although we have not studied it in this paper.
\item The deviation of Higgs VEVs from $S$-symmetric point is important in this model.
Higgs $\mu$ term leading to such deviation is a key issue.
The moduli stabilization is also an issue.
\end{itemize}

\item 4D modular symmetric flavor models (quark flavor model with $A_4\times A_4\times A_4$ modular symmetry in Subsection \ref{subsubsec:non-universal})
\begin{itemize}
\item The moduli $\tau_1$, $\tau_2$ and $\tau_3$ are assumed to be in the vicinity of $ST$ or $T$-symmetric points.
\item There are a pair of Higgs fields.
\item Yukawa couplings have ambiguilties of coupling constants.
We have treated them as ${\cal O}(1)$ parameters since we expect hierarchical structures do not originate from them.
\item Large mass hierarchies originate from the deviation of the moduli from the modular symmetric points.
\item CP violation originates from non-universal moduli on CP violating point.
\item The deviation of the moduli from the modular or CP symmetric points are important in the model.
The moduli stabilization leading to such deviations is a key issue.
Also the origin of the values of the coupling constants should be identified.
\end{itemize}
\end{itemize}

The results on 4D modular symmetric flavor models imply that large mass hierarchies and CP phase require multi moduli parameters.
As we have seen in Chapter \ref{sec:Constructing Siegel modular forms}, the magnetized $T^6$ models lead to zero-modes which behave as the Siegel modular forms.
Moreover, it can be expected that three-generation models with a pair of Higgs fields are obtained on the magnetized $T^6$ models.
In such models, we do not need to identify the lightest direction of Higgs fields and only the moduli stabilization is an unsolved issue.
Then large mass hierarchies and CP phase would be obtained by the deviation of the moduli form the (modular or CP) symmetric points as same as 4D modular symmetric flavor models.
Note that ambiguilties of coupling constants in mass matrices do not exist on the magnetized orbifold models except for the overall factors.
Therefore, the magnetized $T^6$ models with a pair of Higgs fields have the possibilities realizing flavor structures by a few moduli parameters without fine-tuning by Higgs VEVs and coupling constants.
Also the magnetized $T^6$ models with multi Higgs fields are interesting.
They may have the leading order Higgs $\mu$ term being able to fix the lightest mass direction uniquely although the magnetized $T^2/\mathbb{Z}_2$ orbifold model does not.
Then the direction of Higgs VEVs depends on the values of the moduli.
Hence the moduli are the unique orign of the flavor structures.
Thus the magnetized $T^6$ models are attractive and realization of the flavor structures is challenging issue.


\appendix


\chapter{Wilson lines and Scherk-Schwarz phases}
\label{app:W-SS}

In this appendix, we show the equivalence of Wilson lines and SS phases on the magnetized $T^2$ model \cite{Abe:2013bca}.
We start from the zero-modes on $T^2$, $\psi^{(j+\alpha_1,\alpha_2),M}(z,\tau)$, which satisfy the boundary conditions in Eq.~(\ref{eq:BC_psi}).
First we consider the following gauge transformation for the zero-modes,
\begin{align}
\psi^{(j+\alpha_1,\alpha_2),M}(z,\tau)
\to \widetilde{\psi}^{(j+\alpha_1,\alpha_2),M}(z,\tau)
= e^{-i\textrm{Re}\bar{\beta}z} \psi^{(j+\alpha_1,\alpha_2),M}(z,\tau), \quad \beta \in \mathbb{C}.
\end{align}
Then the vector potential $A(z)$ is transformed as
\begin{align}
\widetilde{A}(z) = A(z) - d[\textrm{Re}\bar{\beta}z]
= \frac{\pi M}{\textrm{Im}\tau} \textrm{Im} \left(\left(\bar{z}-\frac{i\textrm{Im}\tau}{\pi M}\bar{\beta}\right)dz\right)
\equiv \frac{\pi M}{\textrm{Im}\tau} \textrm{Im} \left(\overline{\left(z+\widetilde{a}_w\right)}dz\right).
\end{align}
Comparing the vector potential in Eq.~(\ref{eq:vector_potential_Wilson}), we can regard $\widetilde{a}_w$ as Wilson line.
Thus $\widetilde{A}(z)$ obeys the boundary conditions,
\begin{align}
&\widetilde{A}(z+1) = \widetilde{A}(z) + d\widetilde{\chi}_{1}(z), \\
&\widetilde{A}(z+\tau) = \widetilde{A}(z) + d\widetilde{\chi}_{2}(z),
\end{align}
with
\begin{align}
&\widetilde{\chi}_{1}(z) = \frac{\pi M}{\textrm{Im}\tau}\textrm{Im}\left(z+\widetilde{a}_w\right) = \chi_1(z) + \textrm{Re}\beta, \\
&\widetilde{\chi}_{2}(z) = \frac{\pi M}{\textrm{Im}\tau}\textrm{Im}\bar{\tau}\left(z+\widetilde{a}_w\right) = \chi_2(z) + \textrm{Re}\bar{\tau}\beta.
\end{align}
In the same way, we obtain the boundary conditions of $\widetilde{\psi}(z,\tau)$ as
\begin{align}
&\widetilde{\psi}^{(j+\alpha_1,\alpha_2),M}(z+1,\tau) = e^{2\pi i\alpha_1-2i\textrm{Re}\beta} e^{i\widetilde{\chi}_1(z)} \widetilde{\psi}^{(j+\alpha_1,\alpha_2),M}, \\
&\widetilde{\psi}^{(j+\alpha_1,\alpha_2),M}(z+\tau,\tau) = e^{2\pi i\alpha_2-2i\textrm{Re}\bar{\tau}\beta} e^{i\widetilde{\chi}_2(z)} \widetilde{\psi}^{(j+\alpha_1,\alpha_2),M}.
\end{align}
Here let us choose
\begin{align}
\beta = -\pi i \frac{\alpha_1\tau-\alpha_2}{\textrm{Im}\tau}.
\end{align}
Then we obtain the gauge transformed zero-modes,
\begin{align}
\widetilde{\psi}^{(j+\alpha_1,\alpha_2),M}(z,\tau) = e^{\pi i \frac{\textrm{Im}(\alpha_1\bar{\tau}-\alpha_2)z}{\textrm{Im}\tau}} \psi^{(j+\alpha_1,\alpha_2),M}(z,\tau),
\end{align} 
which satisfy the boundary conditions,
\begin{align}
&\widetilde{\psi}^{(j+\alpha_1,\alpha_2),M}(z+1,\tau) = e^{i\widetilde{\chi}_1(z)} \widetilde{\psi}^{(j+\alpha_1,\alpha_2),M}, \\
&\widetilde{\psi}^{(j+\alpha_1,\alpha_2),M}(z+\tau,\tau) = e^{i\widetilde{\chi}_2(z)} \widetilde{\psi}^{(j+\alpha_1,\alpha_2),M},
\end{align}
with
\begin{align}
\widetilde{\chi}_{1}(z) = \frac{\pi M}{\textrm{Im}\tau}\textrm{Im}\left(z+\frac{\alpha_1\tau-\alpha_2}{M}\right), \quad
\widetilde{\chi}_{2}(z) = \frac{\pi M}{\textrm{Im}\tau}\textrm{Im}\bar{\tau}\left(z+\frac{\alpha_1\tau-\alpha_2}{M}\right).
\end{align}
This implies that the gauge transformed zero-modes $\widetilde{\psi}(z,\tau)$ have Wilson line $\zeta=(\alpha_1\tau-\alpha_2)/M$ but vanishing SS phases $(\widetilde{\alpha}_1,\widetilde{\alpha}_2)=(0,0)$.
Thus SS phases can be absorbed into Wilson line through the gauge transformation.


\chapter{Flavor models on $T^2/\mathbb{Z}_2$}
\label{app:Flavor models}


\section{Favorable models}
\label{app:Favorable_models}

Here, we show quark and lepton flavor models satisfying the phenomenologically favorable conditions for quark and lepton flavors in Table \ref{tab:classification}.
\begin{table}[H]
  \caption{Quark and lepton flavor models satisfying the phenomenologically favorable conditions for quark and lepton flavors.
  The first to eighth columns show the flux, $\mathbb{Z}_2$ parity (even, odd = 0, 1) and SS phases $(\alpha_1,\alpha_2)$ of the zero-modes of the fields.
  $g_H$ denotes number of Higgs fields and $p$ denotes the fixed points satisfying Eq.~(\ref{eq:consistent_p}).}
  \label{tab:classification}
\begin{align}
\begin{array}{c|c|c|c|c|c|c|c|c|c} \hline
    B_Q & B_{u_R} & B_{d_R} & B_L & B_{\nu_R} & B_{e_R} & B_{H_u} & B_{H_d} & g_H & p \\ \hline \hline
    5,0,0,0 & 6,0,\frac{1}{2},0 & 6,0,0,\frac{1}{2} & 6,0,\frac{1}{2},\frac{1}{2} & 5,0,0,\frac{1}{2} & 5,0,\frac{1}{2},0 & 11,0,\frac{1}{2},0 & 11,0,0,\frac{1}{2} & 6 & \frac{\tau}{2} \\
5,0,0,0 & 6,0,\frac{1}{2},0 & 6,1,\frac{1}{2},\frac{1}{2} & 6,1,0,\frac{1}{2} & 5,1,\frac{1}{2},\frac{1}{2} & 5,0,\frac{1}{2},0 & 11,0,\frac{1}{2},0 & 11,1,\frac{1}{2},\frac{1}{2} & 6 & 0 \\
5,0,0,0 & 6,0,\frac{1}{2},0 & 7,1,\frac{1}{2},0 & 6,0,\frac{1}{2},\frac{1}{2} & 5,0,0,\frac{1}{2} & 6,1,0,\frac{1}{2} & 11,0,\frac{1}{2},0 & 12,1,\frac{1}{2},0 & 6 & \frac{\tau}{2} \\
5,0,0,0 & 6,0,\frac{1}{2},0 & 7,1,\frac{1}{2},0 & 6,1,0,\frac{1}{2} & 5,1,\frac{1}{2},\frac{1}{2} & 6,0,\frac{1}{2},\frac{1}{2} & 11,0,\frac{1}{2},0 & 12,1,\frac{1}{2},0 & 6 & 0 \\
5,0,0,0 & 6,0,0,\frac{1}{2} & 6,0,\frac{1}{2},0 & 6,0,\frac{1}{2},\frac{1}{2} & 5,0,\frac{1}{2},0 & 5,0,0,\frac{1}{2} & 11,0,0,\frac{1}{2} & 11,0,\frac{1}{2},0 & 6 & \frac{1}{2} \\
5,0,0,0 & 6,0,0,\frac{1}{2} & 6,1,\frac{1}{2},\frac{1}{2} & 6,1,\frac{1}{2},0 & 5,1,\frac{1}{2},\frac{1}{2} & 5,0,0,\frac{1}{2} & 11,0,0,\frac{1}{2} & 11,1,\frac{1}{2},\frac{1}{2} & 6 & 0 \\
5,0,0,0 & 6,0,0,\frac{1}{2} & 7,1,0,\frac{1}{2} & 6,0,\frac{1}{2},\frac{1}{2} & 5,0,\frac{1}{2},0 & 6,1,\frac{1}{2},0 & 11,0,0,\frac{1}{2} & 12,1,0,\frac{1}{2} & 6 & \frac{1}{2} \\
\hline
  \end{array} \notag
\end{align}
\end{table}
\begin{table}[H]
\begin{align}
  \begin{array}{c|c|c|c|c|c|c|c|c|c} \hline
    B_Q & B_{u_R} & B_{d_R} & B_L & B_{\nu_R} & B_{e_R} & B_{H_u} & B_{H_d} & g_H & p \\ \hline \hline
5,0,0,0 & 7,0,\frac{1}{2},\frac{1}{2} & 7,1,\frac{1}{2},0 & 7,1,0,0 & 5,1,\frac{1}{2},\frac{1}{2} & 5,0,\frac{1}{2},0 & 12,0,\frac{1}{2},\frac{1}{2} & 12,1,\frac{1}{2},0 & 6 & 0 \\
5,0,0,0 & 7,0,\frac{1}{2},\frac{1}{2} & 7,1,0,\frac{1}{2} & 6,1,\frac{1}{2},0 & 6,1,0,\frac{1}{2} & 6,0,\frac{1}{2},\frac{1}{2} & 12,0,\frac{1}{2},\frac{1}{2} & 12,1,0,\frac{1}{2} & 6 & 0 \\
5,0,0,0 & 7,0,\frac{1}{2},\frac{1}{2} & 7,1,0,\frac{1}{2} & 7,1,0,0 & 5,1,\frac{1}{2},\frac{1}{2} & 5,0,0,\frac{1}{2} & 12,0,\frac{1}{2},\frac{1}{2} & 12,1,0,\frac{1}{2} & 6 & 0 \\
5,0,0,0 & 7,1,\frac{1}{2},0 & 6,0,\frac{1}{2},0 & 6,0,\frac{1}{2},\frac{1}{2} & 6,1,0,\frac{1}{2} & 5,0,0,\frac{1}{2} & 12,1,\frac{1}{2},0 & 11,0,\frac{1}{2},0 & 6 & \frac{1}{2} \\
5,0,0,0 & 7,1,\frac{1}{2},0 & 6,0,\frac{1}{2},0 & 6,1,0,\frac{1}{2} & 6,0,\frac{1}{2},\frac{1}{2} & 5,1,\frac{1}{2},\frac{1}{2} & 12,1,\frac{1}{2},0 & 11,0,\frac{1}{2},0 & 6 & \frac{1}{2} \\
5,0,0,0 & 7,1,\frac{1}{2},0 & 7,0,\frac{1}{2},\frac{1}{2} & 6,1,0,\frac{1}{2} & 6,0,\frac{1}{2},\frac{1}{2} & 6,1,\frac{1}{2},0 & 12,1,\frac{1}{2},0 & 12,0,\frac{1}{2},\frac{1}{2} & 6 & \frac{1}{2} \\
5,0,0,0 & 7,1,\frac{1}{2},0 & 7,0,\frac{1}{2},\frac{1}{2} & 7,1,0,0 & 5,0,\frac{1}{2},0 & 5,1,\frac{1}{2},\frac{1}{2} & 12,1,\frac{1}{2},0 & 12,0,\frac{1}{2},\frac{1}{2} & 6 & \frac{1}{2} \\
\hline
  \end{array} \notag
\end{align}
\end{table}
\begin{table}[H]
\begin{align}
  \begin{array}{c|c|c|c|c|c|c|c|c|c} \hline
    B_Q & B_{u_R} & B_{d_R} & B_L & B_{\nu_R} & B_{e_R} & B_{H_u} & B_{H_d} & g_H & p \\ \hline \hline
5,0,\frac{1}{2},0 & 7,1,0,\frac{1}{2} & 7,1,0,0 & 6,0,0,\frac{1}{2} & 6,1,\frac{1}{2},0 & 6,1,\frac{1}{2},\frac{1}{2} & 12,1,\frac{1}{2},\frac{1}{2} & 12,1,\frac{1}{2},0 & 6 & \frac{\tau}{2} \\
5,0,\frac{1}{2},0 & 7,1,0,\frac{1}{2} & 7,1,0,0 & 7,1,\frac{1}{2},0 & 5,0,0,\frac{1}{2} & 5,0,0,0 & 12,1,\frac{1}{2},\frac{1}{2} & 12,1,\frac{1}{2},0 & 6 & \frac{\tau}{2} \\
5,0,\frac{1}{2},0 & 7,1,0,0 & 6,0,\frac{1}{2},0 & 6,0,0,\frac{1}{2} & 6,1,\frac{1}{2},\frac{1}{2} & 5,0,0,\frac{1}{2} & 12,1,\frac{1}{2},0 & 11,0,0,0 & 6 & \frac{1+\tau}{2} \\
5,0,\frac{1}{2},0 & 7,1,0,0 & 6,0,\frac{1}{2},0 & 6,1,\frac{1}{2},\frac{1}{2} & 6,0,0,\frac{1}{2} & 5,1,\frac{1}{2},\frac{1}{2} & 12,1,\frac{1}{2},0 & 11,0,0,0 & 6 & \frac{1+\tau}{2} \\
5,0,\frac{1}{2},0 & 7,1,0,0 & 7,0,\frac{1}{2},\frac{1}{2} & 6,1,\frac{1}{2},\frac{1}{2} & 6,0,0,\frac{1}{2} & 6,1,\frac{1}{2},0 & 12,1,\frac{1}{2},0 & 12,0,0,\frac{1}{2} & 6 & \frac{1+\tau}{2} \\
5,0,\frac{1}{2},0 & 7,1,0,0 & 7,0,\frac{1}{2},\frac{1}{2} & 7,1,\frac{1}{2},0 & 5,0,0,0 & 5,1,\frac{1}{2},\frac{1}{2} & 12,1,\frac{1}{2},0 & 12,0,0,\frac{1}{2} & 6 & \frac{1+\tau}{2} \\
5,0,\frac{1}{2},0 & 7,1,0,0 & 7,1,0,\frac{1}{2} & 6,0,0,\frac{1}{2} & 6,1,\frac{1}{2},\frac{1}{2} & 6,1,\frac{1}{2},0 & 12,1,\frac{1}{2},0 & 12,1,\frac{1}{2},\frac{1}{2} & 6 & \frac{1+\tau}{2} \\
\hline
  \end{array} \notag
\end{align}
\end{table}
\begin{table}[H]
\begin{align}
  \begin{array}{c|c|c|c|c|c|c|c|c|c} \hline
    B_Q & B_{u_R} & B_{d_R} & B_L & B_{\nu_R} & B_{e_R} & B_{H_u} & B_{H_d} & g_H & p \\ \hline \hline
5,0,0,\frac{1}{2} & 7,1,0,0 & 7,1,\frac{1}{2},0 & 7,1,0,\frac{1}{2} & 5,0,0,0 & 5,0,\frac{1}{2},0 & 12,1,0,\frac{1}{2} & 12,1,\frac{1}{2},\frac{1}{2} & 6 & \frac{1+\tau}{2} \\
5,1,\frac{1}{2},\frac{1}{2} & 6,1,\frac{1}{2},0 & 6,1,0,\frac{1}{2} & 6,0,\frac{1}{2},\frac{1}{2} & 5,0,\frac{1}{2},0 & 5,0,0,\frac{1}{2} & 11,0,0,\frac{1}{2} & 11,0,\frac{1}{2},0 & 6 & \frac{1}{2} \\
5,1,\frac{1}{2},\frac{1}{2} & 6,1,\frac{1}{2},0 & 6,1,\frac{1}{2},\frac{1}{2} & 6,0,0,\frac{1}{2} & 5,0,0,0 & 5,0,0,\frac{1}{2} & 11,0,0,\frac{1}{2} & 11,0,0,0 & 6 & \frac{1+\tau}{2} \\
5,1,\frac{1}{2},\frac{1}{2} & 6,1,\frac{1}{2},0 & 7,1,0,\frac{1}{2} & 6,0,0,\frac{1}{2} & 5,0,0,0 & 6,0,\frac{1}{2},\frac{1}{2} & 11,0,0,\frac{1}{2} & 12,0,\frac{1}{2},0 & 6 & \frac{1+\tau}{2} \\
5,1,\frac{1}{2},\frac{1}{2} & 6,1,\frac{1}{2},0 & 7,1,0,\frac{1}{2} & 6,0,\frac{1}{2},\frac{1}{2} & 5,0,\frac{1}{2},0 & 6,0,0,\frac{1}{2} & 11,0,0,\frac{1}{2} & 12,0,\frac{1}{2},0 & 6 & \frac{1}{2} \\
5,1,\frac{1}{2},\frac{1}{2} & 6,1,0,\frac{1}{2} & 6,1,\frac{1}{2},0 & 6,0,\frac{1}{2},\frac{1}{2} & 5,0,0,\frac{1}{2} & 5,0,\frac{1}{2},0 & 11,0,\frac{1}{2},0 & 11,0,0,\frac{1}{2} & 6 & \frac{\tau}{2} \\
5,1,\frac{1}{2},\frac{1}{2} & 6,1,0,\frac{1}{2} & 6,1,\frac{1}{2},\frac{1}{2} & 6,0,\frac{1}{2},0 & 5,0,0,0 & 5,0,\frac{1}{2},0 & 11,0,\frac{1}{2},0 & 11,0,0,0 & 6 & \frac{1+\tau}{2} \\
\hline
  \end{array} \notag
\end{align}
\end{table}
\begin{table}[H]
\begin{align}
  \begin{array}{c|c|c|c|c|c|c|c|c|c} \hline
    B_Q & B_{u_R} & B_{d_R} & B_L & B_{\nu_R} & B_{e_R} & B_{H_u} & B_{H_d} & g_H & p \\ \hline \hline
6,0,\frac{1}{2},0 & 5,0,\frac{1}{2},0 & 7,1,\frac{1}{2},0 & 6,1,\frac{1}{2},\frac{1}{2} & 5,1,\frac{1}{2},\frac{1}{2} & 7,0,\frac{1}{2},\frac{1}{2} & 11,0,0,0 & 13,1,0,0 & 6 & 0 \\
6,0,\frac{1}{2},0 & 6,0,0,\frac{1}{2} & 6,1,0,\frac{1}{2} & 6,1,\frac{1}{2},0 & 6,1,0,\frac{1}{2} & 6,0,0,\frac{1}{2} & 12,0,\frac{1}{2},\frac{1}{2} & 12,1,\frac{1}{2},\frac{1}{2} & 6 & 0 \\
6,0,\frac{1}{2},0 & 6,0,0,\frac{1}{2} & 6,1,0,\frac{1}{2} & 6,1,0,\frac{1}{2} & 6,1,\frac{1}{2},0 & 6,0,\frac{1}{2},0 & 12,0,\frac{1}{2},\frac{1}{2} & 12,1,\frac{1}{2},\frac{1}{2} & 6 & 0 \\
6,0,\frac{1}{2},0 & 6,0,0,\frac{1}{2} & 6,1,\frac{1}{2},\frac{1}{2} & 6,1,\frac{1}{2},0 & 6,1,0,\frac{1}{2} & 6,0,\frac{1}{2},\frac{1}{2} & 12,0,\frac{1}{2},\frac{1}{2} & 12,1,0,\frac{1}{2} & 6 & 0 \\
6,0,\frac{1}{2},0 & 6,0,0,\frac{1}{2} & 6,1,\frac{1}{2},\frac{1}{2} & 7,1,0,0 & 5,1,\frac{1}{2},\frac{1}{2} & 5,0,0,\frac{1}{2} & 12,0,\frac{1}{2},\frac{1}{2} & 12,1,0,\frac{1}{2} & 6 & 0 \\
6,0,\frac{1}{2},0 & 6,0,0,\frac{1}{2} & 7,1,0,0 & 6,1,0,\frac{1}{2} & 6,1,\frac{1}{2},0 & 7,0,\frac{1}{2},\frac{1}{2} & 12,0,\frac{1}{2},\frac{1}{2} & 13,1,\frac{1}{2},0 & 6 & 0 \\
6,0,\frac{1}{2},0 & 6,0,0,\frac{1}{2} & 7,1,0,0 & 7,1,0,0 & 5,1,\frac{1}{2},\frac{1}{2} & 6,0,\frac{1}{2},0 & 12,0,\frac{1}{2},\frac{1}{2} & 13,1,\frac{1}{2},0 & 6 & 0 \\
\hline
  \end{array} \notag
\end{align}
\end{table}
\begin{table}[H]
\begin{align}
  \begin{array}{c|c|c|c|c|c|c|c|c|c} \hline
    B_Q & B_{u_R} & B_{d_R} & B_L & B_{\nu_R} & B_{e_R} & B_{H_u} & B_{H_d} & g_H & p \\ \hline \hline
6,0,0,\frac{1}{2} & 6,0,\frac{1}{2},0 & 6,1,\frac{1}{2},0 & 6,1,\frac{1}{2},0 & 6,1,0,\frac{1}{2} & 6,0,0,\frac{1}{2} & 12,0,\frac{1}{2},\frac{1}{2} & 12,1,\frac{1}{2},\frac{1}{2} & 6 & 0 \\
6,0,0,\frac{1}{2} & 6,0,\frac{1}{2},0 & 6,1,\frac{1}{2},0 & 6,1,0,\frac{1}{2} & 6,1,\frac{1}{2},0 & 6,0,\frac{1}{2},0 & 12,0,\frac{1}{2},\frac{1}{2} & 12,1,\frac{1}{2},\frac{1}{2} & 6 & 0 \\
6,0,0,\frac{1}{2} & 6,0,\frac{1}{2},0 & 6,1,\frac{1}{2},\frac{1}{2} & 6,1,0,\frac{1}{2} & 6,1,\frac{1}{2},0 & 6,0,\frac{1}{2},\frac{1}{2} & 12,0,\frac{1}{2},\frac{1}{2} & 12,1,\frac{1}{2},0 & 6 & 0 \\
6,0,0,\frac{1}{2} & 6,0,\frac{1}{2},0 & 6,1,\frac{1}{2},\frac{1}{2} & 7,1,0,0 & 5,1,\frac{1}{2},\frac{1}{2} & 5,0,\frac{1}{2},0 & 12,0,\frac{1}{2},\frac{1}{2} & 12,1,\frac{1}{2},0 & 6 & 0 \\
6,0,0,\frac{1}{2} & 6,0,\frac{1}{2},0 & 7,1,0,0 & 6,1,\frac{1}{2},0 & 6,1,0,\frac{1}{2} & 7,0,\frac{1}{2},\frac{1}{2} & 12,0,\frac{1}{2},\frac{1}{2} & 13,1,0,\frac{1}{2} & 6 & 0 \\
6,0,0,\frac{1}{2} & 6,0,\frac{1}{2},0 & 7,1,0,0 & 7,1,0,0 & 5,1,\frac{1}{2},\frac{1}{2} & 6,0,0,\frac{1}{2} & 12,0,\frac{1}{2},\frac{1}{2} & 13,1,0,\frac{1}{2} & 6 & 0 \\
6,0,0,\frac{1}{2} & 6,0,\frac{1}{2},\frac{1}{2} & 6,1,\frac{1}{2},0 & 6,1,0,\frac{1}{2} & 6,1,\frac{1}{2},\frac{1}{2} & 6,0,\frac{1}{2},0 & 12,0,\frac{1}{2},0 & 12,1,\frac{1}{2},\frac{1}{2} & 6 & 0 \\
\hline
  \end{array} \notag
\end{align}
\end{table}
\begin{table}[H]
\begin{align}
  \begin{array}{c|c|c|c|c|c|c|c|c|c} \hline
    B_Q & B_{u_R} & B_{d_R} & B_L & B_{\nu_R} & B_{e_R} & B_{H_u} & B_{H_d} & g_H & p \\ \hline \hline
6,0,\frac{1}{2},\frac{1}{2} & 6,0,\frac{1}{2},0 & 6,1,0,\frac{1}{2} & 6,1,\frac{1}{2},\frac{1}{2} & 6,1,\frac{1}{2},0 & 6,0,0,\frac{1}{2} & 12,0,0,\frac{1}{2} & 12,1,\frac{1}{2},0 & 6 & 0 \\
6,0,\frac{1}{2},\frac{1}{2} & 6,0,\frac{1}{2},0 & 6,1,0,\frac{1}{2} & 7,1,\frac{1}{2},0 & 5,1,\frac{1}{2},\frac{1}{2} & 5,0,0,0 & 12,0,0,\frac{1}{2} & 12,1,\frac{1}{2},0 & 6 & 0 \\
6,0,\frac{1}{2},\frac{1}{2} & 6,0,\frac{1}{2},0 & 7,1,\frac{1}{2},0 & 6,1,\frac{1}{2},0 & 6,1,\frac{1}{2},\frac{1}{2} & 7,0,\frac{1}{2},\frac{1}{2} & 12,0,0,\frac{1}{2} & 13,1,0,\frac{1}{2} & 6 & 0 \\
6,0,\frac{1}{2},\frac{1}{2} & 6,0,\frac{1}{2},0 & 7,1,\frac{1}{2},0 & 7,1,\frac{1}{2},0 & 5,1,\frac{1}{2},\frac{1}{2} & 6,0,\frac{1}{2},\frac{1}{2} & 12,0,0,\frac{1}{2} & 13,1,0,\frac{1}{2} & 6 & 0 \\
6,0,\frac{1}{2},\frac{1}{2} & 6,0,0,\frac{1}{2} & 6,1,\frac{1}{2},0 & 6,1,\frac{1}{2},\frac{1}{2} & 6,1,0,\frac{1}{2} & 6,0,\frac{1}{2},0 & 12,0,\frac{1}{2},0 & 12,1,0,\frac{1}{2} & 6 & 0 \\
6,0,\frac{1}{2},\frac{1}{2} & 6,0,0,\frac{1}{2} & 6,1,\frac{1}{2},0 & 7,1,0,\frac{1}{2} & 5,1,\frac{1}{2},\frac{1}{2} & 5,0,0,0 & 12,0,\frac{1}{2},0 & 12,1,0,\frac{1}{2} & 6 & 0 \\
6,0,\frac{1}{2},\frac{1}{2} & 6,0,0,\frac{1}{2} & 6,1,0,\frac{1}{2} & 6,1,0,\frac{1}{2} & 6,1,\frac{1}{2},\frac{1}{2} & 6,0,\frac{1}{2},\frac{1}{2} & 12,0,\frac{1}{2},0 & 12,1,\frac{1}{2},0 & 6 & 0 \\
\hline
  \end{array} \notag
\end{align}
\end{table}
\begin{table}[H]
\begin{align}
  \begin{array}{c|c|c|c|c|c|c|c|c|c} \hline
    B_Q & B_{u_R} & B_{d_R} & B_L & B_{\nu_R} & B_{e_R} & B_{H_u} & B_{H_d} & g_H & p \\ \hline \hline
6,1,\frac{1}{2},0 & 6,0,0,\frac{1}{2} & 7,1,0,\frac{1}{2} & 6,1,0,\frac{1}{2} & 6,0,\frac{1}{2},0 & 7,1,\frac{1}{2},0 & 12,1,\frac{1}{2},\frac{1}{2} & 13,0,\frac{1}{2},\frac{1}{2} & 6 & \frac{1}{2} \\
6,1,\frac{1}{2},0 & 6,0,0,\frac{1}{2} & 7,1,0,\frac{1}{2} & 7,1,0,\frac{1}{2} & 5,0,\frac{1}{2},0 & 6,1,\frac{1}{2},0 & 12,1,\frac{1}{2},\frac{1}{2} & 13,0,\frac{1}{2},\frac{1}{2} & 6 & \frac{1}{2} \\
6,1,\frac{1}{2},0 & 6,0,\frac{1}{2},\frac{1}{2} & 6,0,0,\frac{1}{2} & 6,0,\frac{1}{2},0 & 6,1,\frac{1}{2},\frac{1}{2} & 6,1,0,\frac{1}{2} & 12,1,0,\frac{1}{2} & 12,1,\frac{1}{2},\frac{1}{2} & 6 & \frac{1+\tau}{2} \\
6,1,\frac{1}{2},0 & 6,0,\frac{1}{2},\frac{1}{2} & 6,0,0,\frac{1}{2} & 7,1,0,\frac{1}{2} & 5,0,0,0 & 5,0,\frac{1}{2},0 & 12,1,0,\frac{1}{2} & 12,1,\frac{1}{2},\frac{1}{2} & 6 & \frac{1+\tau}{2} \\
6,1,\frac{1}{2},0 & 6,0,\frac{1}{2},\frac{1}{2} & 6,1,\frac{1}{2},\frac{1}{2} & 6,0,\frac{1}{2},0 & 6,1,\frac{1}{2},\frac{1}{2} & 6,0,\frac{1}{2},\frac{1}{2} & 12,1,0,\frac{1}{2} & 12,0,0,\frac{1}{2} & 6 & \frac{1+\tau}{2} \\
6,1,\frac{1}{2},0 & 6,0,\frac{1}{2},\frac{1}{2} & 6,1,\frac{1}{2},\frac{1}{2} & 6,1,\frac{1}{2},\frac{1}{2} & 6,0,\frac{1}{2},0 & 6,1,\frac{1}{2},0 & 12,1,0,\frac{1}{2} & 12,0,0,\frac{1}{2} & 6 & \frac{1+\tau}{2} \\
6,1,\frac{1}{2},0 & 6,0,\frac{1}{2},\frac{1}{2} & 7,1,0,\frac{1}{2} & 6,1,\frac{1}{2},\frac{1}{2} & 6,0,\frac{1}{2},0 & 7,1,0,0 & 12,1,0,\frac{1}{2} & 13,0,\frac{1}{2},\frac{1}{2} & 6 & \frac{1+\tau}{2} \\
\hline
  \end{array} \notag
\end{align}
\end{table}
\begin{table}[H]
\begin{align}
  \begin{array}{c|c|c|c|c|c|c|c|c|c} \hline
    B_Q & B_{u_R} & B_{d_R} & B_L & B_{\nu_R} & B_{e_R} & B_{H_u} & B_{H_d} & g_H & p \\ \hline \hline
6,1,0,\frac{1}{2} & 6,0,\frac{1}{2},\frac{1}{2} & 6,0,\frac{1}{2},0 & 6,0,0,\frac{1}{2} & 6,1,\frac{1}{2},\frac{1}{2} & 6,1,\frac{1}{2},0 & 12,1,\frac{1}{2},0 & 12,1,\frac{1}{2},\frac{1}{2} & 6 & \frac{1+\tau}{2} \\
6,1,0,\frac{1}{2} & 6,0,\frac{1}{2},\frac{1}{2} & 6,0,\frac{1}{2},0 & 7,1,\frac{1}{2},0 & 5,0,0,0 & 5,0,0,\frac{1}{2} & 12,1,\frac{1}{2},0 & 12,1,\frac{1}{2},\frac{1}{2} & 6 & \frac{1+\tau}{2} \\
6,1,0,\frac{1}{2} & 6,0,\frac{1}{2},\frac{1}{2} & 6,1,\frac{1}{2},\frac{1}{2} & 6,0,0,\frac{1}{2} & 6,1,\frac{1}{2},\frac{1}{2} & 6,0,\frac{1}{2},\frac{1}{2} & 12,1,\frac{1}{2},0 & 12,0,\frac{1}{2},0 & 6 & \frac{1+\tau}{2} \\
6,1,0,\frac{1}{2} & 6,0,\frac{1}{2},\frac{1}{2} & 6,1,\frac{1}{2},\frac{1}{2} & 6,1,\frac{1}{2},\frac{1}{2} & 6,0,0,\frac{1}{2} & 6,1,0,\frac{1}{2} & 12,1,\frac{1}{2},0 & 12,0,\frac{1}{2},0 & 6 & \frac{1+\tau}{2} \\
6,1,0,\frac{1}{2} & 6,0,\frac{1}{2},\frac{1}{2} & 7,1,\frac{1}{2},0 & 6,1,\frac{1}{2},\frac{1}{2} & 6,0,0,\frac{1}{2} & 7,1,0,0 & 12,1,\frac{1}{2},0 & 13,0,\frac{1}{2},\frac{1}{2} & 6 & \frac{1+\tau}{2} \\
6,1,0,\frac{1}{2} & 6,0,\frac{1}{2},\frac{1}{2} & 7,1,\frac{1}{2},0 & 7,1,\frac{1}{2},0 & 5,0,0,0 & 6,1,0,\frac{1}{2} & 12,1,\frac{1}{2},0 & 13,0,\frac{1}{2},\frac{1}{2} & 6 & \frac{1+\tau}{2} \\
6,1,0,\frac{1}{2} & 6,1,\frac{1}{2},0 & 6,0,\frac{1}{2},0 & 6,0,\frac{1}{2},0 & 6,0,0,\frac{1}{2} & 6,1,0,\frac{1}{2} & 12,0,\frac{1}{2},\frac{1}{2} & 12,1,\frac{1}{2},\frac{1}{2} & 6 & \frac{1+\tau}{2} \\
\hline
  \end{array} \notag
\end{align}
\end{table}
\begin{table}[H]
\begin{align}
  \begin{array}{c|c|c|c|c|c|c|c|c|c} \hline
    B_Q & B_{u_R} & B_{d_R} & B_L & B_{\nu_R} & B_{e_R} & B_{H_u} & B_{H_d} & g_H & p \\ \hline \hline
6,1,\frac{1}{2},\frac{1}{2} & 6,0,0,\frac{1}{2} & 6,1,0,\frac{1}{2} & 6,0,\frac{1}{2},\frac{1}{2} & 6,1,0,\frac{1}{2} & 6,0,0,\frac{1}{2} & 12,1,\frac{1}{2},0 & 12,0,\frac{1}{2},0 & 6 & \frac{1}{2} \\
6,1,\frac{1}{2},\frac{1}{2} & 6,0,0,\frac{1}{2} & 6,1,0,\frac{1}{2} & 6,1,0,\frac{1}{2} & 6,0,\frac{1}{2},\frac{1}{2} & 6,1,\frac{1}{2},\frac{1}{2} & 12,1,\frac{1}{2},0 & 12,0,\frac{1}{2},0 & 6 & \frac{1}{2} \\
6,1,\frac{1}{2},\frac{1}{2} & 6,0,0,\frac{1}{2} & 7,1,0,0 & 6,1,0,\frac{1}{2} & 6,0,\frac{1}{2},\frac{1}{2} & 7,1,\frac{1}{2},0 & 12,1,\frac{1}{2},0 & 13,0,\frac{1}{2},\frac{1}{2} & 6 & \frac{1}{2} \\
6,1,\frac{1}{2},\frac{1}{2} & 6,0,0,\frac{1}{2} & 7,1,0,0 & 7,1,0,0 & 5,0,\frac{1}{2},0 & 6,1,\frac{1}{2},\frac{1}{2} & 12,1,\frac{1}{2},0 & 13,0,\frac{1}{2},\frac{1}{2} & 6 & \frac{1}{2} \\
6,1,\frac{1}{2},\frac{1}{2} & 6,1,\frac{1}{2},0 & 6,0,\frac{1}{2},0 & 6,0,\frac{1}{2},0 & 6,0,\frac{1}{2},\frac{1}{2} & 6,1,\frac{1}{2},\frac{1}{2} & 12,0,0,\frac{1}{2} & 12,1,0,\frac{1}{2} & 6 & \frac{1}{2} \\
6,1,\frac{1}{2},\frac{1}{2} & 6,1,\frac{1}{2},0 & 6,0,\frac{1}{2},0 & 6,0,\frac{1}{2},\frac{1}{2} & 6,0,\frac{1}{2},0 & 6,1,\frac{1}{2},0 & 12,0,0,\frac{1}{2} & 12,1,0,\frac{1}{2} & 6 & \frac{1}{2} \\
6,1,\frac{1}{2},\frac{1}{2} & 6,1,\frac{1}{2},0 & 6,1,0,\frac{1}{2} & 6,0,\frac{1}{2},\frac{1}{2} & 6,0,\frac{1}{2},0 & 6,0,0,\frac{1}{2} & 12,0,0,\frac{1}{2} & 12,0,\frac{1}{2},0 & 6 & \frac{1}{2} \\
\hline
  \end{array} \notag
\end{align}
\end{table}
\begin{table}[H]
\begin{align}
  \begin{array}{c|c|c|c|c|c|c|c|c|c} \hline
    B_Q & B_{u_R} & B_{d_R} & B_L & B_{\nu_R} & B_{e_R} & B_{H_u} & B_{H_d} & g_H & p \\ \hline \hline
7,1,\frac{1}{2},0 & 5,0,0,0 & 6,0,\frac{1}{2},\frac{1}{2} & 6,0,\frac{1}{2},\frac{1}{2} & 6,1,0,\frac{1}{2} & 7,1,\frac{1}{2},0 & 12,1,\frac{1}{2},0 & 13,1,0,\frac{1}{2} & 6 & \frac{1}{2} \\
7,1,\frac{1}{2},0 & 5,0,0,0 & 6,0,\frac{1}{2},\frac{1}{2} & 7,1,0,0 & 5,0,\frac{1}{2},0 & 6,0,0,\frac{1}{2} & 12,1,\frac{1}{2},0 & 13,1,0,\frac{1}{2} & 6 & \frac{1}{2} \\
7,1,\frac{1}{2},0 & 5,0,0,0 & 6,1,0,\frac{1}{2} & 6,1,0,\frac{1}{2} & 6,0,\frac{1}{2},\frac{1}{2} & 7,1,\frac{1}{2},0 & 12,1,\frac{1}{2},0 & 13,0,\frac{1}{2},\frac{1}{2} & 6 & \frac{1}{2} \\
7,1,\frac{1}{2},0 & 5,0,0,0 & 6,1,0,\frac{1}{2} & 7,1,0,0 & 5,0,\frac{1}{2},0 & 6,1,\frac{1}{2},\frac{1}{2} & 12,1,\frac{1}{2},0 & 13,0,\frac{1}{2},\frac{1}{2} & 6 & \frac{1}{2} \\
7,1,\frac{1}{2},0 & 5,0,0,\frac{1}{2} & 6,0,\frac{1}{2},0 & 6,0,\frac{1}{2},0 & 6,1,0,\frac{1}{2} & 7,1,\frac{1}{2},0 & 12,1,\frac{1}{2},\frac{1}{2} & 13,1,0,0 & 6 & \frac{1}{2} \\
7,1,\frac{1}{2},0 & 5,0,0,\frac{1}{2} & 6,0,\frac{1}{2},0 & 7,1,0,\frac{1}{2} & 5,0,\frac{1}{2},0 & 6,0,0,\frac{1}{2} & 12,1,\frac{1}{2},\frac{1}{2} & 13,1,0,0 & 6 & \frac{1}{2} \\
7,1,\frac{1}{2},0 & 5,0,0,\frac{1}{2} & 6,1,0,\frac{1}{2} & 6,1,0,\frac{1}{2} & 6,0,\frac{1}{2},0 & 7,1,\frac{1}{2},0 & 12,1,\frac{1}{2},\frac{1}{2} & 13,0,\frac{1}{2},\frac{1}{2} & 6 & \frac{1}{2} \\
\hline
  \end{array} \notag
\end{align}
\end{table}
\begin{table}[H]
\begin{align}
  \begin{array}{c|c|c|c|c|c|c|c|c|c} \hline
    B_Q & B_{u_R} & B_{d_R} & B_L & B_{\nu_R} & B_{e_R} & B_{H_u} & B_{H_d} & g_H & p \\ \hline \hline
7,1,0,0 & 5,0,\frac{1}{2},0 & 6,1,\frac{1}{2},\frac{1}{2} & 7,1,\frac{1}{2},0 & 5,0,0,0 & 6,1,0,\frac{1}{2} & 12,1,\frac{1}{2},0 & 13,0,\frac{1}{2},\frac{1}{2} & 6 & \frac{1+\tau}{2} \\
7,1,0,0 & 5,0,0,\frac{1}{2} & 6,0,\frac{1}{2},0 & 6,0,\frac{1}{2},0 & 6,1,\frac{1}{2},\frac{1}{2} & 7,1,0,0 & 12,1,0,\frac{1}{2} & 13,1,\frac{1}{2},0 & 6 & \frac{1+\tau}{2} \\
7,1,0,0 & 5,0,0,\frac{1}{2} & 6,0,\frac{1}{2},0 & 7,1,0,\frac{1}{2} & 5,0,0,0 & 6,0,\frac{1}{2},\frac{1}{2} & 12,1,0,\frac{1}{2} & 13,1,\frac{1}{2},0 & 6 & \frac{1+\tau}{2} \\
7,1,0,0 & 5,0,0,\frac{1}{2} & 6,1,\frac{1}{2},\frac{1}{2} & 6,1,\frac{1}{2},\frac{1}{2} & 6,0,\frac{1}{2},0 & 7,1,0,0 & 12,1,0,\frac{1}{2} & 13,0,\frac{1}{2},\frac{1}{2} & 6 & \frac{1+\tau}{2} \\
7,1,0,0 & 5,0,0,\frac{1}{2} & 6,1,\frac{1}{2},\frac{1}{2} & 7,1,0,\frac{1}{2} & 5,0,0,0 & 6,1,\frac{1}{2},0 & 12,1,0,\frac{1}{2} & 13,0,\frac{1}{2},\frac{1}{2} & 6 & \frac{1+\tau}{2} \\
7,1,0,0 & 5,1,\frac{1}{2},\frac{1}{2} & 6,0,\frac{1}{2},0 & 6,0,\frac{1}{2},0 & 6,0,0,\frac{1}{2} & 7,1,0,0 & 12,0,\frac{1}{2},\frac{1}{2} & 13,1,\frac{1}{2},0 & 6 & \frac{1+\tau}{2} \\
7,1,0,0 & 5,1,\frac{1}{2},\frac{1}{2} & 6,0,\frac{1}{2},0 & 7,0,\frac{1}{2},\frac{1}{2} & 5,0,0,0 & 6,1,0,\frac{1}{2} & 12,0,\frac{1}{2},\frac{1}{2} & 13,1,\frac{1}{2},0 & 6 & \frac{1+\tau}{2} \\
\hline
  \end{array} \notag
\end{align}
\end{table}


\section{Yukawa couplings on the flavor model in Table \ref{tab:ex-model}}
\label{App:Yukawa-couplings}

Here we show Yukawa couplings of up-sector quarks, down-sector quarks, neutrinos and charged leptons, $Y_u^{jk\ell}$, $Y_{d}^{jk\ell}$, $Y_\nu^{jk\ell}$ and $Y_e^{jk\ell}$ on the flavor model in Table \ref{tab:ex-model}.

\paragraph{Up sector quark: $B_Q{\text -}B_{u_R}{\text -}B_{H_u}=(6,0,\frac{1}{2},0){\text -}(6,0,0,\frac{1}{2}){\text -}(12,0,\frac{1}{2},\frac{1}{2})$}~\\
Table \ref{tab:three-model-UpQuarks} shows the zero-mode assignments for quark doublets $Q^j$, right-handed up-sector quarks $u_R^k$ and up type Higgs fields $H_u^\ell$.
\begin{table}[H]
\begin{center}
\renewcommand{\arraystretch}{1.2}
\begin{tabular}{c|c|c|c}
& $Q^j$ & $u_R^k$ & $H^\ell_u$ \\ \hline
0 & $\frac{1}{\sqrt{2}}(\psi_{T^2}^{(1/2,0),6}+\psi_{T^2}^{(11/2,0),6})$ & $\psi_{T^2}^{(0,1/2),6}$ & $\frac{1}{\sqrt{2}}(\psi_{T^2}^{(1/2,1/2),12}-\psi_{T^2}^{(23/2,1/2),12})$ \\
1 & $\frac{1}{\sqrt{2}}(\psi_{T^2}^{(3/2,0),6}+\psi_{T^2}^{(9/2,0),6})$ & $\frac{1}{\sqrt{2}}(\psi_{T^2}^{(1,1/2),6}-\psi_{T^2}^{(5,1/2),6})$ & $\frac{1}{\sqrt{2}}(\psi_{T^2}^{(3/2,1/2),12}-\psi_{T^2}^{(21/2,1/2),12})$ \\
2 & $\frac{1}{\sqrt{2}}(\psi_{T^2}^{(5/2,0),6}+\psi_{T^2}^{(7/2,0),6})$ & $\frac{1}{\sqrt{2}}(\psi_{T^2}^{(2,1/2),6}-\psi_{T^2}^{(4,1/2),6})$ & $\frac{1}{\sqrt{2}}(\psi_{T^2}^{(5/2,1/2),12}-\psi_{T^2}^{(19/2,1/2),12})$ \\
3 & & & $\frac{1}{\sqrt{2}}(\psi_{T^2}^{(7/2,1/2),12}-\psi_{T^2}^{(17/2,1/2),12})$ \\
4 & & & $\frac{1}{\sqrt{2}}(\psi_{T^2}^{(9/2,1/2),12}-\psi_{T^2}^{(15/2,1/2),12})$ \\
5 & & & $\frac{1}{\sqrt{2}}(\psi_{T^2}^{(11/2,1/2),12}-\psi_{T^2}^{(13/2,1/2),12})$ \\
\end{tabular}
\end{center}
\caption{The zero-modes on $T^2/\mathbb{Z}_2$ for up-sector quarks and up type Higgs fields in Table \ref{tab:ex-model}.}
\label{tab:three-model-UpQuarks}
\end{table}
Yukawa couplings are given by
\begin{align}
  Y^{jk\ell}_{u}H_u^\ell = Y^{jk0}_{u}H_u^0+Y^{jk1}_{u}H_u^1+Y^{jk2}_{u}H_u^2+Y^{jk3}_{u}H_u^3+Y^{jk4}_{u}H_u^4+Y^{jk5}_{u}H_u^5,
\end{align}
with
\begin{align}
&Y_u^{jk0} = c_{(6{\text -}6{\text -}12)}
\begin{pmatrix}
  X_0 & \frac{1}{\sqrt{2}}X_1 & 0 \\
  0 & \frac{1}{\sqrt{2}}X_2 & \frac{1}{\sqrt{2}}X_3 \\
  0 & 0 & \frac{1}{\sqrt{2}}X_4 \\
\end{pmatrix}, \quad
Y_u^{jk1} = c_{(6{\text -}6{\text -}12)}
\begin{pmatrix}
  0 & \frac{1}{\sqrt{2}}X_0 & \frac{1}{\sqrt{2}}X_2 \\
  X_1 & 0 & 0 \\
  0 & \frac{1}{\sqrt{2}}X_3 & \frac{1}{\sqrt{2}}X_5 \\
\end{pmatrix},
\notag \\
&Y_u^{jk2} = c_{(6{\text -}6{\text -}12)}
\begin{pmatrix}
  0 & 0 & \frac{1}{\sqrt{2}}X_1 \\
  0 & \frac{1}{\sqrt{2}}X_0 & -\frac{1}{\sqrt{2}}X_5 \\
  X_2 & \frac{1}{\sqrt{2}}X_4 & 0 \\
\end{pmatrix},
\quad
Y_u^{jk3} = c_{(6{\text -}6{\text -}12)}
\begin{pmatrix}
  0 & 0 & -\frac{1}{\sqrt{2}}X_4 \\
  0 & \frac{1}{\sqrt{2}}X_5 & \frac{1}{\sqrt{2}}X_0 \\
  X_3 & \frac{1}{\sqrt{2}}X_1 & 0 \\
\end{pmatrix},
\notag \\
&Y_u^{jk4} = c_{(6{\text -}6{\text -}12)}
\begin{pmatrix}
  0 & -\frac{1}{\sqrt{2}}X_5 & -\frac{1}{\sqrt{2}}X_3 \\
  X_4 & 0 & 0 \\
  0 & \frac{1}{\sqrt{2}}X_2 & \frac{1}{\sqrt{2}}X_0 \\
\end{pmatrix},
\quad
Y_u^{jk5} = c_{(6{\text -}6{\text -}12)}
\begin{pmatrix}
  X_5 & -\frac{1}{\sqrt{2}}X_4 & 0 \\
  0 & \frac{1}{\sqrt{2}}X_3 & -\frac{1}{\sqrt{2}}X_2 \\
  0 & 0 & \frac{1}{\sqrt{2}}X_1 \\
\end{pmatrix},
\notag
\end{align}
where
\begin{align}
  X_N \equiv \sum_{n=0}^5 (-1)^n \eta_{6(N+1/2)+72n}, \quad
  \eta_N \equiv \vartheta
  \begin{bmatrix}
    \frac{N}{432} \\ 0 \\
  \end{bmatrix}
  (0,432\tau).
\end{align}

\paragraph{Down sector quark: $B_Q{\text -}B_{d_R}{\text -}B_{H_d}=(6,0,\frac{1}{2},0){\text -}(6,1,0,\frac{1}{2}){\text -}(12,1,\frac{1}{2},\frac{1}{2})$}~\\
Table \ref{tab:three-model-DownQuarks} shows the zero-mode assignments for quark doublets $Q^j$, right-handed down-sector quarks $d_R^k$ and down type Higgs fields $H_d^\ell$.
\begin{table}[H]
\begin{center}
\renewcommand{\arraystretch}{1.2}
\begin{tabular}{c|c|c|c}
& $Q^j$ & $d_R^k$ & $H^\ell_d$ \\ \hline
0 & $\frac{1}{\sqrt{2}}(\psi_{T^2}^{(1/2,0),6}+\psi_{T^2}^{(11/2,0),6})$ & $\frac{1}{\sqrt{2}}(\psi_{T^2}^{(1,1/2),6}+\psi_{T^2}^{(5,1/2),6})$ & $\frac{1}{\sqrt{2}}(\psi_{T^2}^{(1/2,1/2),12}+\psi_{T^2}^{(23/2,1/2),12})$ \\
1 & $\frac{1}{\sqrt{2}}(\psi_{T^2}^{(3/2,0),6}+\psi_{T^2}^{(9/2,0),6})$ & $\frac{1}{\sqrt{2}}(\psi_{T^2}^{(2,1/2),6}+\psi_{T^2}^{(4,1/2),6})$ & $\frac{1}{\sqrt{2}}(\psi_{T^2}^{(3/2,1/2),12}+\psi_{T^2}^{(21/2,1/2),12})$ \\
2 & $\frac{1}{\sqrt{2}}(\psi_{T^2}^{(5/2,0),6}+\psi_{T^2}^{(7/2,0),6})$ & $\psi_{T^2}^{(3,1/2),6}$ & $\frac{1}{\sqrt{2}}(\psi_{T^2}^{(5/2,1/2),12}+\psi_{T^2}^{(19/2,1/2),12})$ \\
3 & & & $\frac{1}{\sqrt{2}}(\psi_{T^2}^{(7/2,1/2),12}+\psi_{T^2}^{(17/2,1/2),12})$ \\
4 & & & $\frac{1}{\sqrt{2}}(\psi_{T^2}^{(9/2,1/2),12}+\psi_{T^2}^{(15/2,1/2),12})$ \\
5 & & & $\frac{1}{\sqrt{2}}(\psi_{T^2}^{(11/2,1/2),12}+\psi_{T^2}^{(13/2,1/2),12})$ \\
\end{tabular}
\end{center}
\caption{The zero-modes on $T^2/\mathbb{Z}_2$ for down-sector quarks and down type Higgs fields in Table \ref{tab:ex-model}.}
\label{tab:three-model-DownQuarks}
\end{table}
Yukawa couplings are given by
\begin{align}
  Y^{jk\ell}_{d}H_d^k = Y^{jk0}_{d}H_d^0+Y^{jk1}_{d}H_d^1+Y^{jk2}_{d}H_d^2+Y^{jk3}_{d}H_d^3+Y^{jk4}_{d}H_d^4+Y^{jk5}_{d}H_d^5,
\end{align}
with
\begin{align}
&Y_{d}^{jk0} = c_{(6{\text -}6{\text -}12)}
\begin{pmatrix}
\frac{1}{\sqrt{2}}X_1 & 0 & 0 \\
-\frac{1}{\sqrt{2}}X_2 & \frac{1}{\sqrt{2}}X_3 & 0 \\
0 & -\frac{1}{\sqrt{2}}X_4 & X_5 \\
\end{pmatrix},
&Y_{d}^{jk1} = c_{(6{\text -}6{\text -}12)}
\begin{pmatrix}
\frac{1}{\sqrt{2}}X_0 & \frac{1}{\sqrt{2}}X_2 & 0 \\
0 & 0 & X_4 \\
-\frac{1}{\sqrt{2}}X_3 & -\frac{1}{\sqrt{2}}X_5 & 0 \\
\end{pmatrix},
\notag \\
&Y_{d}^{jk2} = c_{(6{\text -}6{\text -}12)}
\begin{pmatrix}
0 & \frac{1}{\sqrt{2}}X_1 & X_3 \\
\frac{1}{\sqrt{2}}X_0 & \frac{1}{\sqrt{2}}X_5 & 0 \\
-\frac{1}{\sqrt{2}}X_4 & 0 & 0 \\
\end{pmatrix},
&Y_{d}^{jk3} = c_{(6{\text -}6{\text -}12)}
\begin{pmatrix}
0 & \frac{1}{\sqrt{2}}X_4 & X_2 \\
-\frac{1}{\sqrt{2}}X_5 & \frac{1}{\sqrt{2}}X_0 & 0 \\
\frac{1}{\sqrt{2}}X_1 & 0 & 0 \\
\end{pmatrix},
\notag \\
&Y_{d}^{jk4} = c_{(6{\text -}6{\text -}12)}
\begin{pmatrix}
\frac{1}{\sqrt{2}}X_5 & \frac{1}{\sqrt{2}}X_3 & 0 \\
0 & 0 & X_1 \\
\frac{1}{\sqrt{2}}X_2 & \frac{1}{\sqrt{2}}X_0 & 0 \\
\end{pmatrix},
&Y_{d}^{jk5} = c_{(6{\text -}6{\text -}12)}
\begin{pmatrix}
\frac{1}{\sqrt{2}}X_4 & 0 & 0 \\
\frac{1}{\sqrt{2}}X_3 & \frac{1}{\sqrt{2}}X_2 & 0 \\
0 & \frac{1}{\sqrt{2}}X_1 & X_0 \\
\end{pmatrix},
\notag
\end{align}
where
\begin{align}
  X_N \equiv \sum_{n=0}^5 (-1)^n \eta_{6(N+1/2)+72n}, \quad
  \eta_N \equiv \vartheta
  \begin{bmatrix}
    \frac{N}{432} \\ 0 \\
  \end{bmatrix}
  (0,432\tau).
\end{align}

\paragraph{Neutrino: $B_L{\text -}B_{\nu_R}{\text -}B_{H_u}=(6,1,0,\frac{1}{2}){\text -}(6,1,\frac{1}{2},0){\text -}(12,0,\frac{1}{2},\frac{1}{2})$}~\\
Table \ref{tab:three-model-Neutrinos} shows the zero-mode assignments for lepton doublets $L^j$, right-handed neutrinos $\nu_R^k$ and up type Higgs fields $H_u^\ell$.
\begin{table}[H]
\begin{center}
\renewcommand{\arraystretch}{1.2}
\begin{tabular}{c|c|c|c}
& $L^j$ & $\nu_R^k$ & $H^\ell_u$ \\ \hline
0 & $\frac{1}{\sqrt{2}}(\psi_{T^2}^{(1,1/2),6}+\psi_{T^2}^{(5,1/2),6})$ & $\frac{1}{\sqrt{2}}(\psi_{T^2}^{(1/2,0),6}-\psi_{T^2}^{(11/2,0),6})$ & $\frac{1}{\sqrt{2}}(\psi_{T^2}^{(1/2,1/2),12}-\psi_{T^2}^{(23/2,1/2),12})$ \\
1 & $\frac{1}{\sqrt{2}}(\psi_{T^2}^{(2,1/2),6}+\psi_{T^2}^{(4,1/2),6})$ & $\frac{1}{\sqrt{2}}(\psi_{T^2}^{(3/2,0),6}-\psi_{T^2}^{(9/2,0),6})$ & $\frac{1}{\sqrt{2}}(\psi_{T^2}^{(3/2,1/2),12}-\psi_{T^2}^{(21/2,1/2),12})$ \\
2 & $\psi_{T^2}^{(3,1/2),6}$ & $\frac{1}{\sqrt{2}}(\psi_{T^2}^{(5/2,0),6}-\psi_{T^2}^{(7/2,0),6})$ & $\frac{1}{\sqrt{2}}(\psi_{T^2}^{(5/2,1/2),12}-\psi_{T^2}^{(19/2,1/2),12})$ \\
3 & & & $\frac{1}{\sqrt{2}}(\psi_{T^2}^{(7/2,1/2),12}-\psi_{T^2}^{(17/2,1/2),12})$ \\
4 & & & $\frac{1}{\sqrt{2}}(\psi_{T^2}^{(9/2,1/2),12}-\psi_{T^2}^{(15/2,1/2),12})$ \\
5 & & & $\frac{1}{\sqrt{2}}(\psi_{T^2}^{(11/2,1/2),12}-\psi_{T^2}^{(13/2,1/2),12})$ \\
\end{tabular}
\end{center}
\caption{The zero-modes on $T^2/\mathbb{Z}_2$ for neutrinos and up type Higgs fields in Table \ref{tab:ex-model}.}
\label{tab:three-model-Neutrinos}
\end{table}
Yukawa couplings are given by
\begin{align}
  Y^{jk\ell}_\nu H_u^\ell = Y^{jk0}_\nu H_u^0+Y^{jk1}_\nu H_u^1+Y^{jk2}_\nu H_u^2+Y^{jk3}_\nu H_u^3+Y^{jk4}_\nu H_u^4+Y^{jk5}_\nu H_u^5,
\end{align}
with
\begin{align}
&Y_\nu^{jk0} = c_{(6{\text -}6{\text -}12)}
\begin{pmatrix}
-\frac{1}{\sqrt{2}}X_1 & -\frac{1}{\sqrt{2}}X_2 & 0 \\
0 & -\frac{1}{\sqrt{2}}X_3 & -\frac{1}{\sqrt{2}}X_4 \\
0 & 0 & -X_5 \\
\end{pmatrix},
&Y_\nu^{jk1} = c_{(6{\text -}6{\text -}12)}
\begin{pmatrix}
\frac{1}{\sqrt{2}}X_0 & 0 & -\frac{1}{\sqrt{2}}X_3 \\
-\frac{1}{\sqrt{2}}X_2 & 0 & \frac{1}{\sqrt{2}}X_5 \\
0 & -X_4 & 0 \\
\end{pmatrix},
\notag \\
&Y_\nu^{jk2} = c_{(6{\text -}6{\text -}12)}
\begin{pmatrix}
0 & \frac{1}{\sqrt{2}}X_0 & \frac{1}{\sqrt{2}}X_4 \\
\frac{1}{\sqrt{2}}X_1 & -\frac{1}{\sqrt{2}}X_5 & 0 \\
-X_3 & 0 & 0 \\
\end{pmatrix},
&Y_\nu^{jk3} = c_{(6{\text -}6{\text -}12)}
\begin{pmatrix}
0 & \frac{1}{\sqrt{2}}X_5 & \frac{1}{\sqrt{2}}X_1 \\
-\frac{1}{\sqrt{2}}X_4 & \frac{1}{\sqrt{2}}X_0 & 0 \\
X_2 & 0 & 0 \\
\end{pmatrix},
\notag \\
&Y_\nu^{jk4} = c_{(6{\text -}6{\text -}12)}
\begin{pmatrix}
-\frac{1}{\sqrt{2}}X_5 & 0 & -\frac{1}{\sqrt{2}}X_2 \\
\frac{1}{\sqrt{2}}X_3 & 0 & \frac{1}{\sqrt{2}}X_0 \\
0 & X_1 & 0 \\
\end{pmatrix},
&Y_\nu^{jk5} = c_{(6{\text -}6{\text -}12)}
\begin{pmatrix}
\frac{1}{\sqrt{2}}X_4 & -\frac{1}{\sqrt{2}}X_3 & 0 \\
0 & \frac{1}{\sqrt{2}}X_2 & -\frac{1}{\sqrt{2}}X_1 \\
0 & 0 & X_0 \\
\end{pmatrix},
\notag
\end{align}
where
\begin{align}
  X_N \equiv \sum_{n=0}^5 (-1)^n \eta_{6(N+1/2)+72n}, \quad
  \eta_N \equiv \vartheta
  \begin{bmatrix}
    \frac{N}{432} \\ 0 \\
  \end{bmatrix}
  (0,432\tau).
\end{align}

\paragraph{Charged lepton: $B_L{\text -}B_{e_R}{\text -}B_{H_d}=(6,1,0,\frac{1}{2}){\text -}(6,0,\frac{1}{2},0){\text -}(12,1,\frac{1}{2},\frac{1}{2})$}~\\
Table \ref{tab:three-model-ChargedLepton} shows the zero-mode assignments for lepton doublets $L^j$, right-handed charged leptons $e_R^k$ and down type Higgs fields $H_d^\ell$.
\begin{table}[H]
\begin{center}
\renewcommand{\arraystretch}{1.2}
\begin{tabular}{c|c|c|c}
& $L^j$ & $e_R^k$ & $H^\ell_d$ \\ \hline
0 & $\frac{1}{\sqrt{2}}(\psi_{T^2}^{(1,1/2),6}+\psi_{T^2}^{(5,1/2),6})$ & $\frac{1}{\sqrt{2}}(\psi_{T^2}^{(1/2,0),6}+\psi_{T^2}^{(11/2,0),6})$ & $\frac{1}{\sqrt{2}}(\psi_{T^2}^{(1/2,1/2),12}+\psi_{T^2}^{(23/2,1/2),12})$ \\
1 & $\frac{1}{\sqrt{2}}(\psi_{T^2}^{(2,1/2),6}+\psi_{T^2}^{(4,1/2),6})$ & $\frac{1}{\sqrt{2}}(\psi_{T^2}^{(3/2,0),6}+\psi_{T^2}^{(9/2,0),6})$ & $\frac{1}{\sqrt{2}}(\psi_{T^2}^{(3/2,1/2),12}+\psi_{T^2}^{(21/2,1/2),12})$ \\
2 & $\psi_{T^2}^{(3,1/2),6}$ & $\frac{1}{\sqrt{2}}(\psi_{T^2}^{(5/2,0),6}+\psi_{T^2}^{(7/2,0),6})$ & $\frac{1}{\sqrt{2}}(\psi_{T^2}^{(5/2,1/2),12}+\psi_{T^2}^{(19/2,1/2),12})$ \\
3 & & & $\frac{1}{\sqrt{2}}(\psi_{T^2}^{(7/2,1/2),12}+\psi_{T^2}^{(17/2,1/2),12})$ \\
4 & & & $\frac{1}{\sqrt{2}}(\psi_{T^2}^{(9/2,1/2),12}+\psi_{T^2}^{(15/2,1/2),12})$ \\
5 & & & $\frac{1}{\sqrt{2}}(\psi_{T^2}^{(11/2,1/2),12}+\psi_{T^2}^{(13/2,1/2),12})$ \\
\end{tabular}
\end{center}
\caption{The zero-modes on $T^2/\mathbb{Z}_2$ for charged leptons and down type Higg fields in Table \ref{tab:ex-model}.}
\label{tab:three-model-ChargedLepton}
\end{table}
Yukawa couplings are given by
\begin{align}
  Y^{jk\ell}_{e} = Y^{jk\ell}_{d}.
\end{align}


\section{Majorana neutrino masses on the flavor model in Table \ref{tab:ex-model}}
\label{App:Majorana}

Majorana masses of right-handed neutrinos $\nu_R$ can be induced by D-brane instanton effects as shown in Appendix \ref{app:Majorana_D-brane}.
For right-handed neutrinos in Table \ref{tab:ex-model}, there are two possible instanton zero-mode configurations, $\beta_1,\gamma_1$ and $\beta_2,\gamma_2$,
\begin{align}
&B_{\beta_1} {\text -} B_{\gamma_1} {\text -} B_{\nu_R} = (3,0,0,\textstyle\frac{1}{2}) {\text -} (3,1,\textstyle\frac{1}{2},\textstyle\frac{1}{2}) {\text -} (6,1,\textstyle\frac{1}{2},0), \\ 
&B_{\beta_2} {\text -} B_{\gamma_2} {\text -} B_{\nu_R} = (2,0,0,0) {\text -} (4,1,\textstyle\frac{1}{2},0) {\text -} (6,1,\textstyle\frac{1}{2},0).
\end{align}
These two configurations are different; nevertheless they give the same Majorana mass matrix up to overall factor.
Therefore we use only the former configuration, $\beta_1,\gamma_1$.
Table \ref{tab:Majorana} shows the zero-mode assignments for instanton zero-modes and right-handed neutrinos.
\begin{table}[H]
\begin{center}
\renewcommand{\arraystretch}{1.2}
\begin{tabular}{c|c|c|c}
& $\beta_1^j$ & $\gamma_1^k$ & $\nu_R^a$ \\ \hline
0 & $\psi_{T^2}^{(0,1/2),3}$ & $\frac{1}{\sqrt{2}}(\psi_{T^2}^{(1/2,1/2),3}+\psi_{T^2}^{(5/2,1/2),3})$ & $\frac{1}{\sqrt{2}}(\psi_{T^2}^{(1/2,0),6}-\psi_{T^2}^{(11/2,0),6})$ \\
1 & $\frac{1}{\sqrt{2}}(\psi_{T^2}^{(1,1/2),3}-\psi_{T^2}^{(2,1/2),3})$ & $\psi_{T^2}^{(3/2,1/2),3}$ & $\frac{1}{\sqrt{2}}(\psi_{T^2}^{(3/2,0),6}-\psi_{T^2}^{(9/2,0),6})$ \\
2 & & & $\frac{1}{\sqrt{2}}(\psi_{T^2}^{(5/2,0),6}-\psi_{T^2}^{(7/2,0),6})$ \\
\end{tabular}
\end{center}
\caption{The zero-modes for instanton zero-modes and right-handed neutrinos.}
\label{tab:Majorana}
\end{table}
The three point couplings $d_a^{jk}$ are given by
\begin{align}
&d_0^{jk} = c_{(3{\text -}3{\text -}6)}
\begin{pmatrix}
  \eta_{1.5}+\eta_{16.5}+\eta_{19.5} & 0 \\
  -\frac{1}{\sqrt{2}}(\eta_{4.5}+\eta_{13.5}+\eta_{22.5}) & \eta_{7.5}+\eta_{10.5}+\eta_{25.5} \\
\end{pmatrix}, \\
&d_1^{jk} = c_{(3{\text -}3{\text -}6)}
\begin{pmatrix}
  0 & \sqrt{2}(\eta_{4.5}+\eta_{13.5}+\eta_{22.5}) \\
  \frac{1}{\sqrt{2}}(\eta_{1.5}+\eta_{16.5}+\eta_{19.5}+\eta_{7.5}+\eta_{10.5}+\eta_{25.5}) & 0 \\
\end{pmatrix}, \\
&d_2^{jk} = c_{(3{\text -}3{\text -}6)}
\begin{pmatrix}
  \eta_{7.5}+\eta_{10.5}+\eta_{25.5} & 0 \\
  -\frac{1}{\sqrt{2}}(\eta_{4.5}+\eta_{13.5}+\eta_{22.5}) & \eta_{1.5}+\eta_{16.5}+\eta_{19.5} \\
\end{pmatrix},
\end{align}
where
\begin{align}
  \eta_N \equiv \vartheta
  \begin{bmatrix}
    \frac{N}{54} \\ 0 \\
  \end{bmatrix}
  (0,54\tau).
\end{align}
Using above $d_a^{jk}$, Majorana masses of right-handed neutrinos in Eq.~(\ref{eq:m_RR}) can be calculated.


\chapter{D-brane instanton effects}
\label{app:D-brane instanton effects}

Here we give a brief review of Majorana neutrino mass terms and Higgs $\mu$ terms induced by D-brane instanton effects \cite{Blumenhagen:2006xt, Ibanez:2006da, Ibanez:2007rs, Antusch:2007jd, Kobayashi:2015siy}.


\section{Majorana neutrino masses}
\label{app:Majorana_D-brane}

In order to obtain Majorana mass of right-handed neutrinos $\nu_R$, let us assume two stacks of D-branes, $D_{N_1}$ and $D_{N_2}$, and the D-brane instanton, $D_{\textrm{inst}}$, with magnetic fluxes.
The D-branes $D_{N_1}$, $D_{N_2}$ and $D_{\textrm{inst}}$ are located to intersect each other.
Then right-handed neutrinos $\nu_R$ can appear as the zero-modes of the open strings between $D_{N_1}$ and $D_{N_2}$.
Similarly, instanton zero-modes $\beta$ ($\gamma$) appear between $D_{N_1}$ ($D_{N_2}$) and $D_{\textrm{inst}}$.
We denote their zero-modes as $\psi_{\nu_R}^a(z)$, $\psi_\beta^j(z)$ and $\psi_\gamma^k(z)$ for right-handed neutrinos $\nu_R^a$, instanton zero-modes $\beta^j$ and $\gamma^k$, respectively.
Right-handed neutrinos and two instanton zero-modes can have non-vanishing three point couplings $d_a^{jk}$,
\begin{align}
d_a^{jk} = g(\textrm{Im}\tau)^{1/2} \int dzd\bar{z} \psi_\beta^j(z) \cdot psi_\gamma^k(z) \cdot (\psi_{\nu_R}^a(z))^*,
\end{align}
and they give the instanton effects,
\begin{align}
\int d^2\beta d^2\gamma e^{-(\textrm{Im}\tau)^{-1/2}d_a^{jk} \beta^j \gamma^k \nu_R^a},
\end{align}
where $\beta$ and $\gamma$ are Grasmannian.
Mass terms do not vanish only if each of $\beta$ and $\gamma$ has two-generation zero-modes.
Implementing Grasmannian integral, we obtain Majorana mass terms of right-handed neutrinos $M_{RR}$,
\begin{align}
\Lambda e^{-S_{\textrm{inst}}} \int d^2\beta d^2\gamma e^{-(\textrm{Im}\tau)^{-1/2} d_a^{jk}\beta^j\gamma^k \nu_R^a}
&= \Lambda e^{-S_{\textrm{inst}}} (\textrm{Im}\tau)^{-1/2} \varepsilon_{jk} \varepsilon_{\ell m} d^{j\ell}_a d^{km}_b \nu_R^a \nu_R^b \\
&\equiv M_{RR}^{ab} \nu_R^a \nu_R^b, \label{eq:m_RR}
\end{align}
where $S_\textrm{inst}$ stands for the instanton action and $\Lambda$ is a typical scale such as the compactification scale.
When we consider magnetized $T^2/\mathbb{Z}_2$ twisted orbifold model, the configurations of the instanton zero-modes leading to non-vanishing three point couplings must satisfy
\begin{align}
\begin{aligned}
&M_{\beta} \pm M_\gamma = \pm M_{\nu_R}, \\
&m_\beta + m_\gamma = m_{\nu_R}~(\textrm{mod}~2), \\
&\alpha_{1\beta} + \alpha_{1\gamma} = \alpha_{1\nu_R} ~ (\textrm{mod}~1), \\
&\alpha_{2\beta} + \alpha_{2\gamma} = \alpha_{2\nu_R} ~ (\textrm{mod}~1), \\
\end{aligned}
\end{align}
where $M$, $m$ and $(\alpha_1,\alpha_2)$ are the flux, $\mathbb{Z}_2$ parity and SS phases of $\beta$, $\gamma$ and $\nu_R$.


\section{Higgs $\mu$ terms}
\label{app:Higgs_mu_terms}

In order to obtain Higgs $\mu$ terms, let us assume three stacks of D-branes, $D_{N_a}$, $D_{N_b}$ and $D_{N_c}$, and the D-brane instanton $D_{\textrm{inst}}$ with magnetic fluxes.
The D-brane $D_b$ is parallel to $D_c$, and $D_a$ are located to intersect them.
Then up (down) type Higgs fields $H_u$ ($H_d$) can appear as the zero-modes of the open strings between $D_a$ and $D_b$ ($D_c$).
Similarly, instanton zero-modes $\alpha$, $\beta$ and $\gamma$ appear as the zero-modes of the open strings between $D_a$ and $D_{\textrm{inst}}$, $D_b$ and $D_{\textrm{inst}}$, and $D_c$ and $D_{\textrm{inst}}$.
We denote their zero-modes as $\psi_{H_u}^j(z)$, $\psi_{H_d}^k(z)$, $\psi_\alpha(z)$, $\psi_\beta(z)$ and $\psi_\gamma(z)$ for up type Higgs fields $H_u^j$, down type Higgs fields $H_d^k$, instanton zero-modes $\alpha$, $\beta$ and $\gamma$, respectively.
Higgs fields and three instanton zero-modes can have non-vanishing three point couplings $Y_u^j$ and $Y_d^k$,
\begin{align}
&Y_u^j = g(\textrm{Im}\tau)^{1/2} \int dzd\bar{z} \psi_\alpha(z) \cdot \psi_\beta(z) \cdot (\psi^j_{H_u}(z))^*, \\
&Y_d^k = g(\textrm{Im}\tau)^{1/2} \int dzd\bar{z} \psi_\alpha(z) \cdot \psi_\gamma(z) \cdot (\psi^k_{H_d}(z))^*,
\end{align}
and they give the instanton effects,
\begin{align}
\int d^2\alpha d\beta d\gamma e^{(\textrm{Im}\tau)^{-1/2}(Y_u^j \alpha \cdot H_u^j\beta + Y_d^k\alpha\cdot H_d^k\gamma)},
\end{align}
where $\alpha$, $\beta$ and $\gamma$ are Grasmannian.
Mass terms do not vanish only if each of $\alpha$, $\beta$ and $\gamma$ has a single zero-mode.
Implementing Grasmannian integral, we obtain Higgs $\mu$ terms,
\begin{align}
\Lambda e^{-S_\textrm{inst}} \int d^2\alpha d\beta d\gamma e^{(\textrm{Im}\tau)^{-1/2}(Y_u^j \alpha \cdot H_u^j\beta + Y_d^k\alpha\cdot H_d^k\gamma)}
&= \Lambda e^{-S_\textrm{inst}} (\textrm{Im}\tau)^{-1} (Y_u^jY_d^k) \varepsilon_{nm} H_{um}^j H_{dn}^k \\
&\equiv \mu^{jk} \varepsilon_{nm} H_{um}^j H_{dn}^k,
\end{align}
where $S_\textrm{inst}$ stands for the instanton action, $\Lambda$ is a typical scale such as the compactification scale and $m,n\in\{0,1\}$ are components of the $SU(2)_L$ doublet.
When we consider magnetized $T^2/\mathbb{Z}_2$ twisted orbifold model, the configurations of the instanton zero-modes leading to non-vanishing three point couplings must satisfy
\begin{align}
\begin{aligned}
&M_{\alpha} \pm M_\beta = \pm M_{H_u}, \\
&m_\alpha + m_\beta = m_{H_u}~(\textrm{mod}~2), \\
&\alpha_{1\alpha} + \alpha_{1\beta} = \alpha_{1H_u} ~ (\textrm{mod}~1), \\
&\alpha_{2\alpha} + \alpha_{2\beta} = \alpha_{2H_u} ~ (\textrm{mod}~1), \\
&M_{\alpha} \pm M_\gamma = \pm M_{H_d}, \\
&m_\alpha + m_\gamma = m_{H_u}~(\textrm{mod}~2), \\
&\alpha_{1\alpha} + \alpha_{1\gamma} = \alpha_{1H_d} ~ (\textrm{mod}~1), \\
&\alpha_{2\alpha} + \alpha_{2\gamma} = \alpha_{2H_d} ~ (\textrm{mod}~1), \\
\end{aligned}
\end{align}
where $M$, $m$ and $(\alpha_1,\alpha_2)$ are the flux, $\mathbb{Z}_2$ parity and SS phases of $H_u$, $H_d$, $\alpha$, $\beta$ and $\gamma$.


\chapter{Three-generation models on $(T^2_1\times T^2_2)/(\mathbb{Z}_2^{(\textrm{t})}\times \mathbb{Z}_2^{(\textrm{per})})$}
\label{app:three_T2xT2dZ2xZ2}

Here we show possible three-generation models with non-vanishing Yukawa couplings on $(T^2_1\times T^2_2)/(\mathbb{Z}_2^{\textrm{(t)}}\times\mathbb{Z}_2^{\textrm{(per)}})$ in Table \ref{tab:three-generation-models_T2xT2/Z2xZ2}.
\begin{table}[H]
\centering
\caption{Possible three-generation models with non-vanishing Yukawa couplings on $(T^2_1\times T^2_2)/(\mathbb{Z}_2^{\textrm{(t)}}\times\mathbb{Z}_2^{\textrm{(per)}})$.
$B_L$, $B_R$ and $B_H$ denote (flux; $\mathbb{Z}_2^{\textrm{(t)}}$ parity; $\mathbb{Z}_2^{\textrm{(per)}}$ parity; SS phases) of $\psi_L$, $\psi_R$ and $\psi_H$, respectively.
$g_H$ denotes the number of Higgs fields.
We omit three-generation models which is equivalent to the model shown in this table by flipping left- and right-handed matter fields.
}
\label{tab:three-generation-models_T2xT2/Z2xZ2}
\begin{tabular}{c|c|c|c||c|c|c|c} \hline
$B_L$ & $B_R$ & $B_H$ & $g_H$ & 
$B_L$ & $B_R$ & $B_H$ & $g_H$ \\ \hline\hline
2;0;0;0,0 & 2;0;0;0,0 & 4;0;0;0,0 & 6 &
2;0;0;0,0 & 3;0;0;0,0 & 5;0;0;0,0 & 6 \\
2;0;0;0,0 & 6;1;0;0,0 & 8;1;0;0,0 & 6 &
2;0;0;0,0 & 5;1;0;0,0 & 7;1;0;0,0 & 6 \\
2;0;0;0,0 & 4;0;0;$\frac{1}{2}$,0 & 6;0;0;$\frac{1}{2}$,0 & 6 &
2;0;0;0,0 & 3;0;0;$\frac{1}{2}$,0 & 5;0;0;$\frac{1}{2}$,0 & 6 \\
2;0;0;0,0 & 4;1;0;$\frac{1}{2}$,0 & 6;1;0;$\frac{1}{2}$,0 & 6 &
2;0;0;0,0 & 5;1;0;$\frac{1}{2}$,0 & 7;1;0;$\frac{1}{2}$,0 & 6 \\
2;0;0;0,0 & 4;0;0;0,$\frac{1}{2}$ & 6;0;0;0,$\frac{1}{2}$ & 6 &
2;0;0;0,0 & 3;0;0;0,$\frac{1}{2}$ & 5;0;0;0,$\frac{1}{2}$ & 6 \\
2;0;0;0,0 & 4;1;0;0,$\frac{1}{2}$ & 6;1;0;0,$\frac{1}{2}$ & 6 &
2;0;0;0,0 & 5;1;0;0,$\frac{1}{2}$ & 7;1;0;0,$\frac{1}{2}$ & 6 \\
2;0;0;0,0 & 4;0;0;$\frac{1}{2}$,$\frac{1}{2}$ & 6;0;0;$\frac{1}{2}$,$\frac{1}{2}$ & 6 &
2;0;0;0,0 & 5;0;0;$\frac{1}{2}$,$\frac{1}{2}$ & 7;0;0;$\frac{1}{2}$,$\frac{1}{2}$ & 6 \\
\hline
\end{tabular}
\end{table}
\begin{table}[H]
\centering
\begin{tabular}{c|c|c|c||c|c|c|c} \hline
$B_L$ & $B_R$ & $B_H$ & $g_H$ & 
$B_L$ & $B_R$ & $B_H$ & $g_H$ \\ \hline\hline
3;0;0;0,0 & 3;0;0;0,0 & 6;0;0;0,0 & 10 &
3;0;0;0,0 & 6;1;0;0,0 & 9;1;0;0,0 & 10 \\
3;0;0;0,0 & 5;1;0;0,0 & 8;1;0;0,0 & 6 &
3;0;0;0,0 & 4;0;0;$\frac{1}{2}$,0 & 7;0;0;$\frac{1}{2}$,0 & 10 \\
3;0;0;0,0 & 3;0;0;$\frac{1}{2}$,0 & 6;0;0;$\frac{1}{2}$,0 & 6 &
3;0;0;0,0 & 4;1;0;$\frac{1}{2}$,0 & 7;1;0;$\frac{1}{2}$,0 & 6 \\
3;0;0;0,0 & 5;1;0;$\frac{1}{2}$,0 & 8;1;0;$\frac{1}{2}$,0 & 10 &
3;0;0;0,0 & 4;0;0;0,$\frac{1}{2}$ & 7;0;0;0,$\frac{1}{2}$ & 10 \\
3;0;0;0,0 & 3;0;0;0,$\frac{1}{2}$ & 6;0;0;0,$\frac{1}{2}$ & 6 &
3;0;0;0,0 & 4;1;0;0,$\frac{1}{2}$ & 7;1;0;0,$\frac{1}{2}$ & 6 \\
3;0;0;0,0 & 5;1;0;0,$\frac{1}{2}$ & 8;1;0;0,$\frac{1}{2}$ & 10 &
3;0;0;0,0 & 4;0;0;$\frac{1}{2}$,$\frac{1}{2}$ & 7;0;0;$\frac{1}{2}$,$\frac{1}{2}$ & 6 \\
3;0;0;0,0 & 5;0;0;$\frac{1}{2}$,$\frac{1}{2}$ & 8;0;0;$\frac{1}{2}$,$\frac{1}{2}$ & 10 &
3;0;0;0,0 & 4;1;0;$\frac{1}{2}$,$\frac{1}{2}$ & 7;1;0;$\frac{1}{2}$,$\frac{1}{2}$ & 10 \\
\hline
\end{tabular}
\end{table}
\begin{table}[H]
\centering
\begin{tabular}{c|c|c|c||c|c|c|c} \hline
$B_L$ & $B_R$ & $B_H$ & $g_H$ & 
$B_L$ & $B_R$ & $B_H$ & $g_H$ \\ \hline\hline
5;1;0;0,0 & 6;0;1;$\frac{1}{2}$,0 & 11;1;1;$\frac{1}{2}$,0 & 10 &
5;1;0;0,0 & 5;0;1;$\frac{1}{2}$,0 & 10;1;1;$\frac{1}{2}$,0 & 10 \\
5;1;0;0,0 & 6;1;1;$\frac{1}{2}$,0 & 11;0;1;$\frac{1}{2}$,0 & 15 &
5;1;0;0,0 & 7;1;1;$\frac{1}{2}$,0 & 12;0;1;$\frac{1}{2}$,0 & 15 \\
5;1;0;0,0 & 6;0;1;0,$\frac{1}{2}$ & 11;1;1;0,$\frac{1}{2}$ & 10 &
5;1;0;0,0 & 5;0;1;0,$\frac{1}{2}$ & 10;1;1;0,$\frac{1}{2}$ & 10 \\
5;1;0;0,0 & 6;1;1;0,$\frac{1}{2}$ & 11;0;1;0,$\frac{1}{2}$ & 15 &
5;1;0;0,0 & 7;1;1;0,$\frac{1}{2}$ & 12;0;1;0,$\frac{1}{2}$ & 15 \\
5;1;0;0,0 & 6;0;1;$\frac{1}{2}$,$\frac{1}{2}$ & 11;1;1;$\frac{1}{2}$,$\frac{1}{2}$ & 15 &
5;1;0;0,0 & 7;0;1;$\frac{1}{2}$,$\frac{1}{2}$ & 12;1;1;$\frac{1}{2}$,$\frac{1}{2}$ & 15 \\
5;1;0;0,0 & 6;1;1;$\frac{1}{2}$,$\frac{1}{2}$ & 11;0;1;$\frac{1}{2}$,$\frac{1}{2}$ & 10 &
5;1;0;0,0 & 5;1;1;$\frac{1}{2}$,$\frac{1}{2}$ & 10;0;1;$\frac{1}{2}$,$\frac{1}{2}$ & 10 \\
4;0;0;$\frac{1}{2}$,0 & 4;0;0;$\frac{1}{2}$,0 & 8;0;0;0,0 & 15 &
4;0;0;$\frac{1}{2}$,0 & 3;0;0;$\frac{1}{2}$,0 & 7;0;0;0,0 & 10 \\
\hline
\end{tabular}
\end{table}
\begin{table}[H]
\centering
\begin{tabular}{c|c|c|c||c|c|c|c} \hline
$B_L$ & $B_R$ & $B_H$ & $g_H$ & 
$B_L$ & $B_R$ & $B_H$ & $g_H$ \\ \hline\hline
4;1;0;$\frac{1}{2}$,0 & 5;0;1;0,0 & 9;1;1;$\frac{1}{2}$,0 & 6 &
4;1;0;$\frac{1}{2}$,0 & 8;1;1;0,0 & 12;0;1;$\frac{1}{2}$,0 & 15 \\
4;1;0;$\frac{1}{2}$,0 & 7;1;1;0,0 & 11;0;1;$\frac{1}{2}$,0 & 15 &
4;1;0;$\frac{1}{2}$,0 & 6;0;1;$\frac{1}{2}$,0 & 10;1;1;0,0 & 6 \\
4;1;0;$\frac{1}{2}$,0 & 5;0;1;$\frac{1}{2}$,0 & 9;1;1;0,0 & 6 &
4;1;0;$\frac{1}{2}$,0 & 6;1;1;$\frac{1}{2}$,0 & 10;0;1;0,0 & 15 \\
4;1;0;$\frac{1}{2}$,0 & 7;1;1;$\frac{1}{2}$,0 & 11;0;1;0,0 & 15 &
4;1;0;$\frac{1}{2}$,0 & 6;0;1;0,$\frac{1}{2}$ & 10;1;1;$\frac{1}{2}$,$\frac{1}{2}$ & 10 \\
4;1;0;$\frac{1}{2}$,0 & 5;0;1;0,$\frac{1}{2}$ & 9;1;1;$\frac{1}{2}$,$\frac{1}{2}$ & 10 &
4;1;0;$\frac{1}{2}$,0 & 6;1;1;0,$\frac{1}{2}$ & 10;0;1;$\frac{1}{2}$,$\frac{1}{2}$ & 10 \\
4;1;0;$\frac{1}{2}$,0 & 7;1;1;0,$\frac{1}{2}$ & 11;0;1;$\frac{1}{2}$,$\frac{1}{2}$ & 10 &
4;1;0;$\frac{1}{2}$,0 & 6;0;1;$\frac{1}{2}$,$\frac{1}{2}$ & 10;1;1;0,$\frac{1}{2}$ & 10 \\
4;1;0;$\frac{1}{2}$,0 & 7;0;1;$\frac{1}{2}$,$\frac{1}{2}$ & 11;1;1;0,$\frac{1}{2}$ & 10 &
4;1;0;$\frac{1}{2}$,0 & 6;1;1;$\frac{1}{2}$,$\frac{1}{2}$ & 10;0;1;0,$\frac{1}{2}$ & 10 \\
\hline
\end{tabular}
\end{table}
\begin{table}[H]
\centering
\begin{tabular}{c|c|c|c||c|c|c|c} \hline
$B_L$ & $B_R$ & $B_H$ & $g_H$ & 
$B_L$ & $B_R$ & $B_H$ & $g_H$ \\ \hline\hline
3;0;0;0,$\frac{1}{2}$ & 7;1;1;$\frac{1}{2}$,0 & 10;1;1;$\frac{1}{2}$,$\frac{1}{2}$ & 10 &
3;0;0;0,$\frac{1}{2}$ & 6;0;1;0,$\frac{1}{2}$ & 9;0;1;0,0 & 10 \\
3;0;0;0,$\frac{1}{2}$ & 5;0;1;0,$\frac{1}{2}$ & 8;0;1;0,0 & 10 &
3;0;0;0,$\frac{1}{2}$ & 6;1;1;0,$\frac{1}{2}$ & 9;1;1;0,0 & 6 \\
3;0;0;0,$\frac{1}{2}$ & 7;1;1;0,$\frac{1}{2}$ & 10;1;1;0,0 & 6 &
3;0;0;0,$\frac{1}{2}$ & 6;0;1;$\frac{1}{2}$,$\frac{1}{2}$ & 9;0;1;$\frac{1}{2}$,0 & 10 \\
3;0;0;0,$\frac{1}{2}$ & 7;0;1;$\frac{1}{2}$,$\frac{1}{2}$ & 10;0;1;$\frac{1}{2}$,0 & 10 &
3;0;0;0,$\frac{1}{2}$ & 6;1;1;$\frac{1}{2}$,$\frac{1}{2}$ & 9;1;1;$\frac{1}{2}$,0 & 6 \\
3;0;0;0,$\frac{1}{2}$ & 5;1;1;$\frac{1}{2}$,$\frac{1}{2}$ & 8;1;1;$\frac{1}{2}$,0 & 6 &
4;1;0;0,$\frac{1}{2}$ & 4;1;0;0,$\frac{1}{2}$ & 8;0;0;0,0 & 15 \\
4;1;0;0,$\frac{1}{2}$ & 5;1;0;0,$\frac{1}{2}$ & 9;0;0;0,0 & 15 &
4;1;0;0,$\frac{1}{2}$ & 4;0;0;$\frac{1}{2}$,$\frac{1}{2}$ & 8;1;0;$\frac{1}{2}$,0 & 10 \\
4;1;0;0,$\frac{1}{2}$ & 5;0;0;$\frac{1}{2}$,$\frac{1}{2}$ & 9;1;0;$\frac{1}{2}$,0 & 10 &
4;1;0;0,$\frac{1}{2}$ & 4;1;0;$\frac{1}{2}$,$\frac{1}{2}$ & 8;0;0;$\frac{1}{2}$,0 & 10 \\
\hline
\end{tabular}
\end{table}
\begin{table}[H]
\centering
\begin{tabular}{c|c|c|c||c|c|c|c} \hline
$B_L$ & $B_R$ & $B_H$ & $g_H$ & 
$B_L$ & $B_R$ & $B_H$ & $g_H$ \\ \hline\hline
5;0;0;$\frac{1}{2}$,$\frac{1}{2}$ & 7;1;1;0,0 & 12;1;1;$\frac{1}{2}$,$\frac{1}{2}$ & 15 &
5;0;0;$\frac{1}{2}$,$\frac{1}{2}$ & 6;0;1;$\frac{1}{2}$,0 & 11;0;1;0,$\frac{1}{2}$ & 15 \\
5;0;0;$\frac{1}{2}$,$\frac{1}{2}$ & 5;0;1;$\frac{1}{2}$,0 & 10;0;1;0,$\frac{1}{2}$ & 10 &
5;0;0;$\frac{1}{2}$,$\frac{1}{2}$ & 6;1;1;$\frac{1}{2}$,0 & 11;1;1;0,$\frac{1}{2}$ & 10 \\
5;0;0;$\frac{1}{2}$,$\frac{1}{2}$ & 7;1;1;$\frac{1}{2}$,0 & 12;1;1;0,$\frac{1}{2}$ & 15 &
5;0;0;$\frac{1}{2}$,$\frac{1}{2}$ & 6;0;1;0,$\frac{1}{2}$ & 11;0;1;$\frac{1}{2}$,0 & 15 \\
5;0;0;$\frac{1}{2}$,$\frac{1}{2}$ & 5;0;1;0,$\frac{1}{2}$ & 10;0;1;$\frac{1}{2}$,0 & 10 &
5;0;0;$\frac{1}{2}$,$\frac{1}{2}$ & 6;1;1;0,$\frac{1}{2}$ & 11;1;1;$\frac{1}{2}$,0 & 10 \\
5;0;0;$\frac{1}{2}$,$\frac{1}{2}$ & 7;1;1;0,$\frac{1}{2}$ & 12;1;1;$\frac{1}{2}$,0 & 15 &
5;0;0;$\frac{1}{2}$,$\frac{1}{2}$ & 6;0;1;$\frac{1}{2}$,$\frac{1}{2}$ & 11;0;1;0,0 & 15 \\
5;0;0;$\frac{1}{2}$,$\frac{1}{2}$ & 7;0;1;$\frac{1}{2}$,$\frac{1}{2}$ & 12;0;1;0,0 & 21 &
5;0;0;$\frac{1}{2}$,$\frac{1}{2}$ & 6;1;1;$\frac{1}{2}$,$\frac{1}{2}$ & 11;1;1;0,0 & 10 \\
5;0;0;$\frac{1}{2}$,$\frac{1}{2}$ & 5;1;1;$\frac{1}{2}$,$\frac{1}{2}$ & 10;1;1;0,0 & 6 &
4;1;0;$\frac{1}{2}$,$\frac{1}{2}$ & 4;1;0;$\frac{1}{2}$,$\frac{1}{2}$ & 8;0;0;0,0 & 15 \\
\hline
\end{tabular}
\end{table}
\begin{table}[H]
\centering
\begin{tabular}{c|c|c|c||c|c|c|c} \hline
$B_L$ & $B_R$ & $B_H$ & $g_H$ & 
$B_L$ & $B_R$ & $B_H$ & $g_H$ \\ \hline\hline
5;0;1;0,0 & 5;1;1;$\frac{1}{2}$,$\frac{1}{2}$ & 10;1;0;$\frac{1}{2}$,$\frac{1}{2}$ & 15 &
8;1;1;0,0 & 8;1;1;0,0 & 16;0;0;0,0 & 45 \\
8;1;1;0,0 & 7;1;1;0,0 & 15;0;0;0,0 & 36 &
8;1;1;0,0 & 6;0;1;$\frac{1}{2}$,0 & 14;1;0;$\frac{1}{2}$,0 & 28 \\
8;1;1;0,0 & 5;0;1;$\frac{1}{2}$,0 & 13;1;0;$\frac{1}{2}$,0 & 21 &
8;1;1;0,0 & 6;1;1;$\frac{1}{2}$,0 & 14;0;0;$\frac{1}{2}$,0 & 28 \\
8;1;1;0,0 & 7;1;1;$\frac{1}{2}$,0 & 15;0;0;$\frac{1}{2}$,0 & 36 &
8;1;1;0,0 & 6;0;1;0,$\frac{1}{2}$ & 14;1;0;0,$\frac{1}{2}$ & 28 \\
8;1;1;0,0 & 5;0;1;0,$\frac{1}{2}$ & 13;1;0;0,$\frac{1}{2}$ & 21 &
8;1;1;0,0 & 6;1;1;0,$\frac{1}{2}$ & 14;0;0;0,$\frac{1}{2}$ & 28 \\
8;1;1;0,0 & 7;1;1;0,$\frac{1}{2}$ & 15;0;0;0,$\frac{1}{2}$ & 36 &
8;1;1;0,0 & 6;0;1;$\frac{1}{2}$,$\frac{1}{2}$ & 14;1;0;$\frac{1}{2}$,$\frac{1}{2}$ & 28 \\
8;1;1;0,0 & 7;0;1;$\frac{1}{2}$,$\frac{1}{2}$ & 15;1;0;$\frac{1}{2}$,$\frac{1}{2}$ & 36 &
8;1;1;0,0 & 6;1;1;$\frac{1}{2}$,$\frac{1}{2}$ & 14;0;0;$\frac{1}{2}$,$\frac{1}{2}$ & 28 \\
\hline
\end{tabular}
\end{table}
\begin{table}[H]
\centering
\begin{tabular}{c|c|c|c||c|c|c|c} \hline
$B_L$ & $B_R$ & $B_H$ & $g_H$ & 
$B_L$ & $B_R$ & $B_H$ & $g_H$ \\ \hline\hline
5;0;1;0,$\frac{1}{2}$ & 5;0;1;0,$\frac{1}{2}$ & 10;0;0;0,0 & 21 &
5;0;1;0,$\frac{1}{2}$ & 6;1;1;0,$\frac{1}{2}$ & 11;1;0;0,0 & 15 \\
5;0;1;0,$\frac{1}{2}$ & 7;1;1;0,$\frac{1}{2}$ & 12;1;0;0,0 & 15 &
5;0;1;0,$\frac{1}{2}$ & 6;0;1;$\frac{1}{2}$,$\frac{1}{2}$ & 11;0;0;$\frac{1}{2}$,0 & 21 \\
5;0;1;0,$\frac{1}{2}$ & 7;0;1;$\frac{1}{2}$,$\frac{1}{2}$ & 12;0;0;$\frac{1}{2}$,0 & 21 &
5;0;1;0,$\frac{1}{2}$ & 6;1;1;$\frac{1}{2}$,$\frac{1}{2}$ & 11;1;0;$\frac{1}{2}$,0 & 15 \\
5;0;1;0,$\frac{1}{2}$ & 5;1;1;$\frac{1}{2}$,$\frac{1}{2}$ & 10;1;0;$\frac{1}{2}$,0 & 15 &
6;1;1;0,$\frac{1}{2}$ & 6;1;1;0,$\frac{1}{2}$ & 12;0;0;0,0 & 28 \\
6;1;1;0,$\frac{1}{2}$ & 7;1;1;0,$\frac{1}{2}$ & 13;0;0;0,0 & 28 &
6;1;1;0,$\frac{1}{2}$ & 6;0;1;$\frac{1}{2}$,$\frac{1}{2}$ & 12;1;0;$\frac{1}{2}$,0 & 21 \\
6;1;1;0,$\frac{1}{2}$ & 7;0;1;$\frac{1}{2}$,$\frac{1}{2}$ & 13;1;0;$\frac{1}{2}$,0 & 21 &
6;1;1;0,$\frac{1}{2}$ & 6;1;1;$\frac{1}{2}$,$\frac{1}{2}$ & 12;0;0;$\frac{1}{2}$,0 & 21 \\
6;1;1;0,$\frac{1}{2}$ & 5;1;1;$\frac{1}{2}$,$\frac{1}{2}$ & 11;0;0;$\frac{1}{2}$,0 & 21 &
7;1;1;0,$\frac{1}{2}$ & 7;1;1;0,$\frac{1}{2}$ & 14;0;0;0,0 & 36 \\
\hline
\end{tabular}
\end{table}


\chapter{Decompositions of zero-modes with higher fluxes}
\label{Decompositions of zero-modes with higher fluxes}

Here we study the decompositions of zero-mode wave functions on the magnetized $T^2/\mathbb{Z}_2$, $T^2/\mathbb{Z}_3$, $T^2/\mathbb{Z}_4$ and $T^2/\mathbb{Z}_6$ orbifolds.
As we have studied in Section \ref{subsec:Torus compactification}, zero-modes on magnetized $T^2$ have degenerate solutions.
The degeneracy number of zero-modes is given by the size of flux.
Then we find that zero-modes on orbifolds with higher fluxes are decomposed into the products of ones with lower fluxes.
Also such decompositions give a simple way counting degeneracy number of zero-modes on magnetized orbifolds.


\section{$T^2/\mathbb{Z}_2$}

First, let us consider zero-modes on $T^2/\mathbb{Z}_2$ with flux $M$, SS phases $(\alpha_1,\alpha_2)$ and $\mathbb{Z}_2$ parity $m$, $\psi^{(j+\alpha_1,\alpha_2),M}_{T^2/\mathbb{Z}_2^m}(z,\tau)$.
On $T^2/\mathbb{Z}_2$ orbifold, the complex coordinate, $z$, is identified as $z\sim -z$.
The fixed points for $\mathbb{Z}_2$ twist $z\to -z$ are:
\begin{align}
z = 0,~\frac{1}{2},~\frac{\tau}{2},~\frac{1+\tau}{2}.
\end{align}
Then SS phases are restricted to $(\alpha_1,\alpha_2)=(0,0)$ and $(1/2,1/2)$.
To express the zero-modes with higher fluxes by ones with lower fluxes, we use following four single zero-modes: zero-mode with flux 1, SS phases (0,0) and $\mathbb{Z}_2$ even parity, $\psi^{(0,0),1}_{T^2/\mathbb{Z}_2^0}$, zero-mode with flux 1, SS phases (1/2,0) and $\mathbb{Z}_2$ even parity, $\psi^{(1/2,0),1}_{T^2/\mathbb{Z}_2^0}$, zero-mode with flux 1, SS phases (0,1/2) and $\mathbb{Z}_2$ even parity, $\psi^{(0,1/2),1}_{T^2/\mathbb{Z}_2^0}$, and zero-mode with flux 1, SS phases (1/2,1/2) and $\mathbb{Z}_2$ odd parity, $\psi^{(1/2,1/2),1}_{T^2/\mathbb{Z}_2^1}$.
We denote them as $\bm{0}_{00}$, $\bm{0}_{10}$, $\bm{0}_{01}$ and $\bm{1}_{11}$, respectively.
As we have discussed in Subsection \ref{subsec:Zero points of zero-modes}, zero-modes on $T^2/\mathbb{Z}_2$ have zero points at fixed points.
We show zero points of $\bm{0}_{00}$, $\bm{0}_{10}$, $\bm{0}_{01}$ and $\bm{1}_{11}$ in Table \ref{tab:zero points of singles T2/Z2}.
\begin{table}[H]
\centering
\begin{tabular}{c|ccccc} \hline
zero-modes & flux & SS phases & $\mathbb{Z}_2$ parity & number of zero-modes & zero points \\ \hline
$\bm{0}_{00}\equiv\psi^{(0,0),1}_{T^2/\mathbb{Z}_2^0}$ & 1 & (0,0) & even & 1 & $\frac{1+\tau}{2}$ \\
$\bm{0}_{10}\equiv\psi^{(1/2,0),1}_{T^2/\mathbb{Z}_2^0}$ & 1 & (1/2,0) & even & 1 & $\frac{1}{2}$ \\
$\bm{0}_{01}\equiv\psi^{(0,1/2),1}_{T^2/\mathbb{Z}_2^0}$ & 1 & (0,1/2) & even & 1 & $\frac{\tau}{2}$ \\
$\bm{1}_{11}\equiv\psi^{(1/2,1/2),1}_{T^2/\mathbb{Z}_2^1}$ & 1 & (1/2,1/2) & odd & 1 & 0 \\ \hline
\end{tabular}
\caption{Zero points of single generation zero-modes on $T^2/\mathbb{Z}_2$.}
\label{tab:zero points of singles T2/Z2}
\end{table}
Using these four zero-modes with flux 1, zero-modes with higher fluxes can be decomposed.
Let us see an example.
We introduce
\begin{align}
Y_0^{(0,0),2} \equiv 
\begin{pmatrix}
\bm{0}_{00}^2 \\
\bm{1}_{11}^2 \\
\end{pmatrix}.
\end{align}
Since $\bm{1}_{11}$ vanishes at $z=0$ while $\bm{0}_{00}$ does not vanish, $Y_0^{(0,0),2}$ spans 2D spaces.
Then, the products of $Y_0^{(0,0),2}$,
\begin{align}
Y_0^{(0,0),4} \equiv
Y_0^{(0,0),2} \otimes Y_0^{(0,0),2}
=
\begin{pmatrix}
\bm{0}_{00}^4 \\
\bm{0}_{00}^2 \bm{1}_{11}^2 \\
\bm{1}_{11}^4 \\
\end{pmatrix},
\end{align}
spans 3D spaces.
This is because $(\bm{0}_{00}^2\bm{1}_{11}^2, \bm{1}_{11}^4)=\bm{1}_{11}^2Y_0^{(0,0),2}$ spanning 2D spaces vanish at $z=0$ while $\bm{0}_{00}^4$ does not vanish.
That is, $\bm{0}_{00}^4$, $\bm{0}_{00}^2$ and $\bm{1}_{11}^2$ are linearly independent.
Note that $Y_0^{(0,0),4}$ obeys the Dirac equation and the boundary conditions for zero-modes with flux 4 and SS phases (0,0).
In addition it has $\mathbb{Z}_2$ even parity.
Meanwhile, we find the product,
\begin{align}
Y_1^{(0,0),4} \equiv \bm{0}_{00}\bm{0}_{10}\bm{0}_{01}\bm{1}_{11},
\end{align}
which obeys the Dirac equation and the boundary conditions for zero-modes with flux 4 and SS phases (0,0) but has $\mathbb{Z}_2$ odd parity.
Notice that sum of degeneracy numbers of $\mathbb{Z}_2$ even and odd modes on $T^2/\mathbb{Z}_2$ is equivalent to degeneracy number of zero-modes on $T^2$.
Since zero-modes on $T^2$ with flux $M$ have $M$ degenerate solutions, $Y_0^{(0,0),4}$ and $Y_1^{(0,0),4}$ can be regarded as the complete basis of even and odd modes with flux 4 and SS phases (0,0), respectively.
Thus even and odd modes with flux $4$ and SS phases (0,0), $\psi_{T^2/\mathbb{Z}_2^0}^{(j,0),4}$ and $\psi_{T^2/\mathbb{Z}_2^1}^{(j,0),4}$, can be expanded by $Y_0^{(0,0),4}$ and $Y_1^{(0,0),4}$, respectively.
This means that $\psi_{T^2/\mathbb{Z}_2^0}^{(j,0),4}$ and $\psi_{T^2/\mathbb{Z}_2^1}^{(0,0),4}$ have 3 and 1 degenerate solutions.
This result is consistent with the number of zero-modes on $T^2/\mathbb{Z}_2$ shown in Table \ref{tab:num_T2/Z2}.

This procedure is generalized as follows.
We introduce
\begin{align}
Y_0^{(0,0),2n} 
\equiv \underbrace{Y_0^{(0,0),2} \otimes Y_0^{(0,0),2} \otimes \cdots \otimes Y_0^{(0,0),2}}_{\textrm{products~of~}n\textrm{~of~}Y_0^{(0,0),2}}
=
\begin{pmatrix}
\bm{0}_{00}^{2n} \\
\bm{0}_{00}^{2n-2} \bm{1}_{11}^2 \\
\vdots \\
\bm{0}_{00}^{2n-2m} \bm{1}_{11}^{2m} \\
\vdots \\
\bm{0}_{00}^{2} \bm{1}_{11}^{2n-2} \\
\bm{1}_{11}^{2n} \\
\end{pmatrix}, \quad m \in \{0,1,...,n\}.
\label{eq:Z2 Y_0^(0,0),2n}
\end{align}
In this stage it is unknown whether $n+1$ components of $Y_0^{(0,0),2n}$ are independent each other.
If $Y_0^{(0,0),2n}$ spans $n+1$-dimensional spaces,
\begin{align}
Y_0^{(0,0),2n+2} \equiv Y_0^{(0,0),2} \otimes Y_0^{(0,0),2n}
=
\begin{pmatrix}
\bm{0}_{00}^{2n+2} \\
\bm{0}_{00}^{2n} \bm{1}_{11}^2 \\
\vdots \\
\bm{0}_{00}^{2n+2-2m} \bm{1}_{11}^{2m} \\
\vdots \\
\bm{0}_{00}^{2} \bm{1}_{11}^{2n} \\
\bm{1}_{11}^{2n+2} \\
\end{pmatrix}
=
\begin{pmatrix}
\bm{0}_{00}^{2n+2} \\
\bm{1}_{11}^2Y_0^{(0,0),2n} \\
\end{pmatrix},
\end{align}
spans $n+2$-dimensional spaces because $\bm{1}_{11}^2Y_0^{(0,0),2n}$ spanning $n+1$-dimensional spaces vanishes at $z=0$ while $\bm{0}_{00}^{2n+2}$ does not vanish.
We have already known that $Y_0^{(0,0),2}$ spans 2D spaces; therefore $Y_0^{(0,0),2n}$ spans $n+1$-dimensional spaces for $n\geq 1$.
Note that $Y_0^{(0,0),2n}$ has flux $2n$, SS phases (0,0) and $\mathbb{Z}_2$ even parity.
Then, we can find the products with flux $2n$, SS phases (0,0) and $\mathbb{Z}_2$ odd parity,
\begin{align}
Y_1^{(0,0),2n} \equiv \bm{0}_{00}\bm{0}_{10}\bm{0}_{01}\bm{1}_{11} Y_0^{(0,0),2n-4},
\end{align}
which spans $n-1$-dimensional spaces for $n\geq 2$.
Since zero-modes on $T^2$ with flux $2n$ have $2n$ degenerate solutions, $Y_0^{(0,0),2n}$ and $Y_1^{(0,0),2n}$ can be regarded as the complete basis of even modes and odd modes on $T^2/\mathbb{Z}_2$, respectively.
Thus even and odd modes with flux $2n$ and SS phases (0,0), $\psi_{T^2/\mathbb{Z}_2^0}^{(j,0),2n}$ and $\psi_{T^2/\mathbb{Z}_2^1}^{(j,0),2n}$, can be expanded by $Y_0^{(0,0),2n}$ and $Y_1^{(0,0),2n}$, respectively.
This means that $\psi_{T^2/\mathbb{Z}_2^0}^{(j,0),2n}$ and $\psi_{T^2/\mathbb{Z}_2^1}^{(j,0),2n}$ span $n+1$ and $n-1$-dimensional spaces.

In the similar way, we can find that even and odd modes with flux $2n+1$ and SS phases (0,0), $\psi_{T^2/\mathbb{Z}_2^0}^{(j+0,0),2n+1}$ and $\psi_{T^2/\mathbb{Z}_2^1}^{(j+0,0),2n+1}$, can be expanded by
\begin{align}
Y_0^{(0,0),2n+1} \equiv \bm{0}_{00} Y_0^{(0,0),2n},
\end{align}
and
\begin{align}
Y_1^{(0,0),2n+1} \equiv \bm{0}_{10}\bm{0}_{01}\bm{1}_{11} Y_0^{(0,0),2n-2},
\end{align}
respectively.
This means that $\psi_{T^2/\mathbb{Z}_2^0}^{(j+0,0),2n+1}$ and $\psi_{T^2/\mathbb{Z}_2^1}^{(j+0,0),2n+1}$ span $n+1$ and $n$-dimensional spaces, respectively.
The sum of degeneracy numbers of each $\mathbb{Z}_2$ mode is equivalent to degeneracy number of zero-modes on $T^2$.
Therefore this result is consistent.

Similar analysis can be applied for zero-modes on $T^2/\mathbb{Z}_2$ with non-vanishing SS phases.
We summarize the results in Table \ref{tab:expansionT2/Z2}.
One can check that basis of decompositions in fourth column has flux, $\mathbb{Z}_2$ parity and SS phases shown in first to third columns.
\begin{table}[H]
\begin{center}
\renewcommand{\arraystretch}{1.0}
\begin{tabular}{ccccc} \hline
flux & $\mathbb{Z}_2$ parity & SS phases & basis of decompositions & number of basis \\ \hline
\multirow{2}{*}{$2n$} & even & (0,0) & $Y_0^{(0,0),2n}$ & $n+1$ \\
& odd & (0,0) & $\bm{0}_{00}\bm{0}_{10}\bm{0}_{01}\bm{1}_{11}Y_0^{(0,0),2n-4}$ & $n-1$ \\ \hline
\multirow{2}{*}{$2n$} & even & (1/2,0) & $\bm{0}_{00}\bm{0}_{10}Y_0^{(0,0),2n-2}$ & $n$ \\
& odd & (1/2,0) & $\bm{0}_{01}\bm{1}_{11}Y_0^{(0,0),2n-2}$ & $n$ \\ \hline
\multirow{2}{*}{$2n$} & even & (0,1/2)  & $\bm{0}_{00}\bm{0}_{01}Y_0^{(0,0),2n-2}$ & $n$\\
& odd & (0,1/2)  & $\bm{0}_{10}\bm{1}_{11}Y_0^{(0,0),2n-2}$ & $n$\\ \hline
\multirow{2}{*}{$2n$} & even & (1/2,1/2)  & $\bm{0}_{10}\bm{0}_{01}Y_0^{(0,0),2n-2}$ & $n$\\
& odd & (1/2,1/2)  & $\bm{0}_{00}\bm{1}_{11}Y_0^{(0,0),2n-2}$ & $n$\\ \hline
\multirow{2}{*}{$2n+1$} & even & (0,0) & $\bm{0}_{00}Y_0^{(0,0),2n}$ & $n+1$ \\
& odd & (0,0) & $\bm{0}_{10}\bm{0}_{01}\bm{1}_{11}Y_0^{(0,0),2n-2}$ & $n$ \\ \hline
\multirow{2}{*}{$2n+1$} & even & (1/2,0) & $\bm{0}_{10}Y_0^{(0,0),2n}$ & $n+1$ \\
& odd & (1/2,0) & $\bm{0}_{00}\bm{0}_{01}\bm{1}_{11}Y_0^{(0,0),2n-2}$ & $n$ \\ \hline
\multirow{2}{*}{$2n+1$} & even & (0,1/2)  & $\bm{0}_{01}Y_0^{(0,0),2n}$ & $n+1$\\
& odd & (0,1/2)  & $\bm{0}_{00}\bm{0}_{10}\bm{1}_{11}Y_0^{(0,0),2n-2}$ & $n$\\ \hline
\multirow{2}{*}{$2n+1$} & even & (1/2,1/2)  & $\bm{0}_{00}\bm{0}_{10}\bm{0}_{01}Y_0^{(0,0),2n-2}$ & $n$\\
& odd & (1/2,1/2)  & $\bm{1}_{11}Y_0^{(0,0),2n}$ & $n+1$\\ \hline
\end{tabular}
\end{center}
\caption{Basis of decompositions of zero-modes on $T^2/\mathbb{Z}_2$ with higher fluxes.
$Y_0^{(0,0),2n}$ is defined by Eq.~(\ref{eq:Z2 Y_0^(0,0),2n}).
In each row, the value of flux is equivalent to the sum of number of basis.
Numbers of basis are consistent with Table \ref{tab:num_T2/Z2}.}
\label{tab:expansionT2/Z2}
\end{table}


\section{$T^2/\mathbb{Z}_3$}

Second, let us consider zero-modes on $T^2/\mathbb{Z}_3$ with flux $M$, SS phases $(\alpha_1,\alpha_2)$ and $\mathbb{Z}_3$ parity $m$, $\psi^{(j+\alpha_1,\alpha_2),M}_{T^2/\mathbb{Z}_3^m}(z,\tau)$.
On $T^2/\mathbb{Z}_3$ orbifold, the complex coordinate, $z$, is identified as $z\sim\omega z$, $\omega=e^{2\pi i/3}$.
Similarly, the complex structure modulus $\tau$ is fixed at $\tau=\omega$.
The fixed points for $\mathbb{Z}_3$ twist $z\to \omega z$ are:
\begin{align}
z = 0,~\frac{2+\omega}{3},~\frac{1+2\omega}{3}.
\end{align}
Then SS phases are restricted to $(\alpha_1,\alpha_2)=(0,0)$, $(1/3,1/3)$ and $(2/3,2/3)$ for $M\in2\mathbb{Z}$ and $(\alpha_1,\alpha_2)=(1/6,1/6)$, $(1/2,1/2)$ and $(5/6,5/6)$ for $M\in2\mathbb{Z}+1$.
To express the zero-modes with higher fluxes by ones with lower fluxes, we use following three single zero-modes: zero-mode with flux 1, SS phases (1/6,1/6) and $\mathbb{Z}_3$ parity 1, $\psi^{(1/6,1/6),1}_{T^2/\mathbb{Z}_3^0}$, zero-mode with flux 1, SS phases (5/6,5/6) and $\mathbb{Z}_3$ parity 1, $\psi^{(5/6,5/6),1}_{T^2/\mathbb{Z}_3^0}$, and zero-mode with flux 1, SS phases (1/2,1/2) and $\mathbb{Z}_3$ parity $\omega$, $\psi^{(1/2,1/2),1}_{T^2/\mathbb{Z}_3^1}$.
We denote them as $\bm{0}_{11}$, $\bm{0}_{55}$ and $\bm{1}_{33}$, respectively.
As same as zero-modes on $T^2/\mathbb{Z}_2$, they have zero points at fixed points.
We show zero points of $\bm{0}_{11}$, $\bm{0}_{55}$ and $\bm{1}_{33}$ in Table \ref{tab:zero points of singles T2/Z3}.
\begin{table}[H]
\centering
\begin{tabular}{c|ccccc} \hline
zero-modes & flux & SS phases & $\mathbb{Z}_3$ parity & number of zero-modes & zero points \\ \hline
$\bm{0}_{11}\equiv\psi^{(1/6,1/6),1}_{T^2/\mathbb{Z}_3^0}$ & 1 & (1/6,1/6) & 1 & 1 & $\frac{2+\omega}{3}$ \\
$\bm{0}_{55}\equiv\psi^{(5/6,5/6),1}_{T^2/\mathbb{Z}_3^0}$ & 1 & (5/6,5/6) & 1 & 1 & $\frac{1+2\omega}{3}$ \\
$\bm{1}_{33}\equiv\psi^{(1/2,1/2),1}_{T^2/\mathbb{Z}_3^1}$ & 1 & (1/2,1/2) & $\omega$ & 1 & 0 \\ \hline
\end{tabular}
\caption{Zero points of single generation zero-modes on $T^2/\mathbb{Z}_3$.}
\label{tab:zero points of singles T2/Z3}
\end{table}
Using these three zero-modes with flux 1, zero-modes with higher fluxes can be decomposed.
Consequently, we obtain the decompositions of zero-modes on $T^2/\mathbb{Z}_3$ as shown in Table \ref{tab:expansionT2/Z3}.
One can check that basis of decompositions in fourth column has flux, $\mathbb{Z}_3$ parity and SS phases shown in first to third columns.
\begin{table}[H]
\caption{Basis of decompositions of zero-modes on $T^2/\mathbb{Z}_3$ with higher fluxes.
$Y_0^{(0,0),6n}$ and $Y_0^{(1/2,1/2),6n+3}$ are defined by Eqs.~(\ref{eq:Z3 Y_0^(0,0),6n}) and (\ref{eq:Z3 Y_0^(1/2,1/2),6n+3}), respectively.
In each row, the value of flux is equivalent to the sum of number of basis.
Numbers of basis are consistent with the results derived before \cite{Kobayashi:2017dyu, Abe:2014noa, Abe:2013bca, Sakamoto:2020pev}.}
\label{tab:expansionT2/Z3}
\begin{center}
\renewcommand{\arraystretch}{1.0}
\begin{tabular}{ccccc} \hline
flux & $\mathbb{Z}_3$ parity & SS phases & basis of decompositions & number of basis \\ \hline
\multirow{3}{*}{$6n$} & 1 & (0,0) & $Y_0^{(0,0),6n}$ & $2n+1$ \\
& $\omega$ & (0,0) & $\bm{0}_{11}\bm{0}_{55}\bm{1}_{33}Y_0^{(1/2,1/2),6n-3}$ & $2n$ \\
& $\omega^2$ & (0,0) & $\bm{0}_{11}^2\bm{0}_{55}^2\bm{1}_{33}^2Y_0^{(0,0),6n-6}$ & $2n-1$ \\ \hline
\multirow{3}{*}{$6n$} & 1 & (1/3,1/3) & $\bm{0}_{11}\bm{0}_{55}^2Y_0^{(1/2,1/2),6n-3}$ & $2n$ \\
& $\omega$ & (1/3,1/3) & $\bm{0}_{11}^2\bm{1}_{33}Y_0^{(1/2,1/2),6n-3}$ & $2n$ \\
& $\omega^2$ & (1/3,1/3) & $\bm{0}_{55}\bm{1}_{33}^2Y_0^{(1/2,1/2),6n-3}$ & $2n$ \\ \hline
\multirow{3}{*}{$6n$} & 1 & (2/3,2/3) & $\bm{0}_{11}^2\bm{0}_{55}Y_0^{(1/2,1/2),6n-3}$ & $2n$ \\
& $\omega$ & (2/3,2/3) & $\bm{0}_{55}^2\bm{1}_{33}Y_0^{(1/2,1/2),6n-3}$ & $2n$ \\
& $\omega^2$ & (2/3,2/3) & $\bm{0}_{11}\bm{1}_{33}^2Y_0^{(1/2,1/2),6n-3}$ & $2n$ \\ \hline
\end{tabular}
\end{center}
\end{table}
\begin{table}[H]
\begin{center}
\renewcommand{\arraystretch}{1.0}
\begin{tabular}{ccccc} \hline
flux & $\mathbb{Z}_3$ parity & SS phases & basis of decompositions & number of basis \\ \hline
\multirow{3}{*}{$6n+1$} & 1 & (1/6,1/6) & $\bm{0}_{11}Y_0^{(0,0),6n}$ & $2n+1$ \\
& $\omega$ & (1/6,1/6) & $\bm{0}_{11}^2\bm{0}_{55}\bm{1}_{33}Y_0^{(1/2,1/2),6n-3}$ & $2n$ \\
& $\omega^2$ & (1/6,1/6) & $\bm{0}_{55}^2\bm{1}_{33}^2Y_0^{(1/2,1/2),6n-3}$ & $2n$ \\ \hline
\multirow{3}{*}{$6n+1$} & 1 & (1/2,1/2) & $\bm{0}_{11}^2\bm{0}_{55}^2Y_0^{(1/2,1/2),6n-3}$ & $2n$ \\
& $\omega$ & (1/2,1/2) & $\bm{1}_{33}Y_0^{(0,0),6n}$ & $2n+1$ \\
& $\omega^2$ & (1/2,1/2) & $\bm{0}_{11}\bm{0}_{55}\bm{1}_{33}^2Y_0^{(1/2,1/2),6n-3}$ & $2n$ \\ \hline
\multirow{3}{*}{$6n+1$} & 1 & (5/6,5/6) & $\bm{0}_{55}Y_0^{(0,0),6n}$ & $2n+1$ \\
& $\omega$ & (5/6,5/6) & $\bm{0}_{11}\bm{0}_{55}^2\bm{1}_{33}Y_0^{(1/2,1/2),6n-3}$ & $2n$ \\
& $\omega^2$ & (5/6,5/6) & $\bm{0}_{11}^2\bm{1}_{33}^2Y_0^{(1/2,1/2),6n-3}$ & $2n$ \\ \hline

\multirow{3}{*}{$6n+2$} & 1 & (0,0) & $\bm{0}_{11}\bm{0}_{55}Y_0^{(0,0),6n}$ & $2n+1$ \\
& $\omega$ & (0,0) & $\bm{0}_{11}^2\bm{0}_{55}^2\bm{1}_{33}Y_0^{(1/2,1/2),6n-3}$ & $2n$ \\
& $\omega^2$ & (0,0) & $\bm{1}_{33}^2Y_0^{(0,0),6n}$ & $2n+1$ \\ \hline
\multirow{3}{*}{$6n+2$} & 1 & (1/3,1/3) & $\bm{0}_{11}^2Y_0^{(0,0),6n}$ & $2n+1$ \\
& $\omega$ & (1/3,1/3) & $\bm{0}_{55}\bm{1}_{33}Y_0^{(0,0),6n}$ & $2n+1$ \\
& $\omega^2$ & (1/3,1/3) & $\bm{0}_{11}\bm{0}_{55}^2\bm{1}_{33}^2Y_0^{(1/2,1/2),6n-3}$ & $2n$ \\ \hline
\multirow{3}{*}{$6n+2$} & 1 & (2/3,2/3) & $\bm{0}_{55}^2Y_0^{(0,0),6n}$ & $2n+1$ \\
& $\omega$ & (2/3,2/3) & $\bm{0}_{11}\bm{1}_{33}Y_0^{(0,0),6n}$ & $2n+1$ \\
& $\omega^2$ & (2/3,2/3) & $\bm{0}_{11}^2\bm{0}_{55}\bm{1}_{33}^2Y_0^{(1/2,1/2),6n-3}$ & $2n$ \\ \hline

\multirow{3}{*}{$6n+3$} & 1 & (1/6,1/6) & $\bm{0}_{11}^2\bm{0}_{55}Y_0^{(0,0),6n}$ & $2n+1$ \\
& $\omega$ & (1/6,1/6) & $\bm{0}_{55}^2\bm{1}_{33}Y_0^{(0,0),6n}$ & $2n+1$ \\
& $\omega^2$ & (1/6,1/6) & $\bm{0}_{11}\bm{1}_{33}^2Y_0^{(0,0),6n}$ & $2n+1$ \\ \hline
\multirow{3}{*}{$6n+3$} & 1 & (1/2,1/2) & $Y_0^{(1/2,1/2),6n+3}$ & $2n+2$ \\
& $\omega$ & (1/2,1/2) & $\bm{0}_{11}\bm{0}_{55}\bm{1}_{33}Y_0^{(0,0),6n}$ & $2n+1$ \\
& $\omega^2$ & (1/2,1/2) & $\bm{0}_{11}^2\bm{0}_{55}^2\bm{1}_{33}^2Y_0^{(1/2,1/2),6n-3}$ & $2n$ \\ \hline
\multirow{3}{*}{$6n+3$} & 1 & (5/6,5/6) & $\bm{0}_{11}\bm{0}_{55}^2Y_0^{(0,0),6n}$ & $2n+1$ \\
& $\omega$ & (5/6,5/6) & $\bm{0}_{11}^2\bm{1}_{33}Y_0^{(0,0),6n}$ & $2n+1$ \\
& $\omega^2$ & (5/6,5/6) & $\bm{0}_{55}\bm{1}_{33}^2Y_0^{(0,0),6n}$ & $2n+1$ \\ \hline
\end{tabular}
\end{center}
\end{table}
The procedure of proof is same as the case of $T^2/\mathbb{Z}_2$.
Here we prove only the decompositions of zero-modes with flux $6n$ and SS phases (0,0).
We introduce
\begin{align}
Y_0^{(1/2,1/2),3} \equiv
\begin{pmatrix}
\bm{0}_{11}^3 \\
\bm{1}_{33}^3 \\
\end{pmatrix}.
\end{align}
Since $\bm{1}_{33}$ vanishes at $z=0$ while $\bm{0}_{11}$ does not vanish, $Y_0^{(1/2,1/2),3}$ spans 2D spaces.
Then, the products of $Y_0^{(1/2,1/2),3}$,
\begin{align}
Y_0^{(0,0),6} \equiv Y_0^{(1/2,1/2),3} \otimes Y_0^{(1/2,1/2),3} =
\begin{pmatrix}
\bm{0}_{11}^6 \\
\bm{0}_{11}^3 \bm{1}_{33}^3 \\
\bm{1}_{33}^6 \\
\end{pmatrix}
=
\begin{pmatrix}
\bm{0}_{11}^6 \\
\bm{1}_{33}^3Y_0^{(1/2,1/2),3}
\end{pmatrix},
\end{align}
spans 3D spaces.
This is because $\bm{1}_{33}^3Y_0^{(1/2,1/2),3}$ vanishes at $z=0$ while $\bm{0}_{11}^6$ does not vanish.
Also we consider the products of $2n$ of $Y_0^{(1/2,1/2),3}$,
\begin{align}
Y_0^{(0,0),6n} 
\equiv \underbrace{Y_0^{(1/2,1/2),3} \otimes Y_0^{(1/2,1/2),3} \otimes \cdots \otimes Y_0^{(1/2,1/2),3}}_{\textrm{products~of~}2n\textrm{~of~}Y_0^{(1/2,1/2),3}}
=
\begin{pmatrix}
\bm{0}_{11}^{6n} \\
\bm{0}_{11}^{6n-3} \bm{1}_{33}^3 \\
\vdots \\
\bm{0}_{11}^{6n-3m} \bm{1}_{33}^{3m} \\
\vdots \\
\bm{0}_{11}^3 \bm{1}_{33}^{6n-3} \\
\bm{1}_{33}^{6n} \\
\end{pmatrix}, \quad m \in \{0,1,...,2n\}.
\label{eq:Z3 Y_0^(0,0),6n}
\end{align}
In this stage it is unknown whether $2n+1$ components of $Y_0^{(0,0),6n}$ are independent each other.
If $Y_0^{(0,0),6n}$ spans $2n+1$-dimensional spaces,
\begin{align}
Y_0^{(1/2,1/2),6n+3} \equiv Y_0^{(1/2,1/2),3} \otimes Y_0^{(0,0),6n}
=
\begin{pmatrix}
\bm{0}_{11}^{6n+3} \\
\bm{0}_{11}^{6n} \bm{1}_{33}^3 \\
\vdots \\
\bm{0}_{11}^{6n+3-3m} \bm{1}_{33}^{3m} \\
\vdots \\
\bm{0}_{11}^3 \bm{1}_{33}^{6n} \\
\bm{1}_{33}^{6n+3} \\
\end{pmatrix}
=
\begin{pmatrix}
\bm{0}_{11}^{6n+3} \\
\bm{1}_{33}^3Y_0^{(0,0),6n} \\
\end{pmatrix},
\label{eq:Z3 Y_0^(1/2,1/2),6n+3}
\end{align}
spans $2n+2$-dimensional spaces because $\bm{1}_{33}^3Y_0^{(0,0),6n}$ spanning $2n+1$-dimensional spaces vanishes at $z=0$ while $\bm{0}_{11}^{6n+3}$ does not vanish.
Moreover, if it is true,
\begin{align}
Y_0^{(0,0),6n+6} \equiv Y_0^{(1/2,1/2),3} \otimes Y_0^{(1/2,1/2),6n+3}
=
\begin{pmatrix}
\bm{0}_{11}^{6n+6} \\
\bm{0}_{11}^{6n+3} \bm{1}_{33}^3 \\
\vdots \\
\bm{0}_{11}^{6n+6-3m} \bm{1}_{33}^{3m} \\
\vdots \\
\bm{0}_{11}^3 \bm{1}_{33}^{6n+3} \\
\bm{1}_{33}^{6n+6} \\
\end{pmatrix}
=
\begin{pmatrix}
\bm{0}_{11}^{6n+6} \\
\bm{1}_{33}^3Y_0^{(1/2,1/2),6n+3} \\
\end{pmatrix},
\end{align}
spans $2n+3$-dimensional spaces because $\bm{1}_{33}^3Y_0^{(0,0),6n+3}$ spanning $2n+2$-dimensional spaces vanishes at $z=0$ while $\bm{0}_{11}^{6n+6}$ does not vanish.
We have already known that $Y_0^{(0,0),6}$ spans 3D spaces; therefore $Y_0^{(0,0),6n}$ spans $2n+1$-dimensional spaces for $n\geq 1$.
In addition it follows from this that $Y_0^{(1/2,1/2),6n+3}$ spans $2n+2$-dimensional spaces for $n\geq 0$.
Note that $Y_0^{(0,0),6n}$ has flux $6n$, SS phases (0,0) and $\mathbb{Z}_3$ parity 1; $Y_0^{(1/2,1/2),6n+3}$ has flux $6n+3$, SS phases (1/2,1/2) and $\mathbb{Z}_3$ parity 1.
Then, we can find the product with flux $6n$, SS phases (0,0) and $\mathbb{Z}_3$ parity $\omega$,
\begin{align}
Y_1^{(0,0),6n} \equiv \bm{0}_{11}\bm{0}_{55}\bm{1}_{33} Y_0^{(1/2,1/2),6n-3},
\end{align}
which spans $2n$-dimensional spaces for $n\geq 1$.
Similarly, we can find the product with flux $6n$, SS phases (0,0) and $\mathbb{Z}_3$ parity $\omega^2$,
\begin{align}
Y_2^{(0,0),6n} \equiv \bm{0}_{11}^2\bm{0}_{55}^2\bm{1}_{33}^2 Y_0^{(1/2,1/2),6n-6},
\end{align}
which spans $2n-1$-dimensional spaces for $n\geq 1$.
The sum of degeneracy numbers of each $\mathbb{Z}_3$ mode is equivalent to degeneracy number of zero-modes on $T^2$.
Therefore $Y_0^{(0,0),6n}$, $Y_1^{(0,0),6n}$ and $Y_2^{(0,0),6n}$ can be regarded as the complete basis of each $\mathbb{Z}_3$ mode with flux $6n$ and SS phases (0,0) on $T^2/\mathbb{Z}_3$.
Thus each $\mathbb{Z}_3$ mode with flux $6n$ and SS phases (0,0), $\psi_{T^2/\mathbb{Z}_2^0}^{(j,0),6n}$, $\psi_{T^2/\mathbb{Z}_2^1}^{(j,0),6n}$ and $\psi_{T^2/\mathbb{Z}_2^2}^{(j,0),6n}$, can be expanded by $Y_0^{(0,0),6n}$, $Y_1^{(0,0),6n}$ and $Y_2^{(0,0),6n}$, respectively.
This means that $\psi_{T^2/\mathbb{Z}_2^0}^{(j,0),6n}$, $\psi_{T^2/\mathbb{Z}_2^1}^{(j,0),6n}$ and $\psi_{T^2/\mathbb{Z}_2^2}^{(j,0),6n}$ span $2n+1$, $2n$ and $2n-1$-dimensional spaces, respectively.


\section{$T^2/\mathbb{Z}_4$}

Third, let us consider zero-modes on $T^2/\mathbb{Z}_4$ with flux $M$, SS phases $(\alpha_1,\alpha_2)$ and $\mathbb{Z}_4$ parity $m$, $\psi^{(j+\alpha_1,\alpha_2),M}_{T^2/\mathbb{Z}_4^m}(z,\tau)$.
On $T^2/\mathbb{Z}_4$ orbifold, the complex coordinate, $z$, is identified as $z\sim iz$.
Similarly, the complex structure modulus $\tau$ is fixed at $\tau=i$.
The fixed points for $\mathbb{Z}_4$ twist $z\to iz$ are:
\begin{align}
z = 0,~\frac{1+i}{2}.
\end{align}
Then SS phases are restricted to $(\alpha_1,\alpha_2)=(0,0)$ and $(1/2,1/2)$.
To express the zero-modes with higher fluxes by ones with lower fluxes, we use following three single zero-modes: zero-mode with flux 1, SS phases (0,0) and $\mathbb{Z}_4$ parity 1, $\psi^{(0,0),1}_{T^2/\mathbb{Z}_4^0}$, zero-mode with flux 1, SS phases (1/2,1/2) and $\mathbb{Z}_4$ parity $i$, $\psi^{(1/2,1/2),1}_{T^2/\mathbb{Z}_4^1}$, and zero-mode with flux 2, SS phases (1/2,1/2) and $\mathbb{Z}_4$ parity 1, $\psi^{(1/2,1/2),2}_{T^2/\mathbb{Z}_4^0}$.
We denote them as $\bm{0}_{00}$, $\bm{1}_{11}$ and $\bm{00}_{11}$, respectively.
As same as zero-modes on $T^2/\mathbb{Z}_2$, they have zero points at fixed points.
We show zero points of $\bm{0}_{00}$, $\bm{1}_{11}$ and $\bm{00}_{11}$ in Table \ref{tab:zero points of singles T2/Z4}.
\begin{table}[H]
\centering
\begin{tabular}{c|ccccc} \hline
zero-modes & flux & SS phases & $\mathbb{Z}_4$ parity & number of zero-modes & zero points \\ \hline
$\bm{0}_{00}\equiv\psi^{(0,0),1}_{T^2/\mathbb{Z}_4^0}$ & 1 & (0,0) & 1 & 1 & $\frac{1+i}{2}$ \\
$\bm{1}_{11}\equiv\psi^{(1/2,1/2),1}_{T^2/\mathbb{Z}_4^1}$ & 1 & (1/2,1/2) & $i$ & 1 & $0$ \\
$\bm{00}_{11}\equiv\psi^{(1/2,1/2),2}_{T^2/\mathbb{Z}_4^0}$ & 1 & (1/2,1/2) & 1 & 1 & None \\ \hline
\end{tabular}
\caption{Zero points of single generation zero-modes on $T^2/\mathbb{Z}_4$.}
\label{tab:zero points of singles T2/Z4}
\end{table}
Using these three zero-modes with flux 1 and 2, zero-modes with higher fluxes can be decomposed.
Consequently, we obtain the decompositions of zero-modes on $T^2/\mathbb{Z}_4$ as shown in Table \ref{tab:expansionT2/Z4}.
One can check that basis of decompositions in fourth column has flux, $\mathbb{Z}_4$ parity and SS phases shown in first to third columns.
\begin{table}[H]
\begin{center}
\renewcommand{\arraystretch}{1.0}
\begin{tabular}{ccccc} \hline
flux & $\mathbb{Z}_4$ parity & SS phases & basis of decompositions & number of basis \\ \hline
\multirow{4}{*}{$4n$} & 1 & (0,0) & $Y_0^{(0,0),4n}$ & $n+1$ \\
& $i$ & (0,0) & $\bm{0}_{00}\bm{00}_{11}\bm{1}_{11}Y_0^{(0,0),4n-4}$ & $n$ \\
& -1 & (0,0) & $\bm{0}_{00}^2\bm{1}_{11}^2Y_0^{(0,0),4n-4}$ & $n$ \\
& $-i$ & (0,0) & $\bm{0}_{00}^3\bm{00}_{11}\bm{1}_{11}^3Y_0^{(0,0),4n-8}$ & $n-1$ \\ \hline
\multirow{4}{*}{$4n$} & 1 & (1/2,1/2)  & $\bm{0}_{00}^2\bm{00}_{11}Y_0^{(0,0),4n-4}$ & $n$\\
& $i$ & (1/2,1/2)  & $\bm{0}_{00}^3\bm{1}_{11}Y_0^{(0,0),4n-4}$ & $n$\\
& -1 & (1/2,1/2)  & $\bm{00}_{11}\bm{1}_{11}^2Y_0^{(0,0),4n-4}$ & $n$\\
& $-i$ & (1/2,1/2)  & $\bm{0}_{00}\bm{1}_{11}^3Y_0^{(0,0),4n-4}$ & $n$\\ \hline
\multirow{4}{*}{$4n+1$} & 1 & (0,0) & $\bm{0}_{00}Y_0^{(0,0),4n}$ & $n+1$ \\
& $i$ & (0,0) & $\bm{0}_{00}^2\bm{00}_{11}\bm{1}_{11}Y_0^{(0,0),4n-4}$ & $n$ \\
& -1 & (0,0) & $\bm{0}_{00}^3\bm{1}_{11}^2Y_0^{(0,0),4n-4}$ & $n$ \\
& $-i$ & (0,0) & $\bm{00}_{11}\bm{1}_{11}^3Y_0^{(0,0),4n-4}$ & $n$ \\ \hline
\multirow{4}{*}{$4n+1$}  & 1 & (1/2,1/2)  & $\bm{0}_{00}^3\bm{00}_{11}Y_0^{(0,0),4n-4}$ & $n$\\
& $i$ & (1/2,1/2)  & $\bm{1}_{11}Y_0^{(0,0),4n}$ & $n+1$\\
& -1 & (1/2,1/2)  & $\bm{0}_{00}\bm{00}_{11}\bm{1}_{11}^2Y_0^{(0,0),4n-4}$ & $n$\\
& $-i$ & (1/2,1/2)  & $\bm{0}_{00}^2\bm{1}_{11}^3Y_0^{(0,0),4n-4}$ & $n$\\ \hline
\multirow{4}{*}{$4n+2$} & 1 & (0,0) & $\bm{0}_{00}^2Y_0^{(0,0),4n}$ & $n+1$ \\
& $i$ & (0,0) & $\bm{0}_{00}^3\bm{00}_{11}\bm{1}_{11}Y_0^{(0,0),4n-4}$ & $n$ \\
& -1 & (0,0) & $\bm{1}_{11}^2Y_0^{(0,0),4n}$ & $n+1$ \\
& $-i$ & (0,0) & $\bm{0}_{00}\bm{00}_{11}\bm{1}_{11}^3Y_0^{(0,0),4n-4}$ & $n$ \\ \hline
\multirow{4}{*}{$4n+2$} & 1 & (1/2,1/2)  & $\bm{00}_{11}Y_0^{(0,0),4n}$ & $n+1$\\
& $i$ & (1/2,1/2)  & $\bm{0}_{00}\bm{1}_{11}Y_0^{(0,0),4n}$ & $n+1$\\
& -1 & (1/2,1/2)  & $\bm{0}_{00}^2\bm{00}_{11}\bm{1}_{11}^2Y_0^{(0,0),4n-4}$ & $n$\\
& $-i$ & (1/2,1/2)  & $\bm{0}_{00}^3\bm{1}_{11}^3Y_0^{(0,0),4n-4}$ & $n$\\ \hline
\multirow{4}{*}{$4n+3$} & 1 & (0,0) & $\bm{0}_{00}^3Y_0^{(0,0),4n}$ & $n+1$ \\
& $i$ & (0,0) & $\bm{00}_{11}\bm{1}_{11}Y_0^{(0,0),4n}$ & $n+1$ \\
& -1 & (0,0) & $\bm{0}_{00}\bm{1}_{11}^2Y_0^{(0,0),4n}$ & $n+1$ \\
& $-i$ & (0,0) & $\bm{0}_{00}^2\bm{00}_{11}\bm{1}_{11}^3Y_0^{(0,0),4n-4}$ & $n$ \\ \hline
\multirow{4}{*}{$4n+3$} & 1 & (1/2,1/2)  & $\bm{0}_{00}\bm{00}_{11}Y_0^{(0,0),4n}$ & $n+1$\\
& $i$ & (1/2,1/2)  & $\bm{0}_{00}^2\bm{1}_{11}Y_0^{(0,0),4n}$ & $n+1$\\
& -1 & (1/2,1/2)  & $\bm{0}_{00}^3\bm{00}_{11}\bm{1}_{11}^2Y_0^{(0,0),4n-4}$ & $n$\\
& $-i$ & (1/2,1/2)  & $\bm{1}_{11}^3Y_0^{(0,0),4n}$ & $n+1$\\ \hline
\end{tabular}
\end{center}
\caption{Basis of decompositions of zero-modes on $T^2/\mathbb{Z}_4$ with higher fluxes.
$Y_0^{(0,0),4n}$ is defined by Eq.~(\ref{eq:Z4 Y_0^(0,0),4n}).
In each row, the value of flux is equivalent to the sum of number of basis.
Numbers of basis are consistent with the results derived before \cite{Kobayashi:2017dyu, Abe:2014noa, Abe:2013bca, Sakamoto:2020pev}.}
\label{tab:expansionT2/Z4}
\end{table}
The procedure of proof is same as the case of $T^2/\mathbb{Z}_2$.
Here we prove only the decompositions of zero-modes with flux $4n$ and SS phases (0,0).
We introduce
\begin{align}
Y_0^{(0,0),4} \equiv
\begin{pmatrix}
\bm{0}_{00}^4 \\
\bm{1}_{11}^4 \\
\end{pmatrix}.
\end{align}
Since $\bm{1}_{11}$ vanishes at $z=0$ while $\bm{0}_{00}$ does not vanish, $Y_0^{(0,0),4}$ spans 2D spaces.
Also we consider the products of $n$ of $Y_0^{(0,0),4}$,
\begin{align}
Y_0^{(0,0),4n} 
\equiv \underbrace{Y_0^{(0,0),4} \otimes Y_0^{(0,0),4} \otimes \cdots \otimes Y_0^{(0,0),4}}_{\textrm{products~of~}n\textrm{~of~}Y_0^{(0,0),4}}
=
\begin{pmatrix}
\bm{0}_{00}^{4n} \\
\bm{0}_{00}^{4n-4} \bm{1}_{11}^4 \\
\vdots \\
\bm{0}_{00}^{4n-4m} \bm{1}_{11}^{4m} \\
\vdots \\
\bm{0}_{00}^4 \bm{1}_{11}^{4n-4} \\
\bm{1}_{11}^{4n} \\
\end{pmatrix}, \quad m \in \{0,1,...,n\}.
\label{eq:Z4 Y_0^(0,0),4n}
\end{align}
In this stage it is unknown whether $n+1$ components of $Y_0^{(0,0),4n}$ are independent each other.
If $Y_0^{(0,0),4n}$ spans $n+1$-dimensional spaces,
\begin{align}
Y_0^{(0,0),4n+4} \equiv Y_0^{(0,0),4} \otimes Y_0^{(0,0),4n} =
\begin{pmatrix}
\bm{0}_{00}^{4n+4} \\
\bm{0}_{00}^{4n} \bm{1}_{11}^4 \\
\vdots \\
\bm{0}_{00}^{4n+4-4m} \bm{1}_{11}^{4m} \\
\vdots \\
\bm{0}_{00}^4 \bm{1}_{11}^{4n} \\
\bm{1}_{11}^{4n+4} \\
\end{pmatrix}
=
\begin{pmatrix}
\bm{0}_{00}^{4n+4} \\
\bm{1}_{11}^4 Y_0^{(0,0),4n} \\
\end{pmatrix},
\end{align}
spans $n+2$-dimensional spaces because $\bm{1}_{11}^4 Y_0^{(0,0),4n}$ spanning $n+1$-dimensional spaces vanishes at $z=0$ while $\bm{0}_{00}^{4n+4}$ does not vanish.
We have already known that $Y_0^{(0,0),4}$ spans 2D spaces; therefore $Y_0^{(0,0),4}$ spans $n+1$-dimensional spaces for $n\geq 0$.
Note that $Y_0^{(0,0),4n}$ has flux $4n$, SS phases (0,0) and $\mathbb{Z}_4$ pairty 1.
Then, we can find the products with flux $4n$, SS phases (0,0) and $\mathbb{Z}_4$ parity $i$, $-1$ and $-i$,
\begin{align}
&Y_1^{(0,0),4n} \equiv \bm{0}_{00}\bm{00}_{11}\bm{1}_{11} Y_0^{(0,0),4n-4}, \quad (\mathbb{Z}_4~\textrm{parity}~i), \\
&Y_2^{(0,0),4n} \equiv \bm{0}_{00}^2\bm{1}_{11}^2 Y_0^{(0,0),4n-4}, \quad (\mathbb{Z}_4~\textrm{parity}~-1), \\
&Y_3^{(0,0),4n} \equiv \bm{0}_{00}^3\bm{00}_{11}\bm{1}_{11}^3 Y_0^{(0,0),4n-8}, \quad (\mathbb{Z}_4~\textrm{parity}~-1).
\end{align}
They span $n$, $n$ and $n-1$-dimensional spaces, respectively.
The sum of degeneracy numbers of each $\mathbb{Z}_4$ mode in equivalent to degeneracy number of zero-modes on $T^2$.
Therefore $Y_0^{(0,0),4n}$, $Y_1^{(0,0),4n}$, $Y_2^{(0,0),4n}$ and $Y_3^{(0,0),4n}$ can be regarded as the complete basis of each $\mathbb{Z}_4$ mode with flux $4n$ and SS phases (0,0) on $T^2/\mathbb{Z}_4$.
Thus each $\mathbb{Z}_4$ mode with flux $4n$ and SS phases (0,0), $\psi_{T^2/\mathbb{Z}_2^0}^{(j,0),4n}$, $\psi_{T^2/\mathbb{Z}_2^1}^{(j,0),4n}$, $\psi_{T^2/\mathbb{Z}_2^2}^{(j,0),4n}$ and $\psi_{T^2/\mathbb{Z}_2^3}^{(j,0),4n}$, can be expanded by $Y_0^{(0,0),4n}$, $Y_1^{(0,0),4n}$, $Y_2^{(0,0),4n}$ and $Y_3^{(0,0),4n}$, respectively.
This means that $\psi_{T^2/\mathbb{Z}_2^0}^{(j,0),4n}$, $\psi_{T^2/\mathbb{Z}_2^1}^{(j,0),4n}$, $\psi_{T^2/\mathbb{Z}_2^2}^{(j,0),4n}$ and $\psi_{T^2/\mathbb{Z}_2^3}^{(j,0),4n}$ span $n+1$, $n$, $n$ and $n-1$-dimensional spaces, respectively.


\section{$T^2/\mathbb{Z}_6$}

Finally, let us consider zero-modes on $T^2/\mathbb{Z}_6$ with flux $M$, SS phases $(\alpha_1,\alpha_2)$ and $\mathbb{Z}_6$ parity $m$, $\psi^{(j+\alpha_1,\alpha_2),M}_{T^2/\mathbb{Z}_6^m}(z,\tau)$.
On $T^2/\mathbb{Z}_6$ orbifold, the complex coordinate, $z$, is identified as $z\sim\omega^{1/2} z$.
Similarly, the complex structure modulus $\tau$ is fixed at $\tau=\omega$.
There is only a single fixed point for $\mathbb{Z}_6$ twist $z\to \omega^{1/2} z$, $z=0$.
Then SS phases are restricted to $(\alpha_1,\alpha_2)=(0,0)$ for $M\in2\mathbb{Z}$ and $(\alpha_1,\alpha_2)=(1/2,1/2)$ for $M\in2\mathbb{Z}+1$.
To express the zero-modes with higher fluxes by ones with lower fluxes, we use following three single zero-modes: zero-mode with flux 2, SS phases (0,0) and $\mathbb{Z}_6$ parity 1, $\psi^{(0,0),2}_{T^2/\mathbb{Z}_6^0}$, zero-mode with flux 1, SS phases (1/2,1/2) and $\mathbb{Z}_6$ parity $\omega^{1/2}$, $\psi^{(1/2,1/2),1}_{T^2/\mathbb{Z}_6^1}$, and zero-mode with flux 3, SS phases (1/2,1/2) and $\mathbb{Z}_6$ parity 1, $\psi^{(1/2,1/2),3}_{T^2/\mathbb{Z}_6^0}$.
We denote them as $\bm{00}_{00}$, $\bm{1}_{11}$ and $\bm{000}_{11}$, respectively.
As same as zero-modes on $T^2/\mathbb{Z}_2$, they have zero points at fixed points.
We show zero points of $\bm{00}_{00}$, $\bm{1}_{11}$ and $\bm{000}_{11}$ in Table \ref{tab:zero points of singles T2/Z6}.
\begin{table}[H]
\centering
\begin{tabular}{c|ccccc} \hline
zero-modes & flux & SS phases & $\mathbb{Z}_6$ parity & number of zero-modes & zero points \\ \hline
$\bm{00}_{00}\equiv\psi^{(0,0),2}_{T^2/\mathbb{Z}_6^0}$ & 2 & (0,0) & 1 & 1 & None \\
$\bm{1}_{11}\equiv\psi^{(1/2,1/2),1}_{T^2/\mathbb{Z}_6^1}$ & 1 & (1/2,1/2) & $\omega^{1/2}$ & 1 & 0 \\
$\bm{000}_{11}\equiv\psi^{(1/2,1/2),3}_{T^2/\mathbb{Z}_6^0}$ & 3 & (1/2,1/2) & 1 & 1 & None \\ \hline
\end{tabular}
\caption{Zero points of single generation zero-modes on $T^2/\mathbb{Z}_6$.}
\label{tab:zero points of singles T2/Z6}
\end{table}
Using these three zero-modes with flux 1, 2 and 3, zero-modes with higher fluxes can be decomposed.
Consequently, we obtain the decompositions of zero-modes on $T^2/\mathbb{Z}_6$ as shown in Table \ref{tab:expansionT2/Z6}.
One can check that basis of decompositions in fourth column has flux, $\mathbb{Z}_6$ parity and SS phases shown in first to third columns.
\begin{table}[H]
\caption{Basis of decompositions of zero-modes on $T^2/\mathbb{Z}_6$ with higher fluxes.
$Y_0^{(0,0),6n}$ is defined by Eq.~(\ref{eq:Z6 Y_0^(0,0),6n}).
In each row, the value of flux is equivalent to the sum of number of basis.
Numbers of basis are consistent with the results derived before \cite{Kobayashi:2017dyu, Abe:2014noa, Abe:2013bca, Sakamoto:2020pev}.}
\label{tab:expansionT2/Z6}
\begin{center}
\renewcommand{\arraystretch}{1.0}
\begin{tabular}{ccccc} \hline
flux & $\mathbb{Z}_6$ parity & SS phases & basis of decompositions & number of basis \\ \hline
\multirow{6}{*}{$6n$} & 1 & (0,0) & $Y_0^{(0,0),6n}$ & $n+1$ \\
& $\omega^{1/2}$ & (0,0) & $\bm{00}_{00}\bm{1}_{11}\bm{000}_{11}Y_0^{(0,0),6n-6}$ & $n$ \\
& $\omega$ & (0,0) & $\bm{00}_{00}^2\bm{1}_{11}^2Y_0^{(0,0),6n-6}$ & $n$ \\
& $\omega^{3/2}$ & (0,0) & $\bm{1}_{11}^3\bm{000}_{11}Y_0^{(0,0),6n-6}$ & $n$ \\
& $\omega^2$ & (0,0) & $\bm{00}_{00}\bm{1}_{11}^4Y_0^{(0,0),6n-6}$ & $n$ \\
& $\omega^{5/2}$ & (0,0) & $\bm{00}_{00}^2\bm{1}_{11}^5\bm{000}_{11}Y_0^{(0,0),6n-12}$ & $n-1$ \\ \hline

\multirow{6}{*}{$6n+1$} & 1 & (1/2,1/2) & $\bm{00}_{00}^2\bm{000}_{11}Y_0^{(0,0),6n-6}$ & $n$ \\
& $\omega^{1/2}$ & (1/2,1/2) & $\bm{1}_{11}Y_0^{(0,0),6n}$ & $n+1$ \\
& $\omega$ & (1/2,1/2) & $\bm{00}_{00}\bm{1}_{11}^2\bm{000}_{11}Y_0^{(0,0),6n-6}$ & $n$ \\
& $\omega^{3/2}$ & (1/2,1/2) & $\bm{00}_{00}^2\bm{1}_{11}^3Y_0^{(0,0),6n-6}$ & $n$ \\
& $\omega^2$ & (1/2,1/2) & $\bm{1}_{11}^4\bm{000}_{11}Y_0^{(0,0),6n-6}$ & $n$ \\
& $\omega^{5/2}$ & (1/2,1/2) & $\bm{00}_{00}\bm{1}_{11}^5Y_0^{(0,0),6n-6}$ & $n$ \\ \hline

\multirow{6}{*}{$6n+2$} & 1 & (0,0) & $\bm{00}_{00}Y_0^{(0,0),6n}$ & $n+1$ \\
& $\omega^{1/2}$ & (0,0) & $\bm{00}_{00}^2\bm{1}_{11}\bm{000}_{11}Y_0^{(0,0),6n-6}$ & $n$ \\
& $\omega$ & (0,0) & $\bm{1}_{11}^2Y_0^{(0,0),6n}$ & $n+1$ \\
& $\omega^{3/2}$ & (0,0) & $\bm{00}_{00}\bm{1}_{11}^3\bm{000}_{11}Y_0^{(0,0),6n-6}$ & $n$ \\
& $\omega^2$ & (0,0) & $\bm{00}_{00}^2\bm{1}_{11}^4Y_0^{(0,0),6n-6}$ & $n$ \\
& $\omega^{5/2}$ & (0,0) & $\bm{1}_{11}^5\bm{000}_{11}Y_0^{(0,0),6n-6}$ & $n$ \\ \hline

\multirow{6}{*}{$6n+3$} & 1 & (1/2,1/2) & $\bm{000}_{11}Y_0^{(0,0),6n}$ & $n+1$ \\
& $\omega^{1/2}$ & (1/2,1/2) & $\bm{00}_{00}\bm{1}_{11}Y_0^{(0,0),6n}$ & $n+1$ \\
& $\omega$ & (1/2,1/2) & $\bm{00}_{00}^2\bm{1}_{11}^2\bm{000}_{11}Y_0^{(0,0),6n-6}$ & $n$ \\
& $\omega^{3/2}$ & (1/2,1/2) & $\bm{1}_{11}^3Y_0^{(0,0),6n}$ & $n+1$ \\
& $\omega^2$ & (1/2,1/2) & $\bm{00}_{00}\bm{1}_{11}^4\bm{000}_{11}Y_0^{(0,0),6n-6}$ & $n$ \\
& $\omega^{5/2}$ & (1/2,1/2) & $\bm{00}_{00}^2\bm{1}_{11}^5Y_0^{(0,0),6n-6}$ & $n$ \\ \hline

\multirow{6}{*}{$6n+4$} & 1 & (0,0) & $\bm{00}_{00}^2Y_0^{(0,0),6n}$ & $n+1$ \\
& $\omega^{1/2}$ & (0,0) & $\bm{1}_{11}\bm{000}_{11}Y_0^{(0,0),6n}$ & $n+1$ \\
& $\omega$ & (0,0) & $\bm{00}_{00}\bm{1}_{11}^2Y_0^{(0,0),6n}$ & $n+1$ \\
& $\omega^{3/2}$ & (0,0) & $\bm{00}_{00}^2\bm{1}_{11}^3\bm{000}_{11}Y_0^{(0,0),6n-6}$ & $n$ \\
& $\omega^2$ & (0,0) & $\bm{1}_{11}^4Y_0^{(0,0),6n}$ & $n+1$ \\
& $\omega^{5/2}$ & (0,0) & $\bm{00}_{00}\bm{1}_{11}^5\bm{000}_{11}Y_0^{(0,0),6n-6}$ & $n$ \\ \hline
\end{tabular}
\end{center}
\end{table}
\begin{table}[H]
\begin{center}
\renewcommand{\arraystretch}{1.0}
\begin{tabular}{ccccc} \hline
flux & $\mathbb{Z}_6$ parity & SS phases & basis of decompositions & number of basis \\ \hline
\multirow{6}{*}{$6n+5$} & 1 & (1/2,1/2) & $\bm{00}_{00}\bm{000}_{11}Y_0^{(0,0),6n}$ & $n+1$ \\
& $\omega^{1/2}$ & (1/2,1/2) & $\bm{00}_{00}^2\bm{1}_{11}Y_0^{(0,0),6n}$ & $n+1$ \\
& $\omega$ & (1/2,1/2) & $\bm{1}_{11}^2\bm{000}_{11}Y_0^{(0,0),6n}$ & $n+1$ \\
& $\omega^{3/2}$ & (1/2,1/2) & $\bm{00}_{00}\bm{1}_{11}^3Y_0^{(0,0),6n}$ & $n+1$ \\
& $\omega^2$ & (1/2,1/2) & $\bm{00}_{00}^2\bm{1}_{11}^4\bm{000}_{11}Y_0^{(0,0),6n-6}$ & $n$ \\
& $\omega^{5/2}$ & (1/2,1/2) & $\bm{1}_{11}^5Y_0^{(0,0),6n}$ & $n+1$ \\ \hline
\end{tabular}
\end{center}
\end{table}
The procedure of proof is same as the case of $T^2/\mathbb{Z}_2$.
Here we prove only the decompositions of zero-modes with flux $6n$ and SS phases (0,0).
We introduce
\begin{align}
Y_0^{(0,0),6} \equiv
\begin{pmatrix}
\bm{00}_{00}^3 \\
\bm{1}_{11}^6\\
\end{pmatrix}.
\end{align}
Since $\bm{1}_{11}$ vanishes at $z=0$ while $\bm{00}_{00}$ does not vanish, $Y_0^{(0,0),6}$ spans 2D spaces.
Also we consider the products of $n$ of $Y_0^{(0,0),6}$,
\begin{align}
Y_0^{(0,0),6n} 
\equiv \underbrace{Y_0^{(0,0),6} \otimes Y_0^{(0,0),6} \otimes \cdots \otimes Y_0^{(0,0),6}}_{\textrm{products~of~}n\textrm{~of~}Y_0^{(0,0),6}}
=
\begin{pmatrix}
\bm{00}_{00}^{3n} \\
\bm{00}_{00}^{3n-3} \bm{1}_{11}^6 \\
\vdots \\
\bm{00}_{00}^{3n-3m} \bm{1}_{11}^{6m} \\
\vdots \\
\bm{00}_{00}^3 \bm{1}_{11}^{6n-6} \\
\bm{1}_{11}^{6n} \\
\end{pmatrix}, \quad m \in \{0,1,...,n\}.
\label{eq:Z6 Y_0^(0,0),6n}
\end{align}
In this stage it is unknown whether $n+1$ components of $Y_0^{(0,0),6n}$ are independent each other.
If $Y_0^{(0,0),6n}$ spans $n+1$-dimensional spaces,
\begin{align}
Y_0^{(0,0),6n+6} \equiv Y_0^{(0,0),6} \otimes Y_0^{(0,0),6n} =
\begin{pmatrix}
\bm{00}_{00}^{3n+3} \\
\bm{00}_{00}^{3n} \bm{1}_{11}^6 \\
\vdots \\
\bm{00}_{00}^{3n+3-3m} \bm{1}_{11}^{6m} \\
\vdots \\
\bm{00}_{00}^3 \bm{1}_{11}^{6n} \\
\bm{1}_{11}^{6n+6} \\
\end{pmatrix}
=
\begin{pmatrix}
\bm{0}_{00}^{3n+3} \\
\bm{1}_{11}^6 Y_0^{(0,0),6n} \\
\end{pmatrix},
\end{align}
spans $n+2$-dimensional spaces because $\bm{1}_{11}^6 Y_0^{(0,0),6n}$ spanning $n+1$-dimensional spaces vanishes at $z=0$ while $\bm{0}_{00}^{3n+3}$ does not vanish.
We have already known that $Y_0^{(0,0),6}$ spans 2D spaces; therefore $Y_0^{(0,0),6}$ spans $n+1$-dimensional spaces for $n\geq 0$.
Note that $Y_0^{(0,0),6n}$ has flux $6n$, SS phases (0,0) and $\mathbb{Z}_6$ pairty 1.
Then, we can find the products with flux $6n$, SS phases (0,0) and $\mathbb{Z}_6$ parity $\omega^{1/2}$, $\omega$, $\omega^{3/2}$, $\omega^2$ and $\omega^{5/2}$,
\begin{align}
&Y_1^{(0,0),6n} \equiv \bm{00}_{00}\bm{1}_{11}\bm{000}_{11}Y_0^{(0,0),6n-6}, \quad (\mathbb{Z}_6~\textrm{parity}~\omega^{1/2}), \\
&Y_2^{(0,0),6n} \equiv \bm{00}_{00}^2\bm{1}_{11}^2 Y_0^{(0,0),6n-6}, \quad (\mathbb{Z}_6~\textrm{parity}~\omega), \\
&Y_3^{(0,0),6n} \equiv \bm{1}_{11}^3\bm{000}_{11}Y_0^{(0,0),6n-6}, \quad (\mathbb{Z}_6~\textrm{parity}~\omega^{3/2}), \\
&Y_4^{(0,0),6n} \equiv \bm{00}_{00}\bm{1}_{11}^4 Y_0^{(0,0),6n-6}, \quad (\mathbb{Z}_6~\textrm{parity}~\omega^2), \\
&Y_5^{(0,0),6n} \equiv \bm{00}_{00}^2\bm{1}_{11}^5\bm{000}_{11} Y_0^{(0,0),6n-12}, \quad (\mathbb{Z}_6~\textrm{parity}~\omega^{5/2}).
\end{align}
They span $n+1$, $n$, $n$, $n$, $n$ and $n-1$-dimensional spaces, respectively.


\chapter{Mass matrix structures of the favorable models in Tables \ref{tab:chi<0.01atinfinite} and \ref{tab:chi<0.01atomega}}
\label{app:Mass matrix structures of the favorable models}

Here we show the phase factors after the basis transformations in Eqs.~(\ref{eq:u_L}) and (\ref{eq:u_R}), and the hierarchical structures of the mass matrices of the favorable models in Tables \ref{tab:chi<0.01atinfinite} and \ref{tab:chi<0.01atomega}.
We express the structures of up and down-sector quark mass matrices by the phase factors $p\equiv\varepsilon/|\varepsilon|$ and powers of $|\varepsilon|\sim 0.15$.
In Table \ref{tab:mass_matrix_structures}, we show the results.
Note that we show only different structures which are not related by unitary transformations for the fields.
In total we find 128 number of different structures.

\begin{table}[H]
  \centering
  \caption{The phase factors after the basis transformations Eqs.~(\ref{eq:u_L}) and (\ref{eq:u_R}), and the hierarchical structures of up and down-sector quark mass matrices of the favorable models in Tables \ref{tab:chi<0.01atinfinite} and \ref{tab:chi<0.01atomega}.
  First row denotes the structure of up-sector quark mass matrix and other rows denote ones of down-sector quark, up to $\langle H_u \rangle$ and $\langle H_d \rangle$.
  $p$ is given by $\varepsilon/|\varepsilon|$ and $|\varepsilon|\sim 0.15$.
  We show only different structures which are not related by unitary transformations for the fields.
  In total we find 128 number of different structures.}
  \label{tab:mass_matrix_structures}
  \renewcommand{\arraystretch}{1.8}
  \begin{tabular}{p{9em}p{9em}p{9em}p{9em}} \hline
\multicolumn{4}{c}{$M_u=\begin{psmallmatrix}
|\varepsilon|^6 & |\varepsilon|^4p^{-1} & |\varepsilon|^6 \\
|\varepsilon|^2 & -|\varepsilon|^3 & -|\varepsilon|^2 \\
1 & |\varepsilon| & -1 \\
\end{psmallmatrix}$} \\ \hline
$\begin{psmallmatrix}
|\varepsilon|^4p^{-1} & |\varepsilon|^3p^{-2} & |\varepsilon|^6 \\
|\varepsilon|^3 & |\varepsilon|^2p^{-1} & |\varepsilon|^2 \\
|\varepsilon| & -|\varepsilon|^3 & -1 \\
\end{psmallmatrix}$
&
$\begin{psmallmatrix}
|\varepsilon|^4p^{-1} & |\varepsilon|^3p^{-2} & |\varepsilon|^6 \\
-|\varepsilon|^3 & |\varepsilon|^2p^{-1} & |\varepsilon|^2 \\
|\varepsilon| & -|\varepsilon|^3 & -1 \\
\end{psmallmatrix}$
&
$\begin{psmallmatrix}
|\varepsilon|^4p^{-1} & |\varepsilon|^3p^{-2} & |\varepsilon|^6 \\
|\varepsilon|^3 & |\varepsilon|^2p^{-1} & |\varepsilon|^2 \\
-|\varepsilon| & -|\varepsilon|^3 & -1 \\
\end{psmallmatrix}$
&
$\begin{psmallmatrix}
|\varepsilon|^4p^{-1} & |\varepsilon|^3p^{-2} & |\varepsilon|^6 \\
-|\varepsilon|^3 & |\varepsilon|^2p^{-1} & |\varepsilon|^2 \\
-|\varepsilon| & -|\varepsilon|^3 & -1 \\
\end{psmallmatrix}$
\\
$\begin{psmallmatrix}
|\varepsilon|^4p^{-1} & |\varepsilon|^3p^{-2} & |\varepsilon|^6 \\
|\varepsilon|^3 & |\varepsilon|^2p^{-1} & |\varepsilon|^2 \\
|\varepsilon| & |\varepsilon|^3 & -1 \\
\end{psmallmatrix}$
&
$\begin{psmallmatrix}
|\varepsilon|^4p^{-1} & |\varepsilon|^3p^{-2} & |\varepsilon|^6 \\
-|\varepsilon|^3 & |\varepsilon|^2p^{-1} & |\varepsilon|^2 \\
|\varepsilon| & |\varepsilon|^3 & -1 \\
\end{psmallmatrix}$
&
$\begin{psmallmatrix}
|\varepsilon|^4p^{-1} & |\varepsilon|^3p^{-2} & |\varepsilon|^6 \\
|\varepsilon|^3 & |\varepsilon|^2p^{-1} & |\varepsilon|^2 \\
-|\varepsilon| & |\varepsilon|^3 & -1 \\
\end{psmallmatrix}$
&
$\begin{psmallmatrix}
|\varepsilon|^4p^{-1} & |\varepsilon|^3p^{-2} & |\varepsilon|^6 \\
-|\varepsilon|^3 & |\varepsilon|^2p^{-1} & |\varepsilon|^2 \\
-|\varepsilon| & |\varepsilon|^3 & -1 \\
\end{psmallmatrix}$
\\
$\begin{psmallmatrix}
|\varepsilon|^4p^{-1} & |\varepsilon|^3p^{-2} & |\varepsilon|^6 \\
|\varepsilon|^3 & |\varepsilon|^2p^{-1} & -|\varepsilon|^2 \\
|\varepsilon| & -|\varepsilon|^3 & 1 \\
\end{psmallmatrix}$
&
$\begin{psmallmatrix}
|\varepsilon|^4p^{-1} & |\varepsilon|^3p^{-2} & |\varepsilon|^6 \\
-|\varepsilon|^3 & |\varepsilon|^2p^{-1} & -|\varepsilon|^2 \\
|\varepsilon| & -|\varepsilon|^3 & 1 \\
\end{psmallmatrix}$
&
$\begin{psmallmatrix}
|\varepsilon|^4p^{-1} & |\varepsilon|^3p^{-2} & |\varepsilon|^6 \\
|\varepsilon|^3 & |\varepsilon|^2p^{-1} & -|\varepsilon|^2 \\
-|\varepsilon| & -|\varepsilon|^3 & 1 \\
\end{psmallmatrix}$
&
$\begin{psmallmatrix}
|\varepsilon|^4p^{-1} & |\varepsilon|^3p^{-2} & |\varepsilon|^6 \\
-|\varepsilon|^3 & |\varepsilon|^2p^{-1} & -|\varepsilon|^2 \\
-|\varepsilon| & -|\varepsilon|^3 & 1 \\
\end{psmallmatrix}$
\\
$\begin{psmallmatrix}
|\varepsilon|^4p^{-1} & |\varepsilon|^3p^{-2} & |\varepsilon|^6 \\
|\varepsilon|^3 & |\varepsilon|^2p^{-1} & -|\varepsilon|^2 \\
|\varepsilon| & |\varepsilon|^3 & 1 \\
\end{psmallmatrix}$
&
$\begin{psmallmatrix}
|\varepsilon|^4p^{-1} & |\varepsilon|^3p^{-2} & |\varepsilon|^6 \\
-|\varepsilon|^3 & |\varepsilon|^2p^{-1} & -|\varepsilon|^2 \\
|\varepsilon| & |\varepsilon|^3 & 1 \\
\end{psmallmatrix}$
&
$\begin{psmallmatrix}
|\varepsilon|^4p^{-1} & |\varepsilon|^3p^{-2} & |\varepsilon|^6 \\
|\varepsilon|^3 & |\varepsilon|^2p^{-1} & -|\varepsilon|^2 \\
-|\varepsilon| & |\varepsilon|^3 & 1 \\
\end{psmallmatrix}$
&
$\begin{psmallmatrix}
|\varepsilon|^4p^{-1} & |\varepsilon|^3p^{-2} & |\varepsilon|^6 \\
-|\varepsilon|^3 & |\varepsilon|^2p^{-1} & -|\varepsilon|^2 \\
-|\varepsilon| & |\varepsilon|^3 & 1 \\
\end{psmallmatrix}$
\\
$\begin{psmallmatrix}
|\varepsilon|^4p^{-1} & |\varepsilon|^3p^{-3} & |\varepsilon|^6 \\
|\varepsilon|^3 & |\varepsilon|^2p^{-2} & |\varepsilon|^2 \\
|\varepsilon| & |\varepsilon|^6 & -1 \\
\end{psmallmatrix}$
&
$\begin{psmallmatrix}
|\varepsilon|^4p^{-1} & |\varepsilon|^3p^{-3} & |\varepsilon|^6 \\
-|\varepsilon|^3 & |\varepsilon|^2p^{-2} & |\varepsilon|^2 \\
|\varepsilon| & |\varepsilon|^6 & -1 \\
\end{psmallmatrix}$
&
$\begin{psmallmatrix}
|\varepsilon|^4p^{-1} & |\varepsilon|^3p^{-3} & |\varepsilon|^6 \\
|\varepsilon|^3 & |\varepsilon|^2p^{-2} & |\varepsilon|^2 \\
-|\varepsilon| & |\varepsilon|^6 & -1 \\
\end{psmallmatrix}$
&
$\begin{psmallmatrix}
|\varepsilon|^4p^{-1} & |\varepsilon|^3p^{-3} & |\varepsilon|^6 \\
-|\varepsilon|^3 & |\varepsilon|^2p^{-2} & |\varepsilon|^2 \\
-|\varepsilon| & |\varepsilon|^6 & -1 \\
\end{psmallmatrix}$
\\
$\begin{psmallmatrix}
|\varepsilon|^4p^{-1} & |\varepsilon|^3p^{-3} & |\varepsilon|^6 \\
|\varepsilon|^3 & |\varepsilon|^2p^{-2} & |\varepsilon|^2 \\
|\varepsilon| & -|\varepsilon|^6 & -1 \\
\end{psmallmatrix}$
&
$\begin{psmallmatrix}
|\varepsilon|^4p^{-1} & |\varepsilon|^3p^{-3} & |\varepsilon|^6 \\
-|\varepsilon|^3 & |\varepsilon|^2p^{-2} & |\varepsilon|^2 \\
|\varepsilon| & -|\varepsilon|^6 & -1 \\
\end{psmallmatrix}$
&
$\begin{psmallmatrix}
|\varepsilon|^4p^{-1} & |\varepsilon|^3p^{-3} & |\varepsilon|^6 \\
|\varepsilon|^3 & |\varepsilon|^2p^{-2} & |\varepsilon|^2 \\
-|\varepsilon| & -|\varepsilon|^6 & -1 \\
\end{psmallmatrix}$
&
$\begin{psmallmatrix}
|\varepsilon|^4p^{-1} & |\varepsilon|^3p^{-3} & |\varepsilon|^6 \\
-|\varepsilon|^3 & |\varepsilon|^2p^{-2} & |\varepsilon|^2 \\
-|\varepsilon| & -|\varepsilon|^6 & -1 \\
\end{psmallmatrix}$
\\
$\begin{psmallmatrix}
|\varepsilon|^4p^{-1} & |\varepsilon|^3p^{-3} & |\varepsilon|^6 \\
|\varepsilon|^3 & |\varepsilon|^2p^{-2} & -|\varepsilon|^2 \\
|\varepsilon| & |\varepsilon|^6 & 1 \\
\end{psmallmatrix}$
&
$\begin{psmallmatrix}
|\varepsilon|^4p^{-1} & |\varepsilon|^3p^{-3} & |\varepsilon|^6 \\
-|\varepsilon|^3 & |\varepsilon|^2p^{-2} & -|\varepsilon|^2 \\
|\varepsilon| & |\varepsilon|^6 & 1 \\
\end{psmallmatrix}$
&
$\begin{psmallmatrix}
|\varepsilon|^4p^{-1} & |\varepsilon|^3p^{-3} & |\varepsilon|^6 \\
|\varepsilon|^3 & |\varepsilon|^2p^{-2} & -|\varepsilon|^2 \\
-|\varepsilon| & |\varepsilon|^6 & 1 \\
\end{psmallmatrix}$
&
$\begin{psmallmatrix}
|\varepsilon|^4p^{-1} & |\varepsilon|^3p^{-3} & |\varepsilon|^6 \\
-|\varepsilon|^3 & |\varepsilon|^2p^{-2} & -|\varepsilon|^2 \\
-|\varepsilon| & |\varepsilon|^6 & 1 \\
\end{psmallmatrix}$
\\
$\begin{psmallmatrix}
|\varepsilon|^4p^{-1} & |\varepsilon|^3p^{-3} & |\varepsilon|^6 \\
|\varepsilon|^3 & |\varepsilon|^2p^{-2} & -|\varepsilon|^2 \\
|\varepsilon| & -|\varepsilon|^6 & 1 \\
\end{psmallmatrix}$
&
$\begin{psmallmatrix}
|\varepsilon|^4p^{-1} & |\varepsilon|^3p^{-3} & |\varepsilon|^6 \\
-|\varepsilon|^3 & |\varepsilon|^2p^{-2} & -|\varepsilon|^2 \\
|\varepsilon| & -|\varepsilon|^6 & 1 \\
\end{psmallmatrix}$
&
$\begin{psmallmatrix}
|\varepsilon|^4p^{-1} & |\varepsilon|^3p^{-3} & |\varepsilon|^6 \\
|\varepsilon|^3 & |\varepsilon|^2p^{-2} & -|\varepsilon|^2 \\
-|\varepsilon| & -|\varepsilon|^6 & 1 \\
\end{psmallmatrix}$
&
$\begin{psmallmatrix}
|\varepsilon|^4p^{-1} & |\varepsilon|^3p^{-3} & |\varepsilon|^6 \\
-|\varepsilon|^3 & |\varepsilon|^2p^{-2} & -|\varepsilon|^2 \\
-|\varepsilon| & -|\varepsilon|^6 & 1 \\
\end{psmallmatrix}$
\\
$\begin{psmallmatrix}
|\varepsilon|^4p^{-2} & |\varepsilon|^3p^{-2} & |\varepsilon|^6 \\
-|\varepsilon|^3p^{-1} & |\varepsilon|^2p^{-1} & |\varepsilon|^2 \\
-|\varepsilon|^4 & -|\varepsilon|^3 & -1 \\
\end{psmallmatrix}$
&
$\begin{psmallmatrix}
|\varepsilon|^4p^{-2} & |\varepsilon|^3p^{-2} & |\varepsilon|^6 \\
-|\varepsilon|^3p^{-1} & |\varepsilon|^2p^{-1} & |\varepsilon|^2 \\
|\varepsilon|^4 & -|\varepsilon|^3 & -1 \\
\end{psmallmatrix}$
&
$\begin{psmallmatrix}
|\varepsilon|^4p^{-2} & |\varepsilon|^3p^{-2} & |\varepsilon|^6 \\
-|\varepsilon|^3p^{-1} & |\varepsilon|^2p^{-1} & |\varepsilon|^2 \\
-|\varepsilon|^4 & |\varepsilon|^3 & -1 \\
\end{psmallmatrix}$
&
$\begin{psmallmatrix}
|\varepsilon|^4p^{-2} & |\varepsilon|^3p^{-2} & |\varepsilon|^6 \\
-|\varepsilon|^3p^{-1} & |\varepsilon|^2p^{-1} & |\varepsilon|^2 \\
|\varepsilon|^4 & |\varepsilon|^3 & -1 \\
\end{psmallmatrix}$
\\
$\begin{psmallmatrix}
|\varepsilon|^4p^{-2} & |\varepsilon|^3p^{-2} & |\varepsilon|^6 \\
-|\varepsilon|^3p^{-1} & |\varepsilon|^2p^{-1} & -|\varepsilon|^2 \\
-|\varepsilon|^4 & -|\varepsilon|^3 & 1 \\
\end{psmallmatrix}$
&
$\begin{psmallmatrix}
|\varepsilon|^4p^{-2} & |\varepsilon|^3p^{-2} & |\varepsilon|^6 \\
-|\varepsilon|^3p^{-1} & |\varepsilon|^2p^{-1} & -|\varepsilon|^2 \\
|\varepsilon|^4 & -|\varepsilon|^3 & 1 \\
\end{psmallmatrix}$
&
$\begin{psmallmatrix}
|\varepsilon|^4p^{-2} & |\varepsilon|^3p^{-2} & |\varepsilon|^6 \\
-|\varepsilon|^3p^{-1} & |\varepsilon|^2p^{-1} & -|\varepsilon|^2 \\
-|\varepsilon|^4 & |\varepsilon|^3 & 1 \\
\end{psmallmatrix}$
&
$\begin{psmallmatrix}
|\varepsilon|^4p^{-2} & |\varepsilon|^3p^{-2} & |\varepsilon|^6 \\
-|\varepsilon|^3p^{-1} & |\varepsilon|^2p^{-1} & -|\varepsilon|^2 \\
|\varepsilon|^4 & |\varepsilon|^3 & 1 \\
\end{psmallmatrix}$
\\
$\begin{psmallmatrix}
|\varepsilon|^4p^{-2} & |\varepsilon|^3p^{-3} & |\varepsilon|^6 \\
-|\varepsilon|^3p^{-1} & |\varepsilon|^2p^{-2} & |\varepsilon|^2 \\
-|\varepsilon|^4 & |\varepsilon|^6 & -1 \\
\end{psmallmatrix}$
&
$\begin{psmallmatrix}
|\varepsilon|^4p^{-2} & |\varepsilon|^3p^{-3} & |\varepsilon|^6 \\
-|\varepsilon|^3p^{-1} & |\varepsilon|^2p^{-2} & |\varepsilon|^2 \\
|\varepsilon|^4 & |\varepsilon|^6 & -1 \\
\end{psmallmatrix}$
&
$\begin{psmallmatrix}
|\varepsilon|^4p^{-2} & |\varepsilon|^3p^{-3} & |\varepsilon|^6 \\
-|\varepsilon|^3p^{-1} & |\varepsilon|^2p^{-2} & |\varepsilon|^2 \\
-|\varepsilon|^4 & -|\varepsilon|^6 & -1 \\
\end{psmallmatrix}$
&
$\begin{psmallmatrix}
|\varepsilon|^4p^{-2} & |\varepsilon|^3p^{-3} & |\varepsilon|^6 \\
-|\varepsilon|^3p^{-1} & |\varepsilon|^2p^{-2} & |\varepsilon|^2 \\
|\varepsilon|^4 & -|\varepsilon|^6 & -1 \\
\end{psmallmatrix}$
\\
$\begin{psmallmatrix}
|\varepsilon|^4p^{-2} & |\varepsilon|^3p^{-3} & |\varepsilon|^6 \\
-|\varepsilon|^3p^{-1} & |\varepsilon|^2p^{-2} & -|\varepsilon|^2 \\
-|\varepsilon|^4 & |\varepsilon|^6 & 1 \\
\end{psmallmatrix}$
&
$\begin{psmallmatrix}
|\varepsilon|^4p^{-2} & |\varepsilon|^3p^{-3} & |\varepsilon|^6 \\
-|\varepsilon|^3p^{-1} & |\varepsilon|^2p^{-2} & -|\varepsilon|^2 \\
|\varepsilon|^4 & |\varepsilon|^6 & 1 \\
\end{psmallmatrix}$
&
$\begin{psmallmatrix}
|\varepsilon|^4p^{-2} & |\varepsilon|^3p^{-3} & |\varepsilon|^6 \\
-|\varepsilon|^3p^{-1} & |\varepsilon|^2p^{-2} & -|\varepsilon|^2 \\
-|\varepsilon|^4 & -|\varepsilon|^6 & 1 \\
\end{psmallmatrix}$
&
$\begin{psmallmatrix}
|\varepsilon|^4p^{-2} & |\varepsilon|^3p^{-3} & |\varepsilon|^6 \\
-|\varepsilon|^3p^{-1} & |\varepsilon|^2p^{-2} & -|\varepsilon|^2 \\
|\varepsilon|^4 & -|\varepsilon|^6 & 1 \\
\end{psmallmatrix}$
\\
$\begin{psmallmatrix}
|\varepsilon|^4p^{-2} & |\varepsilon|^3p^{-2} & |\varepsilon|^6 \\
-|\varepsilon|^6 & |\varepsilon|^2p^{-1} & |\varepsilon|^2 \\
-|\varepsilon|^4 & -|\varepsilon|^3 & -1 \\
\end{psmallmatrix}$
&
$\begin{psmallmatrix}
|\varepsilon|^4p^{-2} & |\varepsilon|^3p^{-2} & |\varepsilon|^6 \\
|\varepsilon|^6 & |\varepsilon|^2p^{-1} & |\varepsilon|^2 \\
-|\varepsilon|^4 & -|\varepsilon|^3 & -1 \\
\end{psmallmatrix}$
&
$\begin{psmallmatrix}
|\varepsilon|^4p^{-2} & |\varepsilon|^3p^{-2} & |\varepsilon|^6 \\
-|\varepsilon|^6 & |\varepsilon|^2p^{-1} & |\varepsilon|^2 \\
|\varepsilon|^4 & -|\varepsilon|^3 & -1 \\
\end{psmallmatrix}$
&
$\begin{psmallmatrix}
|\varepsilon|^4p^{-2} & |\varepsilon|^3p^{-2} & |\varepsilon|^6 \\
|\varepsilon|^6 & |\varepsilon|^2p^{-1} & |\varepsilon|^2 \\
|\varepsilon|^4 & -|\varepsilon|^3 & -1 \\
\end{psmallmatrix}$
\\
$\begin{psmallmatrix}
|\varepsilon|^4p^{-2} & |\varepsilon|^3p^{-2} & |\varepsilon|^6 \\
-|\varepsilon|^6 & |\varepsilon|^2p^{-1} & |\varepsilon|^2 \\
-|\varepsilon|^4 & |\varepsilon|^3 & -1 \\
\end{psmallmatrix}$
&
$\begin{psmallmatrix}
|\varepsilon|^4p^{-2} & |\varepsilon|^3p^{-2} & |\varepsilon|^6 \\
|\varepsilon|^6 & |\varepsilon|^2p^{-1} & |\varepsilon|^2 \\
-|\varepsilon|^4 & |\varepsilon|^3 & -1 \\
\end{psmallmatrix}$
&
$\begin{psmallmatrix}
|\varepsilon|^4p^{-2} & |\varepsilon|^3p^{-2} & |\varepsilon|^6 \\
-|\varepsilon|^6 & |\varepsilon|^2p^{-1} & |\varepsilon|^2 \\
|\varepsilon|^4 & |\varepsilon|^3 & -1 \\
\end{psmallmatrix}$
&
$\begin{psmallmatrix}
|\varepsilon|^4p^{-2} & |\varepsilon|^3p^{-2} & |\varepsilon|^6 \\
|\varepsilon|^6 & |\varepsilon|^2p^{-1} & |\varepsilon|^2 \\
|\varepsilon|^4 & |\varepsilon|^3 & -1 \\
\end{psmallmatrix}$
\\
$\begin{psmallmatrix}
|\varepsilon|^4p^{-2} & |\varepsilon|^3p^{-2} & |\varepsilon|^6 \\
-|\varepsilon|^6 & |\varepsilon|^2p^{-1} & -|\varepsilon|^2 \\
-|\varepsilon|^4 & -|\varepsilon|^3 & 1 \\
\end{psmallmatrix}$
&
$\begin{psmallmatrix}
|\varepsilon|^4p^{-2} & |\varepsilon|^3p^{-2} & |\varepsilon|^6 \\
|\varepsilon|^6 & |\varepsilon|^2p^{-1} & -|\varepsilon|^2 \\
-|\varepsilon|^4 & -|\varepsilon|^3 & 1 \\
\end{psmallmatrix}$
&
$\begin{psmallmatrix}
|\varepsilon|^4p^{-2} & |\varepsilon|^3p^{-2} & |\varepsilon|^6 \\
-|\varepsilon|^6 & |\varepsilon|^2p^{-1} & -|\varepsilon|^2 \\
|\varepsilon|^4 & -|\varepsilon|^3 & 1 \\
\end{psmallmatrix}$
&
$\begin{psmallmatrix}
|\varepsilon|^4p^{-2} & |\varepsilon|^3p^{-2} & |\varepsilon|^6 \\
|\varepsilon|^6 & |\varepsilon|^2p^{-1} & -|\varepsilon|^2 \\
|\varepsilon|^4 & -|\varepsilon|^3 & 1 \\
\end{psmallmatrix}$
\\
$\begin{psmallmatrix}
|\varepsilon|^4p^{-2} & |\varepsilon|^3p^{-2} & |\varepsilon|^6 \\
-|\varepsilon|^6 & |\varepsilon|^2p^{-1} & -|\varepsilon|^2 \\
-|\varepsilon|^4 & |\varepsilon|^3 & 1 \\
\end{psmallmatrix}$
&
$\begin{psmallmatrix}
|\varepsilon|^4p^{-2} & |\varepsilon|^3p^{-2} & |\varepsilon|^6 \\
|\varepsilon|^6 & |\varepsilon|^2p^{-1} & -|\varepsilon|^2 \\
-|\varepsilon|^4 & |\varepsilon|^3 & 1 \\
\end{psmallmatrix}$
&
$\begin{psmallmatrix}
|\varepsilon|^4p^{-2} & |\varepsilon|^3p^{-2} & |\varepsilon|^6 \\
-|\varepsilon|^6 & |\varepsilon|^2p^{-1} & -|\varepsilon|^2 \\
|\varepsilon|^4 & |\varepsilon|^3 & 1 \\
\end{psmallmatrix}$
&
$\begin{psmallmatrix}
|\varepsilon|^4p^{-2} & |\varepsilon|^3p^{-2} & |\varepsilon|^6 \\
|\varepsilon|^6 & |\varepsilon|^2p^{-1} & -|\varepsilon|^2 \\
|\varepsilon|^4 & |\varepsilon|^3 & 1 \\
\end{psmallmatrix}$
\\ \hline
  \end{tabular}
\end{table}
\begin{table}[H]
  \centering
  \renewcommand{\arraystretch}{1.8}
  \begin{tabular}{p{9em}p{9em}p{9em}p{9em}} \hline
\multicolumn{4}{c}{$M_u=\begin{psmallmatrix}
|\varepsilon|^6 & |\varepsilon|^4p^{-1} & |\varepsilon|^6 \\
|\varepsilon|^2 & |\varepsilon|^3 & -|\varepsilon|^2 \\
1 & -|\varepsilon| & -1 \\
\end{psmallmatrix}$} \\ \hline
$\begin{psmallmatrix}
|\varepsilon|^4p^{-1} & |\varepsilon|^3p^{-2} & |\varepsilon|^6 \\
|\varepsilon|^3 & -|\varepsilon|^2p^{-1} & |\varepsilon|^2 \\
|\varepsilon| & -|\varepsilon|^3 & -1 \\
\end{psmallmatrix}$
&
$\begin{psmallmatrix}
|\varepsilon|^4p^{-1} & |\varepsilon|^3p^{-2} & |\varepsilon|^6 \\
-|\varepsilon|^3 & -|\varepsilon|^2p^{-1} & |\varepsilon|^2 \\
|\varepsilon| & -|\varepsilon|^3 & -1 \\
\end{psmallmatrix}$
&
$\begin{psmallmatrix}
|\varepsilon|^4p^{-1} & |\varepsilon|^3p^{-2} & |\varepsilon|^6 \\
|\varepsilon|^3 & -|\varepsilon|^2p^{-1} & |\varepsilon|^2 \\
-|\varepsilon| & -|\varepsilon|^3 & -1 \\
\end{psmallmatrix}$
&
$\begin{psmallmatrix}
|\varepsilon|^4p^{-1} & |\varepsilon|^3p^{-2} & |\varepsilon|^6 \\
-|\varepsilon|^3 & -|\varepsilon|^2p^{-1} & |\varepsilon|^2 \\
-|\varepsilon| & -|\varepsilon|^3 & -1 \\
\end{psmallmatrix}$
\\
$\begin{psmallmatrix}
|\varepsilon|^4p^{-1} & |\varepsilon|^3p^{-2} & |\varepsilon|^6 \\
|\varepsilon|^3 & -|\varepsilon|^2p^{-1} & |\varepsilon|^2 \\
|\varepsilon| & |\varepsilon|^3 & -1 \\
\end{psmallmatrix}$
&
$\begin{psmallmatrix}
|\varepsilon|^4p^{-1} & |\varepsilon|^3p^{-2} & |\varepsilon|^6 \\
-|\varepsilon|^3 & -|\varepsilon|^2p^{-1} & |\varepsilon|^2 \\
|\varepsilon| & |\varepsilon|^3 & -1 \\
\end{psmallmatrix}$
&
$\begin{psmallmatrix}
|\varepsilon|^4p^{-1} & |\varepsilon|^3p^{-2} & |\varepsilon|^6 \\
|\varepsilon|^3 & -|\varepsilon|^2p^{-1} & |\varepsilon|^2 \\
-|\varepsilon| & |\varepsilon|^3 & -1 \\
\end{psmallmatrix}$
&
$\begin{psmallmatrix}
|\varepsilon|^4p^{-1} & |\varepsilon|^3p^{-2} & |\varepsilon|^6 \\
-|\varepsilon|^3 & -|\varepsilon|^2p^{-1} & |\varepsilon|^2 \\
-|\varepsilon| & |\varepsilon|^3 & -1 \\
\end{psmallmatrix}$
\\
$\begin{psmallmatrix}
|\varepsilon|^4p^{-1} & |\varepsilon|^3p^{-2} & |\varepsilon|^6 \\
|\varepsilon|^3 & -|\varepsilon|^2p^{-1} & -|\varepsilon|^2 \\
|\varepsilon| & -|\varepsilon|^3 & 1 \\
\end{psmallmatrix}$
&
$\begin{psmallmatrix}
|\varepsilon|^4p^{-1} & |\varepsilon|^3p^{-2} & |\varepsilon|^6 \\
-|\varepsilon|^3 & -|\varepsilon|^2p^{-1} & -|\varepsilon|^2 \\
|\varepsilon| & -|\varepsilon|^3 & 1 \\
\end{psmallmatrix}$
&
$\begin{psmallmatrix}
|\varepsilon|^4p^{-1} & |\varepsilon|^3p^{-2} & |\varepsilon|^6 \\
|\varepsilon|^3 & -|\varepsilon|^2p^{-1} & -|\varepsilon|^2 \\
-|\varepsilon| & -|\varepsilon|^3 & 1 \\
\end{psmallmatrix}$
&
$\begin{psmallmatrix}
|\varepsilon|^4p^{-1} & |\varepsilon|^3p^{-2} & |\varepsilon|^6 \\
-|\varepsilon|^3 & -|\varepsilon|^2p^{-1} & -|\varepsilon|^2 \\
-|\varepsilon| & -|\varepsilon|^3 & 1 \\
\end{psmallmatrix}$
\\
$\begin{psmallmatrix}
|\varepsilon|^4p^{-1} & |\varepsilon|^3p^{-2} & |\varepsilon|^6 \\
|\varepsilon|^3 & -|\varepsilon|^2p^{-1} & -|\varepsilon|^2 \\
|\varepsilon| & |\varepsilon|^3 & 1 \\
\end{psmallmatrix}$
&
$\begin{psmallmatrix}
|\varepsilon|^4p^{-1} & |\varepsilon|^3p^{-2} & |\varepsilon|^6 \\
-|\varepsilon|^3 & -|\varepsilon|^2p^{-1} & -|\varepsilon|^2 \\
|\varepsilon| & |\varepsilon|^3 & 1 \\
\end{psmallmatrix}$
&
$\begin{psmallmatrix}
|\varepsilon|^4p^{-1} & |\varepsilon|^3p^{-2} & |\varepsilon|^6 \\
|\varepsilon|^3 & -|\varepsilon|^2p^{-1} & -|\varepsilon|^2 \\
-|\varepsilon| & |\varepsilon|^3 & 1 \\
\end{psmallmatrix}$
&
$\begin{psmallmatrix}
|\varepsilon|^4p^{-1} & |\varepsilon|^3p^{-2} & |\varepsilon|^6 \\
-|\varepsilon|^3 & -|\varepsilon|^2p^{-1} & -|\varepsilon|^2 \\
-|\varepsilon| & |\varepsilon|^3 & 1 \\
\end{psmallmatrix}$
\\
$\begin{psmallmatrix}
|\varepsilon|^4p^{-1} & |\varepsilon|^3p^{-3} & |\varepsilon|^6 \\
|\varepsilon|^3 & -|\varepsilon|^2p^{-2} & |\varepsilon|^2 \\
|\varepsilon| & |\varepsilon|^6 & -1 \\
\end{psmallmatrix}$
&
$\begin{psmallmatrix}
|\varepsilon|^4p^{-1} & |\varepsilon|^3p^{-3} & |\varepsilon|^6 \\
-|\varepsilon|^3 & -|\varepsilon|^2p^{-2} & |\varepsilon|^2 \\
|\varepsilon| & |\varepsilon|^6 & -1 \\
\end{psmallmatrix}$
&
$\begin{psmallmatrix}
|\varepsilon|^4p^{-1} & |\varepsilon|^3p^{-3} & |\varepsilon|^6 \\
|\varepsilon|^3 & -|\varepsilon|^2p^{-2} & |\varepsilon|^2 \\
-|\varepsilon| & |\varepsilon|^6 & -1 \\
\end{psmallmatrix}$
&
$\begin{psmallmatrix}
|\varepsilon|^4p^{-1} & |\varepsilon|^3p^{-3} & |\varepsilon|^6 \\
-|\varepsilon|^3 & -|\varepsilon|^2p^{-2} & |\varepsilon|^2 \\
-|\varepsilon| & |\varepsilon|^6 & -1 \\
\end{psmallmatrix}$
\\
$\begin{psmallmatrix}
|\varepsilon|^4p^{-1} & |\varepsilon|^3p^{-3} & |\varepsilon|^6 \\
|\varepsilon|^3 & -|\varepsilon|^2p^{-2} & |\varepsilon|^2 \\
|\varepsilon| & -|\varepsilon|^6 & -1 \\
\end{psmallmatrix}$
&
$\begin{psmallmatrix}
|\varepsilon|^4p^{-1} & |\varepsilon|^3p^{-3} & |\varepsilon|^6 \\
-|\varepsilon|^3 & -|\varepsilon|^2p^{-2} & |\varepsilon|^2 \\
|\varepsilon| & -|\varepsilon|^6 & -1 \\
\end{psmallmatrix}$
&
$\begin{psmallmatrix}
|\varepsilon|^4p^{-1} & |\varepsilon|^3p^{-3} & |\varepsilon|^6 \\
|\varepsilon|^3 & -|\varepsilon|^2p^{-2} & |\varepsilon|^2 \\
-|\varepsilon| & -|\varepsilon|^6 & -1 \\
\end{psmallmatrix}$
&
$\begin{psmallmatrix}
|\varepsilon|^4p^{-1} & |\varepsilon|^3p^{-3} & |\varepsilon|^6 \\
-|\varepsilon|^3 & -|\varepsilon|^2p^{-2} & |\varepsilon|^2 \\
-|\varepsilon| & -|\varepsilon|^6 & -1 \\
\end{psmallmatrix}$
\\
$\begin{psmallmatrix}
|\varepsilon|^4p^{-1} & |\varepsilon|^3p^{-3} & |\varepsilon|^6 \\
|\varepsilon|^3 & -|\varepsilon|^2p^{-2} & -|\varepsilon|^2 \\
|\varepsilon| & |\varepsilon|^6 & 1 \\
\end{psmallmatrix}$
&
$\begin{psmallmatrix}
|\varepsilon|^4p^{-1} & |\varepsilon|^3p^{-3} & |\varepsilon|^6 \\
-|\varepsilon|^3 & -|\varepsilon|^2p^{-2} & -|\varepsilon|^2 \\
|\varepsilon| & |\varepsilon|^6 & 1 \\
\end{psmallmatrix}$
&
$\begin{psmallmatrix}
|\varepsilon|^4p^{-1} & |\varepsilon|^3p^{-3} & |\varepsilon|^6 \\
|\varepsilon|^3 & -|\varepsilon|^2p^{-2} & -|\varepsilon|^2 \\
-|\varepsilon| & |\varepsilon|^6 & 1 \\
\end{psmallmatrix}$
&
$\begin{psmallmatrix}
|\varepsilon|^4p^{-1} & |\varepsilon|^3p^{-3} & |\varepsilon|^6 \\
-|\varepsilon|^3 & -|\varepsilon|^2p^{-2} & -|\varepsilon|^2 \\
-|\varepsilon| & |\varepsilon|^6 & 1 \\
\end{psmallmatrix}$
\\
$\begin{psmallmatrix}
|\varepsilon|^4p^{-1} & |\varepsilon|^3p^{-3} & |\varepsilon|^6 \\
|\varepsilon|^3 & -|\varepsilon|^2p^{-2} & -|\varepsilon|^2 \\
|\varepsilon| & -|\varepsilon|^6 & 1 \\
\end{psmallmatrix}$
&
$\begin{psmallmatrix}
|\varepsilon|^4p^{-1} & |\varepsilon|^3p^{-3} & |\varepsilon|^6 \\
-|\varepsilon|^3 & -|\varepsilon|^2p^{-2} & -|\varepsilon|^2 \\
|\varepsilon| & -|\varepsilon|^6 & 1 \\
\end{psmallmatrix}$
&
$\begin{psmallmatrix}
|\varepsilon|^4p^{-1} & |\varepsilon|^3p^{-3} & |\varepsilon|^6 \\
|\varepsilon|^3 & -|\varepsilon|^2p^{-2} & -|\varepsilon|^2 \\
-|\varepsilon| & -|\varepsilon|^6 & 1 \\
\end{psmallmatrix}$
&
$\begin{psmallmatrix}
|\varepsilon|^4p^{-1} & |\varepsilon|^3p^{-3} & |\varepsilon|^6 \\
-|\varepsilon|^3 & -|\varepsilon|^2p^{-2} & -|\varepsilon|^2 \\
-|\varepsilon| & -|\varepsilon|^6 & 1 \\
\end{psmallmatrix}$
\\
$\begin{psmallmatrix}
|\varepsilon|^4p^{-2} & |\varepsilon|^3p^{-2} & |\varepsilon|^6 \\
|\varepsilon|^3p^{-1} & -|\varepsilon|^2p^{-1} & |\varepsilon|^2 \\
-|\varepsilon|^4 & -|\varepsilon|^3 & -1 \\
\end{psmallmatrix}$
&
$\begin{psmallmatrix}
|\varepsilon|^4p^{-2} & |\varepsilon|^3p^{-2} & |\varepsilon|^6 \\
|\varepsilon|^3p^{-1} & -|\varepsilon|^2p^{-1} & |\varepsilon|^2 \\
|\varepsilon|^4 & -|\varepsilon|^3 & -1 \\
\end{psmallmatrix}$
&
$\begin{psmallmatrix}
|\varepsilon|^4p^{-2} & |\varepsilon|^3p^{-2} & |\varepsilon|^6 \\
|\varepsilon|^3p^{-1} & -|\varepsilon|^2p^{-1} & |\varepsilon|^2 \\
-|\varepsilon|^4 & |\varepsilon|^3 & -1 \\
\end{psmallmatrix}$
&
$\begin{psmallmatrix}
|\varepsilon|^4p^{-2} & |\varepsilon|^3p^{-2} & |\varepsilon|^6 \\
|\varepsilon|^3p^{-1} & -|\varepsilon|^2p^{-1} & |\varepsilon|^2 \\
|\varepsilon|^4 & |\varepsilon|^3 & -1 \\
\end{psmallmatrix}$
\\
$\begin{psmallmatrix}
|\varepsilon|^4p^{-2} & |\varepsilon|^3p^{-2} & |\varepsilon|^6 \\
|\varepsilon|^3p^{-1} & -|\varepsilon|^2p^{-1} & -|\varepsilon|^2 \\
-|\varepsilon|^4 & -|\varepsilon|^3 & 1 \\
\end{psmallmatrix}$
&
$\begin{psmallmatrix}
|\varepsilon|^4p^{-2} & |\varepsilon|^3p^{-2} & |\varepsilon|^6 \\
|\varepsilon|^3p^{-1} & -|\varepsilon|^2p^{-1} & -|\varepsilon|^2 \\
|\varepsilon|^4 & -|\varepsilon|^3 & 1 \\
\end{psmallmatrix}$
&
$\begin{psmallmatrix}
|\varepsilon|^4p^{-2} & |\varepsilon|^3p^{-2} & |\varepsilon|^6 \\
|\varepsilon|^3p^{-1} & -|\varepsilon|^2p^{-1} & -|\varepsilon|^2 \\
-|\varepsilon|^4 & |\varepsilon|^3 & 1 \\
\end{psmallmatrix}$
&
$\begin{psmallmatrix}
|\varepsilon|^4p^{-2} & |\varepsilon|^3p^{-2} & |\varepsilon|^6 \\
|\varepsilon|^3p^{-1} & -|\varepsilon|^2p^{-1} & -|\varepsilon|^2 \\
|\varepsilon|^4 & |\varepsilon|^3 & 1 \\
\end{psmallmatrix}$
\\
$\begin{psmallmatrix}
|\varepsilon|^4p^{-2} & |\varepsilon|^3p^{-3} & |\varepsilon|^6 \\
|\varepsilon|^3p^{-1} & -|\varepsilon|^2p^{-2} & |\varepsilon|^2 \\
-|\varepsilon|^4 & |\varepsilon|^6 & -1 \\
\end{psmallmatrix}$
&
$\begin{psmallmatrix}
|\varepsilon|^4p^{-2} & |\varepsilon|^3p^{-3} & |\varepsilon|^6 \\
|\varepsilon|^3p^{-1} & -|\varepsilon|^2p^{-2} & |\varepsilon|^2 \\
|\varepsilon|^4 & |\varepsilon|^6 & -1 \\
\end{psmallmatrix}$
&
$\begin{psmallmatrix}
|\varepsilon|^4p^{-2} & |\varepsilon|^3p^{-3} & |\varepsilon|^6 \\
|\varepsilon|^3p^{-1} & -|\varepsilon|^2p^{-2} & |\varepsilon|^2 \\
-|\varepsilon|^4 & -|\varepsilon|^6 & -1 \\
\end{psmallmatrix}$
&
$\begin{psmallmatrix}
|\varepsilon|^4p^{-2} & |\varepsilon|^3p^{-3} & |\varepsilon|^6 \\
|\varepsilon|^3p^{-1} & -|\varepsilon|^2p^{-2} & |\varepsilon|^2 \\
|\varepsilon|^4 & -|\varepsilon|^6 & -1 \\
\end{psmallmatrix}$
\\
$\begin{psmallmatrix}
|\varepsilon|^4p^{-2} & |\varepsilon|^3p^{-3} & |\varepsilon|^6 \\
|\varepsilon|^3p^{-1} & -|\varepsilon|^2p^{-2} & -|\varepsilon|^2 \\
-|\varepsilon|^4 & |\varepsilon|^6 & 1 \\
\end{psmallmatrix}$
&
$\begin{psmallmatrix}
|\varepsilon|^4p^{-2} & |\varepsilon|^3p^{-3} & |\varepsilon|^6 \\
|\varepsilon|^3p^{-1} & -|\varepsilon|^2p^{-2} & -|\varepsilon|^2 \\
|\varepsilon|^4 & |\varepsilon|^6 & 1 \\
\end{psmallmatrix}$
&
$\begin{psmallmatrix}
|\varepsilon|^4p^{-2} & |\varepsilon|^3p^{-3} & |\varepsilon|^6 \\
|\varepsilon|^3p^{-1} & -|\varepsilon|^2p^{-2} & -|\varepsilon|^2 \\
-|\varepsilon|^4 & -|\varepsilon|^6 & 1 \\
\end{psmallmatrix}$
&
$\begin{psmallmatrix}
|\varepsilon|^4p^{-2} & |\varepsilon|^3p^{-3} & |\varepsilon|^6 \\
|\varepsilon|^3p^{-1} & -|\varepsilon|^2p^{-2} & -|\varepsilon|^2 \\
|\varepsilon|^4 & -|\varepsilon|^6 & 1 \\
\end{psmallmatrix}$
\\
$\begin{psmallmatrix}
|\varepsilon|^4p^{-2} & |\varepsilon|^3p^{-2} & |\varepsilon|^6 \\
-|\varepsilon|^6 & -|\varepsilon|^2p^{-1} & |\varepsilon|^2 \\
-|\varepsilon|^4 & -|\varepsilon|^3 & -1 \\
\end{psmallmatrix}$
&
$\begin{psmallmatrix}
|\varepsilon|^4p^{-2} & |\varepsilon|^3p^{-2} & |\varepsilon|^6 \\
|\varepsilon|^6 & -|\varepsilon|^2p^{-1} & |\varepsilon|^2 \\
-|\varepsilon|^4 & -|\varepsilon|^3 & -1 \\
\end{psmallmatrix}$
&
$\begin{psmallmatrix}
|\varepsilon|^4p^{-2} & |\varepsilon|^3p^{-2} & |\varepsilon|^6 \\
-|\varepsilon|^6 & -|\varepsilon|^2p^{-1} & |\varepsilon|^2 \\
|\varepsilon|^4 & -|\varepsilon|^3 & -1 \\
\end{psmallmatrix}$
&
$\begin{psmallmatrix}
|\varepsilon|^4p^{-2} & |\varepsilon|^3p^{-2} & |\varepsilon|^6 \\
|\varepsilon|^6 & -|\varepsilon|^2p^{-1} & |\varepsilon|^2 \\
|\varepsilon|^4 & -|\varepsilon|^3 & -1 \\
\end{psmallmatrix}$
\\
$\begin{psmallmatrix}
|\varepsilon|^4p^{-2} & |\varepsilon|^3p^{-2} & |\varepsilon|^6 \\
-|\varepsilon|^6 & -|\varepsilon|^2p^{-1} & |\varepsilon|^2 \\
-|\varepsilon|^4 & |\varepsilon|^3 & -1 \\
\end{psmallmatrix}$
&
$\begin{psmallmatrix}
|\varepsilon|^4p^{-2} & |\varepsilon|^3p^{-2} & |\varepsilon|^6 \\
|\varepsilon|^6 & -|\varepsilon|^2p^{-1} & |\varepsilon|^2 \\
-|\varepsilon|^4 & |\varepsilon|^3 & -1 \\
\end{psmallmatrix}$
&
$\begin{psmallmatrix}
|\varepsilon|^4p^{-2} & |\varepsilon|^3p^{-2} & |\varepsilon|^6 \\
-|\varepsilon|^6 & -|\varepsilon|^2p^{-1} & |\varepsilon|^2 \\
|\varepsilon|^4 & |\varepsilon|^3 & -1 \\
\end{psmallmatrix}$
&
$\begin{psmallmatrix}
|\varepsilon|^4p^{-2} & |\varepsilon|^3p^{-2} & |\varepsilon|^6 \\
|\varepsilon|^6 & -|\varepsilon|^2p^{-1} & |\varepsilon|^2 \\
|\varepsilon|^4 & |\varepsilon|^3 & -1 \\
\end{psmallmatrix}$
\\
$\begin{psmallmatrix}
|\varepsilon|^4p^{-2} & |\varepsilon|^3p^{-2} & |\varepsilon|^6 \\
-|\varepsilon|^6 & -|\varepsilon|^2p^{-1} & -|\varepsilon|^2 \\
-|\varepsilon|^4 & -|\varepsilon|^3 & 1 \\
\end{psmallmatrix}$
&
$\begin{psmallmatrix}
|\varepsilon|^4p^{-2} & |\varepsilon|^3p^{-2} & |\varepsilon|^6 \\
|\varepsilon|^6 & -|\varepsilon|^2p^{-1} & -|\varepsilon|^2 \\
-|\varepsilon|^4 & -|\varepsilon|^3 & 1 \\
\end{psmallmatrix}$
&
$\begin{psmallmatrix}
|\varepsilon|^4p^{-2} & |\varepsilon|^3p^{-2} & |\varepsilon|^6 \\
-|\varepsilon|^6 & -|\varepsilon|^2p^{-1} & -|\varepsilon|^2 \\
|\varepsilon|^4 & -|\varepsilon|^3 & 1 \\
\end{psmallmatrix}$
&
$\begin{psmallmatrix}
|\varepsilon|^4p^{-2} & |\varepsilon|^3p^{-2} & |\varepsilon|^6 \\
|\varepsilon|^6 & -|\varepsilon|^2p^{-1} & -|\varepsilon|^2 \\
|\varepsilon|^4 & -|\varepsilon|^3 & 1 \\
\end{psmallmatrix}$
\\
$\begin{psmallmatrix}
|\varepsilon|^4p^{-2} & |\varepsilon|^3p^{-2} & |\varepsilon|^6 \\
-|\varepsilon|^6 & -|\varepsilon|^2p^{-1} & -|\varepsilon|^2 \\
-|\varepsilon|^4 & |\varepsilon|^3 & 1 \\
\end{psmallmatrix}$
&
$\begin{psmallmatrix}
|\varepsilon|^4p^{-2} & |\varepsilon|^3p^{-2} & |\varepsilon|^6 \\
|\varepsilon|^6 & -|\varepsilon|^2p^{-1} & -|\varepsilon|^2 \\
-|\varepsilon|^4 & |\varepsilon|^3 & 1 \\
\end{psmallmatrix}$
&
$\begin{psmallmatrix}
|\varepsilon|^4p^{-2} & |\varepsilon|^3p^{-2} & |\varepsilon|^6 \\
-|\varepsilon|^6 & -|\varepsilon|^2p^{-1} & -|\varepsilon|^2 \\
|\varepsilon|^4 & |\varepsilon|^3 & 1 \\
\end{psmallmatrix}$
&
$\begin{psmallmatrix}
|\varepsilon|^4p^{-2} & |\varepsilon|^3p^{-2} & |\varepsilon|^6 \\
|\varepsilon|^6 & -|\varepsilon|^2p^{-1} & -|\varepsilon|^2 \\
|\varepsilon|^4 & |\varepsilon|^3 & 1 \\
\end{psmallmatrix}$
\\ \hline
  \end{tabular}
\end{table}


\chapter{Group theory and modular forms of $\Gamma_6$}
\label{app:Gamma_6}


\section{Group theoretical aspects}
\label{app:gt_Gamma_6}

Here, we review the group theoretical aspects of $\Gamma_6$.
$\Gamma_6$ group is generated by two generators, $S$ and $T$-transformations which satisfy the following algebraic relations:
\begin{align}
  S^2 = (ST)^3 = T^6 = ST^2ST^3ST^4ST^3 = \mathbb{I}.
\end{align}
In $\Gamma_6$ group, there are 12 irreducible representations, six singlets $\bm{1}^0_0$, $\bm{1}^0_1$, $\bm{1}^0_2$, $\bm{1}^1_0$, $\bm{1}^1_1$ and $\bm{1}^1_2$, three doublets $\bm{2}_0$, $\bm{2}_1$ and $\bm{2}_2$, two triplets $\bm{3}^0$ and $\bm{3}^1$ and one 6D representation $\bm{6}$.
Each irreducible representation is given by
\begin{align}
\begin{aligned}
  &\bm{1}^r_k:~ S=(-1)^r, \quad T=(-1)^r\omega^k, \\
  &\bm{2}_k:~  S=\frac{1}{2}\begin{pmatrix}-1&\sqrt{3}\\\sqrt{3}& 1\\\end{pmatrix}, \quad T = \omega^k\begin{pmatrix}1&0\\0&-1\\\end{pmatrix}, \\
  &\bm{3}^r:~ (-1)^r\bm{a}_3, \quad (-1)^r\bm{b}_3, \\
  &\bm{6}:~\frac{1}{2}\begin{pmatrix}-\bm{a}_3 & \sqrt{3}\bm{a}_3 \\ \sqrt{3}\bm{a}_3 & \bm{a}_3\\\end{pmatrix}, \quad T=\begin{pmatrix}\bm{b}_3 & \bm{0} \\ \bm{0} & -\bm{b}_3\\\end{pmatrix},
\end{aligned} \label{eq:irreps_6}
\end{align}
where $r=0,1$, $k=0,1,2$ and
\begin{align}
  {\bm a}_3 = \frac{1}{3}
  \begin{pmatrix}
    -1 & 2 & 2 \\
    2 & -1 & 2 \\
    2 & 2 & -1 \\
  \end{pmatrix}, \quad
  {\bm b}_3 =
  \begin{pmatrix}
    1 & 0 & 0 \\
    0 & \omega & 0 \\
    0 & 0 & \omega^2 \\
  \end{pmatrix}.
\end{align}
In this basis, the Kronecker products between irreducible representations are:
\begin{align}
  &\bm{1}^r_i \otimes \bm{1}^s_j = \bm{1}^t_m, \quad
  \bm{1}^r_i \otimes \bm{2}_j = \bm{2}_m, \quad
  \bm{1}^r_i \otimes \bm{3}^s = \bm{3}^t, \quad
  \bm{1}^r_i \otimes \bm{6} = \bm{6}, \\
  &\bm{2}_i \otimes \bm{2}_j = \bm{1}^0_m \oplus \bm{1}^1_m \oplus \bm{2}_m, \quad
  \bm{2}_i \otimes \bm{3}^r = \bm{6}, \quad
  \bm{2}_i \otimes \bm{6} = \bm{3}^0 \oplus \bm{3}^1 \oplus \bm{6}, \\
  &\bm{3}^r \otimes \bm{3}^s = \bm{1}^t_0 \oplus \bm{1}^t_1 \oplus \bm{1}^t_2 \oplus \bm{3}^t_1 \oplus \bm{3}^t_2, \quad
  \bm{3}^r \otimes \bm{6} = \bm{2}_0 \oplus \bm{2}_1 \oplus \bm{2}_2 \oplus \bm{6} \oplus \bm{6}, \\
  &\bm{6} \otimes \bm{6} = \bm{1}^0_0 \oplus \bm{1}^0_1 \oplus \bm{1}^0_2 \oplus \bm{1}^1_0 \oplus \bm{1}^1_1 \oplus \bm{1}^1_2 \oplus \bm{2}_0 \oplus \bm{2}_1 \oplus \bm{2}_2 \oplus \bm{3}^0 \oplus \bm{3}^0 \oplus \bm{3}^1 \oplus \bm{3}^1 \oplus \bm{6} \oplus \bm{6},
\end{align}
where $i,j=0,1,2$, $r,s=0,1$, $m = i+j~(\textrm{mod}~3)$ and $t = r+s~(\textrm{mod}~2)$.
In the following, we show the Clebsch-Gordon (CG) coefficients of these products,
\begin{align}
  &(\alpha_1)_{{\bm{1}^r_i}} \otimes \begin{pmatrix}\beta_1\\\beta_2\\\end{pmatrix}_{{\bm{2}_j}} = \alpha_1P^r_2\begin{pmatrix}\beta_1\\\beta_2\\\end{pmatrix}_{{\bm{2}}_m}, \quad
  (\alpha_1)_{{\bm{1}^r_i}} \otimes \begin{pmatrix}\beta_1\\\beta_2\\\beta_3\\\end{pmatrix}_{{\bm{3}^s}} = \alpha_1 P^i_3\begin{pmatrix}\beta_1\\\beta_2\\\beta_3\\\end{pmatrix}_{{\bm{3}^t}}, \notag \\
  &(\alpha_1)_{{\bm{1}^r_i}} \otimes \begin{pmatrix}\beta_1\\ \beta_2\\ \beta_3\\ \beta_4\\ \beta_5\\ \beta_6\\\end{pmatrix}_{\bm{6}} = \alpha_1 P_6(r,i)\begin{pmatrix}\beta_1\\ \beta_2\\ \beta_3\\ \beta_4\\ \beta_5\\ \beta_6\\\end{pmatrix}_{\bm{6}}, \notag 
\end{align}
\begin{align}
  &\begin{pmatrix}\alpha_1\\\alpha_2\\\end{pmatrix}_{{\bm{2}_i}} \otimes \begin{pmatrix}\beta_1\\\beta_2\\\end{pmatrix}_{{\bm{2}_j}} = \frac{1}{\sqrt{2}}(\alpha_1\beta_1+\alpha_2\beta_2)_{{\bm{1}^0_m}} \oplus \frac{1}{\sqrt{2}}(\alpha_1\beta_2-\alpha_2\beta_1)_{{\bm{1}^1_m}} \oplus \frac{1}{\sqrt{2}}\begin{pmatrix}\alpha_1\beta_1-\alpha_2\beta_2\\-\alpha_1\beta_2-\alpha_2\beta_1\\\end{pmatrix}_{{\bm{2}_m}}, \notag \\
  &\begin{pmatrix}\alpha_1\\\alpha_2\\\end{pmatrix}_{{\bm{2}_i}} \otimes \begin{pmatrix}\beta_1\\\beta_2\\\beta_3\\\end{pmatrix}_{{\bm{3}^r}} = P_6(r,i)\begin{pmatrix}\alpha_1\beta_1\\ \alpha_1\beta_2\\ \alpha_1\beta_3\\ \alpha_2\beta_1\\ \alpha_2\beta_2\\ \alpha_2\beta_3\\\end{pmatrix}_{\bm{6}}, \notag \\
  &\begin{pmatrix}\alpha_1\\\alpha_2\\\end{pmatrix}_{{\bm{2}_i}} \otimes \begin{pmatrix}\beta_1\\ \beta_2\\ \beta_3\\ \beta_4\\ \beta_5\\ \beta_6\\\end{pmatrix}_{\bm{6}} = 
  \frac{P^i_3}{\sqrt{2}}\begin{pmatrix}\alpha_1\beta_1+\alpha_2\beta_4\\\alpha_1\beta_2+\alpha_2\beta_5\\\alpha_1\beta_3+\alpha_2\beta_6\\\end{pmatrix}_{{\bm{3}^0}} \oplus \frac{P^i_3}{\sqrt{2}}\begin{pmatrix}\alpha_1\beta_4-\alpha_2\beta_1\\\alpha_1\beta_5-\alpha_2\beta_2\\\alpha_1\beta_6-\alpha_2\beta_3\\\end{pmatrix}_{{\bm{3}^1}} \oplus \frac{P_6(0,i)}{\sqrt{2}}\begin{pmatrix}\alpha_1\beta_1-\alpha_2\beta_4\\\alpha_1\beta_2-\alpha_2\beta_5\\\alpha_1\beta_3-\alpha_2\beta_6\\-\alpha_1\beta_4-\alpha_2\beta_1\\-\alpha_1\beta_5-\alpha_2\beta_2\\-\alpha_1\beta_6-\alpha_2\beta_3\\\end{pmatrix}_{\bm{6}}, \notag 
\end{align}
\begin{align}
  &\begin{pmatrix}\alpha_1\\\alpha_2\\\alpha_3\\\end{pmatrix}_{{\bm{3}^r}} \otimes \begin{pmatrix}\beta_1\\\beta_2\\\beta_3\\\end{pmatrix}_{{\bm{3}^s}} 
  = \frac{1}{\sqrt{3}}(\alpha_1\beta_1+\alpha_2\beta_3+\alpha_3\beta_2)_{{\bm{1}^t_0}} 
  \oplus \frac{1}{\sqrt{3}}(\alpha_1\beta_2+\alpha_2\beta_1+\alpha_3\beta_3)_{{\bm{1}^t_1}} \notag \\
  &\oplus \frac{1}{\sqrt{3}}(\alpha_1\beta_3+\alpha_2\beta_2+\alpha_3\beta_1)_{{\bm{1}^t_2}} 
  \oplus \frac{1}{\sqrt{3}}\begin{pmatrix}2\alpha_1\beta_1-\alpha_2\beta_3-\alpha_3\beta_2 \\-\alpha_1\beta_2-\alpha_2\beta_1+2\alpha_3\beta_3 \\-\alpha_1\beta_3+2\alpha_2\beta_2-\alpha_3\beta_1 \\\end{pmatrix}_{{\bm{3}^t_1}} 
  \oplus \frac{1}{\sqrt{2}}\begin{pmatrix}-\alpha_2\beta_3+\alpha_3\beta_2 \\-\alpha_1\beta_2+\alpha_2\beta_1 \\\alpha_1\beta_3-\alpha_3\beta_1 \\\end{pmatrix}_{{\bm{3}^t_2}}, \notag \\
  &\begin{pmatrix}\alpha_1\\\alpha_2\\\alpha_3\\\end{pmatrix}_{{\bm{3}^r}} \otimes \begin{pmatrix}\beta_1\\ \beta_2\\ \beta_3\\ \beta_4\\ \beta_5\\ \beta_6\\\end{pmatrix}_{\bm{6}} 
  = \frac{P_2^r}{\sqrt{3}} \begin{pmatrix} \alpha_1\beta_1+\alpha_2\beta_3+\alpha_3\beta_2 \\\alpha_1\beta_4+\alpha_2\beta_6+\alpha_3\beta_5 \\ \end{pmatrix}_{{\bm{2}_0}} 
  \oplus \frac{P_2^r}{\sqrt{3}} \begin{pmatrix} \alpha_1\beta_2+\alpha_2\beta_1+\alpha_3\beta_3 \\\alpha_1\beta_5+\alpha_2\beta_4+\alpha_3\beta_6 \\ \end{pmatrix}_{{\bm{2}_1}} \notag \\
  &\oplus \frac{P_2^r}{\sqrt{3}} \begin{pmatrix} \alpha_1\beta_3+\alpha_2\beta_2+\alpha_3\beta_1 \\\alpha_1\beta_6+\alpha_2\beta_5+\alpha_3\beta_4 \\ \end{pmatrix}_{{\bm{2}_2}} 
  \oplus \frac{P_6(r,0)}{\sqrt{2}} \begin{pmatrix}  
  \alpha_1 \beta_1 - \alpha_3 \beta_2 \\
-\alpha_2 \beta_1 + \alpha_3 \beta_3 \\
-\alpha_1 \beta_3 + \alpha_2 \beta_2 \\
\alpha_1 \beta_4 - \alpha_3 \beta_5 \\
-\alpha_2 \beta_4 + \alpha_3 \beta_6 \\
-\alpha_1 \beta_6 + \alpha_2 \beta_5 \\  
  \end{pmatrix}_{\bm{6}} 
  \oplus \frac{P_6(r,0)}{\sqrt{2}} \begin{pmatrix}  
  \alpha_2 \beta_3 - \alpha_3 \beta_2 \\
\alpha_1 \beta_2 - \alpha_2 \beta_1 \\
-\alpha_1 \beta_3 + \alpha_3 \beta_1 \\
\alpha_2 \beta_6 - \alpha_3 \beta_5 \\
\alpha_1 \beta_5 - \alpha_2 \beta_4 \\
-\alpha_1 \beta_6 + \alpha_3 \beta_4 \\
  \end{pmatrix}_{\bm{6}}, \notag 
\end{align}
\begin{align}
  \begin{pmatrix}\alpha_1\\ \alpha_2\\ \alpha_3\\ \alpha_4\\ \alpha_5\\ \alpha_6\\\end{pmatrix}_{\bm{6}} \otimes \begin{pmatrix}\beta_1\\ \beta_2\\ \beta_3\\ \beta_4\\ \beta_5\\ \beta_6\\\end{pmatrix}_{\bm{6}}
  =~& 
  \begin{matrix}
  \frac{1}{\sqrt{6}}(\alpha_1 \beta_1 + \alpha_2 \beta_3 + \alpha_3 \beta_2 + \alpha_4 \beta_4 + \alpha_5 \beta_6 + \alpha_6 \beta_5)_{{\bm{1}^0_0}}~~ \\
  \oplus \frac{1}{\sqrt{6}}(\alpha_1 \beta_2 + \alpha_2 \beta_1 + \alpha_3 \beta_3 + \alpha_4 \beta_5 + \alpha_5 \beta_4 + \alpha_6 \beta_6)_{{\bm{1}^0_1}} \\
  \oplus \frac{1}{\sqrt{6}}(\alpha_1 \beta_3 + \alpha_2 \beta_2 + \alpha_3 \beta_1 + \alpha_4 \beta_6 + \alpha_5 \beta_5 + \alpha_6 \beta_4)_{{\bm{1}^0_2}} \\
  \oplus \frac{1}{\sqrt{6}}(\alpha_1 \beta_4 + \alpha_2 \beta_6 + \alpha_3 \beta_5 - \alpha_4 \beta_1 - \alpha_5 \beta_3 - \alpha_6 \beta_2)_{{\bm{1}^1_0}} \\
  \oplus \frac{1}{\sqrt{6}}(\alpha_1 \beta_5 + \alpha_2 \beta_4 + \alpha_3 \beta_6 - \alpha_4 \beta_2 - \alpha_5 \beta_1 - \alpha_6 \beta_3)_{{\bm{1}^1_1}} \\
  \oplus \frac{1}{\sqrt{6}}(\alpha_1 \beta_6 + \alpha_2 \beta_5 + \alpha_3 \beta_4 - \alpha_4 \beta_3 - \alpha_5 \beta_2 - \alpha_6 \beta_1)_{{\bm{1}^1_2}} \\
  \end{matrix} \notag \\
  &\oplus \textstyle\frac{1}{\sqrt{6}}\begin{pmatrix}\alpha_1 \beta_1 + \alpha_2 \beta_3 + \alpha_3 \beta_2 - \alpha_4 \beta_4 - \alpha_5 \beta_6 - \alpha_6 \beta_5 \\ -(\alpha_1 \beta_4 + \alpha_2 \beta_6 + \alpha_3 \beta_5 + \alpha_4 \beta_1 + \alpha_5 \beta_3 + \alpha_6 \beta_2)\end{pmatrix}_{{\bm{2}_0}} \notag \\
  &\oplus \textstyle\frac{1}{\sqrt{6}}\begin{pmatrix}\alpha_1 \beta_2 + \alpha_2 \beta_1 + \alpha_3 \beta_3 - \alpha_4 \beta_5 - \alpha_5 \beta_4 - \alpha_6 \beta_6 \\ -(\alpha_1 \beta_5 + \alpha_2 \beta_4 + \alpha_3 \beta_6 + \alpha_4 \beta_2 + \alpha_5 \beta_1 + \alpha_6 \beta_3 )\end{pmatrix}_{{\bm{2}_1}} \notag \\
  &\oplus \textstyle\frac{1}{\sqrt{6}}\begin{pmatrix} \alpha_1 \beta_3 + \alpha_2 \beta_2 + \alpha_3 \beta_1 - \alpha_4 \beta_6 - \alpha_5 \beta_5 - \alpha_6 \beta_4 \\ -(\alpha_1 \beta_6 + \alpha_2 \beta_5 + \alpha_3 \beta_4 + \alpha_4 \beta_3 + \alpha_5 \beta_2 + \alpha_6 \beta_1 ) \\ \end{pmatrix}_{{\bm{2}_2}} \notag \\ 
  &\oplus \textstyle\frac{1}{2\sqrt{3}}\begin{pmatrix} 2\alpha_1 \beta_1 - \alpha_2 \beta_3 - \alpha_3 \beta_2 + 2\alpha_4 \beta_4 - \alpha_5 \beta_6 - \alpha_6 \beta_5 \\
2\alpha_3 \beta_3 - \alpha_1 \beta_2 - \alpha_2 \beta_1 + 2\alpha_6 \beta_6 - \alpha_4 \beta_5 - \alpha_5 \beta_4 \\
2\alpha_2 \beta_2 - \alpha_1 \beta_3 - \alpha_3 \beta_1 + 2\alpha_5 \beta_5 - \alpha_4 \beta_6 - \alpha_6 \beta_4 \\ \end{pmatrix}_{{\bm{3}^0}} \notag \\ 
  &\oplus \textstyle\frac{1}{2}\begin{pmatrix} \alpha_2 \beta_3 - \alpha_3 \beta_2 + \alpha_5 \beta_6 - \alpha_6 \beta_5 \\
\alpha_1 \beta_2 - \alpha_2 \beta_1 + \alpha_4 \beta_5 - \alpha_5 \beta_4 \\
-\alpha_1 \beta_3 + \alpha_3 \beta_1 - \alpha_4 \beta_6 + \alpha_6 \beta_4 \\ \end{pmatrix}_{{\bm{3}^0}} \notag \\ 
  &\oplus \textstyle\frac{1}{2}\begin{pmatrix} \alpha_2 \beta_6 - \alpha_3 \beta_5 - \alpha_5 \beta_3 + \alpha_6 \beta_2 \\
\alpha_1 \beta_5 - \alpha_2 \beta_4 - \alpha_4 \beta_2 + \alpha_5 \beta_1 \\
-\alpha_1 \beta_6 + \alpha_3 \beta_4 + \alpha_4 \beta_3 - \alpha_6 \beta_1 \\ \end{pmatrix}_{{\bm{3}^1}} \notag \\ 
  &\oplus \textstyle\frac{1}{2\sqrt{3}}\begin{pmatrix} 2\alpha_1 \beta_4 - \alpha_2 \beta_6 - \alpha_3 \beta_5 - 2\alpha_4 \beta_1 + \alpha_5 \beta_3 + \alpha_6 \beta_2 \\
-\alpha_1 \beta_5 - \alpha_2 \beta_4 + 2\alpha_3 \beta_6 + \alpha_4 \beta_2 + \alpha_5 \beta_1 - 2\alpha_6 \beta_3 \\
-\alpha_1 \beta_6 + 2\alpha_2 \beta_5 - \alpha_3 \beta_4 + \alpha_4 \beta_3 - 2\alpha_5 \beta_2 + \alpha_6 \beta_1 \\ \end{pmatrix}_{{\bm{3}^1}} \notag \\ 
  &\oplus \textstyle\frac{1}{2\sqrt{3}}\begin{pmatrix} 2\alpha_1 \beta_1 - \alpha_2 \beta_3 - \alpha_3 \beta_2 - 2\alpha_4 \beta_4 + \alpha_5 \beta_6 + \alpha_6 \beta_5 \\
-\alpha_1 \beta_2 - \alpha_2 \beta_1 + 2\alpha_3 \beta_3 + \alpha_4 \beta_5 + \alpha_5 \beta_4 - 2\alpha_6 \beta_6 \\
-\alpha_1 \beta_3 + 2\alpha_2 \beta_2 - \alpha_3 \beta_1 + \alpha_4 \beta_6 - 2\alpha_5 \beta_5 + \alpha_6 \beta_4 \\
-2\alpha_1 \beta_4 + \alpha_2 \beta_6 + \alpha_3 \beta_5 - 2\alpha_4 \beta_1 + \alpha_5 \beta_3 + \alpha_6 \beta_2 \\
\alpha_1 \beta_5 + \alpha_2 \beta_4 - 2\alpha_3 \beta_6 + \alpha_4 \beta_2 + \alpha_5 \beta_1 - 2\alpha_6 \beta_3 \\
\alpha_1 \beta_6 - 2\alpha_2 \beta_5 + \alpha_3 \beta_4 + \alpha_4 \beta_3 - 2\alpha_5 \beta_2 + \alpha_6 \beta_1 \\\end{pmatrix}_{\bm{6}} \notag \\
  &\oplus \textstyle\frac{1}{2}\begin{pmatrix} \alpha_2 \beta_3 - \alpha_3 \beta_2 - \alpha_5 \beta_6 + \alpha_6 \beta_5 \\
\alpha_1 \beta_2 - \alpha_2 \beta_1 - \alpha_4 \beta_5 + \alpha_5 \beta_4 \\
-\alpha_1 \beta_3 + \alpha_3 \beta_1 + \alpha_4 \beta_6 - \alpha_6 \beta_4 \\
-\alpha_2 \beta_6 + \alpha_3 \beta_5 - \alpha_5 \beta_3 + \alpha_6 \beta_2 \\
-\alpha_1 \beta_5 + \alpha_2 \beta_4 - \alpha_4 \beta_2 + \alpha_5 \beta_1 \\
\alpha_1 \beta_6 - \alpha_3 \beta_4 + \alpha_4 \beta_3 - \alpha_6 \beta_1 \\ \end{pmatrix}_{\bm{6}}, \notag
\end{align}
where we have used the notations,
\begin{align}
  P_2 = 
  \begin{pmatrix}
    0 & 1 \\
    -1 & 0 \\
  \end{pmatrix}, \quad
  P_3 = 
  \begin{pmatrix}
    0 & 0 & 1 \\
    1 & 0 & 0 \\
    0 & 1 & 0 \\
  \end{pmatrix}, \quad
  P_6(r,i) =
  \begin{pmatrix}
    \bm{0}_3 & \bm{1}_3 \\
    -\bm{1}_3 & \bm{0}_3 \\
  \end{pmatrix}^r
  \begin{pmatrix}
    P_3 & \bm{0}_3 \\
    \bm{0}_3 & P_3 \\
  \end{pmatrix}^i.
\end{align}
Further details can be found in Ref.~\cite{Li:2021buv}.


\section{Modular forms}
\label{app:modular_forms_6}

Here we review the modular forms of $\Gamma_6$.
The modular forms of level 6 of even weights can be constructed from the products of the Dedekind eta function \cite{Li:2021buv},
\begin{align}
  \eta(\tau) = q^{1/24} \prod_{n=1}^\infty (1-q^n), \quad q = e^{2\pi i\tau}.
\end{align}
Using $\eta$, we introduce following functions,
\begin{align}
  &Y_1(\tau) = 3\frac{\eta^3(3\tau)}{\eta(\tau)}+\frac{\eta^3(\tau/3)}{\eta(\tau)}, \\
  &Y_2(\tau) = 3\sqrt{2} \frac{\eta^3(3\tau)}{\eta(\tau)}, \\
  &Y_3(\tau) = 3\sqrt{2}\frac{\eta^3(6\tau)}{\eta(2\tau)}, \\
  &Y_4(\tau) = -3\frac{\eta^3(6\tau)}{\eta(2\tau)}-\frac{\eta^3(2\tau/3)}{\eta(2\tau)}, \\
  &Y_5(\tau) = \sqrt{6}\frac{\eta^3(6\tau)}{\eta(2\tau)}-\sqrt{6}\frac{\eta^3(3\tau/2)}{\eta(\tau/2)}, \\
  &Y_6(\tau) = -\sqrt{3}\frac{\eta^3(6\tau)}{\eta(2\tau)}+\frac{1}{\sqrt{3}}\frac{\eta^3(\tau/6)}{\eta(\tau/2)}-\frac{1}{\sqrt{3}}\frac{\eta^3(2\tau/3)}{\eta(2\tau)}+\sqrt{3}\frac{\eta^3(3\tau/2)}{\eta(\tau/2)}.
\end{align}
Then we can find four linearly independent modular forms of weight 2:
\begin{align}
  &Y^{(2)}_{{\bm{3}^0}}(\tau) = 
  \begin{pmatrix}
    -Y_1^2 \\ \sqrt{2}Y_1Y_2 \\ Y_2^2 \\
  \end{pmatrix}, \quad
  Y^{(2)}_{{\bm{1}^1_2}}(\tau) = Y_3Y_6-Y_4Y_5, \quad
  Y^{(2)}_{{\bm{2}_0}}(\tau) = \frac{1}{\sqrt{2}}
  \begin{pmatrix}
    Y_1Y_4-Y_2Y_3 \\ Y_1Y_6-Y_2Y_5 \\
  \end{pmatrix}, \\
  &Y^{(2)}_{\bm{6}}(\tau) = \frac{1}{\sqrt{2}}
  \begin{pmatrix}
    Y_1Y_4+Y_2Y_3 \\ \sqrt{2}Y_2Y_4 \\ -\sqrt{2}Y_1Y_3 \\ Y_1Y_6+Y_2Y_5 \\ \sqrt{2}Y_2Y_6 \\ -\sqrt{2}Y_1Y_5 \\
  \end{pmatrix},
\end{align}
Taking tensor products of these modular forms, higher weight modular forms can be constructed.
The modular forms of weight 4 are:
\begin{align}
  &Y^{(4)}_{{\bm{1}^0_0}}(\tau) = \left(Y^{(2)}_{{\bm{2}_0}} Y^{(2)}_{{\bm{2}_0}}\right)_{{\bm{1}^0_0}}, \quad
  Y^{(4)}_{{\bm{1}^0_1}}(\tau) = \left(Y^{(2)}_{{\bm{1}^1_2}} Y^{(2)}_{{\bm{1}^1_2}}\right)_{{\bm{1}^0_1}}, \quad
  Y^{(4)}_{{\bm{2}_0}}(\tau) = \left(Y^{(2)}_{{\bm{2}_0}} Y^{(2)}_{{\bm{2}_0}}\right)_{{\bm{2}_0}}, \\
  &Y^{(4)}_{{\bm{2}_2}}(\tau) = \left(Y^{(2)}_{{\bm{1}^1_2}} Y^{(2)}_{{\bm{2}_0}}\right)_{{\bm{2}_2}}, \quad
  Y^{(4)}_{{\bm{3}^0}}(\tau) = \left(Y^{(2)}_{{\bm{2}_0}} Y^{(2)}_{\bm{6}}\right)_{{\bm{3}^0}}, \quad
  Y^{(4)}_{{\bm{3}^1}}(\tau) = \left(Y^{(2)}_{{\bm{1}^1_2}} Y^{(2)}_{{\bm{3}^0}}\right)_{{\bm{3}^1}}, \\
  &Y^{(4)}_{\bm{6i}}(\tau) = \left(Y^{(2)}_{{\bm{1}^1_2}} Y^{(2)}_{\bm{6}}\right)_{\bm{6}}, \quad
  Y^{(4)}_{\bm{6ii}}(\tau) = \left(Y^{(2)}_{{\bm{2}_0}} Y^{(2)}_{{\bm{3}^0}}\right)_{\bm{6}},
\end{align}
where $Y^{(4)}_{\bm{6i}}$ and $Y^{(4)}_{\bm{6ii}}$ denote two linearly independent 6D modular forms of weight 4.
In what follows we use the same convention for other modular forms.
In the same way, we can construct the modular forms of weight 6 as
\begin{align}
  &Y^{(6)}_{{\bm{1}^0_0}}(\tau) = \left(Y^{(2)}_{{\bm{2}_0}} Y^{(4)}_{{\bm{2}_0}}\right)_{{\bm{1}^0_0}}, \quad
  Y^{(6)}_{{\bm{1}^1_0}}(\tau) = \left(Y^{(2)}_{{\bm{1}^1_2}} Y^{(4)}_{{\bm{1}^0_1}}\right)_{{\bm{1}^1_0}}, \quad
  Y^{(6)}_{{\bm{1}^1_2}}(\tau) = \left(Y^{(2)}_{{\bm{1}^1_2}} Y^{(4)}_{{\bm{1}^0_0}}\right)_{{\bm{1}^1_2}}, \\
  &Y^{(6)}_{{\bm{2}_0}}(\tau) = \left(Y^{(2)}_{{\bm{2}_0}} Y^{(4)}_{{\bm{1}^0_0}}\right)_{{\bm{2}_0}}, \quad
  Y^{(6)}_{{\bm{2}_1}}(\tau) = \left(Y^{(2)}_{{\bm{2}_0}} Y^{(4)}_{{\bm{1}^0_1}}\right)_{{\bm{2}_1}}, \quad
  Y^{(6)}_{{\bm{2}_2}}(\tau) = \left(Y^{(2)}_{{\bm{1}^1_2}} Y^{(4)}_{{\bm{2}_0}}\right)_{{\bm{2}_2}}, \\
  &Y^{(6)}_{{\bm{3}^0i}}(\tau) = \left(Y^{(2)}_{{\bm{3}^0}} Y^{(4)}_{{\bm{1}^0_1}}\right)_{{\bm{3}^0}}, \quad
  Y^{(6)}_{{\bm{3}^0ii}}(\tau) = \left(Y^{(2)}_{{\bm{3}^0}} Y^{(4)}_{{\bm{1}^0_0}}\right)_{{\bm{3}^0}}, \quad
  Y^{(6)}_{{\bm{3}^1}}(\tau) = \left(Y^{(2)}_{{\bm{1}^1_2}} Y^{(4)}_{{\bm{3}^0}}\right)_{{\bm{3}^1}}, \\
  &Y^{(6)}_{\bm{6i}}(\tau) = \left(Y^{(2)}_{{\bm{2}_0}} Y^{(4)}_{{\bm{3}^1}}\right)_{\bm{6}}, \quad
  Y^{(6)}_{\bm{6ii}}(\tau) = \left(Y^{(2)}_{\bm{6}} Y^{(4)}_{{\bm{1}^0_0}}\right)_{\bm{6}}, \quad
  Y^{(6)}_{\bm{6iii}}(\tau) = \left(Y^{(2)}_{{\bm{3}^0}} Y^{(4)}_{{\bm{2}_0}}\right)_{\bm{6}}.
\end{align}
In Table \ref{tab:even_weight_modular_forms} we summarize the modular forms of level 6 of even weights up to 6.
\begin{table}[H]
 \centering
 \renewcommand{\arraystretch}{1.4}
  \begin{tabular}{c|c}
   \hline
   Weights & Modular forms $Y^{(k_Y)}_{\bm{r}}$ \\ \hline
   $k_Y=2$ & $Y_{{\bm{1}^1_2}}^{(2)}$, $Y_{{\bm{2}_0}}^{(2)}$, $Y_{{\bm{3}^0}}^{(2)}$, $Y_{\bm{6}}^{(2)}$ \\ \hline
   $k_Y=4$ & $Y_{{\bm{1}^0_0}}^{(4)}$, $Y_{{\bm{1}^0_1}}^{(4)}$, $Y_{{\bm{2}_0}}^{(4)}$, $Y_{{\bm{2}_2}}^{(4)}$, $Y_{{\bm{3}^0}}^{(4)}$, $Y_{{\bm{3}^1}}^{(4)}$, $Y_{\bm{6i}}^{(4)}$, $Y_{\bm{6ii}}^{(4)}$ \\ \hline
   $k_Y=6$ & $Y_{{\bm{1}^0_0}}^{(6)}$, $Y_{{\bm{1}^1_0}}^{(6)}$, $Y_{{\bm{1}^1_2}}^{(6)}$, $Y_{{\bm{2}_0}}^{(6)}$, $Y_{{\bm{2}_1}}^{(6)}$, $Y_{{\bm{2}_2}}^{(6)}$, $Y_{{\bm{3}^0i}}^{(6)}$, $Y_{{\bm{3}^0ii}}^{(6)}$, $Y_{{\bm{3}^1}}^{(6)}$, $Y_{\bm{6i}}^{(6)}$, $Y_{\bm{6ii}}^{(6)}$, $Y_{\bm{6iii}}^{(6)}$ \\ \hline
  \end{tabular}
  \caption{The modular forms of level 6 of even weights up to 6.}
  \label{tab:even_weight_modular_forms}
\end{table}
Furthermore we construct the singlet modular forms of weights 8, 10, 12 and 14 which we have used in Section \ref{subsec:Gamma_6}.
The singlet modular forms of weight 8 are given by
\begin{align}
  Y_{{\bm{1}^0_0}}^{(8)} = \left(Y_{{\bm{1}^0_0}}^{(4)} Y_{{\bm{1}^0_0}}^{(4)}\right)_{{\bm{1}^0_0}}, \quad
  Y_{{\bm{1}^0_1}}^{(8)} = \left(Y_{{\bm{1}^0_0}}^{(4)} Y_{{\bm{1}^0_1}}^{(4)}\right)_{{\bm{1}^0_1}}, \quad
  Y_{{\bm{1}^0_2}}^{(8)} = \left(Y_{{\bm{1}^0_1}}^{(4)} Y_{{\bm{1}^0_1}}^{(4)}\right)_{{\bm{1}^0_2}}, \quad
  Y_{{\bm{1}^1_2}}^{(8)} = \left(Y_{{\bm{2}_0}}^{(4)} Y_{{\bm{2}_2}}^{(4)}\right)_{{\bm{1}^1_2}}.
\end{align}
The singlet modular forms of weight 10 are given by
\begin{align}
  &Y_{{\bm{1}^0_0}}^{(10)} = \left(Y_{{\bm{1}^0_0}}^{(4)} Y_{{\bm{1}^0_0}}^{(6)}\right)_{{\bm{1}^0_0}}, \quad
  Y_{{\bm{1}^0_1}}^{(10)} = \left(Y_{{\bm{1}^0_1}}^{(4)} Y_{{\bm{1}^0_0}}^{(6)}\right)_{{\bm{1}^0_1}}, \quad
  Y_{{\bm{1}^1_0}}^{(10)} = \left(Y_{{\bm{1}^0_0}}^{(4)} Y_{{\bm{1}^1_0}}^{(6)}\right)_{{\bm{1}^1_0}}, \\
  &Y_{{\bm{1}^1_1}}^{(10)} = \left(Y_{{\bm{1}^0_1}}^{(4)} Y_{{\bm{1}^1_0}}^{(6)}\right)_{{\bm{1}^1_1}}, \quad
  Y_{{\bm{1}^1_2}}^{(10)} = \left(Y_{{\bm{1}^0_0}}^{(4)} Y_{{\bm{1}^1_2}}^{(6)}\right)_{{\bm{1}^1_2}}.
\end{align}
The singlet modular forms of weight 12 are given by
\begin{align}
  &Y_{{\bm{1}^0_0i}}^{(12)} = \left(Y_{{\bm{1}^0_0}}^{(6)} Y_{{\bm{1}^0_0}}^{(6)}\right)_{{\bm{1}^0_0}}, \quad
  Y_{{\bm{1}^0_0ii}}^{(12)} = \left(Y_{{\bm{1}^1_0}}^{(6)} Y_{{\bm{1}^1_0}}^{(6)}\right)_{{\bm{1}^0_0}}, \quad
  Y_{{\bm{1}^0_1}}^{(12)} = \left(Y_{{\bm{1}^1_2}}^{(6)} Y_{{\bm{1}^1_2}}^{(6)}\right)_{{\bm{1}^0_1}}, \\
  &Y_{{\bm{1}^0_2}}^{(12)} = \left(Y_{{\bm{1}^1_0}}^{(6)} Y_{{\bm{1}^1_2}}^{(6)}\right)_{{\bm{1}^0_2}}, \quad
  Y_{{\bm{1}^1_0}}^{(12)} = \left(Y_{{\bm{1}^0_0}}^{(6)} Y_{{\bm{1}^1_0}}^{(6)}\right)_{{\bm{1}^1_0}}, \quad
  Y_{{\bm{1}^1_2}}^{(12)} = \left(Y_{{\bm{1}^0_0}}^{(6)} Y_{{\bm{1}^1_2}}^{(6)}\right)_{{\bm{1}^1_2}}.
\end{align}
The singlet modular forms of weight 14 are given by
\begin{align}
  &Y_{{\bm{1}^0_0}}^{(14)} = \left(Y_{{\bm{1}^0_0}}^{(6)} Y_{{\bm{1}^0_0}}^{(8)}\right)_{{\bm{1}^0_0}}, \quad
  Y_{{\bm{1}^0_1}}^{(14)} = \left(Y_{{\bm{1}^0_0}}^{(6)} Y_{{\bm{1}^0_1}}^{(8)} \right)_{{\bm{1}^0_1}}, \quad
  Y_{{\bm{1}^0_2}}^{(14)} = \left(Y_{{\bm{1}^0_0}}^{(6)} Y_{{\bm{1}^0_2}}^{(8)} \right)_{{\bm{1}^0_2}}, \\
  &Y_{{\bm{1}^1_0}}^{(14)} = \left(Y_{{\bm{1}^1_0}}^{(6)} Y_{{\bm{1}^0_0}}^{(8)} \right)_{{\bm{1}^1_0}}, \quad
  Y_{{\bm{1}^1_1}}^{(14)} = \left(Y_{{\bm{1}^1_0}}^{(6)} Y_{{\bm{1}^0_1}}^{(8)} \right)_{{\bm{1}^1_1}}, \quad
  Y_{{\bm{1}^1_2i}}^{(14)} = \left(Y_{{\bm{1}^1_2}}^{(6)} Y_{{\bm{1}^0_0}}^{(8)} \right)_{{\bm{1}^1_2}}, \\
  &Y_{{\bm{1}^1_2ii}}^{(14)} = \left(Y_{{\bm{1}^1_0}}^{(6)} Y_{{\bm{1}^0_2}}^{(8)} \right)_{{\bm{1}^1_2}}.
\end{align}
Table \ref{tab:even_weight_singlet_modular_forms} summarizes the singlet modular forms of weights 8, 10, 12 and 14.
\begin{table}[H]
 \centering
 \renewcommand{\arraystretch}{1.4}
  \begin{tabular}{c|c}
   \hline
   Weights & Modular forms $Y^{(k_Y)}_{\bm{r}}$ \\ \hline
   $k_Y=8$ & $Y_{{\bm{1}^0_0}}^{(8)}$, $Y_{{\bm{1}^0_1}}^{(8)}$, $Y_{{\bm{1}^0_2}}^{(8)}$, $Y_{{\bm{1}^1_2}}^{(8)}$ \\ \hline
   $k_Y=10$ & $Y_{{\bm{1}^0_0}}^{(10)}$, $Y_{{\bm{1}^0_1}}^{(10)}$, $Y_{{\bm{1}^1_0}}^{(10)}$, $Y_{{\bm{1}^1_1}}^{(10)}$, $Y_{{\bm{1}^1_2}}^{(10)}$ \\ \hline
   $k_Y=12$ & $Y_{{\bm{1}^0_0i}}^{(12)}$, $Y_{{\bm{1}^0_0ii}}^{(12)}$, $Y_{{\bm{1}^0_1}}^{(12)}$, $Y_{{\bm{1}^0_2}}^{(12)}$, $Y_{{\bm{1}^1_0}}^{(12)}$, $Y_{{\bm{1}^1_2}}^{(12)}$ \\ \hline
   $k_Y=14$ & $Y^{(14)}_{{\bm{1}^0_0}}$, $Y^{(14)}_{{\bm{1}^0_1}}$, $Y^{(14)}_{{\bm{1}^0_2}}$, $Y^{(14)}_{{\bm{1}^1_0}}$, $Y^{(14)}_{{\bm{1}^1_1}}$, $Y^{(14)}_{{\bm{1}^1_2i}}$, $Y^{(14)}_{{\bm{1}^1_2ii}}$ \\ \hline
  \end{tabular}
  \caption{The singlet modular forms of level 6 of weights 8, 10, 12 and 14.}
  \label{tab:even_weight_singlet_modular_forms}
\end{table}


\chapter{Group theory and modular forms of $A_4$}
\label{app:A_4}


\section{Group theoretical aspects}
\label{app:gt_A_4}

Here, we review the group theoretical aspects of $A_4$.
$A_4$ group is generated by two generators, $S$ and $T$-transformations which satisfy the following algebraic relations:
\begin{align}
S^2 = (ST)^3 = T^3 = \mathbb{I}.
\end{align}
In $A_4$ group, there are four irreducible representations, three singlets $\bm{1}$, $\bm{1}'$ and $\bm{1}''$ and one triplet $\bm{3}$.
Each irreducible representation is given by
\begin{align}
  &\bm{1}\quad \rho(S)=1, ~\rho(T)=1, \\
  &\bm{1}' \quad \rho(S)=1,~\rho(T)=\omega, \\
  &\bm{1}'' \quad \rho(S)=1, ~\rho(T)=\omega^2, \\
  &\bm{3} \quad 
  \rho(S) = \frac{1}{3}
  \begin{pmatrix}
    -1 & 2 & 2 \\
    2 & -1 & 2 \\
    2 & 2 & -1 \\
  \end{pmatrix},\quad
  \rho(T) =
  \begin{pmatrix}
    1 & 0 & 0 \\
    0 & \omega & 0 \\
    0 & 0 & \omega^2 \\
  \end{pmatrix}.
\end{align}
The Kronecker products between irreducible representations and the CG coefficients are shown in Table \ref{tab:MultiRuleinA4}.
\begin{table}[H]
\begin{center}
\renewcommand{\arraystretch}{1}
\begin{tabular}{c|c} \hline
  The Kronecker products  & the CG coefficients \\ \hline
  $\bm{1}'' \otimes \bm{1}'' = \bm{1}'$ & \multirow{3}{*}{$a^1b^1$} \\
  $~~~~~~~~~~\bm{1}' \otimes \bm{1}' = \bm{1}''$ ~~$(a^1 b^1)$ & \\
  $\bm{1}'' \otimes \bm{1}' = \bm{1}$ & \\ \hline
  \multirow{2}{*}{$\bm{1}'' \otimes \bm{3} = \bm{3}$ ~~$(a^1 b^i)$} & \multirow{2}{*}{$\left(\begin{smallmatrix} a^1b^3\\ a^1b^1\\ a^1b^2\\\end{smallmatrix}\right)$} \\
  & \\ \hline
  \multirow{2}{*}{$\bm{1}' \otimes \bm{3} = \bm{3}$~~$(a^1 b^i)$} & \multirow{2}{*}{$\left(\begin{smallmatrix} a^1b^2\\ a^1b^3\\ a^1b^1\\ \end{smallmatrix}\right)$} \\
  & \\ \hline
  \multirow{5}{*}{$\bm{3}\otimes \bm{3}=\bm{1}\oplus \bm{1}'' \oplus \bm{1}' \oplus \bm{3} \oplus \bm{3}$} & $\begin{smallmatrix}(a^1b^1+a^2b^3+a^3b^2)\end{smallmatrix}$ \\
  & $\oplus\begin{smallmatrix}(a^1b^2+a^2b^1+a^3b^3)\end{smallmatrix}$ \\
  & $\oplus\begin{smallmatrix}(a^1b^3+a^2b^2+a^3b^1)\end{smallmatrix}$ \\
 \multirow{2}{*}{$(a^ib^j)$}  & \multirow{2}{*}{$\oplus\frac{1}{3}\left(\begin{smallmatrix} 2a^1b^1-a^2b^3-a^3b^2\\ -a^1b^2-a^2b^1+2a^3b^3\\ -a^1b^3+2a^2b^2-a^3b^1\\ \end{smallmatrix}\right)$} \\
  & \\
  & \multirow{2}{*}{$\oplus\frac{1}{2}\left(\begin{smallmatrix} a^2b^3-a^3b^2\\ a^1b^2-a^2b^1\\ -a^1b^3+a^3b^1\\ \end{smallmatrix}\right)$} \\
  & \\ \hline
\end{tabular}
\end{center}
\caption{The Kronecker products between irreducible representations of $A_4$ and the CG coefficients.}
\label{tab:MultiRuleinA4}
\end{table}


\section{Modular forms}
\label{app:modular_forms_A_4}

Here we review the modular forms of $\Gamma_3 \simeq A_4$.
The modular forms of level 3 of even weights can be constructed from the products of the Dedekind eta function $\eta(\tau)$ and its derivative,
\begin{align}
  &\eta(\tau) = q^{1/24} \prod_{n=1}^{\infty} (1-q^n), \quad q = e^{2\pi i\tau}, \\
  &\eta'(\tau) \equiv \frac{d}{d\tau} \eta(\tau).
\end{align}
Using $\eta$ and $\eta'$, the modular forms of weight 2 belonging to $A_4$ triplet $\bm{3}$ can be obtained as 
\cite{Feruglio:2017spp}
\begin{align}
  Y^{(2)}_{\bm{3}}(\tau) = 
  \begin{pmatrix}
    Y_1 \\ Y_2 \\ Y_3 \\
  \end{pmatrix},
\end{align}
where
\begin{align}
  &Y_1(\tau) = \frac{i}{2\pi} \left(\frac{\eta'(\tau/3)}{\eta(\tau/3)} + \frac{\eta'((\tau+1)/3)}{\eta((\tau+1)/3)} + \frac{\eta'((\tau+2)/3)}{\eta((\tau+2)/3)} - \frac{27\eta'(3\tau)}{\eta(3\tau)}\right), \\
  &Y_2(\tau) = \frac{-i}{\pi} \left(\frac{\eta'(\tau/3)}{\eta(\tau/3)} + \omega^2\frac{\eta'((\tau+1)/3)}{\eta((\tau+1)/3)} + \omega \frac{\eta'((\tau+2)/3)}{\eta((\tau+2)/3)}\right), \\
  &Y_3(\tau) = \frac{-i}{\pi} \left(\frac{\eta'(\tau/3)}{\eta(\tau/3)} + \omega\frac{\eta'((\tau+1)/3)}{\eta((\tau+1)/3)} + \omega^2\frac{\eta'((\tau+2)/3)}{\eta((\tau+2)/3)}\right).
\end{align}
Taking the products of these modular forms, higher weight modular forms can be constructed.
The modular forms of weight 4 are:
\begin{align}
  &Y_{\bm{1}}^{(4)}(\tau) = Y^2_1+2Y_2Y_3, \quad Y_{\bm{1'}}^{(4)}(\tau) = Y^2_3+2Y_1Y_2, \notag \\
  &Y_{\bm{3}}^{(4)}(\tau) 
  =
  \begin{pmatrix}
    Y^2_1-Y_2Y_3 \\ Y^2_3-Y_1Y_2 \\ Y^2_2-Y_1Y_3 \\
  \end{pmatrix}.
\end{align}
In the same way, we can construct the modular forms of weight 6 as
\begin{align}
  &Y^{(6)}_{\bm{1}}(\tau) = Y^3_1+Y^3_2+Y^3_3-3Y_1Y_2Y_3, \notag \\
  &Y_{\bm{3}}^{(6)}(\tau) = (Y^2_1+2Y_2Y_3)
  \begin{pmatrix}
    Y_1 \\ Y_2 \\ Y_3 \\
  \end{pmatrix}, \quad
  Y_{\bm{3'}}^{(6)}(\tau) = (Y^2_3+2Y_1Y_2)
  \begin{pmatrix}
    Y_3 \\ Y_1 \\ Y_2 \\
  \end{pmatrix}.
\end{align}
The modular forms of weight 8 are given by
\begin{align}
  &Y_{\bm{1}}^{(8)}(\tau) = (Y^2_1+2Y_2Y_3)^2, \quad Y_{\bm{1'}}^{(8)}(\tau) = (Y^2_1+2Y_2Y_3)(Y^2_3+2Y_1Y_2), \quad Y_{\bm{1''}}^{(8)}(\tau) = (Y^2_3+2Y_1Y_2)^2, \notag \\
  &Y_{\bm{3}}^{(8)}(\tau) = (Y^2_1+2Y_2Y_3)
  \begin{pmatrix}
    Y^2_1-Y_2Y_3 \\ Y^2_3-Y_1Y_2 \\ Y^2_2-Y_1Y_3 \\
  \end{pmatrix}, \quad
  Y_{\bm{3'}}^{(8)}(\tau) = (Y^2_3+2Y_1Y_2)
  \begin{pmatrix}
    Y^2_2-Y_1Y_3 \\
    Y^2_1-Y_2Y_3 \\
    Y^2_3-Y_1Y_2 \\
  \end{pmatrix}.
\end{align}


\chapter{Group theory and Siegel modular forms of $\widetilde{\Delta}(96)$}
\label{app:Delta96}


\section{Group theoretical aspects}
\label{app:gt_Delta96}

Here, we review the group theoretical aspects of $\widetilde{\Delta}(96)$.
$\widetilde{\Delta}(96)$ group is generated by two generators, $S$ and $T$-transformations which satisfy the following algebraic relations:
\begin{align}
S^2=-i\mathbb{I}, ~(ST)^3 = \mathbb{I},~T^{8} = (S^{-1}T^{-1}ST)^3 = \mathbb{I}, ~S^2T = TS^2. \label{eq:algebra_S_T}
\end{align}
In Ref.~\cite{Kikuchi:2021ogn}, it was shown that these algebraic relations correspond to $\widetilde{\Delta}(96)$.
In $\widetilde{\Delta}(96)$ group, there are 40 irreducible representations as shown in Table \ref{tab:num_dim_irr_reps}.
\begin{table}[H]
\centering
\begin{tabular}{c|ccccc}\hline
Dimensions & 1 & 2 & 3 & 6 & Total \\ \hline
Number of irr. reps. &  8 & 4 & 24 & 4 & 40 \\ \hline
\end{tabular}
\caption{Number of irreducible representations of $\widetilde{\Delta}(96)$.}
\label{tab:num_dim_irr_reps}
\end{table}
Each irreducible representation is shown in Table \ref{tab:irr_reps}.
\begin{table}[H]
\centering
\begin{tabular}{c|c|c} \hline
$\bm{r}$ & $\rho(S)$ & $\rho(T)$ \\ \hline \hline
$\begin{matrix}\bm{1}_q \\(q=0,1,...,7) \\\end{matrix}$
 & $e^{-\frac{q\pi i}{4}}$
 & $e^{\frac{q\pi i}{4}}$\\ \hline

$\begin{matrix}\bm{2}_q \\(q=0,1,2,3) \\\end{matrix}$
 & $e^{-\frac{q\pi i}{4}}\begin{pmatrix}
-\frac{1}{2} & \frac{\sqrt{3}}{2} \\ 
\frac{\sqrt{3}}{2} & \frac{1}{2} \\ 
\end{pmatrix}$
 & $e^{\frac{q\pi i}{4}}\begin{pmatrix}
1 & 0 \\ 
0 & -1 \\ 
\end{pmatrix}$\\ \hline

$\begin{matrix}\bm{3}_q \\(q=0,1,...,7) \\\end{matrix}$
 & $e^{-\frac{q\pi i}{4}}\begin{pmatrix}
0 & \frac{1}{\sqrt{2}} & \frac{1}{\sqrt{2}} \\ 
\frac{1}{\sqrt{2}} & -\frac{1}{2} & \frac{1}{2} \\ 
\frac{1}{\sqrt{2}} & \frac{1}{2} & -\frac{1}{2} \\ 
\end{pmatrix}$
 & $e^{\frac{q\pi i}{4}}\begin{pmatrix}
1 & 0 & 0 \\ 
0 & 1i & 0 \\ 
0 & 0 & -1i \\ 
\end{pmatrix}$\\ \hline

$\begin{matrix}\bm{3}'_q \\(q=0,1,...,7) \\\end{matrix}$
 & $e^{-\frac{q\pi i}{4}}\begin{pmatrix}
0 & \frac{1}{\sqrt{2}}i & \frac{1}{\sqrt{2}}i \\ 
\frac{1}{\sqrt{2}}i & \frac{1}{2}i & -\frac{1}{2}i \\ 
\frac{1}{\sqrt{2}}i & -\frac{1}{2}i & \frac{1}{2}i \\ 
\end{pmatrix}$
 & $e^{\frac{q\pi i}{4}}\begin{pmatrix}
1 & 0 & 0 \\ 
0 & e^{\frac{\pi i}{4}} & 0 \\ 
0 & 0 & e^{\frac{5\pi i}{4}} \\ 
\end{pmatrix}$\\ \hline

$\begin{matrix}\bm{\hat{3}}_q \\(q=0,1,...,7) \\\end{matrix}$
 & $e^{-\frac{q\pi i}{4}}\begin{pmatrix}
0 & \frac{1}{\sqrt{2}}i & \frac{1}{\sqrt{2}}i \\ 
\frac{1}{\sqrt{2}}i & -\frac{1}{2}i & \frac{1}{2}i \\ 
\frac{1}{\sqrt{2}}i & \frac{1}{2}i & -\frac{1}{2}i \\ 
\end{pmatrix}$
 & $e^{\frac{q\pi i}{4}}\begin{pmatrix}
1 & 0 & 0 \\ 
0 & e^{\frac{3\pi i}{4}} & 0 \\ 
0 & 0 & e^{\frac{7\pi i}{4}} \\ 
\end{pmatrix}$\\
\hline 
\end{tabular}
\caption{Irreducible representations of $\widetilde{\Delta}(96)$.
Number of irreducible representations is 40.}
\label{tab:irr_reps}
\end{table}
The Kronecker products between irreducible representations and the CG coefficients are shown in Table \ref{tab:tensor}.
We have used the following notations,
\begin{align}
P_2 =
%
.
\end{align}
We also use a notation $int(x)$ to express a integral part of real number $x$.
\small
\begin{longtable}{c} 
\caption{The Kronecker products between irreducible representations of $\widetilde{\Delta}(96)$ and the CG coefficients.}
\label{tab:tensor} \\\hline
$\begin{matrix}
\bm{1}_p \oplus \bm{1}_q = \bm{1}_r \\ 
\alpha^1\beta^1 \\ 
\end{matrix}$ \\ \hline
$\begin{matrix}
\bm{2}_p \otimes \bm{1}_q = \bm{2}_r \\
P_2^{int(\frac{r}{4})}
\begin{pmatrix}
\alpha^1\beta^1 \\ \alpha^2\beta^1 \\
\end{pmatrix} \\ 
\end{matrix}$ \\ \hline
$\begin{matrix}
\bm{3}_p \otimes \bm{1}_q = \bm{3}_r \\
\bm{3}'_p \otimes \bm{1}_q = \bm{3}'_r \\
\bm{\hat{3}}_p \otimes \bm{1}_q = \bm{\hat{3}}_r \\
\begin{pmatrix}
\alpha^1\beta^1 \\ \alpha^2\beta^1 \\ \alpha^3\beta^1 \\
\end{pmatrix} \\
\end{matrix}$ \\ \hline
$\begin{matrix}
\bm{6}_p \otimes \bm{1}_q = \bm{6}_r \\
P_6^{int(\frac{r}{4})}\begin{pmatrix}
\alpha^1\beta^1 \\ \alpha^2\beta^1 \\ \alpha^3\beta^1 \\ \alpha^4\beta^1 \\ \alpha^5\beta^1 \\ \alpha^6\beta^1 \\
\end{pmatrix} \\
\end{matrix}$ \\ \hline
$\begin{matrix}
\bm{2}_p\otimes \bm{2}_q = \bm{1}_r \oplus \bm{1}_{r+4} \oplus \bm{2}_r \\
\frac{1}{\sqrt{2}}(\alpha^1\beta^1+\alpha^2\beta^2) 
\oplus\frac{1}{\sqrt{2}}(\alpha^1\beta^2-\alpha^2\beta^1) 
\oplus P_2^{int(\frac{r}{4})}\frac{1}{\sqrt{2}}\begin{pmatrix}\alpha^1\beta^1-\alpha^2\beta^2 \\
-\alpha^1\beta^2-\alpha^2\beta^1 \\
\end{pmatrix} \\
\end{matrix}$ \\ \hline 
$\begin{matrix}
\bm{3}_p\otimes \bm{2}_q = \bm{3}_{r} \oplus \bm{3}_{r+4} \\ 
\bm{\hat{3}}_p\otimes \bm{2}_q = \bm{\hat{3}}_{r} \oplus \bm{\hat{3}}_{r+4} \\
\bm{3}'_p\otimes \bm{2}_q = \bm{3}'_{r} \oplus \bm{3}'_{r+4} \\
\begin{pmatrix}
\alpha^1\beta^1 \\
-\frac{1}{2}\alpha^2\beta^1+\frac{\sqrt{3}}{2}\alpha^3\beta^2 \\
\frac{\sqrt{3}}{2}\alpha^2\beta^2-\frac{1}{2}\alpha^3\beta^1 \\
\end{pmatrix}
\oplus\begin{pmatrix}
\alpha^1\beta^2 \\
-\frac{1}{2}\alpha^2\beta^2-\frac{\sqrt{3}}{2}\alpha^3\beta^1 \\
-\frac{\sqrt{3}}{2}\alpha^2\beta^1-\frac{1}{2}\alpha^3\beta^2 \\
\end{pmatrix} \\
\end{matrix}$ \\
\hline
\end{longtable}
\normalsize


\section{Siegel modular forms}
\label{app:mf_Delta96}

Here we review the Siegel modular forms of $\widetilde{\Delta}(96)$ up to weight 5.
The Siegel modular forms of weight 1/2 are given by
\begin{align*}
&Y^{(1/2)}_{\bm{3}'_7} = 
%
.
\end{align*}
Taking the products of this, the Siegel modular forms of higher weights can be constructed.
The Siegel modular forms of weight 1 are:
\begin{longtable}{l}
$Y^{(1)}_{\bm{3}_6} =
\begin{pmatrix}
Y^{(1/2)}_{\bm{3}'_7}Y^{(1/2)}_{\bm{3}'_7}
\end{pmatrix}_{\bm{3}_6} =
\begin{pmatrix}
\zeta^{2} \\
-\frac{1}{\sqrt{2}}\theta^{2}-\frac{1}{\sqrt{2}}\eta^{2} \\
-\sqrt{2}\eta\theta \\
\end{pmatrix}$, \\
$Y^{(1)}_{\bm{\hat{3}}_0} =
\begin{pmatrix}
Y^{(1/2)}_{\bm{3}'_7}Y^{(1/2)}_{\bm{3}'_7}
\end{pmatrix}_{\bm{\hat{3}}_0} =
\begin{pmatrix}
-\frac{1}{\sqrt{2}}\theta^{2}+\frac{1}{\sqrt{2}}\eta^{2} \\
\sqrt{2}\zeta\theta \\
-\sqrt{2}\zeta\eta \\
\end{pmatrix}$.
\end{longtable}
In the same way, we can construct the modular forms of weight 3/2 as
\begin{longtable}{l}
$Y^{(3/2)}_{\bm{1}_7} =
\begin{pmatrix}
Y^{(1)}_{\bm{\hat{3}}_0}Y^{(1/2)}_{\bm{3}'_7}
\end{pmatrix}_{\bm{1}_7} =
\begin{pmatrix}
-\frac{\sqrt{6}}{2}\zeta\theta^{2}+\frac{\sqrt{6}}{2}\zeta\eta^{2} \\
\end{pmatrix}$, \\
$Y^{(3/2)}_{\bm{\hat{3}}_5} =
\begin{pmatrix}
Y^{(1)}_{\bm{3}_6}Y^{(1/2)}_{\bm{3}'_7}
\end{pmatrix}_{\bm{\hat{3}}_5} =
\begin{pmatrix}
\zeta^{3} \\
-\frac{3}{2}\eta\theta^{2}-\frac{1}{2}\eta^{3} \\
-\frac{1}{2}\theta^{3}-\frac{3}{2}\eta^{2}\theta \\
\end{pmatrix}$, \\
$Y^{(3/2)}_{\bm{6}_3} =
\begin{pmatrix}
Y^{(1)}_{\bm{3}_6}Y^{(1/2)}_{\bm{3}'_7}
\end{pmatrix}_{\bm{6}_3} =
\begin{pmatrix}
-\sqrt{2}\zeta\eta\theta \\
-\frac{1}{2}\theta^{3}+\frac{1}{2}\eta^{2}\theta \\
\zeta^{2}\eta \\
\frac{1}{\sqrt{2}}\zeta\theta^{2}+\frac{1}{\sqrt{2}}\zeta\eta^{2} \\
\frac{1}{2}\eta\theta^{2}-\frac{1}{2}\eta^{3} \\
\zeta^{2}\theta \\
\end{pmatrix}$.
\end{longtable}
The Siegel modular forms of weight 2 are:
\begin{longtable}{l}
$Y^{(2)}_{\bm{1}_4} =
\begin{pmatrix}
Y^{(1)}_{\bm{3}_6}Y^{(1)}_{\bm{3}_6}
\end{pmatrix}_{\bm{1}_4} =
\begin{pmatrix}
\frac{2\sqrt{3}}{3}\eta\theta^{3}+\frac{2\sqrt{3}}{3}\eta^{3}\theta+\frac{1}{\sqrt{3}}\zeta^{4} \\
\end{pmatrix}$, \\
$Y^{(2)}_{\bm{2}_0} =
\begin{pmatrix}
Y^{(1)}_{\bm{3}_6}Y^{(1)}_{\bm{3}_6}
\end{pmatrix}_{\bm{2}_0} =
\begin{pmatrix}
\frac{1}{2\sqrt{2}}\theta^{4}+\frac{3\sqrt{2}}{2}\eta^{2}\theta^{2}+\frac{1}{2\sqrt{2}}\eta^{4} \\
\frac{2}{\sqrt{6}}\eta\theta^{3}+\frac{2}{\sqrt{6}}\eta^{3}\theta-\frac{2}{\sqrt{6}}\zeta^{4} \\
\end{pmatrix}$, \\
$Y^{(2)}_{\bm{3}_0} =
\begin{pmatrix}
Y^{(1)}_{\bm{3}_6}Y^{(1)}_{\bm{3}_6}
\end{pmatrix}_{\bm{3}_0} =
\begin{pmatrix}
\frac{1}{2\sqrt{2}}\theta^{4}-\frac{1}{\sqrt{2}}\eta^{2}\theta^{2}+\frac{1}{2\sqrt{2}}\eta^{4} \\
2\zeta^{2}\eta\theta \\
-\zeta^{2}\theta^{2}-\zeta^{2}\eta^{2} \\
\end{pmatrix}$, \\
$Y^{(2)}_{\bm{3}'_6} =
\begin{pmatrix}
Y^{(1)}_{\bm{3}_6}Y^{(1)}_{\bm{\hat{3}}_0}
\end{pmatrix}_{\bm{3}'_6} =
\begin{pmatrix}
-\frac{1}{\sqrt{2}}\zeta^{2}\theta^{2}+\frac{1}{\sqrt{2}}\zeta^{2}\eta^{2} \\
-\frac{1}{\sqrt{2}}\zeta\eta\theta^{2}+\frac{1}{\sqrt{2}}\zeta\eta^{3} \\
-\frac{1}{\sqrt{2}}\zeta\theta^{3}+\frac{1}{\sqrt{2}}\zeta\eta^{2}\theta \\
\end{pmatrix}$, \\
$Y^{(2)}_{\bm{6}_0} =
\begin{pmatrix}
Y^{(1)}_{\bm{3}_6}Y^{(1)}_{\bm{\hat{3}}_0}
\end{pmatrix}_{\bm{6}_0} =
\begin{pmatrix}
\frac{1}{2}\theta^{4}-\frac{1}{2}\eta^{4} \\
-\sqrt{2}\zeta^{3}\theta \\
-\frac{1}{\sqrt{2}}\zeta\theta^{3}-\frac{3\sqrt{2}}{2}\zeta\eta^{2}\theta \\
\eta\theta^{3}-\eta^{3}\theta \\
\sqrt{2}\zeta^{3}\eta \\
\frac{3\sqrt{2}}{2}\zeta\eta\theta^{2}+\frac{1}{\sqrt{2}}\zeta\eta^{3} \\
\end{pmatrix}$.
\end{longtable}
The Siegel modular forms of weight 5/2 are:
\begin{longtable}{l}
$Y^{(5/2)}_{\bm{3}_5} =
\begin{pmatrix}
Y^{(3/2)}_{\bm{\hat{3}}_5}Y^{(1)}_{\bm{\hat{3}}_0}
\end{pmatrix}_{\bm{3}_5} =
\begin{pmatrix}
-\frac{1}{\sqrt{2}}\zeta^{3}\theta^{2}+\frac{1}{\sqrt{2}}\zeta^{3}\eta^{2} \\
\frac{1}{2}\zeta\theta^{4}-\frac{1}{2}\zeta\eta^{4} \\
\zeta\eta\theta^{3}-\zeta\eta^{3}\theta \\
\end{pmatrix}$, \\
$Y^{(5/2)}_{\bm{3}'_{3a}} =
\begin{pmatrix}
Y^{(3/2)}_{\bm{\hat{3}}_5}Y^{(1)}_{\bm{\hat{3}}_0}
\end{pmatrix}_{\bm{3}'_3} =
\begin{pmatrix}
-2\zeta\eta\theta^{3}-2\zeta\eta^{3}\theta \\
\frac{1}{4}\theta^{5}+\frac{1}{2}\eta^{2}\theta^{3}-\frac{3}{4}\eta^{4}\theta-\zeta^{4}\eta \\
-\frac{3}{4}\eta\theta^{4}+\frac{1}{2}\eta^{3}\theta^{2}+\frac{1}{4}\eta^{5}-\zeta^{4}\theta \\
\end{pmatrix}$, \\
$Y^{(5/2)}_{\bm{3}'_{3b}} =
\begin{pmatrix}
Y^{(3/2)}_{\bm{\hat{3}}_5}Y^{(1)}_{\bm{3}_6}
\end{pmatrix}_{\bm{3}'_3} =
\begin{pmatrix}
\zeta^{5} \\
\frac{1}{4}\theta^{5}+\frac{5}{2}\eta^{2}\theta^{3}+\frac{5}{4}\eta^{4}\theta \\
\frac{5}{4}\eta\theta^{4}+\frac{5}{2}\eta^{3}\theta^{2}+\frac{1}{4}\eta^{5} \\
\end{pmatrix}$, \\
$Y^{(5/2)}_{\bm{3}'_7} =
\begin{pmatrix}
Y^{(3/2)}_{\bm{\hat{3}}_5}Y^{(1)}_{\bm{\hat{3}}_0}
\end{pmatrix}_{\bm{3}'_7} =
\begin{pmatrix}
\frac{1}{2}\zeta\theta^{4}+3\zeta\eta^{2}\theta^{2}+\frac{1}{2}\zeta\eta^{4} \\
\frac{3}{4}\eta\theta^{4}-\frac{1}{2}\eta^{3}\theta^{2}-\frac{1}{4}\eta^{5}-\zeta^{4}\theta \\
-\frac{1}{4}\theta^{5}-\frac{1}{2}\eta^{2}\theta^{3}+\frac{3}{4}\eta^{4}\theta-\zeta^{4}\eta \\
\end{pmatrix}$, \\
$Y^{(5/2)}_{\bm{\hat{3}}_7} =
\begin{pmatrix}
Y^{(3/2)}_{\bm{6}_3}Y^{(1)}_{\bm{\hat{3}}_0}
\end{pmatrix}_{\bm{\hat{3}}_7} =
\begin{pmatrix}
-\frac{1}{2}\zeta\theta^{4}+\zeta\eta^{2}\theta^{2}-\frac{1}{2}\zeta\eta^{4} \\
\zeta^{2}\theta^{3}-\zeta^{2}\eta^{2}\theta \\
-\zeta^{2}\eta\theta^{2}+\zeta^{2}\eta^{3} \\
\end{pmatrix}$, \\
\end{longtable}
The Siegel modular forms of weight 3 are:
\begin{longtable}{l}
$Y^{(3)}_{\bm{1}_6} =
\begin{pmatrix}
Y^{(3/2)}_{\bm{1}_7}Y^{(3/2)}_{\bm{1}_7}
\end{pmatrix}_{\bm{1}_6} =
\begin{pmatrix}
\frac{3}{2}\zeta^{2}\theta^{4}-3\zeta^{2}\eta^{2}\theta^{2}+\frac{3}{2}\zeta^{2}\eta^{4} \\
\end{pmatrix}$, \\
$Y^{(3)}_{\bm{3}_{2a}} =
\begin{pmatrix}
Y^{(3/2)}_{\bm{\hat{3}}_5}Y^{(3/2)}_{\bm{\hat{3}}_5}
\end{pmatrix}_{\bm{3}_2} =
\begin{pmatrix}
\zeta^{6} \\
-\frac{3\sqrt{2}}{4}\eta\theta^{5}-\frac{5\sqrt{2}}{2}\eta^{3}\theta^{3}-\frac{3\sqrt{2}}{4}\eta^{5}\theta \\
-\frac{1}{4\sqrt{2}}\theta^{6}-\frac{15\sqrt{2}}{8}\eta^{2}\theta^{4}-\frac{15\sqrt{2}}{8}\eta^{4}\theta^{2}-\frac{1}{4\sqrt{2}}\eta^{6} \\
\end{pmatrix}$, \\
$Y^{(3)}_{\bm{3}_{2b}} =
\begin{pmatrix}
Y^{(3/2)}_{\bm{6}_3}Y^{(3/2)}_{\bm{6}_3}
\end{pmatrix}_{\bm{3}_2} =
\begin{pmatrix}
-\sqrt{2}\zeta^{2}\eta\theta^{3}-\sqrt{2}\zeta^{2}\eta^{3}\theta \\
\frac{1}{4}\eta\theta^{5}-\frac{1}{2}\eta^{3}\theta^{3}+\frac{1}{4}\eta^{5}\theta+\frac{1}{2}\zeta^{4}\theta^{2}+\frac{1}{2}\zeta^{4}\eta^{2} \\
-\frac{1}{8}\theta^{6}+\frac{1}{8}\eta^{2}\theta^{4}+\frac{1}{8}\eta^{4}\theta^{2}-\frac{1}{8}\eta^{6}+\zeta^{4}\eta\theta \\
\end{pmatrix}$, \\
$Y^{(3)}_{\bm{3}_6} =
\begin{pmatrix}
Y^{(3/2)}_{\bm{6}_3}Y^{(3/2)}_{\bm{6}_3}
\end{pmatrix}_{\bm{3}_6} =
\begin{pmatrix}
\frac{1}{2\sqrt{2}}\zeta^{2}\theta^{4}+\frac{3\sqrt{2}}{2}\zeta^{2}\eta^{2}\theta^{2}+\frac{1}{2\sqrt{2}}\zeta^{2}\eta^{4} \\
\frac{1}{8}\theta^{6}-\frac{1}{8}\eta^{2}\theta^{4}-\frac{1}{8}\eta^{4}\theta^{2}+\frac{1}{8}\eta^{6}+\zeta^{4}\eta\theta \\
-\frac{1}{4}\eta\theta^{5}+\frac{1}{2}\eta^{3}\theta^{3}-\frac{1}{4}\eta^{5}\theta+\frac{1}{2}\zeta^{4}\theta^{2}+\frac{1}{2}\zeta^{4}\eta^{2} \\
\end{pmatrix}$, \\
\end{longtable}
The Siegel modular forms of weight 7/2 are:
\begin{longtable}{l}
$Y^{(7/2)}_{\bm{1}_3} =
\begin{pmatrix}
Y^{(2)}_{\bm{1}_4}Y^{(3/2)}_{\bm{1}_7}
\end{pmatrix}_{\bm{1}_3} =
\begin{pmatrix}
-\sqrt{2}\zeta\eta\theta^{5}+\sqrt{2}\zeta\eta^{5}\theta-\frac{1}{\sqrt{2}}\zeta^{5}\theta^{2}+\frac{1}{\sqrt{2}}\zeta^{5}\eta^{2} \\
\end{pmatrix}$, \\
$Y^{(7/2)}_{\bm{2}_3} =
\begin{pmatrix}
Y^{(3/2)}_{\bm{\hat{3}}_5}Y^{(2)}_{\bm{3}'_6}
\end{pmatrix}_{\bm{2}_3} =
\begin{pmatrix}
\frac{1}{\sqrt{3}}\zeta\eta\theta^{5}-\frac{1}{\sqrt{3}}\zeta\eta^{5}\theta-\frac{1}{\sqrt{3}}\zeta^{5}\theta^{2}+\frac{1}{\sqrt{3}}\zeta^{5}\eta^{2} \\
-\frac{1}{4}\zeta\theta^{6}-\frac{5}{4}\zeta\eta^{2}\theta^{4}+\frac{5}{4}\zeta\eta^{4}\theta^{2}+\frac{1}{4}\zeta\eta^{6} \\
\end{pmatrix}$ \\
$Y^{(7/2)}_{\bm{3}_7} =
\begin{pmatrix}
Y^{(2)}_{\bm{6}_0}Y^{(3/2)}_{\bm{6}_3}
\end{pmatrix}_{\bm{3}_7} =
\begin{pmatrix}
\frac{1}{4}\zeta\theta^{6}-\frac{3}{4}\zeta\eta^{2}\theta^{4}+\frac{3}{4}\zeta\eta^{4}\theta^{2}-\frac{1}{4}\zeta\eta^{6} \\
\sqrt{2}\zeta^{3}\eta\theta^{3}-\sqrt{2}\zeta^{3}\eta^{3}\theta \\
-\frac{1}{\sqrt{2}}\zeta^{3}\theta^{4}+\frac{1}{\sqrt{2}}\zeta^{3}\eta^{4} \\
\end{pmatrix}$, \\
$Y^{(7/2)}_{\bm{3}'_5} =
\begin{pmatrix}
Y^{(3/2)}_{\bm{6}_3}Y^{(2)}_{\bm{3}_0}
\end{pmatrix}_{\bm{3}'_5} =
\begin{pmatrix}
\frac{1}{2}\zeta^{3}\theta^{4}-\zeta^{3}\eta^{2}\theta^{2}+\frac{1}{2}\zeta^{3}\eta^{4} \\
\frac{1}{2}\zeta^{2}\eta\theta^{4}-\zeta^{2}\eta^{3}\theta^{2}+\frac{1}{2}\zeta^{2}\eta^{5} \\
\frac{1}{2}\zeta^{2}\theta^{5}-\zeta^{2}\eta^{2}\theta^{3}+\frac{1}{2}\zeta^{2}\eta^{4}\theta \\
\end{pmatrix}$, \\
\end{longtable}
The Siegel modular forms of weight 4 are:
\begin{longtable}{l}
$Y^{(4)}_{\bm{1}_{0a}} =
\begin{pmatrix}
Y^{(2)}_{\bm{6}_0}Y^{(2)}_{\bm{6}_0}
\end{pmatrix}_{\bm{1}_0} =
\begin{pmatrix}
\frac{1}{4\sqrt{6}}\theta^{8}-\frac{1}{\sqrt{6}}\eta^{2}\theta^{6}+\frac{3}{2\sqrt{6}}\eta^{4}\theta^{4}-\frac{1}{\sqrt{6}}\eta^{6}\theta^{2}+\frac{1}{4\sqrt{6}}\eta^{8}-\frac{4\sqrt{6}}{3}\zeta^{4}\eta\theta^{3}-\frac{4\sqrt{6}}{3}\zeta^{4}\eta^{3}\theta \\
\end{pmatrix}$, \\
$Y^{(4)}_{\bm{1}_{0b}} =
\begin{pmatrix}
Y^{(2)}_{\bm{1}_4}Y^{(2)}_{\bm{1}_4}
\end{pmatrix}_{\bm{1}_0} =
\begin{pmatrix}
\frac{4}{3}\eta^{2}\theta^{6}+\frac{8}{3}\eta^{4}\theta^{4}+\frac{4}{3}\eta^{6}\theta^{2}+\frac{4}{3}\zeta^{4}\eta\theta^{3}+\frac{4}{3}\zeta^{4}\eta^{3}\theta+\frac{1}{3}\zeta^{8} \\
\end{pmatrix}$, \\
$Y^{(4)}_{\bm{2}_{0a}} =
\begin{pmatrix}
Y^{(2)}_{\bm{2}_0}Y^{(2)}_{\bm{1}_4}
\end{pmatrix}_{\bm{2}_0} =
\begin{pmatrix}
\frac{4}{3\sqrt{2}}\eta^{2}\theta^{6}+\frac{4\sqrt{2}}{3}\eta^{4}\theta^{4}+\frac{4}{3\sqrt{2}}\eta^{6}\theta^{2}-\frac{2}{3\sqrt{2}}\zeta^{4}\eta\theta^{3}-\frac{2}{3\sqrt{2}}\zeta^{4}\eta^{3}\theta-\frac{2}{3\sqrt{2}}\zeta^{8} \\
-\frac{1}{\sqrt{6}}\eta\theta^{7}-\frac{7}{\sqrt{6}}\eta^{3}\theta^{5}-\frac{7}{\sqrt{6}}\eta^{5}\theta^{3}-\frac{1}{\sqrt{6}}\eta^{7}\theta-\frac{1}{2\sqrt{6}}\zeta^{4}\theta^{4}-\frac{\sqrt{6}}{2}\zeta^{4}\eta^{2}\theta^{2}-\frac{1}{2\sqrt{6}}\zeta^{4}\eta^{4} \\
\end{pmatrix}$, \\
$Y^{(4)}_{\bm{2}_{0b}} =
\begin{pmatrix}
Y^{(2)}_{\bm{6}_0}Y^{(2)}_{\bm{6}_0}
\end{pmatrix}_{\bm{2}_0} =
\begin{pmatrix}
\frac{1}{4\sqrt{3}}\theta^{8}-\frac{1}{\sqrt{3}}\eta^{2}\theta^{6}+\frac{3}{2\sqrt{3}}\eta^{4}\theta^{4}-\frac{1}{\sqrt{3}}\eta^{6}\theta^{2}+\frac{1}{4\sqrt{3}}\eta^{8}+\frac{4\sqrt{3}}{3}\zeta^{4}\eta\theta^{3}+\frac{4\sqrt{3}}{3}\zeta^{4}\eta^{3}\theta \\
\zeta^{4}\theta^{4}+6\zeta^{4}\eta^{2}\theta^{2}+\zeta^{4}\eta^{4} \\
\end{pmatrix}$, \\
\end{longtable}
The Siegel modular forms of weight 9/2 are:
\begin{longtable}{l}
$Y^{(9/2)}_{\bm{1}_5} =
\begin{pmatrix}
Y^{(5/2)}_{\bm{\hat{3}}_7}Y^{(2)}_{\bm{3}'_6}
\end{pmatrix}_{\bm{1}_5} =
\begin{pmatrix}
\frac{3}{2\sqrt{6}}\zeta^{3}\theta^{6}-\frac{3\sqrt{6}}{4}\zeta^{3}\eta^{2}\theta^{4}+\frac{3\sqrt{6}}{4}\zeta^{3}\eta^{4}\theta^{2}-\frac{3}{2\sqrt{6}}\zeta^{3}\eta^{6} \\
\end{pmatrix}$, \\

$Y^{(9/2)}_{\bm{3}_{1a}} =
\begin{pmatrix}
Y^{(5/2)}_{\bm{3}'_{3b}}Y^{(2)}_{\bm{3}'_6}
\end{pmatrix}_{\bm{3}_1} =
\begin{pmatrix}
-\frac{1}{\sqrt{2}}\zeta^{7}\theta^{2}+\frac{1}{\sqrt{2}}\zeta^{7}\eta^{2} \\
\frac{3}{4}\zeta\eta\theta^{7}+\frac{7}{4}\zeta\eta^{3}\theta^{5}-\frac{7}{4}\zeta\eta^{5}\theta^{3}-\frac{3}{4}\zeta\eta^{7}\theta \\
\frac{1}{8}\zeta\theta^{8}+\frac{7}{4}\zeta\eta^{2}\theta^{6}-\frac{7}{4}\zeta\eta^{6}\theta^{2}-\frac{1}{8}\zeta\eta^{8} \\
\end{pmatrix}$, \\
$Y^{(9/2)}_{\bm{3}_{1b}} =
\begin{pmatrix}
Y^{(5/2)}_{\bm{3}'_{3a}}Y^{(2)}_{\bm{3}'_6}
\end{pmatrix}_{\bm{3}_1} =
\begin{pmatrix}
\sqrt{2}\zeta^{3}\eta\theta^{5}-\sqrt{2}\zeta^{3}\eta^{5}\theta \\
-\frac{1}{4}\zeta\eta\theta^{7}+\frac{3}{4}\zeta\eta^{3}\theta^{5}-\frac{3}{4}\zeta\eta^{5}\theta^{3}+\frac{1}{4}\zeta\eta^{7}\theta-\frac{1}{2}\zeta^{5}\theta^{4}+\frac{1}{2}\zeta^{5}\eta^{4} \\
\frac{1}{8}\zeta\theta^{8}-\frac{1}{4}\zeta\eta^{2}\theta^{6}+\frac{1}{4}\zeta\eta^{6}\theta^{2}-\frac{1}{8}\zeta\eta^{8}-\zeta^{5}\eta\theta^{3}+\zeta^{5}\eta^{3}\theta \\
\end{pmatrix}$, \\
\end{longtable}
The Siegel modular forms of weight 5 are:
\begin{longtable}{l}
$Y^{(5)}_{\bm{1}_2} =
\begin{pmatrix}
Y^{(5/2)}_{\bm{3}_5}Y^{(5/2)}_{\bm{3}_5}
\end{pmatrix}_{\bm{1}_2}$\\$ =
\begin{pmatrix}
\frac{1}{\sqrt{3}}\zeta^{2}\eta\theta^{7}-\frac{1}{\sqrt{3}}\zeta^{2}\eta^{3}\theta^{5}-\frac{1}{\sqrt{3}}\zeta^{2}\eta^{5}\theta^{3}+\frac{1}{\sqrt{3}}\zeta^{2}\eta^{7}\theta+\frac{1}{2\sqrt{3}}\zeta^{6}\theta^{4}-\frac{1}{\sqrt{3}}\zeta^{6}\eta^{2}\theta^{2}+\frac{1}{2\sqrt{3}}\zeta^{6}\eta^{4} \\
\end{pmatrix}$, \\

$Y^{(5)}_{\bm{2}_2} =
\begin{pmatrix}
Y^{(5/2)}_{\bm{\hat{3}}_7}Y^{(5/2)}_{\bm{3}'_7}
\end{pmatrix}_{\bm{2}_2}$\\$ =
\begin{pmatrix}
\frac{1}{\sqrt{2}}\zeta^{2}\eta\theta^{7}-\frac{1}{\sqrt{2}}\zeta^{2}\eta^{3}\theta^{5}-\frac{1}{\sqrt{2}}\zeta^{2}\eta^{5}\theta^{3}+\frac{1}{\sqrt{2}}\zeta^{2}\eta^{7}\theta-\frac{1}{\sqrt{2}}\zeta^{6}\theta^{4}+\sqrt{2}\zeta^{6}\eta^{2}\theta^{2}-\frac{1}{\sqrt{2}}\zeta^{6}\eta^{4} \\
-\frac{3}{4\sqrt{6}}\zeta^{2}\theta^{8}-\frac{\sqrt{6}}{2}\zeta^{2}\eta^{2}\theta^{6}+\frac{5\sqrt{6}}{4}\zeta^{2}\eta^{4}\theta^{4}-\frac{\sqrt{6}}{2}\zeta^{2}\eta^{6}\theta^{2}-\frac{3}{4\sqrt{6}}\zeta^{2}\eta^{8} \\
\end{pmatrix}$, \\

$Y^{(5)}_{\bm{3}_{2a}} =
\begin{pmatrix}
Y^{(5/2)}_{\bm{3}'_{3a}}Y^{(5/2)}_{\bm{3}'_7}
\end{pmatrix}_{\bm{3}_2}$\\$ =
\begin{pmatrix}
-\zeta^{2}\eta\theta^{7}-7\zeta^{2}\eta^{3}\theta^{5}-7\zeta^{2}\eta^{5}\theta^{3}-\zeta^{2}\eta^{7}\theta \\
-\frac{3}{8\sqrt{2}}\eta\theta^{9}-\frac{1}{2\sqrt{2}}\eta^{3}\theta^{7}+\frac{7\sqrt{2}}{8}\eta^{5}\theta^{5}-\frac{1}{2\sqrt{2}}\eta^{7}\theta^{3}-\frac{3}{8\sqrt{2}}\eta^{9}\theta-\sqrt{2}\zeta^{8}\eta\theta \\
\frac{1}{16\sqrt{2}}\theta^{10}+\frac{13}{16\sqrt{2}}\eta^{2}\theta^{8}-\frac{7}{8\sqrt{2}}\eta^{4}\theta^{6}-\frac{7}{8\sqrt{2}}\eta^{6}\theta^{4}+\frac{13}{16\sqrt{2}}\eta^{8}\theta^{2}+\frac{1}{16\sqrt{2}}\eta^{10}-\frac{1}{\sqrt{2}}\zeta^{8}\theta^{2}-\frac{1}{\sqrt{2}}\zeta^{8}\eta^{2} \\
\end{pmatrix}$, \\
$Y^{(5)}_{\bm{3}_{2b}} =
\begin{pmatrix}
Y^{(5/2)}_{\bm{6}_1}Y^{(5/2)}_{\bm{3}'_{3a}}
\end{pmatrix}_{\bm{3}_2}$\\
\end{longtable}
We summarize the Siegel modular forms up to weight 5 in Table \ref{tab:Siegel_Delta96_up_to_5}.
\begin{table}[H]
\centering
\renewcommand{\arraystretch}{1.3}
\begin{tabular}{c|c|c} \hline
Weights & Dimensions & Irreducible representations of the Siegel modular forms \\ \hline
$1/2$ & 3 & $\bm{3}'_7$ \\ \hline
$1$ & 6 & $\bm{3}_6$, $\bm{\hat{3}}_0$ \\ \hline
$3/2$ & 10 & $\bm{1}_7$, $\bm{\hat{3}}_5$, $\bm{6}_3$ \\ \hline
$2$ & 15 & $\bm{1}_4$, $\bm{2}_0$, $\bm{3}_0$, $\bm{3}'_6$, $\bm{6}_0$ \\ \hline
$5/2$ & 21 & $\bm{3}_5$, $\bm{3}'_{3a}$, $\bm{3}'_{3b}$, $\bm{3}'_7$, $\bm{\hat{3}}_7$, $\bm{6}_1$ \\ \hline
$3$ & 28 & $\bm{1}_6$, $\bm{3}_{2a}$, $\bm{3}_{2b}$, $\bm{3}_6$, $\bm{3}'_0$, $\bm{\hat{3}}_0$, $\bm{\hat{3}}_{4a}$, $\bm{\hat{3}}_{4b}$, $\bm{6}_2$ \\ \hline
$7/2$ & 36 & $\bm{1}_3$, $\bm{2}_3$, $\bm{3}_7$, $\bm{3}'_5$, $\bm{\hat{3}}_{1a}$, $\bm{\hat{3}}_{1b}$, $\bm{\hat{3}}_5$, $\bm{6}_{3a}$, $\bm{6}_{3b}$, $\bm{6}_{3c}$ \\ \hline
$4$ & 45 & $\bm{1}_{0a}$, $\bm{1}_{0b}$, $\bm{2}_{0a}$, $\bm{2}_{0b}$, $\bm{3}_0$, $\bm{3}_{4a}$, $\bm{3}_{4b}$, $\bm{3}'_{2a}$, $\bm{3}'_{2b}$, $\bm{3}'_6$, $\bm{\hat{3}}_6$, $\bm{6}_{0a}$, $\bm{6}_{0b}$, $\bm{6}_{0c}$ \\ \hline
$9/2$ & 55 & $\bm{1}_5$, $\bm{3}_{1a}$, $\bm{3}_{1b}$, $\bm{3}_5$, $\bm{3}'_{3a}$, $\bm{3}'_{3b}$, $\bm{3}'_{7a}$, $\bm{3}'_{7b}$, $\bm{3}'_{7c}$, $\bm{3}'_{7d}$, $\bm{\hat{3}}_{3a}$, $\bm{\hat{3}}_{3b}$, $\bm{\hat{3}}_7$, $\bm{6}_{1a}$, $\bm{6}_{1b}$, $\bm{6}_{1c}$ \\ \hline
$5$ & 66 & 
$\begin{matrix}
\bm{1}_2, \bm{2}_2, \bm{3}_{2a}, \bm{3}_{2b}, \bm{3}_{6a}, \bm{3}_{6b}, \bm{3}_{6c}, \bm{3}_{6d}, \bm{3}'_0,  \bm{3}'_{4a}, \bm{3}'_{4b}, \bm{\hat{3}}_{0a}, \bm{\hat{3}}_{0b}, \bm{\hat{3}}_{0c}, \bm{\hat{3}}_{0d}, \bm{\hat{3}}_{4a}, \bm{\hat{3}}_{4b}, \\ \bm{6}_{2a}, \bm{6}_{2b}, \bm{6}_{2c} 
\end{matrix}$
\\ \hline
\end{tabular}
\caption{The Siegel modular forms of $\widetilde{\Delta}(96)$.
The dimensions of the Siegel modular forms of weight $k$ are $\frac{1}{2}(2k+2)(2k+1)$.}
\label{tab:Siegel_Delta96_up_to_5}
\end{table}


\end{document}